UNIVERITÉ DE GENÈVE

Section de Physique

Département de Physique Théorique

FACULTÉ DES SCIENCES

Professeur Antonio RIOTTO

# Primordial Black Holes:

## from Theory to Gravitational Wave Observations

THÈSE

présentée à la Faculté des sciences de l'Université de Genève
pour obtenir le grade de Docteur ès sciences, mention physique

par

## Gabriele Franciolini

de

Jesi (Italie)

Thèse N° – 5599

GENÈVE

Atelier de reproduction de la Section de Physique

2021

# Abstract

Primordial Black Holes (PBH) can form in the early universe and might comprise a significant fraction of the dark matter. Interestingly, they are accompanied by the generation of Gravitational Wave (GW) signals and they could contribute to the merger events currently observed by the LIGO/Virgo Collaboration (LVC). In this thesis, we study the PBH scenario, addressing various properties at the formation epoch and the computation of abundance beyond the Gaussian paradigm, while also developing the theoretical description of PBH evolution through accretion and mergers, with particular focus on modelling their GW signatures. In a second part, we compare the primordial scenario with current GW data, seizing the possible contribution of PBH binaries to LVC signals and forecasting the potential of future GW detectors, such as Einstein Telescope and LISA, to detect mergers of primordial binaries and the stochastic GW background induced at second order by the PBH formation mechanism.

# Résumé

Les trous noirs primordiaux (PBH) peuvent se former dans l'univers primitif et pourraient constituer une fraction importante de la matière noire. Il est intéressant de noter qu'ils sont accompagnés de la génération de signaux d'ondes gravitationnelles (GW) et qu'ils pourraient contribuer aux événements actuellement observés par la collaboration LIGO/Virgo (LVC).

Dans cette thèse, nous abordons divers aspects liés à la physique des PBH. Nous analysons la dépendance du seuil de formation de PBH sur les propriétés statistiques des perturbations de sur-densité, tout en étendant également le calcul de l'abondance de PBH au-delà du paradigme Gaussien en incluant à la fois des non-Gaussianités intrinsèques et non linéaires. Nous décrivons ensuite les propriétés fondamentales d'une population de PBH à l'époque de la formation en discutant la distribution de masse, de spin des PBH à la naissance et du regroupement spatial des PBH. Par conséquent, l'évolution de la population de PBH est abordée, en se concentrant à la fois sur les PBH isolés et les PBH dans les systèmes binaires. Nous identifions et caractérisons l'accrétion comme l'acteur clé dans la modification des masses et des spins de PBH, tandis que l'évolution du regroupement spatial est décrite à travers un traitement analytique qui correspond aux simulations numérique existantes à N-corps limitées à un décalage vers le rouge élevé. Nous discutons enfin de la formation et de l'évolution des systèmes binaires de PBH conduisant à l'état de l'art actuel de la paramétrisation du taux de fusion des PBH.

Dans la deuxième partie de cette thèse, nous nous concentrons sur la comparaison du scénario PBH avec les observations GW. Tout d'abord, nous nous concentrons sur le dernier catalogue d'événements GW publié par la collaboration LVC. Nous montrons que les PBH peuvent avoir un taux de fusion suffisant pour produire un certain nombre de signaux détectables compatibles avec les observations sans violer aucune limite sur l'abondance de PBH. Ensuite, nous comparons et mélangeons le scénario PBH avec des canaux astrophysiques pour saisir la contribution potentielle de PBH aux données GW actuelles.

Compte tenu des connaissances actuelles dans la communauté scientifique des scénarios astro-physiques et primordiaux, les PBH sont compatibles avec le fait d'être une sous-population des données GW actuelles. En particulier, ils offrent une explication compétitive des événements dans l'écart de masse tels que GW190521. Nous mettons en évidence quelques stratégies pour rechercher des signatures incontestables du modèle PBH. Celles-ci reposent sur la recherche d'événements à décalage vers le rouge élevé sur les détecteurs de troisième génération et sur des études à événement unique ou de population recherchant les caractéristiques clés du scénario PBH, telles que les corrélations de rapport de masse - spin effectives induites par les effets d'accrétion.

Enfin, nous nous concentrons sur la caractérisation du fond stochastique GW induit au second ordre par les perturbations scalaires responsables de la formation de PBH. Après une discussion détaillée sur la forme spectrale attendue et la Gaussianité du signal après propagation GW dans l'univers perturbé, nous comparons les signatures de formation de PBH aux expériences GW actuelles et futures. Nous montrons comment LISA sera capable de rechercher des signatures de production de PBH dans la fenêtre de masse où les PBH sont encore autorisés à être l'intégralité de la matière noire. De plus, en supposant que les données publiés par la collaboration NANOGrav soient le premier indice d'un signal GW, nous concevons un scénario dans lequel les PBH peuvent comprendre la totalité de la matière noire et, en même temps, être responsables de la génération d'un fond GW stochastique compatible avec le observations de NANOGrav.

# Members of the Jury

- **Prof. Christian T. Byrnes**
  - *Department of Physics and Astronomy,*
  *University of Sussex, Brighton BN1 9QH, United Kingdom*

- **Prof. Bernard Carr**
  - *School of Physics and Astronomy,*
  *Queen Mary University of London, Mile End Road, London E1 4NS, UK*
  - *Research Center for the Early Universe,*
  *University of Tokyo, Tokyo 113-0033, Japan*

- **Prof. Michele Maggiore**
  - *Département de Physique Théorique and Centre for Astroparticle Physics (CAP),*
  *Université de Genève, 24 quai E. Ansermet, CH-1211 Geneva, Switzerland*

- **Prof. Antonio Riotto**
  - *Département de Physique Théorique and Centre for Astroparticle Physics (CAP),*
  *Université de Genève, 24 quai E. Ansermet, CH-1211 Geneva, Switzerland*

# List of publications

## Papers appearing in this thesis

[1] V. De Luca, G. Franciolini, P. Pani and A. Riotto, *"The Minimum Testable Abundance of Primordial Black Holes at Future Gravitational-Wave Detectors"*, [arXiv preprint]

[2] M. Biagetti, V. De Luca, G. Franciolini, A. Kehagias and Riotto, *"The Formation Probability of Primordial Black Holes"*, Phys. Lett. B **820** (2021), 136602.

[3] G. Franciolini, V. Baibhav, V. De Luca, K. K. Y. Ng, K. W. K. Wong, E. Berti, P. Pani, A. Riotto and S. Vitale, *"Evidence for primordial black holes in LIGO/Virgo gravitational-wave data"*. [arXiv preprint]

[4] V. De Luca, G. Franciolini and A. Riotto, *"Constraining the Initial Primordial Black Hole Clustering with CMB-distortion"*, Phys. Rev. D **104** (2021) no.6, 063526.

[5] V. De Luca, G. Franciolini and A. Riotto, *"Bayesian Evidence for Both Astrophysical and Primordial Black Holes: Mapping the GWTC-2 Catalog to Third-Generation Detectors"*, JCAP **05** (2021), 003.

[6] I. Musco, V. De Luca, G. Franciolini and A. Riotto, *"The Threshold for Primordial Black Hole Formation: a Simple Analytic Prescription"*, Phys. Rev. D **103** (2021), 063538.

[7] K. W. K. Wong, G. Franciolini, V. De Luca, V. Baibhav, E. Berti, P. Pani and A. Riotto, *"Constraining the primordial black hole scenario with Bayesian inference and machine learning: the GWTC-2 gravitational wave catalog"*, Phys. Rev. D **103** (2021) no.2, 023026.

[8] V. De Luca, G. Franciolini and A. Riotto, *"NANOGrav Data Hints at Primordial Black Holes as Dark Matter"*, Phys. Rev. Lett. **126** (2021) no.4, 041303.

[9] V. De Luca, V. Desjacques, G. Franciolini and A. Riotto, *"The clustering evolution of primordial black holes"*, JCAP **11** (2020), 028.

[10] S. Bhagwat, V. De Luca, G. Franciolini, P. Pani and A. Riotto, *"The importance of priors on LIGO-Virgo parameter estimation: the case of primordial black holes"*, JCAP **01** (2021), 037.

[11] V. De Luca, V. Desjacques, G. Franciolini, P. Pani and A. Riotto, *"GW190521 Mass Gap Event and the Primordial Black Hole Scenario"*, Phys. Rev. Lett. **126** (2021) no.5, 051101.

[12] V. De Luca, G. Franciolini, P. Pani and A. Riotto, *"Primordial Black Holes Confront LIGO/Virgo data: Current situation"*, JCAP **06** (2020), 044.

[13] V. De Luca, G. Franciolini, P. Pani and A. Riotto, *"Constraints on Primordial Black Holes: the Importance of Accretion"*, Phys. Rev. D **102** (2020) no.4, 043505.

[14] V. De Luca, G. Franciolini, P. Pani and A. Riotto, *"The evolution of primordial black holes and their final observable spins"*, JCAP **04** (2020), 052.

[15] V. De Luca, G. Franciolini and A. Riotto, *"On the Primordial Black Hole Mass Function for Broad Spectra"*, Phys. Lett. B **807** (2020), 135550.

[16] V. De Luca, G. Franciolini, A. Kehagias and A. Riotto, *"On the Gauge Invariance of Cosmological Gravitational Waves"*, JCAP **03** (2020), 014.

[17] N. Bartolo, D. Bertacca, V. De Luca, G. Franciolini, S. Matarrese, M. Peloso, A. Ricciardone, A. Riotto and G. Tasinato, *"Gravitational wave anisotropies from primordial black holes"*, JCAP **02** (2020), 028.

[18] A. Moradinezhad Dizgah, G. Franciolini and A. Riotto, *"Primordial Black Holes from Broad Spectra: Abundance and Clustering"*, JCAP **11** (2019), 001.

[19] V. De Luca, G. Franciolini, A. Kehagias, M. Peloso, A. Riotto and C. Ünal, *"The Ineludible non-Gaussianity of the Primordial Black Hole Abundance"*, JCAP **07** (2019), 048.

[20] V. De Luca, V. Desjacques, G. Franciolini, A. Malhotra and A. Riotto, *"The initial spin probability distribution of primordial black holes"*, JCAP **05** (2019), 018.

[21] N. Bartolo, V. De Luca, G. Franciolini, M. Peloso, D. Racco and A. Riotto, *"Testing primordial black holes as dark matter with LISA"*, Phys. Rev. D **99** (2019) no.10, 103521.

[22] N. Bartolo, V. De Luca, G. Franciolini, A. Lewis, M. Peloso and A. Riotto, *"Primordial Black Hole Dark Matter: LISA Serendipity"*, Phys. Rev. Lett. **122** (2019) no.21, 211301.

[23] M. Biagetti, G. Franciolini, A. Kehagias and A. Riotto, *"Primordial Black Holes from Inflation and Quantum Diffusion"*, JCAP **07** (2018), 032.

[24] G. Franciolini, A. Kehagias, S. Matarrese and A. Riotto, *"Primordial Black Holes from Inflation and non-Gaussianity"*, JCAP **03** (2018), 016.

## Other Journal Papers

[25] S. S. Bavera, G. Franciolini, G. Cusin, A. Riotto, M. Zevin and T. Fragos, *"Stochastic gravitational-wave background as a tool to investigate multi-channel astrophysical and primordial black-hole mergers"*, *[arXiv preprint]*.

[26] V. De Luca, G. Franciolini, A. Kehagias and A. Riotto, *"Standard Model Baryon Number Violation Seeded by Black Holes"*, Phys. Lett. B **819** (2021), 136454.

[27] K. Kritos, V. De Luca, G. Franciolini, A. Kehagias and A. Riotto, *"The Astro-Primordial Black Hole Merger Rates: a Reappraisal"*, JCAP **05** (2021), 039.

[28] V. De Luca, G. Franciolini, A. Kehagias, A. Riotto and M. Shiraishi, *"Constraining graviton non-Gaussianity through the CMB bispectra"*, Phys. Rev. D **100** (2019) no.6, 063535.

[29] V. De Luca, V. Desjacques, G. Franciolini and A. Riotto, *"Gravitational Waves from Peaks"*, JCAP **09** (2019), 059.

[30] D. Anninos, V. De Luca, G. Franciolini, A. Kehagias and A. Riotto, *"Cosmological Shapes of Higher-Spin Gravity"*, JCAP **04** (2019), 045.

[31] G. Franciolini, G. F. Giudice, D. Racco and A. Riotto, *"Implications of the detection of primordial gravitational waves for the Standard Model"*, JCAP **05** (2019), 022.

[32] G. Franciolini, L. Hui, R. Penco, L. Santoni and E. Trincherini, *"Stable wormholes in scalar-tensor theories"*, JHEP **01** (2019), 221.

[33] G. Franciolini, L. Hui, R. Penco, L. Santoni and E. Trincherini, *"Effective Field Theory of Black Hole Quasinormal Modes in Scalar-Tensor Theories"*, JHEP **02** (2019), 127.

[34] A. Moradinezhad Dizgah, G. Franciolini, A. Kehagias and A. Riotto, *"Constraints on long-lived, higher-spin particles from galaxy bispectrum"*, Phys. Rev. D **98** (2018) no.6, 063520.

[35] G. Franciolini, A. Kehagias, A. Riotto and M. Shiraishi, *"Detecting higher spin fields through statistical anisotropy in the CMB bispectrum"*, Phys. Rev. D **98** (2018) no.4, 043533.

[36] G. Franciolini, A. Kehagias and A. Riotto, *"Imprints of Spinning Particles on Primordial Cosmological Perturbations"*, JCAP **02** (2018), 023.

## White Papers

[37] E. Barausse, E. Berti, T. Hertog, S. A. Hughes, P. Jetzer, P. Pani, T. P. Sotiriou, N. Tamanini, H. Witek and K. Yagi, *et al.* *"Prospects for Fundamental Physics with LISA"*, Gen. Rel. Grav. **52** (2020) no.8, 81.

[38] A. Kashlinsky, Y. Ali-Haimoud, S. Clesse, J. Garcia-Bellido, L. Wyrzykowski, A. Achucarro, L. Amendola, J. Annis, A. Arbey and R. G. Arendt, *et al.* *"Electromagnetic probes of primordial black holes as dark matter"*. arXiv: 1903.04424 [astro-ph.CO]

*Salviati: [...] Or vedete come è facile da intendersi.*
*Sagredo: Tali sono tutte le cose vere, doppo che son trovate;*
*ma il punto sta nel saperle trovare.*

*- Galileo Galilei -*
*Dialogo sopra i due massimi sistemi del mondo (1632)*

*(Salviati: [...] Now you see how easy it is to understand.*
*Sagredo: So are all truths, once they are discovered;*
*the point is in being able to discover them. [39])*

# Contents

# IV    Conclusions                                                                                          193

# 7    Conclusions and outlook                                                                               194

# V    Appendices                                                                                            197

# A    Peak theory of gaussian fields                                                                        198

# B    A general formula for PBH clustering induced by non-Gaussianities                                     202

# C    Bayesian inference framework                                                                          204

# D    Leading astrophysical models of black hole binaries                                                   210

# Part I

# Introduction

# Chapter 1

# The dark matter problem and the primordial black hole opportunity

Our journey begins at the frontiers of human knowledge. The majority of matter filling our universe appears to be "invisible" to us but its presence is required to explain the formation of galaxies and the large-scale structures we observe. As a direct consequence of the recent development of astronomical and cosmological observations, substantial evidence for the existence of such "dark" matter was gathered, but the nature of its constituents is still, however, one of the longstanding mysteries of physics.

In this chapter, we start by reviewing the main observations pointing towards the existence of dark matter on various scales, from galactic up to cosmological scales. These results have paved the way for a plethora of studies, all addressing the same question: *What is the dark matter made of?* We will then proceed by introducing the main character of this thesis, i.e. Primordial Black Holes (PBHs). PBHs were first hypothesised in the late '60s and they have been investigated for more than half a century as a potential solution to the dark matter problem.

We will briefly describe the main characteristics of PBH models and their connection to the physics of the early universe. We will show how recent studies were able to constrain the PBH abundance in most of (but not all) the phenomenologically interesting range of masses. Still, we will argue how important and consequential the study of PBHs may be, even if they represent only a tiny fraction of the dark matter.

Finally, we will introduce the last ingredient in our discussion, being one of the most promising new probes which have revolutionised our ability to observe the universe: Gravitational Waves (GWs). As we will discuss at length in this thesis, GWs will be used in the next decades to search for PBHs in a wide region of the parameter space which is mostly inaccessible otherwise. This journey has already started with the observation of GW events made by the LIGO and Virgo collaborations. Also, PBHs could provide the leading explanation for some GW observations which are currently challenging other astrophysical models, as we will see.

## 1.1 Evidence for dark matter

The history of how scientists proved that dark matter plays a crucial role in the current cosmological paradigm is fascinating. We refer to Refs. [40, 41] for historical reviews.

In this section, we give a rough sketch of the main compelling arguments in favour of the existence of dark matter. We divide the discussion depending on the scales involved in each observation.

### Galactic scales

At the beginning of the 20th century, following some pioneering speculations of W. Kelvin [42] and J. Kapteyn, a first account of the discrepancy between the amount of luminous matter and the total matter in our galaxy was presented by J. Oort in Ref. [43]. It was only in 1939, with the PhD dissertation of H. Babcock [44], that a measurement of the circular velocity of stars and gas at different radial distances from the galactic center (i.e. rotation curves) of the Andromeda Galaxy M31 was reported, showing a rising behaviour extending until about 20 kpc away from its center. This observation would require the existence of large amounts of mass in the outer parts of the galaxy. Then, at the beginning of the 1970s, the seminal work of V. Rubin and K. Ford [45] reported substantially improved measurements of the rotation curves of M31 through observations of the Doppler shift of the Hydrogen 21 cm line, marking a decisive confirmation of this idea. See Fig. 1.1, also reporting subsequent dataset from Refs. [46, 47]. As a consequence, the first explicit statements on the necessity of additional non-luminous mass in the galaxies started to appear (see, for example, Refs. [48–51]).

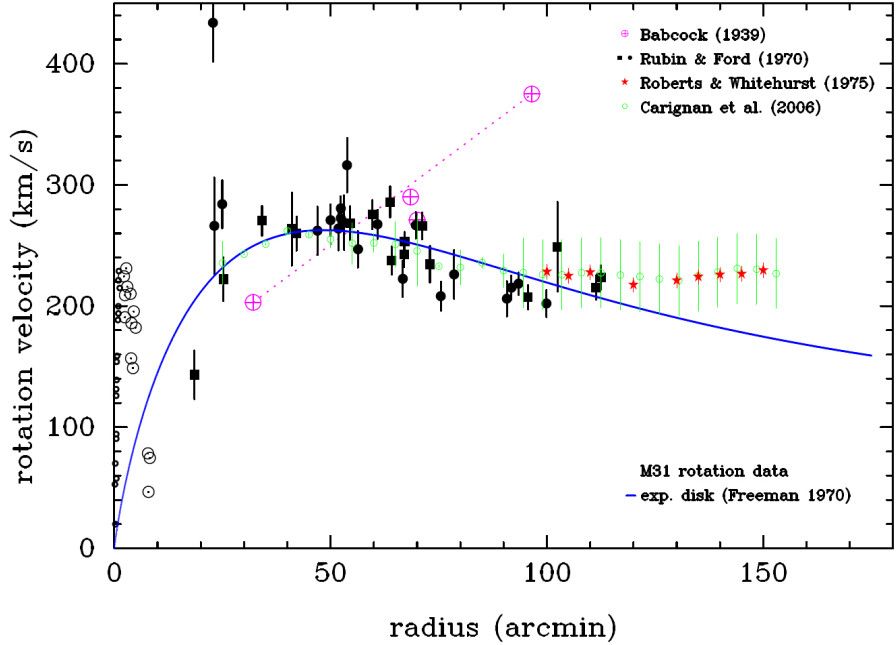

Figure 1.1: *Rotation curve data for the Andromeda galaxy M31. Fig. taken from Ref. [41].*

This conclusion can be understood using simple physical arguments. Newtonian dynamics predicts that the star circular velocity follows from the relation

$$v(r) = \sqrt{\frac{GM(r)}{r}}, \tag{1.1.1}$$

where $r$ is the radial distance from the galactic center, $M(r)$ is the mass contained within the radius $r$ and $G = 6.674 \cdot 10^{-11}$ m$^3$ kg$^{-1}$ s$^{-2}$ is the Newton's gravitational constant. To obtain a flat behaviour of the curve at large distances, one has to assume a halo with $M(r) \propto r$ (i.e. a density profile $\rho(r) \propto 1/r^2$). When compared to the luminous matter of galaxies which only covers a limited region in space, one concludes that some missing "dark" mass extending up to the outer regions of galaxies must be present.

**Galaxy cluster scales**

In a slightly different context, independent hypotheses appeared about the existence of dark matter. It was F. Zwicky who, after noticing large apparent velocities of galaxies within the Coma clusters, first adopted the virial theorem to estimate the mass of the galaxy cluster. In 1933, he found that the total mass of the cluster under consideration must have been hundreds of times larger than the luminous matter [52]. He was also the first to use the terms "dark matter" with the meaning they acquired in recent years. Since then, many additional observations proved that much of the mass within and surrounding galaxies must not interact electromagnetically for it to remain "unseen" by astronomers.

Probably, the most fascinating piece of empirical evidence for dark matter on intergalactic scales comes from the observations of the Bullet Cluster [53–55], see Fig. 1.2. This special system is the result of a collision of two galaxy clusters that passed through each other in a direction roughly perpendicular to our line of sight around 150 million years ago. By using independent probes, researchers were able to show that different constituents have undergone drastically distinct interactions. In particular, galaxies (acting in this context as point-like particles interacting gravitationally) and the dark matter did not sway from their trajectories, while the intergalactic gas, accounting for most of the baryonic mass of the clusters and characterised by a relatively large self-interaction cross-section, was displaced from the barycenters of the corresponding halos.

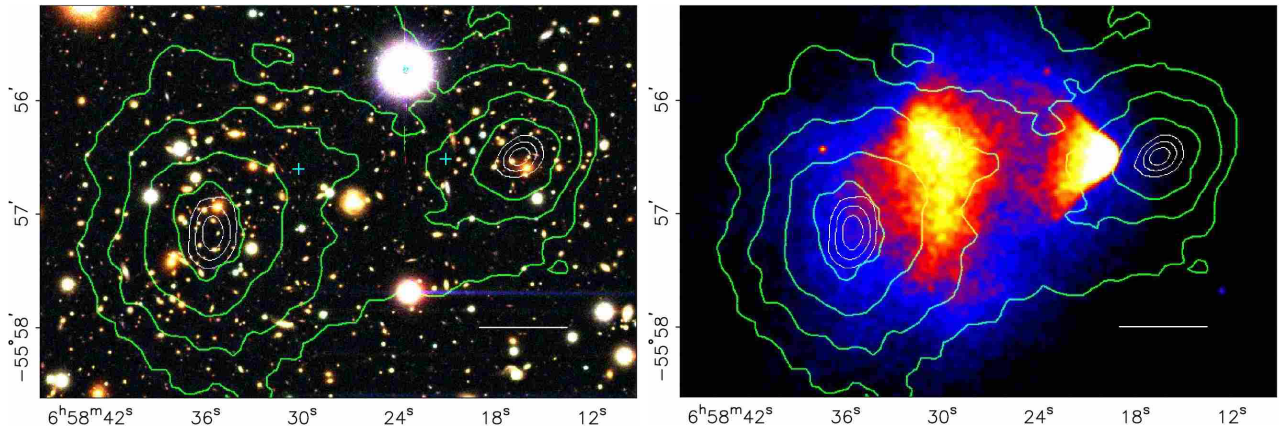

Figure 1.2: *Observations of the Bullet Cluster reported in Ref. [55].* **Left:** *Optical image of the system.* **Right:** *X-ray image of the gas distribution. In both plots, the green contour lines indicate the distribution of mass as inferred from gravitational lensing observations.*

The gravitational lensing maps show, therefore, that the gravitational potential does not trace the plasma distribution contributing to the dominant baryonic mass component, but rather approximately traces the distribution of galaxies. This proves, therefore, that the majority of the gravitating matter in the system is unseen, i.e. "dark". These observations were afterwards extended to similar colliding galaxy cluster systems [56, 57].

**Large-scale structure**

The large scale structure of our universe appears as being organised hierarchically in halos of increasingly larger sizes (starting from galaxies up to galaxy superclusters), see Fig. 1.3 (left panel). Our best theory for explaining the large scale observations describes the formation of structures via gravitational enhancement of small initial fluctuations. In this description, a crucial role is played by assuming a large portion of the energy budget of the universe is in the form of a cold and collisionless dark matter fluid, whose only interaction is gravitational. The cold dark matter model, in other words, has become the leading theoretical paradigm describing the formation of structures in the Universe.

In particular, large-scale structure surveys such as the Field Galaxy Redshift Survey (2dFGRS) [58] and the Sloan Digital Sky Survey (SDSS) [59] inferred properties of the galaxy distribution (such as the galaxy correlation function) which are matched incredibly well by N-body numerical simulations such as the Millennium simulation [60], see Fig. 1.3 (right panel).

In qualitative terms, dark matter can be regarded as a necessary ingredient allowing for the formation of galaxies, such as the one we live in. Before baryons can decouple from photons, around the epoch of recombination at redshift $z \simeq 1100 \div 1400$, their density fluctuations oscillate without growing in amplitude. Dark matter, on the other hand, decouples at much earlier times (depending on the formation mechanism) and its perturbations start growing as soon as they re-enter the cosmological horizon. This allows for the formation of gravitational potential wells tracing the dark matter protostructures where baryons can fall after they have decoupled from the thermal bath. Without the enhancing effect of dark matter, baryon perturbations would not have grown enough to allow for the formation of galaxies before the epoch of dark energy domination, when perturbations stop developing [62].

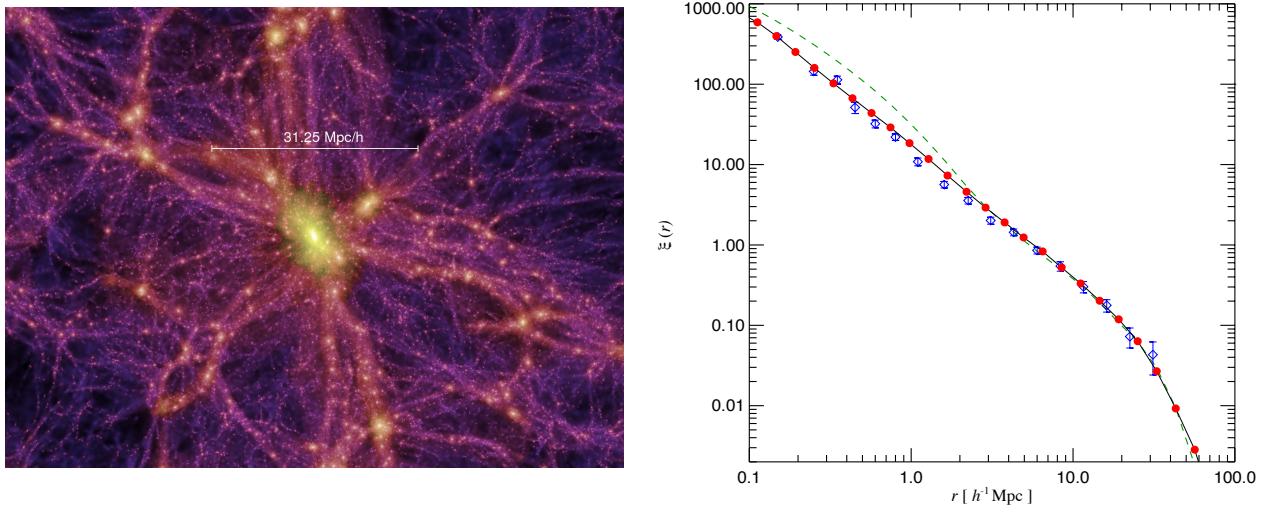

Figure 1.3: ***Left:*** *Snapshot of the Millenium simulation at redshift $z = 0$ taken from Ref. [61].* ***Right:*** *Galaxy 2-point correlation function at the present epoch (red dots) as found performing the N-body Millennium simulation [60] and compared to the measurements of 2dFGRS [58] (blue marks) (with the SDSS [59] survey giving a similar agreement). Fig. taken from Ref. [60].*

## Cosmological scales

Last but not least, another piece of evidence in favour of the existence of dark matter comes from observations on the largest scales accessible to us, being a significant fraction of the current horizon of our universe. Those scales are probed through observations of the Cosmic Microwave Background (CMB) radiation.

The CMB photons were emitted at the epoch of last scattering and bring important information about the matter content in our universe. In particular, the CMB spectrum can be efficiently described as the one of a black body at the temperature of $2.7255\,\mathrm{K}$ with fluctuations which are of the order of one part in $10^5$. By studying the anisotropies of the temperature fluctuations (i.e. their dependence on the direction on the sky), the Planck collaboration was able to infer the free parameters of the $\Lambda$CDM model (see, for example, Ref. [62]), which fits the observations with incredible precision in terms of only a very small set of parameters, see Fig. 1.4.

In particular, the anisotropies are found to be Gaussian and characterised by a power-spectrum which we report in Fig. 1.4 as a function of the angular scale (or multipole $\ell$ in the spherical harmonics decomposition). It is interesting to notice that the position and height of the peaks observed in the power spectrum provide precise information about the amount of dark matter compared to its baryonic counterpart.

The oscillating pattern of the spectrum can be explained with the standard cosmological model as follows. At very large scales, perturbations are still super-Hubble at the time of last scattering and the power spectrum is described by the Sachs-Wolfe term [63] predicting a flat shape whose amplitude is fixed by that of perturbations imprinted to super-horizon modes during inflation. The various peaks are, however, bringing further information. The matter-radiation plasma presents acoustic oscillations ignited by initial density perturbations and sustained by the interplay of radiation pressure and gravitational attraction when modes re-enter the cosmological horizon (see, for example, the discussion in Ref. [64]). Therefore, modes corresponding to maxima (and minima) of the plasma perturbations can be mapped to peaks in the power spectrum. The largest peak in Fig. 1.4 is generated by modes that entered the horizon and only had time to complete half a cycle (the gravitational compression) before photons decoupled from the plasma. The second peak corresponds to the first complete cycle, i.e. compression and rarefaction due to pressure reaction, and so on. Thus, in practical terms, the energy budget composed of baryons affects the ratios between the height of odd and even peaks in the spectrum, while the total matter density composed of both the dark matter and baryons is determined by the overall peak amplitudes.

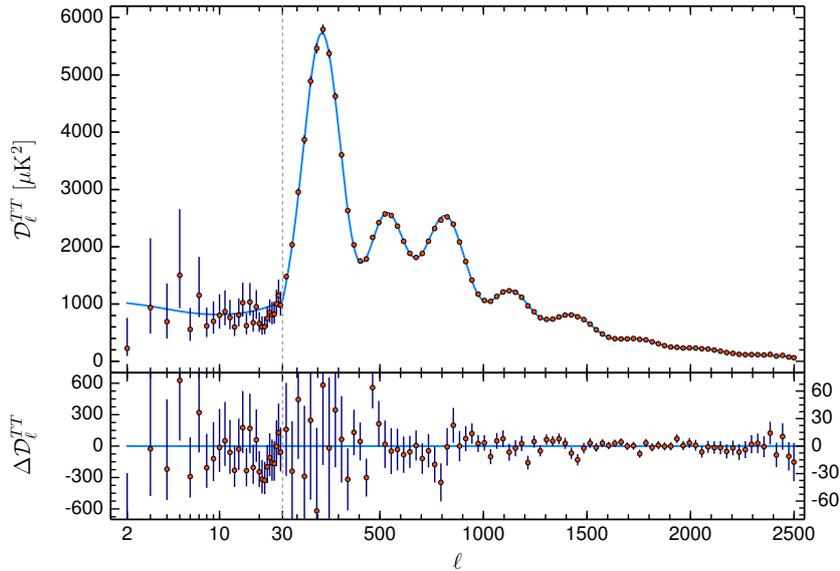

Figure 1.4: *Planck 2018 temperature power spectrum. In the bottom panel, the residuals with respect to the $\Lambda CDM$ model are shown. Figure taken from Ref. [65].*

The measurement made by the Planck collaboration [65] give a precise determination of the dark matter abundance, which is found to be

$$\Omega_{\mathrm{DM}} h^2 = 0.120 \pm 0.001, \tag{1.1.2}$$

when expressed in units of the critical density $\rho_c = 3H_0^2/8\pi G$ as $\Omega_{\mathrm{DM}} \equiv \rho_{\mathrm{DM}}/\rho_c$, and where $h$ indicates the Hubble parameter $H_0$ in units of 100 km s$^{-1}$ Mpc$^{-1}$ and it was measured to be $h = 67.4 \pm 0.5$. Analogously, the baryonic abundance is constrained at

$$\Omega_{\mathrm{b}} h^2 = 0.0224 \pm 0.0001. \tag{1.1.3}$$

This fixes the dark matter energy density in our universe to be around the 26% of the total budget, with a contribution that is roughly five times larger than the one from baryons. In all parts of this thesis, we will follow the literature and define the abundance of PBHs by normalising their energy density with respect to the dark matter one reported above.

## 1.2 Primordial black holes

Having set the stage in the previous section, we can now introduce the main character in this thesis: *Primordial Black Holes.*

The idea that PBHs could form in the early universe first appeared in the literature with the works of Y. Zel'dovich and I. Novikov [66] (1967) and S. Hawking [67] (1971). It is historically rather interesting to note that the discovery of the famous BH evaporation [68, 69] took place during a period when Hawking was focusing his studies on PBHs. It was soon realised that PBHs could contribute to the dark matter in our universe in the works by B. Carr [70, 71] and G. Chapline [72], generate the large-scale structure through Poisson fluctuations [73] while also potentially providing the seeds for the supermassive BHs populating our universe [74]. As PBHs are supposed to form in the early epochs preceding by far the matter-radiation equality and that on cosmological scales PBHs would behave like a cold and collisionless fluid, they are a perfect candidate to be the dark matter, provided they are more massive than $\simeq 10^{-19} M_\odot$ to have a lifetime longer than the age of the Universe [75, 76]. A particularly appealing property of PBH dark matter is that this explanation does not necessarily require any physics beyond the standard model of particle physics, provided one modifies the (mostly unconstrained) early universe description at small scales to account for the generation of large density perturbations responsible for the PBH formation. A considerable amount of work was conducted to search and constrain the abundance of PBHs by leveraging on their peculiar properties at small scales, potentially leading to visible effects such as lensing, electromagnetic emission from accretion processes, gravitational waves, just to name a few.

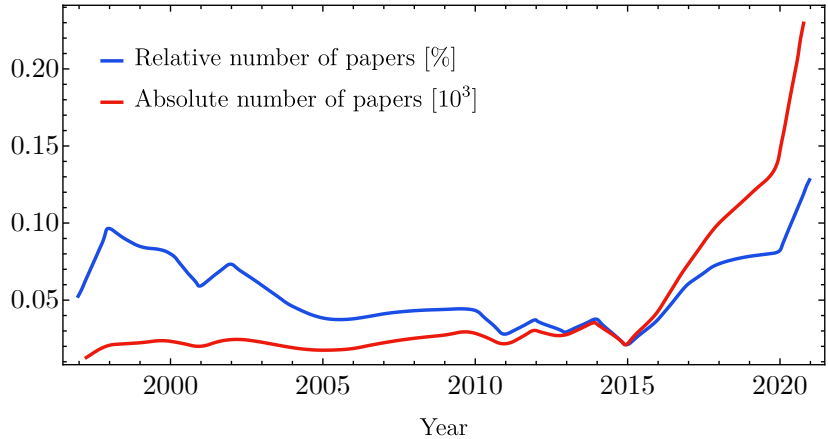

Figure 1.5: *Both absolute and relative number of papers posted on the arXiv with the phrase "Primordial black holes" either in the title or abstract. One can recognise how the GW detection in 2015 jump-started the current wave of interest in PBHs. Data from Ref. [163].*

Starting from the 1980's, many formation mechanisms were devised. The most studied scenario predicts that PBHs come from large fluctuations generated at small scales by the inflationary dynamics [77–107], while many other models envision PBHs as a consequence of an early matter era [108–112], modified gravity [113], scalar field instabilities [114–116], collapse of cosmic strings [117–123] and domain walls [124–130], phase transitions [131–133], bubble collisions [134–136] and standard model Higgs instability [137, 138]. At same time, the phenomenon of PBH formation started being investigated with the aid of dedicated numerical relativity simulations showing it follows the properties of a critical collapse [139–151].

In recent years, the scientific community has experienced multiple waves of interest for PBHs, see Fig. 1.5. The first spark of interest was ignited in the late 1990s as a consequence of the reported detection by the MACHO collaboration 2yr results [152, 153] of multiple Large Magellanic Cloud microlensing events. These events, if interpreted as due to PBHs in our Milky Way, would suggest a significant fraction of the mass in our galaxy to be composed of subsolar compact objects. This suggestive result was, however, outdated by the EROS [154] and OGLE [155–158] results, finding only a reduced fraction of the milky way mass could be in the form of subsolar PBHs, setting more stringent constraints on the PBH abundance in this mass range.

A second wave, which we are still experiencing today, was ignited by the first detection of GWs coming from a black hole merger performed by the LIGO/Virgo collaboration [159]. Indeed, it was soon realised that such a signal would be compatible with a merger of PBHs [160–162]. Also, those groups showed that PBH models could have a merger rate compatible with the GW observation without violating the obvious bound requiring PBHs to be at most as abundant as the dark matter in our universe. Since then, many subsequent works have tried to address the question of whether PBHs could be responsible for all, or a part of, the GW events observed by the LIGO/Virgo collaboration, an endeavour which will also be pursued in this thesis.

Before entering in some details of the PBH model, it is interesting to stress that PBHs, if they were discovered, could have numerous consequences on our current understanding of the universe even if they comprise only a small portion of the dark matter. Here are a few points motivating this statement.

- PBHs are a unique probe to test the universe at the very small scales, leading to the current most stringent constraint on the amplitude of early universe perturbations [164–167] at subparsec comoving scales. Being formed much before the emission of the CMB radiation, PBHs could thus provide information on the physics governing the small scales/ early epochs of our universe. PBHs could also be a test of primordial non-Gaussianities (see for example Ref. [168]).

- PBHs models may produce merger events detectable by the LIGO/Virgo collaboration even if they compose a small fraction (below 0.1%) of the dark matter in the mass range around $30 M_\odot$ [3, 169, 170].

- Even a small fraction of solar mass PBHs can heat the galaxy cores, possibly solving the cusp-core problem (see Ref. [171] for a review) without the need of baryonic processes [172], as well as other dwarf galaxy anomalies [173].

- A detection of PBHs would tell us what (at least a fraction of) the dark matter is made of, and what is not. This is because the coexistence of PBHs and other particle dark matter candidates (such as WIMPs) would lead to a copious particle dark matter annihilation in halos accumulated around individual PBHs in the local universe. This would lead to the production of a detectable gamma-ray signal, strongly constraining the coexistence scenario [174–178].

- Even a tiny abundance of PBHs could produce the primordial seeds for the formation of the supermassive BHs observed at high redshift [74, 179–183] (see Ref. [184] for a review).

### 1.2.1 Primordial black hole preliminaries

PBHs can form in the early universe from the collapse of large density perturbations. Even though different formation mechanisms were devised in the literature, we will always consider this standard scenario unless stated otherwise. Notice, however, that many of the results contained in this thesis are independent of the PBH formation mechanism.

The scenario can be summarised in the following way. We assume there is an enhancement of perturbations at small scales above the value required to match CMB observations at cosmological scales. This can be achieved, for example, by modifying the slow roll paradigm close to the end of inflation (see the original proposal in Ref. [81]). The perturbations, at super-horizon scales by the end of inflation, are then transferred to the radiation during the reheating phase. As long as perturbations are in the super-horizon regime, they are frozen and the gauge-invariant comoving curvature $\zeta$ is constant. As the Hubble horizon is growing compared to the comoving scales during the radiation phase, at a certain epoch the characteristic scale of the perturbations become comparable to the horizon scale. When this happens, gravitational forces become active, initiating the contraction of the overdense regions, which can collapse and form PBHs if they are dense enough. Soon after horizon crossing, radiation pressure can rapidly disperse the overdensity peaks, and therefore the fate of perturbations is decided at horizon crossing. One finds also that there is only a negligible contribution from PBHs formed from the collapse of sub-horizon modes. This scenario is sketched in Fig. 1.6.

One can use Jean length arguments in Newtonian gravity, as first estimated by Ref. [71], to derive the threshold above which an overdensity can collapse. It is found that $\delta_c \sim c_s^2$, where $c_s^2 = 1/3$ is the sound speed of the radiation fluid. This estimate is sufficiently accurate for a rough understanding of the characteristic amplitude of perturbations that can produce PBHs. We will present the value of the threshold $\delta_c$ as dictated by state-of-the-art relativistic numerical simulations in the first chapter of this thesis.

The characteristic mass of PBHs is related to the mass contained in the horizon at the time of formation through the efficiency factor which takes values of the order $\gamma \simeq 0.2$ [71]. Therefore, one finds

$$M_{\rm PBH} = \gamma M_H(t_{\rm form}) = \frac{\gamma}{2} \left( G H_{\rm form} \right)^{-1}.$$  (1.2.1)

We will discuss how the critical collapse modifies this picture and reshapes the mass distribution produced by perturbations on a given scale. Therefore, PBH masses can span an incredibly large range (contrarily to astrophysical black holes) as they are related to the size and density of the cosmological horizon at formation, with a minimum mass that can be, in principle, as small as the Planck mass $\sim 10^{-5}$g (see *e.g.* [185] and references therein).

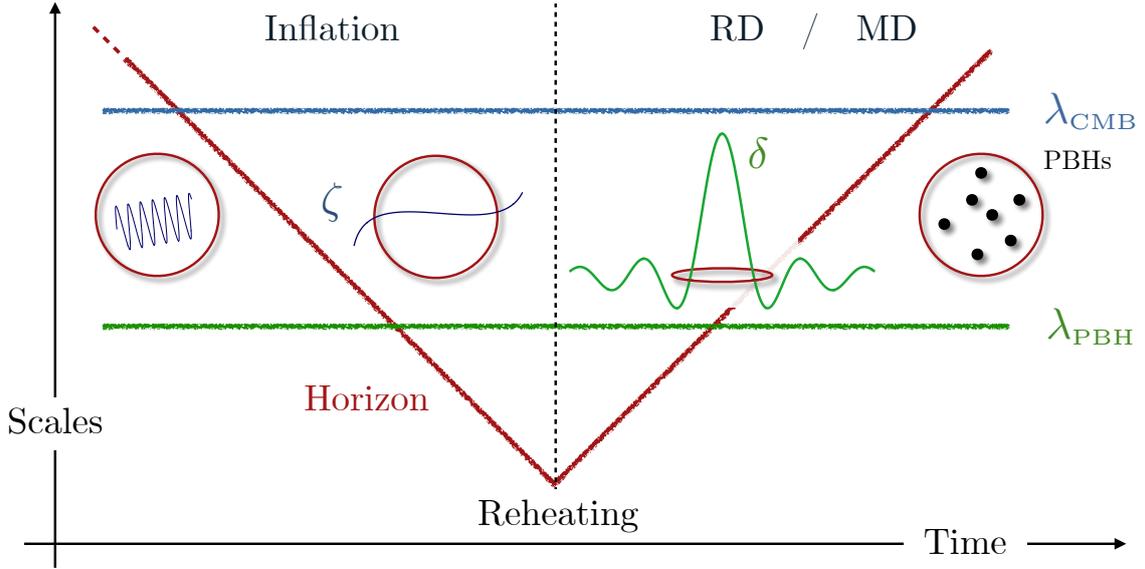

Figure 1.6: *A schematic representation of the standard PBH formation scenario. In red we indicate the comoving horizon, decreasing (increasing) during inflation (radiation/matter domination) with respect to comoving lengths. The green line indicates the comoving scale of perturbations generated during inflation responsible for the PBH formation, much smaller than the CMB scales indicated in blue. Perturbations become super-horizon and stop evolving during inflation, while they can collapse to form PBHs in the subsequent radiation dominated expansion of the universe upon horizon crossing.*

The abundance of PBHs is typically expressed in terms of the ratio between the mean energy density in the form of PBHs normalised to the dark matter in our universe as

$$f_{\rm PBH} \equiv \frac{\Omega_{\rm PBH}}{\Omega_{\rm DM}}, \tag{1.2.2}$$

where $\Omega_{\rm DM}$ is currently best constrained by the Planck measurement as reported in Eq. 1.1.2. Given the mass distribution, one can translate $f_{\rm PBH}$ into a mean PBH number density $n_{\rm PBH}$, which will often be used when discussing the clustering properties of PBHs. As the formation of PBHs in the standard scenario takes place in a radiation dominated universe, their abundance compared to the total energy density gets enhanced until the matter-radiation equality epoch is reached.[1] Therefore, the mass fraction at formation time $\beta$ can be defined as

$$\beta \equiv \frac{\rho_{\rm PBH}}{\rho_{\rm tot}}\bigg|_{\rm form} = \left(\frac{H_0}{H_{\rm form}}\right)^2 \left(\frac{a_{\rm form}}{a_0}\right)^{-3} \Omega_{\rm DM}\, f_{\rm PBH}, \tag{1.2.3}$$

where $H \equiv \dot{a}/a$ is a Hubble parameter defined in terms of the scale factor $a$. Using the mass-horizon relation (1.2.1), one finds

$$\beta \simeq 6.6 \cdot 10^{-9} \left(\frac{\gamma}{0.2}\right)^{-1/2} \left(\frac{g_*(t_{\rm form})}{106.75}\right)^{1/4} \left(\frac{M_{\rm PBH}}{M_\odot}\right)^{1/2} f_{\rm PBH}, \tag{1.2.4}$$

where $g_*$ is a number of relativistic degrees of freedom. This tells us that the mass fraction $\beta$ takes very small values for the interesting range of masses, requiring the PBH collapse to be an extremely rare event not to exceed the obvious limit of $f_{\rm PBH}$ being at most equal to unity. Also, $\beta$ can be interpreted as the number of horizons collapsing to form PBHs. To give an order of magnitude estimate of the characteristic perturbation amplitude giving rise to a phenomenologically viable PBH population, let us estimate $\beta$ assuming Gaussian perturbations and threshold statistics (we will discuss in detail how to go beyond this simple estimate by including both intrinsic and non-linearly induced non-Gaussianities in Chapter 2). In this way, $\beta$ is just the probability that the density contrast overcomes the threshold $\delta_c$, which means

$$\beta \simeq \int_{\delta_c} \frac{\mathrm{d}\delta}{\sqrt{2\pi}\sigma_\delta} \exp\left(-\frac{\delta^2}{2\sigma_\delta^2}\right) \simeq \frac{1}{\sqrt{2\pi}\nu} \exp\left(-\frac{\nu^2}{2}\right) \tag{1.2.5}$$

where we defined the rescaled threshold $\nu \equiv \delta_c/\sigma_\delta$. In order to have a PBH population with $M = 30M_\odot$ and $f_{\mathrm{PBH}} = 10^{-3}$, one finds $\nu = 6.5$, corresponding to a density variance $\sigma_\delta \simeq 0.08$, while a PBH population with $M = 10^{-12}M_\odot$ and $f_{\mathrm{PBH}} = 1$ requires $\nu = 7.7$ and $\sigma_\delta \simeq 0.06$. We adopted a threshold of order $\delta_c = 0.5$ to perform these estimates. These two scenarios will be particularly motivated by the discussion in the next chapters of this thesis. Given these estimates, we see that an enhancement of the perturbations from their amplitude of the order of $\simeq 10^{-5}$ at CMB scales is required.

---

[1]Recall that in a Friedmann-Lemaître-Robertson-Walker universe, the radiation (matter) energy density scales like $a^{-4}$ $(a^{-3})$ as a function of the scale factor $a$ [186].

Finally, let us compute the characteristic size of perturbations giving rise to PBHs. As we already discussed, the PBH mass is related to the size of the horizon at the time of formation, corresponding to the horizon crossing time. This means the comoving perturbation wavenumber is $k_{\mathrm{PBH}} = a(t_{\mathrm{form}})H(t_{\mathrm{form}})$ and can be expressed as a function of the PBH mass by

$$k_{\mathrm{PBH}}(M_{\mathrm{PBH}}) \simeq 2.4 \cdot 10^5 \, \mathrm{Mpc}^{-1} \left(\frac{\gamma}{0.2}\right)^{1/2} \left(\frac{g_*(t_{\mathrm{form}})}{106.75}\right)^{-1/12} \left(\frac{M_{\mathrm{PBH}}}{30M_\odot}\right)^{-1/2}. \tag{1.2.6}$$

Notice that the primordial power spectrum of curvature perturbations $\mathcal{P}_\zeta$ in the interesting range of scales is mostly unconstrained [167]. The most stringent bounds at the scales relevant for PBH formation come from the COBE/FIRAS limits from spectral distortions [187, 188] and Pulsar Timing Array (PTA) [189–191]. Those limits can be schematically summarised as

$$\mathcal{P}_\zeta(k) \lesssim \begin{cases} 10^{-4} & \text{for} \quad k \in [10 \div 10^4] \, \mathrm{Mpc}^{-1} & \text{COBE/FIRAS,} \\ 10^{-2} & \text{for} \quad k \in [10^6 \div 10^7] \, \mathrm{Mpc}^{-1} & \text{PTA,} \end{cases} \tag{1.2.7}$$

while at even smaller scales the most stringent bound is given by requiring no overproduction of PBHs [165, 167, 192]. We will come back to this discussion in the next section dedicated to PBH constraints.

To conclude, in this brief preliminary section we estimated the perturbation scale and amplitude producing a PBH population with mass $M_{\mathrm{PBH}}$ and abundance $f_{\mathrm{PBH}}$. It is important to stress that there exists a plethora of different models which are predicting specific properties of the perturbations, thus affecting the resulting PBH population. We will not review those models here, but the interested reader can find more details in the excellent reviews in Refs. [193–198]. This wide variety of predictions, along with the lack of constraints on most of the parameter space accessible to PBHs (i.e. mass distribution and overall abundance) provides a tremendous opportunity for researchers, meaning there is still a lot to be learned about PBHs. On the other hand, in few cases, the absence of specific predictions for some of the PBH population properties makes the comparison with current experiments and cosmological scenarios harder. In this thesis, we try to address this problem by assuming a model-independent approach as much as possible, while trying to refine the general predictions of the PBH scenario before attempting a comparison with GW data.

In the next section, we will proceed and discuss the current constraints on the PBH abundance in the phenomenologically interesting range of masses.

## 1.2.2   Constraints on the primordial black hole abundance

In this section, we review the constraints on the PBH abundance in the phenomenologically interesting mass range. We warn the reader that few constraints reported in the literature were derived based on a very specific set of assumptions (i.e. all of them are reported for a monochromatic mass function and unclustered PBHs) and may be relaxed or strengthened with model-dependent modifications of the formation mechanism (i.e. strongly non-Gaussian fluctuations or the effect of accretion for $M_{\rm PBH} \gtrsim \mathcal{O}(10) M_\odot$ [13] and clustering). This is a rapidly evolving topic in the literature even though we regard the constraints we report here as well established, see Ref. [199] for the most recent and complete review. A comprehensive summary plot is shown in Fig. 1.7, where only the main (and most stringent) constraints on the abundance are shown for clarity.

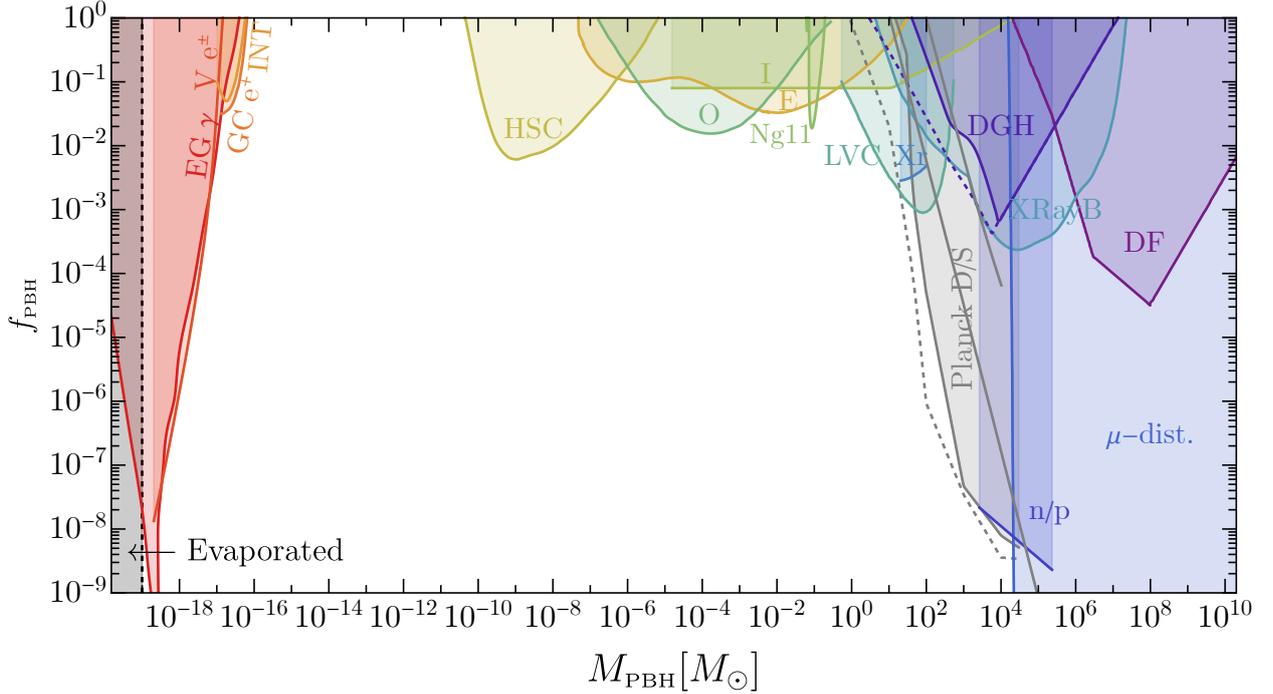

Figure 1.7: *A comprehensive plot of the constraints on the PBH abundance for a monochromatic population of PBHs with mass $M_{\rm PBH}$. For clarity, we decided to show only the most stringent/relevant constraints out of the ones discussed in the main text. In the leftmost portion of the plot, the shaded region indicates PBHs that evaporate in a timescale smaller than the age of the universe and do not constitute a viable dark matter candidate. The version of DGH and Planck (D) shown with dashed lines are not shaded as corresponding to an optimistic (i.e. most constraining) choice of assumptions.*

We divide the discussion based on the kind of physical phenomenon used to constrain the PBH population.

**Hawking radiation**

PBHs with $M_{\rm PBH} \lesssim 10^{-16} M_\odot$ evaporate in a time-scale comparable with the age of the universe. As the energy emitted through the Hawking evaporation becomes increasingly larger for lighter BHs, this process may lead to the production of detectable signatures in the case of ultralight PBHs. The emitted radiation has a black-body spectrum with a temperature $T \propto 1/M_{\rm PBH}$. Indeed, in the lightest portion of Fig. 1.7, the most stringent constraints come from extra-galactic gamma-rays (EG $\gamma$) [200], $e^\pm$ observations by Voyager 1 (V $e^\pm$) [201], positron annihilations in the Galactic Center (GC $e^+$) [202] and gamma-ray observations by INTEGRAL (INT) [203]. We also point out other analogous constraints in this mass window (not shown in Fig. 1.7) were derived in Refs. [194, 204–216].

**Gravitational lensing**

PBHs, being very compact objects, can lead to significant lensing signatures on the electromagnetic radiation reaching the detectors from background sources. Thus, the most stringent constraints in the mass range $10^{-10} \lesssim M_{\rm PBH} \lesssim 10 M_\odot$ come from microlensing searches by Subaru HSC [217, 218], lensing searches of massive compact halo objects towards the Large Magellanic Clouds (EROS, E) [219, 220], fast transient events near-critical curves of massive galaxy cluster Icarus (I) [221], and observations of stars in the Galactic bulge by the Optical Gravitational Lensing Experiment (Ogle,

O) [222]. Finally, Ref. [223] constrained $f_{PBH} \lesssim 0.4$ for $M_{PBH} \gtrsim 10^{-2} M_\odot$ from gravitational lensing of type Ia supernovae. This is not shown in the plot as it would fall below other more stringent bounds.

**Gravitational waves**

PBHs can assemble in binaries both in the early- and late-time universe, leading to observable signals at current LIGO and Virgo detectors. As we will discuss in the next chapters, the former channel predicts the dominant merger rate scaling as $R_{PBH} \approx f_{PBH}^2$. By requiring the number of detectable events per year not to exceed the rate observed by the LIGO/Virgo collaboration (LVC), one can set a constraint on the PBH abundance [7, 169, 170, 224]. For large values of $f_{PBH}$ close to unity, PBH clustering evolution can reduce the merger rate by enhancing binary interactions in dense environments. However, this effect is not sufficient to reduce the rate to a level that would be compatible with LVC observations in the standard scenario (i.e. Poisson distributed PBHs at formation) [14, 224]. This conclusion may not apply to the extreme scenario of strongly clustered PBHs at formation.

Additionally, the NANOGrav experiment searching for a stochastic GW background in the frequency range close to $f \simeq 1\,\mathrm{yr}^{-1}$ would be able to detect the GWs induced at second order by the curvature perturbations responsible for PBH formation. The null observation in the 11-yr dataset was translated into a constraint on the PBH abundance by Ref. [191] (see also [225]). As the PBH abundance is exponentially sensitive to the curvature perturbation amplitude constrained by NANOGrav, the current upper bound has large uncertainties when translated in terms of $f_{PBH}$. Therefore, we take the opportunity to stress that the constraint from the NANOGrav 11-yr data from Ref. [191] has large systematic uncertainties, above all in their choice of the threshold for PBH formation. In Fig. 1.7 we report this constraint (Ng11) but obtained by choosing a threshold on the curvature perturbation at collapse $\zeta_c = 0.6$ (in contrast with the choice $\zeta_c = 1$ made in Ref. [191]) motivated by state-of-art numerical simulations [226] (see also the discussion in Refs. [7, 167, 198]).

We stress that this is only applicable for PBHs formed from the collapse of density perturbations and in the absence of non-Gaussianities (see [100, 227–230]). Also, the NANOGrav collaboration has released the new 12.5 yrs dataset and claimed that the previous constraint should be relaxed due to improved treatment of the intrinsic pulsar red noise [231]. Additionally, the NANOGrav collaboration has reported evidence for a stochastic common process that could be explained by a SGWB signal. We will discuss this possibility in a dedicated chapter of this thesis. For our purposes, as the new constraint would have a similar impact on the mass range of interest, here we choose to show the re-analysis of the 11-yr constraint reported in Ref. [191] as a reference.

**PBH accretion**

Accretion of baryonic matter on PBHs leads to various effects both in the early- and late-time universe. As PBH accretion of gas generates the emission of ionizing radiation in the *early-time* universe, a PBH population would have an impact on the CMB temperature and polarization anisotropies [183, 232, 233]. In particular, Ref. [183] also accounts for the catalysing effect of the early DM halo forming around individual PBHs, if $f_{PBH} < 1$. Due to uncertainties in the accretion physics, Ref. [183] considered two opposite scenarios: the accreting gas is characterised by either a disk or a spherical geometry (Planck D/S respectively). We recall that the relevant electromagnetic emission takes place in the redshift range $300 \lesssim z \lesssim 600$. Therefore, we expect the spherical model (Planck S) to be more accurate, as a thin accretion disk could form only at a much smaller redshift [234]. We, therefore, plot Planck D as a dashed line in Fig. 1.7. We also stress that, since the relevant emission takes place at high redshift, its physics is independent of uncertainties in the accretion model due to the onset of structure formation and, in the standard scenario, PBH clustering is not expected to play a major role based on N-body simulations performed in Ref. [235] (see also [236]). In Fig. 1.7, we additionally plot as grey lines (to the right side of Planck S) the conservative constraints derived in Ref. [232] neglecting the role of (non-PBH) DM halos surrounding PBHs either assuming photoionization or collisional ionization [232] (from left to right). These can be considered as the most conservative bounds on PBHs coming from CMB observations.

Other constraints come from comparing the *late-time* emission of electromagnetic signals from interstellar gas accretion onto PBHs with observations of galactic radio and X-ray isolated sources (XRay) [237, 238] (see also Ref. [239]) and X-ray binaries (XRayB) [240]. Also, the effect of PBH interaction with the interstellar medium compared to data from Leo T dwarf galaxy observations leads to Dwarf Galaxy Heating (DGH) constraints [241, 242].

**Dynamical effects**

The most stringent constraint we report in Fig. 1.7 from dynamical effects is the one denoted as dynamical friction (DF) [243, 244]. If the galactic halo has a large fraction of massive PBHs, some of them must cluster close to the galactic center due to dynamical friction from stars or lighter PBHs making them lose kinetic energy. Such high concentration would not be allowed by the upper limit on the mass contained in the Galactic center. Other dynamical limits for $M_{\mathrm{PBH}} \gtrsim M_\odot$ come from the disruption of wide binaries [245], and survival of star clusters in Eridanus II [246] and Segue I [247, 248]. Those are not shown in Fig. 1.7 as they would be covered by other constraints.

**Indirect constraints**

We define "indirect" constraints the ones that only probe the standard formation mechanism for PBHs, i.e. limiting the amplitude of primordial curvature perturbations.[2] Those limits at small scales can then be translated into a bound on the PBH abundance within a given set of assumptions. From the effect of scalar perturbations on the big-bang nucleosynthesis (i.e. a modification of the neutron-to-proton ratio), the constraint (n/p) was derived in Ref. [249]. Similarly, limits on CMB $\mu$-distortions [250] strongly constrain PBHs above $M_{\mathrm{PBH}} \approx 10^4 M_\odot$. Those limits would be, however, evaded if PBHs are generated from highly non-Gaussian perturbations boosting the PBH production by enhancing the tail of the perturbation distribution while leaving the mean perturbations well below the bounds. Finally, we do not plot other constraints coming from Lyman-$\alpha$ forest observations [251, 252] which would be impacted by Poisson fluctuations in the PBH number density inducing an enhancement of power at small scales when compared to the standard cold dark matter scenario.

**Open opportunity: the asteroidal mass range**

We conclude by stressing the absence of bounds in the asteroidal mass range $M_{\mathrm{PBH}} \in [10^{-16} \div 10^{-10}] M_\odot$. This open window still allows PBHs to explain the dark matter. Limits that were previously set in this window, by using femtolensing effects [253], an extended original version of HSC microlensing searches [217], dynamical constraints derived requiring the survival of compact objects such as white dwarfs [254] and neutron stars [255], were later on relaxed in the re-analysis performed in Refs. [218, 256, 257]. To date, there is general agreement in the literature on the absence of constraints in this interesting mass range. In the final chapters of this thesis, we will show how GW observations at LISA could discover (or rule out) PBHs as dark matter in this mass window, at least if coming from the standard PBH formation scenario.

Notice that the bounds, typically derived for a monochromatic PBH population, can be adapted to extended mass functions using the techniques described in Refs. [258, 259]. One should also bear in mind the difference between constraints applying to high-redshift abundance and masses (such as those coming from CMB and NANOGrav observations, probing the early Universe physics) and the ones constraining late-time Universe quantities. The evolution of PBH masses and abundance with accretion requires constraints to be treated as described in detail in Ref. [13]. We will discuss this effect in detail in this thesis. The main effect of accretion is to alleviate early Universe constraints by shifting them to a higher *late-time* mass range and making them weaker due to the growth of $f_{\mathrm{PBH}}$.

---

[2]Also the NANOGrav constraint falls in the category of indirect constraints. However, we arbitrarily decided to put it in the GWs section due to the nature of the probe used to set the constraint.

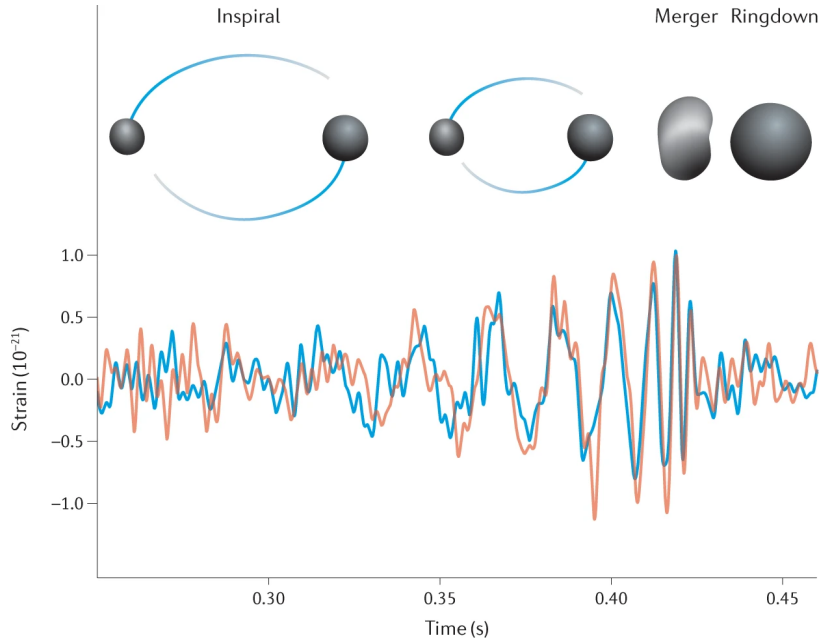

Figure 1.8: *GW signals from GW150914 detected at LIGO Hanford and Livingston observatories on September 14, 2015. In the top row, the result of a numerical relativity simulation showing the evolution of the event horizons during the merger. Figure taken from Refs. [159, 271]*

## 1.3 The gravitational wave revolution

There is only one last ingredient to be introduced, that is: *Gravitational waves.*

The history of GWs starts in the year 1916 when Einstein first predicted that accelerated matter would produce undulations of spacetime which could be interpreted as waves [260]. See, for example, Ref. [261] for a historical account. There was an initial scepticism of Einstein's result, partially based on a technical mistake in the original computations which was pointed out by Eddington a few years later with a very critical remark "the only speed of propagation relevant to them is the speed of thought"[262]. Their existence was still debated up until the late 1950s, when the work of F.Pirani, H. Bondi, R. Feynman, suggested GWs had physical effects on matter, while finally J. Wheeler and J. Weber argued that these waves could be, indeed, measurable [263].

The first "indirect" confirmation of the existence of GWs came after the discovery of pulsars, which are rotating neutron stars emitting radio pulses [264, 265]. It was soon realised that pulsar in binaries would emit GWs [266, 267], making the orbital period evolve as dictated by the properties of GW emission predicted by Einstein's theory of general relativity. An example of such a system was finally observed in 1974 by R. Hulse and J. Taylor [268] and goes under the name of PSR B1913+16. Both received the 1993 Nobel prize in physics thanks to this discovery. Later on, only in 1981, the orbital-period decay of PSR B1913+16 was measured [269] with good precision, showing agreement with the prediction of general relativity.

We then needed to wait until September 2015 to have the first "direct" detection of GWs. The signal GW150914, observed at LIGO facilities [159] and shown in Fig. 1.8, was coming from a BH binary that merged around a billion years ago. This observation also showed the potential of GW signals to give us information on the most violent events in our universe. R. Weiss, B. Barish and K. Thorne were awarded the Nobel prize in physics in 2017 for their decisive contribution leading to this discovery.

Furthermore, it is remarkable that PBHs could explain the properties of such event and produce it with a rate consistent with observations without violating the bound on their abundance $f_{\rm PBH} \lesssim 1$. *Did this event present us with the first direct detection of dark matter [160, 162]?* The jury is still out and we will pursue this question, among others, in this thesis. Since then, the era of gravitational wave astronomy has flourished [270].

GWs can be described as propagating deformations of the spacetime geometry. We will discuss in more detail how the GW measurement is performed in Part III of this thesis. To gain a feeling of how GWs are emitted, we provide here a rough estimate of the GW sourced by a binary system. Let us consider a system of two compact bodies moving in a circular orbit. Einstein, by solving the equations of general relativity, found that the GW signal produced by a system at luminosity distance $D_{\rm L}$ follows the quadrupole formula [272]

$$h_{ij} = \frac{2G}{D_{\rm L}} \frac{{\rm d}^2 Q_{ij}}{{\rm d}t^2}, \tag{1.3.1}$$

where $Q_{ij}$ is the time-dependent quadrupole of the binary. Using Eq. (1.3.1), one can easily find an order of magnitude estimate for the GW amplitude $h_0$, which is defined as the typical value for the components $h_{ij}$. As the quadrupole moment scales as $Q \approx Mr^2$, where $M$ and $r$ are the total mass and size of the system respectively, and since the time derivatives can be estimated through the velocity $v$ as $1/t \approx v/r$, one finds

$$h_0 \approx \frac{GM}{D_{\rm L}} v^2 \simeq 5 \cdot 10^{-19} \left( \frac{M}{M_\odot} \right) \left( \frac{D_{\rm L}}{\rm Mpc} \right)^{-1} \left( \frac{v}{c} \right)^2. \tag{1.3.2}$$

On the other hand, the emission of GWs makes the binary system lose energy until the two compact objects eventually merge. A similar back of the envelope argument leads to the estimate for the maximum GW frequency. This can be estimated by computing the orbital frequency related to the Innermost Stable Circular Orbit (ISCO) after which the binary compact objects plunge and merge. As the ISCO orbit for Schwarzschild BHs is found to be $r_{\rm ISCO} = 6GM$, using the Kepler's law, one obtains

$$f_{\rm ISCO} \approx \frac{1}{6\sqrt{6}(2\pi)GM} = 2.2\,{\rm kHz} \left( \frac{M}{M_\odot} \right)^{-1}. \tag{1.3.3}$$

In more detail, it can be easily shown that the characteristic GW frequency evolves as [273]

$$f_{\rm GW}^{-8/3}(t) = \frac{1}{\pi} \left( \frac{5}{256} \right)^{3/8} (G\mathcal{M})^{-5/8} \left( t_{\rm merger} - t \right)^{-3/8}, \tag{1.3.4}$$

where we defined the so called "chirp mass"

$$\mathcal{M} = \frac{(m_1 m_2)^{3/5}}{(m_1 + m_2)^{1/5}}. \tag{1.3.5}$$

From Eq. (1.3.4), we see that a binary with $m_1 = m_2 = 30 M_\odot$, one obtains a frequency $f_{\rm GW} \simeq 40 {\rm Hz}$ around $\Delta t \simeq 0.1 {\rm s}$ before the merger. For example, the LIGO and Virgo experiments are indeed able to search for GW signals in the frequency range $f \in [10 \div 10^3]$ Hz, where compact binaries from solar to intermediate-mass BHs are expected to produce detectable signals.

In practice, a much wider range of frequencies and more distant sources are expected to be probed with current and future experiments. Pulsar Timing Array (PTA) experiments (discussed in more details in Sec. 6.7) like Parkes Pulsar Timing Array (PPTA) [280], the North American Nanohertz Observatory for Gravitational Waves (NANOGrav) [281], and the European Pulsar Timing Array (EPTA) [282], can search for GWs with frequencies close to nHz. In the range of frequencies around mHz, GWs will be observed by the planned space-based Laser Interferometer Space Antenna (LISA) [277]. Finally, in the range around $10^2$ Hz, current ground-based LIGO/Virgo detectors [283] will be superseded by third generation GW detectors Einstein Telescope (ET) [284] and Cosmic Explorer (CE) [285] in the next decade, greatly extending their reach in terms of maximum distance and detectable mass range.

GW signals can be more broadly distinguished in two main categories. First, individual detections of resolved sources. Those events are associated with strain signals in the detector's response which are recognised as coming from compact binary mergers, analogously to Fig. 1.8. On the other hand, individual mergers falling below the detection threshold will in any case contribute to a Stochastic Gravitational Wave Background (SGWB), potentially observable as a superposition of many undetected signals. A SGWB could also be generated in the early universe by mechanisms different from binary coalescences, such as during inflation, sourced at second order by scalar perturbations, from phase transitions, cosmic strings and many others (see Refs. [286–288] for reviews).

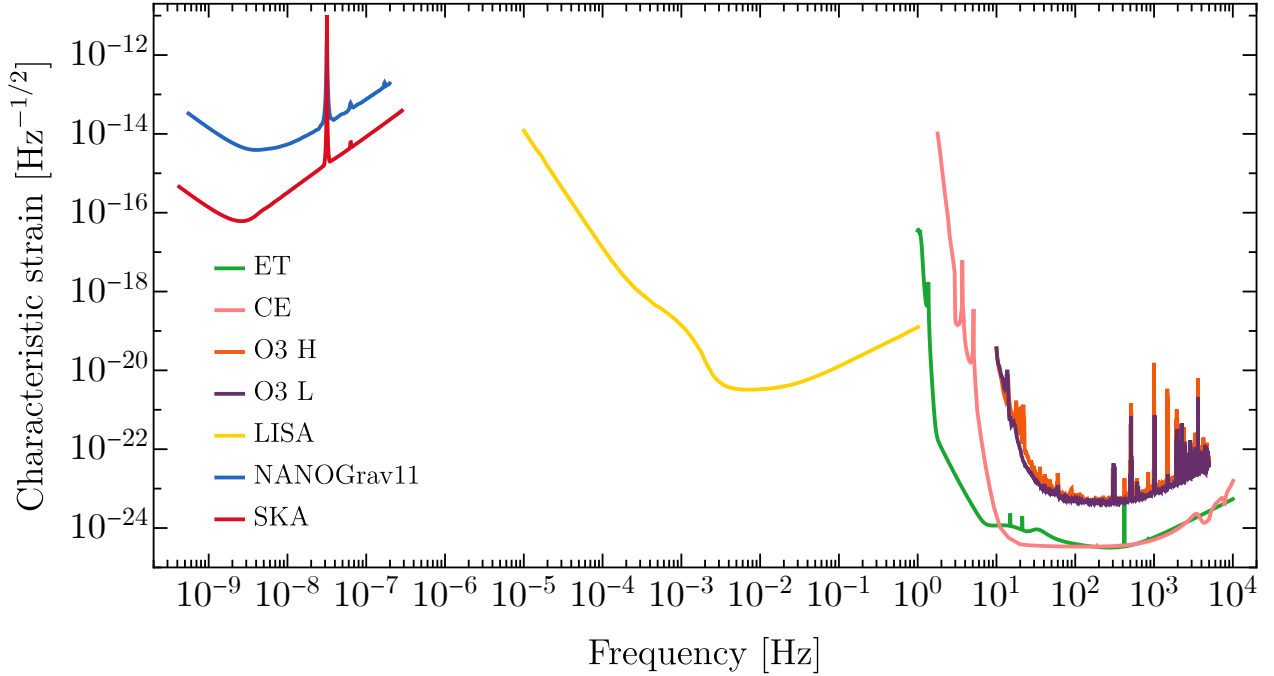

Figure 1.9:  *Noise curves for the LIGO Hanford (H) and Livingston (L) detector during the O3 run [274], the 3G detectors Einstein Telescope (ET) [275] and Cosmic Explorer (CE) [276], and the LISA experiment [277]. As representative for current and future PTA experiments, we show the NANOGrav 11yr sensitivity [278] as well as the future SKA sensitivity [279].*

The SGWB is characterised in terms of the dimensionless energy density $\Omega_{\rm GW} \equiv \rho_{\rm GW}/\rho_c$, similarly to the way we defined $\Omega_{\rm DM}$. The energy density in the SGWB can also be decomposed in its fundamental frequencies, by defining the fractional energy abundance per logarithmic frequency interval as [286]

$$\Omega_{\rm GW}(f) = \frac{1}{\rho_c} \frac{{\rm d}\rho_{\rm GW}}{{\rm d}\ln f}, \tag{1.3.6}$$

where the GW energy density is defined in terms of the time average of the 00-component of the energy-momentum tensor, which is [286]

$$\rho_{\rm GW} = \frac{1}{32\pi G} \left\langle \dot{h}_{ab} \dot{h}^{ab} \right\rangle. \tag{1.3.7}$$

As the standard model producing PBHs is characterised by large curvature perturbations, we can expect a sizeable SGWB generated at second order by the PBH formation mechanism. With the aim of providing an order of magnitude computation, the associated GWs amplitude at emission can be estimated to be $h_0 \simeq \zeta^2$ (see more details in Chapter 6), while the energy density becomes $\rho_{\rm GW} \simeq k_{\rm PBH}^2 \zeta^4 / 32\pi G$. Using the fact that at horizon crossing, when most of GWs are emitted within this mechanism, one has $k_{\rm PBH} \simeq aH$ and that the GWs behave as radiation in the expanding universe after their emission, one finds

$$\Omega_{\rm GW}^{\rm PBH} \approx \Omega_r \zeta^4 \approx 10^{-9}, \tag{1.3.8}$$

where $\Omega_r = 5.38 \cdot 10^{-5}$ is the current abundance of radiation and we chose a perturbation amplitude $\zeta^2 \simeq 10^{-2}$ motivated by Sec. 1.2.1. With a similar back of the envelope estimate, one can relate the frequency $f = k/2\pi$ of modes producing PBHs in the mass window where they are still allowed to be the dark matter, i.e. with masses around $M_{\rm PBH} \approx 10^{-12} M_\odot$, using Eq. (1.2.6), which translates into $f \simeq 3\,{\rm nHz}(M_{\rm PBH}/M_\odot)^{-1/2}$. It is serendipitous to discover that the corresponding frequencies lie in the range that will be probed by LISA, i.e around mHz. Translating the SGWB abundance in terms of a characteristic strain, we find

$$S_h(f) \approx \left[ \frac{3}{4\pi^2} \left( \frac{H_0^2}{f^3} \right) \Omega_{\rm GW}(f) \right]^{1/2} \approx 10^{-18} \, {\rm Hz}^{-1/2} \left( \frac{f}{\rm mHz} \right)^{-3/2} \left( \frac{\zeta}{0.1} \right)^2 . \tag{1.3.9}$$

By comparing this result with Fig. 1.9, we see that such a signal falls within the LISA reach. This shows that *LISA will be able to discover PBH dark matter by looking for the GWs associated to the PBH formation mechanism.*

It is exciting that PBHs could give rise to both types of signal we discussed. We dedicate Part III of this thesis to this topic and show how GWs could provide a unique probe to search for PBHs by looking for the GW signals of PBH mergers and the SGWB related to the PBH formation mechanism. Now, in Part II, we take a step back and start again from the beginning, that is the PBH formation.

# Part II

# Primordial black hole phenomenology

# Chapter 2

# Primordial black hole abundance

In this part of the thesis, we focus on the PBH formation in its various aspects. Let us stress that we are going to assume the "standard" formation scenario in which PBHs form out of the collapse of density perturbations in the early universe, when the energy density was dominated by a relativistic fluid. Some of the conclusions drawn in this part may change if one considered more "exotic" formation scenarios.

In this chapter, in particular, we focus on the theory describing the amount of PBHs produced in the early universe. We are going to describe both the criterion for collapse which accounts for the results of relativistic numerical simulations and the computation of the PBH abundance including both intrinsic and non-linearly induced non-Guassianities.

## 2.1 Threshold for PBH formation

PBHs can arise from the collapse of large density perturbations in the early universe [143, 146, 289, 290]. Such a violent event takes place when the gravitational force overcomes the radiation pressure. A simple order of magnitude estimate, based on a Jeans length argument in Newtonian gravity [71], suggests a density contrast larger than $\delta_c \sim c_s^2$, where $c_s^2 = 1/3$ is the sound speed of the radiation fluid, is sufficient to cause the formation of a PBH. A generalisation of this argument including the effects of general relativity gives $\delta_c \simeq 0.4$ for a radiation dominated Universe [150].

A full characterisation of this phenomenon, including the effect of pressure gradients, requires the description of the evolution of perturbations in the non-linear regime after they re-enter the cosmological horizon. This can be done by performing dedicated numerical simulations enabling us to define a threshold value above which a PBH can form. The threshold value of the density contrast is found to be dependent on the shape of the density peak and falls within the range $0.4 \le \delta_c \le 2/3$, see Refs. [291, 292]. In this section, we present its precise computation, for a given perturbation power spectrum, based on the results of general relativistic numerical simulations. We closely follow the analysis published in Ref. [6].

### 2.1.1 Collapse of density perturbations in the early universe

The initial conditions adopted by numerical simulations of PBH formation are fixed on superhorizon scales when the curvature perturbations are still time independent [293]. In other words, the initial energy density and velocity field are defined only in terms of a time independent curvature profile [147, 294], which can be derived from the shape of the inflationary power spectrum of cosmological perturbations measured on superhorizon scales [295, 296].

It follows from the theory of peaks of random fields in the gaussian approximation [297] that extreme peaks tend to be spherical (see discussion in App. A). Therefore, it is standard practice to assume spherical symmetry on superhorizon scales. An effort to generalise the analysis to non-spherical initial conditions was performed in Ref. [298], which however only found negligible corrections to the value of the threshold in a radiation dominated universe.

The local region characterised by a spherical perturbation is described by the metric written as

$$\begin{aligned} ds^2 &= -\,dt^2 + a^2(t)e^{2\zeta(\hat{r})}\left[d\hat{r}^2 + \hat{r}^2 d\Omega^2\right] \\ &= -\,dt^2 + a^2(t)\left[\frac{dr^2}{1 - K(r)r^2} + r^2 d\Omega^2\right], \end{aligned} \tag{2.1.1}$$

where $a(t)$ is the scale factor, while $K(r)$ and $\zeta(\hat{r})$ are the conserved comoving curvature perturbations defined on a super-Hubble scales. Both quantities vanish at large distances from the perturbation where the Universe is assumed to be the spatially flat Friedmann-Robertson-Walker (FRW) spacetime. The relation between the perturbation in the two metrics is given by

$$K(r)r^2 = -\hat{r}\zeta'(\hat{r})\left[2 + \hat{r}\zeta'(\hat{r})\right], \tag{2.1.2}$$

showing that $K(r)$ is defined in terms of spatial derivatives of the curvature perturbation $\zeta(r)$.

On superhorizon scales one can adopt the gradient expansion [143, 147, 299, 300]. This approximation only accounts for the first non-vanishing terms in the expansion in powers of the small parameter $\epsilon \ll 1$, where $\epsilon$ is identified with the ratio between the Hubble radius and the length scale of the perturbation. In this limit, the energy density profile can be written as [291, 294]

$$\frac{\delta\rho}{\rho_b} \equiv \frac{\rho(r,t) - \rho_b(t)}{\rho_b(t)} = -\frac{1}{a^2 H^2}\frac{4(1+w)}{5+3w}e^{-5\zeta(\hat{r})/2}\nabla^2 e^{\zeta(\hat{r})/2}, \tag{2.1.3}$$

where $H(t) = \dot{a}(t)/a(t)$ is the Hubble parameter, $\rho_b$ is the mean background energy density and $\nabla$ denote differentiation with respect to $\hat{r}$. The parameter $w$ is the coefficient of the equation of state $p = w\rho$ relating the total (isotropic) pressure $p$ to the total energy density $\rho$, which takes the value $w = 1/3$ in a radiation dominated universe.

Let us provide a simplified analytical argument showing the characteristic value of the threshold for PBH collapse [71], see also Ref. [196]. We can write the time-time component of the Einstein equations neglecting the spatial derivatives of $K$ (which appear at higher orders in gradient expansion) as

$$H^2 = \frac{8\pi G}{3}\rho(t) - \frac{K}{a^2}\,. \tag{2.1.4}$$

This is equivalent to the Friedmann equation describing the evolution of a spatially curved universe. In this picture, we can also define the density contrast as

$$\frac{\delta\rho}{\rho_b} = \frac{3K}{8\pi G\rho_b a^2} = \frac{K}{H^2 a^2}\,. \tag{2.1.5}$$

As a consequence of Eq. (2.1.4), a region with $K > 0$ will eventually stop expanding and collapse to form a PBH. This leads to a break down of the separate universe approximation, precisely when the right-hand-side of Eq. (2.1.4) becomes negative, i.e. when $3K/a^2 = 8\pi G\rho$ or $\delta\rho/\rho_b = 1$ at the time called $t_c$. As only modes corresponding to scales larger than the Jeans length can collapse, we can identify $k^2 = a^2 H^2/c_s^2$. This means that, at the horizon crossing time $t_k$ of the relevant perturbation (i.e. when $k^2 = a^2 H^2$), one finds that the only perturbations which are able to collapse should have a density contrast larger than

$$\left.\frac{\delta\rho(t_k)}{\rho_b(t_k)}\right|_c = \frac{K}{H^2(t_k)a^2(t_k)} = \frac{c_s^2 k^2}{H^2(t_k)a^2(t_k)} = c_s^2. \tag{2.1.6}$$

This rough estimate was then refined with dedicated relativistic numerical simulations following the evolution of perturbations until they eventually collapse (or not) after horizon crossing. By measuring the necessary amplitude leading to the PBH formation, the PBH threshold is defined. In the following, after having defined the relevant physical quantity used to measure the perturbation amplitude, we will report the result of numerical simulations performed in Ref. [6].

**The compaction function and the threshold**

It was shown that a convenient way to define a criterion for collapse is in terms of the so called "compaction function" [143, 291], constructed as the mass excess inside the length scale of the perturbation

$$\mathcal{C} \equiv 2 \frac{\delta M(r,t)}{R(r,t)}, \tag{2.1.7}$$

where $R(r,t)$ is the areal radius and $\delta M(r,t)$ is the difference between the Misner-Sharp mass within a sphere of radius $R(r,t)$, and the background mass $M_b(r,t) = 4\pi \rho_b(r,t) R^3(r,t)/3$ within the same areal radius but calculated with respect to a spatially flat background FRW metric.

In the superhorizon regime, the compaction function is time independent and takes the form

$$\mathcal{C} = \frac{2}{3} K(r) r^2 = -\frac{2}{3} \hat{r} \zeta'(\hat{r}) \left[2 + \hat{r} \zeta'(\hat{r})\right] \tag{2.1.8}$$

and the characteristic comoving length scale of the perturbation, defined as the distance $r_m$ at which the compaction function reaches its maximum [291], is found by imposing

$$\mathcal{C}'(r_m) = 0: \qquad K(r_m) + \frac{r_m}{2} K'(r_m) = 0 \qquad \text{or} \qquad \zeta'(\hat{r}_m) + \hat{r}_m \zeta''(\hat{r}_m) = 0. \tag{2.1.9}$$

Therefore, the parameter $\epsilon$ dictating the gradient expansion is defined as

$$\epsilon \equiv \frac{R_H(t)}{R_b(r_m,t)} = \frac{1}{aHr_m} = \frac{1}{aH\hat{r}_m e^{\zeta(\hat{r}_m)}}, \tag{2.1.10}$$

where $R_H = 1/H$ is the cosmological horizon and $R_b(r,t) = a(t)r$ is the background definition of the areal radius. The cosmological horizon crossing time $t_H$ can therefore be defined as the time when

$$\epsilon(t_H) = 1/aHr_m = 1. \tag{2.1.11}$$

The amplitude of the perturbation at $t_H$, which we refer to as $\delta_m \equiv \delta(r_m, t_H)$, is given by the excess of mass averaged over a spherical volume of radius $R_m$, as

$$\delta_m = \frac{4\pi}{V_{R_m}} \int_0^{R_m} \frac{\delta\rho}{\rho_b} R^2 \mathrm{d}R = \frac{3}{r_m^3} \int_0^{r_m} \frac{\delta\rho}{\rho_b} r^2 \mathrm{d}r, \tag{2.1.12}$$

where $V_{R_m} = 4\pi R_m^3/3$. With simple manipulations, see Ref. [291] for details, one can show that $\delta_m = \mathcal{C}(r_m)$ and therefore we conclude that imposing the compaction function to be above the threshold for collapse at the time of horizon crossing can be conveniently translated into the requirement that the smoothed density contrast over a scale $r_m$ is $\delta_m \geq \delta_c$. We will present the values of $\delta_c$ found using dedicated numerical simulations [6] in the following.

It is also interesting to notice that the resulting mass spectrum of PBHs has a behaviour described by the scaling law of critical collapse [291]

$$M_{\mathrm{PBH}} = \mathcal{K}(\delta_m - \delta_c)^\gamma M_H, \tag{2.1.13}$$

where $\gamma \simeq 0.36$ for a radiation dominated universe, $M_H$ indicates the mass of the cosmological horizon at horizon crossing time and $\mathcal{K}$ is an efficiency factor depending on the shape of $\delta\rho/\rho_b$ which was shown to fall in the range $1 \lesssim \mathcal{K} \lesssim 10$. We will use this relation in the following to compute the PBH mass distribution at formation.

**The shape parameter**

To investigate the dependence of the threshold on the shape of the overdensity profile, numerical simulations were performed to study the collapse of a basis of profiles of the curvature perturbation $K(r)$ given by

$$K(r) = \mathcal{A} \exp\left[-\frac{1}{\alpha}\left(\frac{r}{r_m}\right)^{2\alpha}\right]. \tag{2.1.14}$$

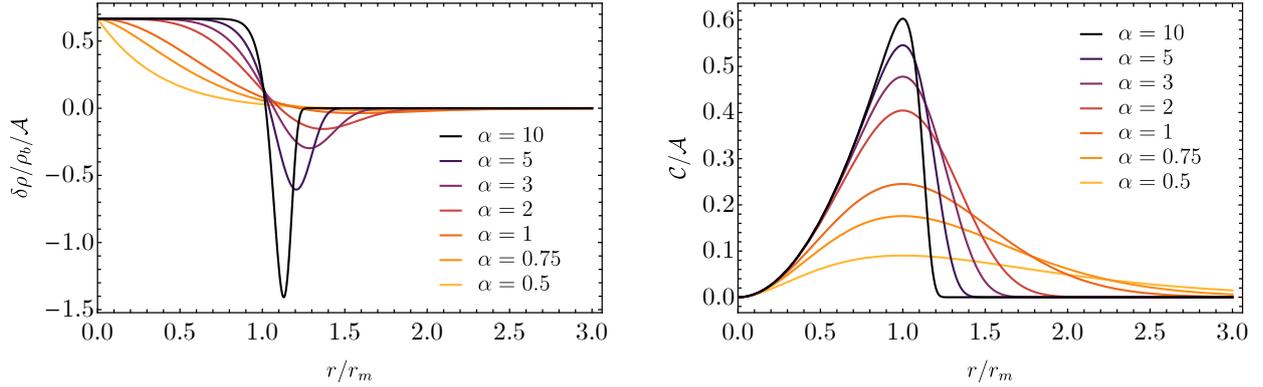

Figure 2.1: **Left:** Rescaled profile of the density contrast, as given by Eq. (2.1.3), for different values of $\alpha$. **Right:** Compaction function corresponding to the density contrast profiles shown in the left panel.

In Figure 2.1 we show the energy density profile $\delta\rho/\rho_b$ and the compaction function $\mathcal{C}$ plotted against $r/r_m$ for different values of $\alpha$. The shape of the density contrast is sharply peaked for $\alpha \ll 1$, corresponding to a broad profile of the compaction function, whereas $\mathcal{C}$ is more peaked for values of $\alpha \gg 1$, corresponding to broader profiles of the density contrast. We highlight here for clarity that the compaction function can be thought of as the curvature of the density profile at $\hat{r}_m$. For the interested reader, we refer to Ref. [291] and Refs. therein for all the details on how the relativistic numerical simulations are performed.

As seen in [291, 292], the threshold for PBH collapse varies depending on the shape of the cosmological perturbation, characterised by the width of the compaction function around its peak. At leading order in the compaction function, the shape is captured by the dimensionless parameter

$$\alpha_{\mathrm{G}} = -\frac{\mathcal{C}''(r_m)r_m^2}{4\mathcal{C}(r_m)}, \tag{2.1.15}$$

which is explicitly independent of the perturbation amplitude. However, going beyond the linear approximation, the shape parameter can be written as [6]

$$\alpha = \frac{2\alpha_{\mathrm{G}}}{1 - 3\mathcal{C}(r_m)/2 + (1 - 3\mathcal{C}(r_m)/2)^{1/2}}. \tag{2.1.16}$$

We note that the correction of the shape parameter turns out to be dependent on the amplitude of the perturbation revealing its non-linear nature.

As we will show in the following with some explicit examples, it is still useful to keep track of the approximated shape parameter $\alpha_{\mathrm{G}}$, since it is possible to find the full shape parameter $\alpha$ by solving [6]

$$F(\alpha)\left[1 + F(\alpha)\right]\alpha = 2\alpha_{\mathrm{G}}, \tag{2.1.17}$$

where

$$F(\alpha) = \sqrt{1 - \frac{2}{5}e^{-1/\alpha}\frac{\alpha^{1-5/2\alpha}}{\Gamma\left(\frac{5}{2\alpha}\right) - \Gamma\left(\frac{5}{2\alpha}, \frac{1}{\alpha}\right)}}. \tag{2.1.18}$$

A numerical fit of the shape parameter $\alpha$ as a function of $\alpha_{\mathrm{G}}$, which represents the solution of Eq. (2.1.17) with an accuracy of few percent, is given by

$$\alpha \simeq \begin{cases} 1.758\,\alpha_{\mathrm{G}}^{2.335} + 1.912\,\alpha_{\mathrm{G}} & 0.1 \lesssim \alpha \lesssim 4.5, \\ 4\,\alpha_{\mathrm{G}}^2 + 3.930\,\alpha_{\mathrm{G}} & \alpha \gtrsim 8. \end{cases} \tag{2.1.19}$$

Finally, as shown in [292], the average value of $\mathcal{C}(r)$ integrated over a volume of comoving radius $r_m$, defined as

$$\bar{\mathcal{C}}(r_m) = \frac{3}{r_m^3}\int_0^{r_m}\mathcal{C}(r)\,r^2\mathrm{d}r, \tag{2.1.20}$$

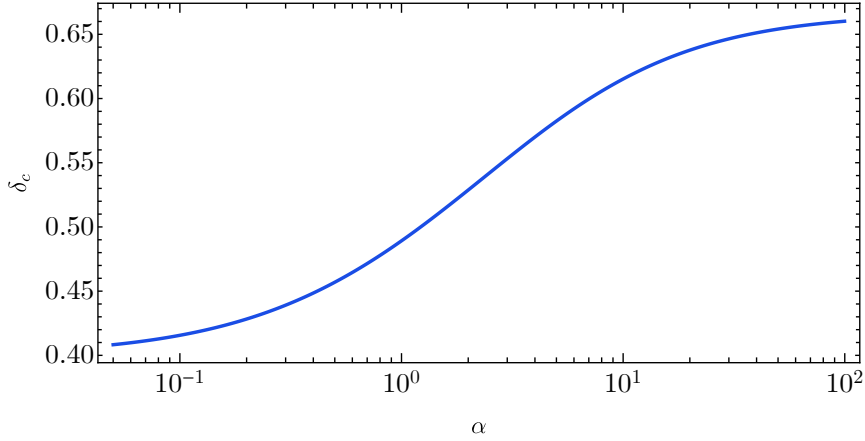

Figure 2.2: *Values of the threshold density contrast $\delta_c$ plotted as a function of the shape parameter $\alpha$.*

has a nearly constant value when computed at the threshold for PBH formation, which is $\bar{\mathcal{C}}_c \simeq 2/5$. This allows to write an analytic expression of the threshold $\delta_c$ as a function of the shape parameter $\alpha$, up to a few percent precision when compared to the results of numerical simulations [292],

$$\delta_c \simeq \frac{4}{15} \frac{e^{-1/\alpha} \alpha^{1-5/2\alpha}}{\Gamma\left(\frac{5}{2\alpha}\right) - \Gamma\left(\frac{5}{2\alpha}, \frac{1}{\alpha}\right)}. \tag{2.1.21}$$

A simple fit of the relation between the threshold and the shape parameter in Eq. (2.1.21) in the relevant range of values of $\alpha$ is given by

$$\delta_c \simeq \begin{cases} \alpha^{0.047} - 0.50 & 0.1 \lesssim \alpha \lesssim 7, \\ \alpha^{0.035} - 0.475 & 7 \lesssim \alpha \lesssim 13, \\ \alpha^{0.026} - 0.45 & 13 \lesssim \alpha \lesssim 30. \end{cases} \tag{2.1.22}$$

It is important to stress that the shape parameter $\alpha$ describes the main features of the profile in the relevant region $0 < r \lesssim r_m$ where PBHs form, while secondary corrections to the profiles in the tails at $r \gtrsim r_m$ only introduce a few percent correction to the threshold $\delta_c$ with respect to the one obtained with (2.1.14). The resulting value of the threshold as a function of $\alpha$ is plotted in Fig. 2.2.

### 2.1.2 The average value of the threshold

In this section, we describe how it is possible to calculate the average value of the shape parameter $\alpha$ by identifying the typical perturbation shape associated with a given cosmological power spectrum. Once $\alpha$ is identified, using the results in the preceding section, one can find the corresponding averaged value of the threshold $\delta_c$.

Assuming Gaussian statistics for the comoving curvature perturbation $\zeta$, we compute the value of $\alpha_{\mathrm{G}}$ from the power spectrum $P_\zeta(k, \eta)$, defined as

$$P_\zeta(k, \eta) = \frac{2\pi^2}{k^3} \mathcal{P}_\zeta(k) T^2(k, \eta), \tag{2.1.23}$$

and computed at the proper time $\eta$ effectively corresponding to the horizon crossing when $\hat{r}_m \sim r_H$, where $r_H = 1/aH$ is the comoving Hubble radius. $\mathcal{P}_\zeta(k)$ is the adimensional form of the power spectrum defined, defined as

$$\langle \zeta(\vec{k}_1)\zeta(\vec{k}_2) \rangle = (2\pi)^3 \, \delta^{(3)}\left(\vec{k}_1 + \vec{k}_2\right) \frac{2\pi^2}{k_1^3} \mathcal{P}_\zeta(k_1), \tag{2.1.24}$$

and the linear transfer function $T(k, \eta)$, given by

$$T(k, \eta) = 3 \frac{\sin(k\eta/\sqrt{3}) - (k\eta/\sqrt{3}) \cos(k\eta/\sqrt{3})}{(k\eta/\sqrt{3})^3}, \qquad (2.1.25)$$

has the effect of cutting subhorizon modes which would not contribute to the dynamics as they are smoothed out as a consequence of pressure gradients during the collapse. See App. E for the derivation of transfer function in linear approximation.

Using Gaussian peak theory to write the characteristic profile of curvature perturbations as (see details in App. A)

$$\zeta(\hat{r}) = \zeta_0 \int dk k^2 \frac{\sin(k\hat{r})}{k\hat{r}} P_\zeta(k, \eta), \qquad (2.1.26)$$

the condition in Eq. (2.1.9), identifying the radius $\hat{r}_m$, becomes

$$\int dk k^2 \left[ (k^2 \hat{r}_m^2 - 1) \frac{\sin(k\hat{r}_m)}{k\hat{r}_m} + \cos(k\hat{r}_m) \right] P_\zeta(k, \eta) = 0. \qquad (2.1.27)$$

The Gaussian shape parameter can be then computed from Eq. (2.2.34) as

$$\alpha_{\mathrm{G}} = -\frac{1}{4} \left[ 1 + \hat{r}_m \frac{\int dk k^4 \cos(k\hat{r}_m) P_\zeta(k, \eta)}{\int dk k^3 \sin(k\hat{r}_m) P_\zeta(k, \eta)} \right], \qquad (2.1.28)$$

showing explicitly that both $\alpha_{\mathrm{G}}$, and consequently $\alpha$ computed using (2.1.19), are varying depending on the shape of the cosmological power spectrum. The same holds for the corresponding value of $\hat{r}_m$ given by the solution of Eq. (2.1.27). Finally, the characteristic threshold can be found using Eq. (2.1.21).

In the following we apply the prescription we presented so far to show how, given a particular form of the power spectrum, the value of the threshold $\delta_c$ is varying. It is interesting to stress here the general trend observed. For increasingly broader spectra, the shape parameter $\alpha$ is decreasing, reflecting the fact that when multiple modes are participating in the collapse, the compaction function becomes flatter up to the relevant scale $\hat{r}_m$. As a consequence, the pressure gradients around the peak of the compaction function are smaller, facilitating the collapse, and the corresponding threshold for PBH formation is reduced.

We will consider various characteristic shapes often use to parametrise realistic cosmological curvature power spectra, which could also serve the reader as a useful reference to find the corresponding threshold for PBH formation.

- **Peaked power spectrum:** we consider a monochromatic power spectrum of the form

$$\mathcal{P}_\zeta(k) = \mathcal{P}_0 k_* \delta_D(k - k_*) \qquad (2.1.29)$$

 where $\delta_D$ indicates the Dirac delta distribution. This scenario corresponds to $k_* \hat{r}_m \simeq 2.74$, $\alpha \simeq 6.33$ and an average threshold $\delta_c \simeq 0.59$. It is interesting to stress that, in this scenario, the threshold is found to be consistent with Ref. [301], where the average profile of $\zeta(r)$ for a peaked power spectrum was directly inserted into Eq. (2.1.3) to set up the initial conditions for the numerical simulations. This is expected as, for a narrow power spectrum parametrised by a Dirac delta function, using peak theory in $\zeta$ or in the density contrast $\delta\rho/\rho_b$ is equivalent [19].

- **Broad power spectrum:** an extended and flat power spectrum of curvature perturbations of the form [15, 18, 295]

$$\mathcal{P}_\zeta(k) \approx \mathcal{P}_0 \Theta(k - k_{\min}) \Theta(k_{\max} - k), \qquad \text{with} \qquad k_{\max} \gg k_{\min}, \qquad (2.1.30)$$

 is also often considered in the context of PBH formation. In this case, from (2.1.27) we have $k_{\max} \hat{r}_m \simeq 4.49$, which gives $\alpha \simeq 3.14$, corresponding to $\delta_c \simeq 0.56$. As one can notice, the threshold for PBH collapse is reduced compared to the monochromatic case. As we discussed, a broader power spectrum generically leads to an increased probability of collapse, as more modes contributes to the dynamics.

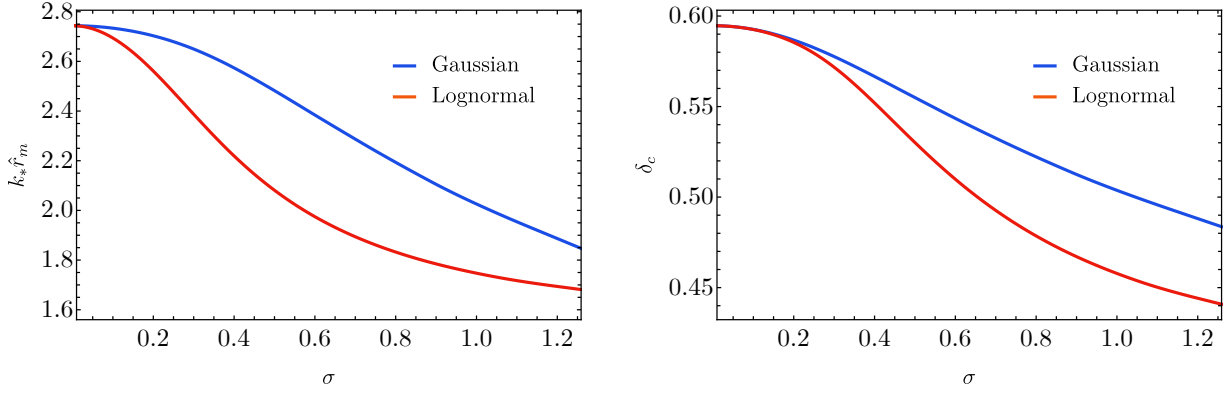

Figure 2.3: **Left:** Peak position of the compaction function $\hat{r}_m$ for both the case of a gaussian and lognormal shape of the power spectrum. **Right:** Average threshold for collapse $\delta_c$ in both cases. In the limit of $\sigma \to 0$, both spectra converge to the monochromatic case.

- **Gaussian power spectrum:** a symmetric spectrum parametrised as a gaussian function

$$\mathcal{P}_\zeta(k) = \mathcal{P}_0 \exp\left[-(k - k_*)^2/2\sigma^2\right], \qquad (2.1.31)$$

depending on the central reference scale $k_*$ and width $\sigma$. We show in Fig. 2.3 the resulting length scale $\hat{r}_m$ and threshold as a function of the spectrum width. In the limit of small $\sigma$, both results converge towards the ones found for the Dirac delta power spectrum while, for increasing values of $\sigma$, the corresponding value of the threshold decreases.

- **Lognormal power spectrum:** a symmetric power spectrum of the form

$$\mathcal{P}_\zeta(k) = \mathcal{P}_0 \exp\left[-\ln^2\left(k/k_*\right)/2\sigma^2\right], \qquad (2.1.32)$$

characterised by a width $\sigma$ and a central scale $k_*$. The results in terms of the characteristic perturbation length and threshold are shown in Fig. 2.3. As one can appreciate, the results have a behaviour similar to the gaussian case. Notice however that the amplified trends observed in this case are due to the fact that, for the lognormal shape, the parameter $\sigma$ identifies a spread in logarithmic space.

- **Cut-Power-Law power spectrum:** this spectrum is given by the functional form

$$\mathcal{P}_\zeta(k) = \mathcal{P}_0 \left(\frac{k}{k_*}\right)^{n_s^*} \exp\left[-(k/k_*)^2\right], \qquad (2.1.33)$$

expressed in terms of a tilt $n_s^*$ and with an exponential cut-off at the momentum scale $k_*$. The relation between the length scale of the overdensity and the scale $k_*$ is shown in Fig. 2.4, along with the corresponding average threshold for PBHs as a function of $n_s^*$. As $n_s^*$ increases, the spectrum becomes narrower and peaks at higher values of $k$. In agreement with the general behaviour seen in the previous examples, as the spectral tilt decreases, a larger number of modes participates in the collapse, resulting in a lower value of the threshold $\delta_c$.

We conclude this section by stressing the importance of the dependence of the threshold on the shape of the collapsing overdensities we have shown. As we will present in the next section, the PBH abundance is typically exponentially sensitive $\delta_c$. As such, it is of crucial importance to include shape effects in the computation of the threshold in a given model as they can lead to significant changes in the overall PBH abundance. Also, we provided a simple analytical prescription to compute the average threshold starting from the power spectrum of comoving curvature perturbations which fully accounts for the results of numerical simulations.

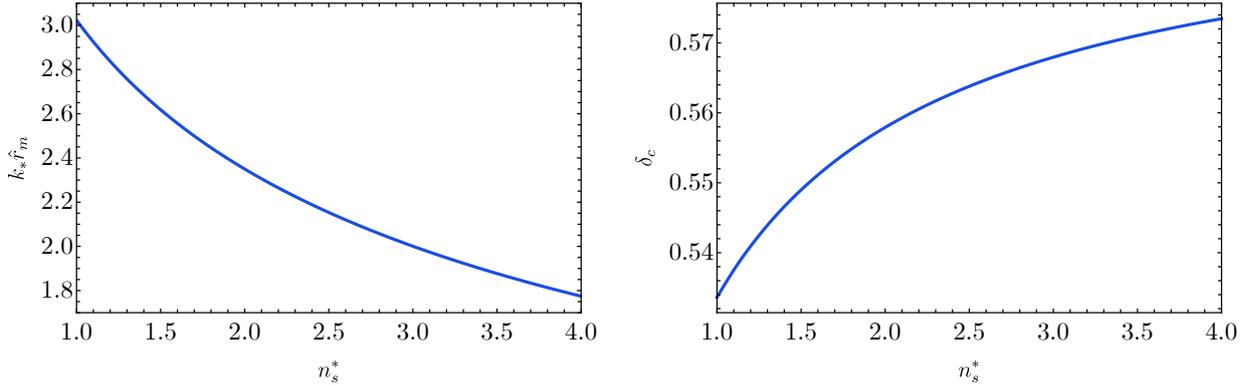

Figure 2.4: *Plots showing the same quantities as in Fig. 2.3, but for the cut-power-law power spectrum characterised by the tilt $n_s^*$.*

**The non linear horizon crossing**

One delicate issue that would need further investigations in the future is the definition of horizon crossing. In this section, we presented the threshold computed at the linearly extrapolated horizon crossing time, defined as in Eq. (2.1.11), which is based on the background evolution of the quantities $a(t)$ and $H(t)$. However, locally around each perturbation, the evolution of the horizon may be modified. This could change, therefore, the exact time of crossing. In general, the cosmological horizon is a marginally trapped surface within an expanding region, which in spherical symmetry is simply defined by the condition $R(r, t) = 2M(r, t)$, where $R(r, t)$ is the areal radius and $M(r, t)$ is the Misner-Sharp mass, i.e. the mass contained within a given sphere of radius $R(r, t)$. Since the region collapsing to form a PBH can be described as a locally closed Universe, the rate of expansion of the cosmological horizon is less than that of the spatially flat background, and this gives rise to an additional growth of the amplitude of the perturbation before reaching the horizon crossing. In the left plot of Fig. 2.5, we show $\epsilon(t_H^{\text{NL}})$ as a function of the shape parameter $\alpha$, with the dashed line fitting the numerical results given by the dots. As one can see, $\epsilon(t_H^{\text{NL}})$ computed at the non-linear horizon crossing is $1.3 \div 1.5$ times larger than one extrapolated linearly ($aHr_m = 1$). In Ref. [6], we provided the values of the threshold computed at the true horizon crossing, as found in the numerical simulations. Those values are reported in the right panel of Fig. 2.5 in terms of the threshold computed at the linearly extrapolated horizon crossing time for various values of $\alpha$. This shows that the threshold value of the density contrast is larger when computed at the true horizon crossing time. This follows from the larger growth of perturbations due to the longer time it takes for the horizon to reach the perturbation scale.

We stress however that, to use this result to compute the PBH abundance, one would need to know also the evolution of the variance of the density contrast up to the real horizon crossing, which would, in turn, require the knowledge of the non-linear transfer function of curvature perturbations. As this issue still needs to be settled in the literature, we decided not to include this effect in the following parts of this thesis and to rely on the criterion for collapse on the linearly extrapolated horizon crossing time.

## 2.2   Computation of the abundance

Computing the abundance of PBHs generated in the early universe in a particular model is a very difficult task. As the collapse is required to be a rare event for PBHs not to exceed the maximum allowed abundance of dark matter in the universe, $f_{\text{PBH}}$ is sensitive to the tails of the probability distribution. Therefore one can potentially encounter large deviations from the gaussian statistics. Also, as the amount of PBHs produced is exponentially dependent on the parameters of the model (i.e. the curvature perturbation power spectrum amplitude), little changes to the techniques adopted to compute $f_{\text{PBH}}$ can cause a variation of many orders of magnitude of the final result. On the other hand, the opposite reasoning is also valid, i.e. large changes in the PBH abundance can be reabsorbed by $\mathcal{O}(1)$ factors in the model. This represents a blessing and a curse for the PBH scenario.

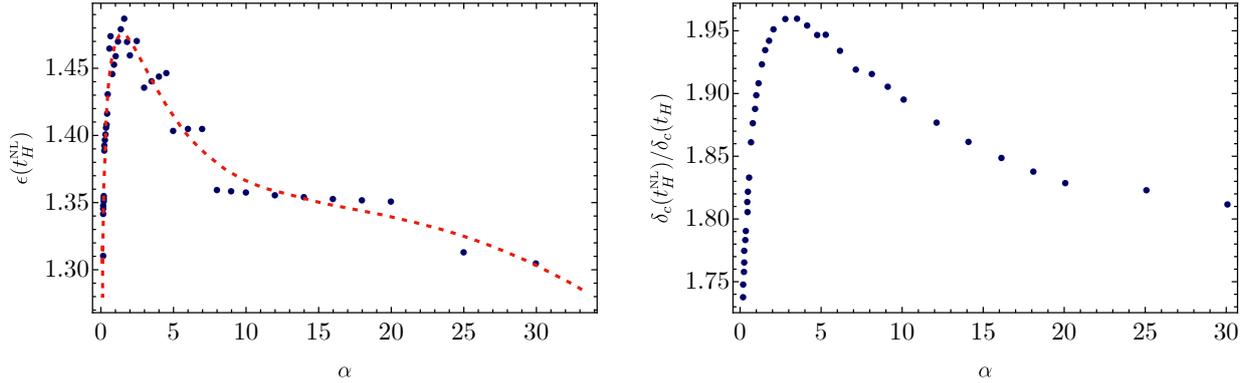

Figure 2.5: **Left:** *Non-linear horizon crossing, measured in terms of $\epsilon(t_H) = 1$, as a function of the shape parameter $\alpha$. Notice that the step-like behaviour of the numerical data is the result of artefacts introduced by the necessary discretization of spacetime performed in the simulation [6].* **Right:** *Ratio of the threshold $\delta_c$ computed at both the non-linear and linearly extrapolated horizon crossings, plotted as a function of the shape parameter $\alpha$.*

It is crucial however to notice that in the case one is interested in looking at the correspondence of multiple observables (e.g. the PBH abundance *and* the amount of GWs emitted at second order) within a given scenario, the freedom to adjust the model parameters is lost and a precise determination of the abundance is of utmost importance. In the recent years there has been a large effort of the scientific community in trying to develop the theory behind the computation of the cosmological abundance of PBHs [295, 296, 302, 303], including both corrections coming from non linearities [19, 166, 304–309] and non-Gaussianities [24, 301, 310–312].

In the following section, we present how the computation of the abundance can be performed, also going beyond the Gaussian approximation, based on Refs. [19, 24].

### 2.2.1 Gaussian approximation

We learned in section 2.1 how the condition for the formation of PBHs can be translated into the requirement that the overdensity field $\delta$, smoothed over the characteristic perturbation scale, is larger than the threshold $\delta_c$. The abundance of PBHs then corresponds to the fraction of local regions with $\delta \geq \delta_c$ evaluated at the time of cosmological horizon crossing of the perturbations.

If we assume the comoving curvature perturbation $\zeta$ to be a gaussian random field and we expand the density perturbation in Eq. (2.1.3) up to linear order as

$$\delta(\vec{x}, t) = -\frac{4}{9a^2H^2}\nabla^2\zeta(\vec{x}) + \ldots,\qquad(2.2.1)$$

we find that also the density contrast follows gaussian statistics. A random gaussian field is completely described by the value of its two point function. We can define the variance $\sigma_\delta^2$ as [313]

$$\sigma_\delta^2 = \int \frac{\mathrm{d}^3k}{(2\pi)^3}\,W^2(k, R_H)\,P_\delta(k),\qquad(2.2.2)$$

where $P_\delta$ is the overdensity power spectrum, defined starting from Eq. (2.1.24) as

$$P_\delta(k) = \frac{16}{81}\left(\frac{k}{aH}\right)^4 P_\zeta(k) + \ldots,\qquad(2.2.3)$$

$R_H = 1/aH$ being the comoving horizon length, $H$ the Hubble rate and $a$ the scale factor, as defined in the preceding section. The quantity $W(k, R_H)$ is a window function corresponding to a top-hat in real space, as used in Eq. (2.2.33). In momentum space it is given by

$$W(k, R) = 3 \frac{\sin(kR) - kR \cos(kR)}{(kR)^3}.$$ (2.2.4)

Two possible prescriptions can be adopted to compute the probability of collapse, as we describe now.

**Threshold statistics**

The simplest way to compute the abundance is using the so-called "threshold statistics", or Press-Schechter theory. In this setup, the primordial mass fraction of the universe stored into PBHs at formation time is predicted to be the integral of the Gaussian probability distribution function falling above $\delta_c$ as[1]

$$P_G(\delta > \delta_c) = \beta_G^{\rm th} = \int_{\delta_c} \frac{{\rm d}\delta}{\sqrt{2\pi}\,\sigma_\delta} e^{-\delta^2/2\sigma_\delta^2}.$$ (2.2.5)

By defining the rescaled threshold parameter

$$\nu = \frac{\delta_c}{\sigma_\delta},$$ (2.2.6)

the Gaussian mass fraction is

$$\beta_G^{\rm th} = \frac{1}{2} {\rm erfc} \left( \frac{\nu}{\sqrt{2}} \right) \simeq \sqrt{\frac{1}{2\pi\nu^2}} e^{-\nu^2/2}$$ (2.2.7)

where the second step is a good approximation only for $\nu \gtrsim 5$. In this result we clearly recognise the exponential dependence of the abundance on the variance $\sigma_\delta^2$.

**Peak theory**

Alternatively, one can draw a correspondence between the PBHs and the local maxima of the overdensity field. Those maxima can be described using peak theory of gaussian random field [297], see details in App. A. As a result, the number density of the maxima with an height overcoming the threshold, directly related to $\beta$, is found to be [303]

$$\beta_G^{\rm pk} \simeq \frac{1}{3\pi} \left( \frac{\langle k^2 \rangle}{3} \right)^{3/2} R_H^3 \left( \nu^2 - 1 \right) e^{-\nu^2/2}$$ (2.2.8)

with $\langle k^2 \rangle$ which is the first moment of the distribution

$$\langle k^2 \rangle = \frac{1}{\sigma_\delta^2} \int \frac{{\rm d}^3 k}{(2\pi)^3} k^2 W^2(k, R_H) P_\delta(k).$$ (2.2.9)

There is a slight discrepancy in the prefactor between the results obtained with threshold statistics and peak theory, which is however minor compared to other systematic uncertainties. Most importantly, both recover the exponential term dominating the parametric dependence of the abundance. We recall here that a characteristic value for $\nu_c$ is of order $\nu \sim 7$ for PBHs in the solar mass range while $\nu \sim 8$ in the asteroidal mass range.

The gaussian expressions (2.2.7) and (2.2.8) therefore make already manifest the essence of the problem we are going to discuss in the following section. PBHs are generated through very large and rare fluctuations. Therefore, their mass fraction at formation is extremely sensitive to changes in the tail of the fluctuation distribution and therefore to any possible non-Gaussianity in the density contrast [23, 24, 88, 168, 310, 311, 314–321]. This implies that non-Gaussianities need to be carefully accounted for. For instance, the presence of a primordial local non-Gaussianity in the comoving curvature perturbation can significantly alter the number density of PBHs through mode coupling, see for example Refs. [322, 323].

---

[1]Notice we are neglecting an additional factor of 2 in Eq. (2.2.5). We will come back to this point in Sec. 2.2.4.

### 2.2.2   The general non-Gaussian threshold statistics

In this section, we present a general description of the computation of the abundance adopting threshold statistics when the density contrast field is non-Gaussian. These formulas are based on a path integral formulation introduced in the context of large scale structure by Ref. [324] and then adapted in the context of PBH models in Ref. [24].

Consider the overdensity field $\delta(\vec{x})$ with a generic probability distribution $P[\delta(\vec{x})]$. The partition function $Z[J]$ in the presence of an external source $J(\vec{x})$ is given by

$$Z[J] = \int [D\delta(\vec{x})] P[\delta(\vec{x})] e^{i \int d^3 x J(\vec{x}) \delta(\vec{x})}, \qquad (2.2.10)$$

where the measure $[D\delta(\vec{x})]$ is normalised such that

$$\int [D\delta(\vec{x})] P[\delta(\vec{x})] = 1. \qquad (2.2.11)$$

As customary in the path integral theory, the functional Taylor expansion of the partition function $Z[J]$ in powers of the source $J(\vec{x})$ specifies the correlators, while the corresponding expansion of $W[J] \equiv \ln Z[J]$ generates only the connected correlation functions $\xi^{(N)} = \xi^{(N)}(\vec{x}_1, \cdots, \vec{x}_N)$ to all orders, which we define as

$$\xi^{(N)} = \frac{\int [D\delta(\vec{x})] P[\delta(\vec{x})]\, \delta(\vec{x}_1) \cdots \delta(\vec{x}_N)}{\int [D\delta(\vec{x})] P[\delta(\vec{x})]} = \frac{(-i)^N}{Z[J]} \frac{\delta^N}{\delta J(\vec{x}_1) \cdots \delta J(\vec{x}_N)} Z[J] \bigg|_{J=0}. \qquad (2.2.12)$$

For later use, we also define the normalised correlators

$$w^{(N)}(\vec{x}_1, \cdots, \vec{x}_N) = \sigma_\delta^{-N} \xi^{(N)}(\vec{x}_1, \cdots, \vec{x}_N) \qquad (2.2.13)$$

where in particular one has $w^{(2)}(0) = 1$. Finally, one can express the functional $W[J]$ in terms of the connected correlation functions $\xi^{(N)}$ as

$$W[J] = \sum_{n=2}^{\infty} \frac{(i)^n}{n!} \int d^3 x_1 \cdots \int d^3 x_n\, \xi^{(n)}(\vec{x}_1, \cdots, \vec{x}_n) J(\vec{x}_1) \cdots J(\vec{x}_n). \qquad (2.2.14)$$

**Threshold statistics**

As done in the previous section, the threshold statistics describes the probability that the smoothed density field $\delta$ exceeds a certain threshold $\delta_c$ at the horizon crossing time $t_H$. Therefore, it is convenient to define a peak overdensity $\rho_\nu(\vec{x})$ as

$$\rho_\nu(\vec{x}) = \Theta(\delta(\vec{x}) - \nu \sigma_\delta), \qquad (2.2.15)$$

where $\Theta(x)$ is the Heaviside step-function.

Let us specialise the computation to a single point statistics. The field $\rho_\nu(\vec{x})$ can be regarded as counting the number of peaks overtaking the threshold. The probability that the overdensity is above the threshold is given by

$$\Pi_\nu(\vec{x}_1) = \Big\langle \rho_\nu(\vec{x}_1) \Big\rangle = \int [D\delta(\vec{x})] P[\delta(\vec{x})] \Theta\Big(\delta(\vec{x}_1) - \nu \sigma_\delta\Big). \qquad (2.2.16)$$

Since $d\Theta(x)/dx = \delta_D(x)$ and $2\pi \delta_D(x) = \int_{-\infty}^{\infty} d\phi\, e^{i\phi x}$, the $\Theta$-function may be expressed as

$$\Theta(\vec{x}) = \int_{-x}^{\infty} da \int_{-\infty}^{\infty} \frac{d\phi}{2\pi} e^{i\phi a}, \qquad (2.2.17)$$

and therefore,

$$\Pi_\nu = \int [D\delta(\vec{x})] P[\delta(\vec{x})] \int_{\nu\sigma_\delta}^{\infty} da \int_{-\infty}^{\infty} \frac{d\phi}{2\pi} e^{i\phi(\delta(\vec{x}) - a)} = \frac{\sigma_\delta}{2\pi} \int_\nu^{\infty} da \int_{-\infty}^{\infty} d\phi\, e^{-i\sigma_\delta \phi\, a} Z[\phi], \qquad (2.2.18)$$

where we identified the partition function and redefined $a \to a/\sigma_\delta$. Using Eq. (2.2.14) we may express $\ln Z[J]$ as (we are dropping the explicit integration over $x_i$ to simplify the notation)

$$\ln Z[J] = \sum_{n=2}^{\infty} \frac{(i)^n}{n!} \xi^{(n)}(\vec{x}_1, \cdots, \vec{x}_n) \prod_{k=1}^{n} \phi_k, \qquad (2.2.19)$$

and since

$$\phi e^{-i\sigma\phi a} = \frac{i}{\sigma_\delta} \frac{\partial}{\partial a} e^{-i\sigma_\delta \phi a}, \qquad (2.2.20)$$

we finally find

$$\Pi_\nu = \frac{\sigma_\delta}{2\pi} \int_\nu^\infty \mathrm{d}a \int_{-\infty}^\infty \mathrm{d}\phi \, \exp\left\{ \sum_{n=3}^\infty \frac{(-1)^n}{n!} \frac{\xi^{(n)}}{\sigma_\delta^n} \prod_{i=1}^n \frac{\partial}{\partial a} \right\} \exp\left( -\frac{1}{2}\sigma_\delta^2 \phi^2 - i\sigma_\delta \phi a \right). \qquad (2.2.21)$$

The last equation can be reduced by performing the Gaussian integration over $\phi_i$ to find

$$\Pi_\nu = (2\pi)^{-1/2} \int_\nu^\infty \mathrm{d}a \, \exp\left\{ \sum_{n=3}^\infty \frac{(-1)^n}{n!} w^{(n)} \prod_{i=1}^n \frac{\partial}{\partial a} \right\} \exp\left( -\frac{1}{2}a^2 \right). \qquad (2.2.22)$$

This expression can be simplified further by expanding the series of Hermite polynomials $H_n$ as

$$\begin{aligned}
\Pi_\nu =& (2\pi)^{-1/2} \int_\nu^\infty \mathrm{d}a \, \left( 1 - \frac{1}{3!} w^{(3)}(0) \frac{\mathrm{d}^3}{\mathrm{d}a^3} + \frac{1}{4!} w^{(4)}(0) \frac{\mathrm{d}^4}{\mathrm{d}a^4} + \cdots \right) \exp\left( -\frac{1}{2}a^2 \right) \\
=& h_0(\nu) + \frac{1}{\sqrt{4\pi}\,3!} w^{(3)}(0) e^{-\nu^2/2} H_2\left( \frac{\nu}{\sqrt{2}} \right) + \frac{1}{\sqrt{16\pi}\,4!} w^{(4)}(0) e^{-\nu^2/2} H_3\left( \frac{\nu}{\sqrt{2}} \right) + \cdots,
\end{aligned} \qquad (2.2.23)$$

or

$$P(\delta > \delta_c) = h_0(\nu) + \frac{e^{-\nu^2/2}}{\sqrt{2\pi}} \sum_{n=3}^\infty \frac{1}{2^{\frac{n}{2}} n!} \Xi_n(0) H_{n-1}\left( \frac{\nu}{\sqrt{2}} \right). \qquad (2.2.24)$$

We stress that the argument (0) in the functional $\Xi_n$ means that all the correlation functions are computed at the same point, i.e. $\Xi_n(0) = \Xi(\vec{x}, \vec{x}, \cdots, \vec{x})$, and the explicit expressions for $\Xi(\vec{x}_1, \vec{x}_2, \cdots, \vec{x}_n)$ is given by Eq. (2.20) in [24]. This is required since we are interested in looking at the effect of higher order correlation functions in the computation of the monopole term, or the single point statistics. For instance, the first order terms are

$$\begin{aligned}
\Xi_3(0) &= w^{(3)}(0), \qquad \Xi_4(0) = w^{(4)}(0), \qquad \Xi_5(0) = w^{(5)}(0), \\
\Xi_6(0) &= w^{(6)}(0) + 10 w^{(3)}(0)^2, \qquad \Xi_7(0) = w^{(7)}(0) + 35 w^{(3)}(0) w^{(4)}(0).
\end{aligned} \qquad (2.2.25)$$

Also, it is important to draw the attention of the reader to the fact that the result in Eq. (2.2.24) is exact and it is not affected by any approximation. Nevertheless, it can be simplified further as in the PBH scenario we are interested in the limit of a large threshold, i.e. $\nu \gg 1$. In this case, using the asymptotic expansion of the Hermite polynomials and the complementary error function when $\nu \gg 1$

$$\begin{aligned}
H_n\left( \frac{\nu}{\sqrt{2}} \right) &= 2^{\frac{n}{2}} \nu^n \Big( 1 + \mathcal{O}(\nu^{-2}) \Big), \\
h_0(\nu) &= \frac{1}{2} \mathrm{Erfc}\left( \frac{\nu}{\sqrt{2}} \right) = \frac{e^{-\nu^2/2}}{\sqrt{2\pi\nu^2}} \Big( 1 + \mathcal{O}(\nu^{-2}) \Big),
\end{aligned} \qquad (2.2.26)$$

we find that Eq. (2.2.24) reduces to

$$\begin{aligned}
P_{\mathrm{NG}}(\delta > \delta_c) &= \frac{e^{-\nu^2/2}}{\sqrt{2\pi\nu^2}} \exp\left\{ \sum_{n=3}^\infty \frac{(-1)^n}{n!} w^{(n)}(0)\, \nu^n \right\} \\
&= \frac{1}{\sqrt{2\pi\nu^2}} \exp\left\{ -\nu^2/2 + \sum_{n=3}^\infty \frac{(-1)^n}{n!} \xi^{(n)}(0) \, \left( \frac{\delta_c}{\sigma_\delta^2} \right)^n \right\},
\end{aligned} \qquad (2.2.27)$$

where the argument (0) in the connected correlation functions means again that all the points have to be taken coinciding with each other. This quantity directly gives the primordial mass fraction $\beta^{\mathrm{th}}$ and is derived with the only approximation of large $\nu$. One can identify the leading order term which gives the gaussian result when all the higher order ($n > 2$) connected correlators vanish.

**Impact of non-Gaussianity**

We can use Eq. (2.2.27) to estimate the impact of non-Gaussianity. Indeed, it is important to estimate if, in a given model, higher order correlators encoding the non-Gaussian corrections are impacting the computation of the abundance in a significant way. To do so, we can define dimensionless quantities, the cumulants, as

$$S_n = \frac{\xi^n(0)}{(\xi^2(0))^{n-1}} = \frac{\overbrace{\langle \delta(\vec{x}) \cdots \delta(\vec{x}) \rangle}^{n-\text{times}}_c}{\sigma_\delta^{2(n-1)}}, \tag{2.2.28}$$

where, with the subscript $c$, we denote the connected contribution to the n-point function. Following Ref. [24], we can define the parameter $\Delta_n$ which describes the logarithmic variation of the PBH abundance after introduction of the $n$-th cumulant as

$$\Delta_n = \frac{d \ln \beta(M)}{d \ln S_n}. \tag{2.2.29}$$

In other terms, the correction parameter $\Delta_n$ is relating the non-Gaussian PBH abundance to the Gaussian result by

$$\frac{\beta_{\text{NG}}^{\text{th}}(M)}{\beta_{\text{G}}^{\text{th}}(M)} = e^{\Delta_n}. \tag{2.2.30}$$

Using Eq. (2.2.30), one can conclude that the resulting PBH abundance is exponentially sensitive to deviation from Gaussianity unless one has $|\Delta_n| \lesssim 1$. Inserting Eqs. (2.2.7) and (2.2.27), we find

$$|\Delta_n| = \frac{1}{n!} \left( \frac{\delta_c}{\sigma_\delta} \right)^2 |S_n| \delta_c^{n-2} \lesssim 1. \tag{2.2.31}$$

This tells us that intrinsic non-Gaussianity of the overdensity field alters exponentially the gaussian prediction for the PBH abundance unless

$$|S_n| \lesssim \left( \frac{\sigma_\delta}{\delta_c} \right)^2 \frac{n!}{\delta_c^{n-2}}. \tag{2.2.32}$$

**Summary**

Let us summarise the results of this section. We devised a general formula in Eq. (2.2.27) giving the PBH abundance when the density contrast field is non-Gaussian. This formula is exact but, in practice, adopting it would require one of the following requirements to be valid:

- the whole set of correlation functions at any non-vanishing order is known and one can perform a re-summation of the series;

- the problem can be approached perturbatively and one can reliably account for the first few non-Gaussian corrections only. In this second case, however, one needs to check using condition (2.2.32), that the higher order terms neglected in the expansion are not giving a sizeable contribution.

### 2.2.3   The ineludible non-Gaussianity

In this section we are going to show that the density contrast is characterised by an intrinsic non-Gaussinity which is unavoidable. Let us start by going back to the relation between the density contrast and the comoving curvature perturbation in Eq. (2.1.3), which can be also written as

$$\delta(\vec{x}, t) = -\frac{4}{9a^2 H^2} e^{-2\zeta(\vec{x})} \left[ \nabla^2 \zeta(\vec{x}) + \frac{1}{2} \partial_i \zeta(\vec{x}) \partial^i \zeta(\vec{x}) \right]. \tag{2.2.33}$$

Going beyond the linear approximation, it is straightforward to notice that even though the comoving curvature perturbation may be assumed to be gaussian, as the density contrast $\delta$ is non-linearly related to $\zeta$, it inevitably inherits deviations from Gaussian statistics. As it was investigated in details in Ref. [19], by going up to 4-th order in the computation of the connected correlation functions of the density contrast beyond the linear approximation, the series (2.2.27) still shows no sign of convergence at low orders. This motivates us to devise a different technique to compute the PBH abundance accounting for the ineludible non-Gaussian nature of the density contrast. We will first do so by showing peaks in the density contrast can be identified with "spiky" peaks of the curvature perturbation and then compute the abundance using peak theory. In the second part, we will adopt threshold statistics to compute the probability that the non-gaussian density contrast is above the threshold.

**Spiky peaks in $\zeta$ versus peaks in $\delta$: an analytical treatment**

There exists a simple argument showing that one can identify the location of the spiky peaks of the curvature perturbation with the position of density peaks. We refer to "spiky" peaks as the ones narrow enough to have a large second derivative at their peaks. This argument, first presented in Ref. [296], goes as follows. We can expand $\zeta(\vec{x})$ around the location $\vec{x}_{\rm pk}$ of the peak of the overdensity $\delta(\vec{x}, t)$ as

$$\zeta(\vec{x}) = \zeta(\vec{x}_{\rm pk}) + \partial_i \zeta(\vec{x}_{\rm pk})(x^i - x^i_{\rm pk}) + \frac{1}{2} \partial_i \partial_j \zeta(\vec{x}_{\rm pk})(x^i - x^i_{\rm pk})(x^j - x^j_{\rm pk}). \tag{2.2.34}$$

We denote $\partial_i \zeta(\vec{x}_{\rm pk})$ as the gradient $\partial_i \zeta(\vec{x})$ evaluated at the peak location $\vec{x}_{\rm pk}$. Around this point we can also approximate

$$\delta(\vec{x}_{\rm pk}, t) \simeq -\frac{4}{9a^2 H^2} e^{-2\zeta(\vec{x}_{\rm pk})} \nabla^2 \zeta(\vec{x}_{\rm pk}), \tag{2.2.35}$$

where we neglected the second term in Eq. (2.2.33) since it only contributes at higher order in $\zeta$ with respect to (2.2.35) and, more importantly, is expected to be small around peaks of $\zeta$.

As the overdensity is required to be above the critical value $\delta^{\rm c}_{\rm pk}$, we find that the curvature of the peak in $\zeta$ should be larger than a given amount, which is

$$-\nabla^2 \zeta(\vec{x}_{\rm pk}) > \frac{9a^2 H^2}{4} e^{2\zeta(\vec{x}_{\rm pk})} \delta^{\rm c}_{\rm pk}. \tag{2.2.36}$$

This explains the requirement of spiky enough peaks in $\zeta$. As the peak in $\zeta$ is centred in $\vec{y}_{\rm pk}$, such that $\partial_i \zeta(\vec{y}_{\rm pk}) = 0$, one can also write

$$\partial_j \zeta(\vec{x}_{\rm pk}) + \partial_i \partial_j \zeta(\vec{x}_{\rm pk})(y^i_{\rm pk} - x^i_{\rm pk}) \simeq 0, \tag{2.2.37}$$

or

$$(y^i_{\rm pk} - x^i_{\rm pk}) \simeq -(\partial_i \partial_j \zeta(\vec{x}_{\rm pk}))^{-1} \partial_j \zeta(\vec{x}_{\rm pk}). \tag{2.2.38}$$

Aligning the principal axes of the constant-$\zeta$ ellipsoids, we find that the eigenvalues of the shear tensor $\zeta_{ij}$ can be written as $-\sigma_2 \lambda_i$, where $\sigma_2$ is the characteristic root-mean-square variance of the components of $\zeta_{ij}$, while $\sigma_1$ being the same quantity related to $\partial_i \zeta$. As routinely done in peak theory, one can also define

$$\lambda_i \simeq \frac{\gamma \nu}{3}, \quad \nu = \frac{\zeta(\vec{x}_{\rm pk})}{\sigma_0}, \quad \gamma = \frac{\sigma_1^2}{\sigma_0 \sigma_2} \tag{2.2.39}$$

where

$$\sigma_j^2 = \int \frac{{\rm d}^3 k}{(2\pi)^3} k^{2j} P_\zeta(k). \tag{2.2.40}$$

The key point is that at the horizon crossing time, one can expect the moments $\sigma_j$ to be much smaller than $(aH)^j$ as the amplitude of the power spectrum enters in the equation for the higher order variances (2.2.40). From Eq. (2.2.36), we deduce that

$$-\nabla^2 \zeta(\vec{x}_{\rm pk}) \sim \lambda_i \sigma_2 > \frac{9a^2 H^2}{4} e^{2\zeta(\vec{x}_{\rm pk})} \delta^{\rm c}_{\rm pk} \gg \sigma_2 \tag{2.2.41}$$

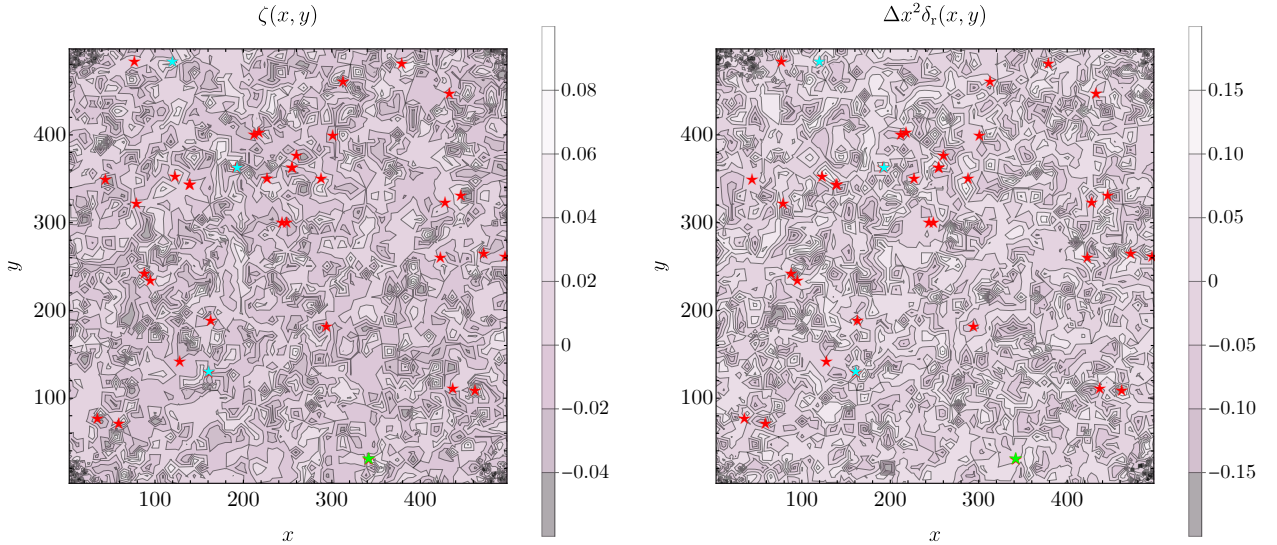

Figure 2.6: *A depiction of the two-dimensional simulation.* **Left:** *gaussian field $\zeta(x, y)$.* **Right:** *density contrast $\delta_{\mathrm{r}}(x, y)$ found from the left panel using the relation in Eq. (2.2.33). The stars indicate the location of the spiky peaks in $\zeta$ and high peaks in $\delta_{\mathrm{r}}$, showing the correspondence between their positions. The colour code identifying the peaks is the same one used in Fig. 2.7.*

and therefore $\lambda_i \sim \gamma\nu \gg 1$. Finally, one finds

$$|y_{\mathrm{pk}}^i - x_{\mathrm{pk}}^i| \simeq |\sigma_1/\sigma_2\lambda_i| \ll |\sigma_1/\sigma_2| \lesssim 1/aH, \qquad (2.2.42)$$

where in the last equality we have used the fact that $\sigma_1/\sigma_2 \simeq k_* r^{-1} \lesssim R_H$ as we evaluate the quantity at the time of PBH formation, which is the horizon crossing time $t_H$ when the relevant mode $k_* \sim 1/R_H$. This shows that large enough peaks of the density contrast field are found close to peaks of the curvature perturbation. In these considerations, we define "close" as being within the Hubble distance between each other. This happens, however, only if $\zeta$ is characterised by large second derivatives at the origin of the peak.

This simple approximate argument can be supported by a numerical simulation that we present in the following.

**Spiky peaks in $\zeta$ versus peaks in $\delta$: a numerical treatment**

We can denote a rescaled density contrast $\delta_{\mathrm{r}}$ defined as

$$\delta(\vec{x}, t) = \frac{4}{9}\frac{1}{a^2 H^2}\delta_{\mathrm{r}}(\vec{x}, t). \qquad (2.2.43)$$

We numerically simulated a random realisation of the gaussian curvature field $\zeta(\vec{x})$ in a $n$-dimensional box of dimensions $N$, discretised using a grid of $N^n$ points with a spacing $\Delta x = 1$ in all directions. We report the result of the analysis in a 2-dimensional space ($N = 2$) for simplicity. The results are consistent with the one found in the realistic case with $N = 3$. We assume a narrow power spectrum of curvature perturbations having the form

$$\mathcal{P}_\zeta(k) = 0.01 \exp\left[-\ln^2(k/k_\star)/0.02\right]. \qquad (2.2.44)$$

The power spectrum amplitude was chosen such that the variance is $\sigma_0^2 = 2.5 \cdot 10^{-3}$, while characteristic scale was fixed at $k_\star = 0.2/\Delta x$. A random realisation of $\zeta(\vec{x})$ and the corresponding density contrast $\delta_{\mathrm{r}}(\vec{x})$ are shown Fig. 5.11. The striking feature of the results shown is the correspondence between the locations of spiky peaks in the curvature perturbations and the density maxima.

In Fig. 2.7 we show a detailed analysis of the simulation. Each point of the plot represents a peak in $\zeta$ with the corresponding values of the rescaled amplitude $\nu = \zeta/\sigma_0$ and the curvature $x = -\nabla^2\zeta/\sigma_2$ shown in both axes. The coloured lines indicate various lower bounds on the curvature $x \gtrsim c \cdot x_\delta^c(\nu)$ (with $c = [1, 0.9, 0.8]$) in terms of the absolute maximum of the density contrast $\delta_{\max} \simeq 0.4$ to be found in the simulation, where

$$x_\delta^c(\nu) \equiv \frac{9a^2H^2}{4\sigma_2} e^{2\sigma_0\nu}\delta_{\max}. \tag{2.2.45}$$

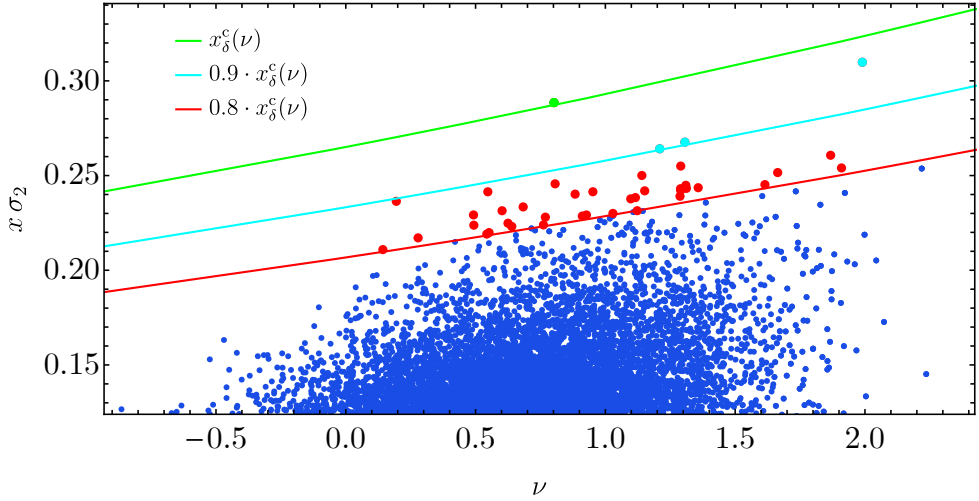

Figure 2.7: *A plot with field values of $\nu = \zeta/\sigma_0$ and $x = -\nabla^2\zeta/sigma_2$ in each position of the grid. All points are peaks in $\zeta$, but only those spiky enough correspond to peaks in $\delta$, as predicted.*

This bound, as we will see, corresponds to the condition to impose on the derivatives of $\zeta$ to find a peak with a local overdensity above the threshold. With red, cyan and green dots we show the points which, at the same location, present peaks in $\delta$ and the condition (2.2.45) is satisfied. Blue dots, on the other hand, are peaks in $\zeta$ not satisfying Eq. (2.2.45). This shows the correspondence between peaks of $\zeta$ and peaks of $\delta$, provided the condition (2.2.45) is met. We generally expect this conclusion to be even stronger when rarer events are simulated. Results in three dimensions, which we do not report here for simplicity, also confirm the present analysis.

#### The non-Gaussian beta from peak theory

Basing our approach on the previous arguments, we can identify the number of peaks of the overdensity field with peaks in the curvature perturbation which are spiky enough, see Eq. (2.2.41). Therefore, we can apply peak theory to the gaussian random field $\zeta$ and expand the density contrast in Eq. (2.2.33) around the peak location $\vec{x}_{\text{pk}}$ of $\zeta$ (such that $\partial_i\zeta(\vec{x}_{\text{pk}}) = 0$) to get

$$\delta(\vec{x}_{\text{pk}}, t) = -\frac{4}{9a^2H^2} e^{-2\zeta(\vec{x}_{\text{pk}})}\nabla^2\zeta(\vec{x}_{\text{pk}}) = \frac{4}{9a^2H^2} e^{-2\sigma_0\nu}x\sigma_2. \tag{2.2.46}$$

We recall that the number density of peaks in $\zeta$ is (see App. A)

$$\mathcal{N}_{\text{pk}}(\nu, x)\mathrm{d}\nu\mathrm{d}x = \frac{e^{-\nu^2/2}}{(2\pi)^2R_*^3} f(x) \frac{\exp[-(x - x_*)^2/2(1 - \gamma^2)]}{[2\pi(1 - \gamma^2)]^{1/2}} \mathrm{d}\nu\mathrm{d}x, \tag{2.2.47}$$

where

$$R_* = \sqrt{3}\frac{\sigma_1}{\sigma_2}, \qquad \gamma = \frac{\sigma_1^2}{\sigma_0\sigma_2}, \qquad x_* = \gamma\nu, \tag{2.2.48}$$

and the function $f(x)$ is given by

$$f(x) = \frac{(x^3 - 3x)}{2}\left[\text{erf}\left(x\sqrt{\frac{5}{2}}\right) + \text{erf}\left(\frac{x}{2}\sqrt{\frac{5}{2}}\right)\right] + \sqrt{\frac{2}{5\pi}}\left[\left(\frac{31x^2}{4} + \frac{8}{5}\right)e^{-\frac{5x^2}{8}} + \left(\frac{x^2}{2} - \frac{8}{5}\right)e^{-\frac{5x^2}{2}}\right]. \tag{2.2.49}$$

Finally, the number density of non-Gaussian peaks of the overdensity above a given threshold $\delta_c^{\rm pk}$ is found by integrating [19]

$$\beta_{\rm NG}^{\rm pk} \simeq 4\pi R_H^3 \int_{-\infty}^{\infty} \mathrm{d}\nu \int_{x_\delta^c(\nu)}^{\infty} \mathrm{d}x\, \frac{e^{-\nu^2/2}}{(2\pi)^2 R_*^3} f(x) \frac{\exp[-(x-x_*)^2/2(1-\gamma^2)]}{[2\pi(1-\gamma^2)]^{1/2}}, \tag{2.2.50}$$

where the lower integration limit on the variable $x$ is set by the condition

$$x_\delta^c(\nu) \simeq \frac{9a^2 H^2}{4\sigma_2} e^{2\sigma_0 \nu} \delta_c^{\rm pk}, \tag{2.2.51}$$

which encodes the requirement that the resulting density contrast peak should be above the threshold $\delta_c^{\rm pk}$. It is also interesting to notice that if we neglected non-linearities, which would mean dropping the exponential in Eq. (2.2.51), we would have automatically recovered the gaussian result in Eq. (2.2.8).

**The non-Gaussian beta from threshold statistics**

We can also present an alternative way to compute the non-Gaussian PBH abundance, which is based on threshold statistics and does not require the identification of peaks of the curvature perturbation with peaks of the density contrast. We start again by considering $\zeta$ as a gaussian random field and, following the notation of the Appendix A of Ref. [297], we define

$$\zeta_i = \partial_i \zeta, \quad \zeta_{ij} = \partial_i \partial_j \zeta. \tag{2.2.52}$$

The cross-correlations of these fields are

$$\langle \zeta\zeta \rangle = \sigma_0^2, \qquad \langle \zeta\zeta_{ij} \rangle = -\frac{\sigma_1^2}{3}\delta_{ij}, \qquad \langle \zeta\zeta_i \rangle = 0,$$

$$\langle \zeta_i\zeta_i \rangle = \frac{\sigma_1^2}{3}\delta_{ij}, \qquad \langle \zeta_i\zeta_{jk} \rangle = 0, \qquad \langle \zeta_{ij}\zeta_{kl} \rangle = \frac{\sigma_2^2}{15}(\delta_{ij}\delta_{kl} + \delta_{ik}\delta_{jl} + \delta_{il}\delta_{jk}). \tag{2.2.53}$$

The variance and higher momenta of the smoothed $\zeta$ distribution can be computed numerically using the Fourier transform of the top-hat window function in real space in Eq. (2.2.4), that is

$$\sigma_j^2 = \int \frac{\mathrm{d}^3 k}{(2\pi)^3} k^{2j} W^2(k, R_H) P_\zeta(k). \tag{2.2.54}$$

The matrix of second derivative $-\zeta_{ij}$ can be diagonalized and written in terms of the eigenvalues $\sigma_2\lambda_i$, ordered such that $\lambda_1 \geq \lambda_2 \geq \lambda_3$. Therefore, we can change variables and define

$$x = -\frac{\nabla^2 \zeta}{\sigma_2} = \lambda_1 + \lambda_2 + \lambda_3, \quad y = \frac{\lambda_1 - \lambda_3}{2}, \quad z = \frac{\lambda_1 - 2\lambda_2 + \lambda_3}{2}, \tag{2.2.55}$$

introducing again the rescaled curvature height $\nu = \zeta(\vec{x})/\sigma_0$ and the non-vanishing correlators become

$$\langle \nu^2 \rangle = 1, \quad \langle x^2 \rangle = 1, \quad \langle x\nu \rangle = \gamma, \quad \langle y^2 \rangle = 1/15, \quad \langle z^2 \rangle = 1/5. \tag{2.2.56}$$

The joint Gaussian probability distribution for these variables is provided by the expression (we define $\eta_i \equiv \zeta_i$)

$$P(\nu, \vec{\eta}, x, y, z)\mathrm{d}\nu \mathrm{d}^3\eta \mathrm{d}x\mathrm{d}y\mathrm{d}z = N|2y(y^2 - z^2)|e^{-Q/2}\mathrm{d}\nu \mathrm{d}x\mathrm{d}y\mathrm{d}z\frac{\mathrm{d}^3\eta}{\sigma_0^3}, \tag{2.2.57}$$

as a function of

$$Q = \nu^2 + \frac{(x - x_*)^2}{(1 - \gamma^2)} + 15y^2 + 5z^2 + \frac{3\vec{\eta} \cdot \vec{\eta}}{\sigma_1^2} \tag{2.2.58}$$

and

$$x_* = \gamma\nu, \quad \gamma = \frac{\sigma_1^2}{\sigma_0\sigma_2}, \quad N = \frac{(15)^{5/2}}{32\pi^3} \frac{6\sigma_0^3}{\sigma_1^3(1 - \gamma^2)^{1/2}}. \tag{2.2.59}$$

We can marginalise over the variables $y \geq 0$ and $z \in [-y, y]$. The resulting probability distribution is therefore given by

$$P(\nu, \vec{\eta}, x)\mathrm{d}\nu\mathrm{d}^3\eta\mathrm{d}x = \frac{6\sqrt{3}}{8\pi^{5/2}\sqrt{2(1-\gamma^2)}\sigma_1^3}e^{-\frac{1}{2}\left(\nu^2 + \frac{(x-x_*)^2}{(1-\gamma^2)} + \frac{3\vec{\eta}\cdot\vec{\eta}}{\sigma_1^2}\right)}\mathrm{d}\nu\mathrm{d}^3\eta\mathrm{d}x. \qquad (2.2.60)$$

We stress that Eq. (2.2.60) completely describes the statistical properties of all the relevant quantities needed to find the density contrast using Eq. (2.2.33). Indeed, we can then write $\delta$ as a function of these variables as

$$\delta(\vec{x}, t) = \frac{4}{9a^2H^2}e^{-2\nu\sigma_0}\left[x\sigma_2 - \frac{1}{2}\vec{\eta}\cdot\vec{\eta}\right]. \qquad (2.2.61)$$

We can pause here and summarise the procedure we are adopting in this section. The essence of the computation of the abundance using threshold statistics resides in integrating the probability distribution function above the value of the threshold. Since the density contrast is non-linearly related to the curvature perturbation, we were forced to construct a non-Gaussian distribution describing the (correlated) behaviour of the quantities entering in Eq. (2.2.61), which are $\nu, \vec{\eta}, x$, in terms of the properties of the curvature field $\zeta$. The final step is therefore simply to integrate probability distribution in the domain where $\delta \geq \delta_c$.

For convenience let us perform the change of variables:

$$x_\delta = x, \quad \vec{\eta}_\delta = \vec{\eta}, \quad \nu = \frac{1}{2\sigma_0}\ln\left[\frac{4\left(x_\delta\sigma_2 - \frac{1}{2}\vec{\eta}_\delta\cdot\vec{\eta}_\delta\right)}{9a^2H^2\delta}\right], \qquad (2.2.62)$$

valid for $x_\delta > \vec{\eta}_\delta\cdot\vec{\eta}_\delta/2\sigma_2$, while the Jacobian of the transformation is given by $J = |1/2\delta\sigma_0|$. The condition requiring the argument of the logarithm to be positive coincides with the condition of having positive overdensities. As we are focusing on the collapse of PBHs, underdense regions where $\delta < 0$ are not relevant. Then, the distribution in terms of the new variables is given by

$$P(\delta, \vec{\eta}_\delta, x_\delta)\mathrm{d}\delta\mathrm{d}^3\eta_\delta\mathrm{d}x_\delta = \frac{6\sqrt{3}}{8\pi^2\sqrt{2\pi(1-\gamma^2)}\sigma_1^3}\left|\frac{1}{2\delta\sigma_0}\right|e^{-Q_3/2}\Theta(x_\delta\sigma_2 - \eta_\delta^2/2)\mathrm{d}\delta\mathrm{d}^3\eta_\delta\mathrm{d}x_\delta \qquad (2.2.63)$$

where

$$Q_3 = \nu^2(\delta, \eta_\delta, x) + \frac{1}{(1-\gamma^2)}\left[x_\delta - \gamma\nu(\delta, \eta_\delta, x)\right]^2 + \frac{3\vec{\eta}_\delta\cdot\vec{\eta}_\delta}{\sigma_1^2}. \qquad (2.2.64)$$

Finally, marginalisying over the directions of the vector $\vec{\eta}_\delta$, we find

$$\beta_{\mathrm{NG}}^{\mathrm{th}} = 4\pi\int_{\delta_c}\mathrm{d}\delta\int_0^\infty\mathrm{d}\eta_\delta\,\eta_\delta^2\int_{\eta_\delta^2/2\sigma_2}^\infty\mathrm{d}x_\delta\,\frac{6\sqrt{3}}{8\pi^2\sqrt{2\pi(1-\gamma^2)}\sigma_1^3}e^{-Q_3/2}. \qquad (2.2.65)$$

Let us stress again here that this is an exact result and which is not affected by any approximations. We also notice that in case one adopted the linear relation between $\delta$ and $\zeta$ as in Eq. (2.2.1), one would recover the Press-Schechter result in Eq. (2.2.5).

In the following we will present some examples of the computation of the abundance for various power spectra:

- **Monochromatic power spectrum:** this form of the power spectrum is found in Eq. (2.1.29). In this case, one finds $\gamma \simeq 1$ and the integral in Eq. (2.2.65) can be performed analytically to get the final abundance as

$$\beta_{\mathrm{NG}}^{\mathrm{th}} \simeq 6\sqrt{\frac{3}{2\pi}}\frac{(1 - 2\sigma_0\,x_c)^{3/2}}{2x_c\,(3 - 5\,\sigma_0\,x_c)^{3/2}}\,\mathrm{e}^{-\frac{x_c^2}{2}}, \qquad (2.2.66)$$

where

$$x_c = -\frac{1}{2\sigma_0}\,W_0\left(-\frac{9a^2H^2\sigma_0\delta_c}{2\sigma_2}\right) \qquad (2.2.67)$$

and $W_0$ is the principal branch of the Lambert function.

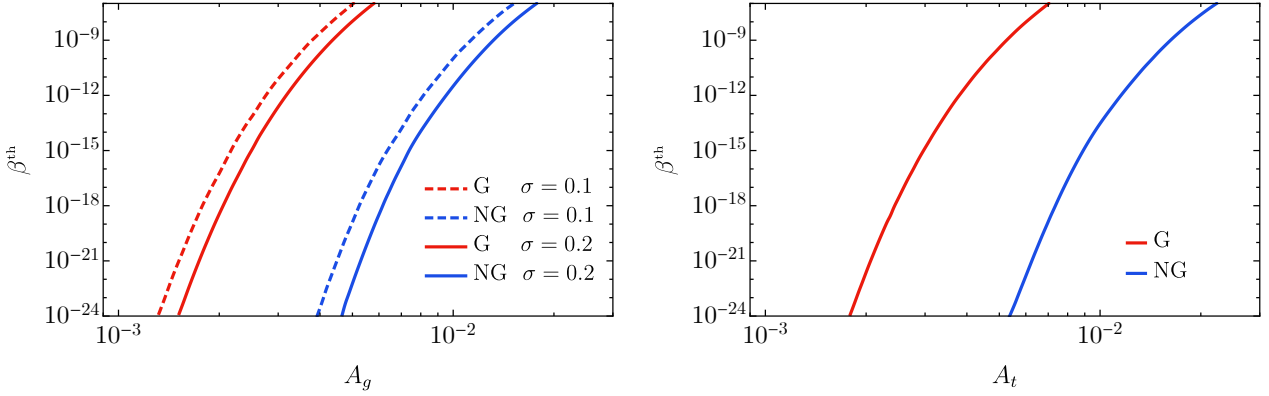

Figure 2.8: $\beta^{\rm th}$ as a function of the amplitude of the power spectrum. For comparison, we show both the Gaussian (red) and non-Gaussian (blue) results. To simplify the comparison between the two methodologies, we plot the result fixing the threshold at $\delta_c = 0.51$. In reality, when different power spectra are considered, the correct threshold should be adopted, see Sec. 2.1. **Left:** Lognormal shape with $A_g \equiv \mathcal{P}_0 \sigma \sqrt{2\pi}$. **Right:** Broad shape with $A_t \equiv \mathcal{P}_0$.

- **Lognormal power spectrum:** defined as in Eq. (2.1.32) with an amplitude $\mathcal{P}_0 = A_g/\sqrt{2\pi\sigma^2}$. This scenario converges towards the monochromatic case for $\sigma \to 0$. The abundance must be computed by integrating Eq. (2.2.65). We show the abundance, compared to the gaussian result which neglected the non-linear relation between the density contrast and the curvature, in Fig. 2.8. As one can notice, non-linearities suppress the abundance by many orders of magnitude. To obtain the same abundance, the amplitude of the power spectrum should be corrected by a factor of $\mathcal{O}(2 \div 3)$, depending on the shape of $\mathcal{P}_\zeta$.

- **Broad power spectrum:** defined in Eq. (2.1.30). Also in this case, the integration must be performed numerically. The resulting abundance is also shown in Fig. 2.8 and shares the same behaviour observed for the lognormal shape. We stress also here that, when computing the variance of the curvature perturbation, we disregarded unphysical long wavelength modes and we set $aH = 1/R_H$ at the horizon crossing time $t_H$ as the infrared cut-off. We will come back to this scenario in the next section, where the so-called "cloud-in-cloud" problem (analogous to the one encountered in the dark matter large scale structure formation theory) will be clarified.

### A note on the inclusion of local non-gaussianity

The threshold statistics approach introduced in the previous section can also be used when both a local non-gaussianity of the curvature perturbation and the non-linear definition of the density contrast are included. This was recently developed further in Ref. [309]. Let us assume the curvature perturbation to be expressed in terms of a local non-gaussianity of the form

$$\zeta = \zeta_g + f_{\rm NL} \left( \zeta_g^2 - \langle \zeta_g^2 \rangle \right). \tag{2.2.68}$$

The gaussian field $\zeta_g$ is completely characterised by the correlators in Eq. (2.2.53). As a consequence, the distribution of the fields $(\nu, \vec{\eta}, x)$, built out of the gaussian field $\zeta_g$, is dictated by Eq. (2.2.60). Therefore, one can define the density contrast by accounting for both its non-linear relation with $\zeta$ and the presence of $f_{\rm NL}$ corrections by combining Eqs. (2.2.33) and (2.2.68). Finally, following the same logic used in the previous section, one finds

$$\beta_{\rm NG}^{\rm th}(f_{\rm NL}) = \gamma \int_{\delta_c} d\delta \int d^3\vec{\eta} \int_{-\infty}^{\infty} d\nu \, J \, \frac{6\sqrt{3}}{8\pi^{5/2}\sqrt{2(1-\gamma^2)}\sigma_1^3} e^{-\frac{1}{2}\left(\nu^2 + \frac{(x_\delta - x_*)^2}{(1-\gamma^2)} + \frac{3\vec{\eta}\cdot\vec{\eta}}{\sigma_1^2}\right)}, \tag{2.2.69}$$

where in this case, as in Ref. [309], we defined

$$x_\delta\left(\delta,\nu,\vec{\eta}\right)=\frac{9(aH)^2}{4\sigma_2}\left[\frac{e^{2\sigma_0\left[\nu+(\nu^2-1)\sigma_0 f_{\text{NL}}\right]}\delta+\left(2+8f_{\text{NL}}+8\nu f_{\text{NL}}\sigma_0+8\nu^2 f_{\text{NL}}^2\sigma_0^2\right)\vec{\eta}\cdot\vec{\eta}}{1+2\nu\,f_{\text{NL}}\,\sigma_0}\right],$$

$$J=\left|\frac{9(aH)^2}{4\sigma_2}\frac{e^{2\sigma_0\left[\nu+(\nu^2-1)\sigma_0 f_{\text{NL}}\right]}}{1+2\nu\,f_{\text{NL}}\,\sigma_0}\right|. \tag{2.2.70}$$

Notice that, when local non-gaussianity is introduced into the picture, also the threshold for PBH formation has to be modified, see Ref. [301] for a detailed investigation of this phenomenon.

**Summary**

We conclude this section by summarising the results we presented. We showed the non-linearities relating the overdensities to the curvature perturbations inevitably introduce non-Gaussian effects that should be carefully treated. As the general formula in Sec. 2.2.2 would require knowing the cumulants of the density contrast up to high orders, we devised new (non-perturbative) methods to compute the PBH abundance. In the first part, after having shown that spiky enough peaks in $\zeta$ correspond to high peaks in $\delta$, we applied peak theory on $\zeta$ to compute the spiky peaks which will collapse, resulting in Eq. (2.2.50). In a second part, we resorted to threshold statistics and wrote a fully non-gaussian distribution for the density contrast, which can be integrated to find the PBH abundance as in Eq. (2.2.65). Both formulations are generally consistent with each other [19] and show that, when non-linearities are included, the PBH abundance is reduced by many orders of magnitude, corresponding to a change of the power spectrum height of a factor $\mathcal{O}(2\div 3)$, depending on the shape. Finally, using the threshold statistics approach, we also presented a formulation of the mass fraction able to capture both the non-linearities and intrinsic local non-Gaussianity of the curvature perturbation.

### 2.2.4   Abundance from broad spectra

When addressing PBH formation from a broad spectrum of curvature perturbations there may be additional subtleties that need to be carefully considered. The basic difference with respect to the narrow spectrum is that one expects many perturbations superimposed on top of each other with varying scales in the allowed range $r\in[1/k_{\text{max}},1/k_{\text{min}}]$. Therefore, there may be peaks with different characteristic sizes $r_m$. One important property to remember though is that the collapse of PBH takes place rapidly around the time of horizon crossing, which is when $r_m\sim R_H$. Therefore, peaks at different sizes collapse at different times.

This poses the issue which goes under the name of the "cloud-in-cloud" problem in the context of halo formation. In the PBH scenario, it corresponds to the question of whether small PBHs are efficiently swallowed by bigger PBHs, thus modifying the overall PBH abundance and mass distribution. In this section, following Ref. [18] and using the excursion set method, we will show that the cloud-in-cloud phenomenon is basically absent for PBHs. The physical motivation behind our findings is the smallness of the probability of forming a PBH (bounded by the requirement of them being at most as abundant as the dark matter in our universe), i.e. PBHs are very rare events. This, in practice, prevents pre-existing PBHs to be swallowed by bigger ones.

**The excursion set method applied to PBH formation**

As already presented in the preceding section, the abundance of PBHs can be computed using the Press-Schechter theory. The same procedure is adopted when computing the mass function of dark matter halos in the context of the large scale structure of the universe. An extension of this formalism, going under the name of excursion set theory [325], corrects the original computation made by Press and Schechter which may miscount the number of virialized dark matter objects. In the dark matter spherical collapse model, one expects a region of radius $R$ and mass $M$, with a smoothed density contrast $\delta(R)$, to collapse and virialize if $\delta(R)$ exceeds a critical value ($\approx 1.68$, see for example

Ref. [326]). However, the same region may be also part of a zone above the threshold for collapse on a scale $R' > R$. Therefore one would expect the smaller volume on scales $R$ to be part of a virialized object of mass $M' > M$.

In the excursion set method, the density perturbations are evolved stochastically with respect to the smoothing scale, and the computation of the probability of collapse is translated into a first-time passage through a barrier problem. The miscounting of collapsed objects is then avoided by only keeping track of trajectories in the smoothing scale passing the threshold for the first time (i.e. only the largest $R$ is counted).

In the following we review the excursion set computation of the abundance and apply it to the PBH scenario. We start by considering the density contrast $\delta(\vec{x})$, smoothed on a particular scale $R$, as

$$\delta(\vec{x}, R) = \int \mathrm{d}^3 x' \, W(|\vec{x} - \vec{x}'|, R) \, \delta(\vec{x}'), \tag{2.2.71}$$

or, analogously,

$$\delta(\vec{k}, R) = \widetilde{W}(\vec{k}, R) \delta(\vec{k}) \tag{2.2.72}$$

where $W(|\vec{x} - \vec{x}'|, R)$ is the window function. As we will show below, an optimal choice for the present discussion is the choice of sharp filter in $k$-space, defined as

$$\widetilde{W}_{\mathrm{sharp}-k}(k, k_f) = \Theta(k_f - k), \tag{2.2.73}$$

with $k_f = 1/R$, $k = |\vec{k}|$. This choice will drastically simplify the computations in the excursion set theory. Notice that the density field in Eq. (2.2.71) is independent of time, and we define its variance as $S$ (not to be confused with $\sigma^2$ which is time-dependent). The formation history of collapsed objects will be tracked by promoting the collapse threshold to be time-dependent and keeping the variance fixed in time. This is possible as the probability of collapse is dependent on the ratio $\delta_c/\sigma$, also called $\nu$ in the preceding sections.

One can evolve the density contrast $\delta(\vec{x}, R)$ in $R$ in a fixed position $\vec{x}$ (set in the origin $\vec{x} = 0$ without loss of generality) with

$$\frac{\partial \delta(\vec{x}, R)}{\partial R} = \Upsilon(R), \qquad \text{with} \qquad \Upsilon(R) \equiv \int \frac{\mathrm{d}^3 k}{(2\pi)^3} \, \frac{\partial \widetilde{W}(k, R)}{\partial R} \delta(\vec{k}). \tag{2.2.74}$$

As the density contrast modes $\delta(\vec{k})$ are stochastic variables, $\Upsilon(R)$ inherits their stochastic properties. Therefore, Eq. (2.2.74) can be interpreted as a Langevin equation, with $R$ playing the role of time, and $\Upsilon(R)$ playing the role of noise. The noise properties are dictated by the two point function (we are going to assume gaussian statistics throughout this section) as

$$\langle \Upsilon(R_1) \Upsilon(R_2) \rangle = \int_{-\infty}^{\infty} \mathrm{d}\ln k \, \mathcal{P}_\delta(k) \frac{\partial \widetilde{W}(k, R_1)}{\partial R_1} \frac{\partial \widetilde{W}(k, R_2)}{\partial R_2}. \tag{2.2.75}$$

Here we see why assuming a sharp filter in momentum space brings significant simplifications in the formalism. In fact, for a generic filter, the right-hand side becomes a function $R_1$ and $R_2$ differing from a Dirac-delta $\delta_D(R_1 - R_2)$. In this setup, Eq. (2.2.74) becomes a Langevin equation with Dirac-delta noise. Then, by changing to the variable $k_f = 1/R$ and defining $Q(k_f) = -(1/k_f)\Upsilon(k_f)$, one gets

$$\frac{\partial \delta(k_f)}{\partial \ln k_f} = Q(k_f), \tag{2.2.76}$$

with

$$\langle Q(k_{f_1}) Q(k_{f_2}) \rangle = \mathcal{P}_\delta(k_{f_1}) \delta_D(\ln k_{f_1} - \ln k_{f_2}). \tag{2.2.77}$$

As a final step, we change the "pseudo-time" variable to the (time-independent) variance $S$ smoothed with a sharp filter in $k$-space as

$$S(k_f) = \int_{-\infty}^{\ln k_f} \mathrm{d}\ln k \, \mathcal{P}_\delta(k), \tag{2.2.78}$$

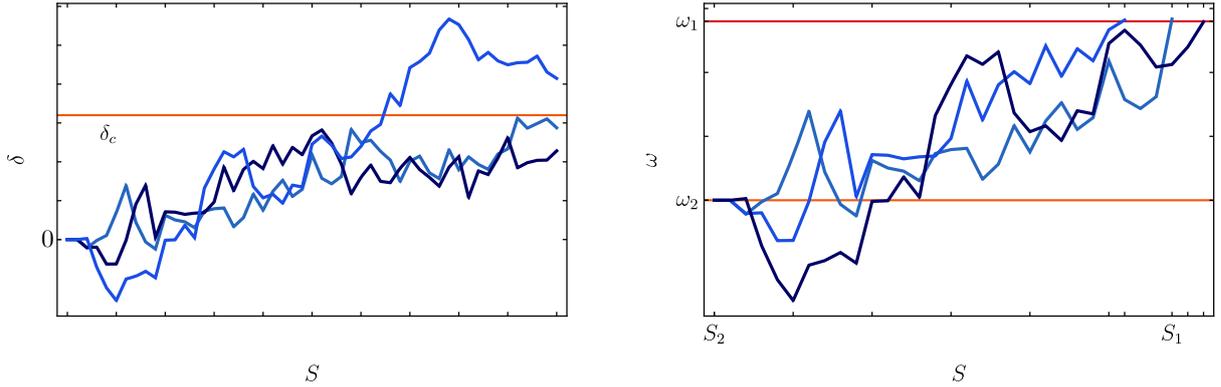

Figure 2.9: *A schematic picture of the random walk performed by the smoothed overdensity for few random realisations.* **Left:** *First passage problem with a barrier.* **Right:** *Two-barrier problem.*

and redefine $\eta(k_F) = Q(k_F)/\Delta^2(k_F)$ to get

$$\frac{\partial \delta(S)}{\partial S} = \eta(S), \tag{2.2.79}$$

with

$$\langle \eta(S_1)\eta(S_2)\rangle = \delta_D(S_1 - S_2). \tag{2.2.80}$$

We recognise in Eq. (2.2.79) a Langevin equation with Gaussian white noise.

The description of this problem is analogous to the one adopted for the phenomenon of the Brownian random walk. In practice, the field $\delta(S)$ performs a stochastic motion with respect to the time variable $S$.[2] Following Ref. [325], we refer to the evolution of $\delta$ as a function of $S$ as a "trajectory" (see a schematic representation of this phenomenon in figure 2.9).

We are now ready to study the solution of the Langevin equation and draw the consequences for the PBH scenario. The solutions of a Langevin equation with white noise can be described by a distribution $P$ obeying the Fokker-Planck (FP) equation. Therefore, $\delta(S)$ is distributed as described by the equation

$$\frac{\partial P}{\partial S} = \frac{1}{2}\frac{\partial^2 P}{\partial \delta^2}. \tag{2.2.81}$$

The Press-Schecter original computation is recovered by solving this equation with boundary conditions requiring the distribution to vanish at $\delta \to \pm\infty$ and accounting for all possible trajectories exceeding the threshold $\delta_c$. This approach makes explicit the so-called "cloud-in-cloud" problem. Indeed, this computation would count as valid all trajectories that make multiple crossings of the threshold at different scales $R$. This issue is resolved in excursion set theory by keeping track of the *first* up-crossing of the trajectories only. This problem is analogous to the study of the Brownian motion with the inclusion of an absorbing barrier (i.e. all the trajectories crossing the barrier for the first time are removed from the subsequent evolution). In practice, this is equivalent to calculating the lowest value of $S$ (or, equivalently, the highest value of $R$) for which the trajectory crosses the threshold. This trajectory will only correspond to the largest virialized object to be formed in that physical location $\vec{x}$. Therefore, one has to require the boundary condition

$$P(\delta, S)\Big|_{\delta=\delta_c} = 0, \tag{2.2.82}$$

---

[2]We could have considered a window function different from a step function in momentum space, (as, for example, the step function in real space adopted in (2.2.4)). In this case, the random walk of the smoothed overdensity would have become non-markovian and retained memory effects between different steps (the opposite happens in the white-noise case where the evolution is markovian and no memory effects are present). Such a more complex setup was addressed in Refs. [327–329] using the path-integral methods. For sake of simplicity, in this paper, we restrict to the markovian case originally studied in Ref. [325].

and the FP equation is solved by

$$P(\delta, S) = \frac{1}{\sqrt{2\pi S}} \left[ e^{-\delta^2/(2S)} - e^{-(2\delta_c - \delta)^2/(2S)} \right].$$ (2.2.83)

Finally, we can convert this result into a fraction of trajectories that have crossed the threshold at "time" smaller or equal to $S$ as

$$P(\delta > \delta_c) = 1 - \int_{-\infty}^{\delta_c} d\delta\, P(\delta, S) = \text{Erfc}\left( \frac{\delta_c}{\sqrt{2S}} \right),$$ (2.2.84)

which is twice the Press-Schecter original prediction.

Excursion set theory is also a powerful tool to study the formation *history* of virialized objects. In the context of the large scale structure of dark matter halos, this was performed in the seminal paper in Ref. [325] and further addressed in Ref. [330]. Following the same logic, but applied to the PBH scenario, we reabsorb the time-dependance into the threshold for collapse and define a time-dependant threshold $\omega(a) \equiv \delta_c/D(a)$. In this case $D(a)$ is the linear growth factor resulting in a threshold scaling like $\omega(a) \equiv \delta_c/a^2$. Therefore, the barriers at two different times are related to one another as

$$\omega(a_1) = \left( \frac{a_2}{a_1} \right)^2 \omega(a_2).$$ (2.2.85)

In other words, the barrier is reduced as time passes, while the variance is constant (intuitively, this is the consequence of the density contrast growing as $a^2$ on super-Hubble scales, while the threshold is fixed at $\delta_c \approx 0.5$). Therefore, the time evolution of the barrier in the PBH scenario is related to the evolution of the comoving Hubble length.

Let us consider the "two-barrier" problem [330], which can be used to find the conditional probability that a trajectory with a first up-crossing of the barrier $\omega_1$ at some $S_1$ will have a secondary upcrossing of $\omega_2$ between $S_2$ and $S_2 + dS_2$ with $S_1 \gg S_2$ and $\omega_1 > \omega_2$ (see again figure 2.9 for a schematic picture). This conditional probability is given by

$$P(S_2, \omega_1 | S_1, \omega_1) dS_2 = \frac{P(S_1, \omega_1 | S_2, \omega_2)}{P(S_1, \omega_1)} dS_2$$

$$= \frac{1}{\sqrt{2\pi}} \left[ \frac{S_1}{S_2(S_1 - S_2)} \right]^{3/2} \frac{\omega_2(\omega_1 - \omega_2)}{\omega_1} \exp\left[ -\frac{(\omega_2 S_1 - \omega_1 S_2)^2}{2S_1 S_2(S_1 - S_2)} \right] dS_2.$$ (2.2.86)

Recall that, in this setup, all the trajectories provide the variation of $\delta$ at a fixed time. However, the formation of PBHs of various masses takes place when the corresponding scale $R$ re-enter the horizon. Therefore, one should find the trajectories at a given fixed time and then match $\delta$ to the corresponding time when the given scale of interest crosses the horizon [331]. In terms of the PBH masses, which is related by a factor $\mathcal{O}(1)$ to the horizon mass $M_H$ associated with a given scale when the latter enters the horizon by

$$R \sim 1/aH \sim a \sim M_H^{1/2}$$ (2.2.87)

in the radiation-dominated case, the time independent variance scales as

$$S(R_1) = S(R_2) \left( \frac{M_{H_1}}{M_{H_2}} \right)^{-(4+n_p)/2}.$$ (2.2.88)

In the previous equation, we defined $n_p$ as the tilt of the broad power spectrum of curvature perturbations. Then, Eq. (2.2.86) estimates the conditional probability that a PBH of mass $M_1$ formed at the time $t_1$ has merged into a PBH of mass between $M_2$ and $M_2 + dM_2$ at time $t_2 > t_1$.

**Consequences for the PBH abundance at various masses**

The quantity described in Eq. (2.2.88) can be used to seize the impact of PBH formation at different scales (time) on the overall abundance. Let us consider various explicit examples. As this issue is only posed when the power spectrum is broad, we consider

$$\mathcal{P}_\zeta \approx \mathcal{P}_0 (k/k_s)^{n_p} \Theta\left( k_s - k \right) \Theta\left( k - k_l \right),$$ (2.2.89)

for various values of the spectral index $n_p$:

- **Broad spectra with a blue tilt:** for which $n_p > 0$. In this case we will find that it is always the smallest scale (the largest momentum) that is the most relevant as far as the dominant portion of PBH population is concerned. Let us recall that fluctuations on small and large scales collapse to form PBHs at the time when the scale factors are $a_s$ and $a_l$ correspondingly. The variances (at fixed time) as a function of the scale are hierarchical, meaning that

$$S_s \simeq \frac{16}{81(n_p+4)} \mathcal{P}_0 k_s^4,$$
$$S_l \simeq \frac{16}{81(n_p+4)} \mathcal{P}_0 \left(\frac{k_l}{k_s}\right)^{n_p} k_l^4 \ll S_s. \qquad (2.2.90)$$

From this point of view, the hierarchy of scales $k_s \gg k_l$ with a blue spectrum translates into a hierarchy for the variances and threshold

$$\omega(a_s) \gg \omega(a_l) \quad \text{and} \quad S_s \gg S_l, \qquad (2.2.91)$$

such that

$$\sqrt{\frac{S_s}{S_l}} = \left(\frac{k_s}{k_l}\right)^{(4+n_p)/2} \gg \frac{\omega(a_s)}{\omega(a_l)} = \left(\frac{k_s}{k_l}\right)^2. \qquad (2.2.92)$$

Therefore, small PBHs are generated with (exponentially) larger abundances than the big ones, as, parametrically, the relative abundances are

$$\beta(M(r_{m,l})) \sim e^{-\omega^2(a_l)/(2S_l)}$$
$$\beta(M(r_{m,s})) \sim e^{-\omega^2(a_s)/(2S_s)} \qquad (2.2.93)$$

and hence,

$$\beta(M(r_{m,l})) \ll \beta(M(r_{m,s})). \qquad (2.2.94)$$

Notice that, in this section, we refer to the characteristic scale of the density perturbations $r_{m,l}$ ($r_{m,s}$) as corresponding to the smallest (largest) mode in the spectrum (2.2.89), namely $k_l$ ($k_s$) respectively. We can consider this limit in Eq. (2.2.86) to find

$$P(S_l, \omega(a_l)|S_s, \omega(a_s)) \mathrm{d}S_l \simeq \frac{\omega(a_l)}{\sqrt{2\pi} \, S_l^{3/2}} \, e^{-\omega^2(a_l)/(2S_l)} \mathrm{d}S_l = \frac{1}{2}\beta(M(r_{m,l})) \frac{\omega^2(a_l)}{S_l^2} \mathrm{d}S_l. \qquad (2.2.95)$$

We can draw two consequences of this result. Firstly, PBHs of higher masses form with a negligible abundance. Moreover, even when they form, the probability of engulfing previously formed lighter PBHs is negligible. In other words, the cloud-in-cloud phenomenon does not exist in practice for broad spectra with a blue tilt and does not alter the overall PBH abundance leaving the final spectrum of masses dominated by the small PBHs. In this sense, broad spectra with a blue tilt give predictions similar to the ones of narrow spectra.

- **Flat broad spectra:** i.e. $n_p = 0$. Also in this a case, the same hierarchy for the thresholds and the variances is maintained

$$\omega(a_s) \gg \omega(a_l) \quad \text{and} \quad S_s \gg S_l. \qquad (2.2.96)$$

However (see Eq. (2.2.92) with $n_p = 0$), the relative ratio of both quantities is constant as

$$\sqrt{\frac{S_s}{S_l}} = \frac{\omega(a_s)}{\omega(a_l)}. \qquad (2.2.97)$$

This condition implies that small PBHs are generated with the same abundance as large ones in terms of $\beta$.

We can compute the probability that a small PBHs is incorporated into a new PBH of a larger mass at $S(< S_s)$ at later times as in Ref. [330]

$$
\begin{aligned}
P(S, \omega(a)|S_s, \omega(a_s)) &= \int_0^S P(S', \omega(a')|S_s, \omega(a_s)) \mathrm{d}S' \\
&= \frac{1}{2} \frac{\omega(a_s) - \omega(a)}{\omega(a_s)} \mathrm{Exp}\left[\frac{2\omega(a)(\omega(a_s) - \omega(a))}{S_s}\right] \mathrm{Erfc}(X) + \frac{1}{2}\mathrm{Erfc}(Y),
\end{aligned}
\tag{2.2.98}
$$

where we approximated the lower limit of integration to zero and defined

$$
\begin{aligned}
X &= \frac{S(\omega(a_s) - 2\omega(a)) + S_s\omega(a)}{[2S_sS(S_s - S)]^{1/2}}, \\
Y &= \frac{S_s\omega(a) - S\omega(a_s)}{[2S_sS(S_s - S)]^{1/2}}.
\end{aligned}
\tag{2.2.99}
$$

Assuming the relations in Eq. (2.2.96), it is easy to see that $X \simeq Y \simeq \omega(a)/\sqrt{2S}$ and the above probability is again found to be negligibly small since

$$
\begin{aligned}
P(S, \omega(a)|S_s, \omega(a_s)) &= \frac{1}{2}\mathrm{Erfc}\left(\frac{\omega(a)}{\sqrt{2S}}\right)\left[1 + \exp\left(2\frac{\omega(a)}{\omega(a_s)}\frac{\omega^2(a_s)}{S_s}\right)\right] \\
&\simeq \beta(M(r_m)) \ll 1.
\end{aligned}
\tag{2.2.100}
$$

To find this result we used the fact that $\omega(a_s)/\sqrt{S_s} = \mathcal{O}(6 \div 8)$ for realistic PBH scenarios, and the argument in the exponential enhancement factor is tamed by the small coefficient $\omega(a)/\omega(a_s) \ll 1$. This again confirms the intuitive picture that, since the formation of a PBH is a rather rare event, having a small PBH swallowed by a larger one is even rarer, scaling like $\beta(M(r_m))\beta(M(r_{m,s})) \simeq \beta^2(M(r_{m,s}))$. This confirms the absence of the cloud-in-cloud problem for broad flat spectra.

It is also interesting to estimate the expected mass distribution in this scenario (we will come back to this issue in Sec. 3.1). At the time when all the PBH population is formed, since PBHs behave like non-relativistic matter and smaller PBHs form earlier in the history of the universe, the abundance of the small PBHs at the formation time $t_l$ is

$$
\frac{\rho_s(t_l)}{\rho_l(t_l)} = \frac{\rho_s(t_s)(a_s/a_l)^3}{\rho_l(t_l)} = \frac{\beta(M(r_{m,s}))}{\beta(M(r_{m,l}))}\frac{a_l}{a_s}\frac{k_s}{k_l} = \left(\frac{M(r_{m,l})}{M(r_{m,s})}\right)^{1/2} \gg 1.
\tag{2.2.101}
$$

In the derivation above we used $\beta = \rho_{\mathrm{PBH}}/\rho_{\mathrm{tot}}$ at the time of the PBH formation. Therefore, even though the probability of forming a PBH is independent of the mass, i.e. $\beta(M(r_{m,s})) = \beta(M(r_{m,l}))$, the PBHs with the smallest mass will give the largest contribution to the mass distribution. In this sense, again, the predictions are similar to the ones of a narrow spectrum. Of course, if the hierarchy between the smallest and largest momenta is not that large, the big PBHs could still have an interesting cosmological role.

- **Broad spectra with a red tilt:** i.e. $n_p < 0$. In this case, from the expression (2.2.90), we find that large-scale variances (corresponding to larger masses) are greater than those at small scales when both are computed at their corresponding horizon crossing time. The formation of PBHs will be therefore dominated by the largest scales and PBHs will have the largest possible mass allowed by the power spectrum $M_l$ (corresponding to $k_l$). This can also be confirmed using analytical arguments analogous to the ones presented above. In such a case, the relative abundance of small PBHs is highly suppressed, and the cloud-in-cloud problem is absent. In this sense, broad spectra with a red tilt also give predictions similar to the peaked power spectra [34].

Let us conclude this discussion with some final remarks. We based our considerations on few simple properties tightly bound to the nature of the PBH scenario, which are:

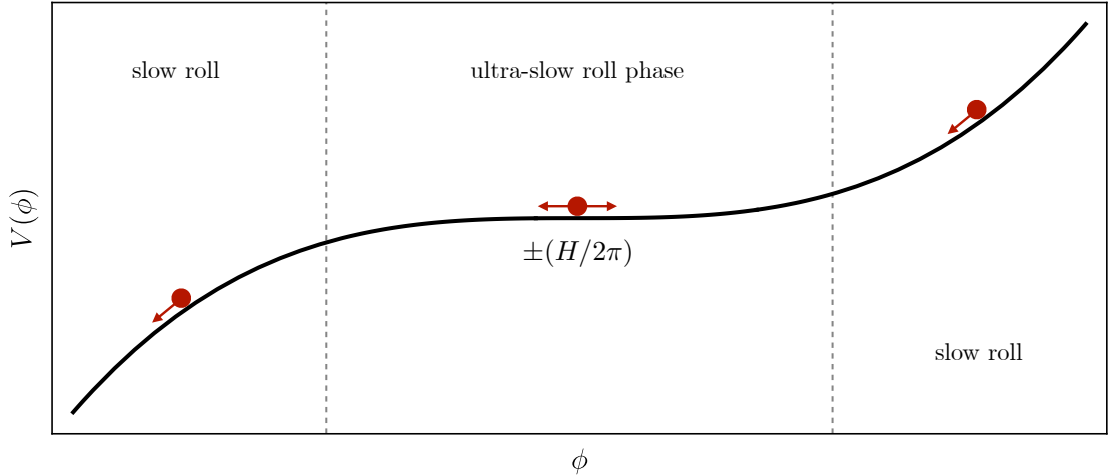

Figure 2.10: *A sketch of the inflaton potential during the phases relevant for the PBH production, highlighting the possibility of quantum jumps during the ultra-slow roll phase.*

- PBH collapse is a rare event;

- the PBH abundance is exponentially sensitive to the amplitude of perturbations;

- PBHs form at horizon crossing, therefore different masses are formed at different times.

As a consequence, we have shown that, even when broad spectra of curvature perturbations are considered, the qualitative features of the resulting PBH population are similar to the one produced by a narrow power spectrum and the cloud-in-cloud problem is absent in the PBH scenario.

### 2.2.5   Impact of quantum diffusion on the abundance in single-field models

In this section, we want to take a little detour and consider the case in which the curvature perturbations responsible for the PBH production are generated in a single-field model of inflation. We will show that, in this specific scenario, the impact of quantum diffusion on the local PBH abundance cannot be neglected. We will closely follow Ref. [23]. [3]

Most of the single field models of inflation are based on the slow-roll paradigm for which the power spectrum of the comoving curvature perturbation can be computed using [340]

$$\mathcal{P}_\zeta^{1/2}(k) = \left(\frac{H}{2\pi\phi'}\right), \qquad \text{with} \qquad \phi' = \frac{\mathrm{d}\phi}{\mathrm{d}N}, \tag{2.2.102}$$

where $N$ is the number of e-folds, the prime denotes differentiation with respect to $N$ and we set the Planck mass to unity for simplicity. To match the right power spectrum amplitude, the generation of PBHs requires an enhancement of the power spectrum of about seven orders of magnitude from its value on CMB scales ($\mathcal{P}_\zeta(k_{\mathrm{CMB}}) \sim 10^{-9}$) within a few e-folds $\Delta N$. In full generality, even without considering a specific model of single-field inflation, one may easily conclude that the slow-roll conditions must be violated as $\phi'$ is required to change rapidly with time [105]. This can be achieved by inserting a feature in the scalar potential causing a non slow-roll phase [341–350], thus producing a sizeable resonance in the power spectrum of curvature perturbations. A typical example of such a feature is given by a plateau, see Fig. 2.10, corresponding to a flatter region where the field reduces its velocity leading to an enhancement of the perturbation at small scales.

---

[3]There still exists an ongoing effort in trying to work out all the consequences of quantum diffusion on the computation of the PBH abundance in single field models of inflation. While, in this section, we want to point out the relevance of the problem, we refer to Refs. [332–339] for some recent progress.

As the derivative of the potential is extremely small in the ultra-slow roll phase, the equation of motion of the inflaton background $\phi$ reduces to

$$\phi'' + 3\phi' + \frac{1}{H^2}\frac{\partial V}{\partial \phi} \simeq \phi'' + 3\phi' = 0. \tag{2.2.103}$$

Using an order of magnitude estimate, the comoving curvature perturbation is expected to grow as

$$\phi' \sim e^{-3N} \qquad \text{and} \qquad \mathcal{P}_\zeta^{1/2} \sim e^{3N}. \tag{2.2.104}$$

This exponential growth makes possible the generation of large fluctuations in the curvature perturbation and the subsequent formation of PBHs upon horizon re-entry during the radiation phase.

Now the issue of quantum diffusion becomes apparent [320]. The reason is the following. The stochastic equation of motion for the classical inflaton field should also account for the possible kicks received by $\phi$ which are of the order of $\pm(H/2\pi)$ [351]. This is done by introducing an effective stochastic force

$$\phi'' + 3\phi' + \frac{1}{H^2}\frac{\partial V}{\partial \phi} = \xi, \tag{2.2.105}$$

where $\xi$ is a Gaussian random noise with

$$\langle \xi(N)\xi(N')\rangle = \frac{9H^2}{4\pi^2}\delta(N - N'). \tag{2.2.106}$$

This stochastic force comes from "integrating out" short modes, typically at scales much shorter than the horizon $k \ll aH$, which are excited due to the quantum effects acting on the inflaton. Typically, this quantum term is negligible if the condition

$$\frac{2\pi}{3H^3}\frac{\partial V}{\partial \phi} \gg 1. \tag{2.2.107}$$

is satisfied. In this case, the classic evolution dominates. During the ultra-slow roll phase, however, $\partial V/\partial \phi$ is required to be small enough so that $\phi'$ decrease rapidly, thus violating slow-roll. For the same reason, (2.2.107) may be violated.

The stochastic equation (2.2.105) can be written as an Ornstein-Uhlenbeck process

$$\frac{d\phi}{dN} = \Pi,$$
$$\frac{d\Pi}{dN} + 3\Pi + \frac{1}{H^2}\frac{\partial V}{\partial \phi} = \xi,$$
$$\langle \xi(N)\xi(N')\rangle = D\delta(N - N'), \tag{2.2.108}$$

where $D = 9H^2/4\pi^2$ is the diffusion coefficient. One can compute the distribution of the field position and velocities by simulating many realisations of the stochastic random evolution, or one can translate the problem into an equation for the probability distribution $P(\phi, \Pi, N)$, which goes under the name of Kramers-Moyal (KM) equation, as [352]

$$\frac{\partial P}{\partial N} = -\frac{\partial}{\partial \phi}(\Pi P) + \frac{\partial}{\partial \Pi}\left[3\Pi P + \frac{V_{,\phi}}{H^2}P\right] + \frac{D}{2}\frac{\partial^2}{\partial \Pi^2}P. \tag{2.2.109}$$

One can set the initial conditions for the probability distribution to be

$$P(\phi, \Pi, 0) = \delta_D(\phi - \phi_0)\delta_D(\Pi - \Pi_0), \tag{2.2.110}$$

where we assumed that $\phi_0$ and $\Pi_0$ are the position and velocity of the inflaton at the onset of the ultra-slow roll phase and that the evolution has been purely classical in the preceding stages of inflation.

Table 2.1: *Parameters adopted for the model [353].*

| $a_w$ | $b_w$ | $c_w$ | $d_w$ | $g_w$ | $r_w$ | $\mathcal{V}$ | $W_0$ | $c_{up}$ |
|-------|-------|-------|-------|-------|-------|------|-------|------|
| 0.02 | 1 | 0.04 | 0 | $3.076278 \cdot 10^{-2}$ | $7.071067 \cdot 10^{-1}$ | $10^3$ | 12.35 | 0.0382 |

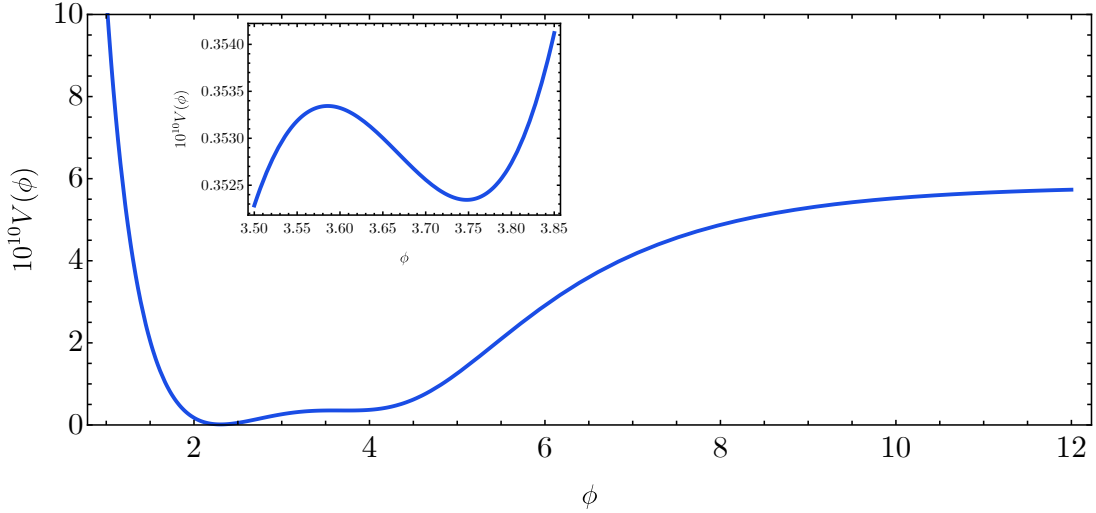

Figure 2.11: *Inflaton potential from Ref. [353] with parameter set defined in Tab 2.1. In the inset, the detail of the local minimum and maximum around the inflection point.*

**Explicit example: model [353]**

To show the impact of quantum diffusion on a realistic example, we will consider the model proposed in Ref. [353]. Analogous conclusions are expected for other single field models. More details can be found in Ref. [23]. The inflaton potential in this model reads

$$V(\phi) = \frac{W_0^2}{\mathcal{V}^3} \left[ \frac{c_{up}}{\sqrt[3]{\mathcal{V}}} + \frac{a_w}{e^{\frac{\phi}{\sqrt{3}}} - b_w} - \frac{c_w}{e^{\frac{\phi}{\sqrt{3}}}} + \frac{e^{\frac{2\phi}{\sqrt{3}}}}{\mathcal{V}} \left( d_w - \frac{g_w}{r_w e^{\sqrt{3}\phi}/\mathcal{V} + 1} \right) \right], \quad (2.2.111)$$

where the parameters used in our analysis can be found in Tab. 2.1. This potential gives rise to the power spectrum matching the CMB observations at large scales (in terms of its amplitude and spectral index) as well as producing a population of PBHs with masses $M \approx 10^{-15} M_\odot$ which comprise the totality of the dark matter. As one can observe in Fig. 2.11, this potential presents an inflection point that leads to a violation of the slow-roll conditions during the inflaton evolution. The ultra-slow roll phase lasts only a few e-folds and a boost in the curvature power spectrum is generated. To compute the perturbations in this model, we numerically solve the Mukhanov-Sasaki equation for the comoving curvature perturbation

$$\zeta''(k) + 2\frac{z'}{z}\zeta'(k) + k^2\zeta(k) = 0, \quad (2.2.112)$$

where, for convenience, the prime denotes conformal time derivative $d/d\eta$ (dot being derivative with respect to the cosmological time) and $z = a\dot{\phi}/H$. The initial conditions for the perturbations in the asymptotic past can be set to coincide with the Bunch-Davies vacuum in the limit $(-k\tau) \gg 1$. The function $z$ satisfies the following equation

$$\frac{z''}{z} = 2a^2 H^2 \left( 1 + \epsilon - \frac{3}{2}\eta + \epsilon^2 - 2\epsilon\eta + \frac{1}{2}\eta^2 + \frac{1}{2}\xi^2 \right) \quad (2.2.113)$$

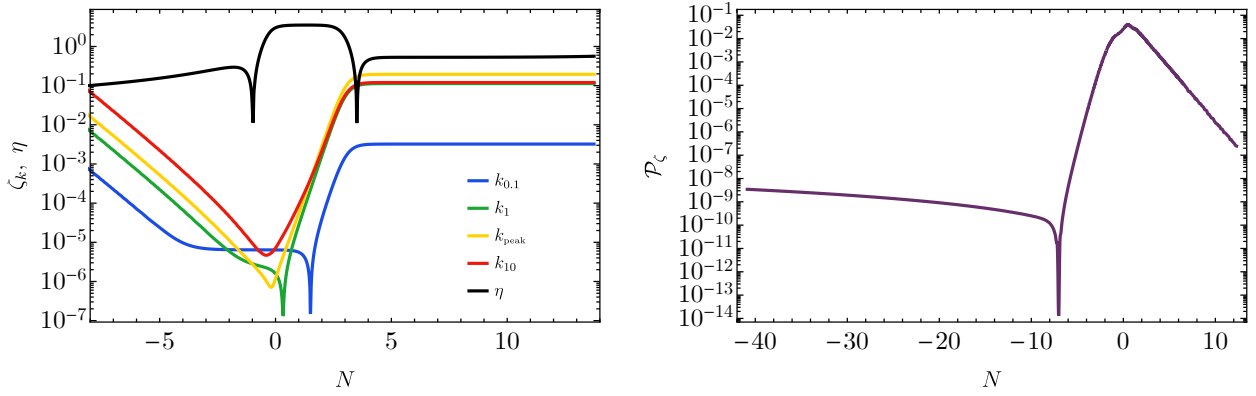

Figure 2.12: **Left:** *Evolution of different modes in time during the ultra-slow roll phase. We define $k_{pk}$ as the mode corresponding to the largest amplification of the spectrum and $k_1$ as the one crossing the horizon at the beginning of the transition (approximately at $N \simeq -1$). Consequently, $k_{0.1}$ ($k_{10}$) corresponds to $k \sim k_1/10$ ($10\,k_1$), respectively.* **Right:** *Power spectrum as a function of the e-fold number $N$ corresponding to the horizon crossing time when $k = aH$. In both figures, we have set $N = 0$ at the time when $\eta$ reaches its plateau value during the ultra-slow roll phase $\eta = 3$.*

where

$$\epsilon = -\frac{\dot{H}}{H^2}, \qquad \eta = -\frac{\ddot{\phi}}{H\dot{\phi}}, \qquad \xi^2 = 3(\epsilon + \eta) - \eta^2 - \frac{1}{H^2}\frac{\partial^2 V}{\partial \phi \partial \phi}. \qquad (2.2.114)$$

The crucial point is that, during the ultra-slow roll phase, the second term in Eq. (2.2.112) becomes positive, driving the growth of perturbations (differently from what happens during the slow roll evolution when it acts as a dumping term, eventually freezing the super-horizon perturbations). Indeed, one finds

$$\frac{z'}{z} = aH\left(1 + \epsilon - \eta\right) < 0, \qquad (2.2.115)$$

when $\eta$ becomes large and positive. See Fig. 2.12 for the evolution of $\eta$ and few modes exiting the horizon during the ultra-slow roll phase. Also, in Fig. 2.12, we show the resulting power spectrum of comoving curvature perturbations computed adopting the mean background evolution without the effect of quantum jumps.

We repeat the computation of the curvature perturbation for various realisations of the background evolution and then feed this spectrum into the machinery to compute the primordial PBH abundance $\beta_{\rm prim}^{\rm qd}$. The physical interpretation of the different enhancements of the power spectrum found in different realisations of the stochastic background evolution is the following. As the inflaton field passes through the plateau and starts experiencing quantum kicks, its trajectory deviates from the classical evolution. When the quantum noise slows down the field, the inflaton remains in the ultra-slow roll phase for a longer period and the perturbations get enhanced further. The opposite happens when quantum jumps make the field go faster through the feature of the inflaton potential. Our results show that $\ln \beta_{\rm prim}^{\rm qd}$ is approximately Gaussian distributed around the value of $\ln \beta_{\rm prim}^{\rm cl}$ computed using the classical inflaton evolution, and with a standard deviation $\sigma_{\beta_{\rm prim}^{\rm qd}} \approx 6$, see Fig. 2.13.

From the results above, one can deduce that in different patches of the universe one expects the value of the PBH local abundance to violently fluctuate around the average abundance of PBHs. This effect can be interpreted as a non-Gaussian signature in the comoving curvature perturbations which modulates the variance of the field (and therefore the resulting PBH abundance) at scales somewhat larger than the PBH characteristic scale. The result shows, therefore, that quantum diffusion effects cannot be neglected when computing the prediction of an ultra-slow roll model of inflation responsible for the production of PBHs.

Notice that, as the ultra-slow roll phase, where the quantum noise is responsible for the fluctuations of the variance, only lasts up to a few ($\lesssim 5$) e-folds, such fluctuations in the beta function take place at very small scales. This is explained as follows. An overdense region comes from trajectories in the field space which have received quantum kicks that have slowed down the field going through the ultra-slow roll phase and vice versa. Therefore, the correlation scale, corresponding to the earliest quantum kick, can only be a number of e-folds $\lesssim 5$ larger than the cosmological horizon at the time of PBH formation. In the case of PBHs in the mass range of interest for LIGO, it is plausible in fact, that the formation of binaries and subsequently the merger rate would not be modified by this effect, as the hierarchy of scales between the PBH formation and binary formation is larger than $\mathcal{O}(10^3)$ (see discussion in Sec. 4.5).

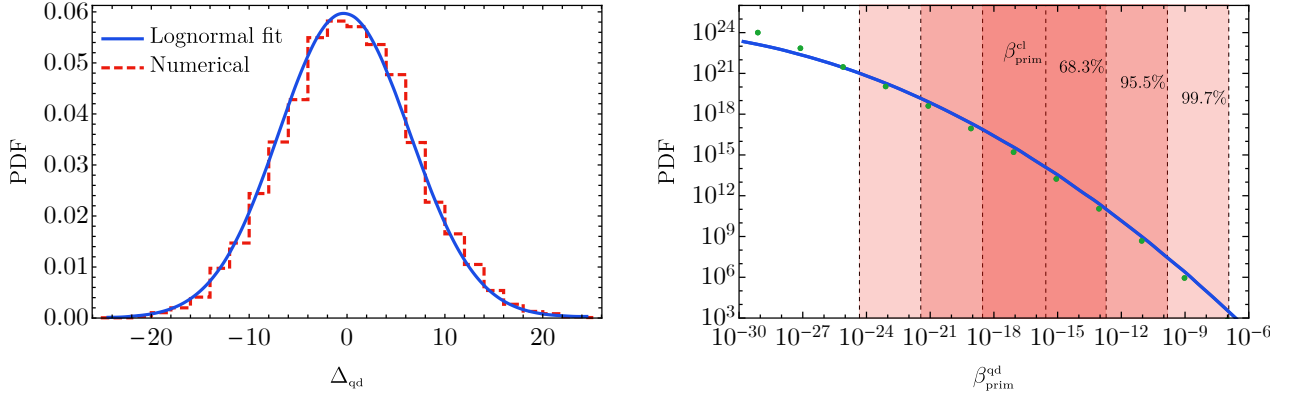

Figure 2.13: **Left:** *The probability density distribution for $\Delta_{\mathrm{qd}} \equiv \ln\left(\beta_{\mathrm{prim}}^{\mathrm{qd}}/\beta_{\mathrm{prim}}^{\mathrm{cl}}\right)$, which is nearly Gaussian distributed around the classical value determined ignoring quantum diffusion, for the model [353].* **Right:** *corresponding $\beta_{\mathrm{prim}}^{\mathrm{qd}}$. The results are derived from $10^4$ realisations of the stochastic evolution.*

# Chapter 3

# Primordial black hole properties at formation

In this chapter, we review the properties of PBHs at the formation epoch by discussing both the mass and spin distributions and how PBHs cluster at high redshift.

## 3.1 Mass distribution

An important property of the PBH population, other than the overall abundance, is given by its mass distribution, $\psi(M)$, representing the mass fraction of PBHs with mass within $(M, M + \mathrm{d}M)$, which is routinely defined by the relation (see [196] for a review)

$$\psi(M) = \frac{1}{\rho_{\mathrm{PBH}}} \frac{\mathrm{d}\rho_{\mathrm{PBH}}(M)}{\mathrm{d}M} \tag{3.1.1}$$

and normalised such that

$$\int \psi(M)\mathrm{d}M = 1. \tag{3.1.2}$$

In order to compute it, starting from the density perturbation distribution, one should additionally take into account two important effects:

- the critical collapse property, which relates the PBH mass $M_{\mathrm{PBH}}$ to the horizon mass at formation depending on the value of the collapsing overdensity by [139, 140, 354] (see Eq. (3.1.6))

$$M_{\mathrm{PBH}}(\delta) = \mathcal{K} \left( \delta - \delta_c \right)^\gamma M_H, \tag{3.1.3}$$

where $\gamma = 0.36$ in a radiation dominated universe [146, 148, 149, 355, 356] and $\mathcal{K}$ depends on the collapsing overdensity shape. As a reference value, we will adopt $\mathcal{K} = 4$ as done in Ref. [15]. Therefore, the critical collapse requires modifying the definition of mass fraction at collapse, as multiple masses are generated at the same time in different Hubble patches. This can be can be computed as

$$\beta = \int_{\delta_c}^{\infty} \frac{M_{\mathrm{PBH}}}{M_H} P(\delta)\mathrm{d}\delta = \mathcal{K} \int_{\delta_c}^{\infty} (\delta - \delta_c)^\gamma P(\delta)\mathrm{d}\delta. \tag{3.1.4}$$

- the evolution with redshift of the abundance of PBH, which is scaling like non-relativistic matter

$$\rho_{\mathrm{PBH}} \sim a^{-3}, \tag{3.1.5}$$

and therefore is enhanced compared to the dominant energy density of the universe behaving as radiation $\rho_{\mathrm{tot}} \sim a^{-4}$ up the epoch of matter-radiation equality at redshift $z_{\mathrm{eq}} = 3402$. This means, in practice, that the abundance of light PBHs formed earlier than larger ones, as the cosmological horizon is growing with time, gets larger enhancement factors. This effect distorts the mass function predicted at redshift $z \lesssim z_{\rm eq}$. More in details, one finds, at the time of formation of the heaviest PBH,

$$\frac{\rho_s(\eta_l)}{\rho_l(\eta_l)} = \frac{\rho_s(\eta_s)(a_s/a_l)^3}{\rho_l(\eta_l)} = \frac{\beta(M_s)}{\beta(M_l)}\frac{a_l}{a_s} = \frac{\beta(M_s)}{\beta(M_l)}\left(\frac{M_l}{M_s}\right)^{1/2}. \tag{3.1.6}$$

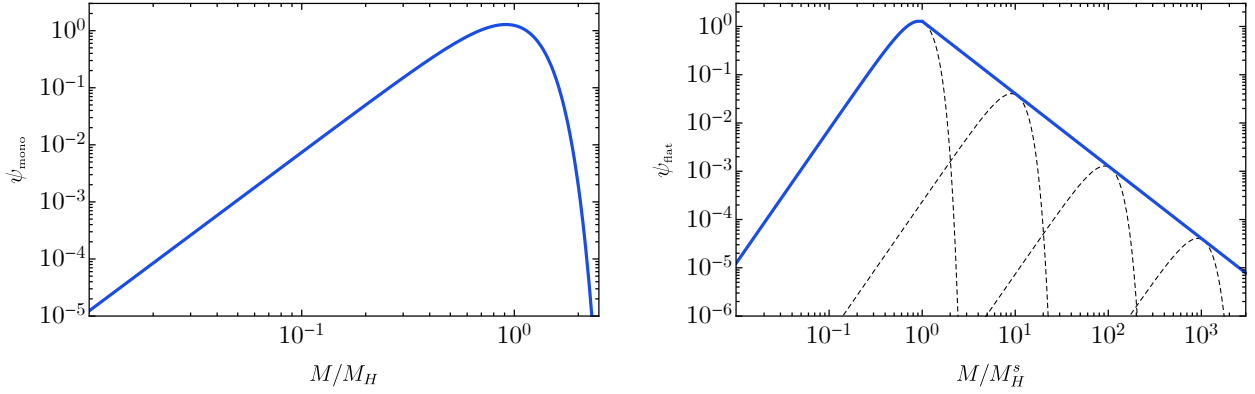

Figure 3.1: ***Left:** Mass function for a monochromatic power spectrum. **Right:** mass function for a broad and flat power spectrum. It presents the characteristic scaling $\psi(M) \propto M^{-3/2}$ due to the redshift effect. It can be viewed as the convolution of different critical mass functions with an amplitude scaling like $\sim (M/M_H)^{-3/2}$. The small mass tail is dominated by the critical collapse giving the characteristic scaling $\sim M^{2.8}$.*

In the following, we will present the mass function obtained in a couple of noticeable examples, namely the monochromatic and flat power spectra, while also defining a useful parametrisation in terms of the lognormal shape which is often used in the literature. The latter was shown to be able to capture a variety of PBH formation models.

- **Monochromatic case:** also referred to as "critical mass function", corresponds to the mass distribution found in the case scalar perturbations are characterised by a single scale $k_* \sim 1/r_m$. Therefore, all PBHs are expected to form around the same time, at cosmological horizon crossing when $r_m \sim R_H$. The fact that the mass also depends on the amplitude of the perturbations creates an effective broadening of the mass function, leading to a shape of the form [140, 357]

$$\psi_{\rm mono}(M) = \frac{1}{\beta}\frac{1}{\sqrt{2\pi}\sigma_\delta\gamma M_H}\left(\frac{M}{\mathcal{K}M_H}\right)^{\frac{1}{\gamma}}\exp\left[-\frac{(\delta_c + (M/\mathcal{K}M_H)^{\frac{1}{\gamma}})^2}{2\sigma_\delta^2}\right], \tag{3.1.7}$$

  and shown in Fig. 3.1.

  This mass function presents a low-mass tail scaling like $\sim M^{2.8}$ and an exponential fall-off after the peak located around $M \sim M_H$. As one can understand intuitively, the exponential cut-off at large masses is due to the exponential dumping of perturbations with amplitude larger than $\delta_c$. In this case, as all the PBH are formed at the same cosmological time, the redshift factor is an overall scaling and does not affect the shape of the mass distribution.

  The result has been derived assuming gaussian statistics. It was however generalised, accounting for both the non-linear relation between the curvature perturbations and the density contrast and possible local non-gaussianities, in Ref. [167, 358]. In practice, this mass function can be viewed as setting the minimum width a PBH mass distribution can posses when formed through the collapse of density perturbations.

- **Flat spectrum:** corresponding to the parametrisation Eq. (2.1.30). In this case, it is important to account for the redshift factor as different modes re-enter the horizon at different times and the PBH formation takes place at different epochs.

As found in the preceding section when considering this flat power spectrum, the probability of collapse $\beta(M)$ is constant in the range $[M_H^s, M_H^l]$, where $M_H^s$ ($M_H^l$) corresponds to the horizon mass at the time of horizon crossing of the smallest (largest) scale in the spectrum $1/k_s$ ($1/k_l$). The fact that the mass fraction $\beta$ at formation is constant over the whole range of masses is a consequence of few key results [15]:

 *1)* the shape of the perturbations is the same for the various peaks produced by the broad spectrum, leading to a constant threshold $\delta_c = 0.56$, see Sec. 2.1.2;

 *2)* the probability of collapse is constant as different scales re-enter the horizon at different times, see Eq. (2.2.97);

 *3)* the "cloud-in-cloud" is absent for PBHs, as shown in Sec. 2.2.4.

By expanding (3.1.6) for a mass $M + \mathrm{d}M$ at the time of matter radiation equality when all PBHs has formed, one finds

$$\frac{\rho_s(M + \mathrm{d}M)}{\rho_l(M_{\mathrm{eq}})} - \frac{\rho_s(M)}{\rho_l(M_{\mathrm{eq}})} = \left(\frac{M_{\mathrm{eq}}}{M + \mathrm{d}M}\right)^{1/2} - \left(\frac{M_{\mathrm{eq}}}{M}\right)^{1/2} \sim -\frac{M_{\mathrm{eq}}^{1/2}}{M^{3/2}}\mathrm{d}M. \tag{3.1.8}$$

Therefore, the mass function can be written as

$$\psi_{\mathrm{flat}}(M) \simeq \frac{1}{f_{\mathrm{PBH}}}\frac{M_{\mathrm{eq}}^{1/2}}{M^{3/2}}\beta(M). \tag{3.1.9}$$

with a constant $\beta(M)$ for $M \in [M_H^s, M_H^l]$ [15]. As shown in Fig. 3.1, this mass function is characterised by a constant slope going like $\sim M^{-3/2}$ [71, 78, 189, 194, 359]. At masses smaller then $M_H^s$, the mass function is dominated by the critical collapse of the first mode re-entering the horizon, thus having a scaling with the mass $\sim M^{2.8}$.

 • **Lognormal parametrisation:** defined as

$$\psi(M) = \frac{1}{\sqrt{2\pi}\sigma M}\exp\left[-\frac{\ln^2(M/M_c)}{2\sigma^2}\right], \tag{3.1.10}$$

in terms of a reference mass $M_c$ and a width $\sigma$. This was shown to provide a very good approximation of the mass function obtained in models where the power spectrum of curvature perturbations possesses a symmetric peak [167, 258, 360–362]. By varying its parameters, one is able to describe a large class of models and it will be considered in the following as the standard parametrisation of the PBH mass function. An alternative to this shape has been considered in Ref. [363], at the cost of introducing an additional skewness parameter.

Notice that another definition of the mass function is also often used in the literature, which corresponds to the mass fraction with respect to the entirety of the dark matter in logarithmic mass interval, and can be written as

$$f(M) = \frac{1}{\rho_{\mathrm{DM}}}\frac{\mathrm{d}\rho_{\mathrm{PBH}}}{\mathrm{d}\ln M}, \tag{3.1.11}$$

normalised such that

$$\int f(M)\mathrm{d}\ln M = f_{\mathrm{PBH}}, \tag{3.1.12}$$

where $\rho_{\mathrm{DM}}$ is the abundance of dark matter in the universe. One can translates this definition in terms of the one we adopted in Eq. (3.1.1) using

$$f(M) = f_{\mathrm{PBH}}\, M\, \psi(M). \tag{3.1.13}$$

For the broad mass distribution, for instance, the fall-off given by $f(M) \propto M^{-1/2}$ is less pronounced and more directly related to the observational constraints as a function of the PBH mass, as we will see in the following chapters.

## 3.2   Natal spin of PBHs

Black holes in general relativity are very simple objects, whose properties are completely described by their mass and spin.[1] Therefore, having analysed the properties of the PBH mass distribution, we now turn our attention to the PBH spin. In this section, we present the computation of the PBH spin at formation, while its evolution will be addressed in the following chapter.

To compare the PBH scenario to current BH observations, having a precise determination of the PBH spin distribution is crucial. We can think about, for example, the measurements of GWs coming from merging of two black holes, where one of the key observables reconstructed from GW waveform is the so-called "effective spin parameter" $\chi_{\text{eff}}$. This is defined as the orbital projection of the two spins $\vec{S_1}$ and $\vec{S_2}$ onto the binary angular momentum $\vec{L}$ as

$$\chi_{\text{eff}} = \frac{M_1 \vec{S_1} + M_2 \vec{S_2}}{M_1 + M_2} \cdot \hat{L}. \tag{3.2.1}$$

Currently, most of the observed GW events are characterised by a $\chi_{\text{eff}}$ which is compatible with zero [274], with only few exceptions. As part of the ongoing effort in trying to understand if PBHs may be compatible with current LIGO/Virgo observations, it is, therefore, crucial to understand the prediction for the spin in the PBH scenario.

In this section, we follow closely Ref. [20] and show a small (albeit non-vanishing) natal spin is expected for PBHs formed out of the collapse of radiation density perturbations. As we discussed in the preceding sections, extreme peaks in density contrast are expected to possess a nearly spherical shape. This property was used to study the collapse of overdensities leading to the generation of PBHs. However, the leading order anisotropy of the collapsing ellipsoid is coupled to the first-order tidal gravitational field and can generate a first-order torque. This result is extremely similar to the renowned "tidal-torque theory" in large-scale structure [364–369]. In the PBH case, however, the shape of the Lagrangian region collapsing to form a PBH is determined when the characteristic scale of the perturbation is still on superhorizon scales. To successfully generate a non zero spin at first order in perturbation theory then, both the following conditions must be satisfied:

- the lengths of the principal axes of the inertia tensor should differ. This happens in the homogeneous ellipsoid approximation, we will adopt following the studies of gravitational collapse of ellipsoidal perturbations in the large scale structure literature [370–374];

- the presence of non-vanishing off-diagonal components of the velocity shear, which means the collapsing ellipsoidal perturbation must possess an inertia tensor that is misaligned with that of the velocity shear.

As we will see, the latter condition cannot be achieved when perturbations are still on superhorizon scales. However, after horizon crossing and prior to decoupling from the Hubble expansion, torques are generated, leading to a small spin-up of the configuration. Lacking a better description of the dynamics after turnaround time, we will assume the angular momentum remains constant, as it happens for non-relativistic dust [366, 367, 369]. Still, this approximation should also be investigated further with the aid of general relativistic 3-dimensional numerical simulations.

### 3.2.1   The spin of PBHs as local density maxima

We start by addressing the issue of how to define angular momentum in general relativity. We refer to Ref. [375] (and more recently [376]) for a detailed discussion. The angular momentum is a conserved quantity that originates from rotational invariance. When the spacetime possesses a symmetry, it is possible to define a coordinate-independent global quantity using the technique described in Ref. [377]. This is done by computing the flux integral of the derivative of the Killing vector associated with the symmetry over closed two-surfaces surrounding the matter sources. If all the matter source is included within the integration surface, it can be shown that such a quantity is conserved and it does not depend on the choice of the particular surface where the flux is measured. One can therefore define the amplitude of the Komar angular momentum on a given time-slicing $\Sigma$ through the Komar integral [375]

$$S(\Sigma) = \frac{1}{16\pi G_N} \int_{\partial\Sigma} \mathrm{d}S_{\mu\nu} D^\mu \phi^\nu, \tag{3.2.2}$$

where $\phi^\nu$ is the rotational Killing vector of the asymptotic flat metric and $G_N$ is the Newton's constant. Equivalently, using both Gauss' law and the Einstein equations, one can also write

$$S(\Sigma) = \int_\Sigma \mathrm{d}S_\mu J^\mu(\phi), \tag{3.2.3}$$

where $J^\mu(\phi) = T^\mu_\nu \phi^\nu - T\phi^\mu/2$ and $T^\mu_\nu$ is energy-momentum tensor. We can focus on the case in which the source is made of a relativistic perfect fluid and simplify the expression to get

$$S(\Sigma) = \int_\Sigma \mathrm{d}V \, T^0_{\ \mu}\phi^\mu = \frac{4}{3} \int_\Sigma \mathrm{d}V \rho \, \vec{v} \cdot \vec{\phi}, \tag{3.2.4}$$

where $\vec{v}$ is the velocity field and $\rho$ is the density field. Finally, as we will be dealing with collapsing overdensity peaks, we can expand the density contrast and the velocity around the peak location $\vec{x}_{\mathrm{pk}}$, so that

$$S_i = \frac{4}{3} a^4(\eta) \epsilon_{ijk} \int \mathrm{d}^3 x \, \rho(\vec{x}, \eta)(x - x_{\mathrm{pk}})^j (v - v_{\mathrm{pk}})^k, \tag{3.2.5}$$

where $\vec{x}$ is a comoving coordinate and we have adopted the conformal time $\eta$. In the setting of cosmological perturbation theory, the velocity is inherently a first order quantity. One should also expand the density contrast, leading to higher-order contributions to the spin. In the next section we will provide the description of those contributions using peak theory [297].

---

[1]We are neglecting the case of electrically charged BHs as one does not expect them to form in the standard PBH scenario.

**First-order description of the PBH spin**

To start, we describe the collapsing overdensities as high peaks of the random density fluid and adopt a triaxial ellipsoid description of the volume $V_e$. To describe the statistics of the (correlated) quantities entering into the definition of the spin, we will use the results of peak theory to be found in Ref. [297]. We expand the overdensity field around its peak up to second order as

$$\delta(\vec{x}) = \frac{\delta\rho(\vec{x})}{\bar{\rho}} \simeq \delta_{\mathrm{pk}} + \frac{1}{2}\zeta_{ij}(x - x_{\mathrm{pk}})^i (x - x_{\mathrm{pk}})^j > f\delta_{\mathrm{pk}}, \qquad \zeta_{ij} = \left.\frac{\partial^2 \delta}{\partial x^i \partial x^j}\right|_{\mathrm{pk}}. \tag{3.2.6}$$

As we are dealing with a peak, the first gradient term around $\vec{x}_{\mathrm{pk}}$ is vanishing. Notice that we adopt $\zeta_{ij}$ to describe the second derivatives acting on the density perturbations, not to be confused with the comoving curvature perturbations symbol used elsewhere. The free parameter $f$ will be fixed in the following by requiring the overdensity peak to be above the threshold for collapse. By rotating the coordinates, we can align them with the principal axes of length $\lambda_i$ of the constant-overdensity ellipsoids and write

$$\delta \simeq \delta_{\mathrm{pk}} - \frac{1}{2}\sigma_\zeta \sum_1^3 \lambda_i (x^i - x_{\mathrm{pk}}^i)^2, \tag{3.2.7}$$

where $\sigma_\zeta$ is the variance of $\zeta_{ij}$. As standard practice, we can define the rescaled peak height $\nu = \delta_{\mathrm{pk}}/\sigma_\delta$ and solve for the positions where $\delta = f\delta_{\mathrm{pk}}$, i.e.

$$2\frac{\sigma_\delta}{\sigma_\zeta}(1 - f)\nu = \sum_1^3 \lambda_i (x^i - x_{\mathrm{pk}}^i)^2. \tag{3.2.8}$$

Those solutions describe the boundary of the integration volume in the integral (3.2.5). They also define the principal semi-axes of the collapsing ellipsoid as

$$a_i^2 = 2\frac{\sigma_\delta}{\sigma_\zeta}\frac{(1 - f)}{\lambda_i}\nu. \tag{3.2.9}$$

In the PBH scenario we can always expand in the limit of large $\nu$ to get

$$\lambda_i = \frac{\gamma\nu}{3}\left(1 + \epsilon_i\right), \tag{3.2.10}$$

where we introduced $\gamma = \sigma_\times^2/\sigma_\delta\sigma_\zeta$ and $\epsilon_i = \mathcal{O}\left(1/\gamma\nu\right)$ corresponds to the small deviations from spherical symmetry, see app. A. Also, $\sigma_\times^2$ is the characteristic cross-correlation between $\delta$ and $\zeta_{ij}$. Parametrically, we find that the characteristic size of the perturbation along the different axes is $a_i^2 \sim 6\sigma_\delta^2/\sigma_\times^2$ or $a_i^2 \sim \sigma_\delta/\sigma_\zeta|\zeta_{ij}| \sim R_*^2$, where $R_*$ is the characteristic perturbation scale. Finally, we stress that the difference $|\lambda_i - \lambda_j| \sim \mathcal{O}(1)$ because the "ellipticity" $\epsilon_i$ scales like $1/(\gamma\nu)$. We can also expand the velocity field analogously to Eq. (3.2.6) as

$$(v - v_{\rm pk})^k = v_l^k(x - x_{\rm pk})^l, \qquad v_l^k = \left.\frac{\partial v^k}{\partial x^l}\right|_{\rm pk}, \tag{3.2.11}$$

to get

$$S_i = \frac{4}{3}a^4(\eta)\epsilon_{ijk}\overline{\rho}_{\rm rad}(\eta)v_l^k\int_{V_e}{\rm d}^3x(x - x_{\rm pk})^j(x - x_{\rm pk})^l. \tag{3.2.12}$$

As the perturbation crosses the horizon, we expect also the volume $V_e$ to be modified. However, those deformations become significant only after turnaround[2], and therefore at times $t < t_{\rm t.a.}$ we approximate the perturbation as a rigid body of volume $V_e$. The boundary of the perturbation is set by the requirement $\delta = f\delta_{\rm pk}$, as discussed above.

In order to perform the integration in Eq. (3.2.12), we set, without loss of generality, $\vec{x}_{\rm pk} = 0$ to find

$$S_i = \frac{4}{3}a^4(\eta)\epsilon_{ijk}\overline{\rho}_{\rm rad}(\eta)g_v(\eta)\widetilde{v}_{kl}\int_{V_e}{\rm d}^3x\ x^jx^l. \tag{3.2.13}$$

The time dependence of the velocity shear has been factorised in terms of the linear factor $g_v$ as

$$v_l^k(\eta) = g_v(\eta)\widetilde{v}_l^k, \tag{3.2.14}$$

while $\widetilde{v}_l^k$ defines the normalised (time independent) velocity shear. Going to the coordinates of the ellipsoidal perturbation, the integration measure becomes

$$\int_{V_e}{\rm d}^3x = a_1a_2a_3\int_0^1 r^2{\rm d}r\int_0^{2\pi}{\rm d}\phi\int_0^\pi {\rm d}\theta\,\sin\theta \tag{3.2.15}$$

and the coordinates are written as

$$x_1 = a_1r\cos\phi\sin\theta, \quad x_2 = a_2r\sin\phi\sin\theta, \quad x_3 = a_3r\cos\theta. \tag{3.2.16}$$

Therefore, one can find that

$$\int_{V_e}{\rm d}^3x\ x^jx^l = \frac{4\pi}{15}a_1a_2a_3\begin{bmatrix} a_1^2 & 0 & 0 \\ 0 & a_2^2 & 0 \\ 0 & 0 & a_3^2 \end{bmatrix}_{jl}. \tag{3.2.17}$$

Finally, Eq. (3.2.13) can be written as [365]

$$\vec{S}^{(1)} = \frac{16\pi}{45}a^4(\eta)\overline{\rho}_{\rm rad}(\eta)g_v(\eta)a_1a_2a_3([a_2^2 - a_3^2]\widetilde{v}_{23}, [a_3^2 - a_1^2]\widetilde{v}_{13}, [a_1^2 - a_2^2]\widetilde{v}_{12}). \tag{3.2.18}$$

We can pause here and stress that Eq. (4.4.29) already contains few important results. The spin is shown to be non-vanishing at first order in perturbation theory if both the following conditions are met. First, the semi-major axes must be different (i.e. the geometry of the collapsing overdensity should depart from spherical symmetry). Second, the off-diagonal components of the velocity shear ($v_{ij}$ for $i \neq j$) should be non-vanishing and, therefore, misaligned with the inertia tensor.

---

[2]The turnaround time $t_{\rm t.a.}$ is defined as the time at which the overdensity decouples from the Hubble flow and starts contracting.

To simplify the notation, Eq. (4.4.29) we can also be rewritten as

$$\vec{S}^{(1)} = \left[\frac{4}{3}a^4(\eta)\overline{\rho}_{\rm rad}(\eta)g_v(\eta)(1-f)^{5/2}R_*^5\right]\frac{16\sqrt{2}\pi}{135\sqrt{3}}\left(\frac{\nu}{\gamma}\right)^{\frac{5}{2}}\frac{1}{\sqrt{\Lambda}}\left(-\alpha_1\widetilde{v}_{23},\alpha_2\widetilde{v}_{13},-\alpha_3\widetilde{v}_{12}\right), \quad (3.2.19)$$

where $\alpha_i$ are defined such that

$$\alpha_1 = \frac{1}{\lambda_3} - \frac{1}{\lambda_2}, \quad \alpha_2 = \frac{1}{\lambda_3} - \frac{1}{\lambda_1}, \quad \alpha_3 = \frac{1}{\lambda_2} - \frac{1}{\lambda_1}, \quad (3.2.20)$$

and $\alpha_i \geq 0$, with $\alpha_2 \geq \alpha_1, \alpha_3$. Also, we introduced

$$\Lambda = \lambda_1\lambda_2\lambda_3, \quad R_* = \sqrt{3}\frac{\sigma_\times}{\sigma_\zeta}. \quad (3.2.21)$$

We now can split the relation in two part as

$$\vec{S}^{(1)}(\eta) = S_{\rm ref}(\eta)\vec{s}_{\rm e}^{(1)}. \quad (3.2.22)$$

where $S_{\rm ref}(\eta)$ identifies the time dependent part of the spin, defined as

$$S_{\rm ref}(\eta) = \frac{4}{3}a^4(\eta)\overline{\rho}_{\rm rad}(\eta)g_v(\eta)R_*^5(1-f)^{5/2}, \quad (3.2.23)$$

while $\vec{s}_{\rm e}^{(1)}$ tracks its geometrical properties. Notice that the amplitude $S_{\rm ref}(\eta)$ represents a common factor for all peaks, while the stochastic shape properties are captured in the remaining term. Also, the time derivative of $S_{\rm ref}(\eta)$ can be thought of as the effective torque acting on the matter in the neighbourhood of local density maxima.

**Second-order description of the PBH spin**

In this section, we derive the second order contribution to the PBH spin. Following the preceding steps, we expand also the density contrast as

$$\rho(\vec{x},\eta) = \overline{\rho}_{\rm rad}(\eta) + \delta\rho = \overline{\rho}_{\rm rad}(\eta)\left(1+\delta_{\rm pk}\right) + \overline{\rho}_{\rm rad}(\eta)\left[\frac{1}{2}\zeta_{ij}(x-x_{\rm pk})^i(x-x_{\rm pk})^j\right]. \quad (3.2.24)$$

This translates into a second order term appearing in addition to Eq. (3.2.12)

$$S_i = \frac{4}{3}a(\eta)^4\epsilon_{ijk}\overline{\rho}_{\rm rad}(\eta)v_l^k\left[(1+\delta_{\rm pk})\int_{V_{\rm e}}{\rm d}^3x\,x^jx^l + \frac{1}{2}\zeta_{mn}\int_{V_{\rm e}}{\rm d}^3x\,x^jx^lx^mx^n\right]. \quad (3.2.25)$$

In this case as well, we integrate the previous equation by choosing the coordinates aligned with the principal axes of the ellipsoid (see Eq. (3.2.15)). By defining

$$I^{jlmn} = \int_{V_{\rm e}}{\rm d}^3x\,x^jx^lx^mx^n \quad (3.2.26)$$

we find that the non zero multipole integrals are

$$I^{1111} = \frac{4\pi}{35}a_1^5a_2a_3, \quad I^{2222} = \frac{4\pi}{35}a_1a_2^5a_3, \quad I^{3333} = \frac{4\pi}{35}a_1a_2a_3^5, \quad (3.2.27)$$

$$I^{2211} = I^{1122} = I^{2121} = I^{2112} = I^{1221} = I^{1212} = \frac{4\pi}{105}a_1^3a_2^3a_3, \quad (3.2.28)$$

$$I^{3311} = I^{1133} = I^{3131} = I^{3113} = I^{1331} = I^{1313} = \frac{4\pi}{105}a_1^3a_2a_3^3, \quad (3.2.29)$$

$$I^{3322} = I^{2233} = I^{2332} = I^{2323} = I^{3232} = I^{3223} = \frac{4\pi}{105}a_1a_2^3a_3^3. \quad (3.2.30)$$

Finally, the spin is given by

$$\vec{S} = \vec{S}^{(1)}(1 + \delta_{\text{pk}}) + \vec{S}^{(2)}, \tag{3.2.31}$$

where $\vec{S}^{(2)}$ is expressed in terms of its components as

$$
\begin{aligned}
S_1^{(2)} = \frac{16\pi}{90} a^4(\eta) \bar{\rho}_{\text{rad}}(\eta) g_v(\eta) a_1 a_2 a_3 \cdot \frac{1}{7} &\left[ 3a_2^4 \widetilde{v}_{23} \zeta_{22} - 3a_3^4 \widetilde{v}_{23} \zeta_{33} + a_1^2 a_2^2 (\widetilde{v}_{23} \zeta_{11} + 2\widetilde{v}_{13} \zeta_{12}) \right. \\
&\left. - a_1^2 a_3^2 (\widetilde{v}_{23} \zeta_{11} + 2\widetilde{v}_{12} \zeta_{13}) + a_2^2 a_3^2 (-\widetilde{v}_{23} \zeta_{22} - 2\widetilde{v}_{22} \zeta_{23} + 2\widetilde{v}_{33} \zeta_{23} + \widetilde{v}_{23} \zeta_{33}) \right],
\end{aligned} \tag{3.2.32}
$$

$$
\begin{aligned}
S_2^{(2)} = \frac{16\pi}{90} a^4(\eta) \bar{\rho}_{\text{rad}}(\eta) g_v(\eta) a_1 a_2 a_3 \cdot \frac{1}{7} &\left[ -3a_1^4 \widetilde{v}_{13} \zeta_{11} + 3a_3^4 \widetilde{v}_{13} \zeta_{33} - a_1^2 a_2^2 (2\widetilde{v}_{23} \zeta_{12} + \widetilde{v}_{13} \zeta_{22}) \right. \\
&\left. + a_2^2 a_3^2 (\widetilde{v}_{13} \zeta_{22} + 2\widetilde{v}_{12} \zeta_{23}) + a_1^2 a_3^2 (\widetilde{v}_{13} \zeta_{11} + 2\widetilde{v}_{11} \zeta_{13} - 2\widetilde{v}_{33} \zeta_{13} - \widetilde{v}_{13} \zeta_{33}) \right],
\end{aligned} \tag{3.2.33}
$$

$$
\begin{aligned}
S_3^{(2)} = \frac{16\pi}{90} a^4(\eta) \bar{\rho}_{\text{rad}}(\eta) g_v(\eta) a_1 a_2 a_3 \cdot \frac{1}{7} &\left[ 3a_1^4 \widetilde{v}_{12} \zeta_{11} - 3a_2^4 \widetilde{v}_{12} \zeta_{22} + a_1^2 a_2^2 (-\widetilde{v}_{12} \zeta_{11} - 2\widetilde{v}_{11} \zeta_{12} \right. \\
&\left. + 2\widetilde{v}_{22} \zeta_{12} + \widetilde{v}_{12} \zeta_{22}) + a_1^2 a_3^2 (2\widetilde{v}_{23} \zeta_{13} + \widetilde{v}_{12} \zeta_{33}) - a_2^2 a_3^2 (2\widetilde{v}_{13} \zeta_{23} + \widetilde{v}_{12} \zeta_{33}) \right].
\end{aligned} \tag{3.2.34}
$$

In the following we will compute the expected PBH spin, focusing on the first order contribution in perturbation theory. In order to do so, we will study first the properties of perturbations when they are still in the super-horizon limit. And then discuss the evolution of the spin under torques up to the turnaround time.

### 3.2.2  The PBH dimensionless Kerr parameter at formation time

Let us divide our analysis in the dynamics before and after horizon crossing, as the properties of the collapsing perturbations change in the two regimes. In the last part of this section we will provide the estimate for the PBH spin at formation time.

#### The PBH spin before horizon crossing

The PBH scenario predicts that PBHs form when the density perturbations cross the horizon and can collapse under gravitational pressure. In the stages preceding the collapse, the density perturbations have characteristic wavelengths much larger than the horizon and the separate universe approach can be applied [294]. Therefore, one performs the gradient expansion to describe, in a simpler form, the relevant observables. At this point, the choice of gauge becomes important. We will adopt the gauge in which numerical simulations are performed, going under the name of constant mean curvature slicing (CMC) [143].

In the CMC slicing, the overdensity is related to the comoving curvature perturbation as [294]

$$\delta(\vec{x}, \eta) = -\frac{4}{3} \frac{1}{\mathcal{H}^2} e^{-5\zeta(\vec{x})/2} \nabla^2 e^{\zeta(\vec{x})/2}, \tag{3.2.35}$$

This expression is fully non-perturbative and only assumes an expansion in gradients. We also defined the Hubble parameter in terms of the conformal time as $\mathcal{H} = aH$. Notice that this quantity is 3/2 times larger than the one in the comoving slicing (CG) (2.1.3), and consequently the threshold for PBH formation is larger than the one presented in Sec.2.1, i.e. $\delta_{\text{cmc}}^c = 3\delta_{\text{CG}}^c/2$.

As the universe expands, the overdensities grow, scaling like $\delta \sim a^2$. In the rare regions where peaks are so high to reach the threshold, the perturbation can collapse after they re-enter the horizon, i.e. when their characteristic size becomes comparable to the cosmological horizon. Even though the gradient expansion becomes increasingly less accurate, it was shown to lead to an acceptable criterion for the PBH formation, as confirmed by nonlinear numerical studies [143]. Therefore, we will adopt a similar strategy in the following. The spin is then expected to grow until the system completely decouples from the background expansion and the rest of the environment leading to a reduction of the torques.

In the gauge where the off-diagonal $_{0i}$ component of the pertubed metric are zero, the velocity is [294]

$$v^i(\vec{x}, \eta) = \frac{1}{12\mathcal{H}} \partial^i \delta(\vec{x}, \eta). \tag{3.2.36}$$

Therefore, the velocity is found to grow more rapidly than the density contrast, evolving as $\delta \sim a^3$ on superhorizon scales. An important consequence of Eq. (3.2.36) is that the velocity shear has vanishing off-diagonal terms. In physical terms, this can be interpreted as the velocity shear $v_l^k$ being always aligned with the principal axis of the inertia tensor, therefore no spin can be generated, independently of the deviation from sphericity of the collapsing region. This conclusion holds for any particular realization of a density peak, regardless of the probability distribution of the random variables $\delta$ and $v^i$. Notice also that the same is true for the contribution of the spin at second order. The relation (3.2.36) only assumes gradient expansion (i.e. $k \ll \mathcal{H}$).

**The PBH spin after horizon crossing**

After the perturbations cross the horizon, the relation between the velocity and the overdensity changes. Therefore, as we will see, the spin of the collapsing overdensity can build up due to the linear tidal torque. This growth stops when the perturbation decouples from the background. We can estimate the time of decoupling as follows. The Jeans criterion, found by equating the sound crossing and free-fall timescale, dictates that gravitational instability occurs when

$$\frac{k}{\mathcal{H}} \lesssim \sqrt{\frac{2\pi}{3}} \frac{(1 + \delta_{\text{pk}})^{1/2}}{c_{\text{S}}} \approx \frac{2}{c_{\text{S}}}. \tag{3.2.37}$$

As a radiation dominated fluid is described by $c_{\text{S}} \simeq 1/\sqrt{3}$, the perturbations must decouple from the background around horizon crossing. If not, the effect of radiation pressure would stabilise them or disperse the overdensity. This is also observed in numerical simulations.

In the following, we will assume therefore that turnaround occurs at horizon crossing and there is only a small time window for acquiring angular momentum through linear tidal-torque during the collapse. This is one of the reasons why, as we will see, the final PBH spin is small.

We describe the evolution of perturbations after horizon crossing adopting the CMC slicing in order to be consistent with the previous section, translating them from the more familiar Newtonian longitudinal gauge. In the following derivation, we will only focus on perturbation with characteristic comoving scale $k \gg k_{\text{eq}}$, where $k_{\text{eq}}$ corresponds to the mode re-entering the horizon at matter radiation equality, which is always satisfied in the PBH scenario, as the collapse takes place deep in radiation domination. Since the mean free path is very short, one can neglect the impact of anisotropic stress and take the Bardeen potentials [378] to be $\Phi = \Psi$. Near horizon crossing, the radiation density perturbation is [379]

$$\delta(\vec{k}, \eta) \simeq -6\Phi(\vec{k}, 0) \cos(kc_{\text{S}}\eta) + 4\Phi(\vec{k}, \eta), \tag{3.2.38}$$

where

$$\Phi(\vec{k}, \eta) = 3\Phi(\vec{k}, 0) \frac{\sin(kc_{\text{S}}\eta) - (kc_{\text{S}}\eta) \cos(kc_{\text{S}}\eta)}{(kc_{\text{S}}\eta)^3}. \tag{3.2.39}$$

At the same time, the radiation velocity modes are expressed as

$$v^i(\vec{k}, \eta) = i\frac{9}{2}\frac{k^i}{k}\Phi(\vec{k}, 0) c_{\text{S}} \sin(kc_{\text{S}}\eta), \tag{3.2.40}$$

while the (physical) velocity shear at horizon crossing is given by

$$v_i^j(\vec{k}, \eta_H) = ik_i v^j(\vec{k}, \eta_H) = -\frac{9}{2}\frac{k_i k^j}{k}\Phi(\vec{k}, 0)c_{\text{S}} \sin(k/k_{\text{S}}) \tag{3.2.41}$$

in terms of $k_{\text{S}} \equiv k_H/c_{\text{S}}$. In the CMC gauge, both the density perturbation and the velocity shear take the form [20]

$$\delta_{\text{cmc}}(\vec{k}, \eta_H) = -6\Phi(\vec{k}, 0) \frac{\left[2\left(3c_{\text{S}}^2 + 1\right) + (k/k_{\text{S}})^2\right]\cos(k/k_{\text{S}}) - 2\left(3c_{\text{S}}^2 + 1\right)(k_{\text{S}}/k)\sin(k/k_{\text{S}})}{6c_{\text{S}}^2 + (k/k_{\text{S}})^2},$$

$$v_{\text{cmc}i}^j(\vec{k}, \eta_H) = -\frac{9}{2}\Phi(\vec{k}, 0)\frac{k_i k^j}{k}c_{\text{S}}\left(\frac{k}{k_{\text{S}}}\right)^2 \frac{\sin(k/k_{\text{S}})}{6c_{\text{S}}^2 + (k/k_{\text{S}})^2}. \tag{3.2.42}$$

With the previous definition, we can introduce the rescaled density $\nu$, shear $\widetilde{v}_{ij}$ and Hessian $\zeta_{ij}$, still adopting the same notation used in [367]:

$$
\begin{aligned}
\sigma_{\delta_{\text{cmc}}} \nu(\vec{x}, \eta_H) &= \frac{V}{(2\pi)^3} \int \mathrm{d}^3 k \left(\frac{k}{k_{\text{S}}}\right)^2 T_\delta(k, \eta_H) \, \Phi(\vec{k}, 0) \, W(k) \, e^{i\vec{k}\cdot\vec{x}}, \\
\zeta_{\text{cmc}ij}(\vec{x}, \eta_H) &= -\frac{V}{(2\pi)^3} \int \mathrm{d}^3 k \, k_i k_j \, \delta_{\text{cmc}}(\vec{k}, \eta_H) \, W(k) \, e^{i\vec{k}\cdot\vec{x}}, \\
v_{\text{cmc}i}^j(\vec{x}, \eta_H) &= -k_H \frac{V}{(2\pi)^3} \int \mathrm{d}^3 k \, \frac{k_i k^j}{k^2} \frac{T_v(k, \eta_H)}{T_\delta(k, \eta_H)} \delta_{\text{cmc}}(\vec{k}, \eta_H) \, W(k) \, e^{i\vec{k}\cdot\vec{x}} \equiv g_v(\eta_H) \widetilde{v}_{\text{cmc}i}^j(\vec{x}, \eta_H),
\end{aligned}
\tag{3.2.43}
$$

where $W(k)$ is the familiar window function defined in Eq. (2.2.4) with a smoothing scale $R = 1/k_H$, while we defined the effective transfer functions

$$
\begin{aligned}
T_\delta(k, \eta_H) &= -6 \left(\frac{k_{\text{S}}}{k}\right)^2 \frac{\left[2\left(3c_{\text{S}}^2+1\right)+(k/k_{\text{S}})^2\right]\cos\left(k/k_{\text{S}}\right) - 2\left(3c_{\text{S}}^2+1\right)(k_{\text{S}}/k)\sin\left(k/k_{\text{S}}\right)}{6c_{\text{S}}^2 + (k/k_{\text{S}})^2}, \\
T_v(k, \eta_H) &= \frac{9}{2} \left(\frac{k}{k_{\text{S}}}\right) \frac{\sin\left(k/k_{\text{S}}\right)}{6c_{\text{S}}^2 + (k/k_{\text{S}})^2}.
\end{aligned}
\tag{3.2.44}
$$

From Eq. (3.2.43), we can immediately read that in the sub-horizon regime the velocity shear is not proportional to the gradients of the density contrast. Therefore, the velocity shear presents non-vanishing off-diagonal terms, as opposed to the super-horizon regime. This leads to the generation of some angular momentum at linear order. The characteristic variances can be computed from the spectral moments as

$$
\sigma_{\delta_{\text{cmc}}} = \sigma_0, \qquad \sigma_{\zeta_{\text{cmc}}} = \sigma_2, \qquad \sigma_{\times_{\text{cmc}}} = \sigma_1
\tag{3.2.45}
$$

where

$$
\sigma_j^2 \equiv \frac{V}{2\pi^2} \int \mathrm{d}k \, k^{2+2j} \left|\delta_{\text{cmc}}(\vec{k}, \eta_H)\right|^2 W^2(k).
\tag{3.2.46}
$$

Finally, the velocity shear time evolution, defined in Eq. (3.2.14), is found to be

$$
\begin{aligned}
g_v^2(\eta_H) &= k_H^2 \frac{V}{2\pi^2} \int \mathrm{d}k \, k^2 \frac{T_v^2(k, \eta_H)}{T_\delta^2(k, \eta_H)} \left|\delta_{\text{cmc}}(\vec{k}, \eta_H)\right|^2 W^2(k) \\
&\sim \frac{T_v^2(k_H, \eta_H)}{T_\delta^2(k_H, \eta_H)} k_H^2 \sigma_{\delta_{\text{cmc}}}^2(\eta_H),
\end{aligned}
\tag{3.2.47}
$$

where in the second step we neglected the time dependence of the transfer functions as, in the relevant cases, the power spectrum is peaked at the reference scale $k = k_H$. The ratio in Eq. (3.2.47) is found to be $T_v(k_H, \eta_H)/T_\delta(k_H, \eta_H) \approx 0.5$. Finally, working with the assumption that turnaround takes place soon after horizon crossing, the linear order reference spin from Eq. (3.2.23) is found to be

$$
S_{\text{ref}}^{(1)}(\eta_H) = \frac{4}{3} a^4(\eta_H) \bar{\rho}_{\text{rad}}(\eta_H) g_v(\eta_H) R_*^5 (1-f)^{5/2}.
\tag{3.2.48}
$$

This expression gives rise to the dominant contribution to the PBH spin at formation. From now on, we will drop the label $^{(1)}$ to simplify the notation.

### An estimate of the PBH dimensionless Kerr parameter at formation time

As usually done to describe the spin of a Kerr BH, we define the dimensionless parameter $\chi$ as

$$
\chi = \frac{S}{G_N M^2}.
\tag{3.2.49}
$$

From the definition of the spin at first order (3.2.22), we write

$$
\chi = \frac{S_{\text{ref}}(\eta_H)}{G_N M^2} s_{\text{e}} \equiv A(\eta_H) s_{\text{e}},
\tag{3.2.50}
$$

where the pre-factor

$$A(\eta_H) = \frac{4}{3} \frac{a^4(\eta_H) g_v(\eta_H) \overline{\rho}_{\rm rad}(\eta_H) R_*^5 (1-f)^{5/2}}{G_N M^2} \qquad (3.2.51)$$

accounts for the time dependent evolution of perturbations while the rescaled spin parameter $s_{\rm e}$ accounts for the statistical variations of the geometry of the collapsing overdensity. We can estimate the factor $f$ using $(1-f) \sim 1/3$ (in the CMC gauge), while and $R_* \sim \sqrt{3} k_H^{-1} \sim \sqrt{3} \mathcal{H}^{-1}(\eta_H)$ [291]. After few careful manipulations [20], the Kerr parameter at the time of formation is given by the simple relation

$$\chi = A(\eta_H) s_{\rm e} \sim \left[ \frac{1}{2\pi} \sigma_{\delta_{\rm cmc}}(\eta_H) \right] s_{\rm e}. \qquad (3.2.52)$$

Since $\chi$ scales linearly with the square root of the variance of the density contrast, it is explicitly given in terms of first-order relation. Notice that a characteristic value for the variance in the PBH scenario can be taken to be $\sigma_{\delta_{\rm cmc}}(\eta_H) = \delta_{\rm cmc}^{\rm c}/\nu \sim 0.08$ giving $\chi \approx 10^{-2} s_{\rm e}$, for $\nu = 8$.

The missing step now is only the characterisation of the distribution of the rescaled spin $s_{\rm e}$. As an order of magnitude estimate, we can take $s_{\rm e}$ from Eq. (3.2.19) to be

$$s_{\rm e} \simeq \frac{16\sqrt{2}\pi}{135\sqrt{3}\Lambda} \nu^{5/2} \widetilde{v}_{ij} \alpha_i \qquad \text{for} \qquad i \neq j. \qquad (3.2.53)$$

In the following section we will show that the velocity shear is scaling like $\sqrt{1-\gamma^2}$, with $\gamma = \sigma_\times^2 / \sigma_\delta \sigma_\zeta$ and therefore the final prediction for the characteristic PBH spin magnitude at formation is

$$\chi \sim 10^{-2} \sqrt{1-\gamma^2}. \qquad (3.2.54)$$

If one considers the limit of monochromatic perturbations, the parameter describing the shape of the power spectrum $\gamma$ tends to unity. In that case, the spin vanishes. This limit can be explained by looking at Eq. (3.2.59). The velocity shear tends to be aligned with the principal axes of the perturbation ellipsoid, due to its correlation with the inertia tensor. In the limit $\gamma \to 1$, one achieves a complete alignment and no spin can be generated.

**The rescaled PBH spin from the statistics of local maxima**

In this section we compute the expectation value of the rescaled spin $s_{\rm e}$, along with its distribution, starting from the probability distributions of the key quantities entering in its definition. Notice that in this section we will adopt the standard approach used in peak theory [297], but extending it to account for the velocity shear in addition to the density field.

The (correlated) variables entering in $s_{\rm e}$ are

$$\delta, \quad \zeta_i = \frac{\partial \delta}{\partial x_i}, \quad \zeta_{ij} = \frac{\partial^2 \delta}{\partial x_i \partial x_j}, \quad v_i^j = \frac{\partial v^j}{\partial x^i}, \qquad (3.2.55)$$

where we are adopting the same notation used in Ref. [365]. Notice the last two matrices only have six independent components as they are both symmetric. We dropped the label CMC in order to simplify the notation in this section. We stack all the variables in a single vector $V$, to get the gaussian joint distribution

$$f(V_i) \mathrm{d}^{16} V_i = \frac{1}{(2\pi)^8 |\mathbf{M}|^{1/2}} e^{-\frac{1}{2}(V_i - \langle V_i \rangle) \mathbf{M}_{ij}^{-1} (V_j - \langle V_j \rangle)} \mathrm{d}^{16} V_i, \qquad (3.2.56)$$

where the covariance matrix $\mathbf{M}$ is defined as

$$\mathbf{M}_{ij} = \langle (V_i - \langle V_i \rangle)(V_j - \langle V_j \rangle) \rangle. \qquad (3.2.57)$$

We can change variables to their dimensionless form by defining

$$\nu = \delta/\sigma_\delta,$$

$$x = -(\zeta_{11} + \zeta_{22} + \zeta_{33})/\sigma_\zeta, \qquad y = -\frac{1}{2}(\zeta_{11} - \zeta_{33})/\sigma_\zeta, \qquad z = -\frac{1}{2}(\zeta_{11} - 2\zeta_{22} + \zeta_{33})/\sigma_\zeta,$$

$$v_A = -(\widetilde{v}_{11} + \widetilde{v}_{22} + \widetilde{v}_{33}), \qquad v_B = -\frac{1}{2}(\widetilde{v}_{11} - \widetilde{v}_{33}), \qquad v_C = -\frac{1}{2}(\widetilde{v}_{11} - 2\widetilde{v}_{22} + \widetilde{v}_{33}),$$

$$w_1 = \widetilde{v}_{23}, \qquad w_2 = \widetilde{v}_{13}, \qquad w_3 = \widetilde{v}_{12},$$

$$\widetilde{\zeta}_1 = \zeta_1/\sigma_\times, \qquad \widetilde{\zeta}_2 = \zeta_2/\sigma_\times, \qquad \widetilde{\zeta}_3 = \zeta_3/\sigma_\times,$$

$$\widetilde{\zeta}_{12} = \zeta_{12}/\sigma_\zeta, \qquad \widetilde{\zeta}_{13} = \zeta_{13}/\sigma_\zeta, \qquad \widetilde{\zeta}_{23} = \zeta_{23}/\sigma_\zeta, \tag{3.2.58}$$

and compute all the non vanishing entries of the covariance matrix by setting the expectation value of the field to zero $\langle V_j \rangle = 0$ to get

$$\langle x^2 \rangle = \langle \nu^2 \rangle = \langle v_A^2 \rangle = 3\langle \widetilde{\zeta}_1^2 \rangle = 15\langle \widetilde{\zeta}_{12}^2 \rangle = 15\langle w_3^2 \rangle = \langle v_A \nu \rangle = ... = 1, \tag{3.2.59}$$

$$\langle x\nu \rangle = \langle x v_A \rangle = 5\langle v_C z \rangle = 15\langle v_B y \rangle = 15\langle \widetilde{\zeta}_{12} w_3 \rangle = ... = \gamma, \tag{3.2.60}$$

$$\langle z^2 \rangle = 3\langle y^2 \rangle = \langle v_C^2 \rangle = 3\langle v_B^2 \rangle = 1/5. \tag{3.2.61}$$

The ellipses indicate the analogous terms with a permutation of the indices and, as we defined in the previous section, $\gamma = \sigma_\times^2/\sigma_\delta\sigma_\zeta$. As $\langle v_A \nu \rangle^2 = \langle v_A^2 \rangle \langle \nu^2 \rangle = 1$, the variables $v_A$ and $\nu$ are fully correlated, so one can be expressed in terms of the other and we can reduce the sample to fifteen independent variables. Here we stress also that, in the limit $\gamma = 1$, the velocity shear is aligned with the inertia tensor, and the expected spin vanishes, as it was found in the preceding section.

The system is independent of three Euler angles describing the orientation of the principal axes of $-\zeta_{ij}/\sigma_\zeta$, so we can integrate over them. Then, we can define the eigenvalues of the matrix $-\zeta_{ij}/\sigma_\zeta$ as

$$x = \lambda_1 + \lambda_2 + \lambda_3, \quad y = \frac{1}{2}(\lambda_1 - \lambda_3), \quad z = \frac{1}{2}(\lambda_1 - 2\lambda_2 + \lambda_3), \tag{3.2.62}$$

to express the distribution of the remaining 12 variables as

$$f(\nu, \widetilde{\zeta}_i, \lambda_i, v_B, v_C, w_i) = \frac{5^5 3^{11/2}}{2(2\pi)^{11/2}} \Gamma^3 |(\lambda_1 - \lambda_2)(\lambda_2 - \lambda_3)(\lambda_1 - \lambda_3)| e^{-Q_2/2}, \tag{3.2.63}$$

with

$$\Gamma = \frac{1}{1 - \gamma^2}, \tag{3.2.64}$$

and

$$Q_2 = \Gamma\nu^2 - 2\gamma\Gamma x\nu + \Gamma x^2 + 15\Gamma y^2 - 30\gamma\Gamma y v_B + 15\Gamma v_B^2 + 5\Gamma z^2$$
$$- 10\gamma\Gamma z v_C + 5\Gamma v_C^2 + 15\Gamma(w_1^2 + w_2^2 + w_3^2) + 3(\widetilde{\zeta}_1^2 + \widetilde{\zeta}_2^2 + \widetilde{\zeta}_3^2). \tag{3.2.65}$$

In order to proceed, and consistently with the peak theory picture, we can simplify the expressions by expanding the derivatives of the density fields around its peak as

$$\zeta_i = \frac{\partial\delta}{\partial x_i} = \left(\frac{\partial^2\delta}{\partial x_i \partial x_j}\right)\bigg|_{\mathrm{pk}} (x - x_{\mathrm{pk}})^j = \zeta_{ij}|_{\mathrm{pk}}(x - x_{\mathrm{pk}})^j, \tag{3.2.66}$$

such that

$$\mathrm{d}^3\widetilde{\zeta}_i = \left(\frac{\sigma_\zeta}{\sigma_\times}\right)^3 |\lambda_1\lambda_2\lambda_3| \mathrm{d}^3 x_i. \tag{3.2.67}$$

We can then marginalise over $\widetilde{\zeta}_i$, $v_B$ and $v_C$ to get the final number density of peaks in the volume element $\mathrm{d}\nu\mathrm{d}^3\lambda_i\mathrm{d}^3 w_i$ as [20]

$$N_{\mathrm{pk}}(\nu, \lambda_i, w_i)\mathrm{d}\nu\mathrm{d}^3\lambda_i\mathrm{d}^3 w_i = \frac{5^4 3^{15/2}}{2^{13/2}\pi^{9/2}} \frac{\Gamma^2}{R_*^3} \lambda_1\lambda_2\lambda_3(\lambda_1 - \lambda_2)(\lambda_2 - \lambda_3)(\lambda_1 - \lambda_3)e^{-Q_4/2}\mathrm{d}\nu\mathrm{d}^3\lambda_i\mathrm{d}^3 w_i, \tag{3.2.68}$$

where

$$Q_4 = \nu^2 + \Gamma(x - x_*)^2 + 15y^2 + 5z^2 + 15\Gamma w^2, \qquad (3.2.69)$$

with $x_* = \gamma\nu$ and $w^2 = w_1^2 + w_2^2 + w_3^2$.

In the end we are interested in computing the expectation value and distribution of the rescaled spin $s_e$, so we can change variable by using Eq. (3.2.19) as

$$s_e = \frac{2^{9/2}\pi\nu^{5/2}w}{5 \times 3^{7/2}\gamma^{5/2}\sqrt{\lambda_1\lambda_2\lambda_3}}\sqrt{\beta^2 + (\alpha_3^2 - \beta^2)u^2}, \qquad (3.2.70)$$

where

$$\beta^2(\lambda_i, \phi) = \alpha_1^2\cos^2\phi + \alpha_2^2\sin^2\phi, \qquad (3.2.71)$$

and use it to rewrite Eq. (3.2.68)

$$N_{\text{pk}}(\nu, \lambda_i, w_i)\text{d}\nu\text{d}^3\lambda_i\text{d}^3w_i = N_{\text{pk}}(\nu, \lambda_i, w, u, \phi)w^2\text{d}w\text{d}\phi\text{d}^3\lambda_i\text{d}\nu\frac{\text{d}u}{\text{d}s_e}\text{d}s_e. \qquad (3.2.72)$$

The total number of peaks of height $\nu$ and rescaled spin $s_e$ are then

$$N_{\text{pk}}(\nu, s_e) = \frac{4Cs_e}{R_*^3\nu^5}\int_0^\infty\text{d}\lambda_1\int_0^{\lambda_1}\text{d}\lambda_2\int_0^{\lambda_2}\text{d}\lambda_3\int_{\alpha_1}^{\alpha_2}\text{d}\beta\frac{e^{-Q_5/2}F(\lambda_i)\Lambda T(\alpha_3, \beta, s_e, \nu)}{\sqrt{|(\alpha_1^2 - \beta^2)(\alpha_2^2 - \beta^2)(\alpha_3^2 - \beta^2)|}}, \qquad (3.2.73)$$

where

$$\Lambda = \lambda_1\lambda_2\lambda_3, \quad C = \frac{3^{11}5^{11/2}\gamma^5\Gamma^{3/2}}{2^{13}\pi^{13/2}}, \quad Q_5 = \nu^2 + \Gamma(x - x_*)^2 + 15y^2 + 5z^2, \qquad (3.2.74)$$

and

$$T = \Theta(\alpha_3^2 - \beta^2)e^{\frac{-15\Gamma w_3^2}{2}}D(X) + \Theta(\beta^2 - \alpha_3^2)\frac{\sqrt{\pi}}{2}e^{\frac{-15\Gamma w_\beta^2}{2}}\text{erf}(X). \qquad (3.2.75)$$

Here the argument of the Dawson's integral $D(X)$ is given by

$$X = \sqrt{\frac{15}{2}\Gamma|w_\beta^2 - w_3^2|}, \qquad (3.2.76)$$

with

$$w_3 = \frac{\sqrt{\Lambda}s_e}{K\nu^{5/2}\alpha_3}, \quad w_\beta = \frac{\sqrt{\Lambda}s_e}{K\nu^{5/2}\beta}, \quad K = \frac{2^{9/2}\pi}{5 \times 3^{7/2}\gamma^{5/2}}. \qquad (3.2.77)$$

Finally, knowing the total number of peaks, we can translate the result as a conditional probability distribution function for $s_e$ given a specific height $\nu$ as [365]

$$P(s_e|\nu)\text{d}s_e = \frac{N_{\text{pk}}(\nu, s_e)}{N_{\text{pk}}(\nu)}\text{d}s_e, \qquad (3.2.78)$$

where the comoving differential peak density as a function of the peak height is given by (see App. A)

$$N_{\text{pk}}(\nu)\text{d}\nu = \frac{1}{(2\pi)^2R_*^3}e^{-\frac{\nu^2}{2}}G(\gamma, x_*)\text{d}\nu. \qquad (3.2.79)$$

In the previous step we introduced the function

$$G(\gamma, x_*) = \int_0^\infty\text{d}xf(x)\sqrt{\frac{\Gamma}{2\pi}}e^{-\frac{\Gamma}{2}(x - x_*)^2} \qquad (3.2.80)$$

and

$$f(x) = \frac{(x^3 - 3x)}{2}\left[\text{erf}\left(x\sqrt{\frac{5}{2}}\right) + \text{erf}\left(\frac{x}{2}\sqrt{\frac{5}{2}}\right)\right] + \sqrt{\frac{2}{5\pi}}\left[\left(\frac{31x^2}{4} + \frac{8}{5}\right)e^{-\frac{5x^2}{8}} + \left(\frac{x^2}{2} - \frac{8}{5}\right)e^{-\frac{5x^2}{2}}\right]. \qquad (3.2.81)$$

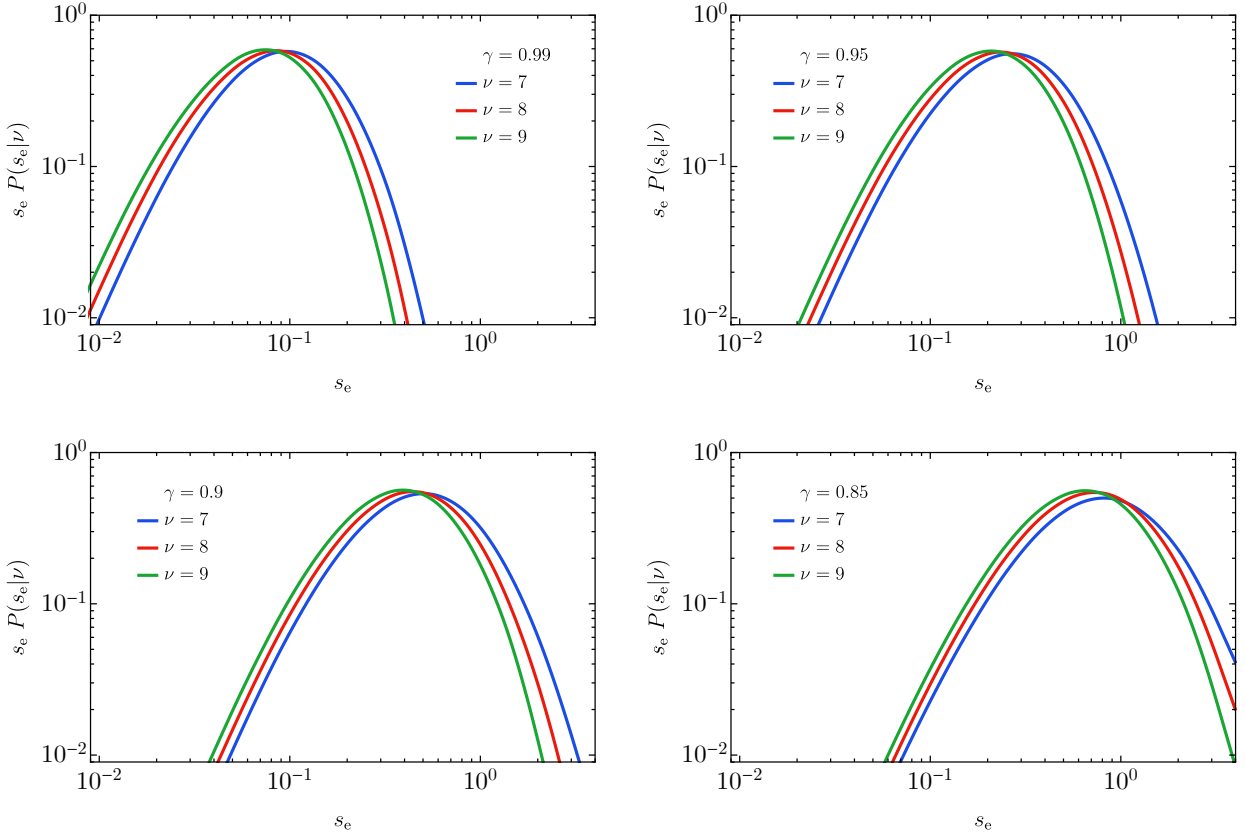

Figure 3.2: *Normalised distribution function for $s_e$.*

Unfortunately, it is not possible to integrate Eq. (3.2.73) to provide an analytical formula for the distribution of the rescaled spin. Such integration must be performed numerically. In Fig. 3.2 we show the resulting distribution $P(s_e|\nu)$ for few characteristic choices of $\nu$. As one can easily appreciate from the figures, the conditional probability distribution shows a systematic shift towards smaller values of the $x$-axis for increasingly higher peaks (i.e. larger values of $\nu$). This can be interpreted as the preference of high peaks to possess increasingly more spherical configurations. Furthermore, at fixed $\nu$, larger values of $\gamma$ translates to smaller characteristic spin values, consistently with the approximation in Eq. (6.3.20) scaling like $s_e \propto \sqrt{1-\gamma^2}$.

### 3.2.3   The analytical spin distribution in the limit of high peaks

Being required to be a rare event, the formation of a PBH is always accompanied by a small value of the variance compared to the threshold, meaning a large value of $\nu$. We can therefore investigate analytically the large $\nu$ limit. In this setup, the probability $P(s_e|\nu)$ can be described in terms of the a rescaled variable $h$ defined as[3]

$$s_e \equiv \frac{2^{9/2}\pi}{5\gamma^6\nu}\frac{h}{\Gamma^{1/2}} = \frac{2^{9/2}\pi}{5\gamma^6\nu}\sqrt{1-\gamma^2}\,h. \tag{3.2.82}$$

The characteristic scaling $\propto \sqrt{1-\gamma^2}$ is dictated by the behaviour of the velocity shear, see Eq. (3.2.53). The distribution of the parameter $h$ can be analytically approximated as

$$P(h)\mathrm{d}h = \exp\left(-2.37 - 4.12\ln h - 1.53\ln^2 h - 0.13\ln^3 h\right)\mathrm{d}h. \tag{3.2.83}$$

This distribution is the result of a numerical fit performed choosing the parameter $\gamma = 0.9$ and $\nu = 8$. We checked that it provides a satisfactory approximation of the distribution in the parameter space relevant for the PBH formation. We also show how the approximation in Eq. (3.2.83), translated in terms of $P(s_e|\nu)$, compares to the numerical results in Fig. 3.3.

---

[3]We highlight to the reader as a note of caution that Eq. (C.7) in Ref. [365] presents an incorrect scaling with $\Gamma$. The factor $\Gamma^{1/2}$ should be $\Gamma^{-1/2}$ instead.

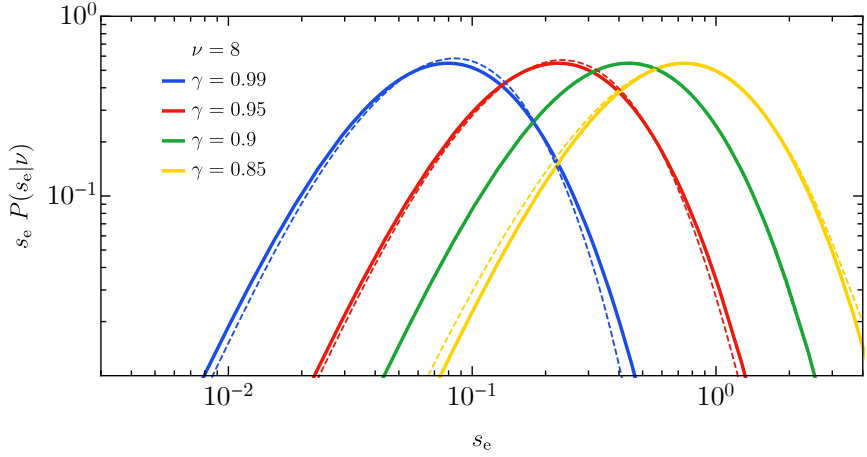

Figure 3.3: *Comparison between the numerical result (dashed line) and the fitted expression (solid lines) for the probability distribution $P(s_e|\nu)$ in the large $\nu$ limit.*

Finally, the distribution of the Kerr parameter $\chi$, in the large $\nu$ limit, is

$$
P(\chi|\nu)\mathrm{d}\chi = \left(\frac{5\gamma^6\nu}{2^{9/2}\pi}\frac{\Gamma^{1/2}}{A(\eta_H)}\right)\exp\left[-2.37 - 4.12\ln\left(\frac{5\gamma^6\nu}{2^{9/2}\pi}\frac{\Gamma^{1/2}}{A(\eta_H)}\chi\right)\right.
$$
$$
\left.-1.53\ln^2\left(\frac{5\gamma^6\nu}{2^{9/2}\pi}\frac{\Gamma^{1/2}}{A(\eta_H)}\chi\right) - 0.13\ln^3\left(\frac{5\gamma^6\nu}{2^{9/2}\pi}\frac{\Gamma^{1/2}}{A(\eta_H)}\chi\right)\right]\mathrm{d}\chi, \qquad (3.2.84)
$$

where $A(\eta_H)$ is defined in Eq. (4.3.2).

### 3.2.4 PBH abundance as a function of spin

In the preceding section, we have given a prescription for describing the distribution of PBH spin at formation. The result, however, is presented as a function of the peak height $\nu$. As the height of the peak is related to the local overdensity, and therefore to the mass of the PBH, we need to account for the critical collapse relation to find the relative abundance as a function of the PBH spin. This exercise was done in Ref. [380] where, however, the authors considered an initial flat distribution of $\chi$.

Adopting the critical collapse theory of radiation fluids, we can write [380]

$$
M = C_M|\delta - \delta_c(q)|^\gamma,
$$
$$
S = c_S|\delta - \delta_c(q)|^{\gamma_S}q,
$$
$$
\delta_c(q) = \delta_{0c} + Kq^2, \qquad (3.2.85)
$$

where $\gamma = 0.3558$, $\gamma_S = (5/2)\gamma = 0.8895$ and $K = 0.005685$ [381]. The parameter $q$ describes the effect of rotation on the threshold and it is related to the Kerr parameter $\chi$ by

$$
q = \frac{C_M^2}{C_J}\left(\frac{M}{C_M}\right)^{-1/2}\chi. \qquad (3.2.86)
$$

As one can intuitively expect, the larger the angular momentum, the higher the threshold for collapse. The parameter $\delta_c(q)$ is related to the density fluctuations $\delta$ as in Eq. (3.2.85), which can be inverted to get

$$
\delta = \delta_c(q) + \left(\frac{M}{C_M}\right)^{1/\gamma} = \delta_{0c} + K\left(\frac{C_M^2}{C_J}\right)^2\left(\frac{M}{C_M}\right)^{-1}\chi^2 + \left(\frac{M}{C_M}\right)^{1/\gamma}. \qquad (3.2.87)
$$

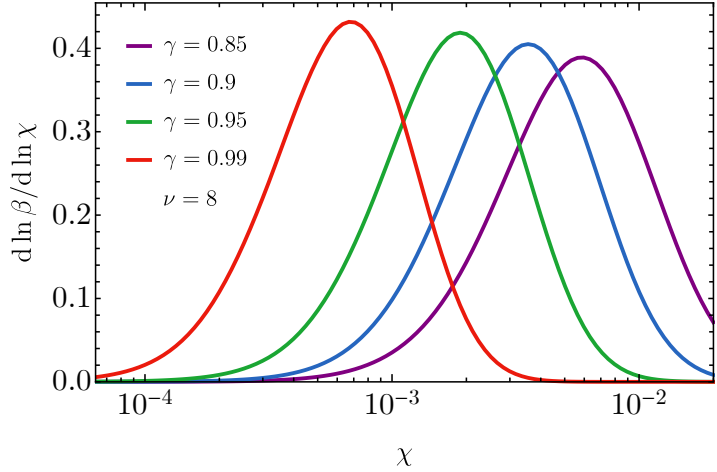

Figure 3.4: *The final relative abundance as a function of the spin. The results are obtained with $C_M = 5.118 M_H$ and $C_J = 26.19 M_H$ [380]. We adopted a value of $\delta_{0c}/\sigma_\delta = 8$.*

For demonstrative purposes, we adopt the linear approximation in which the distribution of the density contrast is gaussian with a characteristic variance $\sigma_\delta$. Therefore, the mass fraction at formation can be found to be

$$\beta = \frac{1}{M_H} \int_0^\infty \mathrm{d}q P(q) \int_{\delta_c(q)} \mathrm{d}\delta M(\delta) P(\delta), \tag{3.2.88}$$

We can change variables from $(q, \delta)$ to $(M, \chi)$, and the measure becomes

$$\mathrm{d}q\, \mathrm{d}\delta = \frac{C_M}{\gamma\, C_J} \left(\frac{M}{C_M}\right)^{-\frac{3}{2}+\frac{1}{\gamma}} \mathrm{d}\chi\, \mathrm{d}M. \tag{3.2.89}$$

Finally, the mass fraction can be written as

$$\beta = \frac{1}{M_H} \frac{1}{\sqrt{2\pi}\sigma_\delta} \frac{C_M^2}{\gamma\, C_J} \int_0^\infty \mathrm{d}\chi P(\chi) \int \mathrm{d}M \left(\frac{M}{C_M}\right)^{-\frac{1}{2}+\frac{1}{\gamma}} e^{-\frac{1}{2\sigma_\delta^2} \left(\delta_{0c} + K\left(\frac{C_M^2}{C_J}\right)^2 \left(\frac{M}{C_M}\right)^{-1} \chi^2 + \left(\frac{M}{C_M}\right)^{1/\gamma}\right)^2}, \tag{3.2.90}$$

where $P(\chi)$ is the distribution of the Kerr parameter $\chi$ we have previously calculated. As one can expect by looking at Eq. (3.2.85) and from the predicted characteristic amplitude of $\chi \lesssim 10^{-2}$, the effect of the spin on the threshold is minor[4] and no significant change on the overall abundance is observed. Also, we do not find significant deviations from the critical mass distribution due to the inclusion of spin effects [20]. On the other hand, by differentiating $\beta$ with respect to $\chi$ and marginalising over the mass, we find the final distribution of the spin of PBHs right after formation. In Fig. 3.4 we plot $\mathrm{d}\ln\beta/\mathrm{d}\ln\chi$ found using the Kerr parameter distribution $P(\chi)$ obtained numerically in the previous section. The trends described in the preceding discussion are also confirmed here.

We conclude with few important comments. First, all the results found in this section were derived assuming the density contrast and the velocity shear are gaussian. It would be interesting to see how the results may change with the inclusion of non-Gaussianities. Secondly, we stress that in this section we derived the expected spin of PBHs coming from the collapse of radiation overdensities and found it to be at the sub-percent level *at formation time*. It is important to keep in mind that at later stages of the universe, the PBH spin can grow under, for example, the effect of baryonic accretion. Therefore, the expected spin distribution at low redshift could be, in fact, very different. This may change drastically the prediction of the PBH model for merger events detectable by LIGO/Virgo experiments. We will come back to this important issue in the following chapter. Finally, we mention that in case one considered PBH formation in a different (non-standard) scenario, the prediction for the PBH spin at formation may be different. While for PBHs produced by long-range scalar-mediated forces [383] the spin at formation may be small [384], other scenarios such as the collapse during an early matter-dominated epoch [385], survival of Planck-scale relics [386, 387] and the collapse of Q-balls [388] may generate a larger initial spin.

---

[4] This result was also confirmed in Ref. [382].

## 3.3   Clustering at formation

As we have seen in the preceding parts of this chapter, PBHs are formed from the collapse of density perturbations in the very early universe at the time $t_H$ of horizon crossing, i.e. when the characteristic scale of the perturbation $r_m$ is comparable to the Hubble horizon $R_H$. After they form, they can be described as discrete objects populating our universe. An important question is how their spatial distribution is characterised. In other words, it is interesting to understand if their spatial distribution follows a Poisson distribution (dictating random positioning in a given volume) or if they are more likely to be close to each other forming clusters of PBHs. We will address this question focusing on the clustering properties *at formation*, while leaving the question of how clustering evolves with redshift for the next chapter.

We can describe the PBH correlation function, which dictates the excess probability of finding PBHs in a given volume with respect to the average, using the peak approach introduced to describe the large scale structure of the universe, see for example Refs. [297, 389]. We define the overdensity of PBH centred at position $\vec{x}_i$ as

$$\delta_{\mathrm{PBH}}(\mathbf{x}) = \frac{1}{n_{\mathrm{PBH}}} \sum_i \delta_D(\vec{x} - \vec{x}_i) - 1 \;, \tag{3.3.1}$$

where $\delta_D(\vec{x})$ identifies the Dirac delta distribution, $n_{\mathrm{PBH}}$ is the average number density of PBH per comoving volume and $i$ indicates the various position of PBHs after formation. One can compute the PBH two point function as

$$\begin{aligned}
\left\langle \delta_{\mathrm{PBH}}(\vec{x})\delta_{\mathrm{PBH}}(0) \right\rangle &= \frac{1}{n_{\mathrm{PBH}}} \delta_D(\vec{x}) - 1 + \frac{1}{n_{\mathrm{PBH}}^2} \left\langle \sum_{i \neq j} \delta_D(\vec{x} - \vec{x}_i)\delta_D(\vec{x}_j) \right\rangle \\
&= \frac{1}{n_{\mathrm{PBH}}} \delta_D(\vec{x}) + \xi_{\mathrm{PBH}}(x)
\end{aligned} \tag{3.3.2}$$

where, in complete analogy with the large scale structure theory (see for example Ref. [390, 391]), $\xi_{\mathrm{PBH}}(x)$ indicates the reduced correlation function. As PBHs come from the collapse of radiation overdensities, one expects them to trace the underlying radiation density fluid and therefore, at large scales,

$$\xi_{\mathrm{PBH}}(x) = b_0^2 \xi_r(x) \;. \tag{3.3.3}$$

In this relation, we denoted $\xi_r(x)$ as the radiation correlation function and defined $b_0$ to be the linear, scale-independent, PBH bias. It is also useful to define the local PBH number density perturbation at the position $\vec{x}$ as

$$\delta_{\mathrm{PBH}}(\vec{x}) \equiv \frac{P_1(> \nu | \vec{x})}{P_1(> \nu)} - 1 \tag{3.3.4}$$

where $P_1$ is the probability distribution function for the density field.

In the following, we will show that PBHs are not clustered at formation if coming from gaussian density perturbations [18, 322, 392–394]. This is a simple consequence of the fact that PBHs form at the scale of the horizon, and it is difficult to generate correlations on scales larger than few Hubble patches, due to the equivalence principle. We then proceed to show that this conclusion is changed when non-Gaussianities are introduced into the model.

### 3.3.1 The (no-go) Gaussian case

We adopt the linear approximation relating the density contrast and the curvature perturbation at super horizon scales as

$$\delta(k, a) \simeq \frac{4}{9} \left( \frac{k}{aH} \right)^2 \zeta(k) \tag{3.3.5}$$

where the prefactor $R_H = 1/aH$ corresponds to the Hubble horizon size. We stress that we are forced to consider the density contrast as the criterion for collapse is expressed in terms of $\delta_c$ as discussed in Sec. 2.1. This is because it is not possible to define consistent a criterion for the collapse in terms of the curvature perturbations, as one would be sensitive to a shift of $\zeta$ by a constant value, which should not affect the local physics due to diffeomorphism invariance [303]. In such a situation, a careful treatment of the super-horizon modes is needed. As the density contrast is defined in terms of the laplacian of the curvature perturbations, it is insensitive to this problem.

In the following, we will adopt a notation similar to the peak-background split routinely adopted in the context of large scale structure theory. We will refer to $k_s$ as the short modes corresponding to the PBH scale where there is the enhancement of the power spectrum. Therefore, at the time of collapse $t_H$, one has $R_H \sim 1/k_s$. We also define $k_l$ as the modes at large scales ($k_l \ll k_s$ corresponding to a hierarchy of characteristic distances $r_l \gg r_s$). The local probability of collapse, in the standard Press-Schechter formalism, can be modulated by the presence of a long mode $\delta_l$. This happens if the collapse takes place in an overdense region modulated by a long perturbation $\delta_l$. In practice, the threshold $\delta_c$ is simply reduced to $\delta_c - \delta_l$, while leaving the variance of the short modes $\sigma_s^2$ unchanged, see a depiction in Fig. 3.5. In such a region, the PBH number density is therefore changed at first order in $\delta_l$ as

$$\delta_{\text{PBH}}(\vec{x}) \simeq \frac{\partial \delta_c}{\partial \delta_l} \frac{\partial \log P_1}{\partial \delta_c} \delta_l(\vec{x}) \simeq \frac{\nu}{\sigma_s} \delta_l(\vec{x}), \tag{3.3.6}$$

where in the last step we have expanded in the high peak limit $\nu \equiv \delta_c/\sigma_s \gg 1$. As $\delta_l$ corresponds to the radiation overdensity fluid, we can identify the scale independent bias given by the Kaiser term [389]

$$b_0 \simeq \frac{\delta_c}{\sigma_s^2} \simeq \frac{\nu}{\sigma_s}, \tag{3.3.7}$$

and the correlation function of PBHs is therefore, on large scales,

$$\xi_{\text{PBH}}(x, a_H) = b_0^2 \xi_r(x, a_H) . \tag{3.3.8}$$

For characteristic numbers involved in models producing stellar mass PBHs, one finds $b_0 \simeq \mathcal{O}(10^2)$, while $a_H$ indicates quantities evaluated at the time of horizon crossing of the small scales $t_H$.

We note that the spatial modulation must be imprinted at the time of PBH collapse, which corresponds to the time of horizon crossing of the short modes $aH \sim k_s$ and therefore although a biasing factor $\nu/\sigma_s$ is of order $b_0 \sim 10^2$, $\delta_{\text{PBH}}$ is still suppressed compared to amplitude of the perturbation of $\zeta_l$. This is because of the large hierarchy of scales between $k_l^{-1}$ and $k_s^{-1}$, which means the long modes are much larger than the PBH scale, and therefore

$$\delta_{\text{PBH}}(k_l) \simeq \frac{4}{9} b_0 \left( \frac{k_l}{k_s} \right)^2 \zeta_l \ll \zeta_l, \tag{3.3.9}$$

where we used that, at the time of formation, the short scale is crossing the horizon $k_s \approx 1/R_H$.

The considerations we made so far indicate that Gaussian perturbations have difficulties in producing clustered PBHs at relatively large scales. However, at sufficiently small scales, the clustering properties may become relevant. It is, therefore, important to understand at which scale PBH clustering dominates over the Poisson noise. This scale can be computed by considering the counts of neighbors. The mean count $\langle N \rangle$ in a cell of volume $V$ centred on a PBH is

$$\langle N \rangle = n_{\text{PBH}} V + n_{\text{PBH}} \int_V \mathrm{d}^3 x \, \xi_{\text{PBH}}(x) . \tag{3.3.10}$$

$\langle N \rangle$ significantly deviates from Poisson if the contribution from the second piece in Eq. (3.3.10) rise above the discreteness noise $\bar{n}_{\rm PBH} V$. This can happen at scales $\lesssim x_\xi(a)$ defined as the characteristic (comoving) clustering length through the relation $\xi_{\rm PBH}(x_\xi(a), a) = 1$, while, by definition, the PBH correlation function can be expressed as

$$\xi_{\rm PBH}(x, a_H) = \frac{1}{2\pi^2} \int_0^\infty dk \, k^2 \, P_{\rm PBH}(k, a_H) \, j_0(kx), \qquad (3.3.11)$$

in terms of the spherical Bessel function $j_0$. In Ref. [393], it was shown that for a narrow spectrum at small scales as defined in Eq. (2.1.29), the dominant contribution to the PBH clustering is given by the small scale perturbations. Thus, the comoving scale $x_\xi$ at which $\xi_{\rm PBH}(r) \simeq 1$ is always smaller than the average comoving separation between two PBHs, $\bar{r} = (3/4\pi\bar{n}_{\rm PBH})^{1/3}$. In more quantitative terms, one finds

$$\frac{x_\xi}{\bar{r}} \simeq 10^{-3} \left(\frac{M_{\rm PBH}}{M_\odot}\right)^{1/6}, \qquad \text{with} \qquad x_\xi \simeq 10^{-4} \, {\rm kpc} \left(\frac{M_{\rm PBH}}{M_\odot}\right)^{1/2}. \qquad (3.3.12)$$

This result is only weakly dependent on the PBH abundance. Therefore, in the Gaussian scenario, PBH clustering on top of Poisson noise is always irrelevant at scales larger than the mean PBH separation.

To get a feeling of the numbers involved, we can consider the case of PBHs with a mass $M_{\rm PBH} \sim 30 M_\odot$. In this case, the characteristic scale at formation is $k_s \sim 2.4 \cdot 10^2 \, {\rm kpc}^{-1}$, see Eq. (1.2.6). On the other hand, in order for the PBH abundance not to exceed the dark matter one, the number density of PBH is

$$\bar{n}_{\rm PBH} \simeq {\rm kpc}^{-3} f_{\rm PBH} \left(\frac{M_{\rm PBH}}{30 M_\odot}\right)^{-1}. \qquad (3.3.13)$$

Therefore, one can infer that the scale related to the average separation of PBHs of mass $M_{\rm PBH} \sim 30 M_\odot$ is $\bar{k} \sim k_l \sim 1.6 \, {\rm kpc}^{-1}$. Having in mind the formation of PBH binaries, for instance, with $M_{\rm PBH} \simeq 30 \, M_\odot$, the initial distances relevant for the present merger rate is $\simeq 4 \cdot 10^{-2} \, {\rm kpc}$, and it was concluded that in the Gaussian scenario clustering is not relevant. We will come back to this point in the following.

One may be tempted to try inducing a significant clustering by enhancing the power spectrum of curvature perturbations at the scale $k_l$. This however would require having a strongly red-tilted power spectrum

$$\langle \zeta_l^2 \rangle / \langle \zeta_s^2 \rangle \gg 1. \qquad (3.3.14)$$

As the PBH abundance is exponentially sensitive to the characteristic perturbation amplitude, this would mean that the small PBHs associated with the scale $k_s$ are an irrelevant portion of the PBH population, which would be dominated instead by masses $M_l$ corresponding to the formation scale $k_l$, see discussion in Sec. 2.2.4 and Ref. [18].

### 3.3.2  The non-Gaussian case

The conclusions drawn in the previous section change drastically when non-Gaussian perturbations are considered. Indeed, non-Gaussianities can generate a coupling between short and long scales which may evade the argument against PBH clustering derived in the previous section.

Let us consider the case in which the curvature presents a local non-Gaussianity parametrised as [300, 395–397]

$$\zeta(x) = \zeta_g(x) + f_{\rm NL} \left[\zeta_g^2(x) - \langle \zeta_g^2(x) \rangle\right]. \qquad (3.3.15)$$

In this case, the long-wavelength density perturbation $\delta_l$ not only effectively reduces the threshold for collapse $\delta_c$, as in the gaussian case, but it is also able to modulate the variance of short-wavelength perturbations, $\sigma_s$. Hence, the bias parameter, computed as

$$b = \frac{\mathrm{d} \log P_1}{\mathrm{d} \delta_l} \bigg|_{\delta_l = 0} \qquad (3.3.16)$$

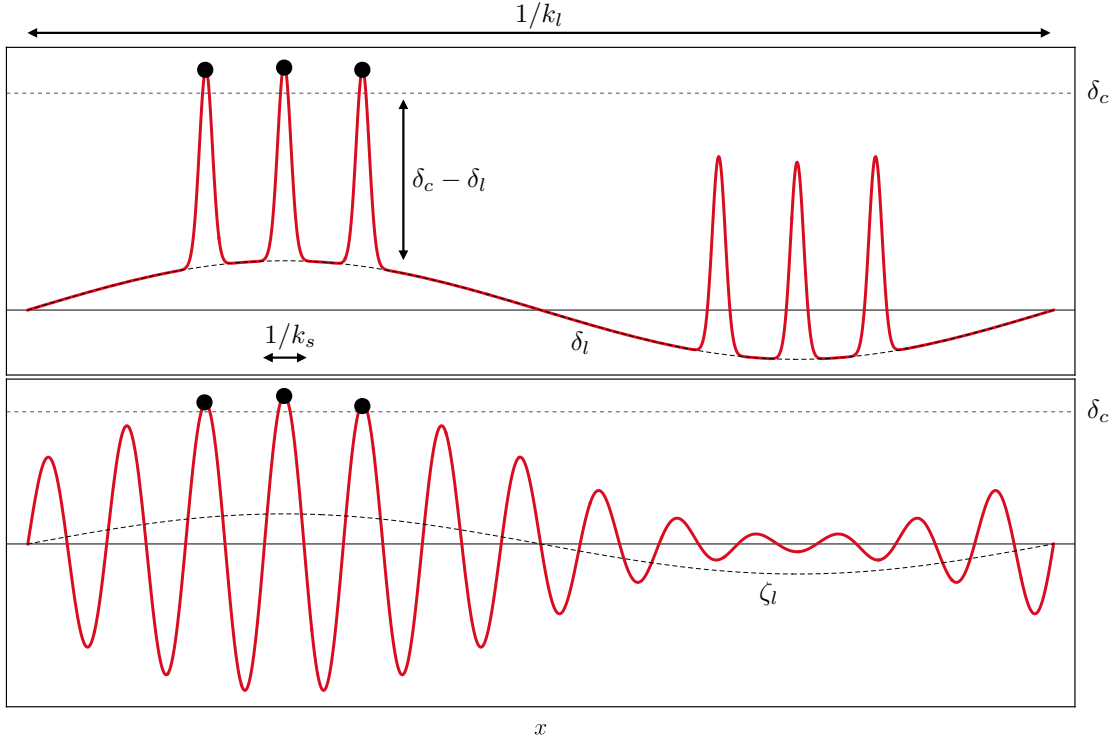

Figure 3.5: *In the top panel, we show the effect of long density contrast modes on the small scale density perturbations. Its effect can be viewed as a local rescaling of the threshold for collapse. This contribution becomes negligible on scales $k_l \ll k_s$ due to the gradients in the density contrast relation with the curvature power spectrum. In the bottom panel, the scenario with local non-Gaussianity is shown. The short mode variance is modulated by $f_{\rm NL}\zeta_l$, leading to a spatial correlation between the peaks which can reach the threshold for collapse $\delta_c$ (black dots indicates the peaks reaching the threshold and collapsing to form a PBH). Figure inspired by Refs. [322, 323].*

is modified by an additional contribution

$$\Delta b = \frac{\partial \log \sigma_s}{\partial \delta_l} \frac{\partial \log P_1}{\partial \log \sigma_s}\bigg|_{\delta_l=0}, \tag{3.3.17}$$

which goes under the name of *scale-dependent bias* (see Ref. [398] for its definition in the context of large scale structure). Again, in the peak-background split picture, we can consider the curvature field as composed by long scale perturbations $\zeta_l$ and short-scale perturbations $\zeta_s$ at the PBH scales. The variance $\sigma_s^2$ is then modulated by the long modes as

$$\sigma_s(\vec{x}) = (1 + 2f_{\rm NL}\zeta_l(\vec{x}))\bar{\sigma}_s, \tag{3.3.18}$$

where $\bar{\sigma}_s^2$ denotes the average variance of $\delta_s$. A pictorial representation of the effect of long modes with and without non-Gaussianities is shown in Fig. 3.5. From Eq. (3.3.17) one finds the scale dependent bias to be

$$\Delta b(k) = \frac{9}{2}\nu^2 f_{\rm NL} \left(k R_{\rm PBH}\right)^{-2}. \tag{3.3.19}$$

Therefore, we observe that the PBH modulation at large scales is linearly related to $\zeta_l$ with no additional suppression factors due to the gradients, as found in the gaussian case, since

$$\delta_{\rm PBH}(k_l) \simeq (b_0 + \Delta b)\,\delta_l \simeq \left[\frac{4}{9}\frac{\nu}{\sigma_{\rm PBH}}(k_l R_{\rm PBH})^2 + 2\nu^2 f_{\rm NL}\right]\zeta_l(k_l) \simeq 2\nu^2 f_{\rm NL}\zeta_l(k_l). \tag{3.3.20}$$

In App. B we report a general formula describing the PBH clustering induced by non-Gaussianities based on the same path integral approach used in Sec. 2.2.2.

To conclude this section, we highlight the fact that, when non-Gaussianities are introduced into the model, one can generate a sizeable PBH clustering (on top of Poisson). This, however, may not come for free if PBHs are a sizeable fraction of the dark matter in the universe, as we now explain.

**Isocurvature constraints on non-Gaussianity and PBH abundance**

If a significant fraction of dark matter is made of PBHs, a sizeable clustering at large scales would be responsible for the production of isocurvature modes in the DM density fluid, which are strongly constrained by CMB observations [322, 323].

The Planck experiments set the most stringent bound on the relative abundance of isocurvature modes. Reported at at 95% CL, this bounds is [399]

$$100\beta_{\rm iso} < 0.095 \qquad \text{for fully correlated,}$$
$$100\beta_{\rm iso} < 0.107 \qquad \text{for fully anti-correlated,} \tag{3.3.21}$$

where by fully correlated (fully anti-correlated) we indicate a positive (negative) $f_{\rm NL}$ and where we defined $\beta_{\rm iso}$ as the relative abundance of isocurvature modes

$$\beta_{\rm iso} \equiv \frac{\mathcal{P}_S}{\mathcal{P}_S + \mathcal{P}_\zeta}. \tag{3.3.22}$$

The isocurvature perturbations can be written in terms of the PBH perturbation as

$$S \equiv \delta_{\rm PBH} - \frac{3}{4}\delta. \tag{3.3.23}$$

In the absence of non-Gaussianities, the isocurvature modes are present at scales incredibly small compared to the one probed by CMB measurements, and therefore can be neglected. On the other hand, in the presence of non-Gaussianities inducing a scale-dependent bias, the isocurvature power spectrum can be written as

$$\mathcal{P}_S(k_{\rm CMB}) \simeq (\Delta b)^2 \mathcal{P}_\zeta(k_{\rm CMB}) \simeq \left(2\nu^2 f_{\rm NL}\right)^2 \mathcal{P}_\zeta(k_{\rm CMB}) \tag{3.3.24}$$

where we used the fact that $\Delta b$ provides the dominant contribution to $\delta_{\rm PBH}$. At this point, we can also reintroduce the abundance $f_{\rm PBH}$. Intuitively, as PBHs make an increasingly smaller fraction of the dark matter in the universe, also the isocurvature modes are scaling like $\propto f_{\rm PBH}$. Finally, combining Eqs. (3.3.21) and (3.3.22), one finds

$$-0.0164 < \nu f_{\rm NL} f_{\rm PBH} < 0.0154. \tag{3.3.25}$$

We stress again that this bound strictly applies to a type of non-Gaussianity relating the PBH scales to the CMB scales, as the CMB observations limiting the isocurvature modes are only reaching very large scales.

### 3.3.3  Constraining the clustering of LIGO/Virgo sources

To conclude this chapter, we address the possibility of constraining the PBH clustering at formation in the mass range currently observed by the LIGO/Virgo collaboration, which is around PBH masses of $M_{\rm PBH} \approx 30 M_\odot$. We follow the discussion presented in Ref. [4].

In this mass window, the range of initial comoving distances relevant for the calculation of the present merger rate is $(4 \cdot 10^{-5} \div 10^{-3})$ Mpc [196]. Indeed, only PBHs separated by a distance smaller than $\sim 10^{-3}$ Mpc can form a binary system, while there is also a minimum separation $\sim 4 \cdot 10^{-5}$ Mpc for which PBHs can form a binary and merge in a large enough timescale allowing for an observation of the emitted signal at GW experiments. We will discuss this process in more details in the next chapter.

The crucial point is that this range of scales strongly overlaps with the window where a sizeable CMB $\mu$-distortion may be produced, that is in the range $(10^{-4} \div 2 \cdot 10^{-2})$ Mpc, which is not testable through CMB anisotropies observations. CMB distortions are caused by the energy injection originated by the dissipation of acoustic waves through the Silk damping as they re-enter the horizon and start oscillating [400, 401]. As PBH clustering can be induced by sizeable curvature perturbations through non-Gaussianities on scales relevant for the merger rate, and those scales overlap with those where CMB μ-distortion would be generated, future μ-distortion experiments have the opportunity to test (and constrain) the amount of PBH clustering in this phenomenologically interesting mass window. Forecasted constraints from PIXIE ($\mu < 3 \cdot 10^{-8}$) [402], SuperPIXIE ($\mu < 7 \cdot 10^{-9}$) [403], Voyage2050 ($\mu < 1.9 \cdot 10^{-9}$) and $10 \times$ Voyage2050 ($\mu < 1.9 \cdot 10^{-10}$) [404] show that future experiments will allow, therefore, to test the hypothesis of large PBH clustering induced by primordial non-Gaussianities.

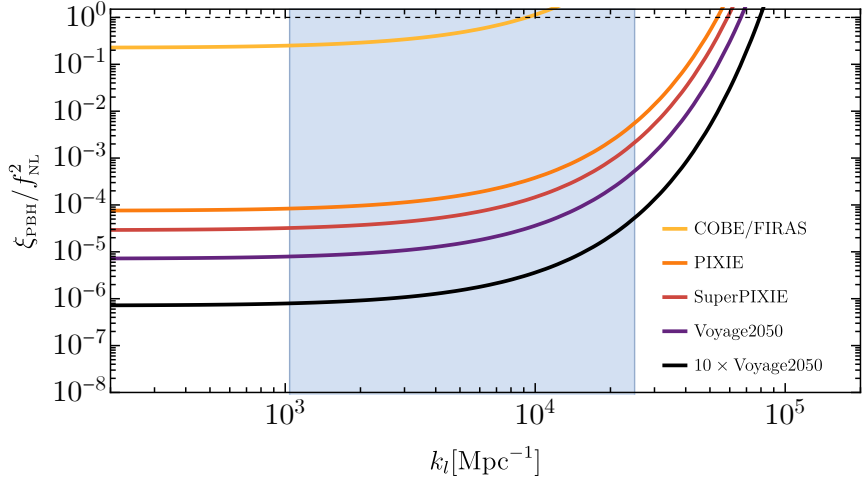

**Figure 3.6:** *Limits on the PBH correlation function from the CMB μ-distortion. The blue band indicates the range of scales relevant for the binary formation.*

For presentation purposes, let us consider the scenario of a local-type non-Gaussian model, as discussed in Sec. 3.3.2. Assuming a monochromatic power spectrum at the clustering scale $k_l$ for simplicity, one obtains an initial PBH correlation function [297]

$$\xi_{\text{PBH}}(x) \simeq 4\nu^4 f_{\text{NL}}^2 A_l j_0(k_l x), \tag{3.3.26}$$

where $A_l$ is the amplitude of the power spectrum at the relevant scales for the μ-distortion measurements.

In Fig. 3.6, we plot the forecasted limits on the PBH correlation function coming from the CMB μ-distortion at the scales relevant for the merger rate. For the computation of the bound we follow Refs. [405, 406]. We also chose a value of the parameter $\nu$ roughly corresponding to $f_{\text{PBH}} = 10^{-3}$, a figure which is compatible with current constraints [199]. A variation of the abundance only requires a fractional change of $\nu$, thus not affecting our results in a significant amount.

As the distortion is directly proportional to the amplitude $A_l$ of the large-scale curvature perturbations, given the current and future constraints, only a large value of $f_{\text{NL}}$ may provide a PBH correlation $\xi_{\text{PBH}} \gtrsim 1$. For instance, if PIXIE does not measure any CMB μ-distortion, generating PBH clustering requires $|f_{\text{NL}}| \gtrsim 10$. The current COBE/FIRAS limit ($\mu < 9 \cdot 10^{-5}$ [187]) imposes $A_l \lesssim 10^{-4}$, which corresponds to $|f_{\text{NL}}| \gtrsim 1$. [5] Finally, we remark that as long as $\xi_{\text{PBH}} \lesssim 1$, the overall PBH abundance is not altered by non-Gaussianities as the short-scale variance is only significantly affected if $f_{\text{NL}} \gtrsim A_l^{-1/2}$.

Other model dependent scenarios are considered in Ref. [4] reaching even more stringent conclusions. These results show that future experiments looking for CMB μ-distortion will be able to significantly constrain the parameter space in which non-Gaussianities are able to induce PBH clustering at formation in the standard scenario where PBHs come from the collapse of density perturbations. This result may have an impact on the interpretation of the merger events in the LIGO/Virgo mass range.

---

[5]We note that our estimate for the minimum $f_{\text{NL}}$ required to produce a significant PBH clustering able to modify the merger rate is consistent with the result reported in Ref. [358]. Looking at their Fig. 6, we infer that the merger rate is impacted by non-Gaussian corrections if $f_{\text{NL}} \zeta_l \gtrsim 10^{-2}$, where $\zeta_l$ is the typical amplitude of the large-scale curvature perturbations. Setting the maximum allowed value $\zeta_l \sim A_l^{1/2} \sim 10^{-2}$, one finds that clustering overcomes the Poisson level for $f_{\text{NL}} \gtrsim 1$.

# Chapter 4

# Primordial black hole evolution

In this chapter, we develop the theory describing the evolution of a PBH population after formation. The dominant effects to be considered are baryonic accretion, mergers and clustering evolution. While the first two can drastically change the predictions of a PBH model in terms of mass and spin distributions, the latter is relevant for the computation of the merger rate for large values of $f_{\rm PBH}$.

To give some motivation for entering in this challenging endeavour, let us consider the following example. If PBHs accretes matter, they become heavier and the mass function naturally shifts. However, also the PBH spin can be significantly impacted. Consider, for example, the effect of spherical accretion on the PBH Kerr parameter, defined as

$$\chi = \frac{|\vec{S}|}{M^2}. \qquad (4.0.1)$$

As spherically infalling gas does not transfer angular momentum to the PBH, while it enhances its mass, $\chi$ is progressively reduced. On the other hand, quite the opposite happens in case a thin accretion disk forms around the PBH, as $\chi$ can be efficiently enhanced up to maximally spinning values $\chi \sim 0.998$.

Merger events can also drastically change the expected mass and spin distributions of the PBH population. Take for example the merger of two black holes with initially small spins. Numerical relativistic simulations provide evidence for the final spin $S_f$ being [407]

$$\chi_f = \frac{S_f}{G_N M_f^2} \simeq 0.69 - 0.56 \left( \frac{M_1 - M_2}{M_1 + M_2} \right)^2, \qquad (4.0.2)$$

where $G_N$ is Newton's constant and $M_f$ is the remnant BH mass, which is expected to be of the order of the sum of the merging BH masses. We see a shift of both masses and spins of the PBH population as a result of mergers. These are just a few examples to show that, depending on the details of PBH evolution after formation, the predicted properties of the PBH population can change drastically.

As our final goal is to describe the properties of binary mergers which could be seen at the current (LIGO/Virgo/KAGRA) and future (Einstein Telescope or Cosmic Explorer) GW detectors, we will focus, in particular, on the mass range between a fraction of the solar mass up to $10^4 M_\odot$.

## 4.1 PBH accretion

Our goal in this section is to quantify the rate of variation of the PBH mass along with describing the geometrical properties of the infalling material. In doing so, we will follow in details the description reported in Refs. [12, 14] and references therein. We also stress that we will adopt the state of the art description of the accretion process as it was introduced in Refs. [169, 234]. We will not however address the computation of the radiation emitted during the accretion process. This is relevant to compute the constraints coming from CMB observations (more details can be found in Refs [183, 232, 233, 361]).

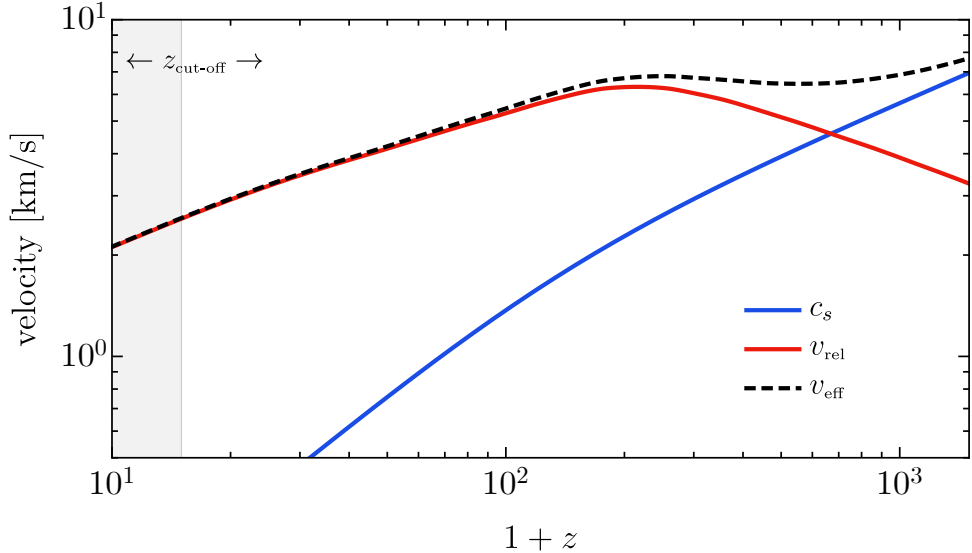

Figure 4.1: *Characteristic relative velocity and sound speed as a function of redshift. Due to the onset of structure formation (going beyond linear theory) and the reionization of the universe, those quantities are not representative below the cut-off redshift $z_{\text{cut-off}}$ (see the main text).*

As we will see shortly, one of the important parameters determining accretion rate is the relative velocity between the accreting system (being a PBH or a PBH binary) and the surrounding gas. Therefore we will consider the two cases separately

- **accretion onto an isolated PBH:** in this case the relative velocity $v_{\text{rel}}$ between dark matter and baryons is typically smaller than or comparable to the speed of sound in the gas $c_s$ at high redshift, while it becomes dominant at redshift $z \lesssim 500$;

- **accretion onto PBHs in binaries:** in this setup, the orbital velocities are dominant and therefore determines the relative motion of the PBH with respect to the fluid. In this case, however, as PBHs reside in a deeper gravitational potential well generated by the binary system, deviations from the mean gas density need to be taken into account.

The accretion rate is found to peak at $z \lesssim 100$ [234, 408], and therefore, the former case is relevant for describing the properties of PBHs in binaries formed in present-day structures at low redshift, while the latter dictates the evolution of PBH in binaries formed in the early universe, which are dominating the merger rate in the standard scenario (more details can be found in Sec. 4.5. In the following, we address both cases separately.

### 4.1.1   Accretion onto isolated PBHs

Let us consider an isolated PBH of mass $M$ surrounded by an intergalactic medium principally composed of hydrogen gas. We define the relative velocity of the compact object with respect to the medium as $v_{\text{rel}}$. This velocity can be computed in linear perturbation theory, following Ref. [234], as $\langle v_{\text{rel}} \rangle \equiv \langle v_{\text{DM}} \rangle - \langle v_{\text{b}} \rangle$, where "DM" and "b" refer to dark matter and baryons respectively. Both quantities are [409]

$$\langle v_i \rangle = \frac{H^2}{2\pi^2} \int_0^\infty \mathrm{d}k f_i^2(k,t) P(k) W^2(k, r_0), \qquad (4.1.1)$$

where $i = \text{DM}$ or b while $P$ is their corresponding power spectrum, $W$ a window function smoothing over a comoving volume $r_0$ and $f_i$ accounts for the faster rate of growth of the baryonic perturbations after decoupling redshift. The resulting relative velocity is shown in Fig. 4.1.

Throughout the evolution of the universe in the matter dominated epoch, the mean cosmic number density $n_{gas}$ of hydrogen particles in the gas is described as

$$n_{gas} \simeq 200 \, \text{cm}^{-3} \left( \frac{1+z}{1000} \right)^3. \tag{4.1.2}$$

The sound speed of the gas in equilibrium at the temperature of the intergalactic medium is given by

$$c_s \simeq 5.7 \, \text{km/s} \left( \frac{1+z}{1000} \right)^{1/2} \left[ \left( \frac{1+z_{dec}}{1+z} \right)^\beta + 1 \right]^{-1/2\beta}, \tag{4.1.3}$$

with $\beta = 1.72$, and $z_{dec} \simeq 130$ being the redshift at which the baryonic matter decouples from the radiation fluid. An increase of the gas temperature is expected during the onset of the reionization epoch, around $z \sim \mathcal{O}(10)$, and consequently the parametrization of the sound speed (4.1.3) is expected to break down. As we shall see, the key quantity affecting the mass accretion rate is the effective velocity $v_{eff}$ defined as the quadratic sum of $v_{rel}$ and $c_s$ as

$$v_{eff} = \sqrt{v_{rel}^2 + c_s^2}. \tag{4.1.4}$$

Finally, the Bondi-Hoyle accretion rate is given by [234, 408, 410–413]

$$\dot{M}_B = 4\pi \lambda m_H n_{gas} v_{eff} r_B^2, \tag{4.1.5}$$

where we introduced the Bondi-Hoyle radius

$$r_B \equiv \frac{M}{v_{eff}^2} \simeq 1.3 \times 10^{-4} \, \text{pc} \left( \frac{M}{M_\odot} \right) \left( \frac{v_{eff}}{5.7 \, \text{km/s}} \right)^{-2}. \tag{4.1.6}$$

This radius can be intuitively considered as the characteristic size of the region where the gravitational attraction can win over the radiation pressure. The accretion eigenvalue $\lambda$ accounts for corrections due to the Hubble expansion, the coupling of the CMB radiation to the gas through Compton scattering, and the gas viscosity. Its analytical expression, to be found in Ref. [408], reads

$$\lambda = \exp \left( \frac{9/2}{3 + \hat{\beta}^{0.75}} \right) x_{cr}^2, \tag{4.1.7}$$

in terms of the sonic radius

$$x_{cr} \equiv \frac{r_{cr}}{r_B} = \frac{-1 + (1 + \hat{\beta})^{1/2}}{\hat{\beta}} \tag{4.1.8}$$

and the gas viscosity parameter $\hat{\beta}$ given by

$$\hat{\beta} = \left( \frac{M}{10^4 M_\odot} \right) \left( \frac{1+z}{1000} \right)^{3/2} \left( \frac{v_{eff}}{5.74 \, \text{km/s}} \right)^{-3} \left[ 0.257 + 1.45 \left( \frac{x_e}{0.01} \right) \left( \frac{1+z}{1000} \right)^{5/2} \right], \tag{4.1.9}$$

as a function of the redshift $z$, the PBH mass $M$, effective velocity $v_{eff}$, and ionization fraction of the cosmic gas $x_e$.

For convenience, one can define a dimensionless accretion rate

$$\dot{m} = \frac{\dot{M}_B}{\dot{M}_{Edd}}, \tag{4.1.10}$$

which is normalised with respect to the Eddington one

$$\dot{M}_{Edd} = 1.44 \times 10^{17} \text{g/s} \left( \frac{M}{M_\odot} \right) = 2.2 M_\odot / \text{Gyr} \left( \frac{M}{M_\odot} \right). \tag{4.1.11}$$

We will denote as "super-Eddington" accretion regimes, the cases in which $\dot{m} > 1$. As we will see more in details in the section dedicated to the evolution of the PBH spin, super-Eddington rates ($\dot{m} \gtrsim 1$) are associated with the formation of a (geometrically) thin accretion disk. We can rewrite the accretion rate using Eq. (4.1.10) to get, see for example Ref. [414],

$$\dot{M} \sim 0.002 \, \dot{m} \left( \frac{M(t)}{10^6 M_{\odot}} \right) M_{\odot} \, \mathrm{yr}^{-1}, \qquad (4.1.12)$$

dictating the evolution of the mass with respect to time.

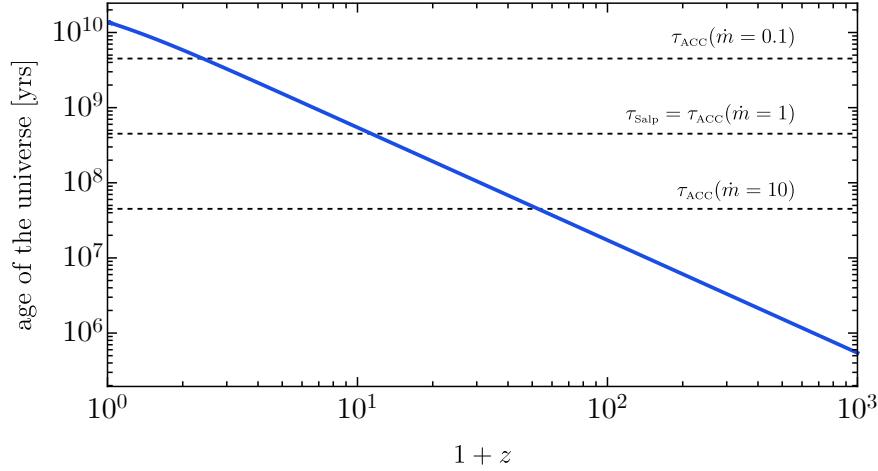

Figure 4.2: *Typical accretion time scale $\tau_{\mathrm{ACC}} \equiv \tau_{\mathrm{Salp}}/\dot{m}$ compared to the age of the universe at a given redshift $z$. The relation between the redshift and the age of the universe $t = t(z)$ has been derived within the $\Lambda$-CDM universe.*

The characteristic time scale $\tau_{\mathrm{ACC}}$ of the process is often expressed in terms of the Salpeter time. This is defined as

$$\tau_{\mathrm{Salp}} = \sigma_{\mathrm{T}}/4\pi m_{\mathrm{p}} = 4.5 \times 10^8 \, \mathrm{yr}, \qquad (4.1.13)$$

where $\sigma_{\mathrm{T}}$ is the Thompson cross section and $m_{\mathrm{p}}$ is the proton mass. Therefore, one has

$$\tau_{\mathrm{ACC}} \equiv \tau_{\mathrm{Salp}}/\dot{m}. \qquad (4.1.14)$$

We show a comparison of this timescale with the age of the universe at a given redshift in Fig. 4.2. As one can see, for an Eddington accretion rate ($\dot{m} = 1$), $\tau_{\mathrm{ACC}}$ becomes comparable with the age of the universe around redshift $z_{\mathrm{cut\text{-}off}} \simeq 10$. Larger accretion rates can be effective at earlier times, provided $\dot{m}$ is large enough.

**The effect of a dark matter halo on the accretion rate**

The PBH abundance is constrained to be below 100% of the dark matter for a wide range of masses [199]. In case $f_{\mathrm{PBH}} < 1$, it is necessary to include the effect of an additional dark matter fluid which can catalyse the accretion process. Indeed, when PBHs are coexisting with a secondary dark matter component, the latter starts accumulating in the PBH gravitational potential well at high redshift. This has also been confirmed by dedicated N-body simulations in Refs. [176, 235]. Therefore, a dark halo of mass $M_h$, truncated at a radius $r_h$, is formed around each PBH. In the following, we will assume a power-law density profile $\rho \propto r^{-\alpha}$, with approximately $\alpha \simeq 2.25$, as found in Refs. [176, 235, 415]. The halo mass and size grow with redshift following the scaling [176, 416]

$$M_h(z) = 3M \left( \frac{1+z}{1000} \right)^{-1}, \qquad (4.1.15)$$

$$r_h = 0.019 \, \mathrm{pc} \left( \frac{M}{M_{\odot}} \right)^{1/3} \left( \frac{1+z}{1000} \right)^{-1}, \qquad (4.1.16)$$

as long as the PBHs are isolated and, eventually, stop when most of the available dark matter in the vicinity of the PBH has been accreted, which corresponds to, approximately, when

$$3f_{\rm PBH}\left(1+\frac{z}{1000}\right)^{-1}=1. \tag{4.1.17}$$

In the following, we will neglect the effect of non-linearities which may modify the growth of the dark matter halo. This would be surely relevant at redshift around $\mathcal{O}(10)$.

The effect of the halo $M_h$ on the accretion rate is to enlarge the gravitational attraction of the gas to the accreting system. Therefore, it acts as a catalyst enhancing the gas accretion rate. We also stress that, on the other hand, the amount of mass which is accreted from the surrounding dark matter halo is negligible [408].

We can identify two regimes. When the characteristic size of the halo is much smaller than the Bondi radius, the gas is not resolving the (halo+PBH) structure and accretion can be described as in Eq. (4.1.5), where however the mass of the PBH $M$ must be substituted with the dominant halo mass $M_h$. In the opposite case, the potential is modified and one should apply corrections to the formula in Eq. (4.1.5). We can define the parameter

$$\kappa\equiv\frac{r_{\rm B}}{r_h}=0.22\left(\frac{1+z}{1000}\right)\left(\frac{M_h}{M_\odot}\right)^{2/3}\left(\frac{v_{\rm eff}}{\rm km/s}\right)^{-2}. \tag{4.1.18}$$

When $\kappa\geq 2$, we are interested in the first case and one can parametrise the accretion rate as

$$\dot{m}\equiv\frac{\dot{M}_{\rm B}}{\dot{M}_{\rm Edd}}=0.023\lambda\left(\frac{1+z}{1000}\right)\left(\frac{M}{M_\odot}\right)\left(\frac{v_{\rm eff}}{5.74\,\rm km/s}\right)^{-3}, \tag{4.1.19}$$

where the sonic radius and viscosity are described as in the naked scenario discussed previously [234]. On the other hand, when $\kappa<2$ one has instead to correct the quantities with respect to the naked case, finding

$$\hat{\beta}^h\equiv\kappa^{\frac{p}{1-p}}\hat{\beta},\quad\lambda^h\equiv\tilde{\Upsilon}^{\frac{p}{1-p}}\lambda(\hat{\beta}^h),\quad r_{\rm cr}^h\equiv\left(\frac{\kappa}{2}\right)^{\frac{p}{1-p}}r_{\rm cr}, \tag{4.1.20}$$

where $p=2-\alpha$ and

$$\tilde{\Upsilon}=\left(1+10\hat{\beta}^h\right)^{\frac{1}{10}}\exp(2-\kappa)\left(\frac{\kappa}{2}\right)^2. \tag{4.1.21}$$

In Fig. 4.3 we show the resulting accretion rate $\dot{m}$ for various PBH masses along with the evolution of the universe for PBHs with a surrounding dark halo.

**The accretion cut-off**

In this section, we define the accretion parameter $z_{\rm cut\text{-}off}$ which captures the uncertainties in the accretion model and that is going to be used in our analysis in the following chapters.

We focus on redshifts smaller than $z\lesssim 100$ as, at higher redshift, the characteristic accretion time scale $\tau_{\rm ACC}$ is orders of magnitude larger than the age of the universe for the largest accretion rate we encounter. We highlight that the accretion rate formula scales like $\dot{m}\sim v_{\rm eff}^{-3}$, where the effective velocity is defined in Eq. 4.1.4. With the onset of structure formation, PBHs and baryonic matter start falling within the potential well of large structures and their characteristic velocities increase significantly. In fact, one expects a large portion of the PBH population to be affected by large-scale structures after redshift around $z\simeq 10$, with an increase of the relative velocity up to one order of magnitude above the one estimated in the previous section. This induces a consequent large suppression of the accretion rate [169, 234, 417]. An additional effect which is reducing the accretion rate is the reionization, as it leads to an enhancement of the temperature of the gas, which causes $c_s$ to increase.

Due to the expected sharp decrease of the accretion efficiency close to the structure formation and reionization epoch, one can confidently stop the evolution of the mass at around $z_{\rm cut\text{-}off}\sim\mathcal{O}(10)$. The exact position of this cut-off is currently highly uncertain and we will consider this as a nuisance parameter in the accretion model. Very low values of $z_{\rm cut\text{-}off}$ would correspond to sustained accretion while a high redshift cut-off would correspond to reduced accretion rates. We will come back to this point, along with few critical remarks about the uncertainties in the accretion model, in Sec. 4.1.3.

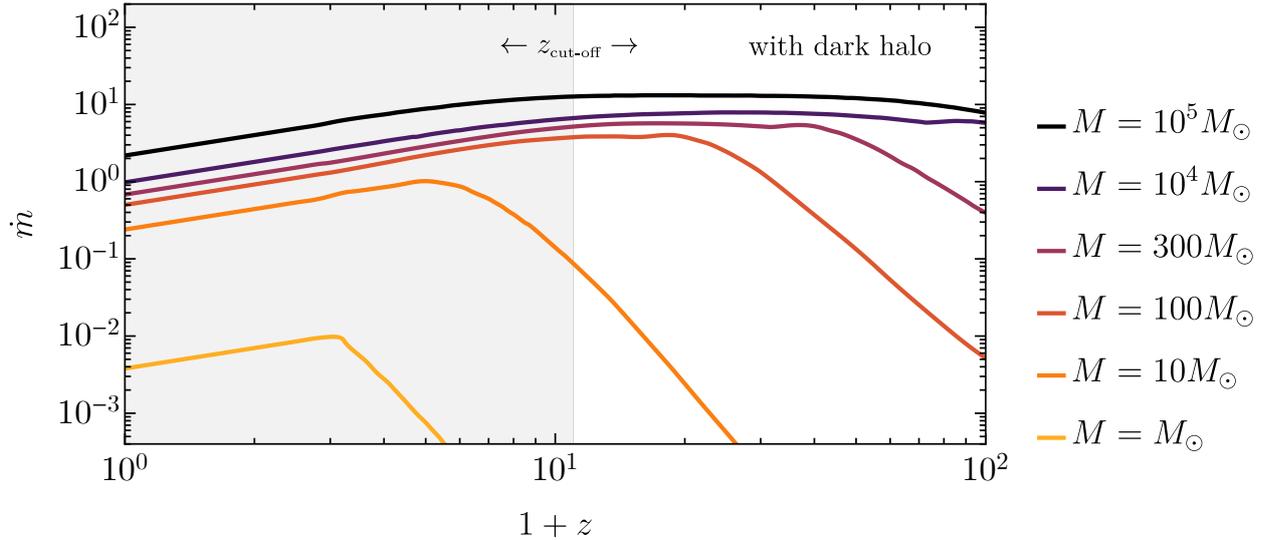

Figure 4.3: *The dimensionless accretion rate $\dot{m}$ as a function redshift for isolated PBHs with a dark matter halo. We show the results for various PBHs masses, see also [234]. The shaded region indicates the critical redshift (e.g. $z = 10$ in the plot) below which the accretion model becomes affected by the structure formation and reionization. For a discussion about $z_{\text{cut-off}}$ see the main text.*

In the following, for presentation purposes, we will consider two scenarios. In one case (which we denote Model I), the accretion rate is drastically decreased after redshift $z \sim 10$. This scenario may be regarded as more realistic. In a second case (called Model II), which is undoubtedly a more extreme scenario, we assume the impact of structure formation and reionization is limited and a moderate accretion is maintained up to very low redshifts. This can be realised if a PBH remains sufficiently isolated and only falls in the large scale structure very late in the evolution of the universe. In the latter case, we follow the procedure of Ref. [234] highlighted in the previous section until the evolution is monotonic, and then extrapolate the behaviour of $\dot{m}$ down to lower redshifts, as shown in Figs. 4.3 and 4.4.

### 4.1.2   Accretion onto binary PBHs

When PBHs are in binaries, the description of the accretion process becomes even more complex. In this section, we describe how we parametrised the mass accretion rate on the two PBHs composing the binary. First, we need to divide the discussion into a *global* accretion processes, which refers to the infall flux of gas onto the binary system as a whole, and *local* accretion processes, i.e. the accretion of material by the two PBHs composing the binary. In this description, we closely follow Ref. [12], where the analysis in Ref. [14] was extended to account for binaries with generic mass ratios and eccentric orbits.

We consider a binary composed by two PBHs with masses $M_1$ and $M_2$. It is going to be useful also to define the total mass of the binary $M_{\text{tot}} = M_1 + M_2$, the reduced mass $\mu = M_1 M_2 / (M_1 + M_2)$ and mass ratio $q = M_2 / M_1 \leq 1$. The geometrical parameters describing the binary motion are the semi-major axis $a$, and eccentricity $e$, see Fig. 4.5 for a depiction of the binary system, along with the relevant scales entering into the accretion model.

We define the Bondi radius of the binary in terms of the effective velocity $v_{\text{eff}} = \sqrt{c_s^2 + v_{\text{rel}}^2}$, constructed adopting the relative velocity $v_{\text{rel}}$ of the binary's center of mass with respect to the gas, as

$$r_{\text{B}}^{\text{bin}} = \frac{M_{\text{tot}}}{v_{\text{eff}}^2}, \tag{4.1.22}$$

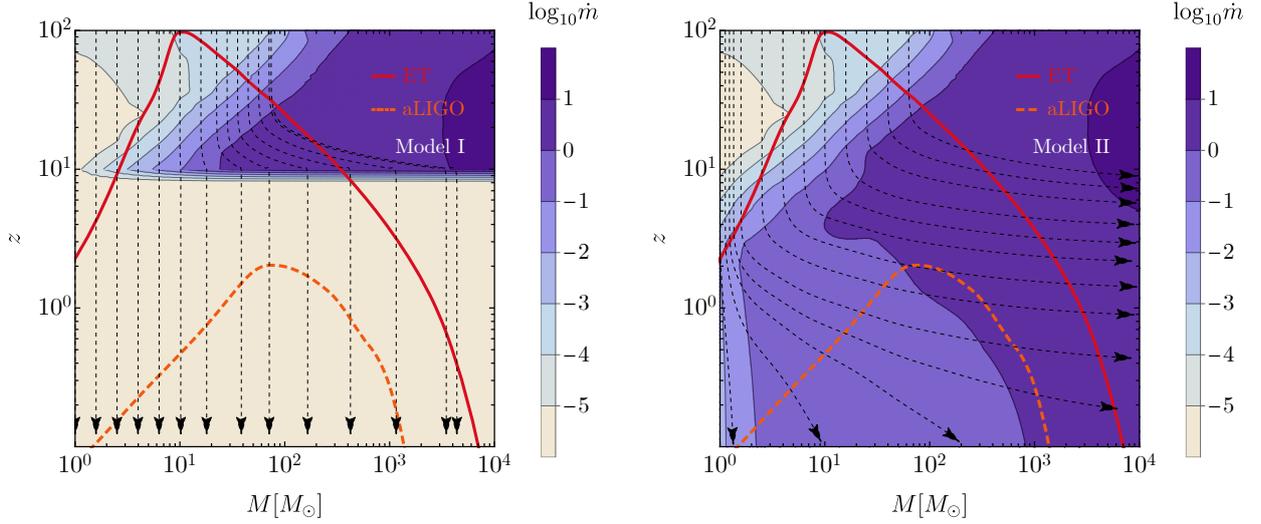

Figure 4.4: *The accretion rate parameter $\dot{m}$ as a function of the mass of PBHs and redshift. We show the evolution of the mass of individual PBHs along with redshift with black dashed lines. **Left:** Model I, where the accretion rate is drastically reduced at redshift $z \simeq 10$ as a consequence of structure formation and reionization. **Right:** Model II, with sustained accretion up to very low redshift.*

Depending on its relation with the binary semi-major axis, we can define two regimes.

- $a \gg r_{\text{B}}^{\text{bin}}$: in this case PBHs are very distant from each other and both accrete independently as parametrised in the previous section. Notice, however, that the characteristic relative velocity of the compact objects with respect to the background gas is affected by the orbital motion.

- $a \ll r_{\text{B}}^{\text{bin}}$: in this case the infalling gas attracted to the center of mass of the binary does not resolve the binary structure from large distances. Therefore, there exists a flux of material on the binary system, which can be described by the Bondi-Hoyle accretion formula (4.1.5) as

$$\dot{M}_{\text{bin}} = 4\pi\lambda m_H n_{\text{gas}} v_{\text{eff}}^{-3} M_{\text{tot}}^2 \,. \tag{4.1.23}$$

Let us focus on the latter case as it corresponds to the relevant configuration for PBH binaries formed in the early universe. Following the standard description of the binary motion, see for example Ref. [418], we can write the PBH positions and velocities with respect to the center of mass as

$$\begin{aligned} r_1 &= \frac{q}{1+q}r, & v_1 &= \frac{q}{1+q}v, \\ r_2 &= \frac{1}{1+q}r, & v_2 &= \frac{1}{1+q}v, \end{aligned} \tag{4.1.24}$$

in terms of their relative distance and velocity defined as

$$r = a(1 - e\cos u), \qquad v = \sqrt{M_{\text{tot}}\left(\frac{2}{r} - \frac{1}{a}\right)} \,. \tag{4.1.25}$$

In the previous equation, we defined the orbital angle $u$, whose time evolution is given by the implicit expression

$$[u(t) - e\sin u(t)]\sqrt{\frac{a^3}{M_{\text{tot}}}} = t - T, \tag{4.1.26}$$

where $T$ is an integration constant. We also define the PBH effective velocities, accounting for the contribution of each orbital motion, as

$$\begin{aligned} v_{\text{eff},1} &= \sqrt{v_{\text{eff}}^2 + v_1^2}, \\ v_{\text{eff},2} &= \sqrt{v_{\text{eff}}^2 + v_2^2}. \end{aligned} \tag{4.1.27}$$

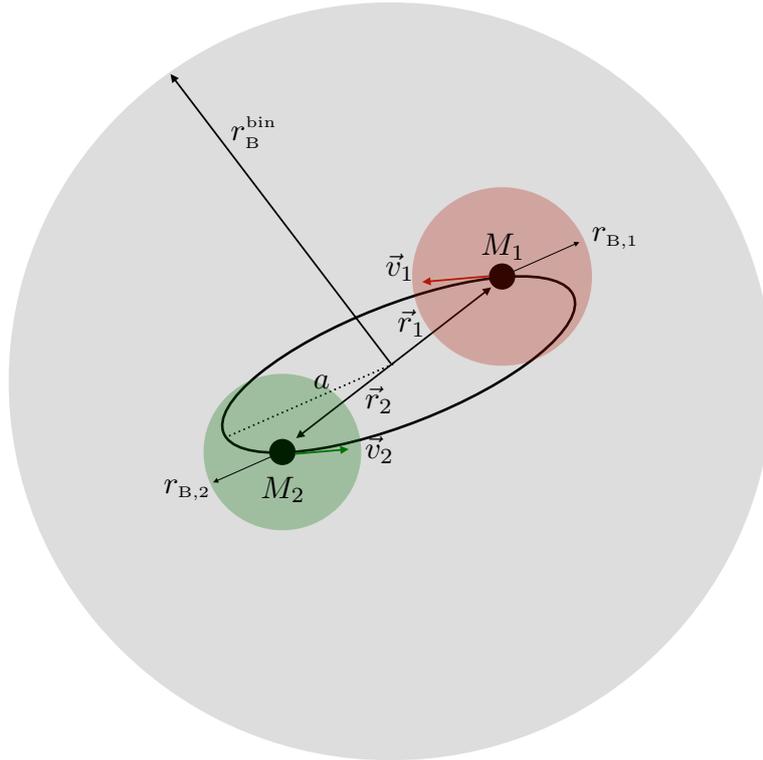

Figure 4.5: *Schematic illustration of the accreting binary system, along with the relevant scales involved. The shaded grey (green/red) region corresponds to the binary (single PBH) Bondi radius. It is found that the hierarchy of scales for binaries which would be observable at the LIGO/Virgo detector always obeys $M_i \ll r_{\mathrm{B,i}} \lesssim a \ll r_{\mathrm{B}}^{\mathrm{bin}}$, and therefore the Bondi radius of the binary is always much bigger than the orbital distance.*

Now we can make use of the fact that we are working in the limit $a \ll r_{\mathrm{B}}^{\mathrm{bin}}$. As the Bondi radius of the binary is much larger than its typical semi-axis, we expect the total flow of baryons towards the binary to be constant, i.e.

$$4\pi m_H n_{\mathrm{gas}}(R) v_{\mathrm{ff}}(R) R^2 = \mathrm{const} = \dot{M}_{\mathrm{bin}}, \qquad (4.1.28)$$

where we defined the free fall velocity of the gas as $v_{\mathrm{ff}}$. This can be computed by requiring that, at large distances $R \sim r_{\mathrm{B}}^{\mathrm{bin}}$, it reduces to the effective velocity $v_{\mathrm{eff}}$, i.e.

$$v_{\mathrm{ff}}(R) = \sqrt{v_{\mathrm{eff}}^2 + \frac{2M_{\mathrm{tot}}}{R} - \frac{2M_{\mathrm{tot}}}{r_{\mathrm{B}}^{\mathrm{bin}}}} \,. \qquad (4.1.29)$$

Also, we identified with $n_{\mathrm{gas}}(R)$ the density profile at a distance $R$ from the center of mass of the binary. Rearranging (4.1.28), we can therefore compute the number density of hydrogen gas at a distance $R$ from the binary center as

$$n_{\mathrm{gas}}(R) = \frac{\dot{M}_{\mathrm{bin}}}{4\pi m_H v_{\mathrm{ff}}(R) R^2}. \qquad (4.1.30)$$

We highlight that the number density scales linearly with the total binary accretion rate $\dot{M}_{\mathrm{bin}}$. This means that the larger the accretion rate of gas towards the binary system, the denser the environment the individual PBHs populate. We can also interpret Eq. (4.1.30) as showing that, being the flow of infalling baryons constant, $n_{\mathrm{gas}}(R)$ increases going towards the binary from its asymptotic value to be found at a distance larger than $r_{\mathrm{B}}^{\mathrm{bin}}$.

The accretion rates for each PBH in the binary can be expressed as

$$\dot{M}_1 = 4\pi m_H n_{\mathrm{gas}}(r_{\mathrm{B,1}}) v_{\mathrm{eff,1}}^{-3} M_1^2,$$
$$\dot{M}_2 = 4\pi m_H n_{\mathrm{gas}}(r_{\mathrm{B,2}}) v_{\mathrm{eff,2}}^{-3} M_2^2, \qquad (4.1.31)$$

in terms of the effective velocity of each PBH as in Eq. (4.1.27) and the (overdense) gas profile (4.1.30). As a consistency check of this setup, one can also estimate the individual Bondi radii $r_{B,i}$ of each accreting PBH, finding it is much smaller than the binary Bondi radius $r_B^{bin}$. This is because the individual PBH velocities induced by the orbital motion are much larger than both $v_{rel}$ and $c_s$ for $a \ll r_B^{bin}$. We also stress that the single PBH accretion rates are computed for two naked PBHs. This choice is justified since, given the large orbital velocities and the small separation, we do not expect the PBHs to be able to retain their individual dark matter halo. We, therefore, took $\lambda \approx 1$, compatible with the naked scenario at low redshift described above. When computing the binary accretion rate, on the other hand, we take into account the dark halo surrounding the system and the eigenvalue $\lambda$ includes corrections in terms of $\kappa$ [234], as discussed above. Finally, one can write the analytical formula for the individual accretion rates as

$$\dot{M}_1 = \dot{M}_{bin}\sqrt{\frac{1 + \zeta + (1-\zeta)\gamma^2}{2(1+\zeta)(1+q) + (1-\zeta)(1+2q)\gamma^2}}\,,$$

$$\dot{M}_2 = \dot{M}_{bin}\sqrt{\frac{(1+\zeta)q + (1-\zeta)q^3\gamma^2}{2(1+\zeta)(1+q) + (1-\zeta)q^2(2+q)\gamma^2}}\,, \tag{4.1.32}$$

where we defined $\zeta = e\cos u$ and $\gamma^2 = av_{eff}^2/\mu q$. As one expects, there exists a periodic modulation of the accretion rate which depends on the binary motion in the overdense environment. The previous equations simplify in the limit of very small mass ratio, $q \to 0$, and become independent from $e$ and $u$ as

$$\dot{M}_1 = \dot{M}_{bin} + \mathcal{O}(q)\,, \qquad \dot{M}_2 = \sqrt{\frac{q}{2}}\dot{M}_{bin} + \mathcal{O}(q^{3/2})\,. \tag{4.1.33}$$

In the general case, we can also choose to average over the orbital motion. This can be done as the characteristic accretion timescale $\tau_{acc}$ is much longer than the characteristic orbital period (we will come back to this point in the section addressing the binary evolution). Therefore, averaging over the angle $u$ in Eq. (4.1.32), one finds that the dependence on the eccentricity becomes negligible [12] and one can write

$$\dot{M}_1 = \dot{M}_{bin}\sqrt{\frac{M_1q^2 + a(1+q)v_{eff}^2}{(1+q)[2M_1q^2 + a(1+2q)v_{eff}^2]}}\,,$$

$$\dot{M}_2 = \dot{M}_{bin}\sqrt{\frac{q[M_1 + a(1+q)v_{eff}^2]}{(1+q)[2M_1 + a(2+q)v_{eff}^2]}}\,. \tag{4.1.34}$$

As a final step, we consider the limit

$$M_1q^2 \gg av_{eff}^2\,. \tag{4.1.35}$$

One enters in this regime when

$$M_1 \gtrsim \mathcal{O}(M_\odot)\,, \qquad \text{and} \qquad a \sim \mathcal{O}(10^6)M_1 \tag{4.1.36}$$

for any $q > 10^{-2}$. Then, the previous expressions reduce to

$$\dot{M}_1 = \dot{M}_{bin}\frac{1}{\sqrt{2(1+q)}}\,,$$

$$\dot{M}_2 = \dot{M}_{bin}\frac{\sqrt{q}}{\sqrt{2(1+q)}}\,. \tag{4.1.37}$$

As an additional consistency check, it is also worth stressing that with this parametrisation one recovers $\dot{M}_1 = \dot{M}_2 = \dot{M}_{bin}/2$ in the case of equal mass binaries $q \to 1$.

In the following sections, where the binary evolution is addressed, we will show explicitly that we are always interested in the regime in which Eq. (4.1.37) provides a good description of the evolution of the masses of each PBH in the binary. This description has also the quality of being independent of the orbital parameters. The mass growth is therefore dictated by the *external* condition given by $\dot{M}_{\mathrm{bin}}$ and the initial mass ratio $q_i$. We can also express Eq. (4.1.37) in terms of the Eddington normalised rates ($\dot{m}_i = \tau_{\mathrm{Salp}} \dot{M}_i/M_i$) as

$$\dot{m}_1 = \dot{m}_{\mathrm{bin}} \sqrt{\frac{1+q}{2}}, \qquad \dot{m}_2 = \dot{m}_{\mathrm{bin}} \sqrt{\frac{1+q}{2q}}. \tag{4.1.38}$$

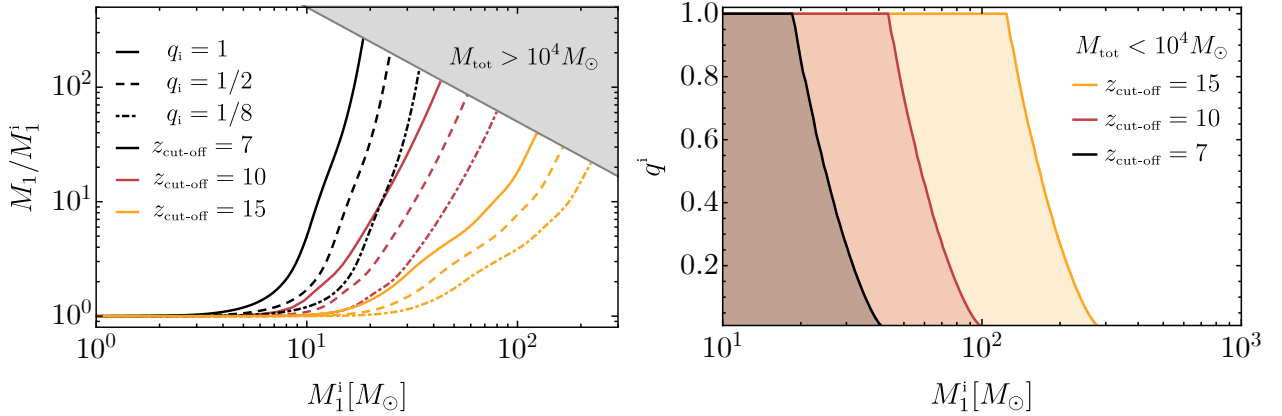

Figure 4.6: **Left:** *Evolution of the primary mass from its initial value at formation to the value after $z_{cut\text{-}off}$, for various initial mass ratios and different cut-off redshifts.* **Right:** *The shaded region indicates the parameter space for which the final total mass $M_{tot}$ is below $M_{tot} = 10^4 M_\odot$.*

Focusing on the mass ratio evolution, we can rearrange Eq. (4.1.37) to show that

$$\dot{q} = q \left( \frac{\dot{M}_2}{M_2} - \frac{\dot{M}_1}{M_1} \right) = \frac{q}{\tau_{\mathrm{Salp}}} \left( \dot{m}_2 - \dot{m}_1 \right). \tag{4.1.39}$$

As the secondary mass evolves faster than the primary, given the fact that it experiences an overdense environment driven by the total binary mass $M_{\mathrm{tot}}$, the mass ratio monotonically grows with time. This growth stops when the mass ratio is close to unity, at which point both masses are growing at a similar rate. The prediction of the effect of accretion onto PBHs in binary is therefore that masses tend to become more similar, within a timescale inversely proportional to the binary mass accretion rate.

**Predictions for the PBH mass evolution in binaries**

In this section, we give a summary of the predictions for the distribution of masses of PBHs in binaries, following the accretion model described in the previous section. We denote $M_j$ as the *final* mass of the $j$-th binary component while $q$ corresponds to the *final* mass ratio of the binary. With final we refer to values after all relevant accretion has taken place, meaning quantities at redshift $z < z_{\mathrm{cut\text{-}off}}$. On the other hand, we will define quantities at formation epoch with a subscript "i". Recall that a PBH with mass $M_{\mathrm{PBH}}$ forms at the characteristic redshift

$$z_i \simeq 2 \cdot 10^{11} \left( \frac{M_{\mathrm{PBH}}}{M_\odot} \right)^{-1/2} \tag{4.1.40}$$

Therefore, the *initial* masses are denoted by $M_j^i$ and the initial mass ratio by $q_i$. We adopt the standard notation in which $M_2 \leq M_1$ (and consequently $q \leq 1$). As the mass ratio is increasing up to the fixed point $q \to 1$, this ordering is preserved during the accretion-driven evolution.

To give a feeling of the numbers involved, in Fig. 4.6 we show the evolution of the primary mass of the binary. We choose three different cut-off redshifts $z_{\mathrm{cut\text{-}off}} = (15, 10, 7)$, corresponding to different accretion strengths and durations. We also choose different values of the initial mass ratio. As one can appreciate, binaries with the primary component lighter than a few solar masses are not modified even for relatively efficient accretion (i.e. small $z_{\text{cut-off}}$). The masses above $M_i \approx 10 M_\odot$, instead, can grow up to two orders of magnitude. For a smaller mass ratio, the change of the primary mass is reduced, as the binary system as a whole is lighter and it can attract less gas in the vicinity of the binary. Let us stress here again that we want to focus on binaries that may be eventually detectable by ground-based experiments. Therefore, as the characteristic frequency of a merger with $M_{\text{tot}} \gtrsim 10^4 M_\odot$ is not observable, for simplicity we disregard the description of all the binaries that are pushed above that value by accretion. Consistently with this choice, in the right panel of Fig. 4.6, we show the parameter space in the $q_i - M_1^i$ plane which is resulting in a binary with final total mass below $M_{\text{tot}} \simeq 10^4 M_\odot$.

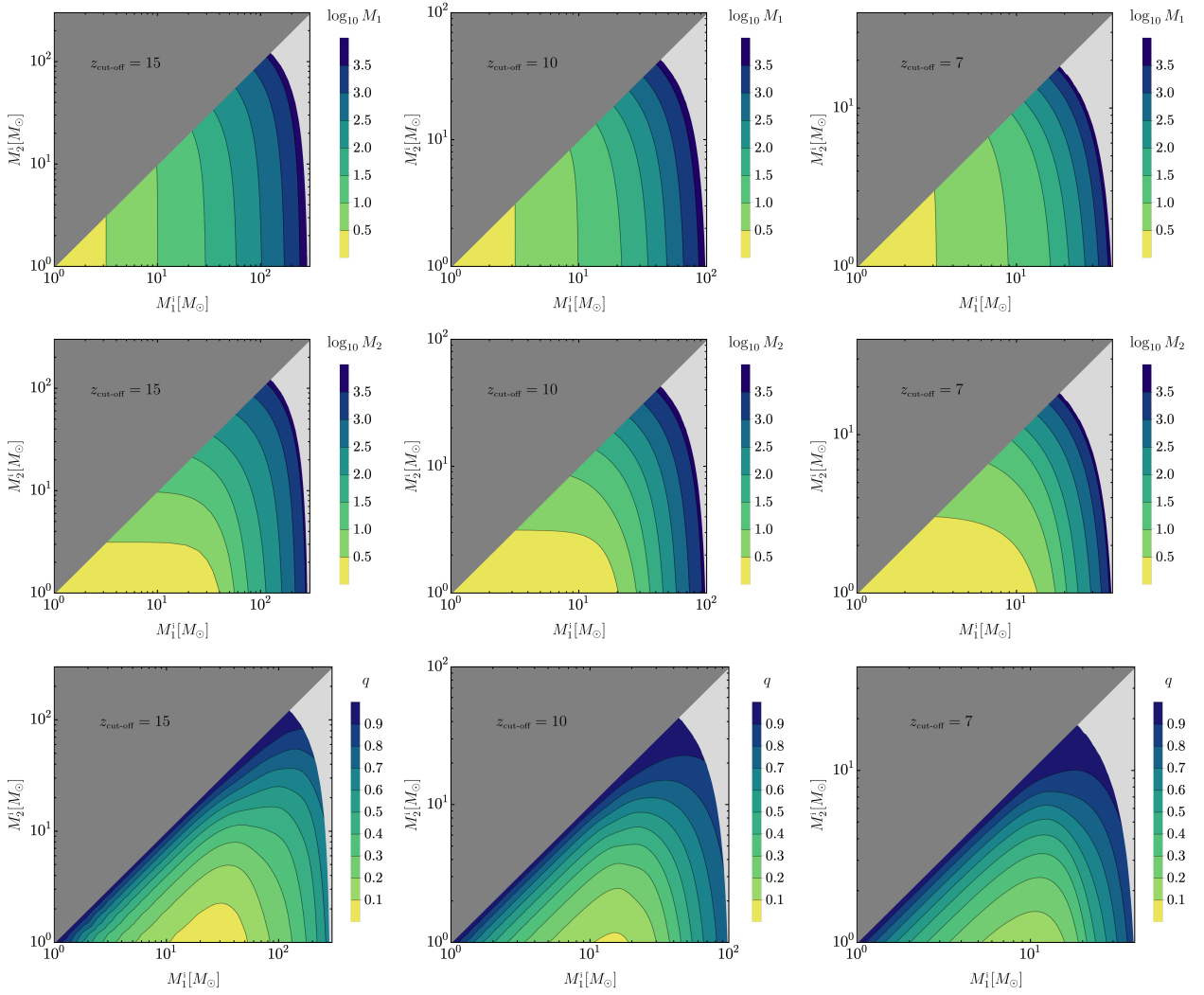

Figure 4.7: **From top to bottom:** *final masses and mass ratio as a function of the corresponding initial masses* $(M_1^i, M_2^i)$. *As in the previous case, we show three accretion scenarios parametrised by different values of the cut-off redshift. We shade the upper part of the corner as we impose $M_2 \leq M_1$, a feature which is maintained by the evolution set by accretion, see discussion in the text.* **Left:** $z_{\text{cut-off}} = 15$. **Center:** $z_{\text{cut-off}} = 10$. **Right:** $z_{\text{cut-off}} = 7$.

Let us finally mention that heavier objects would be natural candidates for intermediate-mass BH binaries, and therefore possible sources for space-based detectors like LISA. The study of accretion onto such heavier systems requires, however, dedicated studies.

To give the connection of the masses at high redshift predicted by a particular formation mechanism to the final masses when accretion has taken place, we also plot the maps determining this correspondence in Fig. 4.7. Indeed, we show the final masses $M_1, M_2$ and their mass ratio $q$ in terms of the initial masses $M_1^i, M_2^i$, for an evolution taking place until the cut-off redshift $z_{\text{cut-off}} = (15, 10, 7)$. As we can see in the bottom plots, the result shows the tendency of accretion to shift the mass ratio towards unity. For any given secondary initial mass $M_2^i$, there exists a point in which the tendency in the plot is inverted and the final mass ratio starts growing with $M_1^i$. This shows the tendency of the PBH model to generate symmetric binaries when accretion is efficient.

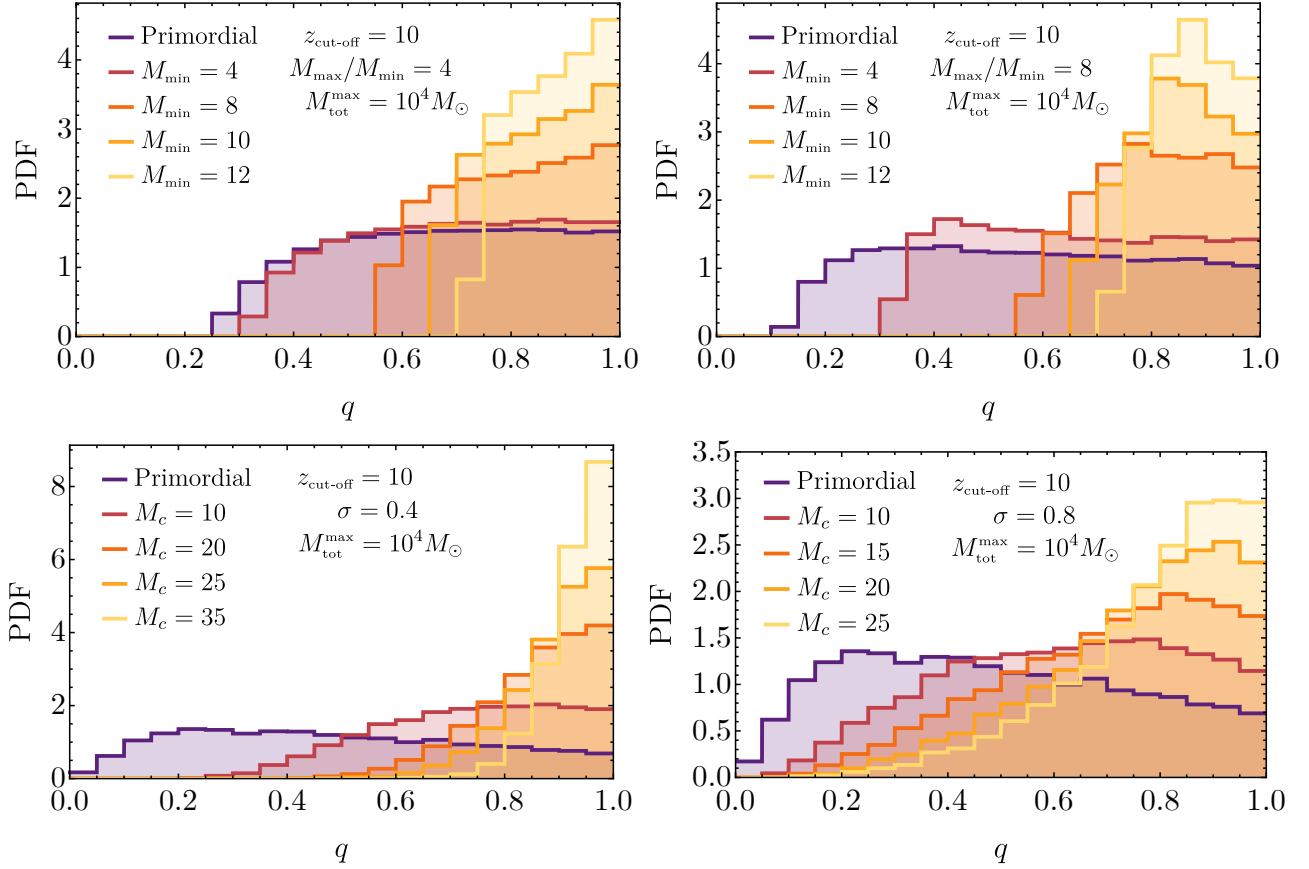

Figure 4.8: *Examples of q distribution evolution for an accretion scenario characterised by $z_{\text{cut-off}} = 10$. The distributions are found neglecting binaries that are pushed to values of $M_{\text{tot}}$ above $10^4 M_\odot$, as they are irrelevant for the ground-based detectors.* **Top:** *Power-law mass function.* **Bottom:** *Lognormal mass function.*

In Fig. 4.8, we further analyse the effect of accretion on the mass ratio distribution expected for a PBH binary population. We consider two characteristic mass functions often used in literature and representative of a wide class of PBH models. We refer to Sec. 3.1 for all the details. Namely, we adopt a power law mass function

$$\psi(M, z_i) = \frac{1}{2} \left( M_{\text{min}}^{-\frac{1}{2}} - M_{\text{max}}^{-\frac{1}{2}} \right)^{-1} M^{-\frac{3}{2}} \qquad (4.1.41)$$

and a lognormal mass function

$$\psi(M, z_i) = \frac{1}{\sqrt{2\pi}\sigma M} \exp\left( -\frac{\log^2(M/M_c)}{2\sigma^2} \right). \qquad (4.1.42)$$

As we can see in the examples shown in Fig. 4.8, mass functions with sizeable support where PBHs experience sizeable accretion rates, give rise to final mass ratio distributions which are strongly peaked around zero. We considered various choices for the mass function parameters for presentation purposes. In the next chapter, we will compare the prediction of the PBH model in a data-driven analysis, where the choice of the parameters of the mass function will be fixed by a best-fit procedure on the GW data.

It is worth mentioning here the importance of selection effects when building such distributions. In the case shown in Fig. 4.8, we define an "ad-hoc" selection by discarding binaries heavier than $10^4 M_\odot$. As accretion becomes increasingly stronger, a larger portion of binaries is expected to surpass this limit, and lost from the "observable population". This reduces the number of symmetric binaries in the count and changes the shape of the observable distribution. In the actual analysis, which is going to be presented in the following chapter, the selection effects will be naturally set by the sensitivity of the LIGO/Virgo detectors.

### 4.1.3   Limitations of the accretion model

As we already expressed in various parts of the text, large uncertainties are affecting the description of accretion of baryonic matter onto compact objects throughout the evolution of the universe. This is because accretion is a complex phenomenon that requires the precise knowledge of various quantities difficult to model throughout cosmic history (one example being the velocity of PBHs, etc.). In the previous discussion, we introduced an effective parameter $z_{\text{cut-off}}$ which determines the epoch at which accretion stops being effective. In practice, as we are going to be interested in the PBH population at low redshift, i.e. after all relevant accretion has taken place, one can also vary $z_{\text{cut-off}}$ to capture scenarios in which the accretion rate is systematically smaller than the one found with this model.

In the following, we collect the critical uncertainties affecting the accretion model we adopted [232, 234, 408].

- **Local feedback:** this corresponds to the thermal feedback induced by the X-ray emission. When accretion onto PBHs takes place, the electromagnetic emission can impact the gas temperature and ionization in the vicinity of each PBH. We neglected this effect in our analysis as the effect of local heating is found to be negligible for the mass range of interest for LIGO/Virgo [232, 408].

- **Global feedback & X-ray pre-heating:** these feedback effects modify the temperature and ionization of the cosmic gas and were included in [408], neglecting sources of X-ray heating other than PBHs (see for example Ref. [419]). We note, however, that the cosmic ionisation computed in Ref. [408] has been revisited in the detailed analysis made in Ref. [232]. The latter showed that the global feedback is far less important for PBHs with masses in the LIGO/Virgo band. In particular, we note that Ref. [232] found an accretion rate which is consistent with the one presented in Ref. [408] *without* the effect of global heating. Still, we stress the crucial importance of better modelling the evolution of the temperature of the intergalactic medium during the epochs leading to the reionization of the universe, i.e. $10 \lesssim z \lesssim 30$. This is due to the dependence of the accretion rate on the effective velocity, which would be impacted by an increase in temperature (corresponding to an increase of the sound speed of the gas $c_s$).

- **DM halo:** as we discussed, while our results are consistent with Ref. [232], the inclusion of the dark matter halo around each PBH enhances the accretion rate. Following the recent development of N-body numerical simulations, it was shown that the formation of a dark halo around a PBH in the early universe is inevitable [183, 235] if $f_{\text{PBH}}$ is smaller than unity. There are, however, uncertainties in the halo growth due to possible non-linear effects modifying the scaling with redshift $M_h \propto 1/z$ close to the epoch of structure formation.

- **Structure formation:** as large scale halos start forming, around redshift $z \simeq 10$, an increasingly larger fraction of PBHs is inevitably captured in the potential well of the large-scale structures. This leads to an increase in the relative velocity of PBHs. In Ref. [417], the evolution of the velocity was estimated, matching the results of large-scale structure numerical simulations, finding an increase of an order of magnitude around $z_{\text{cut-off}} = 10$ with respect to the linear prediction. As a result, the accretion rate is expected to drop sharply, due to its non-linear dependence on the velocity [169, 234, 236]. On the other hand, deviations from this value can still be expected due to the uncertain effect of the global thermal feedback on the speed of sound in the medium.

- **Spherical accretion & disk geometry**: the investigations of accretion onto compact objects which we adopted as a base for our modelling of the PBH accretion are mostly based on semi-analytical studies assuming a quasi-spherical flow of matter. It was shown in Ref. [420] that there may be outflows that make the quasi-spherical approximation break down. The final accretion rate expected in such a situation is however uncertain as it also depends on the exact geometry of the accretion flow and the direction of the outflows relative to the motion of the compact object with respect to the surrounding gas.

- **Angular momentum transfer**: as we will see in details in the section dedicated to the evolution of the PBH spin, when $\dot{m} \sim 1$ one expects the formation of a geometrically thin accretion disk. In this case, the transfer of angular momentum can be described with the geodesic model introduced in Ref. [421]. However, for $\dot{m} \ll 1$ or $\dot{m} \gg 1$, deviations from this geometrical configuration are expected, and therefore modifications to the efficiency of the angular momentum transfer, as well as to the accretion luminosity and feedbacks are possible. For example, in the latter case, corresponding to the super-Eddington regime, the disk is expected to become geometrically thicker. Even though this configuration is more complex, numerical simulations suggest that the characteristic spin evolution time scale is not altered significantly, see Ref. [422]. This discussion will be of particular interest in the following sections where the effect of accretion on the PBH spin will be addressed.

All the points raised in the previous list require complex and model-dependent simulations to make significant improvements with respect to the current state of the art in the literature. As we discussed above, we will capture modifications of the accretion efficiency due to systematic uncertainties in the model by allowing $z_{\text{cut-off}}$ to vary in an agnostic way.

## 4.2 Mass function evolution

In this section, we present the effect of accretion on the PBH mass function, along with the change of the overall PBH abundance. In doing so, we will closely follow Ref. [14]. In the second part of this section, following Ref. [13], we will also discuss what are the consequences of the modification of the PBH population when constraints at different epochs are considered.

To make an example of those effects, let us consider the idealised case of a monochromatic population of PBHs with mass $M_{\text{PBH}} \simeq \mathcal{O}(30) M_{\odot}$. Depending on the strength of accretion, we saw that $M_{\text{PBH}}$ can grow up to orders of magnitude. This implies they must contribute to a larger portion of the energy density budget of the universe, and therefore a larger $f_{\text{PBH}}$ is expected at low redshift. On the other hand, let us assume a particular constraint limits the abundance of PBHs of a given (high redshift) mass. This does not translate into an identical bound on $f_{\text{PBH}}$ at low redshift. As PBHs grow throughout cosmic history, the high redshift bound gets shifted to higher masses. This effect is very important when considering bounds coming from CMB observations, as we shall see. Notice also that, in fact, all constraints are sensitive to the PBH number density $n_{\text{PBH}}$ and are only translated in terms of $f_{\text{PBH}}$ when the mass distribution is considered. This procedure should be performed by accounting for accretion effects on an experiment-by-experiment basis.

### 4.2.1 Accretion effects on the PBH mass function and abundance

Following the notation adopted in Sec. 3.1, we define the (redshift-dependent) mass function $\psi(M, z)$ as corresponding to the relative fraction of PBHs characterised by a mass falling in the interval $(M, M + \mathrm{d}M)$. As in the previous section, we will refer to the initial mass function at formation as $\psi(M_i, z_i)$. Its evolution is then governed by the equation [13, 14]

$$\psi(M(M_i, z), z)\mathrm{d}M = \psi(M_i, z_i)\mathrm{d}M_i. \tag{4.2.1}$$

We stress again here our choice of notation for which $M(M_i, z)$ is the final mass at redshift $z < z_{\text{cut-off}}$ reached by an initial PBH with mass $M_i$ at redshift $z_i$.

As PBH binaries are expected to be a subdominant population with respect to the total amount of PBHs (scaling as $f_{\text{PBH}}^2$, see next sections), one needs to consider the evolution of the mass function as dominantly given by accretion onto isolated PBHs. Therefore, when considering the impact of accretion on constraints at low redshift, accretion on isolated PBHs will provide the leading effect.

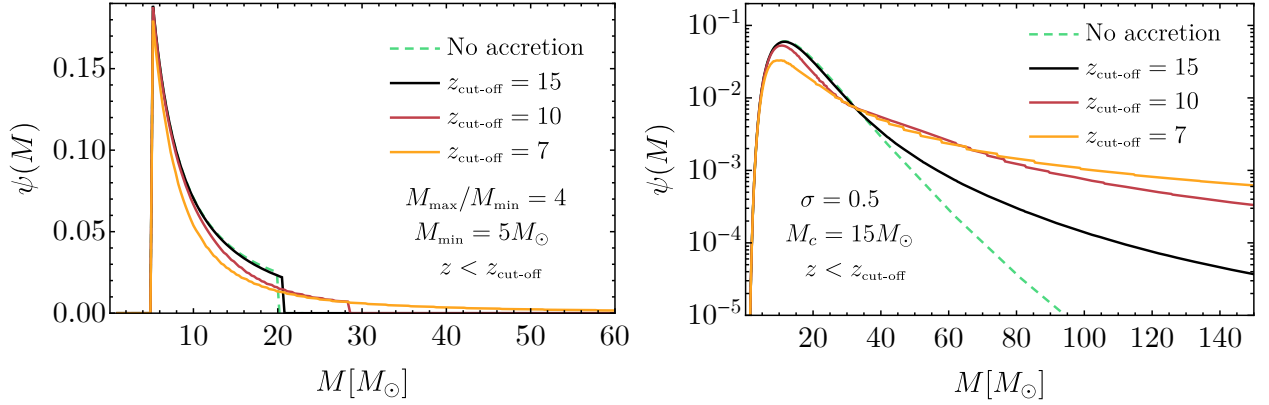

Figure 4.9: *Example of the mass function evolution for three representative values of $z_{cut\text{-}off}$.* **Left:** *Power-law mass function.* **Right:** *Lognormal mass function.*

As the accretion rate does not scale linearly with the mass (being relatively larger for heavier PBHs), one also expects the mass distribution to be distorted, acquiring a larger tail at high masses. This produces an enhancement of the high-mass tail that can reach values orders of magnitude larger than its corresponding value at formation [13]. We show a couple of representative examples of this effect in Fig. 4.9.

Finally, also the overall value of $f_{PBH}$ is affected by accretion. For simplicity, we are going to assume a non-relativistic dominant dark matter component, while PBHs only accounts for a small fraction of the total dark matter budget. Then, one can show that [13]

$$f_{PBH}(z) = \frac{\rho_{PBH}}{(\rho_{DM} - \rho_{PBH}) + \rho_{PBH}} = \frac{\langle M(z)\rangle}{\langle M(z_i)\rangle (f_{PBH}^{-1}(z_i) - 1) + \langle M(z)\rangle}, \qquad (4.2.2)$$

defined in terms of the average mass

$$\langle M(z)\rangle = \int \mathrm{d}M M \psi(M, z). \qquad (4.2.3)$$

In Fig. 4.10 we show that, due to accretion effects, $f_{PBH}(z)$ at low redshift ($z \ll z_{cut\text{-}off}$) can be significantly larger than $f_{PBH}(z_i)$.

One may worry that this growth of $f_{PBH}$ consequently leads to a net growth of the total dark matter abundance in the universe, potentially violating the observed ratio between the energy density of baryons compared to the dark matter one. However, we stress again that such a growth is only observed as long as PBHs remain a *subdominant* component of the dark matter, as the accretion rate is computed with the inclusion of the dark halo. The overall dark matter abundance is, therefore, mainly unchanged. If PBHs were a sizeable fraction of the dark matter, the accretion rate would be drastically reduced, and no significant modification of $f_{PBH}$ would be observed.

### 4.2.2 Impact on constraints at different redshifts

In this section, we address the issue of comparing constraints sensitive to the PBH population at different epochs. As we discussed in the previous section, accretion distorts the mass distribution of the PBH population giving relatively more importance on the high-mass tail as well as augmenting the overall value of $f_{PBH}$.

To perform this task, we need to introduce the technique routinely used in the literature to translate constraints derived assuming monochromatic PBH populations to extended ones. Following the prescription described in Refs. [258, 259] for extended mass distributions, and Ref. [13] with the inclusion of accretion effects, we estimate the bound on the fraction $f_{PBH}(z_\delta)$ of PBHs at the redshift $z_\delta$ from

$$f_{PBH}(z_\delta) \lesssim \left( \int_{M_{min}(z_\delta)}^{M_{max}(z_\delta)} \mathrm{d}M \frac{\psi(M, z_\delta)}{f_{max}(M, z_\delta)} \right)^{-1}, \qquad (4.2.4)$$

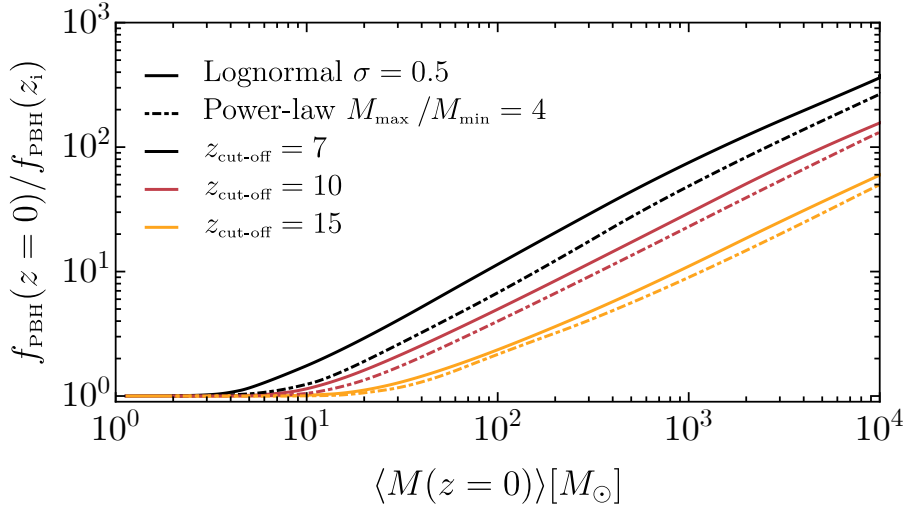

Figure 4.10: *Example of evolution of the PBH abundance for both power-law and lognormal mass functions.*

where $M_{min}(z_\delta)$ and $M_{max}(z_\delta)$ identify the range of masses affected by the given constraint, and $f_{max}(M, z_\delta)$ represents the maximum allowed fraction for a monochromatic mass function at the redshift of interest for a given experiment [199, 258]. As the mass function depends on the redshift, the constraint will be computed using the quantities at redshift $z_\delta$. Then, the result will be translated into a constraint applying at low redshift quantities, therefore the bound $f_{PBH}(z_\delta)$ is mapped to $f_{PBH}(z = 0)$ using Eq. (4.2.2). This procedure will be performed for all experiments sensitive to high redshift quantities. On the other hand, constraints like the one coming from lensing observations at the current epoch, will be computed with the present (evolved) mass function while they are not evolved further.

For simplicity, for any given PBH mass range, we consider only the most stringent constraint (see, e.g., Ref. [199] for a recent review on the topic). In Fig. 4.11, we show a sketch of how the constraints are sensitive to quantities at different redshifts. As PBH accretion is surely negligible for masses below $\mathcal{O}(1)M_\odot$, only constraint in the massive portion of the plot should be carefully addressed with the procedure described in this section.

In the range of masses of interest, i.e. $M \in (1 \div 10^4)M_\odot$, the relevant constraints come from lensing, dynamical processes, formation of structures, and accretion related phenomena. In particular, the bounds coming from lensing observations, are the ones inferred from Supernovae [223], the MACHO and EROS experiments [219, 220], ICARUS (I) [221] and radio [424] observations. They all depend on lensing sources at low redshift $z \ll z_{cut-off}$. Dynamical constraints come from disruption of wide binaries [245] and the requirement of survival of star clusters in Eridanus II [246] and Segue I [247]. All those referred to small redshifts observations. Bounds coming from observations of the Lyman-$\alpha$ forest are sensitive to physical phenomena happening at a redshift larger than $z \approx 4$ [252]. The most stringent constraint in the region above $\sim \mathcal{O}(10)M_\odot$ comes from CMB observations. In particular, Planck data on CMB anisotropies constrain the emission of X-rays due to spherical or disk (Planck D) [183, 232] accretion. This process can impact the CMB only at high redshifts when $z \gtrsim 300$. On the other hand, the observed number of X-ray (Xr) [237, 238] and X-ray binaries (XrB) at low redshifts [240] put additional constraint in this heavy portion of the mass window of interest. Finally, we also consider constraints on the primordial abundance of sub-solar PBHs as set by the LIGO-Virgo observations [423]. We stress that a dedicated discussion of the constraints coming from GW observation of BH mergers will be performed in the following chapter. A comprehensive plot of all the constraints can be found in Fig. 10 of Ref. [199] for a monochromatic PBH mass function.

We decided to present the results in terms of quantities in the present day universe, i.e. $f_{PBH}(z = 0)$ and $\langle M(z = 0) \rangle$. This will be particularly useful when comparing constraints to the recent observations of events by the LIGO/Virgo detectors. When dealing with future experiments like the Einstein

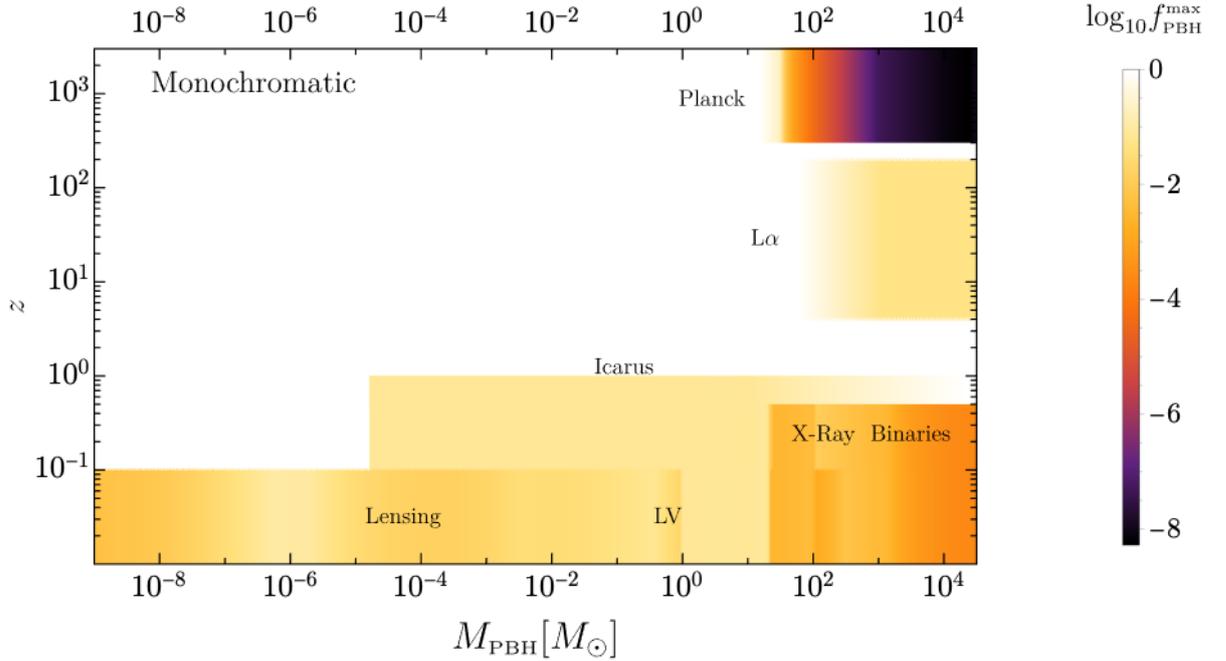

Figure 4.11: *Representative sketch of how different constraints are intrinsically dependent on redshift, see also Ref. [197]. This is of crucial importance when comparing different constraints to the low redshift PBH population, as we will do in the chapter dedicated to the confrontation of the PBH scenario with the LIGO/Virgo measurements. For simplicity here, the constraint coming from the LIGO/Virgo (LV) observations is only plotted in the sub-solar mass range [423]. More details on the constraint in all the observable window of LVC detectors will be presented in the following chapter.*

Telescope [275], which will be able to reach mergers at redshift higher than $z_{\text{cut-off}}$, the corresponding constraint must be evaluated the redshift of interest.

In Fig. 4.12, we show how the constraint on the PBH abundance at low redshift changes depending on the strength of accretion for a monochromatic and lognormal mass function. We choose to plot a single envelope resulting from the most stringent bound for any value of $\langle M(z=0) \rangle$, while the labels identify the corresponding experiment dominating each portion of the graphs. We show the results for various accretion strengths, corresponding to $z_{\text{cut-off}} = 15$, 10 and 7. As the dominant constraint for masses larger than $\sim \mathcal{O}(10)M_\odot$ is coming from CMB observations (dependent on high redshift phenomena), the bound at low redshift gets increasingly less stringent as accretion is shifting the constraint to larger masses, with respect to the original constraint which neglected accretion.

To summarise this section, we have described how accretion onto PBHs may change the interpretation of the observational bounds on the current fraction of PBHs for a given mass range. Our findings are particularly relevant when trying to assess the possibility that the BH mergers observed by current GW experiments could be ascribed to PBH binaries, as we will see in the following chapter. We stress that the effect of accretion on PBHs is intrinsically redshift dependent, and therefore it is also of crucial importance when considering the forecasts of future experiments, like the Einstein Telescope, which will probe higher redshifts and higher PBH masses. Also, the present discussion may play a crucial role when trying to understand if PBHs may provide an explaining the supermassive black holes observed at $z \gtrsim 6$ [425], as it shows that if accretion is strong enough, one may still generate large BH seeds by evading the CMB bounds set at high redshift, see also discussion in Ref. [183]. Finally, we acknowledge that the present analysis is sensitive to the same uncertainties affecting the accretion dynamics, as we discussed in the previous section. This is the reason why we decided to present different scenarios, depending on the parameter $z_{\text{cut-off}}$.

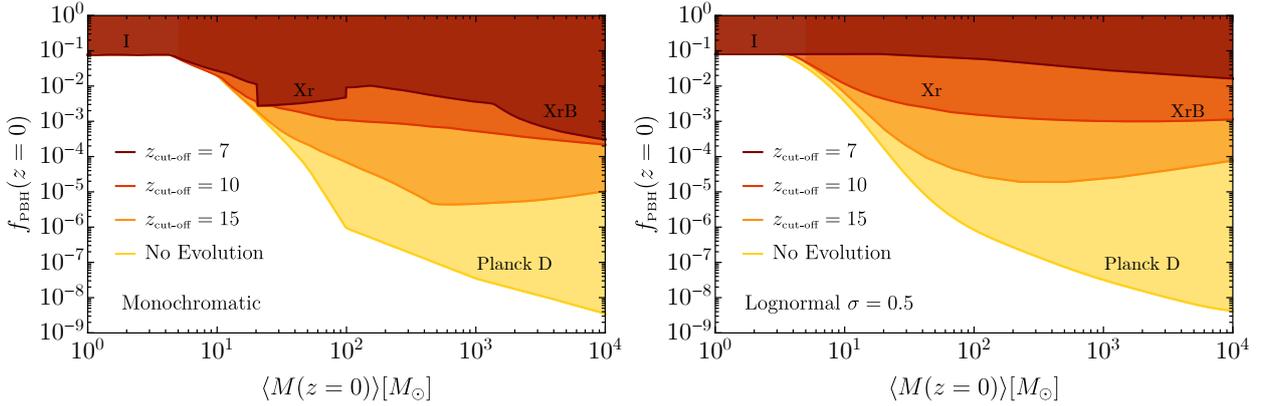

Figure 4.12: *Combined constraints on $f_{PBH}(z = 0)$ today as a function of the average mass $\langle M(z=0)\rangle$ for different accretion strengths, corresponding to $z_{cut\text{-}off} = 15$, 10 and 7. For comparison, we also show the original bound which neglected accretion ("No Evolution"). As the dominant bound coming from CMB is set at high redshift, stronger accretion leads to a larger shift of this bound on the heavier portion of the parameter space.* **Left:** *monochromatic mass distribution.* **Right:** *lognormal mass distribution with $\sigma = 0.5$ at formation.*

## 4.3 Spin evolution

In this section, we study the evolution of PBH spin by considering both isolated PBHs and PBHs in binary systems. We adopt the modelling of accretion presented in the previous section and follow closely Ref. [14].

The crucial property of the accretion phenomenon affecting the evolution of the PBH angular momentum is the geometry of the accretion flow. The infalling accreting gas can carry angular momentum, which crucially determines the evolution of the PBH spin (see, for example, Ref. [426]).[1]

### 4.3.1 Spin evolution due to thin-disk accretion

Before laying down the consequences of accretion from a thin disk, we set up the condition for such a configuration to be realised.

**Conditions for a thin-disk formation**

Focusing on an isolated PBH, the gas possesses enough angular momentum if its characteristic velocity (which can be estimated using the baryon velocity variance [234]) is larger than the Keplerian velocity at the vicinity of the PBH. If such a condition is met, one expects the flow to develop a disk configuration. Therefore, the minimum PBH mass at a given redshift $z$ for which accretion proceeds in a non-spherical regime is [14, 234]

$$M \gtrsim 6 \times 10^2 M_\odot \, D^{1.17} \xi^{4.33}(z) \frac{(1 + z/1000)^{3.35}}{\left[1 + 0.031 \left(1 + z/1000\right)^{-1.72}\right]^{0.68}}, \tag{4.3.1}$$

where $\xi(z) = \text{Max}[1, \langle v_{\text{eff}}\rangle/c_s]$ parametrises the effect of the PBH proper velocity in reducing of the Bondi radius and the additional factor $D \sim \mathcal{O}(1 \div 10)$ introduces corrections coming from relativistic effects.

If condition (4.3.1) is met, there exists enough angular momentum in the system to allow for the formation of a disk. However, we will make the conservative assumption that an efficient transmission of angular momentum to the PBH is present only when a *thin* accretion disk is attained. If $\dot{m} < 1$ and accretion is non-spherical, an advection-dominated accretion flow (ADAF) may form [432]. On the other hand, $\dot{m} \approx 1$, the non-spherical accretion can proceed by acquiring a geometrically thin accretion disk [433]. Finally, in the regime characterised by super-Eddington accretion rates $\dot{m} \gg 1$, the luminosity of the gas might distort the disk forcing it to become thicker. As the accretion rates in the mass range of interest for the LIGO/Virgo observations are not reaching levels above the Eddington rate, following Ref. [234], we assume that a thin disk forms when both Eq. (4.3.1) and

$$\dot{m} \gtrsim 1 \tag{4.3.2}$$

are satisfied. In practice, it can be shown that the latter condition is always a sufficient condition for the formation of a thin accretion disk in the parameter space we are interested in [14].

---

[1]It was shown that PBH spin may change under the effect of "plasma-driven" superradiant instabilities [427–429] extracting angular momentum from the compact object. This phenomenon is, however, dependent upon the geometry of the surrounding plasma and was shown to be negligible for realistic systems [430, 431].

The setup may be different when PBHs are bounded in binary systems. It turns out that, in this case, the transfer of angular momentum to each PBH is more efficient [14]. This is because each binary component has a velocity that is dominated by the orbital motion, and therefore in this case condition (4.3.1) is absent.

To summarize, for both isolated and binary PBHs we can assume that a thin accretion disk forms whenever $\dot{m} \gtrsim 1$ along the cosmic history. We stress again here, that in the parameter space we are interested in, $\dot{m}$ never exceeds unit significantly [14] and the thin disk regime can be regarded as a reliable approximation in the super-Eddington regime.

**The geodesic model of thin disks**

In the presence of a thin accretion disk, it is natural to expect an efficient transfer of angular momentum from the gas to the PBH and, therefore, a subsequent spin-up. We recall that PBHs are born with very small angular momentum, see Sec. 3.2. We can conservatively take (almost) spinless PBHs as our initial condition. As spin is build up from the accretion disk itself, we expect it to be aligned perpendicularly to the accretion disk [433, 434].

In this configuration, we adopt the geodesic model to describe the angular-momentum accretion as presented in the seminal work in Ref. [421]. For circular disk motion the rate of change of the PBH angular momentum magnitude

$$J \equiv |\vec{J}| \equiv \chi M^2 \tag{4.3.3}$$

is related to the mass accretion rate (see also Refs. [421, 435–437])

$$\dot{J} = \frac{L(M, J)}{E(M, J)} \dot{M}, \tag{4.3.4}$$

where we defined

$$E(M, J) = \sqrt{1 - 2\frac{M}{3r_{\mathrm{ISCO}}}}$$
$$L(M, J) = \frac{2M}{3\sqrt{3}} \left( 1 + 2\sqrt{3\frac{r_{\mathrm{ISCO}}}{M} - 2} \right), \tag{4.3.5}$$

and the Innermost Stable Circular Orbit (ISCO) radius reads

$$r_{\mathrm{ISCO}}(M, J) = M \left[ 3 + Z_2 - \sqrt{(3 - Z_1)(3 + Z_1 + 2Z_2)} \right], \tag{4.3.6}$$

with

$$Z_1 = 1 + \left(1 - \chi^2\right)^{1/3} \left[ (1 + \chi)^{1/3} + (1 - \chi)^{1/3} \right],$$
$$Z_2 = \sqrt{3\chi^2 + Z_1^2}. \tag{4.3.7}$$

Finally, Eq. (4.3.4) can be used to describe the time evolution of the dimensionless Kerr parameter $\chi(t)$. After few manipulations, one finds

$$\dot{\chi} = (\mathcal{F}(\chi) - 2\chi) \frac{\dot{M}}{M}, \tag{4.3.8}$$

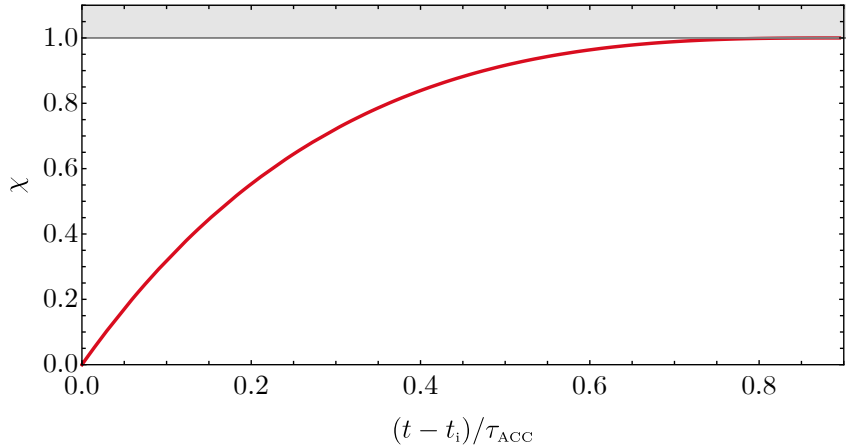

Figure 4.13: *Evolution of the dimensionless Kerr parameter $\chi$ as a function of time as predicted by the thin-disk accretion model. We rescale the time interval with respect to the accretion timescale $\tau_{ACC}$. The grey band shows the upper bound due to radiation effects at $\chi = 0.998$, see Ref. [435]. We recall that the effective timescale of the evolution is set by the accretion rate through the relation $\tau_{ACC} \equiv \tau_{Salp}/\dot{m}$.*

where we have defined the combination $\mathcal{F}(\chi) \equiv L(M, J)/ME(M, J)$, which can be shown to be only a function of the adimensional Kerr parameter $\chi$.

The equation dictating the spin evolution (4.3.8) predicts that the PBH spin grows in a typical accretion time scale $\tau_{ACC}$ until it reaches extremality, $\chi \simeq 1$, see for example Fig. 4.13. To be precise, theoretical arguments prevents the value of the spin to be accreted above the limit $\chi_{max} = 0.998$ [435] due to radiation effects. This limit was reduced in Ref. [422] where magnetohydrodynamic simulations of accretion disks around Kerr BHs were performed, even though it is not clear if such a result could be applied to thin accretion disks. More importantly, Ref. [422] showed that the characteristic spin evolution time scale does not change significantly in more realistic accretion models [422], such as the one we expect for larger values of the accretion rates accompanied by geometrically thicker accretion disks.

To summarize our findings, we stress that a key prediction of the PBH scenario is that the spin of sufficiently massive PBHs is large at small redshift. One expects, therefore, that above a given mass, for which the PBHs can experience a phase of super-Eddington accretion during the cosmic evolution, the PBH spin is large and close to extremality. We also stress that condition (4.3.2) can be satisfied more easily by PBHs in binaries as their accretion rates are enhanced by the binary system as a whole, see related discussion in the previous section.

### 4.3.2   Prediction for the spin of isolated PBHs

In the previous section we showed that, whenever an isolated PBH is accreting with a rate close to the Eddington one, we expect its spin to evolve and eventually reach extremality within a characteristic accretion timescale $\tau_{Salp}$. This gives rise to a rapid transition between the two regimes characterised by small and large PBH spin, depending on whether the conditions for thin-disk accretion are satisfied during the cosmological evolution of the isolated PBHs.

To present the prediction for where such a transition may happen in the plane $(M, z)$, we consider the two scenarios presented in Sec. 4.1.1. Namely, we consider Model I, where accretion takes place up to redshift $z_{cut-off} \sim 10$ and a more extreme scenario, called Model II, in which accretion is sustained up to much lower redshifts. The corresponding mass accretion rate of both models can be found in Fig. 4.4.

In Fig. 4.14 we show the evolution of the PBH spin in the $(M, z)$ plane for Model I and Model II, respectively. We start the evolution with initial conditions set at high redshifts. We conservatively choose $z \sim 100$ as one does not expect relevant changes before that epoch due to the relatively long accretion timescales compared to the age of the universe even for the largest accretion rate we encounter, see Fig. 4.2. As a reference, we superimpose the expected horizon redshift of current and future GW experiments like aLIGO and Einstein Telescope (ET) [275]. We note, however, that the prediction for the PBH spins in binaries formed in the early universe will be presented in the next section.

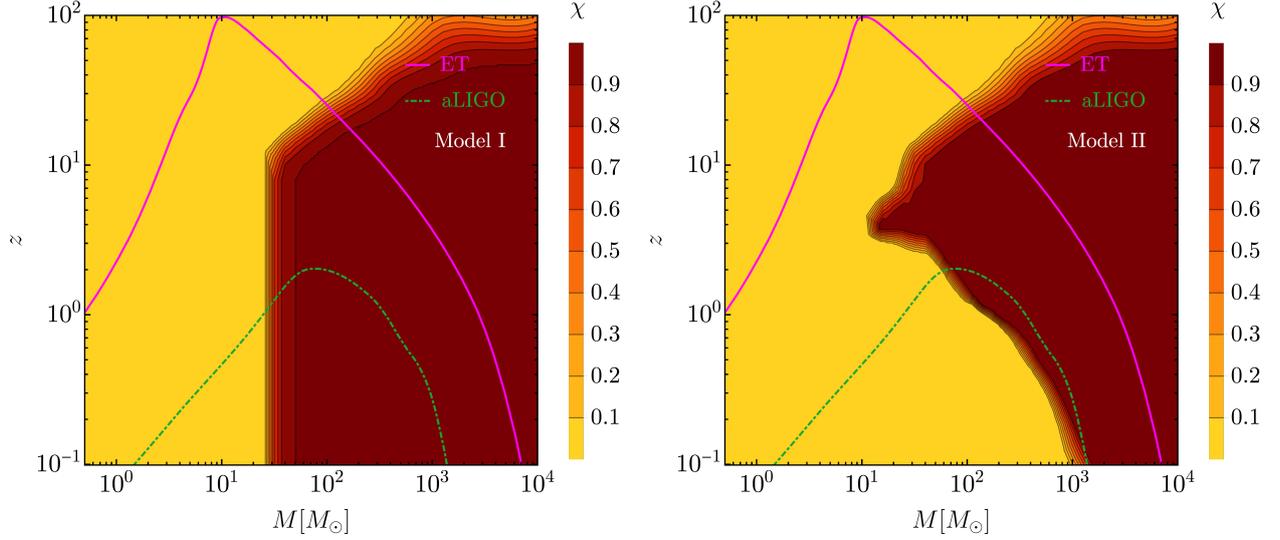

Figure 4.14: *Evolution of the spin χ as a function of the PBH mass M and the redshift z for isolated objects. As a reference, in both planes, we show the corresponding horizons redshift for current and future ground-based experiments like LIGO/Virgo and Einstein Telescope.* **Left:** *Model I.* **Right:** *Model II.*

We can list now the important features of the results shown in Fig. 4.14.

- In both models, we do not observe any significant spin evolution for PBH with masses below $\lesssim \mathcal{O}(10) M_\odot$. This is because, for lighter PBHs, the accretion rate is never strong enough to be accompanied by the formation of a geometrically thin accretion disk.

- Once the condition $\dot{m} \geq 1$ is met, the spin is rapidly brought extremal values, and one can see a sharp transition to $\chi \sim 1$ in few units of $z$.

- In Model I, where there exists an accretion cut-off redshift at around $z_{\text{cut-off}} \simeq 10$, we see that the PBH spin is maintained constant after $z_{\text{cut-off}}$ as no significant accretion is expected at lower redshift. Therefore, if we focus on the prediction of the spin for isolated PBHs at low redshift, we see the presence of a sharp transition between spinless PBHs and maximally spinning PBHs at around $M \sim \mathcal{O}(30) M_\odot$. Apart from its precise location, this transition is an inevitable prediction of the PBH model.

- The exact position of the transition between the two regimes at low redshift, as discussed in the preceding point, depends on the value of $z_{\text{cut-off}}$. By reducing (increasing) its value, one would observe the transition shifting at smaller (larger) values of the mass $M$.

- In Model II, where accretion is supposed to last until very small redshift, we observe a shift for such a transition at even higher masses. This is explained by the fact that, at low redshift, one finds significant accretion which is however taking place with rates below $\dot{m} \sim 1$. This brings PBHs to increasingly higher masses without producing thin accretion disks, see also the evolution of the PBH masses in Fig. 4.4.

- In general, for masses smaller than $\sim 10\, M_\odot$, one expects spherical accretion to take place. This setup, while leaving $|\vec{J}|$ unaffected, is increasing the PBH mass. Therefore, the parameter $\chi$ is decreased and any memory of the initial spin is erased.

We conclude this section by highlighting two important remarks. On the one hand, we have shown that accretion is, in fact, able to spin up PBHs which are massive enough. This goes against the common lore dictating the PBHs are generically expected to be almost spinless objects. On the other hand, as accretion is not efficient for light PBHs, one does not expect any spin to be accreted for PBHs lighter than few solar masses.

The predictions in this section may be relevant for binaries formed in the late-time universe as they are composed of PBHs which have evolved in isolation (i.e. not in a binary system) throughout the cosmic evolution.

### 4.3.3 Prediction for the spin of PBHs in binaries

In this section, we present the prediction for the spin of PBH in binaries formed in the early universe. In Fig. 4.15, we show the final spins of PBH binary components as a function of their final masses for three different choices of $z_{\text{cut-off}}$. While the exact position of the transition between low spin and high spin depends on the exact value of the cut-off chosen, the qualitative behaviour is consistent between the various scenarios plotted. First, one expects the transition to happen at smaller masses compared to what was found in the isolated evolution. This is because PBHs in binaries experience slightly larger accretion rates compared to isolated PBHs. Second, a sharp prediction of the PBH model is that the secondary object in the binaries is accreting more than the primary. Starting from a negligible spin at formation, one, therefore, expects the secondary PBH to spin faster than the primary. We stress here that the possibility of having mergers of highly-spinning PBHs components may increase the stochastic GW background signal resulting from unresolved sources, see Ref. [438] for details about the radiated energy from a merging event in terms of the BHs spin. This may potentially affect the bounds deduced from the non-observation of a SGWB at LIGO/Virgo [12].

We now describe what is the expectation for the PBH spin directions relative to the binary angular momentum. As, at formation, PBHs possess a negligible spin and can only spin up due to baryonic accretion from a thin disk, the expected PBH spin direction is related to the disk orientation. Crucially, the Bondi radii of the individual PBHs in the binary are comparable to or slightly smaller than the characteristic orbital distance [12]. Therefore, the case at hand is different from the common-envelope astrophysical scenario and the two disks do not influence each other, although the geometry of the accretion flow is complicated due to the binary motion. The orientation of the accretion disks are therefore independent and randomly distributed with respect to the orbital angular momentum. We expect, therefore, the spin directions to be uncorrelated and evenly distributed on the 2-sphere.

### 4.3.4 Implications for GW events

Ultimately, we are interested in providing the prediction for the key observables which can be measured in GW merger events with current (LIGO/Virgo) and future (e.g., ET) detectors.

The spin of BHs in the binary mostly affects the gravitational waveform at leading post-Newtonian order through the combination called effective spin parameter. This is defined as the mass weighted projection of the individual spins onto the orbital angular momentum as

$$\chi_{\text{eff}} = \frac{M_1 |\boldsymbol{\chi_1}| \cos\beta + M_2 |\boldsymbol{\chi_2}| \cos\gamma}{M_1 + M_2} = \frac{|\boldsymbol{\chi_1}| \cos\beta + q\, |\boldsymbol{\chi_2}| \cos\gamma}{1 + q}, \qquad (4.3.9)$$

where $M_1$ and $M_2$ are the individual BH masses, $\beta$ and $\gamma$ the angles between each spin and the binary angular momentum and $q = M_2/M_1$. As $|\boldsymbol{\chi_i}| < 1$, the possible range of values for the effective spin parameter is $|\chi_{\text{eff}}| < 1$. Following Ref. [14], we have averaged over the angles between the total angular momentum and the individual PBH spins and we have plotted, in Fig. 4.16, the expected distribution of $\chi_{\text{eff}}$ as a function of the final PBH mass $M_1$ for different fixed values of the final mass ratio parameter $q$. We stress that we expect a symmetric distribution around zero due to the averaging procedure over all possible spin orientations. On the other hand, the width of the distribution is dictated by the values of both spin magnitudes $\chi_1$ and $\chi_2$.

The characteristic feature of the PBH model is the presence of a transition between small and large values of the effective spin parameter, whose precise location depends on the value of $z_{\text{cut-off}}$. As a reference, we also plot the scenario expected by neglecting accretion in the PBH evolution. Also, as a reference, we show the dataset from the first two LIGO/Virgo observation runs. We will come back to the proper confrontation of the PBH model with the GW data in the following chapter. We note here that future upgraded detectors will be able to measure the individual spins with 30% accuracy [444], therefore alleviating the degeneracy between the individual spins and other binary parameters such as the mass ratio in the parameter estimation.

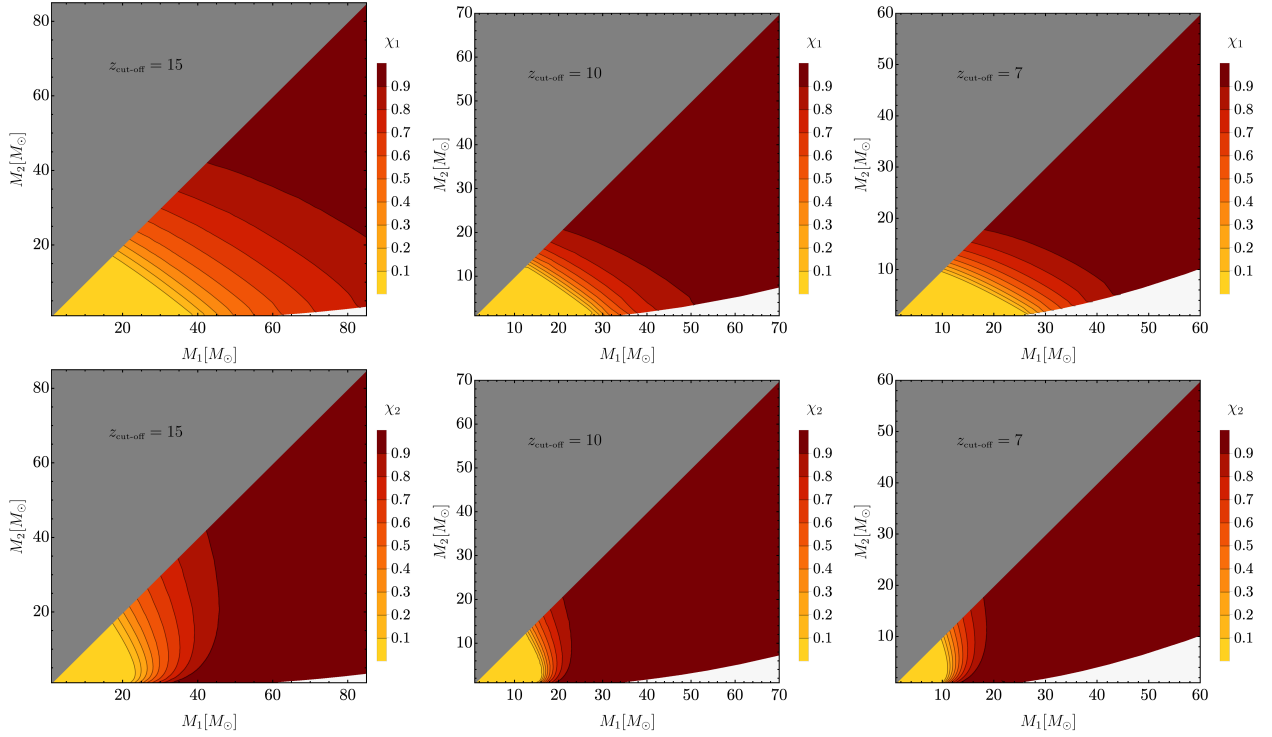

Figure 4.15: *Final spins as a function of the final masses* $(M_1, M_2)$ *for different values of the cut-off redshift, as in Fig. 4.7. We stress that the white region in the plot corresponds to the parameter space where you do not expect PBH binaries to be produced. As we are plotting with respect to the final masses and accretion naturally makes the mass ratio grow with time, there is a minimum mass ratio below which final binaries cannot fall in the PBH model. We also note that, as expected in the PBH model, the secondary spin is always larger than the primary, as a consequence of the larger relative accretion experience by the light PBH in the binary. **Top:** Final spin of the primary object* $\chi_1$. ***Bottom:** Final spin of the secondary object* $\chi_2$.

We can also study the expected final mass and spin of the remnant BH produced after the merger event. The final spin can be found following Ref. [407, 445–447] as

$$\chi_f = \frac{1}{(1+q)^2} \Big[ |\boldsymbol{\chi}_1|^2 + |\boldsymbol{\chi}_2|^2 q^4 + 2|\boldsymbol{\chi}_2||\boldsymbol{\chi}_1|q^2 \cos\alpha + |\boldsymbol{\ell}|^2 q^2$$
$$+ 2\left(|\boldsymbol{\chi}_1|\cos\beta + |\boldsymbol{\chi}_2|q^2\cos\gamma\right)|\boldsymbol{\ell}|q \Big]^{1/2}, \tag{4.3.10}$$

with

$$|\boldsymbol{\ell}| = 2\sqrt{3} + t_2\nu + t_3\nu^2 + \frac{s_4}{(1+q^2)^2}\left(|\boldsymbol{\chi}_1|^2 + |\boldsymbol{\chi}_2|^2 q^4 + 2|\boldsymbol{\chi}_1||\boldsymbol{\chi}_2|q^2\cos\alpha\right)$$
$$+ \left(\frac{s_5\nu + t_0 + 2}{1+q^2}\right)\left(|\boldsymbol{\chi}_1|\cos\beta + |\boldsymbol{\chi}_2|q^2\cos\gamma\right), \tag{4.3.11}$$

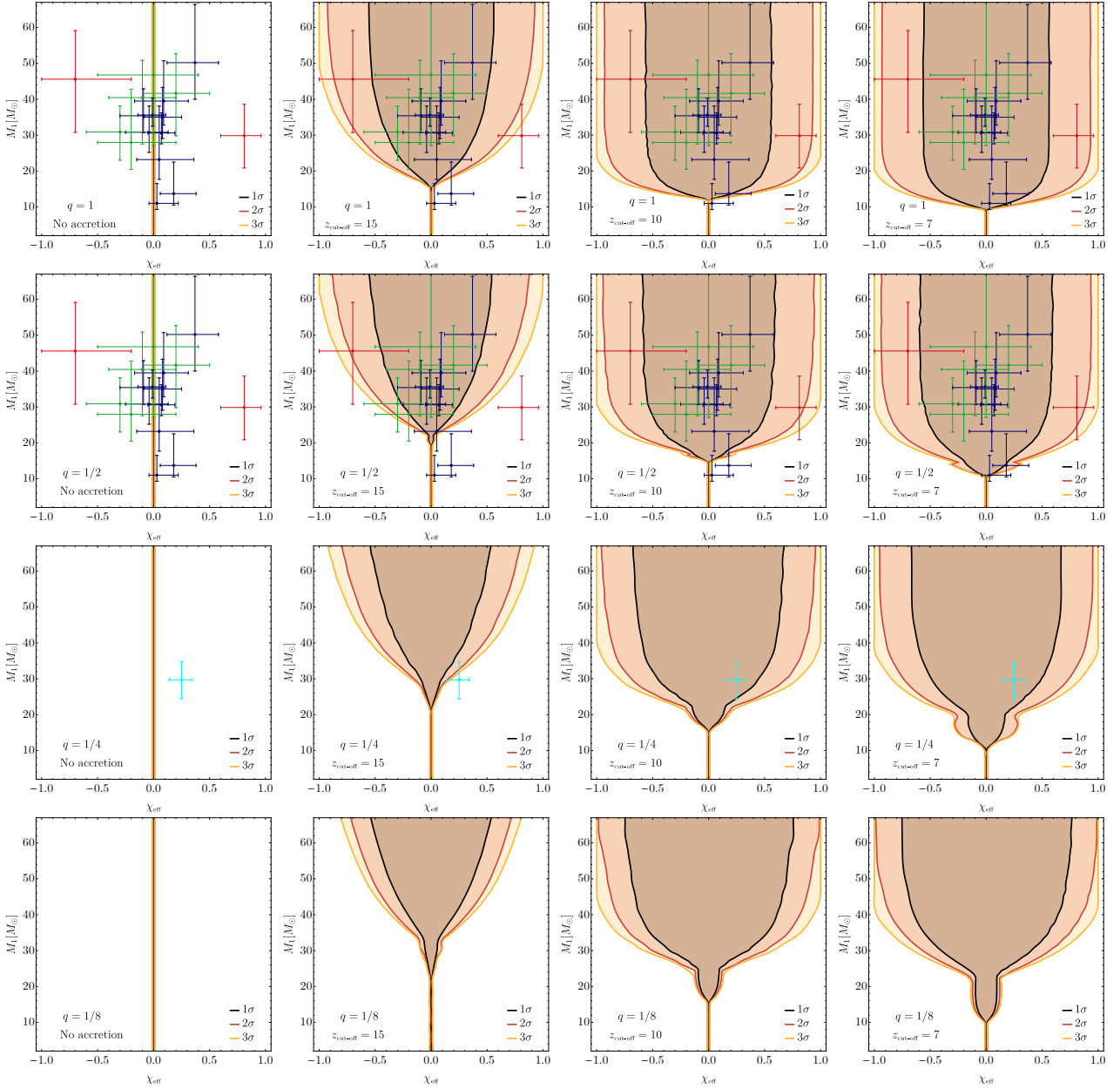

**Figure 4.16:** *The distribution of $\chi_{\text{eff}}$ as a function of the final primary mass $M_1$ for different values of the parameter $q$. As a reference, we show in blue the events listed in the GWTC-1 catalogue in Ref. [439], whereas green and red data points refer to the events discovered in Refs. [440, 441]. We note however that the red data points (referring to the events GW151216 and GW170403) may report measurements of the effective spin parameter which are significantly affected by the choice of prior on the spin angles [442]. We will address the sensitivity of spin measurements to the priors assumed in the next chapter. Finally, the cyan point corresponds to GW190412, see Ref. [443].*

in terms of the numerical parameters

$$s_4 = -0.1229 \pm 0.0075,$$
$$s_5 = 0.4537 \pm 0.1463,$$
$$t_0 = -2.8904 \pm 0.0359,$$
$$t_2 = -3.5171 \pm 0.1208,$$
$$t_3 = 2.5763 \pm 0.4833.$$

$$(4.3.12)$$

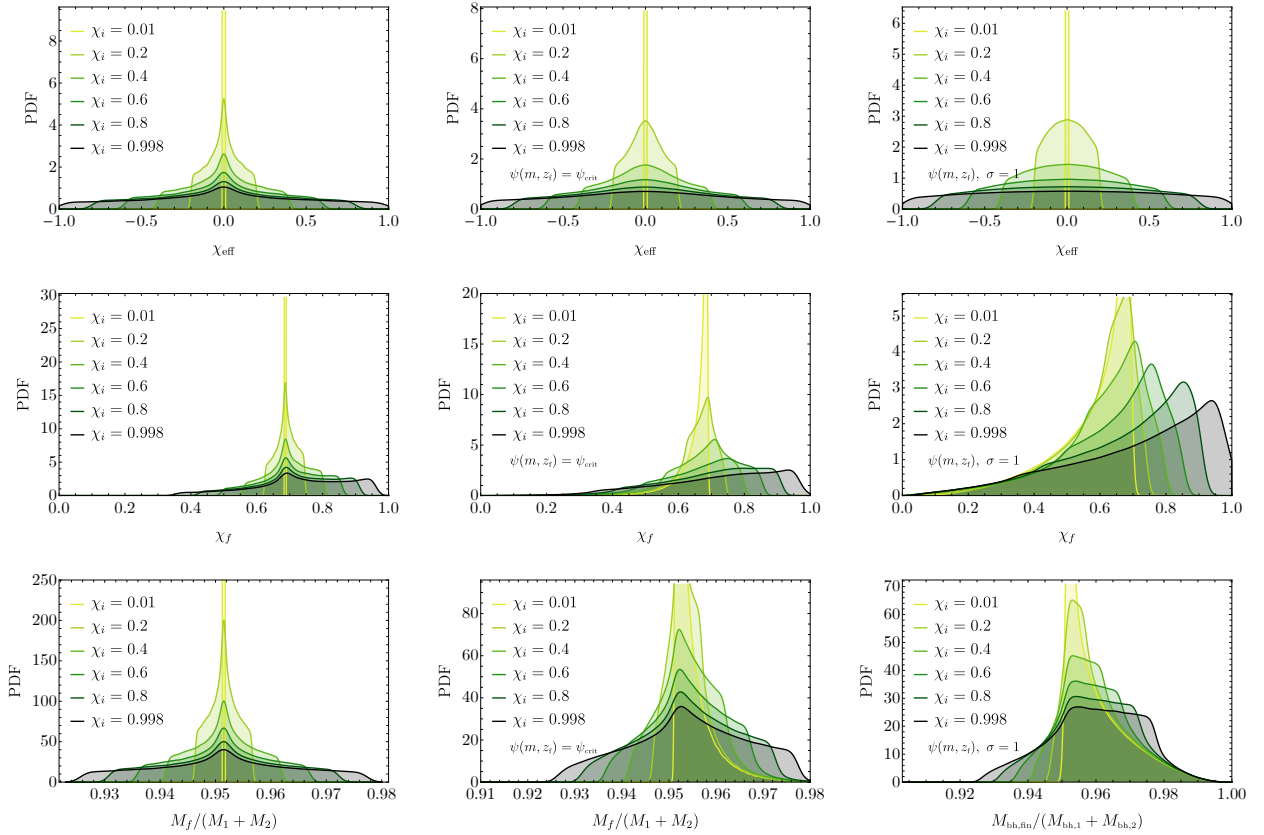

Figure 4.17: *Distribution of the parameters dependent on the individual spins of the binary components, assuming $\chi_1 = \chi_2 = \chi_i$ at the time of merger.* **Top:** *effective spin parameter distribution of the binary.* **Center:** *distribution of the spin of the final BH remnant.* **Bottom:** *distribution of the final remnant mass.* **From left to right:** *monochromatic mass function ($q = 1$), critical mass function $\psi_{\rm crit}$, and lognormal mass function with width $\sigma = 1$.*

Also, we defined $\nu = q/(1+q)^2$ as the symmetric mass ratio and

$$\cos\alpha = \hat{\boldsymbol{\chi}}_2 \cdot \hat{\boldsymbol{\chi}}_1, \qquad \cos\beta = \hat{\boldsymbol{\chi}}_1 \cdot \hat{\boldsymbol{L}}, \qquad \cos\gamma = \hat{\boldsymbol{\chi}}_2 \cdot \hat{\boldsymbol{L}} \qquad (4.3.13)$$

are the angles (at large separation) between each individual spin and the direction of the orbital angular momentum $\hat{\boldsymbol{L}}$, respectively.

The final mass of the remnant BH is [448]

$$M_f = (M_1 + M_2) \times [1 + 4\nu(m_0 - 1) + 16m_1\nu^2(|\boldsymbol{\chi}_1|\cos\beta + |\boldsymbol{\chi}_2|\cos\gamma)], \qquad (4.3.14)$$

as a function of the numerical coefficients

$$\begin{aligned} m_0 &= 0.9515 \pm 0.001, \\ m_1 &= -0.013 \pm 0.007. \end{aligned} \qquad (4.3.15)$$

To obtain the probability distribution functions (PDFs) of the final spin (4.3.10) and mass (4.3.14) of the merger remnant, along with the effective spin of the binary (4.3.9), one has to perform a statistical ensemble average over the masses of the binary components and the allowed angles of the spin vectors. As predicted by the PBH model, we used a uniform distribution for both spin vector orientations on the unit two-sphere. Also, for presentation purposes, we adopt various shapes of the mass functions $\psi(M)$, see Sec. 3.1. Results are shown in Fig. 4.17, where, for simplicity, we have assumed that both components of the binary have the same spin before merger, $\chi_1 = \chi_2 = \chi_i$. More details can be found in Ref. [14]. As the distributions shown in Fig. 4.17 are only a function of the individual BH masses and spins, they are derived with the same physical prescriptions adopted in other astrophysical scenarios [426, 449, 450]. However, the crucial difference to be found in the PBH scenario is the presence of the peculiar correlation between the PBH masses, spin and redshift, along with the particular mass functions considered, which are motivated by the PBH formation mechanism.

The first column of Fig. 4.17 shows the PDFs for a monochromatic PBH population, giving rise to symmetric binaries with $q = 1$, while the second and third columns show the case of broader shapes of the mass distribution, namely a critical and lognormal (with width $\sigma = 1$) mass functions, respectively. As the individual spins are isotropically oriented, the PDF of $\chi_{\rm eff}$ of the binary (first row) is symmetric around 0 and only sharply peaked at $\chi_{\rm eff} \simeq 0$ for small individual spins. For larger values of $\chi_i$, the distribution of $\chi_{\rm eff}$ gets much broader. The PDF of the final spin $\chi_f$ (second row) is peaked around 0.68, which is the value found for vanishing individual spins, and the distribution becomes broader for larger values of the PBH spins and broader mass functions. Finally, the probability distribution for the final mass peaks at the value $M_f \simeq 0.96(M_1 + M_2)$ for low initial spins, while becomes broader with larger support at higher values as spins are increased or the mass function gets broader.

## 4.4 Clustering evolution

In this section, we present a description of the evolution of clustering of PBHs. We will closely follow the results presented in Ref. [9]. Starting from the PBH clustering properties at formation provided in Sec. 3.3, we build an analytical picture heavily based on techniques developed in the context of large-scale structure formation theory. As we shall see, with such insights one can match the results of N-body simulations describing the PBH spatial clustering.

We warn the reader though that, as those are very expensive simulations to perform, in terms of computational power needed to follow the evolution of multiple scales involved in the problem, the results available in the literature are only limited to redshift higher than $z \simeq 100$. Therefore any result at lower redshift is based on an extrapolation of our analytical description, and therefore assumes that no new physical effects enter into the picture, potentially changing the results. As this is a quickly developing field, we envisage further improvements to be made in consolidating this analytical description as well as the advent of pioneering low redshift cosmological simulations including PBHs in the near future.

### Initial conditions

PBHs are born out of the collapse of rare overdensity peaks in the early universe. In the following, we are going to focus on how PBH are spatially distributed and what the evolution of their correlation function is. Following the notation introduced in Sec. 3.3 we define the overdensity of PBHs with respect to the total background dark matter energy density as

$$\frac{\delta\rho_{\rm PBH}(\vec{x}, z)}{f_{\rm PBH}\bar{\rho}_{\rm DM}} = \frac{1}{\bar{n}_{\rm PBH}} \sum_i \delta_D(\vec{x} - \vec{x}_i(z)) - 1, \qquad (4.4.1)$$

where $\delta_D(\vec{x})$ is the three-dimensional Dirac distribution which accounts for the discreteness of PBHs and the index $i$ indicates the various positions of each PBH. In the previous equation, we also introduced the average number density of PBH per comoving volume

$$\bar{n}_{\rm PBH} \simeq 3.2\, f_{\rm PBH} \left( \frac{20\, M_\odot/h}{M_{\rm PBH}} \right) (h/{\rm kpc})^3. \qquad (4.4.2)$$

Consequently, the two point function can be written in the general form

$$\left\langle \frac{\delta\rho_{\rm PBH}(\vec{x}, z)}{\bar{\rho}_{\rm DM}} \frac{\delta\rho_{\rm PBH}(0, z)}{\bar{\rho}_{\rm DM}} \right\rangle = \frac{f_{\rm PBH}^2}{\bar{n}_{\rm PBH}} \delta_{\rm D}(\vec{x}) + \xi(x, z), \qquad (4.4.3)$$

in terms of the the so-called reduced PBH correlation function $\xi(x, z)$.

We stress that the if PBHs were distributed à la Poisson, the correlation function would be negligible and the two point function in real space would only be dictated by the uncorrelated random positioning of PBHs in the universe. We stress that the Poisson term is always present, as it is inevitably related to the discreteness of PBHs. We will refer to PBHs as being clustered if they possess a non-vanishing reduced correlation function dominating over the Poisson term.

In this section, we will consider the scenario in which PBHs are not initially clustered at formation, as it is the case in the absence of non-gaussianities correlating short a large scales, see discussion in Sec. 3.3. Therefore, we can assume a Poisson distribution at the formation redshift $z_i$ and approximate the initial PBH power spectrum as done in Ref. [235]. To simplify the discussion, we will consider a PBH population dominated by a single PBH mass. This can be regarded as a good approximation not only if the power spectrum of the curvature perturbation is peaked around a single comoving momentum, but also when it is broad, see discussion in Sec. 2.2.4 as, in such a case, the mass function would be inevitably peaked at the smallest PBH which can be formed upon horizon re-entry [15]. We therefore consider the following initial power spectrum at the formation redshift $z_i$

$$\Delta_i^2(k) = \frac{k^3}{2\pi^2} \int d^3x \, e^{i\vec{k}\cdot\vec{x}} \left\langle \frac{\delta\rho_{\text{PBH}}(\vec{x}, z_i)}{\bar{\rho}_{\text{DM}}} \frac{\delta\rho_{\text{PBH}}(0, z_i)}{\bar{\rho}_{\text{DM}}} \right\rangle \approx f_{\text{PBH}}^2 \left(\frac{k}{k_*}\right)^3, \quad (4.4.4)$$

normalised in terms of the characteristic wavenumber

$$k_* = (2\pi^2 \bar{n}_{\text{PBH}})^{1/3} \simeq 4 \, f_{\text{PBH}}^{1/3} \left(\frac{20 \, M_\odot/h}{M_{\text{PBH}}}\right)^{1/3} h/\text{kpc}. \quad (4.4.5)$$

Practically, the characteristic wavenumber is inversely proportional to the mean separation between PBHs. We stress that, being interested in the clustering behaviour at small scales where the PBH Poisson noise dominates the (sub-)structure formation, we are going to neglect the contribution of adiabatic perturbations, which are instead the dominant player when the formation of the large scale structure of the universe is considered.

We will describe the evolution of PBH clustering by dividing the discussion into three parts, categorised depending on the characteristic size of the overdensity $\Delta$: the linear, quasi-linear and non-linear regimes.

### 4.4.1  The linear regime

In the linear regime, the PBH density contrast only starts evolving after matter-radiation equality $z_{\text{eq}}$ according to [235]

$$\Delta_L^2(k, z) \simeq \left(1 + \frac{3}{2} f_{\text{PBH}} \frac{1 + z_{\text{eq}}}{1 + z}\right)^2 \Delta_i^2(k). \quad (4.4.6)$$

In the previous relation, we imposed the matter-dominated epoch behaviour of density perturbations scaling like $\approx (1 + z)^{-1}$.

One only expects a deviation from this behaviour when the PBH number density enters the quasi-linear (QL) regime. This happens at the scale at which the density contrast is of order unity, i.e.

$$\Delta_L^2(k = k_{\text{L-QL}}(z), z) \simeq 1, \quad (4.4.7)$$

which implies

$$k_{\text{L-QL}}(z) \simeq \frac{4}{f_{\text{PBH}}^{1/3}} \left(\frac{20 \, M_\odot/h}{M_{\text{PBH}}}\right)^{1/3} \left[1 + 26 f_{\text{PBH}} \left(\frac{100}{1 + z}\right)\right]^{-2/3} h/\text{kpc}. \quad (4.4.8)$$

Therefore, depending on the epoch of interest, the transition between the linear and quasi-linear regime varies with redshift, as in general perturbation grows in time in the matter dominated phase.

For comparison, in Fig. 4.18, we show the PBH power spectra extracted from the N-body simulations of Ref. [235][2] at $z = 99$, for various values of the PBH abundance $f_{\text{PBH}}$. The stars are located at the values of $k_{\text{L-QL}}$, which is found to fit sufficiently well the numerical results. As one can appreciate, the smaller $f_{\text{PBH}}$, the smaller the PBH power spectrum for a given scale $k$. This happens since, when $f_{\text{PBH}}$ is reduced, the number density of PBH decreases (i.e. PBHs are rarer and their average separation becomes larger) and therefore the Poisson noise is only relevant at smaller scales.

---

[2]We thank D. Inman and Y. Ali-Haïmoud for sharing their results with us.

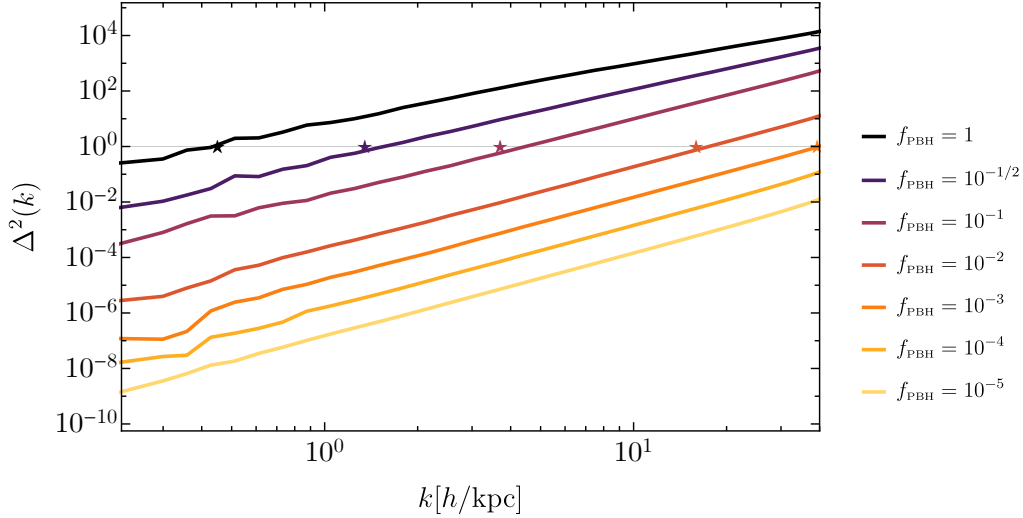

Figure 4.18: *The PBH power spectra at $z = 99$ for different values of $f_{PBH}$ and $M_{PBH} = 20 M_\odot/h$. The lines represent the data taken from Ref. [235]. The stars indicate the scale where the transition to the quasi-linear regime is predicted to happen following Eq. (4.4.8). In the linear regime ($\Delta^2 \lesssim 1$), the power spectrum is found to be scaling like $\propto k^3$ and follows the characteristic Poisson behaviour.*

### 4.4.2 The quasi-linear regime

When PBH overdensities enter the quasi-linear regime, they start forming dense structures that can decouple from the Hubble flow, start collapsing and virialize to form halos. The characteristic "virial" density is predicted to be roughly 200 times the background density at the time of virialization. We will assume that the formation of PBH halos is hierarchical like in a standard cold dark matter cosmology: the smaller PBH halos form first and act as progenitors of more massive halos virializing at a later epoch.

We can regard the PBH perturbations as quasi-linear until the power spectrum $\Delta^2$ at a given wavenumber $k$ reaches $\sim 200$. This corresponds to a linearly extrapolated two-point correlator reaching the value of $\approx 5.85$ [451]. The redshift of collapse can be found by requiring that the local overdensity averaged over a comoving radius $R$ reaches the critical value $\delta_c \simeq 1.68$. It is useful to define the volume averaged correlation function

$$\bar{\xi}(R, z) = \frac{3}{4\pi R^3} \int_0^R \mathrm{d}s \, 4\pi s^2 \xi(s, z), \qquad (4.4.9)$$

where

$$\xi(x, z) \simeq \int \frac{\mathrm{d}k}{k} \, e^{i\vec{k}\cdot\vec{x}} \Delta^2(k, z). \qquad (4.4.10)$$

This quantity can be regarded as an indicator of the squared overdensity within the radius $R$. As we shall see, these quantities will be related to the number of PBHs in a given volume with respect to the average. In Eq. (4.4.10) we neglected the contribution from the Poisson piece in the correlation function, which is subdominant in the quasi-linear regime. It can be shown that the volume averaged correlation function is related to the power at some effective wavenumber as [451]

$$\Delta^2(k, z) \simeq \bar{\xi}(1/k, z). \qquad (4.4.11)$$

We want to find the evolution of the correlation function in this intermediate regime. To do so, one can use the conservation of particle pairs (in this context we refer to PBHs as being the individual components of the halo) to write down an equation satisfied by $\xi(x, z)$. We will momentarily neglect the effect of two-body relaxation and the consequent evaporation of PBHs from the clusters, which will be addressed later. We define the mean number of neighbours in a region of size $x$ as [452]

$$N(x, z) = \bar{n}_{\text{PBH}} \int_0^x \mathrm{d}s \, 4\pi s^2 \left[1 + \xi(s, z)\right] \tag{4.4.12}$$

and the conservation of neighbours implies [452]

$$\frac{\partial \xi}{\partial t} + \frac{1}{ax^2} \frac{\partial}{\partial x} \left[x^2 (1 + \xi) v\right] = 0, \tag{4.4.13}$$

where $a$ is the scale factor and $v(x, t)$ denotes the mean relative velocity of pairs at separation $x$ and time $t$. Therefore, the pair conservation equation yields the (mass conservation) relation

$$x^3 (1 + \bar{\xi}) = R^3, \tag{4.4.14}$$

in terms of the initial shell radius $R$.

Eq. (4.4.14) is the fundamental relation dictating the behaviour of the correlation function in the quasi-linear regime. As long as the evolution is linear and $\bar{\xi} \ll 1$, one finds that $R \sim x$. On the other hand, as clustering grows, $\bar{\xi}$ increases and the scale $x$ becomes smaller than $R$. As a consequence, the correlation function in the quasi-linear regime $\bar{\xi}_{\text{QL}}(x)$ can be expressed in terms of the linearly extrapolated expression given by $\bar{\xi}_{\text{L}}(R)$, as we shall see.

Let us consider a region surrounding a (linear) density peak which is expected to grow and to seed the formation of a cluster. The density profile around a peak is expected to be proportional to the underlying correlation function [297] (neglecting the contribution from gradients). Therefore one can write

$$\rho(x) \simeq \rho_{\text{bkg}} \left(1 + \xi(x)\right) \tag{4.4.15}$$

The mean density contrast is found to scale with the initial shell radius $R$ as $\bar{\xi}_{\text{L}}(R)$, so long as the linear regime is maintained. Following the standard spherical collapse model, the perturbation expands up to a maximum radius $x_{\text{max}}$ proportional to $x_{\text{max}} \approx R/\bar{\xi}_{\text{L}}(R)$ [452]. Taking the effective radius proportional to $x_{\text{max}}$ and considering a halo of mass $M$, we have

$$\bar{\xi}_{\text{QL}}(x) \sim \rho(x) \sim \frac{M}{x^3} \sim \frac{R^3}{(R/\bar{\xi}_{\text{L}}(R))^3} \sim \bar{\xi}_{\text{L}}^3(R), \tag{4.4.16}$$

where the correlation function must be evaluated at the position

$$R^3 \sim x^3 \bar{\xi}_{\text{L}}^3 \sim x^3 \bar{\xi}_{\text{QL}}. \tag{4.4.17}$$

As the linear behaviour is dictated by the Poisson scaling $\bar{\xi}_{\text{L}}(x) \sim x^{-3}$, one finds

$$\bar{\xi}_{\text{QL}}(x) \sim x^{-9/4}. \tag{4.4.18}$$

We show this predicted scaling in Fig. 4.19, where one can observe an excellent agreement with the results of the N-body simulation in Ref. [235].

The evolution in time of the correlation function can found by imposing the total dark matter energy density dominates the energy budget in the universe, which is an adequate approximation after $z_{\text{eq}}$ and until the late time universe when the dark energy takes over. In this setup, the evolution has to be self-similar if the initial power spectrum is a power-law [452] and the Boltzmann equation for the self-gravitating PBHs admits a self-similar solution of the form $\xi(x, t) = f(x/t^\alpha)$ (for more details, see Ref. [453]). This behaviour is consistent with the linear scaling of the correlation function only if $\alpha = 4/9$ is considered. As the quasi-linear correlation function $\bar{\xi}_{\text{QL}}(x, z)$ can only depend upon the combination $x(1+z)^{2/3}$, one finds $\Delta_{\text{QL}}^2(k)$ to be scaling like $(1+z)^{-3/2}$. Notice that the dependence on time is weaker than in the linear regime. This scaling is also apparent in Eq. (4.4.19) if one considers the regime $26 f_{\text{PBH}}(10^2/(1+z)) \gtrsim 1$. Finally, we can write

$$\Delta_{\text{QL}}^2(k) \simeq \left(\frac{k}{k_{\text{L-QL}}(z)}\right)^{9/4}$$

$$\simeq 0.04 \, f_{\text{PBH}}^{3/4} \left(\frac{20 \, M_\odot/h}{M_{\text{PBH}}}\right)^{-3/4} \left[1 + 26 f_{\text{PBH}} \left(\frac{100}{1+z}\right)\right]^{3/2} \left(\frac{k}{h/\text{kpc}}\right)^{9/4}. \tag{4.4.19}$$

### 4.4.3 The non-linear regime

The truly non-linear regime needs careful separate treatment. Indeed, as the power spectrum reaches values above $\sim 200$, the description of the non-linear physics changes. To track the evolution of PBH perturbations in the non-linear regime, we consider the stable clustering hypothesis. This means we are assuming that, even though the separation between clusters may be altered by the expansion of the Universe, their internal structure is completely decoupled from the surrounding environment and remains constant with time. In other words, to maintain a stable structure, the relative velocity $v(x, z)$ between two objects separated by a distance $x$ within the cluster should balance the Hubble flow ($Hr = \dot{a}x$). Therefore, one should impose $v(x, z) = -\dot{a}x$.

Adopting the the stable clustering hypothesis, the pair conservation equation (4.4.13) in the non-linear regime $\xi \gg 1$ can be recast into [452]

$$\frac{\partial}{\partial t}(1 + \xi) = \frac{H}{x^2}\frac{\partial}{\partial t}\left[a^3(1 + \xi)\right].$$ (4.4.20)

The previous equation admits a power-law solution of the form

$$\xi_{\mathrm{NL}}(x, z) \sim \frac{x^{-m}}{(1 + z)^{3-m}}.$$ (4.4.21)

As done in the previous section, the index $m$ can be determined from self-similarity considerations. As the correlation function must be of the form $\xi(x, t) = f(x/t^\alpha)$ with $\alpha = 4/9$ to consistently reproduce the linear behaviour or, equivalently, $\xi(x, z) \sim f(x(1 + z)^{2/3})$, Eq. (4.4.21) shows that one must obtain

$$\xi_{\mathrm{NL}}(x, z) \sim \left[x(1 + z)^{2/3}\right]^{-m}.$$ (4.4.22)

Therefore, one finds $m = 9/5$. Again, in the previous considerations, we have assumed that the universe is dominated by the dark matter energy density.

The transition between the quasi-linear and non-linear regime can be found upon requiring $\bar{\Delta}^2 \sim \xi \sim 200$, or $(k_{\mathrm{L\text{-}QL}}/k_{\mathrm{QL\text{-}NL}})^{-9/4} \sim 200$. This gives the non-linear scale

$$k_{\mathrm{QL\text{-}NL}}(z) \simeq 42 f_{\mathrm{PBH}}^{-1/3}\left(\frac{M_{\mathrm{PBH}}}{20 M_\odot/h}\right)^{-1/3}\left[1 + 26 f_{\mathrm{PBH}}\left(\frac{100}{1 + z}\right)\right]^{-2/3} h/\mathrm{kpc}.$$ (4.4.23)

Finally, the corresponding power spectrum in the non-linear regime reads

$$\Delta_{\mathrm{NL}}^2(k) \simeq 200\left(\frac{k}{k_{\mathrm{QL\text{-}NL}}(z)}\right)^{9/5}$$

$$\simeq 0.2 f_{\mathrm{PBH}}^{3/5}\left(\frac{M_{\mathrm{PBH}}}{20 M_\odot/h}\right)^{3/5}\left[1 + 26 f_{\mathrm{PBH}}\left(\frac{100}{1 + z}\right)\right]^{6/5}\left(\frac{k}{h/\mathrm{kpc}}\right)^{9/5}.$$ (4.4.24)

We therefore highlight that the non-linear PBH power spectrum scales with redshift as $(1 + z)^{-6/5}$ for $26 f_{\mathrm{PBH}}(10^2/(1 + z)) \gtrsim 1$.

We show a comparison between this prediction and the results of the N-body simulation performed in Ref. [235] in Fig. 4.19. We indicate with dots the location of $k_{\mathrm{QL\text{-}NL}}$, while the straight dashed lines represent the various power-laws predicted in the quasi-linear and non-linear regime. Our findings are in reasonable agreement with the numerical N-body data. Notice that the mismatch grows when increasingly smaller $f_{\mathrm{PBH}}$ are considered, as we do not account for the back-reaction of the dark matter fluid not in the form of PBHs. This feedback is absent when $f_{\mathrm{PBH}} = 1$ where we see that the agreement between our prediction and the numerical data is very good.

### 4.4.4 The PBH halo profile at small scales

Another prediction of the previous description of PBH clustering in the non-linear regime is related to the density profile of clusters at small scales. The density profile is defined as the (conditional) correlation function subject to the constraint that one member of each particle pair certainly is at the center of a halo, while the (unconditional) correlation function is obtained when the location of each pair member is unconstrained.

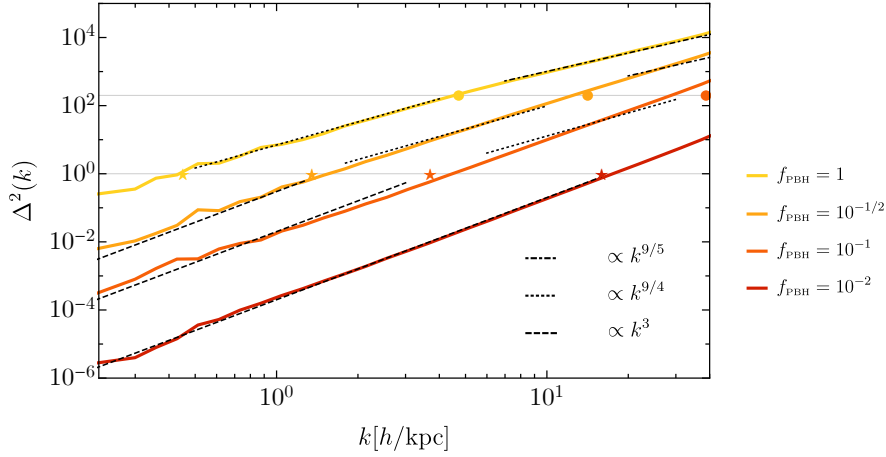

Figure 4.19: *Same as Fig. 4.18 but focusing on the lines reaching the quasi-linear regime. The dots indicate the predictions of the expression (4.4.23) and the predicted power laws are indicated by straight lines.*

The prediction obtained with the stable clustering hypothesis is the following. We restrict our focus to the small scales which are in the non-linear regime. In such a case, if we consider a PBH pair, both objects are most likely coming from the same PBH halo, or cluster. In this limit, if the PBH density profile is characterised by a power-law behaviour $\rho_{\mathrm{PBH}}(x) \sim x^{-\epsilon_{\mathrm{PBH}}}$, then it follows that the two-point correlation function must scale like $\xi_{\mathrm{NL}} \sim x^{-2\epsilon_{\mathrm{PBH}}+3}$ [454, 455]. This is because one can write the correlation function as a convolution of two PBH overdensities as

$$\xi_{\mathrm{NL}}(x) \approx \int \mathrm{d}^3 s\, \rho(\vec{s})\rho(|\vec{s} - \vec{x}|).  \qquad (4.4.25)$$

This corresponds to the "1-halo" term in the large scale structure context. Therefore, by imposing $(-2\epsilon_{\mathrm{PBH}} + 3) = -9/5$, we infer that the PBH density profile should satisfy

$$\rho_{\mathrm{PBH}}(x) \sim x^{-12/5}.  \qquad (4.4.26)$$

In Fig. 4.20, we plot the PBH cluster density profile as found in the N-body simulation performed in Ref. [170], along with our result, showing the adequacy of our analytical description.[3]

We can also work out the characteristic halo size contributing to the PBH correlation function. To do so, we apply Press-Schechter theory [456] to the initial Poisson power spectrum. The resulting number density of PBH halos with mass between $M$ and $(M + \mathrm{d}M)$ reads

$$\frac{\mathrm{d}n(M, z)}{dM} = \frac{\overline{\rho}_{\mathrm{PBH}}}{\sqrt{\pi}} \left(\frac{M}{M_*(z)}\right)^{1/2} \frac{e^{-M/M_*(z)}}{M^2},  \qquad (4.4.27)$$

where $\overline{\rho}_{\mathrm{PBH}}$ is the average PBH energy density and the characteristic halo mass at redshift $z$ is defined as (see also Ref. [236])

$$M_*(z) = N_*(z) \cdot M_{\mathrm{PBH}} \simeq f_{\mathrm{PBH}}^2 \left(\frac{2600}{1+z}\right)^2 M_{\mathrm{PBH}}.  \qquad (4.4.28)$$

In the halo model framework [457], the correlation function in the non-linear regime may also be written as [455, 458]

$$\xi(x, z) = \frac{1}{\overline{\rho}_{\mathrm{DM}}^2} \int \mathrm{d}M \frac{\mathrm{d}n(M, z)}{dM} M^2 \lambda_M(x, z),  \qquad (4.4.29)$$

---

[3]We stress that Fig. 7 of Ref. [170] shows the properties of the PBHs surrounding a *central* binary at $z \simeq 1100$. This plot is therefore representing the PBH density profile rather than a correlation function as stated in Ref. [170]. We thank M. Raidal and H. Veermäe for clarifying this point.

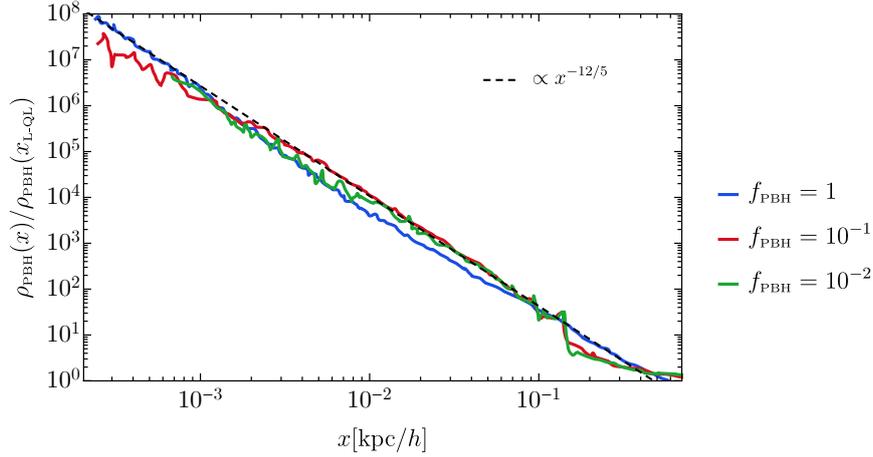

Figure 4.20: *The PBH density profile at redshift $z \simeq 1100$ for different values of $f_{\mathrm{PBH}}$ as obtained in Ref. [170] (see their Fig. 7). The prediction given by our analytical description in Eq. (4.4.26) is shown as a black dashed line.*

where

$$\lambda_M(x, z) = \int \mathrm{d}^3 s \, \rho_{\mathrm{PBH}}(s, M, z)\rho_{\mathrm{PBH}}(|\vec{s} + \vec{x}|, M, z) \simeq \frac{1.22}{4\pi R_{\mathrm{vir}}^3} \left( \frac{x}{R_{\mathrm{vir}}} \right)^{-9/5}. \tag{4.4.30}$$

The previous equation made use of the average density profile of a halo of mass $M$ [458]

$$\rho_{\mathrm{PBH}}(x, M, z) = \left( \frac{3M}{5 \cdot 4\pi R_{\mathrm{vir}}^3} \right) \left( \frac{x}{R_{\mathrm{vir}}} \right)^{-12/5}, \tag{4.4.31}$$

and the virial radius $R_{\mathrm{vir}}$ defined as

$$R_{\mathrm{vir}}^3 = \left( \frac{3M}{4\pi \cdot 200 \, \bar{\rho}_{\mathrm{PBH}}} \right), \tag{4.4.32}$$

assuming an average overdensity $\sim 200$ within each virialized halo. The largest contribution to Eq. (4.4.29) is given by halos with masses around $(11/10)M_*(z)$.

Ref. [459] found a steeper PBH profile, going like $\rho_{\mathrm{PBH}}(r) \sim r^{-2.8}$ between $(10^{-3} \div 10)\,\mathrm{pc}$. We stress however that their simulations only involve the evolution of a single cluster (as also done in Ref. [460]) and assume a cluster of about $10^3$ PBHs with $f_{\mathrm{PBH}} = 1$ already at redshift $z \sim 10^3$ and with a kpc characteristic comoving size. This is in contrast with the expectation given by the Press-Schechter theory for initially Poisson distributed objects, confirmed also with the N-body simulation in Ref. [235], which predicts that the typical halo has only a few PBHs at that redshift.

We conclude this section by highlighting the limitations of our results:

- We have employed the stable clustering hypothesis which is however only valid before the cosmological constant- or dark energy-dominated period, as it assumes the dark matter to be the dominant component in the energy budget of the universe.

- Binaries tend to sink towards the center of halos since they are heavier than single PBHs. This segregation process can catalyse binary-PBHs interactions and heat up the cluster core, possibly modifying its shape.

- Similarly to the previous point, other phenomena can occur at very small scales, including encounters with massive PBHs in the core which cause the lighter PBHs to be ejected and form their own, albeit shallower, profile [459].

- We only consider a monochromatic population of PBHs. With an extended mass function, one can expect the largest PBHs in the spectrum to act as additional seeds of substructures on top of Poisson fluctuations.

- We did not include the backreaction of a secondary dark matter component when $f_{\mathrm{PBH}} \neq 1$.

#### 4.4.5   Halo evaporation and characteristic survival time

In our description of the PBH correlation function, we did not include the possible evaporation of PBHs from the clusters. In this section, we argue this effect is likely not relevant due to the competing effect of the inclusion of clusters in larger halos. We show this by assuming the PBH population makes the entirety of the dark matter, that is $f_{\rm PBH} = 1$.

The formation redshift of a cluster of $N = M/M_{\rm PBH}$ PBHs can be estimated from Eq. (4.4.28), which gives

$$1 + z_{\rm form} = \frac{2600}{\sqrt{N}}. \tag{4.4.33}$$

Interactions between PBHs that randomly encounter each other can give a PBH enough energy to escape from the cluster. The characteristic evaporation time of a system of $N = M/M_{\rm PBH}$ PBHs, clustered in a region of size $R$ and subject to the gravitational force, is given by [461]

$$t_{\rm ev} \simeq 14 \frac{N}{\log N} \frac{R}{v}, \tag{4.4.34}$$

where

$$v \simeq \sqrt{\frac{GNM_{\rm PBH}}{R}}. \tag{4.4.35}$$

Therefore, for the typical cluster virialization radius $R_{\rm vir}$, one finds

$$t_{\rm ev} \simeq 2.5 \cdot 10^4 \, {\rm Gyr} \, (\log N)^{-1} \left(\frac{N}{100}\right)^{1/2} \left(\frac{M_{\rm PBH}}{20 M_\odot/h}\right)^{-1/2} \left(\frac{R_{\rm vir}}{{\rm kpc}/h}\right)^{3/2}. \tag{4.4.36}$$

There exists, however, a competing effect, which may modify the evolution of clusters. As clustering proceeds hierarchically, the larger clusters may end up including smaller ones. The survival time of a given halo of mass $M$ can be computed by resorting again to the Press-Schechter formalism. Following Ref. [330], we define a time dependent threshold for collapse of an overdensity as $\omega(z) \equiv \delta_c/a = \delta_c(1 + z)$ (we assume to be in the matter dominated phase of the universe when the density contrast evolves linearly with the scale factor), where $\delta_c \simeq 1.68$. [4] The time independent variance of the PBH perturbations becomes

$$S(R) = (1 + z)^2 \int \frac{{\rm d}k}{k} \Delta_{\rm L}^2(k, z) W^2(kR), \tag{4.4.37}$$

where $W(kR)$ is the Fourier transform of a top-hat window function. Using the linear power spectrum defined in Eq. (4.4.4), one finds

$$\frac{\omega}{\sqrt{S}} = \left(\frac{M}{M_*}\right)^{1/2} = 2 \cdot 10^{-4} \left(\frac{M}{M_{\rm PBH}}\right)^{1/2} \delta_c \, (1 + z), \tag{4.4.38}$$

from which we derive

$$S(M) = 2.5 \cdot 10^7 \left(\frac{M}{M_{\rm PBH}}\right)^{-1} = 2.5 \cdot 10^7/N. \tag{4.4.39}$$

Let us consider two halos with masses $M_1$ and $M_2$ (while their corresponding thresholds and variances are $S_i$ and $\omega_i$). The probability that the former (smaller) one, formed at redshift $z_{\rm form}(M_1)$ is incorporated into the latter is given by [330]

$$g(S_2, \omega_2 | S_1, \omega_1){\rm d}\omega_2 = \sqrt{\frac{2}{\pi}} \frac{1}{\omega_1} \sqrt{\frac{S_1}{S_2(S_1 - S_2)}} \exp\left[\frac{2\omega_2(\omega_1 - \omega_2)}{S_1}\right] \left\{\frac{-S_2(\omega_1 - 2\omega_2) - S_1(\omega_1 - \omega_2)}{S_1 e^{X^2}}\right.$$
$$\left. + \sqrt{\frac{\pi}{2}} \sqrt{\frac{S_2(S_1 - S_2)}{S_1}} \left[1 - \frac{(\omega_1 - 2\omega_2)^2}{S_1}\right] [1 - {\rm erf}(-X)]\right\} {\rm d}\omega_2, \tag{4.4.40}$$

---

[4]We stress for clarity that, in this section, we are considering the formation of clusters (or halos) of PBHs. We use the term *collapse* to denote the decoupling of a system of $N$ PBHs from the Hubble flow and its consequent virialisation. This should not be confused with the collapse of overdensities leading to the generation of PBHs studied in Chapter 3.

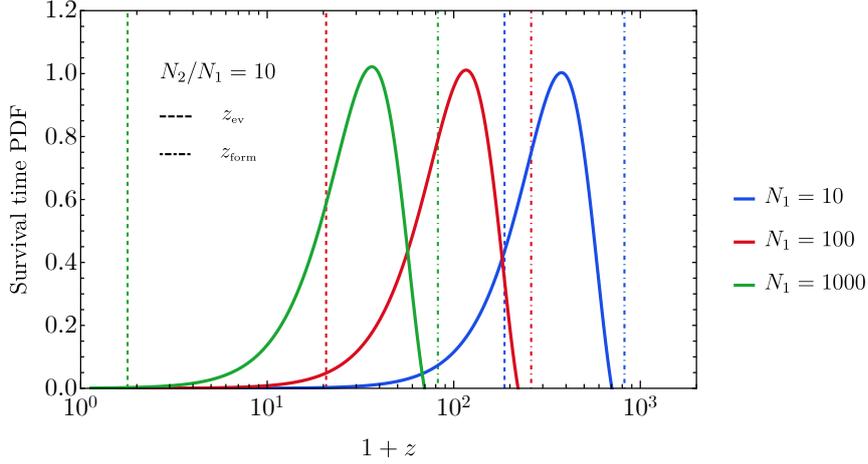

Figure 4.21: *Probability distribution for the survival time of a halo with $N_1$ PBHs at $z = z_{\mathrm{form}}$ before it gets included into a halo with $N_2 = 10N_1$ PBHs. Dashed (dot-dashed) vertical lines indicate the characteristic evaporation (formation) redshift of the cluster of $N_1$ objects.*

where $\omega_2 < \omega_1$, $S_2 < S_1$ and

$$X = \frac{S_2(\omega_2 - 2\omega_1) + S_1\omega_2}{\sqrt{2S_1S_2(S_1 - S_2)}}. \tag{4.4.41}$$

From this relation, one can show that the evaporation time of PBH halos is typically larger than their survival time, which implies that PBH halos are, in practice, stable against evaporation [9]. We support this finding by showing, in Fig. 4.21, the probability that a halo containing $N_1$ PBHs at redshift $z_{\mathrm{form}}$ is incorporated into a bigger halo containing $N_2 = 10N_1$ PBHs at redshift $z < z_{\mathrm{form}}$. In all cases, the peak of the distribution occurs before the evaporation redshift $z_{\mathrm{ev}}$ shown as the vertical dashed line.

We conclude that the correlation function is not altered by evaporation at least in the case we considered with $f_{\mathrm{PBH}} = 1$. For a smaller PBH abundance, the dynamics may be more complicated due to the presence of an additional dark matter fluid, while, however, PBH clustering is increasingly less relevant [235].

## 4.5 Binary formation and evolution

In this section, we focus on the description of the mechanism of PBH binary formation. We will address both the formation in the early universe, typically happening before matter radiation equality, and in the late-time universe taking place in present-day dark matter halos, see [196] for a review. For simplicity, we provide the estimates for equal mass binaries, while the complete description accounting for extended mass functions will be adopted in the next section dedicated to the computation of the merger rate. In the second part, we will study the evolution of binaries given by the loss of energy radiated through GWs while also commenting on the effect of accretion in hardening the binaries. We will follow the description presented in Ref. [12].

### 4.5.1 Binary formation in the early universe

A pair of PBHs of mass $M$ can decouple from the Hubble expansion if their mutual gravitational attraction starts dominating over the Hubble flow. If they are separated by a physical distance $R$, the condition for decoupling becomes

$$MR^{-3}(z) > \rho(z) \tag{4.5.1}$$

where $\rho$ represents the background cosmic energy density. We conveniently normalise each quantity to their value at the epoch of matter-radiation equality $z_{\mathrm{eq}}$. Then, a PBH binary system decouples

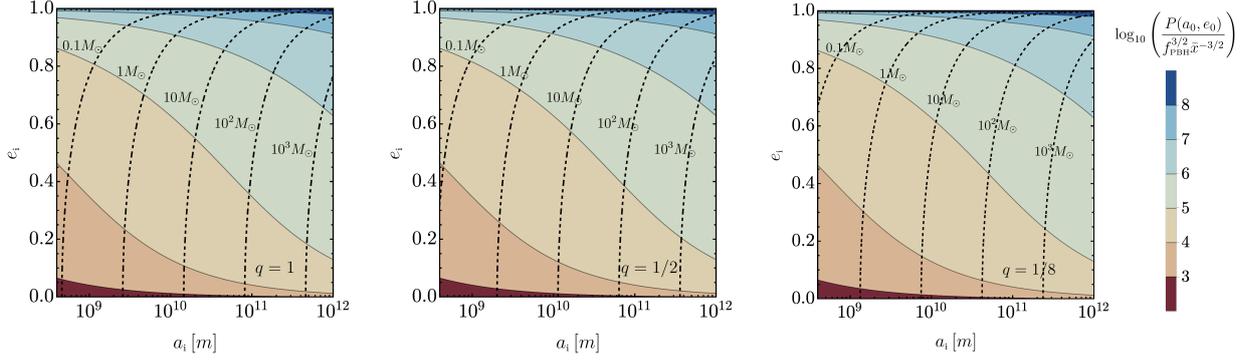

Figure 4.22: *Distribution of the binary parameters $e_i$ and $a_i$. For presentation purposes, we also show with black dashed contour lines the combination of parameters $e_i$ and $a_i$ giving a coalescence time equal to the age of the universe $t_c(a_i, e_i) = t_0$.* **Left:** *Mass ratio $q = 1$.* **Center:** *$q = 1/2$.* **Right:** *$q = 1/8$*

at $z_{dec}$ if

$$\frac{1 + z_{dec}}{1 + z_{eq}} = f_{PBH}\left(\frac{\bar{x}}{x}\right)^3 - 1 > 0, \tag{4.5.2}$$

where $f_{PBH}$ indicates the usual PBHs abundance in terms of the dark matter one at that epoch and the mean physical separation $\bar{x}$ is defined as

$$\bar{x}(z_{eq}) = \left(\frac{M}{\rho_{PBH}(z_{eq})}\right)^{1/3} = \frac{1}{(1 + z_{eq})f_{PBH}^{1/3}}\left(\frac{8\pi G}{3H_0^2}\frac{M}{\Omega_{DM}}\right)^{1/3}. \tag{4.5.3}$$

From Eq. (4.5.2) one can find that the characteristic formation redshift is of the order $z_{dec} > 10^4$ for the masses relevant for current ground-based detectors and $f_{PBH} \approx 10^{-3}$ required to match the observed merger rate [169]. We will come back to this point in the following.

Initially, the motion of two PBHs bounded in the binary system is radial. However, due to the action of tidal torques from the surrounding radiation overdensities and neighbouring isolated PBHs, the binary acquires angular momentum. Following the simplified description given in Refs. [162, 462], one can describe the initial semi major axis $a_{eq}$ and eccentricity $e_{eq}$ of the binary as

$$a_{eq} = \frac{\alpha}{f_{PBH}}\frac{x^4}{\bar{x}^3}, \qquad e_{eq} = \sqrt{1 - \beta^2\left(\frac{x}{y}\right)^6}, \tag{4.5.4}$$

where $y$ is the physical distance to the third PBH at $z_{eq}$ and $\alpha$ and $\beta$ are $\mathcal{O}(1)$ coefficients to be fixed numerically. Following Ref. [463], we take $\alpha = 0.4$ and $\beta = 0.8$. Imposing the geometrical condition $x < y < \bar{x}$, one gets an upper bound on the eccentricity as

$$e_{eq,\,max} = \sqrt{1 - f_{PBH}^{3/2}\left(\frac{\beta^2}{\alpha^{3/2}}\right)\left(\frac{a_{eq}}{\bar{x}}\right)^{3/2}}. \tag{4.5.5}$$

Assuming a uniform probability distribution for both $x$ and $y$ in three dimensional space, one can derive a distribution for the expected semi-major axis and eccentricities at formation, given by

$$dP = \frac{3}{4}f_{PBH}^{3/2}\bar{x}^{-3/2}\left(\frac{\beta^{1/2}}{\alpha^{3/2}}\right)a_{eq}^{1/2}e_{eq}(1 - e_{eq}^2)^{-3/2}da_{eq}de_{eq}. \tag{4.5.6}$$

In Fig. 4.22, we plot the distribution of the parameters $a_{eq}$ and $e_{eq}$. We can see that the PBHs binaries are expected to be predominantly eccentric, with a distribution peaking at large values of $e_i$.

After formation, the binary shrinks due to the loss of energy through GW emission. We will discuss this process in details in the following section. We stress that the simplified description of the binary formation process given in this section has been improved both with analytical [169] and numerical [170] analyses. In the section dedicated to the PBH merger rate, we will adopt the *state of the art* description of the binary formation given in Ref. [170]. We also mention here that the formation of binaries in the early universe provides the dominant channel like the ones formed in the late-time universe give rise to smaller merger rates. Therefore, this channel dictates the properties of the PBH binaries to be compared with GW observations. We will come back to this point later on.

### 4.5.2 Binary formation in the late-time universe

An additional formation mechanism of PBH binaries is due to dynamical captures in present-day halos [160, 464]. We again consider two PBHs with equal mass $M$. If a PBH moving at a given velocity $v$ passes close to another PBH, the energy loss due to the GW emission can reduce the kinetic energy of the first objects which becomes bound to the latter. In this process, one can estimate the energy loss as [196]

$$\Delta E = \frac{85\pi\sqrt{GM}G^3M^4}{12r_{\rm p}^{7/2}}.$$ (4.5.7)

In the newtonian approximation, the impact parameter is estimated to be [464, 465]

$$b^2(r_{\rm p}) = r_{\rm p}^2 + 2GMr_{\rm p}/v^2,$$ (4.5.8)

where

$$r_{\rm p} = \left(\frac{85\pi}{3\sigma_v^2}\right)^{2/7}GM$$ (4.5.9)

and $\sigma_v$ represents the characteristic rms relative velocity between the two objects. Therefore, the cross section for a binary formation simplifies to become

$$\sigma_{\rm bin} \simeq \left(\frac{85\pi}{3}\right)^{2/7}\frac{\pi\,(2GM)^2}{v^{18/7}}.$$ (4.5.10)

By defining $n_{\rm PBH}^{\rm cl}$ as the number density of PBHs with mass $M$ and considering an environment with volume $V_{\rm cl}$, we can finally write the differential rate of binary formation in a dense environment as

$$\gamma_{\rm LU} \simeq 17 \cdot 2^{10/7}\,(n_{\rm PBH}^{\rm cl})^2 V_{\rm cl}G^2\frac{M^{12/7}}{\sigma_v^{11/7}}.$$ (4.5.11)

This formula was obtained by performing a statistical average assuming a Maxwellian distribution of the velocities. It is evident by looking at Eq. (4.5.11) that only dense environments are contributing predominantly to this channel, due to the scaling $\gamma_{\rm LU} \sim (n_{\rm PBH}^{\rm cl})^2$.

As binaries formed through this process are highly eccentric [466–468], their characteristic merger time is very small. As such, the formation rate in Eq. 4.5.11 can be considered a proxy for the actual merger rate in the particular environment considered. To extract the cosmological merger rate, however, one needs to integrate the contribution from all PBH clusters, distributed as predicted by the halo mass function of the large scale structure. We will come back to this point in the next section dedicated to the merger rate computation.

### 4.5.3 GW-driven evolution

First, we consider the case in which accretion is absent or inefficient throughout the binary history. In this case, the binary evolves only through GW radiation emission. We review this process by following Sec. 4.1 of Ref. [273].

Neglecting the emission of GWs, one can show that the eccentricity $e$ and semi-major axis $a$ are constants of motion. Therefore, they can be expressed in terms of the (conserved) angular momentum $L$ and energy $E$ as

$$e^2 = 1 + \frac{2EL^2}{M_{\rm tot}^2\mu^3}, \qquad a = \frac{GM_{\rm tot}\mu}{2|E|},$$ (4.5.12)

where $\mu = M_1 M_2 / M_{\text{tot}}$.

To compute the energy and angular momentum lost through the GW emission, one can adopt the weak-field/slow-motion approximation and use the quadrupole formula. Following Refs. [266, 267], the evolution of energy and angular momentum of the binary is dictated by the equations

$$\frac{dE}{dt} = -\frac{32}{5}\frac{\mu^2 M_{\text{tot}}^3}{a^5}\frac{1}{(1-e^2)^{7/2}}\left(1 + \frac{73}{24}e^2 + \frac{37}{96}e^4\right),$$

$$\frac{dL}{dt} = -\frac{32}{5}\frac{\mu^2 M_{\text{tot}}^{5/2}}{a^{7/2}}\frac{1}{(1-e^2)^2}\left(1 + \frac{7}{8}e^2\right), \tag{4.5.13}$$

or, in terms of $e$ and $a$, by

$$\frac{da}{dt} = -\frac{64}{5}\frac{\mu M_{\text{tot}}^2}{a^3}\frac{1}{(1-e^2)^{7/2}}\left(1 + \frac{73}{24}e^2 + \frac{37}{96}e^4\right),$$

$$\frac{de}{dt} = -\frac{304}{15}\frac{\mu M_{\text{tot}}^2}{a^4}\frac{e}{(1-e^2)^{5/2}}\left(1 + \frac{121}{304}e^2\right). \tag{4.5.14}$$

This system of equations can be integrated to find the merging time $t_c$, defined for simplicity as $a(t_c) = 0$. For an initial orbit with $e(t_i) = e_i$ and $a(t_i) = a_i$ one finds

$$t_c(a_i, e_i) = t_c(a_i)\frac{48}{19}\frac{1}{g^4(e_i)}\int_0^{e_i} de\, \frac{g^4(e)(1-e^2)^{5/2}}{e(1 + 121e^2/304)}, \tag{4.5.15}$$

where we defined the function

$$g(e) = \frac{e^{12/19}}{1-e^2}\left(1 + \frac{121}{304}e^2\right)^{870/2299}. \tag{4.5.16}$$

In the limit of circular orbits, Eq. (4.5.15) simplifies to become

$$t_c(a_i, e_i = 0) \equiv t_c(a_i) = \frac{5}{256}\frac{a_i^4}{M_{\text{tot}}^2 \mu} \tag{4.5.17}$$

while in the highly eccentric $e_i \to 1$ limit, one finds

$$t_c(a_i, e_i \to 1) \simeq t_c(a_i)\frac{768}{429}(1 - e_i^2)^{7/2}. \tag{4.5.18}$$

The parameters leading to a coalescence time equal to the age of the universe, $t_0 = 13.7\,\text{Gyr}$, are shown in Fig. 4.23.

In the PBH model, binaries are typically formed with high eccentricities [162, 196] and, therefore, we will adopt Eq. (4.5.18) to compute the characteristic merger time.

### 4.5.4  Accretion-driven evolution

Here, we modify the evolution of binaries by also accounting for the effect of baryonic mass accretion, as modelled in the previous sections. We extend the results of Ref. [469] where the accretion-driven inspiral was only described for circular and symmetric binaries with $q \simeq 1$.

In general, accretion introduces an additional secular change to the orbital parameters, on top of the inevitable GW radiation reaction [414, 469, 470]. Due to the different timescales involved in the problem, we can study the accretion-driven phase and the GW-driven phase separately. This is because we are mostly interested in mergers taking place in the late time-universe (as the detection redshift is very small for the current horizons of LIGO and Virgo experiments) and accretion is drastically suppressed for $z < z_{\text{cut-off}}$. Therefore, in the first phase, the binary evolves following accretion effects, while closer to the merger, the GW-driven inspiral takes over.

We can support this picture by comparing the characteristic evolution time-scales for both the GW-driven and accretion-driven evolution. From Eq. (4.5.15), a merger occurring at $z \approx 0$ corresponds to a binary with orbital separation $a = \mathcal{O}(10^{11} \, \mathrm{m})$ at $z \sim z_{\text{cut-off}} = 10$. Therefore, the characteristic time scale of variation of the semi-major axis due to GW emission is

$$T_{\text{GW}} \sim \frac{a}{\dot{a}}\Big|_{\text{GW}} = 4 \times 10^{17} \, \mathrm{s} \left(\frac{a}{1.4 \times 10^{11} \mathrm{m}}\right)^4 \left(\frac{M}{30 M_\odot}\right)^{-3} \left(\frac{1-e^2}{0.1}\right)^{7/2}. \tag{4.5.19}$$

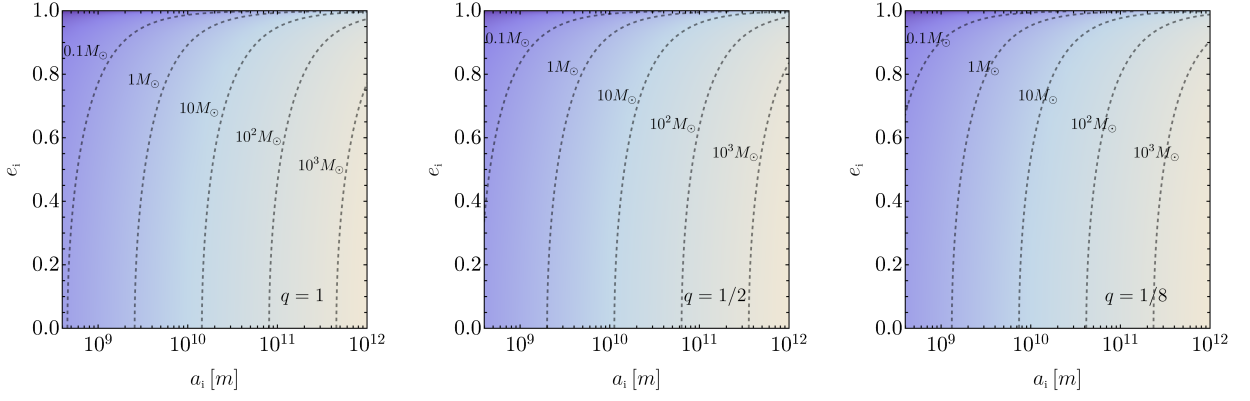

**Figure 4.23:** *The contour lines indicate the combination of parameters $e_i$ and $a_i$ giving a coalescence time equal to the age of the universe $t_c(a_i, e_i) = t_0$. The labels indicate $M_1$ and the mass ratio considered ($q = 1$, $1/2$ and $1/8$ respectively).*

We stress that the normalisation $a = 1.4 \times 10^{11} \mathrm{m}$ was chosen to match the value giving a merger in a time equal to the age of the universe for binaries with equal massed $M = 30 M_\odot$ and $e = 0.95$. As $T_{\text{GW}} \propto a^4$ and we require the merger to take place at small redshift, for $z > z_{\text{cut-off}}$ the typical GW time scale is much larger than the accretion one given in Eq. (4.1.14). Therefore, we can assume that the inspiral is driven solely by accretion when $z > z_{\text{cut-off}}$ and solely by GW radiation-reaction when $z < z_{\text{cut-off}}$. In the case of 3G detectors (such as the Einstein Telescope [471] or the space mission LISA [277]) searching for events at $z > z_{\text{cut-off}}$, one would need to conduct a more in-depth analysis as this hierarchy of scales may not be respected making the evolution more complex.

We also stress that, for the parameter space we are interested in, the mass accretion time scale (4.1.14) is always much larger than the orbital period

$$T_{\text{orbital}} \sim \left(\frac{M_{\text{tot}}}{a^3}\right)^{-1/2} \sim 8 \times 10^5 \, \mathrm{s} \left(\frac{M}{30 M_\odot}\right)^{-1/2} \left(\frac{a}{1.4 \cdot 10^{11} \mathrm{m}}\right)^{3/2}. \tag{4.5.20}$$

Thus, the mass accretion phenomenon can be considered, in first approximation, an adiabatic process.

If the masses vary adiabatically, one can compute the action variables associated to the elliptic motion and treat them as adiabatic invariants. In particular, the adiabatic invariants for the Keplerian two-body problem are [472]

$$I_\phi = \frac{1}{2\pi} \int_0^{2\pi} p_\phi \mathrm{d}\phi = L_z, \tag{4.5.21}$$

$$I_r = \frac{1}{2\pi} \int_{r_{\min}}^{r_{\max}} p_r \mathrm{d}r = -L_z + \sqrt{M_{\text{tot}} \mu^2 a}, \tag{4.5.22}$$

where the energy and angular momentum of the binary are defined as

$$E = -\frac{\mu M_{\text{tot}}}{2a}, \qquad L_z = \sqrt{\frac{M_{\text{tot}}^2 \mu^3 (e^2 - 1)}{2E}} = \sqrt{1 - e^2} \sqrt{M_{\text{tot}} \mu^2 a}. \tag{4.5.23}$$

The adiabatic invariance of $I_\phi$ and $I_r$ can be expresses as

$$\frac{\mathrm{d}I_\phi}{\mathrm{d}t} = \frac{\partial L_z}{\partial e} \frac{\partial e}{\partial t} + \frac{\partial L_z}{\partial a} \frac{\partial a}{\partial t} + \frac{\partial L_z}{\partial \mu} \frac{\partial \mu}{\partial t} + \frac{\partial L_z}{\partial M_{\text{tot}}} \frac{\partial M_{\text{tot}}}{\partial t} = 0,$$
$$\frac{\mathrm{d}I_r}{\mathrm{d}t} = \frac{\partial I_r}{\partial e} \frac{\partial e}{\partial t} + \frac{\partial I_r}{\partial a} \frac{\partial a}{\partial t} + \frac{\partial I_r}{\partial \mu} \frac{\partial \mu}{\partial t} + \frac{\partial I_r}{\partial M_{\text{tot}}} \frac{\partial M_{\text{tot}}}{\partial t} = 0. \tag{4.5.24}$$

After few manipulations, one can recast the previous conditions as a system of equations dictating the evolution of the binary parameters

$$\frac{\partial a}{\partial t} = -\frac{2a}{\mu}\frac{\partial \mu}{\partial t} - \frac{a}{M_{\text{tot}}}\frac{\partial M_{\text{tot}}}{\partial t},$$
$$\frac{\partial e}{\partial t} = 0\,. \tag{4.5.25}$$

The last equation shows that the eccentricity is a constant of motion for an accretion-driven inspiral. This property can also be directly derived by expressing the eccentricity in terms of the adiabatic invariants as [472]

$$e = \sqrt{1 - \left(\frac{I_\phi}{I_\phi + I_r}\right)^2}\,. \tag{4.5.26}$$

It follows that, in the adiabatic approximation, accretion is only able to affect the semi-major axis, whose evolution follows the equation

$$\frac{\dot{a}}{a} + 2\frac{\dot{\mu}}{\mu} + \frac{\dot{M}_{\text{tot}}}{M_{\text{tot}}} = 0. \tag{4.5.27}$$

This is consistent with the findings in Ref. [469] in the absence of GW emission and for circular binaries. However, it also extends to binaries with a generic eccentricity, which remains a constant of motion.

Finally, Eq. (4.5.27) can be recast in terms of the individual PBH accretion rates as

$$\frac{\dot{a}}{a} + \frac{M_2(M_1 + 2M_2)\dot{M}_1 + M_1(2M_1 + M_2)\dot{M}_2}{M_1 M_2(M_1 + M_2)} = 0\,. \tag{4.5.28}$$

Therefore, within the aforementioned approximations, one can evolve the PBH masses $M_i$ using Eq. (4.1.37) and then account for the evolution of the semi-major axis using Eq. (4.5.28). In the limit of symmetric binaries (i.e. $q \to 1$), the evolution in Eq. (4.5.28) simplifies to become

$$\frac{\dot{a}}{a} + 3\frac{\dot{M}_1}{M_1} = 0\,. \tag{4.5.29}$$

## 4.6   Merger rate

In the previous section, we have described how PBH binaries are formed and how they evolve through accretion-driven and GW-driven phases. Those are the basic ingredients to compute the number of mergers expected from a given population of PBHs. As the merger rate is the crucial quantity that is needed to understand if the currently available GW detections could be of primordial origin, we dedicate this section to its computation, also accounting for the possible suppression coming from the interaction of binaries with other compact objects in dense environments.

We will divide the discussion into different parts. First, we will address the merger rate of binaries formed in the early universe, while also discussing two possible suppression effects. Then, we will show the computation of the merger rate coming from late-time universe binaries, and demonstrate it only provides a subdominant contribution. Finally, we will present the complete parameterisation of the merger rate which can be used to compare the prediction of the PBH model with the data.

The considerations contained in this section strictly apply to the vanilla scenario in which PBHs come from the collapse of overdensities and are not initially clustered above Poisson at formation. This is the scenario we will also adopt in the following chapter where the PBH model will be confronted with the GW data.

### 4.6.1 Merger rate of early universe binaries including accretion effects

The computation of the merger rate is straightforward. In general, it is equivalent to estimating the number of binaries that are merging at the time of observation $t_{obs}$. Given the characteristic GW evolution timescale (4.5.18), the computation can be translated into evaluating the fraction of primordial binaries with geometric parameters $a$ and $e$ allowing them to merge in a time $t_{obs} - t_{form} \simeq t_{obs}$. Neglecting the effect of accretion for the time being, one can compute the rate as

$$dR_0(t) = \int dn_b de \frac{dP}{de} \delta\left(t - t_c(\mu, M_{tot}, a, e)\right),$$ (4.6.1)

where $dn_b$ is the differential number of binaries and the coalescence time takes the compact form

$$t_c = \frac{3}{85} \frac{a_i^4 (1 - e_i^2)^{7/2}}{\eta(z_i) M_{tot}^3(z_i)}.$$ (4.6.2)

In this relation, we stressed that, if one neglects accretion, all quantities are defined at the formation epoch $z_i$. We use the same notation adopted in the section dedicated to accretion 4.1. Following Ref. [170], one finds

$$dR_0 = \frac{1.6 \times 10^6}{\text{Gpc}^3 \, \text{yr}} f_{PBH}^{\frac{53}{37}}(z_i) \left(\frac{t}{t_0}\right)^{-\frac{34}{37}} \eta_i^{-\frac{34}{37}} \left(\frac{M_{tot}^i}{M_\odot}\right)^{-\frac{32}{37}} \psi(M_1^i, z_i)\psi(M_2^i, z_i) dM_1^i dM_2^i,$$ (4.6.3)

where $f_{PBH}(z_i)$ is the initial abundance of PBHs, $\eta_i = \mu^i/M_{tot}^i$ is the initial symmetric mass ratio, defined in terms of the reduced mass $\mu^i = M_1^i M_2^i/M_{tot}^i$ and total mass $M_{tot}^i = M_1^i + M_2^i$ of the binary components at the formation time. We also stress that we adopt the PBH mass function $\psi(M^i, z_i)$ at the formation time $z_i$ normalised as in Eq. (3.1.2). With the subscript 0 we denote the merger rate without suppression factors which will be introduced in the following.

**Effect of accretion on the merger rate**

In the presence of accretion both the masses and the semi-major axis change in time. As a consequence, the characteristic coalescence time is modified and, therefore, the merger rate is affected.

As we discussed in the previous section, the accretion-driven phase occurs earlier and independently from the GW emission. Therefore, as accretion effectively stops at the cut-off redshift $z \sim z_{cut\text{-}off}$, one can compute the subsequent GW evolution with the masses and semi-major axis computed at $z \sim z_{cut\text{-}off}$. In practice, the coalescence time, after all the relevant accretion has taken place, is

$$t_c^{acc} = \frac{3}{85} \frac{\mathcal{N}^4 a_i^4 (1 - e_i^2)^{7/2}}{\eta(z_{cut\text{-}off}) M_{tot}^3(z_{cut\text{-}off})} \equiv \frac{\mathcal{N}^4}{\mathcal{S}} t_c(M_j^i).$$ (4.6.4)

In details, the factor

$$\mathcal{S} = \frac{\eta(z_{cut\text{-}off}) M_{tot}^3(z_{cut\text{-}off})}{\eta(z_i) M_{tot}^3(z_i)}$$ (4.6.5)

accounts for the fact that the GW emission is stronger as the masses are larger after accretion has taken place. The second factor

$$\mathcal{N} \equiv \frac{a(z_{cut\text{-}off})}{a_i} = \exp\left[-\int_{t_i}^{t_{cut\text{-}off}} dt \left(\frac{\dot{M}_{tot}}{M_{tot}} + 2\frac{\dot{\mu}}{\mu}\right)\right],$$ (4.6.6)

is introduced to correct the semi-major axis that has shrunk during the accretion-driven phase. Finally, including accretion effects and written in terms of the final masses, the merger rate is

$$dR_0^{acc}(t, M_j, f_{PBH}(z_i)) = \frac{\mathcal{S}}{\mathcal{N}^4} dR_0\left(t\mathcal{S}/\mathcal{N}^4, M_j^i, f_{PBH}(z_i)\right)$$
$$= \mathcal{N}^{-12/37} \mathcal{S}^{3/37} dR_0\left(t, M_j^i, f_{PBH}(z_i)\right)$$ (4.6.7)

where with $M_j^i$ and $M_j$ we identify the masses $(M_1, M_2)$ at formation and final time, respectively. Notice that the merger rate is entirely expressed in terms of the high redshift quantities and the mass accretion maps dictate the values of the correction factors $\mathcal{N}$ and $\mathcal{S}$.

### 4.6.2 Early-time universe suppression

As PBH binaries in the early universe are not formed in isolation but are surrounded by other PBHs and matter fluctuations, the real distribution of eccentricities $e$ or, equivalently, the dimensionless angular momenta $j \equiv \sqrt{1 - e^2}$, is modified [169, 170, 473–475]. We will introduce such corrections as an additional suppression factor to the merger rate, following Ref. [170]. We stress that this effect is genuinely a high redshift phenomenon and, as such, it is independent from the subsequent development of PBH clustering and mass evolution due to accretion.

One can assume the masses $M_i$ and positions $x_i$ of the surrounding PBHs and the matter density perturbations to be statistically independent. Then, the binary angular momentum induced by such environmental effects can be written as

$$\vec{J} = \vec{J}_{\text{PBH}} + \vec{J}_{\text{M}}. \tag{4.6.8}$$

As those quantities are statistically independent, the corresponding variances can be summed to get

$$\sigma_{\vec{J}} = \sigma_{\vec{J}_{\text{PBH}}} + \sigma_{\vec{J}_{\text{M}}} \tag{4.6.9}$$

where [170]

$$\sigma_{\vec{J}_{\text{PBH}}} = \frac{6}{5} j_0^2 \frac{\langle M^2 \rangle / \langle M \rangle^2}{\bar{N}(y)}, \tag{4.6.10}$$

$$\sigma_{\vec{J}_{\text{M}}} = \frac{6}{5} j_0^2 \frac{\sigma_{\text{M}}^2}{f_{\text{PBH}}^2}. \tag{4.6.11}$$

In the previous relations, we introduced the characteristic angular momentum $j_0 \simeq 0.4 f_{\text{PBH}}/\delta_b$ in terms of the binary overdensity with respect to the background at formation $\delta_b \approx a_{\text{eq}}/a_{\text{dec}}$. Also, $\sigma_{\text{M}}^2 \simeq 3.6 \cdot 10^{-5}$ identifies the the matter perturbation variance at the time at which the binary is formed, the mass distribution momenta are defined as

$$\langle X \rangle = \int X \psi(M, z_i) \mathrm{d}M, \tag{4.6.12}$$

and, finally, the expected number of PBHs within a sphere of comoving radius $y$ around the initial PBH is

$$\bar{N}(y) = \frac{M_{\text{tot}}^i}{\langle M \rangle} \frac{f_{\text{PBH}}(z_i)}{f_{\text{PBH}}(z_i) + \sigma_{\text{M}}}. \tag{4.6.13}$$

By comparing Eq. (4.6.10) with (4.6.11), one finds that the effect of surrounding PBHs is equal to the one of matter perturbations with a variance

$$(\sigma_{\text{M}}^{\text{PBH}})^2 \approx \frac{f_{\text{PBH}}^2 \langle M^2 \rangle / \langle M \rangle^2}{\bar{N}(y)}, \tag{4.6.14}$$

which is consistent with the Poisson nature of the PBH distribution at formation assumed throughout. By computing the final distribution of PBH binary angular momentum $j \mathrm{d}P/\mathrm{d}j$, Ref. [170] derived the suppression factor

$$S = \frac{e^{-\bar{N}(y)}}{\Gamma(21/37)} \int \mathrm{d}v v^{-\frac{16}{37}} \exp\left[ -\bar{N}(y) \langle M \rangle \int \frac{\mathrm{d}M}{M} \psi(M, z_i) F\left( \frac{M}{\langle M \rangle} \frac{v}{\bar{N}(y)} \right) - \frac{3\sigma_{\text{M}}^2 v^2}{10 f_{\text{PBH}}^2(z_i)} \right] \tag{4.6.15}$$

in terms of the generalised hypergeometric function

$$F(z) = {}_1F_2\left( -\frac{1}{2}; \frac{3}{4}; \frac{5}{4}; -\frac{9z^2}{16} \right) - 1. \tag{4.6.16}$$

In Fig. 4.24, we show a few examples of the suppression factor $S$ as a function of the PBH abundance. Few key properties can be highlighted:

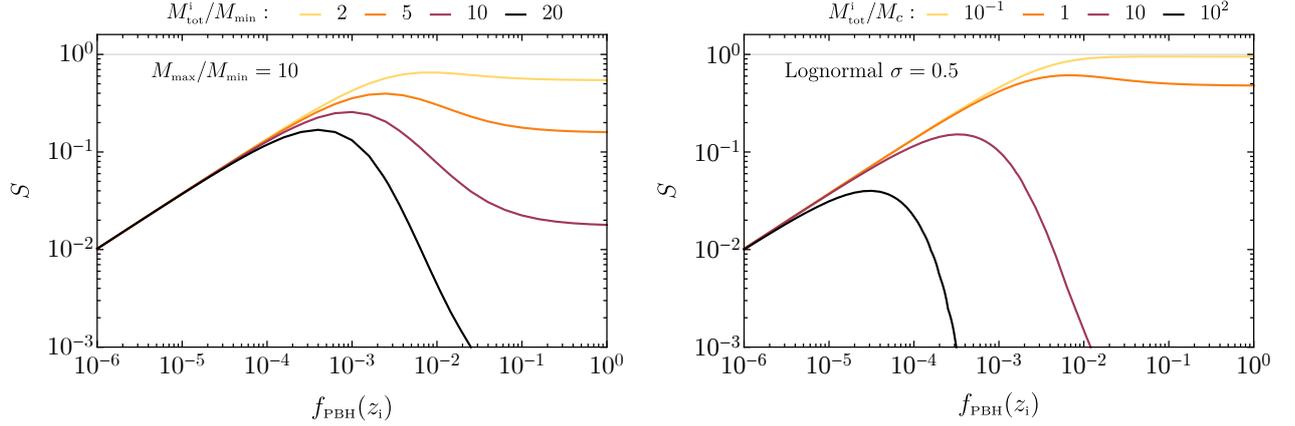

**Figure 4.24:** *Left:  Suppression factor for a power-law mass function with $M_{max} = 10 M_{min}$ and various $M^i_{tot}$.* ***Right:*** *Suppression factor for a lognormal mass function with $\sigma = 0.5$.*

- The suppression factor only depends on the value of the total mass of the binary $M_{tot}$ through the factor $\bar{N}(y)$.

- for a small enough PBH abundance, the suppression factor becomes independent of the total mass of the binary. This can be explained as follows. As $f_{PBH}$ is reduced, it becomes increasingly harder to find neighbouring PBHs able to torque the binary. Therefore, as the torque is only generated by matter perturbations, by looking at Eq. (4.6.11), we see that the dependence on $M_{tot}$ is lost.

- For a sizeable enough abundance, there is an exponential suppression of the rate of massive binaries with $M_{tot} \gg \langle M \rangle$. This is because the most massive PBHs in the spectrum are affecting gravitationally a large region of the universe, thus the expected number of neighbouring PBHs $\bar{N}(y)$ increases.

In the following section, we will also provide a simpler analytical parametrisation of the suppression factor in Eq. (4.6.15) by following [476].

### 4.6.3   Late-time universe suppression

There is another phenomenon that is able to suppress the merger rate of PBH binaries. After binaries are formed (before matter radiation equality), they can enter in clusters that start to form throughout cosmic history. As the interaction of binaries with other compact objects leads to a variation of the semi-major axis and eccentricity, impacting, therefore, the coalesce time, the merger rate gets modified.

Recall that the coalescence time of symmetric binaries ($q \simeq 1$) due to the emission of GWs is

$$t_{GW} \simeq \frac{3}{170\, G M_{PBH}} a^4 (1 - e^2)^{7/2}. \tag{4.6.17}$$

If clustering enhances PBH interactions, generating departures from the initially high eccentricity regime, it inevitably increase significantly the coalescence time of binaries.

As we argued in the section dedicated to PBH clustering at formation, in the vanilla scenario, PBHs are only distributed à la Poisson. This property at formation has been already incorporated into the formula describing the merger rate. However, closer to the matter radiation equality, clusters with an increasingly large number of objects start to form, and PBH binaries can end up populating those dense environments. It was shown in Refs. [170, 460] with dedicated numerical simulations that the geometry of the binaries gets modified after multiple interactions with other PBHs in the clusters. Therefore, we need to account for such a suppression of the merger rate in a redshift dependent manner.

We stress here, however, that this effect is only relevant for large values of the PBH abundance, as in the opposite case dense PBH clusters are much more rare [9, 170, 224].

We will estimate the size of the suppression by assuming $f_{\rm PBH} \simeq 1$ in this section. We choose to be conservative and consider the largest possible suppression by focusing on the effects which may increase the disruption of PBH binaries and, consequently, decrease the merger rate. First of all, binaries are typically heavier than single PBHs and therefore tend to sink towards the halo center and concentrate within a smaller segregation radius. Secondly, the halo core may be subject to the so-called gravothermal instability which arises from the negative heat capacity of self-gravitating systems [477]. As PBHs with large velocity may evaporate from the cluster core, this reduces the characteristic velocity of PBHs, leading to a collapse of the core. As a consequence, the central density increases enhancing the frequency of binary-PBH encounters. Following Ref. [224], we conservatively assume that all binaries entering into an unstable halo core are perturbed so much that their characteristic coalescence time exceeds the age of the universe.

The gravothermal instability timescale is defined as [478]

$$t_{\rm GI} = \frac{v^3(r)}{G^2 M_{\rm PBH} \rho_{\rm PBH}(r) \log(M(<r)/M_{\rm PBH})},  \tag{4.6.18}$$

where we adopt the PBH density halo profile as derived in the non-linear regime in the previous section. For a cluster (or halo) with mass $M_h$, the PBH density profile can be therefore written as (see Eq. (4.4.26))

$$\rho_{\rm PBH}(r) \simeq \frac{3M_h}{20\pi} R_{\rm vir}^{-3/5} r^{-12/5},  \tag{4.6.19}$$

in terms of the virial radius $R_{\rm vir}$ found by imposing

$$\rho_{\rm PBH}(<R_{\rm vir}) = 200 M_{\rm PBH} \bar{n}_{\rm PBH}.  \tag{4.6.20}$$

The characteristic PBH velocity at a given radius $r$ from the cluster center is then given by

$$v(r) = \sqrt{\frac{GM(<r)}{r}} \simeq \sqrt{GM_h R_{\rm vir}^{-3/5} r^{-1/5}}.  \tag{4.6.21}$$

We finally need to require the gravothermal timescale to be smaller than the Hubble time. Focusing, for instance, on $z = 0$ and the characteristic halo mass $M_*(z=0)$ given by Eq. (4.4.28), we find the critical radius for the gravothermal instability to occur to be $\sim 3 \cdot 10^{-3}$ kpc$/h$, corresponding to a critical number of PBHs $N_c \sim 4.6 \cdot 10^4$.

The missing ingredient which is required to estimate the final merger rate after the suppression is the fraction of initial PBH binaries which are contained in gravothermally unstable cores. For an initially Poisson distribution, the probability of finding a PBH within a halo made up of $N$ PBHs at redshift $z$ is [479] (see also [480])

$$p_N(z) \propto N^{-1/2} e^{-N/N_*(z)}.  \tag{4.6.22}$$

Then, the probability of finding a binary within a cluster of $N$ PBHs is approximately proportional to $p_N$, while the probability of finding a binary in a sub-halo of $N$ PBHs embedded in a larger cluster of $N' > N$ PBHs is proportional to $p_N \cdot p_{N'}$ [224]. The conservative fraction of unperturbed binaries at the present time is bounded to be larger than

$$P_{\rm np} \gtrsim 1 - \sum_{N=3}^{N_c} \bar{p}_N(z_{\rm form}^c) - \sum_{N'>N_c} \left[ \sum_{N=3}^{N_c} \widetilde{p}_N(z_{\rm form}^c) \right] \bar{p}_{N'}(z_{\rm form}^c),  \tag{4.6.23}$$

where $z_{\rm form}^c$ is the formation time of the halo with $N_c$ PBHs and

$$\sum_{N \geq 2} \bar{p}_N = 1 \quad \text{and} \quad \sum_{N=2}^{N'} \widetilde{p}_N = 1.  \tag{4.6.24}$$

Focusing on the present day merger rate, we find $P_{\rm np} \simeq 10^{-2}$, which can be interpreted as the probability that a binary is not disrupted by dynamical suppressions in clusters $S_{\rm cl} \equiv P_{\rm np}$. A quantitatively similar conclusion, without however accounting for the corrections due to the cluster profile in the non-linear regime, was found in Ref. [224].

We conclude this discussion by stressing that, in practice, this suppression will not play a crucial role in the computation of the merger rate. This is because the rate inferred for large values of the abundance $f_{\rm PBH}$ is still above the bound set by the LIGO/Virgo observations, even when accounting for the suppression factor coming from disruptions in clusters [9, 224, 481]. This also justifies our procedure which only focused on the conservative estimate of the merger rate, i.e. accounting for the largest possible suppression effect. Therefore, one is forced to consider a smaller PBH abundance $f_{\rm PBH} \ll 1$ to be compatible with the observations. In such a case, the suppression discussed in this section is not present as it becomes increasingly harder for PBH binaries to fall in the rare PBH clusters.

### 4.6.4   Merger rate of late-time binaries

In this section, we focus on the merger rate of binaries formed in the late-time universe. We will proceed with the computation for the largest possible abundance $f_{\rm PBH} = 1$ and show the contribution from this channel is largely subdominant with respect to the one coming from primordial binaries [9, 169], even when accounting for the enhancement due to clustering induced by initial Poisson fluctuations.

Using Eq. (4.5.10), one can compute the merger rate of PBH binaries formed in a halo of mass $M_h$ as [160][5]

$$R_h(M_h) = 2\pi \int_0^{R_{\rm vir}} {\rm d}r\, r^2 \left( \frac{\rho_{\rm PBH}(r)}{M_{\rm PBH}} \right)^2 \langle \sigma_{\rm bin} v \rangle, \tag{4.6.25}$$

where the brackets indicates the thermally averaged cross-section, i.e. the mean of the combination $\sigma_{\rm bin} v$ averaged over the velocities distributed following a Maxwell-Boltzmann distribion. For simplicity, we assume that the PBH density profile is constant within the core of size $r_s$. We will discuss below the nature of this density plateau in the inner region of the halo. Outside this inner region when $r > r_s$, we adopt the halo profile $\rho_{\rm PBH}(r) \sim r^{-12/5}$ as derived in Eq. (4.4.26). The resulting present-day merger rate reads

$$\mathcal{R}_h(M_h) \simeq 22 \left( G M_h^{4/5} M_{\rm PBH}^{1/5} \bar{n}_{\rm PBH}^{1/5} \right)^{17/14} r_s^{-52/35}. \tag{4.6.26}$$

To compute the cosmological merger rate, one needs to convolute the resulting rate $\mathcal{R}_h$ with the halo mass function ${\rm d}n/{\rm d}M_h$ describing the distribution in mass of the halos. Adopting ${\rm d}n/{\rm d}M_h$ as derived in Eq. (4.4.27) using the Press-Schechter formalism, one finds

$$R_{\rm LU} = \int {\rm d}M_h \frac{{\rm d}n}{{\rm d}M_h} \mathcal{R}_h(M_h)$$
$$\simeq 1.5 \cdot 10^3\, G^{17/14} (M_{\rm PBH} \bar{n}_{\rm PBH})^{73/42} M_*^{-11/21} \mathcal{R}_{\rm cl}^{52/35}, \tag{4.6.27}$$

where we introduced the the dimensionless "cluster factor" $\mathcal{R}_{\rm cl} = R_*/r_s$ describing the hierarchy of scales between the inner region of the clusters and the characteristic scale

$$R_* \simeq 9 \left( h M_{\rm PBH}/20 M_\odot \right)^{1/3}\ {\rm kpc}/h \tag{4.6.28}$$

identifying the virial radius of an halo of mass $M_*$. Inserting the value for the mean present halo mass $M_* = 6.8 \cdot 10^6 M_{\rm PBH}$, we arrive at

$$R_{\rm LU} \simeq 10^{-4} \left( \frac{M_{\rm PBH}}{20 M_\odot/h} \right)^{-11/21} \mathcal{R}_{\rm cl}^{52/35} {\rm Gpc}^{-3} {\rm yr}^{-1}. \tag{4.6.29}$$

The clustering factor $\mathcal{R}_{\rm cl}$ can be computed by estimating the size of the inner core radius. One expects each cluster to develop a core of radius $r_s$ when the gravitational PBH interactions equalise the kinetic energies [483]. Imposing the relaxation time of the core to be smaller than the age of the Universe, one obtains $\mathcal{R}_{\rm cl} \simeq 10^2$.

---

[5]We point out that Ref. [482] claimed, motivated by the results of N-body simulations, that the realistic merger rate of PBH in clusters to be smaller than what estimated in Ref. [160].

As one can appreciate by comparing Eqs. (4.6.29) with (4.6.36), this channel gives a subdominant contribution to the total merger rate at the present time coming from a PBH population. This conclusion, explicitly derived in the case of $f_{\rm PBH} \simeq 1$, gets even stronger for smaller values of the abundance. This is because, as PBHs become increasingly less clustered, the merger rate of early universe binaries is less suppressed by the dynamical disruption of binaries. On the other hand, the merger rate of late-time binaries decreases as the chance of encounters in halos is diminished by the reduction of PBH clustering for smaller $f_{\rm PBH}$. We will, therefore, neglect this channel in the following chapters.

### 4.6.5  Second-generation mergers

In this section, we provide a simplified argument in favour of neglecting second-generation mergers in the PBH scenario. A more refined description of the second generation merger rate would require dedicated numerical simulations.

We define a second-generation merger as the one involving a BH remnant of previous black-hole mergers. We report here the formalism to compute the secondary merger rates following the approach adopted in Ref. [484, 485] We stress here that, for simplicity, in the present analysis we neglect both suppression factors coming from the disruption of binaries. Taking them into account would render our conclusions slightly stronger. Also, this treatment of the second generation mergers completely neglects the impact of peculiar velocities, which could be transferred to the remnant PBH as a result of the emission of GWs during the merger [486–488] (see, for example, Refs. [489–491] to see the result of this effect on the astrophysical second-generation merger rate). Again, this estimate can be regarded as a conservative upper bound on the fraction of second-generation mergers.

Given a PBH mass function $\psi(M)$, normalised to unity as in Eq. (3.1.2), we can define the fraction of the present number density of PBHs with mass $M$ with respect to the total average as given by the expression

$$F(M) = \frac{\psi(M)}{M} \left[ \int {\rm dln} M' \, \psi(M') \right]^{-1}.$$

(4.6.30)

As found in Ref. [484], the fraction of PBHs which have undergone a merger event before time $t$ is

$$P_{\rm PBH}^{(1)}(t) = 1.34 \times 10^{-2} \left( \frac{M_c}{M_\odot} \right)^{\frac{5}{37}} \left( \frac{t}{t_0} \right)^{\frac{3}{37}} f_{\rm PBH}^{\frac{16}{37}} \Upsilon_1,$$

(4.6.31)

where the dependence on the shape of the mass function is captured in the factor

$$\Upsilon_1 = \left( \int {\rm dln} x \, \tilde{\psi}(x) \right)^{\frac{16}{37}} \int (x_1 + x_2)^{\frac{36}{37}} x_1^{\frac{3}{37}} x_2^{\frac{3}{37}} x_3^{-\frac{21}{37}} \prod_{i=1}^{3} {\rm d} x_i \, \tilde{F}(x_i),$$

(4.6.32)

where we rescaled the dimensional quantities with respect to the mass scale $M_c$ as $\tilde{F}(x = M/M_c) = M_c F(M, M_c)$ and $\tilde{\psi}(x = M/M_c) = M_c \psi(M, M_c)$. Then, it was shown that the fraction of PBHs that have merged in a second-merger process at time $t$ is instead given by [484]

$$P_{\rm PBH}^{(2)}(t) = 1.21 \times 10^{-4} \left( \frac{M_c}{M_\odot} \right)^{\frac{10}{37}} \left( \frac{t}{t_0} \right)^{\frac{6}{37}} f_{\rm PBH}^{\frac{32}{37}} \Upsilon_2,$$

(4.6.33)

in terms of the factor

$$\Upsilon_2 = \left( \int {\rm dln} x \, \tilde{\psi}(x) \right)^{\frac{32}{37}} \int (x_1 + x_2)^{\frac{6}{37}} x_3^{\frac{6}{37}} x_4^{-\frac{42}{37}} (x_1 + x_2 + x_3)^{\frac{72}{37}} \prod_{i=1}^{4} {\rm d} x_i \, \tilde{F}(x_i).$$

(4.6.34)

The conditional probability that an observable PBH merger is the result of a second-merger process at time $t$ is finally given by

$$P_{\rm PBH}^{(2|1)}(t) = 9 \times 10^{-3} \left( \frac{M_c}{M_\odot} \right)^{\frac{5}{37}} \left( \frac{t}{t_0} \right)^{\frac{3}{37}} f_{\rm PBH}^{\frac{16}{37}} \Upsilon \qquad \text{with} \qquad \Upsilon = \frac{\Upsilon_2}{\Upsilon_1}.$$

(4.6.35)

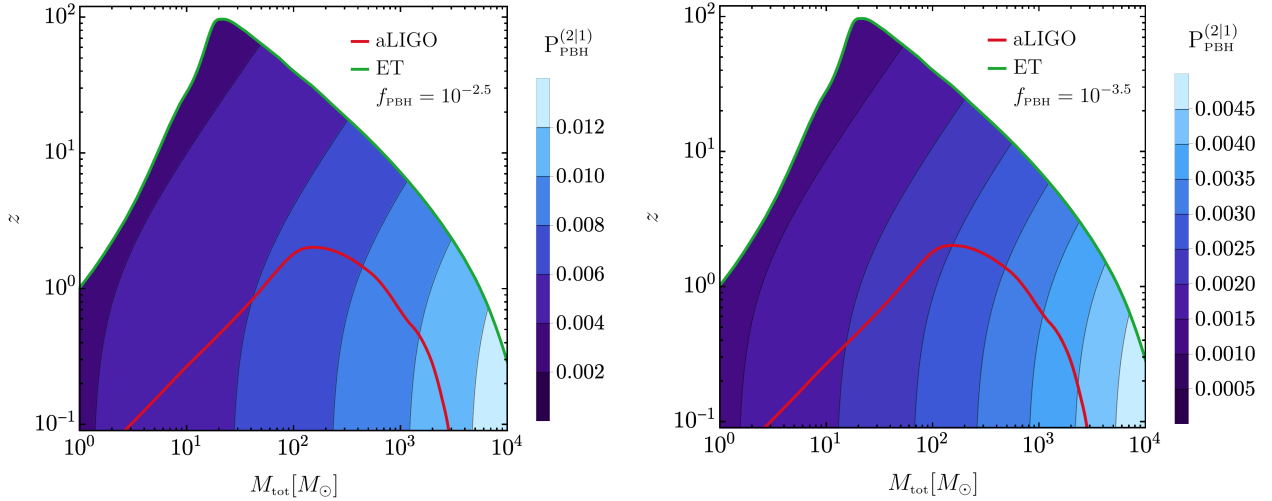

Figure 4.25: *Total fraction of secondary mergers at redshift $z$ in terms of the total (source frame) mass $M_{tot}$. We also show, as a reference, the horizons for aLIGO (red line) and ET (green line), see Refs. [471, 492].* **Left:** $f_{\mathrm{PBH}} = 10^{-2.5}$. **Right:** $f_{\mathrm{PBH}} = 10^{-3.5}$.

Let us consider different mass functions, motivated by the PBH formation mechanism in Sec. 3.1. For the critical mass function, defined in Eq. (3.1.7), one finds $\Upsilon_{\mathrm{crit}} = 6$. For a lognormal mass function, the resulting value for the rescaled merger fraction is $\Upsilon \sim (5 \div 15)$, depending on the width of the distribution $\sigma$ as defined in Eq. (3.1.10). Finally, for a broad mass function as in Eq. (3.1.9), the resulting value for the rescaled merger fraction is $\Upsilon = 4.75$.

In Fig. 4.25, we show the fraction of second generation mergers $P_{\mathrm{PBH}}^{(2|1)}(z)$ in terms of the total (source frame) mass $M_{\mathrm{tot}}$ of the observed binary and the redshift $z$. We consider two representative values for the PBH abundance. The results shown in Fig. 4.25 assume a critical mass function, while the other choices of $\psi(M)$ can be adopted by rescaling $\Upsilon$ with respect to $\Upsilon_{\mathrm{crit}}$. One can safely conclude that only a tiny fraction (at the sub-percent level) of BHs may come from a previous merger.

### 4.6.6 The final merger rate parametrisation

In this section, we just collect all the results described previously to derive an analytical parametrisation of the PBH merger rate which can be adopted to compare the PBH model with the GW observations.

The merger rate is therefore written as [5]

$$
\begin{aligned}
\frac{\mathrm{d}R_{\mathrm{PBH}}}{\mathrm{d}m_1^{\mathrm{i}}\mathrm{d}m_2^{\mathrm{i}}} = {}& 1.6 \times 10^6 \mathrm{Gpc}^{-3}\,\mathrm{yr}^{-1} f_{\mathrm{PBH}}^{\frac{53}{37}}\, \eta_{\mathrm{i}}^{-\frac{34}{37}} \left(\frac{t}{t_0}\right)^{-\frac{34}{37}} \left(\frac{M_{\mathrm{tot}}^{\mathrm{i}}}{M_{\odot}}\right)^{-\frac{32}{37}} \psi_{\mathrm{i}}(m_1^{\mathrm{i}}, z_{\mathrm{i}})\psi_{\mathrm{i}}(m_2^{\mathrm{i}}, z_{\mathrm{i}}) \\
& \times \exp\left[\frac{12}{37}\int_{t_{\mathrm{i}}}^{t_{\mathrm{cut\text{-}off}}}\mathrm{d}t\left(\frac{\dot{M}_{\mathrm{tot}}}{M_{\mathrm{tot}}} + 2\frac{\dot{\mu}}{\mu}\right)\right]\left(\frac{\eta(z_{\mathrm{cut\text{-}off}})}{\eta(z_{\mathrm{i}})}\right)^{3/37}\left(\frac{M_{\mathrm{tot}}(z_{\mathrm{cut\text{-}off}})}{M_{\mathrm{tot}}(z_{\mathrm{i}})}\right)^{9/37} \\
& \times S\left(M_{\mathrm{tot}}^{\mathrm{i}}, f_{\mathrm{PBH}}, \psi_{\mathrm{i}}\right) S_{\mathrm{cl}}(t/t_0, f_{\mathrm{PBH}}),
\end{aligned}
\tag{4.6.36}
$$

where, in order, one can identify the factors coming from the unsuppressed rate (4.6.3), the accretion enhancement (4.6.7), the suppression factor due to early universe dynamics (4.6.15) and the suppression factor due to interactions in clusters (4.6.23). For convenience, we introduced the time $t_{\mathrm{cut\text{-}off}}$ corresponding to the age of the universe at redshift $z = z_{\mathrm{cut\text{-}off}}$ and, following Ref. [476],

$$
S \approx \frac{\sqrt{\pi}(5/6)^{21/74}}{\Gamma(29/37)}\left[\frac{\langle m^2\rangle/\langle m\rangle^2}{\bar{N}(y) + \mathcal{C}} + \frac{\sigma_{\mathrm{M}}^2}{f_{\mathrm{PBH}}^2}\right]^{-\frac{21}{74}}e^{-\bar{N}(y)},
\tag{4.6.37}
$$

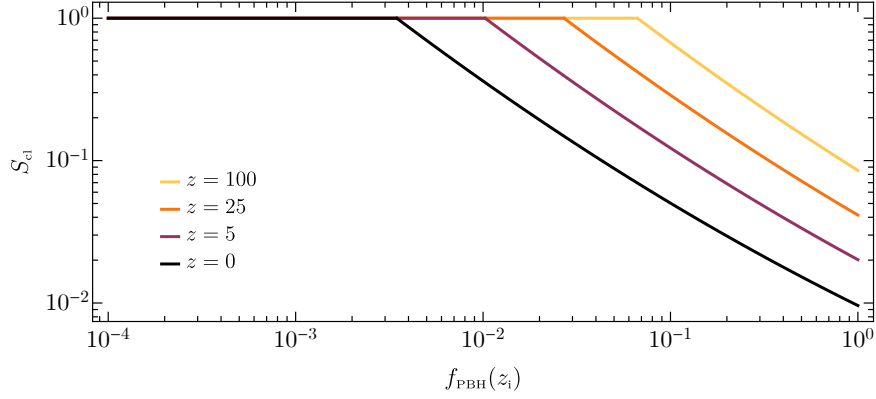

Figure 4.26: *Suppression factor $S_{cl}$ coming from disruptions of binaries in clusters as a function of the initial PBH abundance and for various values of redshift $z$.*

where

$$\mathcal{C} = f_{\mathrm{PBH}}^2 \frac{\langle m^2 \rangle / \langle m \rangle^2}{\sigma_{\mathrm{M}}^2} \left\{ \left[ \frac{\Gamma(29/37)}{\sqrt{\pi}} U \left( \frac{21}{74}, \frac{1}{2}, \frac{5 f_{\mathrm{PBH}}^2}{6 \sigma_{\mathrm{M}}^2} \right) \right]^{-\frac{74}{21}} - 1 \right\}^{-1}. \quad (4.6.38)$$

In Ref. [476] it was found that this approximation is accurate within 7% for a lognormal mass function with $\sigma \leq 2$. Finally, again following Ref. [476], we parametrise the suppression factor due to interactions in clusters as

$$S_{\mathrm{cl}}(x) \approx \min \left[ 1, 9.6 \cdot 10^{-3} x^{-0.65} \exp \left( 0.03 \ln^2 x \right) \right] \qquad \text{with} \qquad x \equiv (t(z)/t_0)^{0.44} f_{\mathrm{PBH}}. \quad (4.6.39)$$

We show a plot of $S_{\mathrm{cl}}$ for various values of the PBH abundance in Fig. 4.26. Notice also that, for $f_{\mathrm{PBH}} \lesssim 0.003$, one always finds $S_2 \simeq 1$, i.e. the suppression of the merger rate due to disruption inside PBH clusters is negligible. This is also supported by the results obtained through a cosmological N-body simulation finding that PBHs are essentially isolated for a small enough abundance [235].

### 4.6.7   SGWB from unresolved mergers

PBH mergers that are not individually resolved by the detector contribute to a SGWB. The non-observation of a SGWB during the first LVC observing runs can be used to constrain the PBHs abundance, see Ref. [170, 493–496].

From the differential merger rate $\mathrm{d}R(z)$ at redshift $z$ as found in Eq. (4.6.36), one can compute the spectrum of the SGWB of frequency $\nu$ as

$$\Omega_{\mathrm{GW}}(\nu) = \frac{\nu}{\rho_0} \int_0^{\frac{\nu_3}{\nu} - 1} \mathrm{d}z \, \mathrm{d}M_1 \, \mathrm{d}M_2 \, \frac{1}{(1+z)H(z)} \frac{\mathrm{d}R(M_1, M_2, z)}{\mathrm{d}M_1 \mathrm{d}M_2} \frac{\mathrm{d}E_{\mathrm{GW}}(\nu_s)}{\mathrm{d}\nu_s}, \quad (4.6.40)$$

where $\rho_0 = 3H_0^2 / 8\pi$ and $\nu_s = \nu(1+z)$ is the redshifted source frequency. When computing the prediction for this SGWB in a given PBH model, one should additionally include a theta function removing the mergers that would be individually resolved by the experiment (i.e. requiring the SNR to be below the threshold). This is, however, in practice, found to be a negligible correction for current GW experiments, see e.g. Ref. [12]. The additional redshift factor $1/(1+z)$ is introduced to account for the difference in the clock rates between the merger and detection times.

For the GW energy spectrum $\mathrm{d}E_{\mathrm{GW}}$ with frequency in between $(\nu, \nu + \mathrm{d}\nu)$, we use a phenomenological expression which, in the non-spinning limit[6], is given by [498, 499]

$$\frac{\mathrm{d}E_{\mathrm{GW}}(\nu)}{\mathrm{d}\nu} = \frac{\pi^{2/3}}{3} M_{\mathrm{tot}}^{5/3} \eta \times \begin{cases} \nu^{-1/3} \left[ 1 + \alpha_2 (\pi M_{\mathrm{tot}} \nu)^{2/3} \right]^2 & \text{for} \quad \nu < \nu_1, \\ w_1 \nu^{2/3} \left[ 1 + \epsilon_1 (\pi M_{\mathrm{tot}} \nu)^{1/3} + \epsilon_2 (\pi M_{\mathrm{tot}} \nu)^{2/3} \right]^2 & \text{for} \quad \nu_1 \leq \nu < \nu_2, \\ w_2 \nu^2 \frac{\sigma^4}{(4(\nu - \nu_2)^2 + \sigma^2)^2} & \text{for} \quad \nu_2 \leq \nu < \nu_3, \end{cases} \quad (4.6.41)$$

---

[6]For a study of the impact of BH spins onto the emitted GWs energy see, for example, Refs. [438, 498].

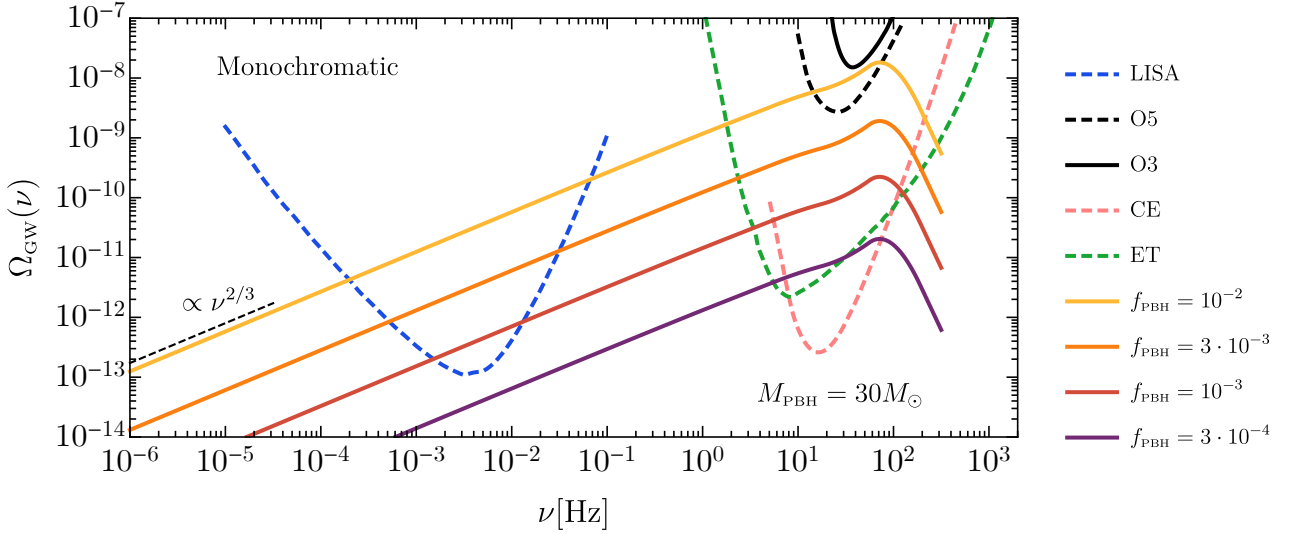

Figure 4.27: *The SGWB generated by unresolved mergers of binaries formed from a monochromatic PBH population with $M_{PBH} = 30M_\odot$. For comparison, we show the current limit from the LIGO/Virgo O3 observational run [497], along with the sensitivity curves of future detectors.*

where $\eta = m_1 m_2/M_{tot}^2$, $\alpha_2 = -323/224 + \eta\, 451/168$, $\epsilon_1 = -1.8897$, $\epsilon_2 = 1.6557$,

$$w_1 = \nu_1^{-1} \frac{[1 + \alpha_2 (\pi M_{tot}\nu_1)^{2/3}]^2}{[1 + \epsilon_1 (\pi M_{tot}\nu_1)^{1/3} + \epsilon_2 (\pi M_{tot}\nu_1)^{2/3}]^2},$$
$$w_2 = w_1 \nu_2^{-4/3} [1 + \epsilon_1 (\pi M_{tot}\nu_2)^{1/3} + \epsilon_2 (\pi M_{tot}\nu_2)^{2/3}]^2, \tag{4.6.42}$$

and

$$\pi M_{tot}\nu_1 = (1 - 4.455 + 3.521) + 0.6437\eta - 0.05822\eta^2 - 7.092\eta^3,$$
$$\pi M_{tot}\nu_2 = (1 - 0.63)/2 + 0.1469\eta - 0.0249\eta^2 + 2.325\eta^3,$$
$$\pi M_{tot}\sigma = (1 - 0.63)/4 - 0.4098\eta + 1.829\eta^2 - 2.87\eta^3,$$
$$\pi M_{tot}\nu_3 = 0.3236 - 0.1331\eta - 0.2714\eta^2 + 4.922\eta^3. \tag{4.6.43}$$

The SGWB is found to have a characteristic low frequency tail scaling like $\Omega_{GW}(\nu) \sim \nu^{2/3}$. For presentation purposes, in Fig. 4.27, we show the SGWB expected from a monochromatic PBH population and compare it to the sensitivity of current and future GW experiments. However, the constraining power on the PBH abundance from the observed merger rate of resolved sources is significantly stronger than the one potentially inferred from non-observation of a SGWB.

It is important to stress though that cross-correlating the future observations of a SGWB with the rate inferred from resolved sources may help to understand the nature of the binary BHs, being of either astrophysical or primordial origin. This is because the merger rate of PBH binaries monotonically grows with redshift, while the astrophysical one is peaking at redshift $z \sim 2$ [500–503]. Therefore, the primordial scenario predicts a larger contribution from high redshift and a stronger SGWB, given the same low redshift resolved merger rate [496].

## Part III

# Gravitational wave signatures of primordial black holes

# Chapter 5

# Gravitational waves from primordial black hole mergers

The recent detection of GWs signals coming from BH mergers made by the LIGO/Virgo Collaboration (LVC) [504, 505] has ignited a renewed interest in PBHs. This is because a population of PBHs can generate binaries in the mass range currently observable by LVC[1]. Also, it can help in explaining mergers in the so-called mass gap, a range of masses the astrophysical channels may struggle to reach. Based on the PBH model analysed in the previous chapters, where we presented an extended description of the formation of PBH binaries, alongside the prediction for their properties and merger rates, we want to compare the PBH model with current data, highlighting the potentials and shortcomings of the primordial model for explaining the origin of the observed binaries.

The final goal of this chapter is to address the following fundamental questions: *Can PBHs contribute to the LIGO/Virgo events? Do current data allow for a confident determination of such a contribution? What is the best strategy to search for PBH mergers and how can one distinguish them from astrophysical sources?* As we will see, addressing them could be too ambitious at the current stage given both the theoretical limitations and the amount of available data, but our results may indicate the path towards answering those questions in the future.

We will proceed with various steps, each one characterised by an incrementally larger complexity of the analysis. As a starting point, we will address the question of whether PBHs can produce observable binaries without violating current constraints on their abundance. To do so, we will adopt the *conservative* hypothesis (which maximises the possible overlap with constraints) that all the events detected so far have a primordial origin. We will also focus, in particular, on the GW190521 "mass gap" event, currently challenging other astrophysical formation pathways. The results show that PBH may contribute to the current GW detection without violating constraints on $f_{\rm PBH}$, a conclusion that is partially dependent on the amount of accretion PBHs experience throughout the universe evolution.

Then, we will check whether the vanilla PBH scenario can reproduce the features of the GWTC-2 catalog [507] (currently containing up to $\sim 50$ events) by making a Bayesian model comparison with a simple astrophysical phenomenological model. This will highlight some shortcomings of the vanilla PBH model while showing a strong preference for a multichannel scenario mixing both astrophysical and primordial models. Finally, we will compare (and mix) the PBH model with state-of-the-art astrophysical models (described in App. D) to perform the most comprehensive population inference on the GWTC-2 catalog including the primordial channel as a subpopulation. This analysis will show that the primordial channel can consistently explain mergers in the upper mass gap such as GW190521, but (depending on the astrophysical channels one considers) a significant fraction of the events could be of primordial origin even if one neglected GW190521. This proves that the PBH model can be competitive with astrophysical models while also being one of the leading explanations for events in the mass gap. Also, our analysis will highlight potential strategies to distinguish astrophysical and primordial origins when a larger dataset will be available in the future.

At the end of the chapter, we will have a section dedicated to the effect of priors in the interpretation of single merger events and we will also discuss the reach of future third-generation (3G)

---
[1]For early speculations on this possibility, see Ref. [506].

detectors such as the Einstein Telescope and Cosmic Explorer when searching for PBHs.

## 5.1 The LIGO/Virgo observations

Even though the LIGO experiment has started taking data already in the year 2002, only 5 years have passed since the very first detection of a GWs signal coming from a binary BH system called GW150914 [508]. Since then, many more detections were made and the era of GW astronomy has begun [270].

Three different stages of the LIGO/Virgo experiments have been performed already (O1, O2 and O3a), with more data from the O3b run about to be released at the time of writing. The latest catalog of compact binary coalescence sources published by LVC reported up to 39 events, most of which are confidently interpreted as binary BHs [274], bringing the number of detections to 47 [507]. We will not consider in our analysis other events independently found by other research groups as they usually have a lower statistical significance (see e.g. [509–511] and references therein).

### 5.1.1 The GWTC-2 catalog

We show a summary plot of the binary parameter for each event detected so far by LVC in Fig. 5.1. The key features of the dataset can be summarised with the following list of points (see Ref. [507] for more details.

- The overall local merger rate density was estimated in Ref. [507] to be $R(z = 0) = 23.9^{+14.3}_{-8.6} \mathrm{Gpc}^{-3} \mathrm{yr}^{-1}$ assuming a constant rate in redshift and a "POWER-LAW + PEAK" parametrisation of the mass function. A lower figure is obtained if the rate is allowed to evolve following a power law $(1+z)^\kappa$, finding $R(z = 0) = 19.3^{+15.1}_{-9.0} \mathrm{Gpc}^{-3} \mathrm{yr}^{-1}$.

- There are events with individual masses above the lower edge of the pair-instability mass gap expected from supernova theory (around $\approx 45 M_\odot$, even though its precise location is still uncertain [513–528]). In particular, the events GW190521 [512], GW190602 and GW190519 have the largest masses in the catalog, and show there is a sizeable merger rate in the mass range $45 M_\odot < m_1 < 100 M_\odot$ (estimated to be around $0.71^{+0.65}_{-0.37} \mathrm{Gpc}^{-3} \mathrm{yr}^{-1}$). This massive portion of the catalog will be of particular interest for the PBH model, as we will discuss in details.

- Even though there are large uncertainties in the estimation of the mass ratio of BH binaries, most of the events has a $q$ compatible with unity. There are two noticeable exceptions given by GW190814 [529] and GW190412 [443]. The former have a secondary mass $m_2 = 2.59^{+0.08}_{-0.09} M_\odot$ potentially interpreted as a neutron star. We will come back to this point in the following. The latter instead have a mass ratio $q = 0.28^{+0.13}_{-0.07}$. The interpretation of these events, however, may change if a prior which is different from the uninformative one adopted by LVC was used [530–532]. We will discuss whether the assumption of a PBH motivated prior is consistent with the event properties in Sec. 5.3.

- The merger rate evolution with redshift is still poorly constrained, being the current detector's reach limited to $z \lesssim 1$. There is, however, a mild preference for a merger rate growing with redshift (85% credibility [507]), with a growth index constrained at $\kappa = 1.3 \pm 2.1$. This result is fully compatible with the monotonic growth of the PBH merger rate at high redshift, even though current measurements are extremely limited in this respect, as the furthest event observed is GW190706_222641 with a source redshift $z = 0.71^{+0.32}_{-0.27}$.

- The majority of signals are characterised by very low spins. There are, however, 9 out of 44 events with a $\chi_{\mathrm{eff}}$ incompatible with 0 and an indication that some events have spins misaligned with the binary's angular momentum (compatible with the PBH scenario predicting spins evenly distributed in all directions).

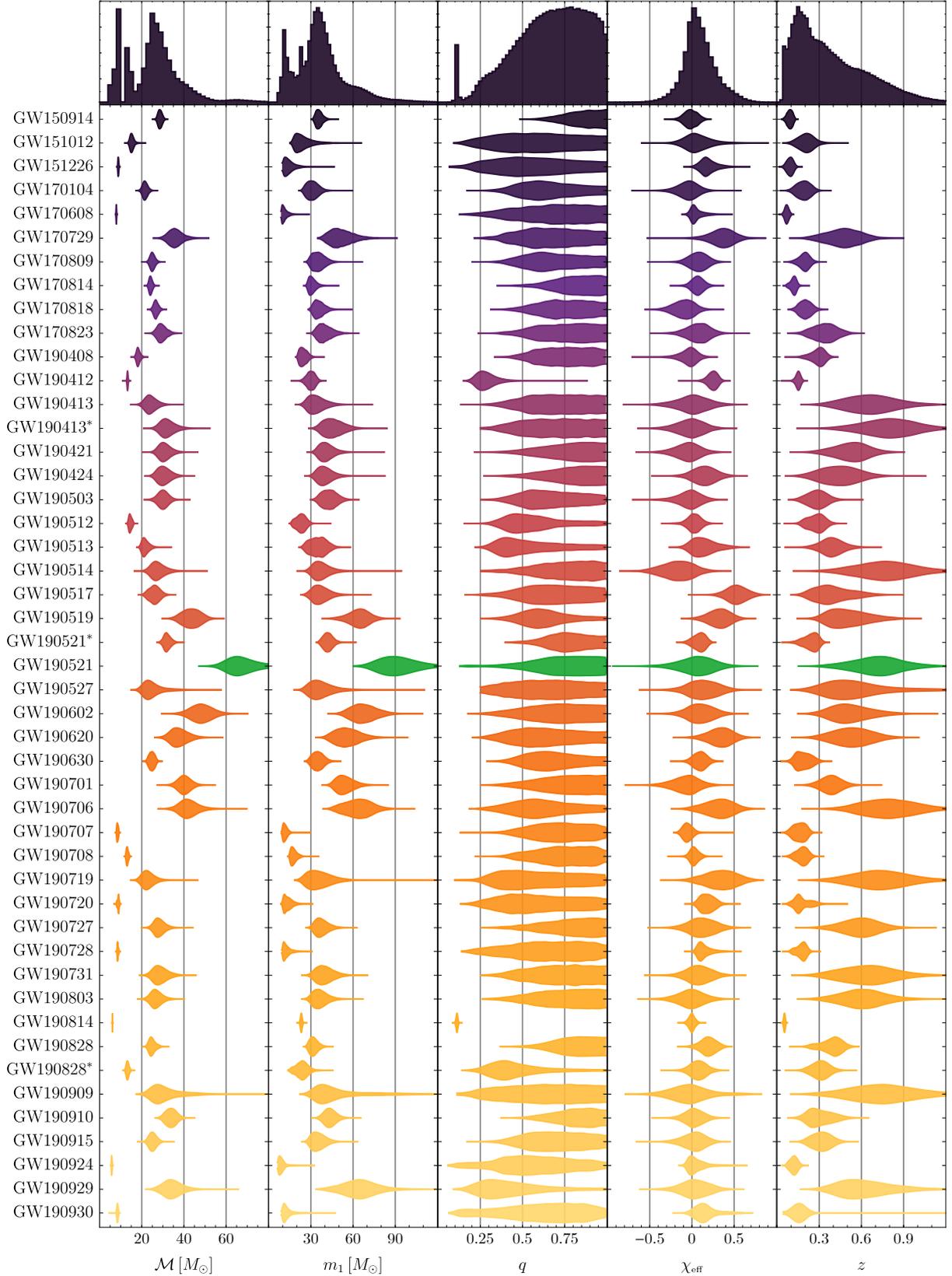

Figure 5.1: *Posterior distributions for the key quantities* $(\mathcal{M}, m_1, q, \chi_{\text{eff}}, z)$ *of the events in the GWTC-1/2 catalogs. Figure made using the data release associated with Refs. [274, 439]. In green we highlight the mass gap event GW190521 [512]. The combined distributions are shown in the top row. The labels with an asterisk correspond to GW190413\*=GW190413\_134308, GW190521\*=GW190521\_074359 and GW190828\*=GW190828\_065509, respectively.*

- There is evidence of features in the primary mass distribution beyond a simple power-law parametrisation. Ref. [507] found that that a power-law plus a Gaussian peak was the phenomenological model which best reproduced the observed data, decisively preferred compared to the power-law model with a Bayes factor of $\log \mathcal{B} = -1.91$.[2] This may hint at multiple populations giving rise to the observed events.

In our analysis, we include the same subset of events selected for the population analysis done by LVC in Ref. [507]. In particular, we exclude events with a large false-alarm rate (GW190426, GW190719, GW190909). Also, there are events with secondary masses $m_2 < 3 M_\odot$ which requires careful treatment. An electromagnetic counterpart was detected and identified with a kilonova only for GW170817 [533]. This confidently shows that at least one of the binary components (and most likely both) is a neutron star. However, two events (GW190425 and GW190814) were not associated with any electromagnetic counterpart, and it is more uncertain to confirm whether their light components are neutron stars instead of, potentially, astrophysical or primordial BHs [529, 534]. Consistently with the LVC population analysis [507], we anyway excluded both events from the dataset. We note that their inclusion is not expected to change our main result while a dedicated analysis would be an interesting extension of our work. [3]

### 5.1.2   The BH binary detectability

When performing a population analysis, it is of crucial importance to account for the detection bias of the experiment [535, 536]. In this section, we present the formalism to compute the probability to observe the GW signal from a BH merger at the LIGO/Virgo experiments.

A compact quasi-circular BH binary is fully characterized by the BHs masses $m_1$ and $m_2$, dimensionless spins $\chi_1$ and $\chi_2$, and by the merger redshift $z$. For simplicity, we will neglect the impact of spins in the BH binary detectability, since the large majority of the GWTC-2 events are compatible with zero spin. Each binary has also a specific position and orientation with respect to the detectors. Those are typically expressed in terms of right ascension $\alpha$, declination $\delta$, orbital-plane inclination $\iota$, and polarization angle $\psi$. To summarise, we will refer to $\theta_i = [m_1, m_2, z]$ as the intrinsic while $\theta_e = [\alpha, \delta, \iota, \psi]$ as the extrinsic binary parameters, respectively.

Given a binary with intrinsic parameters $\theta_i$, one can compute the detectability (i.e. the probability that the detector will be able to discover the GWs coming from its associated merger event) by averaging over the extrinsic parameters $\theta_e$. This amounts to compute detection probability by solving the integral

$$p_{det}(\theta_i) = \int p(\theta_e) \, \Theta[\rho(\theta_i, \theta_e) - \rho_{thr}] \, d\theta_e \,, \tag{5.1.1}$$

where $p(\theta_e)$ is the probability distribution function of $\theta_e$, $\Theta$ indicates the Heaviside step function and $\rho$ is the Signal-to-Noise-Ratio (SNR). In the case of the GWTC-1 catalog, it was shown that $p_{det}$ can be computed using the single-detector semianalytic approximation introduced in Refs. [537, 538] with a SNR threshold $\rho_{thr} = 8$. This procedure does not lead to significant departures from the large-scale injection campaigns performed by LVC for the O1 [539] and O2 [540] runs. We will also adopt this approach when computing the detectability of binaries during the O3 run.

In the case of a single detector, non-precessing sources and focusing on the dominant quadrupole moment only, one can factor out the SNR dependency on $\theta_e$ to obtain $\rho(\theta_i, \theta_e) = \omega(\theta_e)\rho_{opt}(\theta_i)$, where $\rho_{opt}$ is the SNR of an "optimal" source located overhead the detector with face-on inclination. One can, therefore, compute $p_{det}(\theta_i)$ by evaluating the cumulative distribution function

$$P(\omega_{thr}) = \int_{\omega_{thr}}^1 p(\omega') d\omega' \tag{5.1.2}$$

---

[2]For clarity, we note that with "log" we indicate the decimal logarithm while with "ln" the natural logarithm.

[3]We note that the Bayesian framework we adopt(see App. C), at variance with Ref. [476], fully accounts for the correlation between the intrinsic event parameters (such as masses and spins) in the posterior data and it is not sensitive to the particular choice of single-event priors used by the LVC when performing the parameter estimation. As shown in Refs. [530–532] (and in Refs. [10] for the PBH scenario), the priors may modify the interpretation of some events, also possibly affecting the maximum likelihood analysis as performed in Ref. [476]. Finally, contrarily to Ref. [476], we also include the spin information in the inference.

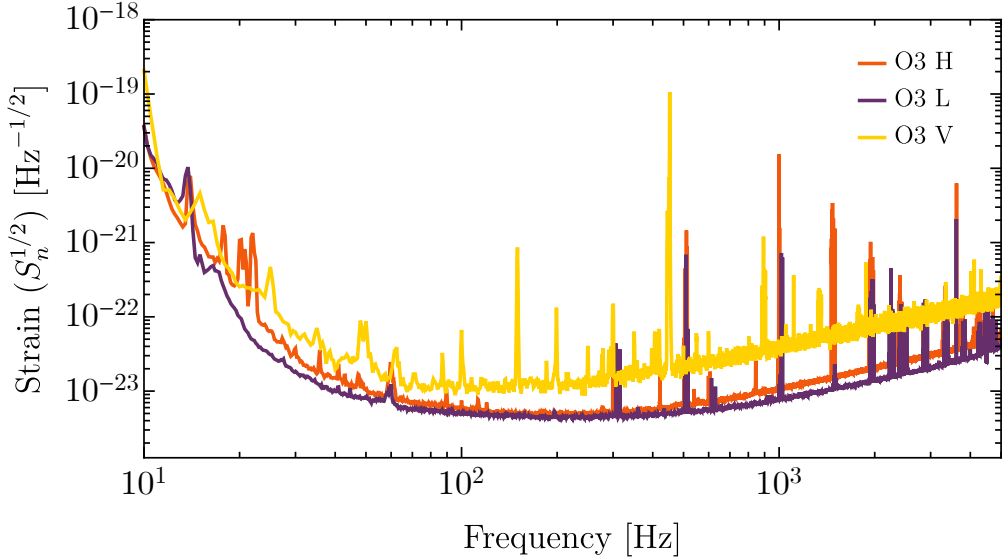

Figure 5.2: *Noise curves of the LIGO Hanford (H), Livingston (L) and Virgo (V) experiments during the O3 run [274].*

at the point $\omega_{\mathrm{thr}} = \rho_{\mathrm{thr}}/\rho_{\mathrm{opt}}(\theta_{\mathrm{i}})$ which is only a function of the optimal SNR and the required threshold. In this work we consider the case of isotropic sources for which $\alpha, \cos\delta, \cos\iota$, and $\psi$ are uniformly distributed on each of their physically allowed ranges. The resulting function $P(\omega_{\mathrm{thr}})$ can be found in Ref. [541].

The optimal SNR of individual GW events is given in terms of the GW waveform in Fourier space $\tilde{h}(\nu)$ by [273]

$$\rho_{\mathrm{opt}}^2(m_1, m_2, z) \equiv \int_0^\infty \frac{4|\tilde{h}(\nu)|^2}{S_n(\nu)}\mathrm{d}\nu, \tag{5.1.3}$$

where $S_n$ is the strain noise. The plot of $S_n$ as a function of frequency during the current LIGO/Virgo observation run can be found in Fig 5.2. In particular, when computing the binary detectability during the O1-O2 (O3a) observation runs, we followed Ref. [7] and adopted `aLIGOEarlyHighSensitivityP1200087` (`aLIGOMidHighSensitivityP1200087`) as the detector noise power spectral densities, as implemented in the publicly available repository `pycbc` [542]. Consistently with the computational framework described to compute the binary detectability, we also adopt the non-precessing waveform model IMRPhenomD [543, 544]. We show two examples of GW signals in Fig. 5.3. Notice, finally, that the optimal SNR scales like the inverse of the luminosity distance, i.e. $\rho \propto 1/D_{\mathrm{L}}(z)$.[4]

In the section dedicated to 3rd generation detectors such as the Einstein Telescope and Cosmic Explorer, we will estimate the detector bias with the same procedure described here, while adopting their respective design sensitivity curves.

## 5.2  Primordial black hole mergers at LIGO/Virgo

As described at the beginning of this chapter, we will confront the PBH model with the LVC data at various levels of complexity. The starting point will be assuming that all the observed events may be ascribed to the PBH channel. This will allow us to maximise the potential overlap with current constraints on the PBH abundance. We will refer to this as the "single PBH population scenario".

We will also focus, in particular, on the mass gap event GW190521 challenging the astrophysical formation scenarios and test whether this event can be produced by the PBH channel with a rate compatible with the available observations. Then, to test how well the PBH model can explain the GWTC-2 events, we will compare it to both an astrophysical phenomenological model and the ab-initio astrophysical isolated and dynamical channels (reviewed in App. D). Finally, we will also consider mixed astrophysical and primordial scenarios.

---

[4]Recent work introduced machine learning techniques to reduce the computational time required to estimate the detector response [545, 546]. Future analysis will require the adoption of these techniques to make the computational run manageable when the number of individual detections will be orders of magnitude larger than what is currently [283].

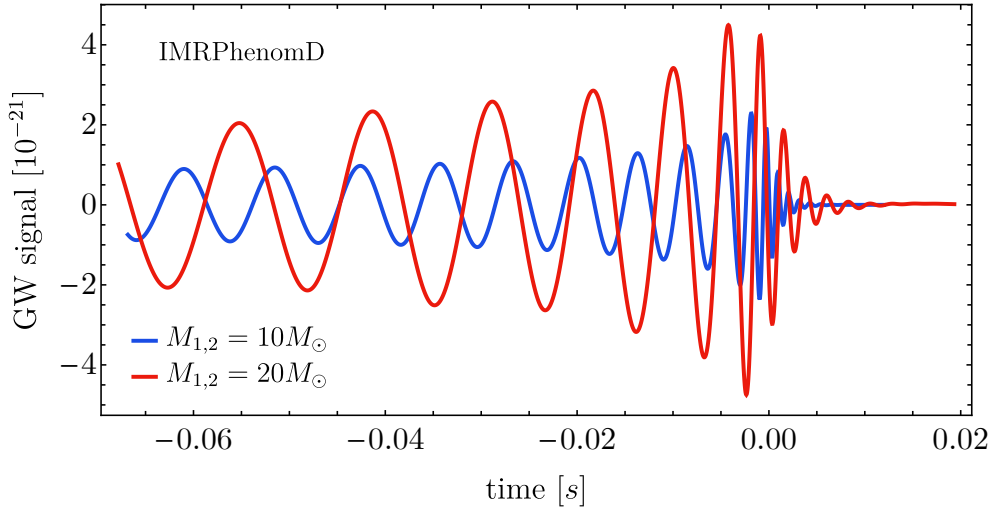

Figure 5.3: *Example of the GW signal coming from the merger of two equal mass BHs located at a distance of $D_L = 100$Mpc (or redshift $z = 0.022$). The plot was done using the IMRPhenomD waveform model [543, 544] we consider in this work to compute the SNR at the LVC experiments.*

In all the following analyses, we will consider the lognormal parameterisation of the PBH mass function, as introduced in Sec. 3.1, which we repeat here for clarity

$$\psi(M) = \frac{1}{\sqrt{2\pi}\sigma M} \exp\left[-\frac{\ln^2(M/M_c)}{2\sigma^2}\right], \tag{5.2.1}$$

where $M_c$ is the central mass scale and $\sigma$ its width. Therefore, the hyperparameters describing the PBH model are $\lambda_{\text{PBH}} = [M_c, \sigma, f_{\text{PBH}}, z_{\text{cut-off}}]$, with $f_{\text{PBH}}$ being the PBH abundance and $z_{\text{cut-off}}$ characterising the accretion model, see Sec. 4.1. Finally, the PBH merger rate parametrisation is provided in Eq. (4.6.36).

To constrain model parameters and perform model comparisons, we adopt a hierarchical bayesian analysis as presented in detail in App. C, where all the relevant event parameters, as well as the population parameters involved, are summarised.

### 5.2.1 Single PBH population scenario

In this section, we present the results based on the GWTC-2 analysis under the assumption that the PBH model is the only binary formation channel at play. The primary aim of this section is to understand if the best fit PBH population would be allowed by current constraints and if its characteristic features may be compatible with the observed set of events. We will closely follow Refs. [12] and [7], where this analysis was carried out focusing on the GWTC-1 and GWTC-2 catalog, respectively.

The result of the inference is shown in Fig. 5.4, while the 90% C.I. for the PBH parameters are shown in Tab. 5.1. We highlight here that the present analysis is characterised by several improvements with respect to the existing literature. Indeed, many attempts to perform population inference neglected the role of accretion (see e.g. [170, 476, 485, 547, 548]), which was shown to be relevant [12] and to have an impact on both the mass and spin distributions and the overall merger rate, see dedicated discussion in Chapter 4. We limit the cut-off redshift to be as high as $z_{\text{cut-off}} \simeq 30$, roughly corresponding to the case in which accretion becomes negligible in the mass range of interest. In that case, and restricting on the GWTC-1 dataset, our results are in general agreement with previous works. It is also important to stress that here, for the first time, the accretion hyperparameter $z_{\mathrm{cut\text{-}off}}$ was allowed to vary and we inferred its posterior distribution.

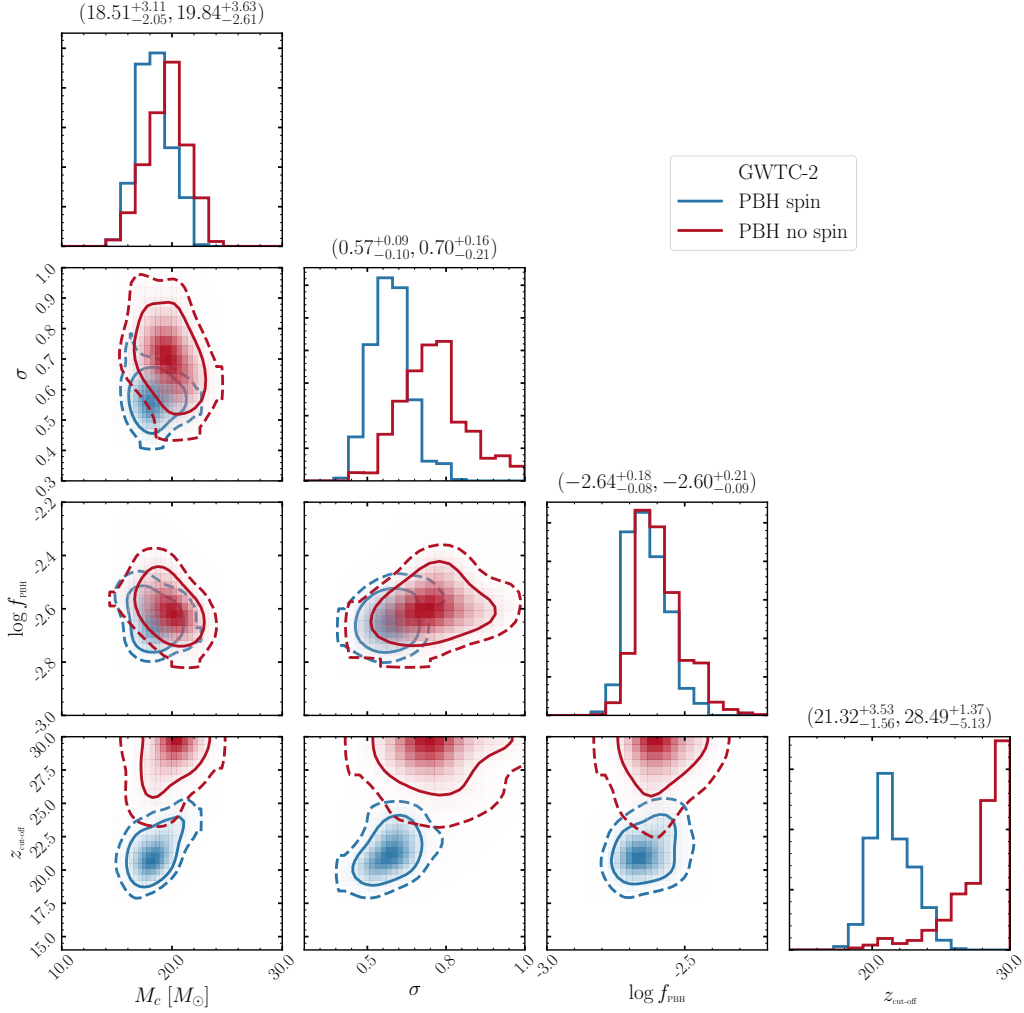

Figure 5.4:  *Population posterior obtained with an inference on the GWTC-2 dataset, under the assumption all events have a primordial origin.  Blue lines are obtained using four observables ($m_1$, $m_2$, $\chi_{\mathrm{eff}}$, $z$), whereas red lines corresponds to the result omitting the effective spin information in the inference.  The first (second) number on top of each column indicate 90% C.I. obtained by including (omitting) spin information.*

Table 5.1:  *Hyperparameters of the PBH model inferred from GWTC-2 data assuming only the PBH channel with a lognormal initial mass function is at play.*

| Parameter | $M_c\,[M_\odot]$ | $\sigma$ | $\log f_{\mathrm{PBH}}$ | $z_{\mathrm{cut\text{-}off}}$ |
|---|---|---|---|---|
| 90% C.I. | $18.51^{+3.11}_{-2.05}$ | $0.57^{+0.09}_{-0.10}$ | $-2.64^{+0.18}_{-0.08}$ | $21.32^{+3.53}_{-1.56}$ |

We show a comparison with the results coming from neglecting the spin information in the inference in Fig. 5.4. The most striking difference between the two results is the change in the posterior of $z_{\mathrm{cut\text{-}off}}$. The other parameters adjust following their correlation with $z_{\mathrm{cut\text{-}off}}$. When $\chi_{\mathrm{eff}}$ is neglected in the inference, the mass distribution of the events in the catalog favours high values of $z_{\mathrm{cut\text{-}off}}$, making accretion less relevant. However, when $\chi_{\mathrm{eff}}$ information is included in the inference, the posterior of $z_{\mathrm{cut\text{-}off}}$ gets narrower and peaks at smaller values. This is because various events in O3a have effective spin incompatible with zero, see Fig. 5.1, and accretion is necessary to spin up PBHs. Correspondingly, the best-fit value of $M_c$ (the characteristic PBH mass at formation) decreases only slightly, while the posterior of $\sigma$ (the initial width of the PBH mass function), gets narrower and peaks at a smaller value, because accretion both shifts and broadens the mass function. As expected, the PBH abundance is relatively stable with respect to changes of the other hyperparameters. The presence of some accretion enhances the rate, which is therefore related to a small decrease of $f_{\rm PBH}$. Overall, we find $f_{\rm PBH} \simeq 3 \cdot 10^{-3}$, indicating that the PBH population inferred from the data comprise at most a subpercent fraction of the totality of the dark matter in our universe.

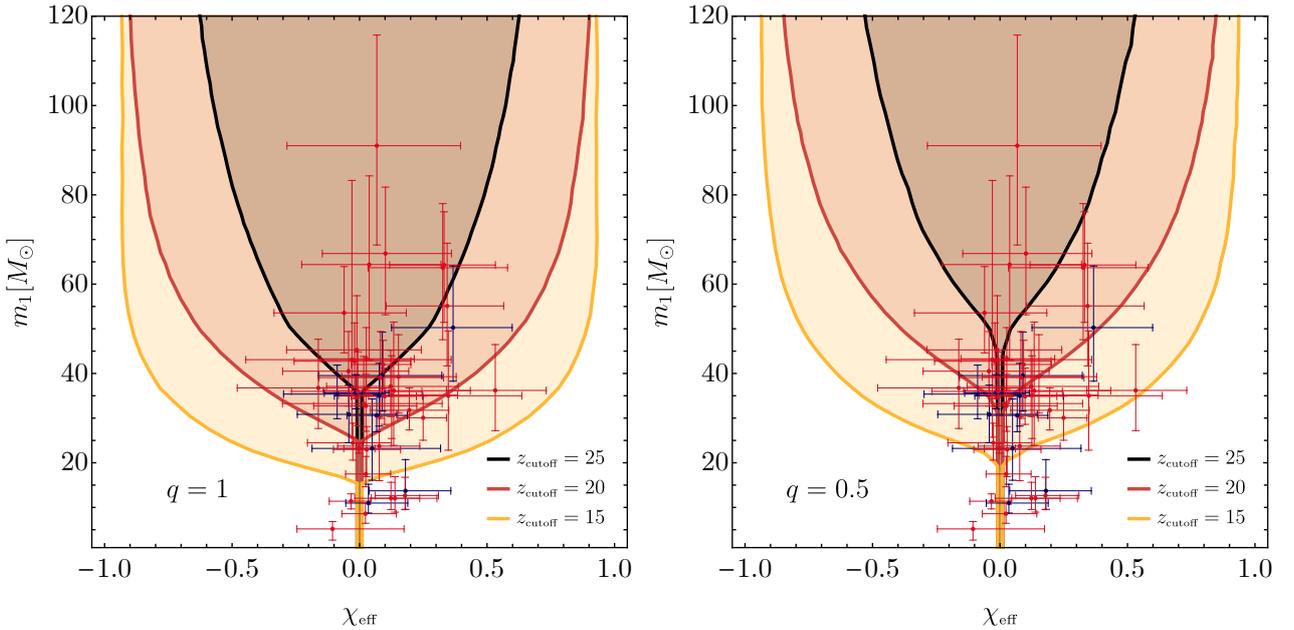

Figure 5.5: *Prediction for the $\chi_{\rm eff}$ distribution as a function of the primary BH mass $m_1$ and cutoff redshift, at $2\sigma$ C.L., for the best PBH scenario inferred from the GWTC-2 dataset. In blue (red) we show the events from the O1/O2 (O3a) catalog. Plot analogous to Fig. 4.16, but done with data driven PBH population parameters. Left: $q = 1$. Right: $q = 1/2$.*

We can investigate further how spin information is crucial in determining $z_{\rm cut\text{-}off}$ by looking at Fig. 5.5, where we show the $2\sigma$ C.I. of the effective spin parameter distribution predicted by the PBH model in terms of the primary component mass. We show different values of the mass ratio and various cutoff redshifts around the $2\sigma$ range obtained from best-fitting the primordial scenario. As predicted by the PBH model, see Sec. 4.3, the individual spin directions with respect to the total angular momentum are isotropic and independently distributed. Therefore, we averaged over all possible angles. For larger binaries' total masses, accretion is stronger and PBH spins grow up to increasingly larger values compared to the one inherited at formation. This only happens after a given threshold which is dependent on the accretion hyperparameter $z_{\rm cut\text{-}off}$. The distributions shown in Fig. 5.5, therefore, highlight the transition from initially vanishing spins to larger values depending on both the binary mass and accretion strength. We also show in blue data from the GWTC-1, while in red new detections from the O3a run, as reported by the LVC analyses of the strain data with agnostic priors. Therefore, we see a $z_{\rm cut\text{-}off}$ around 20 is necessary to reach the few spinning and moderately massive events in the GWTC-2 catalog.

In Fig. 5.6, we show the intrinsic distribution of the most relevant binary parameters, i.e. primary mass $m_1$, mass ratio $q$, effective spin $\chi_{\rm eff}$, and precession spin $\chi_{\rm p}$, as inferred from our best-fit model. For visual comparison, we show the marginalized posterior distribution for the primary mass $m_1$ in the PBH scenario along with the corresponding preferred result found in Ref. [507] assuming a "Power-law + Peak" mass function. We stress that Fig. 5.6 reports the intrinsic population distribution, which does not account for selection effects. Those are not the observable distributions accounting for the fact that current detectors favour the observation of larger masses. A striking property of the PBH model, at least in the vanilla scenario characterised by a lognormal initial mass distribution, is that it is not able to produce the multiple features observed in the primary mass distribution. This will serve as a motivation to extend the present analysis considering a mixture of astrophysical and primordial models, as we will see.

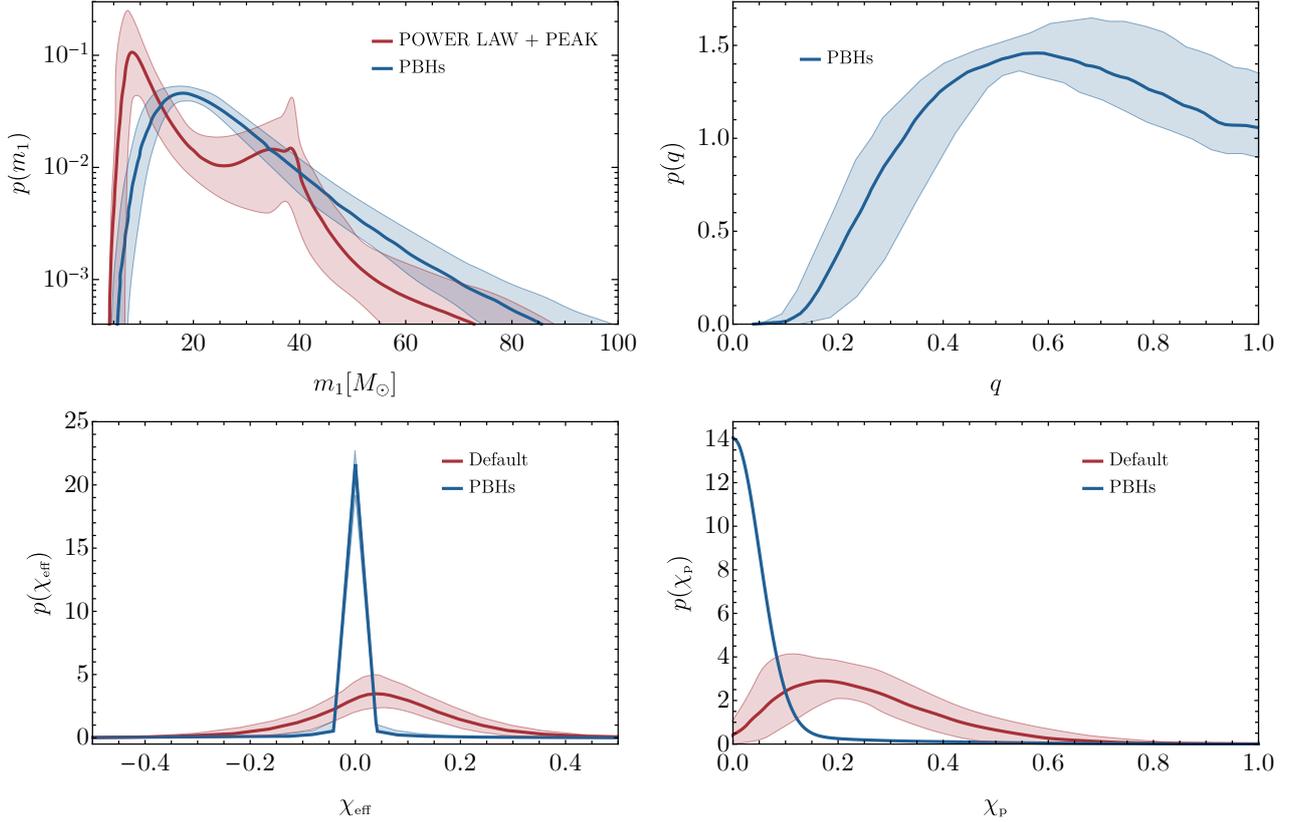

Figure 5.6: ***Top-left:*** *Distribution of the primary mass $m_1$;* ***Top-right:*** *mass ratio $q$;* ***Bottom-left:*** *effective spin $\chi_{\rm eff}$;* ***Bottom-right:*** *precession spin $\chi_{\rm p}$. The distributions in blue correspond to the best-fit PBH model including $\chi_{\rm eff}$ (but not $\chi_{\rm p}$) in the inference. For visual comparison, we show the 90% C.L. distributions found by the LVC in Ref. [507].*

On the top right, we plot the marginalized distribution for the mass ratio in the PBH case. The distribution peaks at a relatively small mass ratio, $q \sim 0.5$ due to the preference for a relatively wide mass function and large values of $z_{\rm cutoff}$. A lower cut-off, related to stronger accretion, would lead to more symmetric binaries [12, 14]. On the bottom panels, we plot the marginalized distributions for the effective spin parameter $\chi_{\rm eff}$ (left) and the precession spin $\chi_{\rm p}$ (right), defined as

$$\chi_{\rm p} = \max\left[\chi_1 \sin\theta_1, \left(\frac{4q+3}{4+3q}\right) q\chi_2 \sin\theta_2\right] \tag{5.2.2}$$

where $\theta_1$ and $\theta_2$ are the angles between the two spin directions and the binary's angular momentum. The latter parametrizes the spin components perpendicular to the binary angular momentum responsible for the precession of the orbital plane. We also show in red the so-called "Default" model adopted in the population analysis performed by LVC, see Appendix D.1 of Ref. [507]. In both cases, the probability distributions inferred from the PBH model show a narrow peak around zero since the best-fit PBH mass function is dominated by relatively small masses, which are correlated with small spins. This explains the difference with the "Default" model, for which masses and spins are not correlated, giving rise to a peak at non-vanishing spins and a broader distribution.

Finally, we compare the evolution of the PBH merger rate density $R(z)$, as described by Eq. (4.6.36), with the power-law evolution model for astrophysical sources found in Ref. [507]. Even though with large uncertainty bands, the latter is slightly less steep than the $R(z) \propto (1+z)^{2.7}$ behaviour predicted from the star formation rate [549]. Overall, the measured merger rate evolution is compatible with the PBH scenario, which predicts fewer mergers compared to the stellar-origin scenario in the high redshift side of the LVC horizon.

Let us stress, however, that the PBH scenario predicts a merger rate that increases monotonically at redshifts beyond the LVC horizon, while in the astrophysical case the rate is expected to decrease soon after $z \approx 2$, where the star formation rate is expected to peak. There may be a secondary peak at higher redshift due to astrophysical Population III binaries. We will come back to this point in the following sections. This difference is of crucial importance for 3G detectors, which may be able to detect mergers up to redshift $z \simeq 50$ [471, 550]. We will come back to this point in Sec. 5.4 dedicated to 3G detectors.

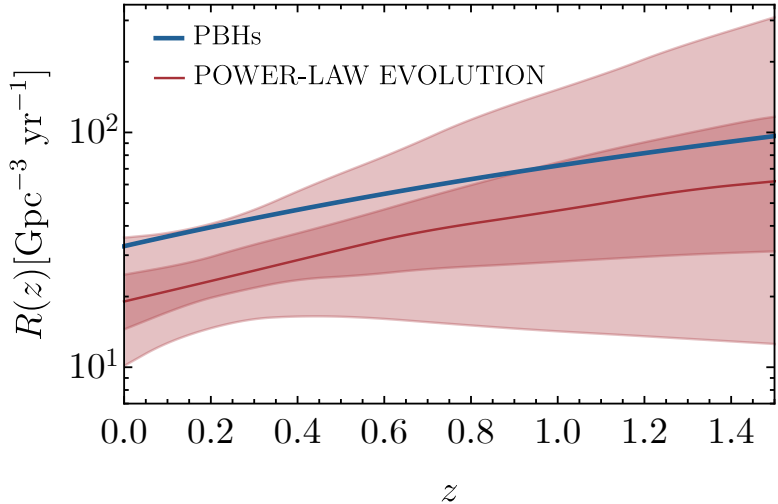

Figure 5.7: *PBH merger rate density evolution given by Eq.* (4.6.36) *for the best-fit population (blue line). We also show in red the 50% (90%) C.I. for the power-law inferred merger rate evolution found by the LVC [507] population analysis.*

**Confrontation with PBH constraints**

In Fig. 5.8 we show the most relevant constraints on the PBH abundance in the mass range of interest (see detailed discussion in Sec. 1.2.2) and compare them to the population inferred from the GWTC-2 dataset. We recall that we re-analyse all constraints taking into account both the width of the PBH mass function and the effect of accretion as described in Sec. 4.2.2.

By looking at Fig. 5.8, we see that if interpreted as coming from the PBH scenario, the GWTC-2 events seem to be in tension with Planck D. However, one should recall that the Planck D constraint is derived assuming thin disk accretion taking place already at high redshift $300 \lesssim z \lesssim 600$ when CMB anisotropies are induced by the electromagnetic emission from accreting PBHs. This is in contrast with the argument presented in Sec. 4.1.1, following Ref. [234], dictating that the formation of a thin accretion disk can only take place at lower redshift. Therefore, at such a high redshift, spherical accretion is regarded as more realistic. The constraint derived with this assumption (Planck S), is compatible with GWTC-2.

We also stress again here that the constraint from the NANOGrav 11-yr data from Ref. [191] has large systematic uncertainties, above all in their choice of the threshold $\zeta_c = 1$ for PBH formation (where $\zeta$ is the curvature perturbation responsible for the creation of PBHs upon collapse). In order to account for these uncertainties, we also plot the constraint derived choosing a threshold $\zeta_c = 0.6$ motivated by state-of-art numerical simulations [226] (see also the discussion in Ref. [198]). As one can appreciate, the exponential dependence of the PBH abundance on the threshold makes the constraint relax by many orders of magnitude. Given the uncertainties discussed above, in Fig. 5.8 we have decided to show the Planck D and the NANOGrav constraints ($\zeta_c = 1$) by dashed lines without filling the corresponding excluded region.

We conclude that the assumption that all the events in the GWTC-2 catalog are originated from PBHs is not in contrast with current observational constraints. However, one should also address the question of whether the PBH model is providing a competitive fit of the data. This can only be quantified by comparing it with other astrophysical binary BH populations to check whether the PBH model can reproduce the features of the mass distribution which starts to emerge in the larger

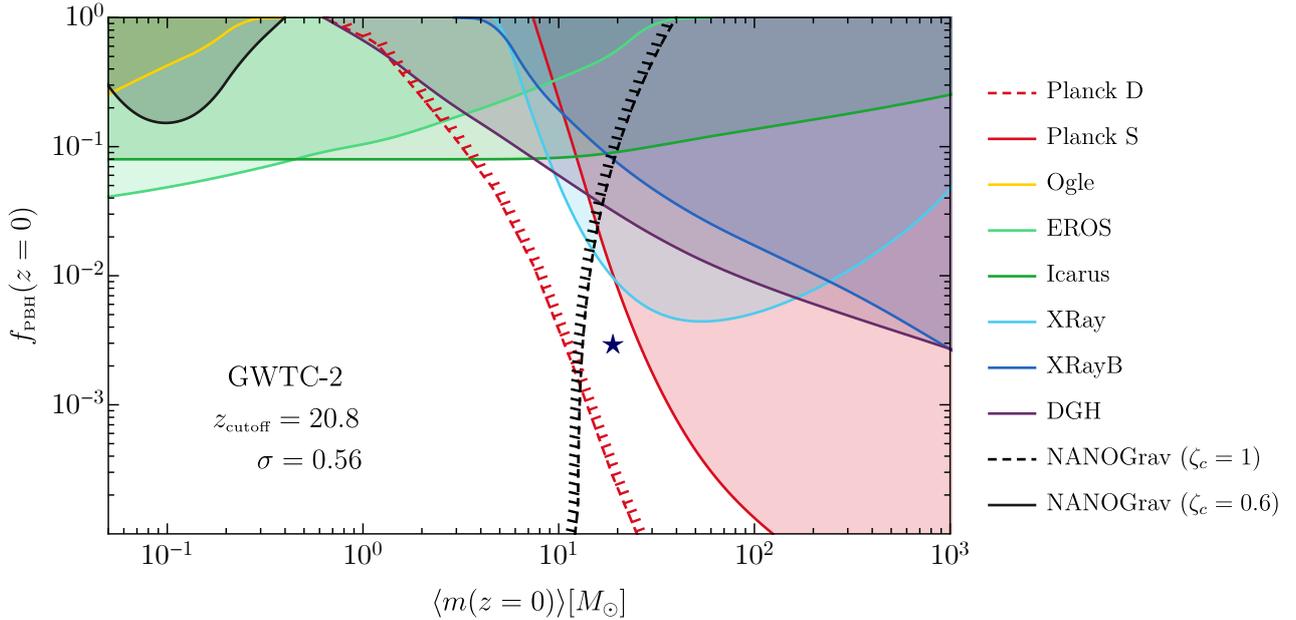

Figure 5.8: *Summary of the most stringent constraints on PBH abundance $f_{\mathrm{PBH}}(z=0)$ as a function of the mean PBH mass $\langle m(z=0)\rangle$, both defined at low redshift. The blue star indicates the median values for the population parameters $[M_c, \sigma, f_{\mathrm{PBH}}, z_{\text{cut-off}}]$ inferred with the GWTC-2 dataset. As discussed in the main text, we have indicated by a red (black) dashed line the bounds from Planck D (NANOGrav) which carry significant uncertainties, while the dashes perpendicular to the line indicate the excluded portion of the plot.*

GWTC-2 dataset. We will address those questions in the next sections. Furthermore, it would be interesting to extend this analysis by considering different PBH mass functions and improved accretion models.

### 5.2.2   The mass gap event GW190521

In this section, we focus our attention on the most massive event observed so far by the LIGO/Virgo collaboration in light of its possible explanation as a primordial binary BH. The discussion will closely follow Ref. [11].

The recent discovery of GW190521 [512], the GW signal originated from the coalescence of two BHs with masses $m_1 = 85^{+21}_{-14}M_\odot$ and $m_2 = 66^{+17}_{-18}M_\odot$, see Fig. 5.1, has focused even more attention on the PBH scenario. This is because astrophysical models are not expected to produce BHs with masses in the so-called mass gap expected between about $\approx 45M_\odot$ and $\approx 135M_\odot$. Such a gap in the allowed range of BH masses produced by astrophysical formation scenarios is due to the pulsational pair-instability phenomenon. If the mass of the progenitor star is large enough, its core may ignite the production of electron-positron pairs at high temperatures. This produces a series of violent contractions and expansion forcing the star to lose mass through solar winds, depending on the metallicity, and, therefore, reducing the final mass of the remnant BH. This may happen in stars with a helium core mass in the range $\sim (32 \div 64)M_\odot$. However, stars with a mass above $\gtrsim 200\ M_\odot$ and very low metallicity (typical of population III) can overcome the pulsation pair instability and form a population of intermediate mass BHs with masses above $\sim 135M_\odot$ [513–515, 517–523]. See Refs. [528, 551–554] for additional discussions on the astrophysical uncertainties of this prediction.

Let us mention, however, that it is not excluded that BHs within the mass gap might have an astrophysical origin due to hierarchical coalescences of smaller BHs [450, 489, 490, 555–558], via direct collapse of a stellar merger between an evolved star and a main sequence companion [559, 560], by rapid mass accretion in primordial dense clusters [561], or by beyond Standard Model physics that reduces the impact of pair instability [551].

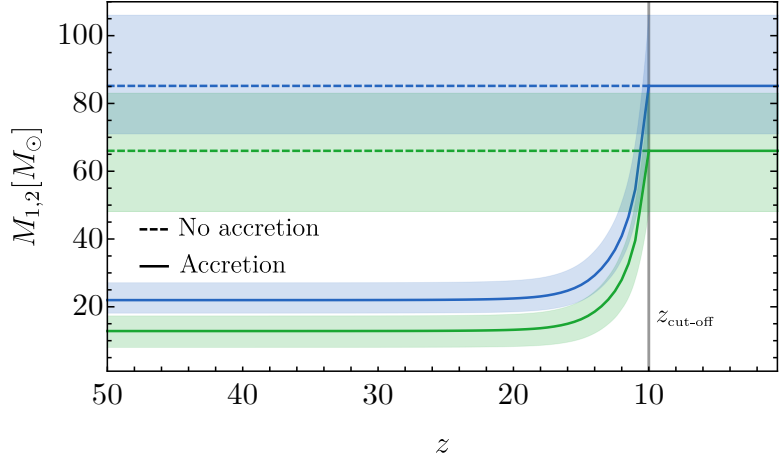

Figure 5.9: *Evolution of the PBH masses for GW190521 without and with accretion ($z_{cut\text{-}off} = 10$). Green (blue) bands correspond to the primary (secondary) PBH mass within 90% C.I. as reported by the LVC parameter estimation.*

The properties of GW190521 are compatible with the primordial scenario. First of all, such a massive binary can be easily produced as a result of the collapse of density perturbations in the early universe, see Chapter 2. Indeed, PBHs can populate the mass gap. Also, if accretion is relevant for PBHs, the mass ratio should be close to unity and the individual spins are expected to be non-vanishing, with the spin of the secondary component always bigger than the one of the primary [12]. Even though no firm conclusion can be drawn due to the small SNR and the large systematic uncertainties coming from waveform modelling [562], this pattern seems to agree with the parameters measured for GW190521. The strain data show evidence for large individual spins, likely lying on the orbital plane. While sizeable spins would be in tension with the non-accreting scenario, they are easily explained when accretion is efficient. Also, as discussed in particular in Sec. 4.1.2, since the Bondi radii of the individual PBHs in the binary are comparable to the characteristic orbital distance, the accretion flow's geometry is complex: the orientations of individual accretion disks are independent and randomly distributed irrespective of the direction of the orbital angular momentum. Thus, the inferred values of the binary components' spins are consistent with the accreting PBH scenario.

To further assess whether GW190521 may have a primordial origin, one has to test if the PBH abundance $f_{PBH}$, necessary to reach the corresponding merger rate inferred by the LVC collaboration, is allowed by the current experimental constraints [199]. We will perform such an analysis for two opposite scenarios:

*i)* the PBH channel is only responsible for the single event GW190521;

*ii)* all the detections by LIGO/Virgo (including GW190521) are originated by PBH mergers, and the corresponding mass function is determined as in the previous section.

In the first case, to bracket the uncertainties on the accretion model, we consider both the case where accretion is completely negligible and the case with efficient accretion $z_{cut\text{-}off} \approx 10$. As the number of mass gap events is expected to increase in the future, with more data one will be able to perform a more proper population study on multiple mass gap events. In the second case, on the other hand, we will infer the value of $z_{cut\text{-}off}$ from the GWTC-2 catalog, as performed in the previous section.

In Fig. 5.9, we show the evolution of both masses in a GW190521-like binary. If accretion with $z_{cut\text{-}off} \approx 10$ is present, the high redshift masses would be $M_1^i \approx 22 M_\odot$ and $M_2^i \approx 13 M_\odot$, respectively. [5]

---

[5] We note that PBH accretion, including relativistic effects, was specifically studied in Ref. [563]. Even though different assumptions compared to the accretion model considered here were taken, the initial masses leading to a GW190521-like event were found to be close to the one reported here.

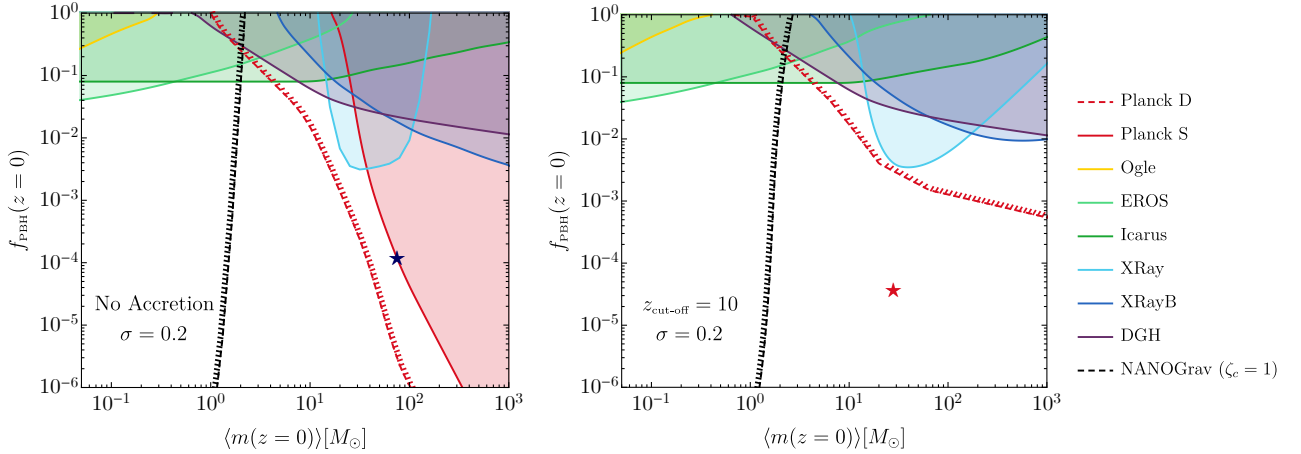

Figure 5.10: *Same as Fig. 5.8 but for the case of a single primordial GW190521-like event.* **Left:** *Scenario with irrelevant accretion. The blues star indicates the PBH abundance needed to explain GW190521 alone within the PBH scenario. A marginal tension with the realistic version of the Planck constraint is present.* **Right:** *Scenario with efficient accretion. The PBH population required to explain the GW190521 event evades any bound on $f_{PBH}$. We stress that the shift of the (early universe) CMB bounds is due to the accretion phenomenon changing the mapping between early- and late-time universe quantities [13], see also detailed discussion Sec. 4.2.2.*

**GW190521 as a single PBH event without accretion**

In this scenario, the mass function is only determined by GW190521. We choose the representative values $M_c = 75 M_\odot$ and $\sigma = 0.2$; different choices would result in a change of the merger rates only by $\mathcal{O}(1)$ factor. Also, since the rate scales indicatively as $R \propto f_{PBH}^2$ and the abundance is found to be relatively stable with respect to reasonable changes of the mass function. From Eq. (4.6.36), we estimate the abundance required to match the observed event rate which, given the current high uncertainties [512, 562], we assume to be $R \simeq 1/\text{yr}$. This yields $f_{PBH} \simeq 1.2 \cdot 10^{-4}$. In Fig. 5.10, we compare this value with constraints on $f_{PBH}$, showing a marginal tension with bounds coming from CMB anisotropies as studied in Ref. [183].

**GW190521 as a single PBH event with accretion.**

In this scenario, the two PBHs evolve from initial masses of approximately $M_1^i \simeq 22 M_\odot$ and $M_2^i \simeq 13 M_\odot$, respectively (Fig. 5.9). We choose a representative narrow initial mass function with $M_c = 18.5 M_\odot$ and $\sigma = 0.2$, related to a PBH population with most of the binaries having final masses in the range identified by GW190521 after accretion has taken place. The abundance corresponding to the observed merger rate is $f_{PBH}(z_i) \simeq 2.5 \cdot 10^{-5}$ ($f_{PBH}(z = 0) \simeq 3.7 \cdot 10^{-5}$), which is smaller than the one found in the previous section since accretion generally leads to an enhancement of the merger rate, see Eq. (4.6.36). As shown in Fig. 5.10, this value of $f_{PBH}$ is allowed by the current constraints, which are strongly relaxed in the presence of accretion [13], see also detailed discussion Sec. 4.2.2.

**GW190521 as an event within the full PBH population.**

Finally, we address the question of whether GW190521 fits into the PBH population inferred in the previous section assuming all the GW events have a primordial origin. [6]

The inference adopting the GWTC-2 catalog was presented in Sec. 5.2.1, where also the hyperparameter describing the accretion phenomenon was constrained to be $z_{\text{cut-off}} = 21.32^{+3.53}_{-1.56}$ to best fit the catalog under the assumption the PBH channel is the only responsible for GW events. Also, the presence of GW190521 with such a high mass causes a preference for a broader mass function, see a more detailed discussion in the following sections.

---

[6]We update the results reported in Ref. [11]. They were found performing an analysis of the GW events available at the time, which only included the GWTC-1 events and GW190412, GW190814 and GW190521. The results of the two analyses are qualitatively consistent with each other.

Fixing the merger rate with the LIGO/Virgo observations restrains the abundance to be log $f_{\mathrm{PBH}}(z_i) = -2.64^{+0.18}_{-0.08}$, which is compatible with current constraints, see Fig. 5.8. The rate of observable events at LIGO/Virgo for binaries in the mass gap or at least as massive as GW190521 is

$$R(M_{1,2} > 65M_\odot) \simeq 0.9/\mathrm{yr},$$
$$R(M_1 > 85M_\odot, M_2 > 65M_\odot) \simeq 0.7/\mathrm{yr}, \tag{5.2.3}$$

which are compatible with the observed rate. We conservatively took $65M_\odot$ as the lower edge of the mass gap. Therefore, we conclude that in the full population scenario with accretion, GW190521 is not only perfectly allowed, but also generally expected. This conclusion is solid against modelling systematics (e.g., changes of the accretion rate) and is also conservative. Indeed, in the more realistic hypothesis where some GW events have an astrophysical origin, removing them from the analysis would make our conclusion on the viability of GW190521 as a primordial binary even more robust. This will become evident in the following sections where the mixed astrophysical and primordial scenario will be considered.

These considerations give a strong motivation to reconsider the parameter estimation of GW190521 incorporating the correct PBH-motivated prior distributions to infer the binary parameters in the PBH scenario [10], and to perform a Bayesian comparison between the primordial and astrophysical scenarios that may explain spinning binaries in the mass gap, like hierarchical mergers [450, 489, 490, 555, 556, 558] or others [559, 560].

### 5.2.3   Mixed scenario: PBHs and the astrophysical phenomenological model

In this section, we make the first step towards trying to perform an analysis of the GWTC-2 based on a mixed model including both primordial and astrophysical channels.

We consider the representative astrophysical phenomenological model, physically equivalent to the one named "Truncated" model and adopted by the LVC in the population analysis of both the GWTC-1 [540] and GWTC-2 [507] catalogs. [7] This model is based on an analytical parameterisation of the merger rate in terms of the local merger rate $R_0$, the mass spectrum slope $\zeta$, both a minimum $m_{\min}$ and maximum $m_{\max}$ mass, and the exponent $\beta$ of the symmetric mass ratio as free parameters. The corresponding differential merger rate can be written as

$$\frac{\mathrm{d}R_{\mathrm{ABH}}}{\mathrm{d}m_1\mathrm{d}m_2} = \mathcal{N}R_0(1+z)^\kappa(m_1+m_2)^\alpha\left[\frac{m_1m_2}{(m_1+m_2)^2}\right]^\beta\psi(m_1)\psi(m_2), \tag{5.2.4}$$

where $\mathcal{N}$ is a normalization factor ensuring the local merger rate is $R(z=0) \equiv R_0$. Also, the mass function is taken to be a power-law with sharp cut-off at both ends as

$$\psi(m|\zeta, m_{\min}, m_{\max}) \propto m^{-\zeta} \qquad \text{for} \qquad m_{\min} < m < m_{\max}, \tag{5.2.5}$$

with an overall constant enforcing unitary normalization as

$$\int \psi(m)\mathrm{d}\ln m = 1. \tag{5.2.6}$$

We stress that this model allows for the existence of both a lower and upper mass gap in the astrophysical BH spectrum through the inclusion of both a minimum and a maximum mass cut-off. On the other hand, it does not introduce any additional feature for the primary mass distribution on top of the power-law scaling. As the GWTC-2 catalog already hints at the existence of some features in the binary BH spectrum, the LVC collaboration introduced more complex phenomenological models having secondary peaks in the spectrum of $m_1$. In practice, within the scenario considered here, we have the aim of testing if such features may be produced by a subdominant PBH population. We will come back to this point in the following.

---

[7]We note that we adopt a symmetric parametrisation of the merger rate with respect to the BH masses, differently from what the LVC collaboration used in Ref. [507].

In the population analysis performed by the LVC collaboration, three different spin models were adopted, going under the name of "Default", "Gaussian" and "Multi-" spin model, respectively, see Appendix D of Ref. [507] for more details. [8] We adopt the "Gaussian" model and, consistently with our Bayesian inference setup described in App. C, we do not include the precession spin parameter $\chi_{\rm p}$ in the model. In the "Gaussian" model, the effective spin parameter $\chi_{\rm eff}$ is normally distributed with mean $\mu_{\chi_{\rm eff}}$ and variance $\sigma_{\chi_{\rm eff}}$ truncated within its physical range $\chi_{\rm eff} \in [-1, 1]$.

The choice of hyperparameters for the phenomenological ABH model is summarized in Tab. C.2, while the corresponding prior ranges considered in the analysis are found in Tab. C.1. Finally, we fix the evolution of the merger rate with redshift to the best fit value inferred in Ref. [507], i.e. $\kappa \simeq 1.5$. We stress however that, being the data only limited to redshifts smaller than unity, the result of the inference is still largely insensitive to the merger rate evolution we assumed. We also fix $\alpha = 0$, without loss of generality, see related discussion in Ref. [476].

**Single phenomenological ABH population**

We first perform the analysis by assuming all events are coming from the ABH sector and we show the resulting posterior Fig. 5.11. We list here the important features of the posterior distributions.

- The inferred local merger rate density $R_0 = 15.97^{+5.02}_{-3.72}{\rm Gpc}^{-3}{\rm yr}^{-1}$ is consistent with the one found by the LVC collaboration.

- There exists a sharp drop of the posterior for values of $m_{\rm max}$ smaller than $\lesssim 60 M_\odot$. This is due to the lack of support of the model where the posterior of $m_1$ for the heaviest event GW190521 falls [512] and the choice of a hard cut-off at the maximum mass in the ABH model. Analogous behaviour is found for the quantity $m_{\rm min}$.

- We observe an anti-correlation between the overall rate $R_0$ and the mass function drop-off $\zeta$. This is explained by the fact that, for a steeper mass function, the binary masses are expected to be on the lighter portion of the observable mass range. Light events, however, are characterised by a smaller SNR. Therefore, the rate must compensate and grow to match the number of observable events predicted by the ABH model to the GWTC-2 catalog.

- An anti-correlation, similar to the previous point, exists between $\zeta$ and $m_{\rm max}$. This is because, as the mass function is allowed to extend to larger masses, the steepness is increased to mitigate the contribution of massive events in the observable mass range.

- As the vast majority of the events in the GWTC-2 catalog are compatible with $q \sim 1$, one finds a preference for a large value of $\beta$, i.e. a tendency towards symmetric binaries.

- The effective spin parameter distribution is found to be narrow and only slightly shifted towards positive values of $\chi_{\rm eff}$.

Even though in the phenomenological model we considered in Eq. (5.2.4) there is no correlation between the spin distribution and the mass distribution, there exists a well-known parameter degeneracy between $(\chi_{\rm eff}, q)$ in the GW data [564–569]. We found that, in this case, the inferred population is only slightly affected by the removal of spin information in the inference.

**Single PBH/ABH populations comparison**

We also perform a comparison of how well the single-population (either ABH or PBH) scenarios can explain the entirety of GW events detected so far. The model comparison is done by computing the Bayes factors, see App. C. In Refs. [548] and [476] using the GWTC-1 and GWTC-2 dataset respectively, it was reported that the LVC phenomenological model was strongly favoured when compared to the PBH scenario. This was explained by the over-production of heavy systems having large values of chirp mass $\mathcal{M}$ and positive skewness over the range of observed masses in the PBH model [548]. This can also be deduced by looking at the distribution of $m_1$ in Fig. 5.6 compared to the best model found by the LVC collaboration, which shows a failure to produce the surplus of events around $m_1 \simeq 40 M_\odot$ without enhancing the tail at large masses. This conclusion is confirmed by our analysis, as the Bayes factor (see Table 5.3) turns out to be $\log \mathcal{B}_{\mathrm{PBH}}^{\mathrm{ABH}} \approx 2.74$, which implies a decisive preference for the ABH model. Furthermore, this shows that the inclusion of accretion in the PBH model, as well as spin information in the inference, both neglected in Refs [476, 548], is not able to reduce the gap between the ABH and PBH models in the attempt to explain all the LVC observation employing a single population.

---

[8]Notice that, as reported by Ref. [507], the population reconstructed using either the "Gaussian" and "Default" models agree with each other.

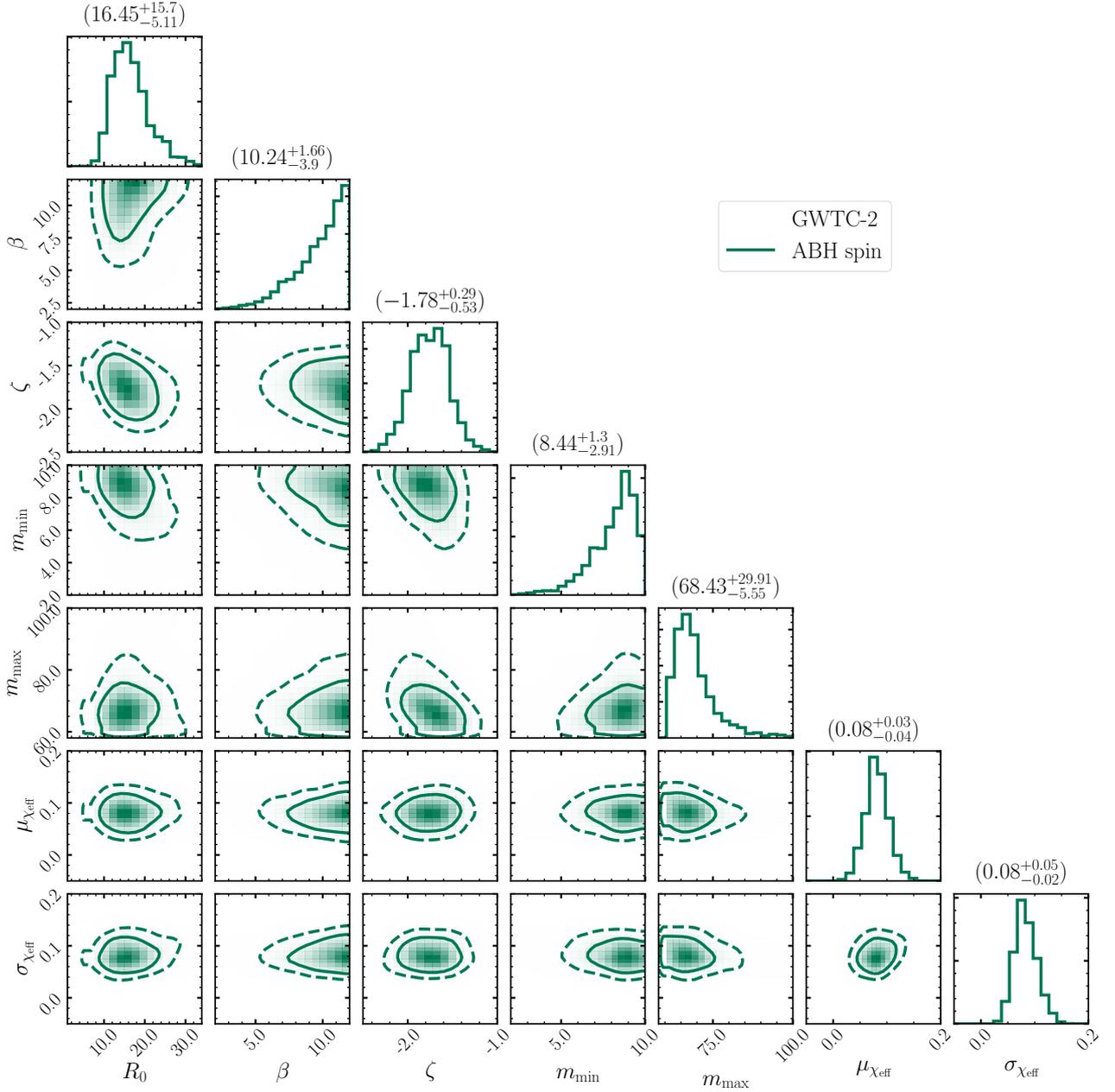

Figure 5.11: *ABH population inferred from the GWTC-2 catalog. The corresponding 90% C.I. are reported on top of each column.*

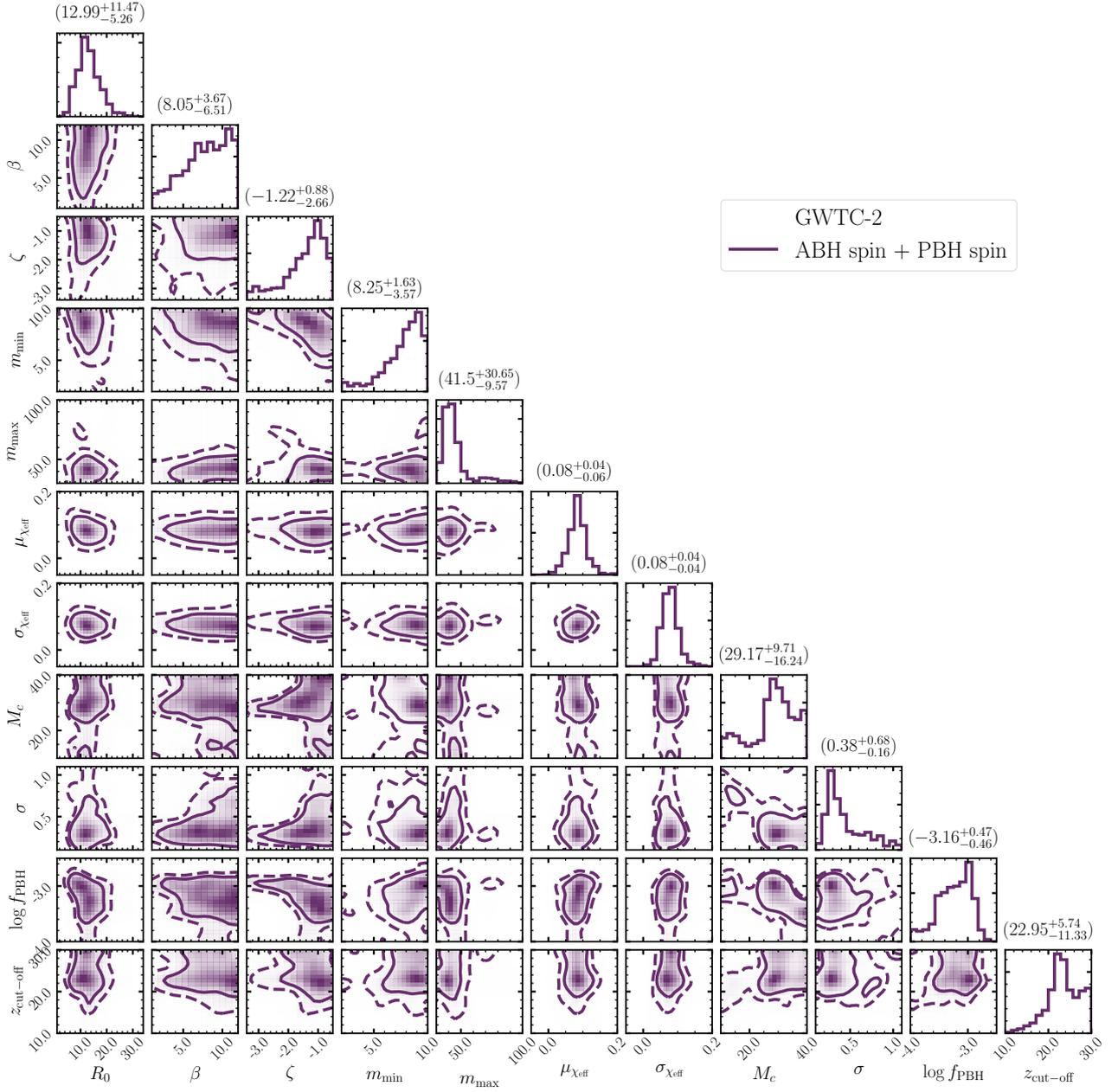

Figure 5.12: *PBH+ABH mixed population inference using the GWTC-2 catalog.*

## Mixed PBH/ABH population

In this section, we relax the hypothesis that all GW events come from a single population and we investigate the mixed scenario where the GWTC-2 population is supposed to contain both ABH and PBH binaries. Similar analyses including multiple astrophysical formation channels, but without accounting for a possible PBH contribution, can be found in Refs. [570–579]. Consistently with the previous sections, the set of hyperparameters of the mixed model contains 7 astrophysical inputs

$$\boldsymbol{\lambda}_{\text{ABH}} = [R_0, \beta, \zeta, m_{\text{min}}, m_{\text{max}}, \mu_{\chi_{\text{eff}}}, \sigma_{\chi_{\text{eff}}}],$$  (5.2.7)

as well as 4 primordial parameters

$$\boldsymbol{\lambda}_{\text{PBH}} = [M_c, \sigma, f_{\text{PBH}}, z_{\text{cut-off}}],$$  (5.2.8)

see App. C. The posterior distribution of $\boldsymbol{\lambda} = \boldsymbol{\lambda}_{\text{ABH}} \cup \boldsymbol{\lambda}_{\text{PBH}}$ resulting from the inference is shown in Fig. 5.12.

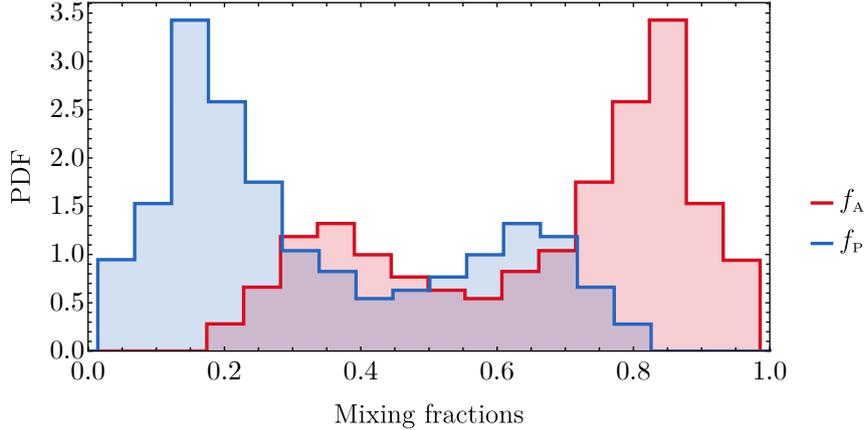

Figure 5.13: *Distribution of the observable mixing fraction $f_P$ and $f_A = 1 - f_P$*

Table 5.2: *Bayesian evidence ratios.*

| Model $\mathcal{M}$ | PBH | ABH | PBH+ABH |
|---|---|---|---|
| $\log_{10} \mathcal{B}^{\mathcal{M}}_{\mathrm{PBH+ABH}} \equiv \log_{10}\left(Z_{\mathcal{M}}/Z_{\mathrm{PBH+ABH}}\right)$ | -4.91 | -2.17 | 0 |

As one expects, both the PBH abundance and the local ABH merger rate density are reduced in the mixed scenario compared to the corresponding single-population analysis. This confirms that only a subdominant portion of the dark matter is allowed to be in the form of PBHs in the mass range probed by LVC. This is a consequence of the fact that a smaller portion of the events are ascribed to either sectors. In particular, we can quantify the mixing fractions in the GWTC-2 events by defining

$$f_P \equiv N^{\mathrm{det}}_{\mathrm{PBH}}/(N^{\mathrm{det}}_{\mathrm{ABH}} + N^{\mathrm{det}}_{\mathrm{PBH}}) = 1 - f_A, \tag{5.2.9}$$

which corresponds to the relative observable fraction of events in both models. In Fig. 5.13 we show the posterior distribution of both mixing fractions. On average, more events are interpreted as having astrophysical origin, with a fraction $f_A = 0.74^{+0.19}_{-0.46}$, corresponding to about three times the number of PBHs in the GWTC-2 catalog. This indicates a preference of the data for having a larger contribution coming from the ABH phenomenological model with only a smaller contribution to the GWTC-2 catalog from PBH binaries. The secondary peak structure in $f_P \simeq 2/3$ is due to a subdominant region in the posterior distribution with small $m_{\max}$ along with large $f_{\mathrm{PBH}}$. Overall, however, it is interesting to notice that there is a statistically significant portion of the posterior for which the mixing fractions are comparable.

We quantify how well the mixed model can perform by computing, also in this case, the corresponding Bayes factors with respect to both isolated scenarios, see Tab. 5.3. We find $\log \mathcal{B}^{\mathrm{ABH+PBH}}_{\mathrm{ABH}} \approx 2.17$, providing decisive evidence (according to Jeffreys' criterion [580]) for our best-fit mixed ABH-PBH model relative to a single-population ABH model. This high significance in favour of the inclusion of the PBH channel is supported by the fact that the posterior distribution for the mixing fraction $f_P$ is incompatible with zero. This implies that the presence of some PBH events in the GWTC-2 catalog seems to be demanded to better fit the data when only a power-law astrophysical model is adopted. Notice that a similar Bayes factor is found in the LVC analysis when comparing the "Truncated" with the "Power-law + Peak" model, showing the primordial channel can efficiently generate the surplus of events in the massive portion of the catalog.

Overall, the parameter distributions found in the mixed population inference are generally similar to the ones observed in the isolated scenarios. One important exception is provided by the preferred value of the upper mass cut-off $m_{\max}$ in the ABH sector, which is found to be significantly reduced with respect to the single population scenario, being $m_{\max} = 41.6^{+30.65}_{-9.57} M_\odot$. This is caused by the strong preference of the inference to interpret massive events as primordial binaries, see Figs. 5.14 and 5.15 and their discussion below. As a consequence of the anti-correlation between $\zeta$ and $m_{max}$, also the mass function is found to be slightly less steep. In the PBH sector, the preferred PBH mass function is shifted towards larger masses (with a larger reference mass scale $M_c$) and a narrower shape, as it only needs to accommodate a smaller portion of the GWTC-2 events compared to the single-population scenario. The local merger rate of the subpopulation of PBH binaries is found to be

$$R_{\mathrm{PBH}}(z=0) = 1.8^{+4.0}_{-1.5}\mathrm{Gpc}^{-3}\mathrm{yr}^{-1}. \tag{5.2.10}$$

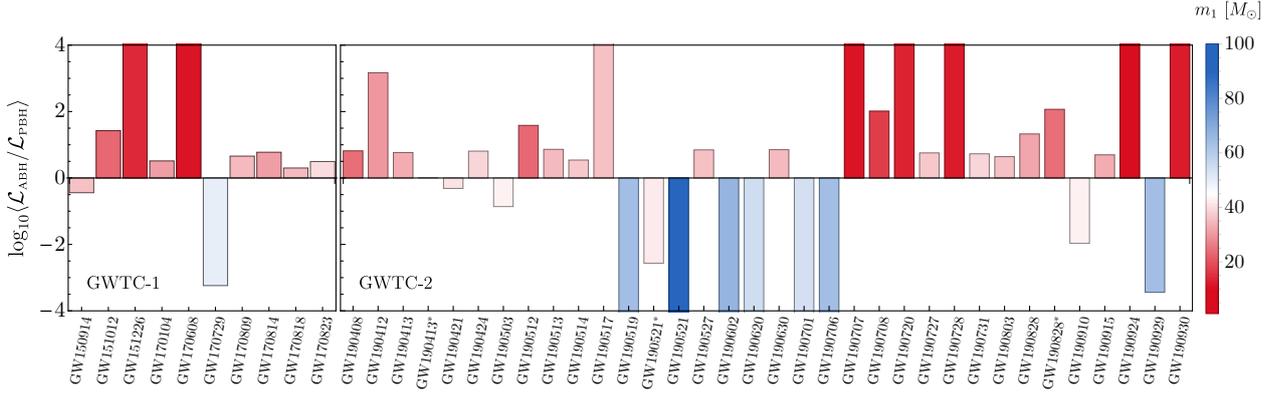

Figure 5.14: *Plot of the ratio between the ABH and PBH likelihood for each event in the GWTC-2 catalog averaged over the population posterior. The labels with an asterisk correspond to GW190413\*=GW190413_134308, GW190521\*=GW190521_074359 and GW190828\*=GW190828_065509, respectively.*

Finally, the mixed population inference confirms a preference for weak accretion with a value of $z_{\mathrm{cut-off}}$ which is compatible with the one found in Sec. 5.2.1. The posterior on $z_{\mathrm{cut-off}}$ is broader than in the single-population PBH case since in the mixed scenarios PBHs are subdominant and the hyperparameters of their population are less constrained by the data.

We can further understand how each GWTC-2 event is interpreted by the inference by looking at Fig. 5.14, where we show the ratio of the contribution to the likelihood function $\mathcal{L} \equiv p(\boldsymbol{\lambda}|\boldsymbol{d})/\pi(\boldsymbol{\lambda})$ in Eq. (C.3) from both the astrophysical and primordial sectors, averaged over the population posterior. The plot shows that events with primary masses with support in the astrophysical mass gap (i.e. above $\mathcal{O}(45)M_{\odot}$) are confidently assigned to the PBH population.

A similar message is contained in Fig. 5.15, where we show the distribution of primary mass $m_1$ and mass ratio $q$ for both sectors separately and combined. The most striking difference between the two channels is the presence of a tail at large masses well within the upper mass gap in the PBH binary $m_1$ distribution. On the other hand, the mass ratio distribution is similar between both sectors, with a slightly enhanced ability of the PBH model to accommodate asymmetric binaries. We again conclude that PBH binaries within the mixed model could be responsible for events in the GWTC-2 catalog with large masses which do not fit well within the ABH model.

It is interesting to compare our findings to the population analysis performed by the LVC collaboration. As also found in this work, Ref. [507] concluded that events in the GWTC-2 catalog are suggesting the presence of an additional distinct population of binaries with primary masses above $\approx 45 M_{\odot}$. Within the standard stellar formation scenario, it is difficult to accommodate massive events due to the pulsational pair supernova instability, preventing the formation of binaries with masses above $\approx 45 M_{\odot}$, even though the precise location of such a cut-off is still uncertain [513–515, 517–528]. One interesting possibility within the astrophysical sector is that massive events in the catalog are coming from second generation mergers in globular clusters or galactic nuclei [450, 489, 490, 555–558]. Our findings show that introducing a PBH population of binaries in the inference naturally leads to the interpretation of those events as coming from primordial binaries. Interestingly, the distribution of the mass ratio of PBH binaries inferred from the data is found to be similar to the one of the astrophysical channel (i.e. both showing the same tendency for symmetric binaries), in contrast to the expectation of second-generation mergers which predicts a smaller mass ratio (see for example Ref. [449, 581]). This difference may be used in the future to disentangle primordial and second-generation mergers in the high-mass portion of the dataset.

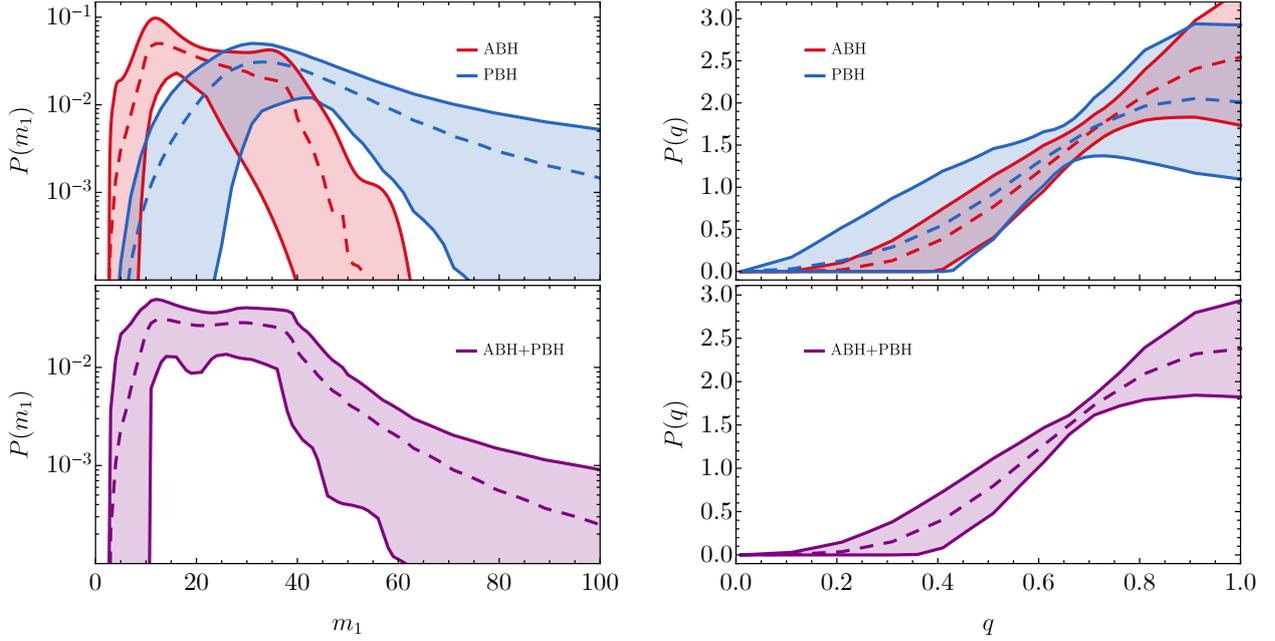

Figure 5.15:   **Top:** *Primary mass (left) and mass ratio (right) distributions for the PBH and ABH sectors, respectively.*   **Bottom:** *Cumulative population distribution. The bands around the dashed lines represent the 90% C.I.*

Finally, in Fig. 5.16 we show the comparison of the PBH population inferred in the mixed scenario with the current constraints on the PBH abundance. This plot indicates there is no tension with current constraints, see also related discussion in Sec. 5.2.1.

Before concluding this section, let us stress the main result of this analysis: a mixed population turns out to be preferred compared to both single PBH (ABH) scenarios at least in the case one assumes a lognormal PBH mass distribution. This is based on a simplified analysis and motivates an effort to consider a mixed astro-primordial scenario that included state-of-the-art astrophysical channels, to allow for a proper comparison and more definitive conclusions. We dedicate the next session to this endeavour.

### 5.2.4   Mixed scenario: PBHs and the astrophysical channels

The analysis presented in the previous section is hinting towards the possibility that a single population (either astrophysical [577] or primordial [5, 476, 548]) would be unable to explain the wealth of GW observations we have so far. In this section, we provide the most comprehensive hierarchical Bayesian inference study of the GWTC-2 catalog, which includes a state-of-the-art PBH model [12, 170] alongside several astrophysical models that can reproduce many features of the observed population. This allows us to quantify the evidence for PBHs in current GW data given the present knowledge of astrophysical formation scenarios. We closely follow the discussion in Ref. [3].

We adopt the astrophysical models collected in Ref. [577], where the most comprehensive astrophysical multichannel inference on the LVC data was performed. We divide the various astrophysical channels into subgroups where binaries formed in isolation or dynamically in dense environments (see dedicated discussion in App. D).

Among the field formation scenarios, we consider a late-phase Common Envelope (CE), binaries with a Stable Mass Transfer (SMT) between the star and the first BH formed, and Chemically Homogeneous Evolution (CHE). Ref. [577] showed that CE and SMT are the dominant binary formation pathways and therefore we will restrict our analysis to those. These two channels were simulated in Ref. [577] using the POSYDON framework [582, 583], which combines the population synthesis code COSMIC [584] and MESA [585] to model the population and evolution of binaries. The hyperparameters affecting the prediction of the CE channel is the CE efficiency, simulated for a discrete set of values $\alpha_{CE} \in [0.2, 0.5, 1.0, 2.0, 5.0]$, with large values of $\alpha_{CE}$ leading to efficient CE evolution. Both channels are also dependent on the assumption of natal BH spin, which was simulated for the discrete values of $\chi_b \in [0, 0.1, 0.2, 0.5]$.

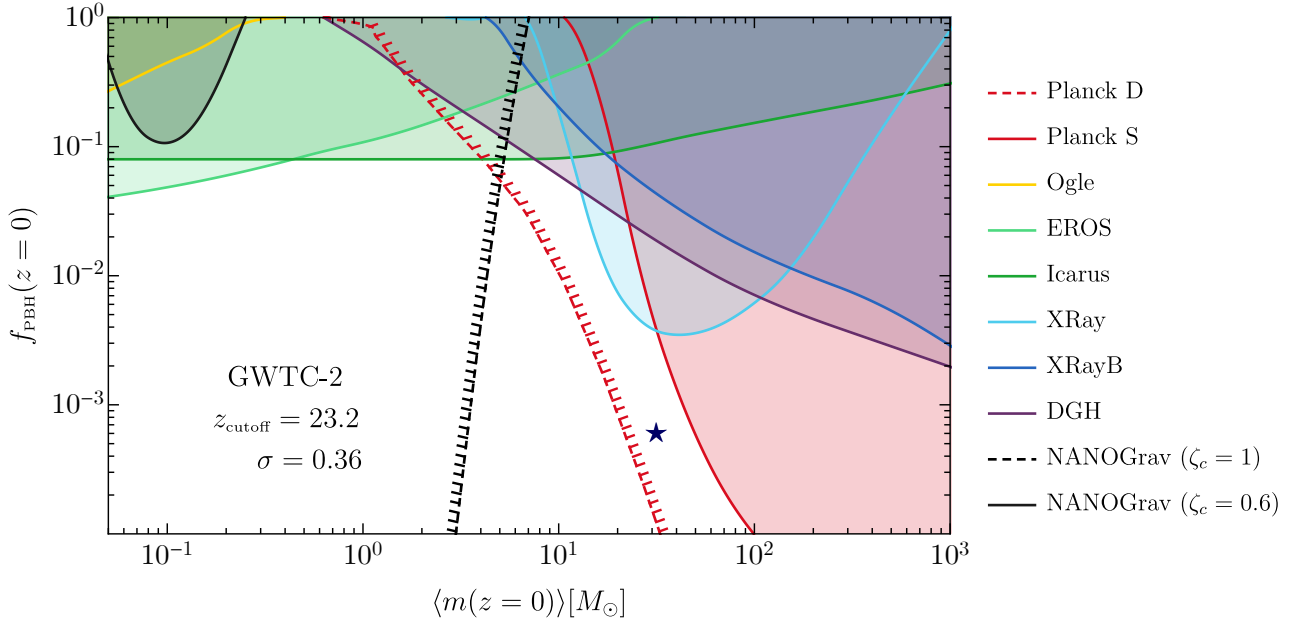

Figure 5.16: *Same as Fig. 5.8 for the PBH population inferred from the mixed PBH+ABH scenario studied in Sec. 5.2.3.*

The two dynamical models predict the binary formation to take place in old, metal-poor globular clusters (GC) and in nuclear star clusters (NSC). The GC models are taken from numerical simulation using the cluster Monte Carlo code CMC [555]. As large natal spins imply larger ejection probabilities of the remnants, large values of $\chi_b$ lead to a smaller probability of repeated mergers. The NSC model uses COSMIC to generate the BH masses from a single stellar population with metallicity $Z = (0.01, 0.1, 1)Z_{\odot}$, and evolve the clusters and their BHs using the semi-analytical approach described in Ref. [586]. This procedure was repeated for all values of $\chi_b$ listed above.

On the PBH side, we adopt the PBH model as described in Chapter 4 as used in the previous sections. We only stress that, as PBHs form from the collapse of radiation density perturbations in the early universe [20, 376], their natal spins are negligible and independent of $\chi_b$. Also, recall that similarly to the dynamical astrophysical channels, the PBH spin orientations with respect to the binary's angular momentum are independent and uniformly distributed on the sphere. This leads to a similar prediction for the distribution of $\chi_{eff}$ for both the PBH and dynamical astrophysical models.

Overall, our astrophysical models depend on the hyperparameters

$$\boldsymbol{\lambda}_{ABH} = [\alpha_{CE}, \chi_b, N_{CE}, N_{SMT}, N_{GC}, N_{NSC}], \tag{5.2.11}$$

where the number of events in each channel $N_i$, following Ref. [577], is assumed to be unconstrained and independent of $\alpha_{CE}$ and $\chi_b$. The PBH channel depends on

$$\boldsymbol{\lambda}_{PBH} = [M_c, \sigma, f_{PBH}, z_{\text{cut-off}}], \tag{5.2.12}$$

with the number of mergers in the PBH model univocally determined by the population parameters $\boldsymbol{\lambda}$ with an overall scaling $N_{PBH} \approx f_{PBH}^2$ [12, 170]. See App. C for more details on the data analysis.

Among the events in the GWTC-2 catalog, we follow the same procedure adopted in the preceding sections, discard those with a large false-alarm rate (GW190426, GW190719, GW190909) [507] and two events involving neutron stars (GW170817, GW190425). We will, for the moment, discard also GW190814 [529] and assume that the secondary component of GW190814 is a neutron star. We will show that the results of this analysis would be mainly unaffected was this event included in the dataset.

Unlike Ref. [577], however, we do not exclude GW190521 [512]. The selected catalog has $N_{\rm obs} = 44$ or 43, depending on whether GW190521 is included or not. The analysis without GW190521 will be useful to understand how much of the evidence for PBHs is relying on the mass gap event. Also, it can be regarded as the "most conservative" analysis as far as PBHs are concerned since it would correspond to the case GW190521 were coming from an additional (and specifically tuned) astrophysical channel not considered here or the reconstruction of the GW190521 source properties were affected by systematical uncertainties, see Refs. [587–597].

**The observable mixing fractions**

The results of our hierarchical Bayesian analysis are summarized in Fig. 5.17, showing the posterior distributions of the detectable mixing fractions

$$\beta_i^{\rm det} = \frac{N_i^{\rm det}}{\sum_j N_j^{\rm det}},  \tag{5.2.13}$$

where $i, j = [{\rm CE, GC, NSC, SMT, PBH}]$ for the different models and $N^{\rm det}$ is the number of *observable* events during the observation time of the O1+O2+O3a LVC runs, computed by accounting for the experimental selection bias. We present various scenarios mixing the PBH population with different combinations of astrophysical channels: a simplified $2 + 1$ multichannel assuming only the two main astrophysical models (CE and GC, left panel), and two 3+1 combinations including three astrophysical channels: CE+GC+NSC and CE+GC+SMT. Tab. 5.3 shows the corresponding Bayes factors for various mixed scenarios with and without the inclusion of a PBH subpopulation.

First of all, we observe that a two-channel CE+GC model is insufficient to explain the data. Either with or without GW190521, Fig. 5.17 (left panel) shows that in the 2+1 (CE+GC+PBH) case the inferred PBH population fraction is approximately one third. Tab. 5.3 confirms this conclusion, by showing that CE+GC has the lowest evidence compared to all scenarios we considered (only comparable with CE+GC+NSC when GW190521 is included) and that CE+GC+PBH is decisively favoured over CE+GC. Also, the inclusion of the NSC channel does not improve the overall fit. This is because the NSC and GC channels compete to explain similar events, whereas the PBH and GC channels produce more complementary populations, even though a certain amount of anti-correlation between the two channels is observed in the posterior distributions.

By comparing the various three-channel scenarios we considered, we conclude that models with a NSC subpopulation are not favoured: NSCs account for some events in the central range of chirp masses, but the relative fraction of NSC events is small (both with and without GW190521). We find that the CE+GC+SMT channel has larger evidence. This is because the SMT channel complements CE and GC by predicting more massive binaries (but see e.g. [598] for a discussion of uncertainties in this prediction). However, even the SMT channel fails to produce binaries in the mass gap.

On the other hand, PBHs can efficiently produce binaries in the mass gap, so the presence of GW190521 in the dataset leads to stronger evidence in favour of mixed astrophysical+PBH models. Furthermore, PBHs can also account for a significant fraction of the heavy events other than GW190521 when the SMT channel is not included. This leads to a sizeable PBH fraction $\beta_{\rm PBH}^{\rm det} = 0.31_{-0.26}^{+0.28}$ ($0.27_{-0.24}^{+0.28}$) at 90% C.I. in the CE+GC+PBH (CE+GC+NSC+PBH) scenarios with GW190521, as shown in Fig. 5.17. This conclusion would be unaffected even if GW190521 were an outlier belonging to a different astrophysical channel as Fig. 5.17 and the Bayes factors show.

Let us finally focus on four-channel scenarios. As shown in the insets of the right panel of Fig. 5.17, when we include GW190521 the posterior of the PBH mixing fraction has vanishing support at $\beta_{\rm PBH}^{\rm det} \approx 0$. The first percentile of $\beta_{\rm PBH}^{\rm det}$ is $(0.022, 0.014, 0.002)$ for the (CE+GC+PBH, CE+GC+NSC+PBH, CE+GC+SMT+PBH) mixed scenarios. This confirms our previous conclusion: the PBH channel is the only one (given the astrophysical models and uncertainties we considered) which can produce binaries in the mass gap with a sufficiently high rate able to predict GW190521. We stress that both GC and NSC have, in fact, a subdominant tail populating the mass gap with multi-generation BH mergers. However, this tail does not provide a rate that is sufficiently high to justify at least 1/44 events in the mass gap during O1/O2/O3a observation time. For this reason the inference always automatically interpret at least GW190521 as a PBH binary, giving a lower bound on the PBH mixing fraction.

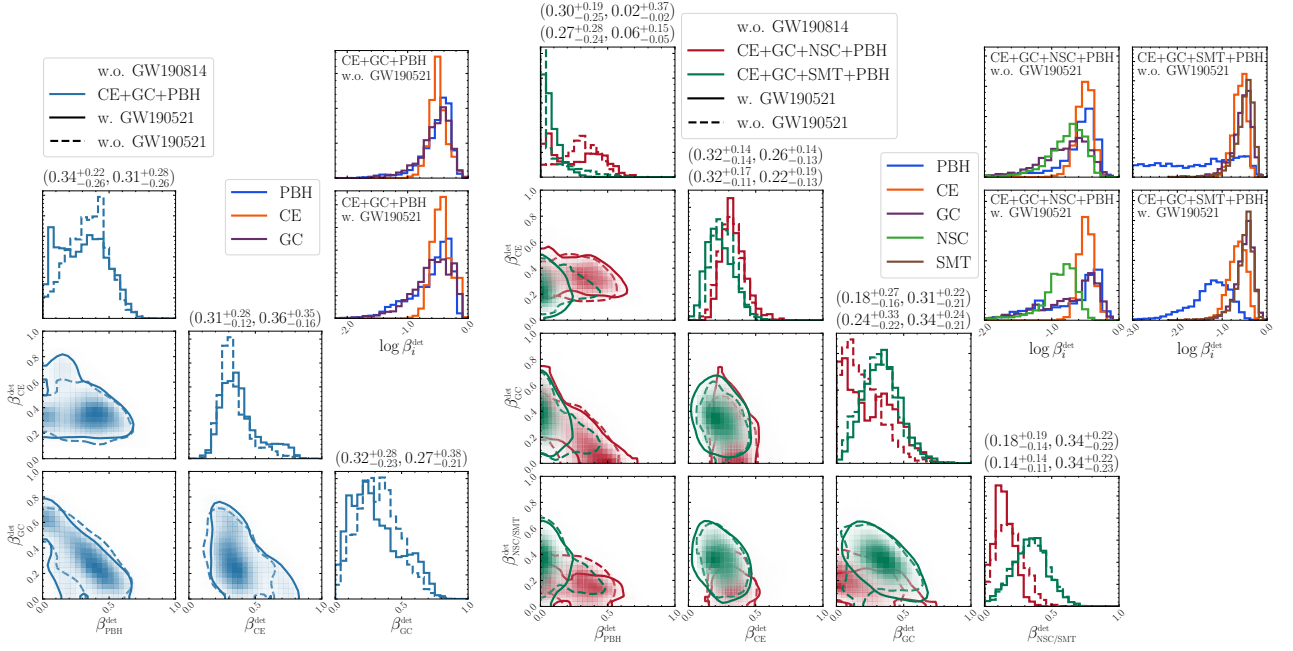

Figure 5.17:  *Posterior distributions of the individual detectable mixing fractions $\beta_i^{det}$ of different populations.* **Left:**  *2+1 (CE+GC+PBH) model.* **Right:**  *3+1 models (CE+GC+NSC+PBH and CE+GC+SMT+PBH). In all cases we excluded GW190814, and we present results with (solid line) and without (dashed line) including GW190521. The insets in the top right of each corner plot show each $\beta_i^{det}$ in logarithmic scale to highlight the monotonic behaviour for small values of $\beta_i^{det}$. In the left panel, the 90% C.I. for $\beta_i^{det}$ reported above each column correspond to the models in the inset (from top to bottom); those on the right panel are sorted in the same way as the four panels in the inset. The corresponding posteriors for the PBH hyperparameters are shown in Fig. 5.18.*

Table 5.3:  *Bayesian evidence ratios for the different mixed astrophysical and primordial populations (normalised with respect to the CE+GC scenario), obtained by marginalising over $\alpha_{CE}$ and $\chi_b$, both with and without GW190521.*

| Model | $\log \mathcal{B}_{CE+GC}^{\mathcal{M}}$: | w.o. GW190521 | w. GW190521 |
|---|---|---|---|
| CE+GC+PBH | | 1.22 | 2.38 |
| CE+GC+NSC | | 0.52 | -0.15 |
| CE+GC+SMT | | 1.39 | 0.72 |
| CE+GC+NSC+PBH | | 1.43 | 2.30 |
| CE+GC+SMT+PBH | | 1.31 | 2.58 |

The smallest PBH fraction ($\beta_{PBH}^{det} = 0.06_{-0.05}^{+0.15}$) corresponds to CE+GC+SMT+PBH: we observe that the SMT channel can reproduce most events below the mass gap. To be even more conservative, we can also exclude GW190521 from the dataset. Then, the posterior distribution of $\beta_{PBH}^{det}$ "flattens out" in the CE+GC+SMT+PBH scenario, becoming compatible with zero (see the blue histogram in the top-right inset of Fig. 5.17), even though significant support at relatively larger fractions is present, due to the degeneracies with both the GC and SMT channels. This very conservative scenario may suggest that the PBH fraction can be compatible with zero if the mass gap event was interpreted in other ways, e.g. by allowing for the large uncertainties in the astrophysical models considered here, or through additional astrophysical channels. For example, heavy binaries in the mass gap could form in AGN disks [599–601]. On the other hand, even in such a case, current data do not possess enough constraining power to rule out a PBH contribution.

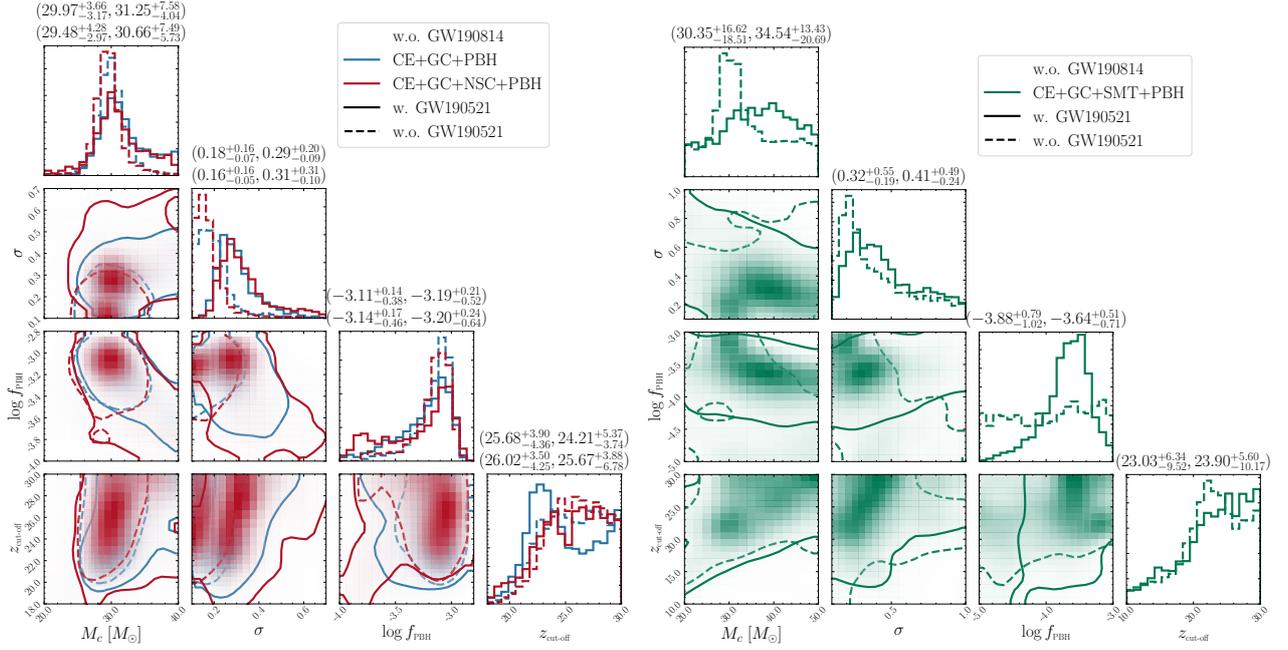

Figure 5.18: *Posterior distributions of the hyperparameters of the PBH model for the CE+GC+PBH, CE+GC+NSC+PBH and CE+GC+SMT+PBH mixed models (blue, red and green, respectively) with (solid line) and without (dashed line) the mass gap event GW190521, and excluding GW190814. The left (right) values on top of each column corresponds to the 90% C.I. in the case without (with) GW190521. In the left plot, the 90% C.I. on top (bottom) correspond to the CE+GC+PBH (CE+GC+NSC+PBH) scenario.*

Finally, we note that $\beta_{PBH}^{det}$ does not depend significantly on the inclusion of the lower mass-gap event GW190814, (see Fig. 5.21 and related discussion).

In the following, we will support our interpretation of the results by showing the observable distributions of the chirp mass and mass ratio of each channel and how they compare to the GWTC-2 distribution of events in the various scenarios analysed.

### The PBH subpopulation

In Fig. 5.18 we show the posterior distributions of the hyperparameters of the PBH model for the CE+GC+PBH, CE+GC+NSC+PBH and CE+GC+SMT+PBH mixed scenarios, both with and without GW190521. These results are complementary to those shown in Fig. 5.17, where we only focused on the mixing fractions.

We note that the posterior distributions in the first two cases (CE+GC+PBH and CE+GC+NSC+PBH) are strikingly similar. The only relevant differences are found when comparing results with and without GW190521. When we include the mass gap event, we obtain a larger width $\sigma$ of the PBH mass function and a higher tail of the posterior distribution of $M_c$ at large values. This happens because the PBH channel is necessary to produce heavy binaries in the mass gap to explain GW190521.

In the CE+GC+PBH mixed model including the mass gap event, the distribution of $z_{cut-off}$ shows two peaks at $z_{cut-off} \approx 23$ and $z_{cut-off} \approx 30$. The first one corresponds to the case where PBH accretion is needed to explain some of the (few) spinning events in the catalog [12, 14], while the second peak corresponds to the case where the observed events associated with PBHs by the inference are mostly nonspinning. A similar behaviour, even though less pronounced, is observed in Fig. 5.12 in the case of a phenomenological ABH model mixed with PBHs. In the other mixed cases, the posterior of $z_{cut-off}$ is approximately flat above $z_{cut-off} \simeq 25$, potentially due to the inference interpreting as PBHs events with smaller spins. As discussed in App. C, we have also checked that the posterior remains flat above $z_{\text{cut-off}} \gtrsim 30$, where accretion is indeed negligible in the mass range of interest and all models become degenerate.

In the two cases when the SMT channel is included, the PBH population parameters are less constrained. Focusing on the case including GW190521 in the catalog, the small observable fraction ascribed to the PBH sector ($\beta_{\text{PBH}}^{\text{det}} = 0.06_{-0.05}^{+0.15}$) implies that the PBH sub-population is only constrained by a small number of events dominated, in particular, by GW190521. As a consequence, the population is in general shifted to heavier masses, but with relatively larger uncertainties on the mass function parameters. Also, $f_{\text{PBH}}$ peaks at slightly smaller values compared to the other scenarios, while still being bounded from below. If instead, one removes GW190521, the posterior of $f_{\text{PBH}}$ possesses a plateau reaching values compatible with zero. This confirms the importance of mass-gap events for assessing the PBH contribution to the observed GW events. As a consequence, all the other PBH hyperparameters are only constrained in a fraction of the posterior (for a sufficiently high $f_{\text{PBH}}$), while they are unconstrained otherwise. This produces a multidimensional tail filling all the prior hyper-space.

In all cases we tested, the inferred observable fraction $\beta_{\text{PBH}}^{\text{det}}$ requires the PBH abundance to be below $f_{\text{PBH}} \lesssim 10^{-3}$, confirming that PBHs can only be a fraction of the dark matter in the mass range currently observed by the LIGO and Virgo experiments. Finally, we note that all the sub-populations we obtained in the various scenarios would pass every constraint on the PBH abundance, being the parameter we found compatible with the scenario obtained in the previous section and confronted with the constraints in Fig. 5.16.

### Observable population distributions

To better understand how the GWTC-2 events are interpreted by the inference, in Figs. 5.19 and 5.20, we plot the contribution of each population to the observed chirp mass and mass ratio distribution for all the scenarios we considered.

Let us focus on results including GW190521 first (i.e. Fig. 5.19). We observe that in all cases, both astrophysical isolated channels (i.e. CE and SMT) are not reaching values of chirp masses larger than $\mathcal{M} \approx 40 M_\odot$ with a significant rate. This is a consequence of the presence of a sharp drop of the binary production close to the mass gap boundary. On the other hand, the CE channel is the only one producing binaries in the range below $\mathcal{O}(10) M_\odot$, making it the only explanation for the lightest events in the catalog. The CE channel also never strongly overlaps with the PBH channel, explaining the negligible correlation between the two mixing fractions in Fig. 5.17. We also note that the SMT channel is complementing the CE channel in covering most of the mass range of the GWTC-2 catalog. Indeed, when SMT is included as a subpopulation as shown in Fig. 5.19c, the PBH fraction is smaller and its relevant contribution to the observable events is relegated to the high mass portion of the plot.

The dynamical channels (i.e. GC and NSC) can reach larger values of chirp mass. This happens thanks to hierarchical mergers producing the "multi-peak" structure at high $\mathcal{M}$. Even though both channels de facto produce binaries in the mass gap, they fail to do so at a sufficiently high rate. For this reason also the subdominant tails at high masses of the GC and NSC channels are not visible in the plot with the current choice of scale. On the contrary, the PBH channel can efficiently produce binaries in the mass gap, giving the leading explanation for GW190521 also following a population analysis (as opposed to the single event considerations adopted in Sec. 5.2.2).

The PBH population overlaps mostly with the GC channel (and with NSC/SMT, when included in the inference) while reaching larger values of $\mathcal{M}$. When the PBH distribution extends to explain the rate of GW190521, it becomes less competitive at explaining the bulk of events in the central range of $\mathcal{M}$. For this reason the posterior of $f_{\text{PBH}}$ has a significant tail at small values (see Fig. 5.18). This does not happen when GW190521 is removed as shown in Fig. 5.20, since then the PBH model can more efficiently reproduce events in the central range of $\mathcal{M}$.

The chirp mass and mass ratio distributions that result from neglecting GW190521 are shown in Fig. 5.20 instead. In particular, by comparing Figs. 5.20 with Fig. 5.19, we see that the removal of GW190521 makes the PBH chirp mass distribution more peaked around $\mathcal{M} \approx 30 M_\odot$. In the most conservative scenario (CE+GC+SMT+PBH without the mass gap event), the fraction $\beta_{\text{PBH}}^{\text{det}}$ is

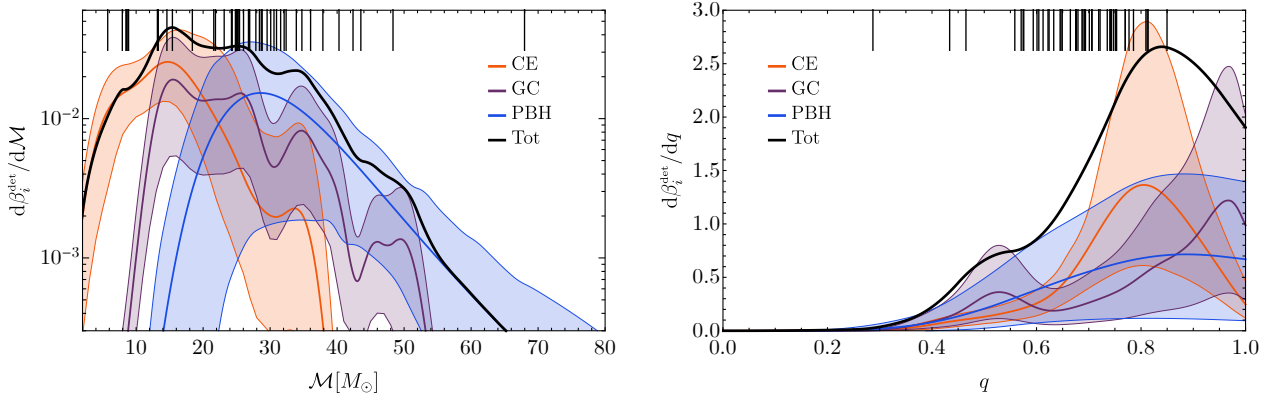

(a) *CE+GC+PBH mixed scenario including GW190521.*

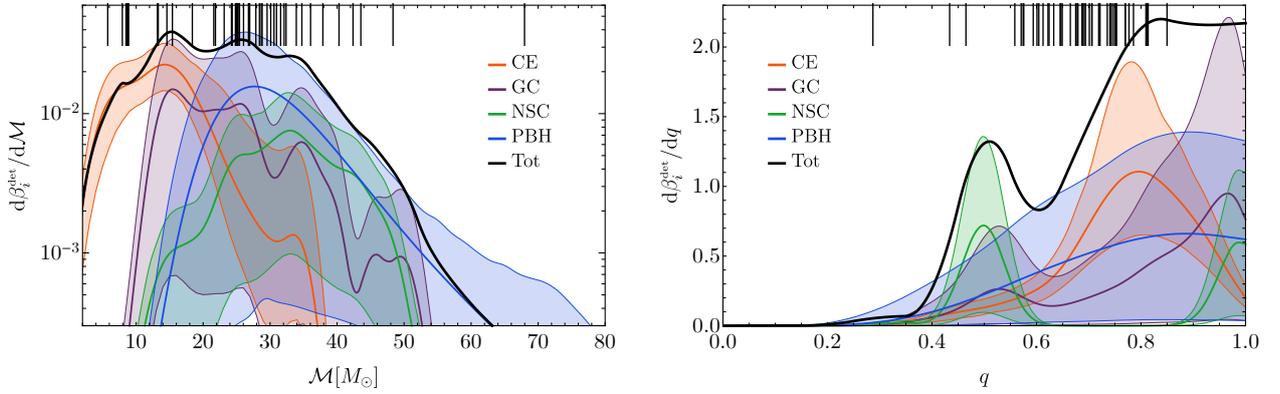

(b) *CE+GC+NSC+PBH mixed scenario including GW190521.*

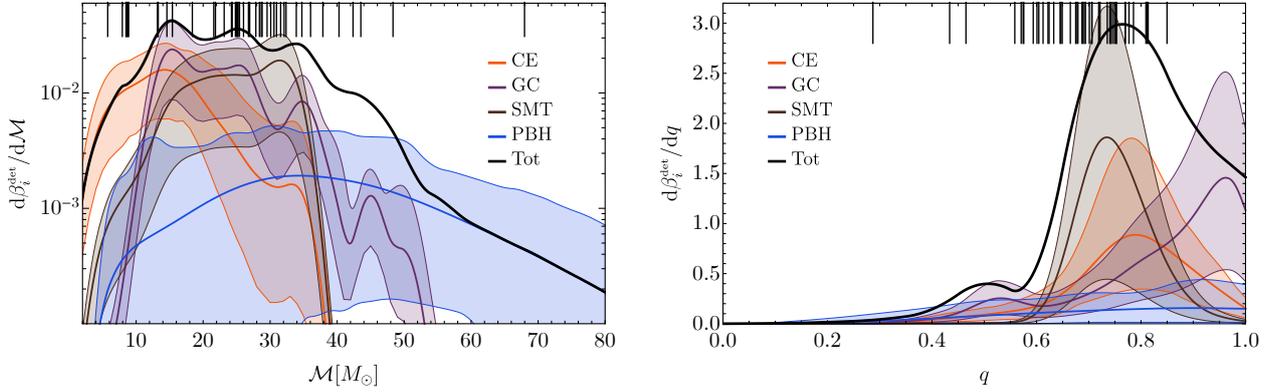

(c) *CE+GC+SMT+PBH mixed scenario including GW190521.*

Figure 5.19: *Individual contribution to the observable distributions for each mixed scenario we considered (CE+GC+PBH, CE+GC+NSC+PBH and CE+GC+SMT+PBH from top to bottom) with the inclusion of GW190521 in the dataset. The bands indicate 90% C.I., while the black line corresponds to the mean total population. Top dashes indicate the mean observed values for the GWTC-2 events.* **Left:** *Chirp mass.* **Right:** *Mass ratio.*

compatible with zero, and we only set an upper bound on the PBH contribution to the observable events. This also implies that the PBH contribution shown in Fig. 5.20c (blue band) is not bounded from below.

Note that the mass ratio distributions are overall very similar to each other, peaking close to $q \simeq 1$. Only the NSC channel, for reasons explained in Ref. [577], is characterized by a bimodal mass ratio

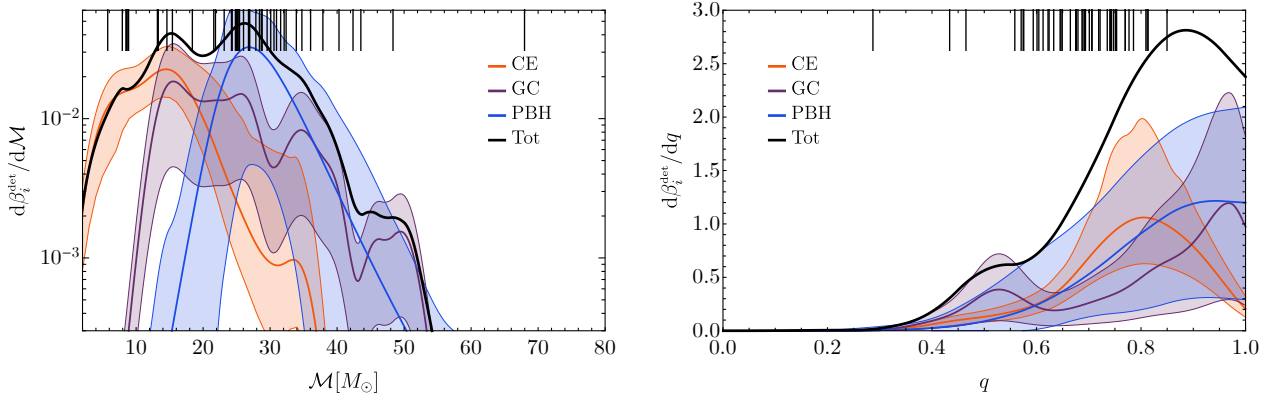

(a) *CE+GC+PBH mixed scenario without GW190521.*

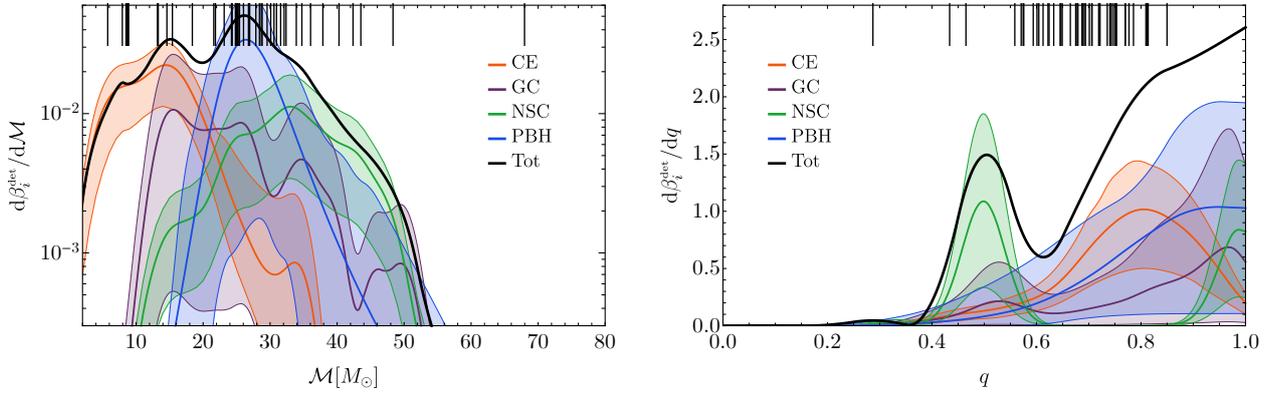

(b) *CE+GC+NSC+PBH mixed scenario without GW190521.*

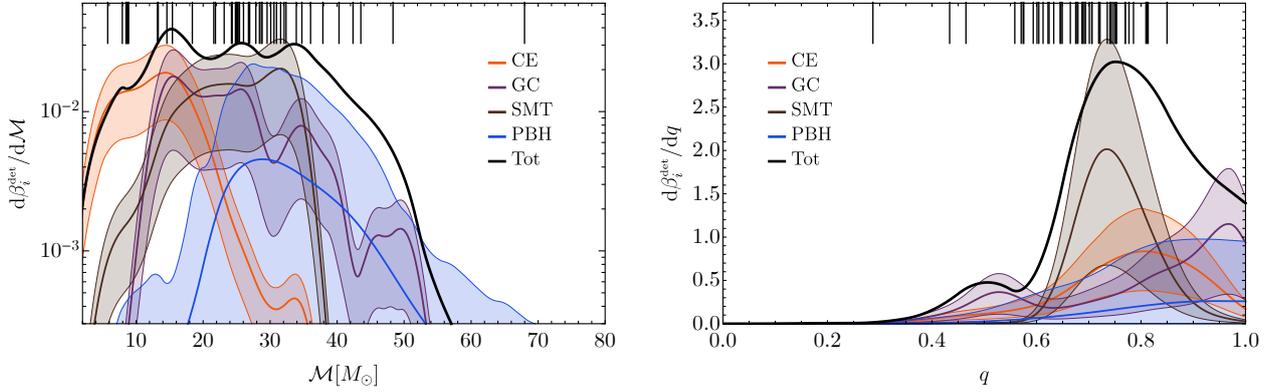

(c) *CE+GC+SMT+PBH mixed scenario without GW190521.*

Figure 5.20: *Same as Fig. 5.19 for the analysis without the inclusion of GW190521 in the dataset. In the bottom panels corresponding to the CE+GC+SMT+PBH case, the PBH contribution is unbounded from below, being the observable fraction $\beta_{PBH}^{det}$, in this analysis without including GW190521 and with SMT as a sub-population, compatible with zero.*

distribution.

Finally, we checked that the inclusion of the asymmetric merger GW190814 does not change our results. In Fig. 5.21, we show the posterior distributions obtained including/excluding GW190521 and GW190814 in various combinations for the CE+GC+PBH mixed model. We observe that this particular event does not lead to a significant modification of either the mixing fractions or the PBH population (comparing blue and green curves). This is because GW190814 is always ascribed to the CE population since the latter has the strongest support at small masses.

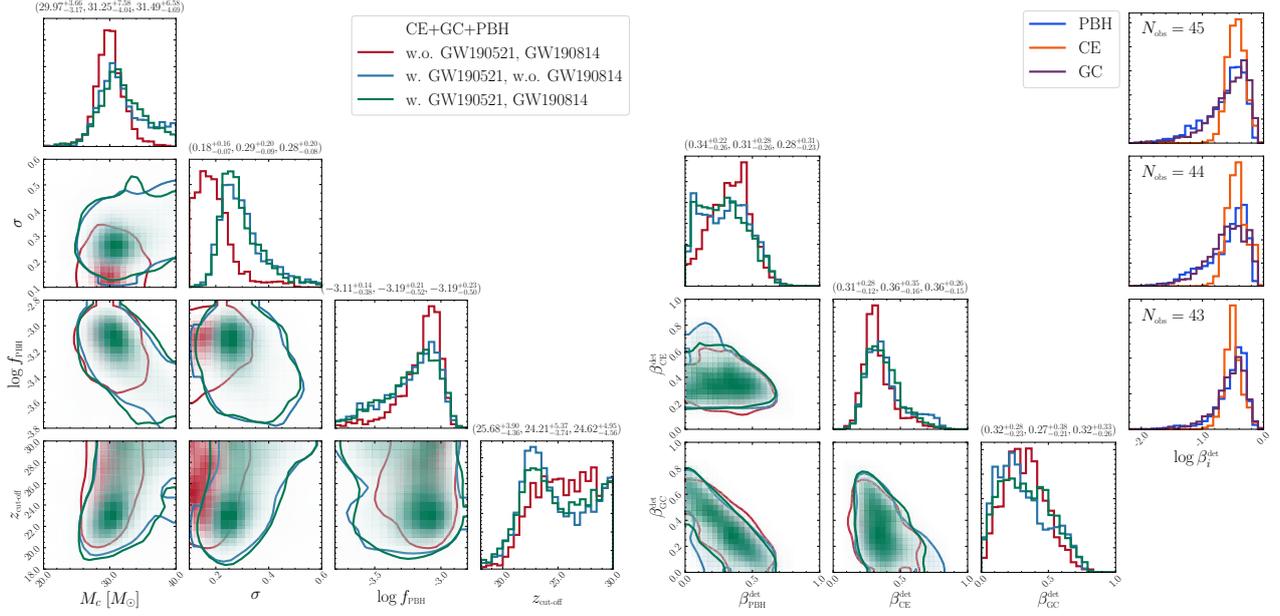

Figure 5.21: **Left:** Posterior distributions of the hyperparameters of the PBH population. **Right:** Individual detectable mixing fractions $\beta_i^{\mathrm{det}}$ of different populations. Both panels show results for the CE+GC+PBH mixed scenario while each color corresponds to the case both GW190521 and GW190814 are neglected (red, $N_{\mathrm{obs}} = 43$), only GW190521 is additionally included (blue, $N_{\mathrm{obs}} = 44$) and both GW190521 and GW190814 are included (green, $N_{\mathrm{obs}} = 45$). The 90% C.I. reported on top of each column correspond to the various cases, following (red, blue, green) ordering. The insets in the top right show the individual $\beta_i^{\mathrm{det}}$ on a logarithmic scale.

### Summary

Let us summarise our findings. First of all, as the mixed population of BBHs from different astrophysical channels considered in this analysis was shown to reproduce most features of the current BBH catalog (except for the upper-mass-gap event GW190521 [577]), it is quite remarkable that the inferred PBH mixing fraction can be significant when the additional primordial channel is added in the multi-population inference. The relative PBH abundance we infer depends on which astrophysical channels are included in the analysis. We find that the PBH channel can explain very massive binaries with a subpopulation consistent with the GWTC-2 catalog, and it is also statistically favoured against competitive dynamical astrophysical population models, such as GC and NSC. We also find that the SMT channel can complement the other astrophysical formation channels in the mass range below the mass gap, drastically reducing the need for PBHs, except for explaining the mass-gap event GW190521. If we further neglect GW190521, the Bayesian evidence for CE+GC+SMT becomes comparable to CE+GC+SMT+PBH, showing that the constraining power of the current data set is not sufficient to draw final conclusions. Note however that the fraction of SMT events inferred from the analysis is $\beta_{\mathrm{SMT}}^{\mathrm{det}} = 0.34 \pm 0.22$, showing SMT would have to be the dominant channel.

We conclude that assessing the evidence for a primordial population in GW data requires a more robust understanding of astrophysical populations. While we have considered four state-of-the-art astrophysical models as in Ref. [577], there might exist others that are competitive against the primordial subpopulation. On the PBH side, we adopted a standard lognormal distribution for the PBH mass function at formation, but it would be important to test the impact of other (model-dependent) assumptions and of different priors on the PBH hyperparameters motivated by specific formation mechanisms (see for example Refs. [167, 363]).

Confidently confirming the primordial nature of a subset of events in current data would require:

- individual-event analyses for large signal-to-noise ratio events, especially by cross-correlating

merger rates with mass, spin, and redshift measurements trying to identify key features of the PBH scenario (see e.g. [10] and discussion in Sec. 5.3);

- population studies focusing on spin distributions [602–604], and accounting for the $q - \chi_{\rm eff}$ correlation introduced by accretion effects in PBH models [12, 14] to break the degeneracy between the PBH and astrophysical channels;

Finally, we stress that a conclusive verdict on the primordial nature of a subpopulation of BBHs will come from third-generation GW detectors such as the Einstein Telescope [275] and Cosmic Explorer [605], able to detect BH mergers up to $z \approx 50$ [471]. We discuss this possibility in detail in Sec. 5.4.

## 5.3  Individual GW event confrontation with the PBH model

A crucial step in the analysis of a GW event is the binary parameter reconstruction. This is performed adopting a Bayesian framework by comparing the signal with theoretical waveform models to obtain credible intervals for parameters (such as masses, spins, binary location and distance, ...) characterising the GW source.

Crucially, it was shown that current analysis may still be sensitive to the choice of priors [530–532]. When a physically motivated prior is adopted, the parameters estimated from the GW signal can change significantly and potentially affect the physical interpretation of the source. This may also have particular implications on theoretical models of binary formation and evolution. In this section, we want to address the question of whether the parameters of some events would change if we considered PBH informed priors.

As we discussed in the previous chapters, the PBH model predicts that, if accretion is not efficient during their cosmic history, PBHs are expected to be almost non-spinning. On the other hand, if accretion is efficient, massive binaries tend to be symmetrical and highly spinning. This prediction may not be necessarily compatible with some observed events. The compatibility of the PBH model with the data can be, in principle, quantified by computing the ratio of evidences (Bayes factors) resulting from the analysis performed with either uninformative (denoted also LVC in this section) or PBH priors.

As a test case, we consider few peculiar events from the first and second GWTC catalogs under the assumption that they come from a PBH population with negligible accretion, i.e. high $z_{\rm cut\text{-}off}$. This is compatible with the inferred properties of the PBH model derived in the mixed population inference performed in the preceding section. See Ref. [10] for additional results including accretion effects. Our results show that adopting PBH priors can significantly change the inferred mass ratio and effective spin of some binary black hole events, especially those identified as high-mass, asymmetrical, or spinning by a standard analysis using agnostic priors. Also, for several events, the Bayes factors are only mildly affected by the new priors, implying that it is hard to distinguish whether current merger signals have primordial or astrophysical origin on an event by event basis. In particular, if binaries identified by LIGO/Virgo as strongly asymmetrical (including GW190412) are of primordial origin, their mass ratio inferred from the data can be closer to unity. For GW190412, the latter property is strongly affected by the inclusion of higher harmonics in the waveform model. We will follow the analysis presented in Ref. [10].

### 5.3.1  Parameter estimation with PBH informed priors

We divide currently detected events [439, 443] into 4 categories and study one event in each category. We will consider both the LVC (agnostic) and PBH priors. In particular, we consider (see Table 5.4):

1. GW150914 [159], as representative for moderate mass, symmetrical, binary BH systems;

2. GW170608 [606], as representative for low mass, symmetrical, binary BH systems;

3. GW170729 [439], as representative for moderate mass, asymmetrical, binary BH systems;

Table 5.4: *Extensive summary of the GW events analyzed with PBH-motivated priors constructed assuming negligible accretion effects. The last two columns refer to the analysis including only the dominant $l = m = 2$ harmonic (GW190412*) and also higher harmonics (GW190412), as discussed in the text. We show the 90% credible intervals of the binary parameters obtained with the standard (flat) priors on the (source-frame) masses and spins (LVC) and with PBH-motivated priors (PBH). The last row reports the (decimal) log Bayes factors comparing the PBH to the LVC cases.*

| | | GW150914 | GW170818 | GW170608 | GW170729 | GW190412* | GW190412 |
|---|---|---|---|---|---|---|---|
| | | Moderate mass, Symmetrical, Non-spinning | Moderate mass, Symmetrical, Non-spinning | Low mass, Symmetrical, Non-spinning | Moderate mass, Asymmetrical, Spinning (?) | Low mass, Asymmetrical, Spinning | Low mass, Asymmetrical, Spinning |
| **Param.** | **Prior** | | | | | | |
| $\mathcal{M}\,[M_\odot]$ | LVC | $27.92^{+1.55}_{-1.37}$ | $25.79^{+2.75}_{-2.26}$ | $7.95^{+0.16}_{-0.17}$ | $39.44^{+7.43}_{-7.00}$ | $13.05^{+0.70}_{-0.37}$ | $13.43^{+0.68}_{-0.49}$ |
| | PBH | $28.51^{+1.01}_{-0.88}$ | $25.90^{+2.45}_{-1.78}$ | $7.90^{+0.21}_{-0.19}$ | $32.90^{+4.18}_{-3.29}$ | $12.94^{+0.87}_{-0.48}$ | $13.22^{+0.65}_{-0.53}$ |
| $m_1\,[M_\odot]$ | LVC | $34.66^{+4.77}_{-2.66}$ | $35.15^{+8.87}_{-5.37}$ | $11.49^{+4.02}_{-2.02}$ | $58.61^{+14.88}_{-11.97}$ | $28.90^{+5.15}_{-5.24}$ | $28.55^{+3.17}_{-3.04}$ |
| | PBH | $35.43^{+4.12}_{-2.55}$ | $34.31^{+7.75}_{-4.61}$ | $9.53^{+0.80}_{-0.48}$ | $56.06^{+11.15}_{-12.07}$ | $18.87^{+3.26}_{-3.37}$ | $22.75^{+1.88}_{-2.43}$ |
| $m_2\,[M_\odot]$ | LVC | $29.86^{+2.87}_{-4.19}$ | $25.30^{+5.17}_{-5.99}$ | $7.35^{+1.43}_{-1.71}$ | $35.79^{+13.58}_{-11.97}$ | $8.40^{+1.65}_{-1.04}$ | $8.93^{+0.93}_{-0.76}$ |
| | PBH | $30.40^{+2.87}_{-4.19}$ | $26.18^{+4.42}_{-5.54}$ | $8.65^{+0.45}_{-0.68}$ | $26.5^{+8.98}_{-5.89}$ | $11.98^{+2.57}_{-1.78}$ | $10.46^{+1.23}_{-0.80}$ |
| $q$ | LVC | $0.86^{+0.12}_{-0.20}$ | $0.72^{+0.25}_{-0.27}$ | $0.64^{+0.28}_{-0.28}$ | $0.6^{+0.35}_{-0.24}$ | $0.29^{+0.13}_{-0.07}$ | $0.31^{+0.07}_{-0.05}$ |
| | PBH | $0.86^{+0.12}_{-0.17}$ | $0.77^{+0.20}_{-0.26}$ | $0.91^{+0.08}_{-0.13}$ | $0.47^{+0.32}_{-0.14}$ | $0.63^{+0.30}_{-0.16}$ | $0.46^{+0.11}_{-0.06}$ |
| $\chi_{\text{eff}}$ | LVC | $-0.06^{+0.10}_{-0.13}$ | $-0.05^{+0.19}_{-0.22}$ | $0.06^{+0.14}_{-0.05}$ | $0.29^{+0.22}_{-0.27}$ | $0.22^{+0.09}_{-0.11}$ | $0.20^{+0.06}_{-0.07}$ |
| | PBH | $0.00^{+0.00}_{-0.04}$ | $0.00^{+0.00}_{-0.00}$ | $0.00^{+0.00}_{-0.00}$ | $0.00^{+0.01}_{-0.00}$ | $0.00^{+0.00}_{-0.01}$ | $0.00^{+0.00}_{-0.00}$ |
| $z$ | LVC | $0.10^{+0.03}_{-0.04}$ | $0.26^{+0.09}_{-0.10}$ | $0.07^{+0.02}_{-0.02}$ | $0.29^{+0.07}_{-0.13}$ | $0.17^{+0.03}_{-0.07}$ | $0.13^{+0.04}_{-0.05}$ |
| | PBH | $0.09^{+0.03}_{-0.04}$ | $0.26^{+0.09}_{-0.11}$ | $0.07^{+0.03}_{-0.03}$ | $0.26^{+0.09}_{-0.13}$ | $0.17^{+0.04}_{-0.07}$ | $0.12^{+0.05}_{-0.05}$ |
| $\log(Z_{\text{PBH}}/Z_{\text{LVC}})$ | | $-0.43 \pm 0.10$ | $-0.30 \pm 0.06$ | $0.61 \pm 0.12$ | $-1.74 \pm 0.06$ | $-0.83 \pm 0.12$ | $-2.26 \pm 0.10$ |

4. GW190412* [443], as representative for low mass, asymmetrical, binary BH systems. As for the previous sources, here in the analysis we considered only the dominant $l = m = 2$ harmonic;

5. GW190412 [443], same as the previous entry, but including also higher harmonics in the waveform model, see details in App. C.3.

We define low-mass and moderate-mass binaries (depending on whether $m_{1,2} \lesssim 20 M_\odot$), between symmetrical and asymmetrical systems (depending on whether $q$ is compatible with unity), and between non-spinning and spinning binaries (depending on whether the posterior distribution of $\chi_{\text{eff}}$ is compatible with zero or not). We stress that these criteria are expressed in terms of parameters originally obtained with the standard priors on the masses and the spins adopted by the LVC (see Tab. 5.4).

As stated before, for simplicity we are going to assume a PBH scenario with negligible accretion and a lognormal mass distribution as compatible with the population inferences performed in the preceding sections. The results of the analysis are summarised in Tab. 5.4, while some details on the setup of the Bayesian parameter estimation are reported in App. C.3.

Let us discuss the result of the analysis for GW150914 [159], whose posterior distribution is shown in Fig. 5.22 (left panel). With the PBH-motivated priors, the effective spin is assumed to be negligible and one obtains a slightly better measurement of the chirp mass since the dimensionality of the waveform parameter space is effectively reduced. However, the new posterior is well within the $1\sigma$ contour of the standard one, whereas the posterior distribution of the mass ratio is almost unaffected. This is simply understood by noticing that the effective spin parameter of GW150914 as measured by the LVC is compatible with zero, so a narrow prior $\chi_{\text{eff}} \approx 0$ is compatible with the event and does not affect the inference on the other parameters. Likewise, the chirp mass estimated for GW150914 is well within the best-fit lognormal distribution inferred for PBHs. Also, the Bayes factors are not favouring either model, showing this event is not incompatible with the primordial hypothesis and the effect of the PBH-motivated priors without accretion is negligible.

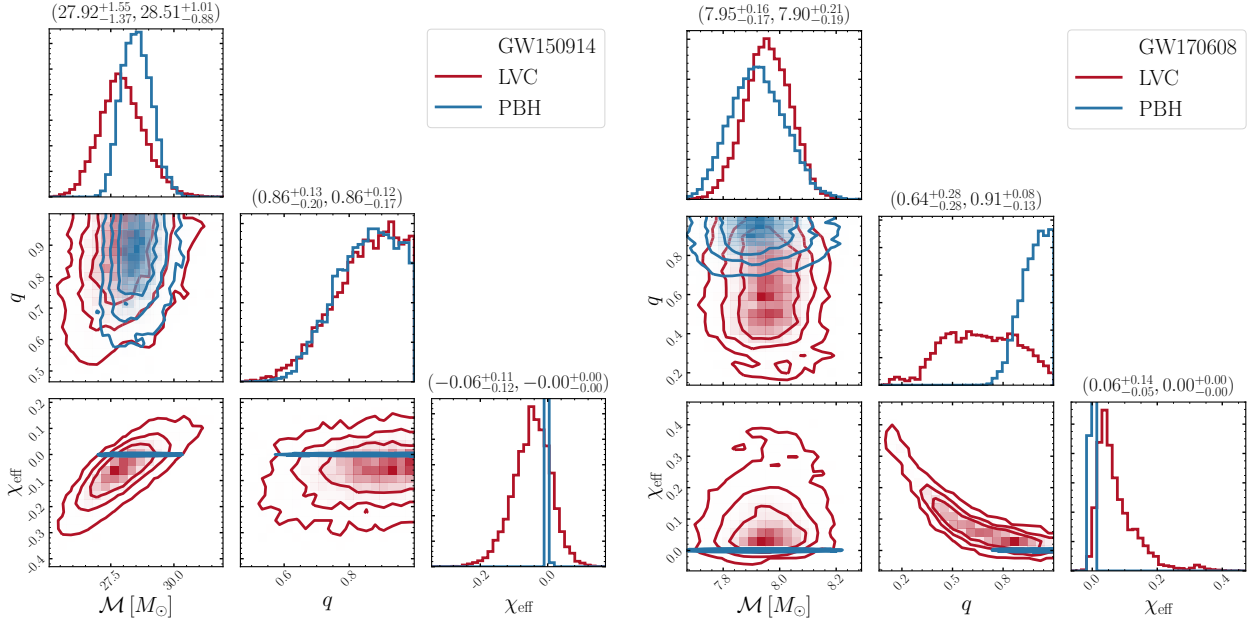

Figure 5.22: **Left:** *Posterior distributions for the binary BH parameters of GW150914. The LVC case is shown in red while the results with PBH-motivated priors correspond to blue contours. The 90% C.I. reported on top of each column follows (LVC,PBH) ordering.* **Right:** *Same as left panel but for GW170608.*

The results of the analysis of GW170608 [606] are shown in Fig. 5.22 (right panel). In this case the PBH prior forces the posterior of $\chi_{eff}$ close to zero, in a small tail of the LVC posterior. This also impacts the posterior distribution of $q$ significantly, allowing for a more accurate measurement which is skewed towards larger values ($q = 0.91^{+0.08}_{-0.13}$), although still compatible with the standard-prior case. Also in this case, the Bayes factor does not sufficiently disfavour either scenario giving a slight advantage to the PBH model.

The results for GW170818 [439] follows the discussion above and does not provide enough evidence to prove/disprove the primordial hypothesis. On the other hand, GW170729, which was found to be a moderately spinning event by the LVC analysis, seems to disfavour the PBH hypothesis in the conservative scenario with negligible accretion, with a Bayes factor $\log \mathcal{B}^{PBH}_{LVC} = -1.74 \pm 0.06$.

Finally, let us focus on GW190412, the first binary BH event published from the O3 run [443]. The results are shown in Fig. 5.23 for both the cases without (left panel) and with (right panel) the inclusion of higher harmonics in the waveform. This event deserves particular attention as it confidently corresponds to a spinning ($\chi_{eff} = 0.22^{+0.09}_{-0.11}$) and asymmetric ($q = 0.29^{+0.13}_{-0.07}$) binary in the LVC analysis. However, different prior assumptions might qualitatively change the inferred binary BH parameters [532, 607, 608].

First, we observe that the posteriors for $\mathcal{M}$ as shown in Fig. 5.23 are very similar, in both cases with (right) and without (left) the inclusion of higher harmonics. However, for this event $\chi_{eff} = 0$ is excluded roughly at $3.3\sigma$ confidence level if one adopts the standard flat priors on the spins. This is in tension with the PBH-motivated prior without accretion which imposes $\chi_{eff} \approx 0$. As a consequence, the Bayes factors weakly (decisively) disfavour the PBH interpretation in the absence of accretion in the analysis excluding (including) higher harmonics.

Also, the measurement of the mass ratio is strongly affected to compensate for this tension, since $q$ and $\chi_{eff}$ are correlated in the waveform. Specifically, the distribution of $q$ broadens up and gains support at higher values. When the higher harmonics are included (right corner plot in Fig. 5.23), the distribution of $q$ is still shifted towards larger values relative to the agnostic-prior case, but it loses support near $q \approx 1$ with respect to the PBH-prior case with the dominant $l = m = 2$ harmonic only. The reason can be understood by the fact that for GW190412 the subdominant $l = m = 3$ mode was confidently detected [443], which implies a certain degree of asymmetry for the system. The latter can arise mainly from $q \neq 1$ and, to a less extent, from misaligned spin vectors. Since our waveform model for this analysis (IMRPhenomHM) assumes aligned spin vectors, the only asymmetry can come from the mass ratio of the system, which cannot be unity. Nonetheless, the qualitative result holds: imposing $\chi_{\mathrm{eff}} \approx 0$ results in a larger inferred value of $q$ compared to the agnostic prior case. We expect this to be a generic result that should apply to any binary for which the distribution of $\chi_{\mathrm{eff}}$ (as inferred using standard priors) is incompatible with zero. Finally, we note that the redshift is almost insensitive to the different choice of priors in all cases considered, see Tab. 5.4.

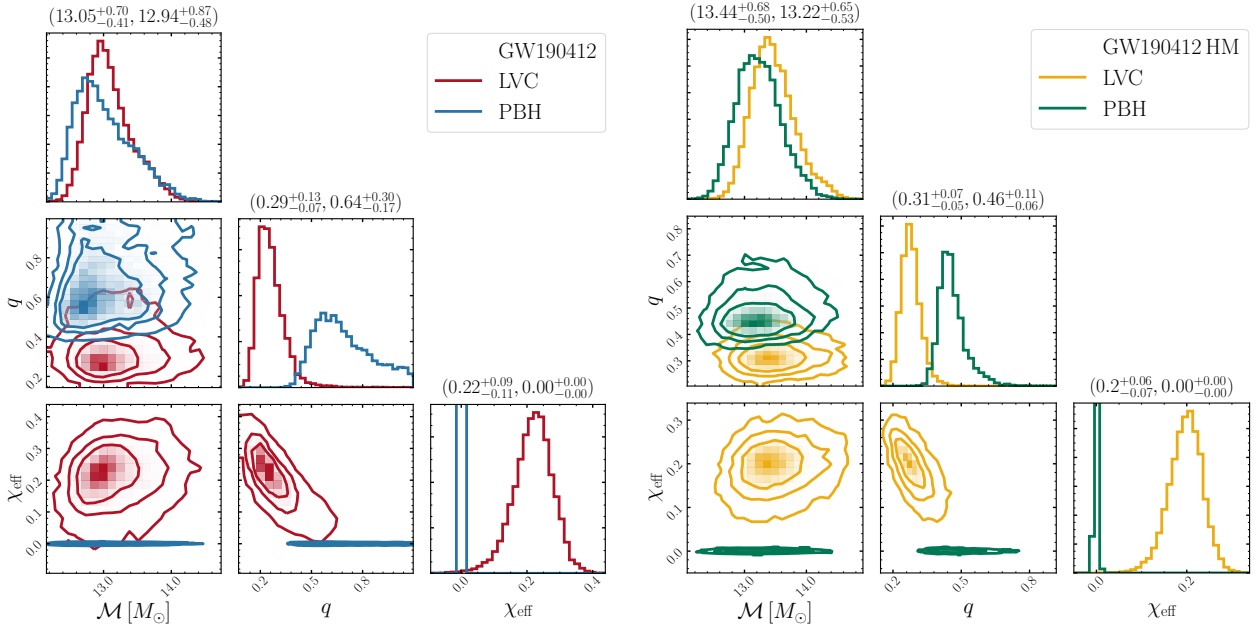

Figure 5.23: *Same as in Fig. 5.22 but for GW190412.* **Left:** *Scenario where only the dominant mode is included.* **Right:** *Scenario including higher harmonics in the waveform model.*

## Summary

Based on the results we presented in the preceding discussion, we can draw few general conclusions which we summarise in the following list:

- The chirp mass and redshift of the binary are almost insensitive to assuming a different prior motivated by the PBH model. On the other hand, the posterior distribution of the mass ratio and the effective spin of some events can be significantly affected.

- A prior predicting negligible spins (as suggested by the PBH scenario without efficient accretion) can strongly affect the parameter estimation of binaries identified as spinning by the LVC analysis. In particular, due to the correlation between mass ratio and spins [564–569], the posterior distribution of the mass ratio gains more support close to unity relative to the LVC case. Thus, binaries identified as (highly) spinning and (highly) asymmetrical using standard agnostic priors might be non-spinning and symmetrical. Interestingly enough, for most of the events considered in this work, the Bayes factors do not strongly favour one hypothesis against the other, putting both interpretations on the same footing.

- As a consequence of the previous point, if GW190412 is assumed to be of primordial origin, its mass ratio inferred from the data is larger than in the standard agnostic-prior case. In particular, neglecting higher harmonics in the waveform even $q = 1$ is compatible within $2\sigma$. Extending the analysis to higher harmonics [9], the posterior of $q$ loses support near $q \approx 1$, since the detection of a $l = m = 3$ mode for GW190412 requires some asymmetry in the system. Finally, also the evidence is affected by the inclusion of the higher harmonics, and the preference for the LVC scenario is increased. Notice that this conclusion is compatible with the results of Sec. 5.2.4 confidently assigning GW190412 to the isolated astrophysical formation channel.

---

[9]We note that the waveform we adopt for this analysis only describes aligned spins, see App. C.3.

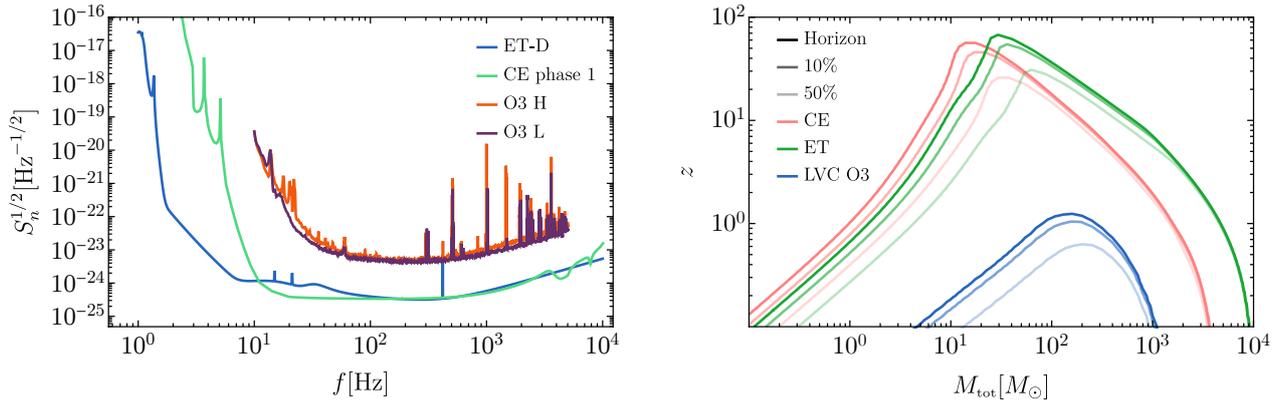

Figure 5.24: *Result for the LIGO Hanford (H) and Livingston (L) during the O3 run [274] and both the 3G detectors Cosmic Explorer (CE) during phase 1 from [276] and ET at design sensitivity from [275].* **Left:** *Noise curves.* **Right:** *Horizon redshift for equal mass binaries. We defined the horizon as the maximum distance at which a binary can in principle be observed if optimally oriented with respect to the detectors, while 10%, 50% as the redshift at which those fractions of binaries are observable, corresponding to $SNR = (8, 10, 19)$, respectively.*

- The latter discussion also shows the relevance of an accurate waveform modelling for GW190412. While the log Bayes factor does not favour either interpretation of the event when only the dominant harmonic in the waveform, the inclusion of higher harmonics breaks this degeneracy. This shows that modelling the waveform accurately is essential to perform model selection.

Many qualitative aspects of this analysis are expected to remain valid beyond the PBH scenario. In particular, the conclusions for the non-accreting PBH case are also valid for other scenarios in which the binary BHs have an astrophysical origin predicting small spins. In this case, we expect the mass ratio of binaries identified as spinning using LVC priors should be strongly affected and grow to compensate for a lacking effective spin.

An interesting extension of this analysis would be to repeat the parameter estimation of GW190412 using a waveform model which included both higher harmonics and precession, to check whether the inclusion of spin misalignment could introduce a sufficient asymmetry into the system and modify the posteriors and the evidence analysis. Finally, it would be interesting to compare the odds of the PBH hypothesis with those of different astrophysical scenarios, e.g. hierarchical mergers [449, 608], within the Bayesian framework used here. This would be of particular interest in light of the result of Sec. 5.2.4 where a degeneracy of the PBH model with the dynamical formation channels was found.

## 5.4  Primordial black hole mergers at 3G detectors

We now focus our attention on the future of GW experiments. New designs, extending the reach of current LIGO/Virgo detectors, are already been considered and planned. A couple of noticeable examples are given by the Einstein Telescope (ET) and Cosmic Explorer (CE) [275]. Those apparatuses will be able to reduce the detector noise by more than one order of magnitude compared to current technology, by also extending the range of frequencies currently observed, see Fig. 1.9. This will allow reaching much further signals coming from binaries with a wider range of masses.

To get an idea of the radical improvement expected within the next two decades, we show in Fig. 5.24 the expected strain noise planned to be reached at ET and CE, along with the horizon redshift for equal mass binaries with a total mass $M_{tot}$ in the source frame. As one can appreciate, 3G detectors will be able to observe local mergers with a much higher SNR, allowing for a more precise determination of the binary parameters [609]. Also, they will be able to observe binaries up to redshift as large as $z \gtrsim 50$ in their optimal mass window. This property, as we will see, can be exploited to unequivocally discover a PBH population of binaries. We will discuss this possibility by following Ref. [5].

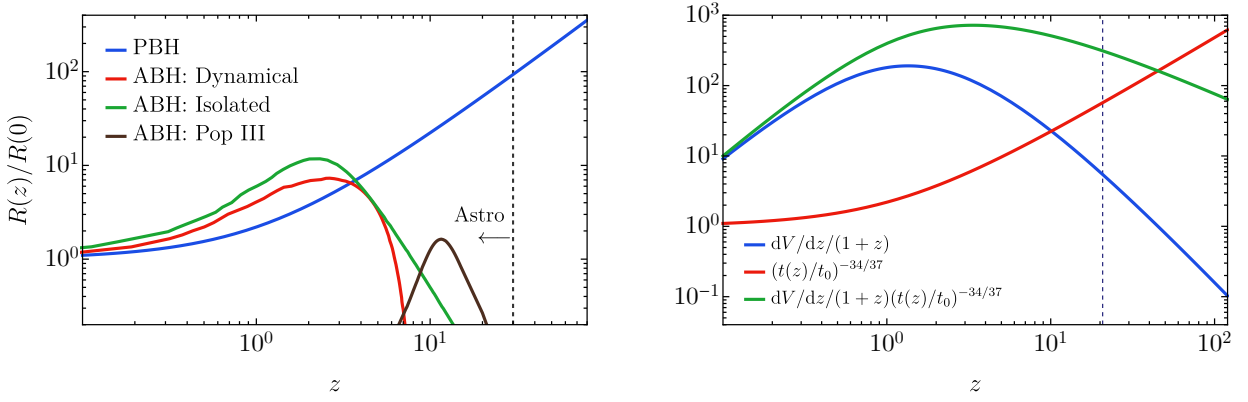

Figure 5.25: **Left:** Merger rate evolution of the PBH model compared to various representative astrophysical channels (dynamical, isolated and Pop III, see Ref. [579] and Refs. therein). The vertical dashed line indicates redshift $z \approx 30$, above which one does not expect any ABH mergers and where the PBH merger rate can still be sizeable. **Right:** Comoving volume factor $dV/dz$ (in units of $Gpc^3$), redshift factor $1/(1+z)$ and their contribution combined with the PBH merger rate.

One of the most prominent differences between the PBH scenario and any astrophysical formation scenario is the predicted redshift evolution of the merger rate. Even though the GWTC-2 catalog is still limited to "local" (i.e. low redshift) GW signals coming from sources with $z \lesssim 1$, future observations with extended horizons will be able to test the merger rate evolution to distinguish between different binary BH populations at current [610–612] and future detectors [579, 613–615].

The redshift evolution of the merger rate density for the PBH model is found to be monotonically increasing as [12, 169, 170]

$$R_{\rm PBH}(z) \approx (t/t_0)^{-34/37} , \qquad (5.4.1)$$

extending up to redshifts $z = \mathcal{O}(10^3)$. Notice that the evolution of the merger rate with time shown in Eq. (5.4.1) is entirely determined by the binary formation mechanism (i.e. how pairs of PBHs decouple from the Hubble flow) before the matter-radiation equality era, see Sec. 4.5 and Eq. (4.6.36). Eq. (5.4.1) is, therefore, a robust prediction of the PBH model assuming the standard formation scenario where PBHs are generated with an initial spatial Poisson distribution. On the other hand, the merger rate of astrophysical binaries is expected to peak at redshift of a few with a possible second peak coming from a Pop III population at redshift $z \sim 10$ [579, 616–618]. In Fig. 5.25, we show a comparison between the merger rate evolution of different astrophysical channels and the PBH model.

The total number of events per unit time produced in the PBH model is also dependent on the volume $(dV/dz)$ and redshift $((1+z)^{-1})$ factors as (see App. C and the discussion around Eq. (C.10))

$$N_{\rm det}/T_{\rm obs} \equiv \int dm_1 dm_2 dz \, p_{\rm det}(m_1, m_2, z) \, \frac{1}{1+z} \frac{dV}{dz} \frac{dR_{\rm PBH}}{dm_1 dm_2}, \qquad (5.4.2)$$

where $p_{\rm det}$ accounts for the detection bias. We show the distribution with redshift of the volume/redshift factors, alongside the PBH intrinsic merger rate, in Fig. 5.25. As the volume accessible in each redshift bin has its maximum value close to $z \approx \mathcal{O}(1)$, the monotonically growing PBH merger rate give rise to a distribution of events in redshift which is peaking at $z \approx \mathcal{O}(\text{few})$, while still admitting a sizeable contribution at very high redshift ($z \gtrsim 30$). In other words, the limiting factor in the search for high-redshift mergers will be the experiment reach.

Astrophysical models predict that the first BHs are born from Pop III stars [619–622]. Their formation should, however, only take place at $z \lesssim 25$ [623]. We will conservatively assume that they form below redshift $z = 30$. Furthermore, their merger time depends on the formation mechanisms of the binaries and could range from $\mathcal{O}(\text{Gyr})$ (in which case they merge at $z \lesssim 6$) to $\mathcal{O}(10 \, \text{Myr})$ if they form dynamically in Pop III clusters, in which case they would merge almost at the redshift of BH formation [623].

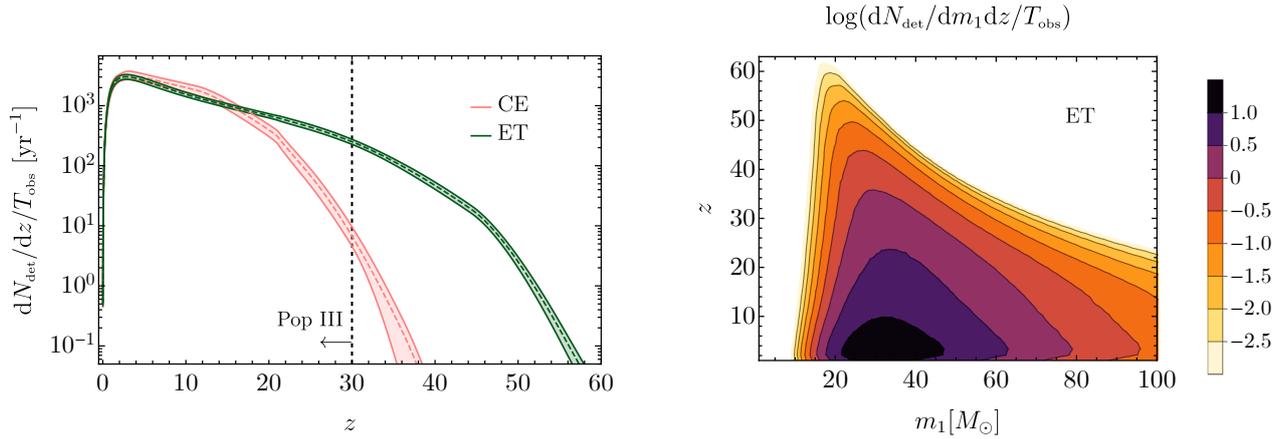

Figure 5.26: ***Left:*** *Distribution of observable events per year as a function of redshift at ET and CE coming from the PBH subpopulation.* ***Right:*** *Corresponding distribution of observable events as a function of both primary mass and redshift at ET.*

Therefore, assuming the most conservative scenario (i.e. including BH mergers formed from Pop III high redshift clusters), the detection of a binary BH with $z \sim 30$ would be a smoking-gun signal in favour of PBHs, as no astrophysical contamination is expected at such high redshift in standard cosmologies, see also Refs. [624, 625]. We can, therefore, make a forecast for the observability of a PBH population of binaries at high redshifts through 3G detectors by assuming the PBH population we found to be compatible with the GWTC-2 analysis in the previous section. For definiteness, we consider the inference performed in Sec. 5.2.3 and follow the results of Ref. [5].

In Fig. 5.26, we show the expected number of observable events per year at ET and CE as a function of redshift. As one can appreciate, both ET and CE are expected to observe a copious number of events per year up to redshift $z \approx 30$ coming from the PBH population we consider. Note that the larger observation rate at high redshift expected for ET comes from superior behaviour of the adopted ET-D sensitivity curve at low-frequencies when compared to the CE phase-1 design (see Fig. 5.2). The low-frequency reach is particularly relevant for heavy and/or high-redshift mergers. In the right panel of Fig. 5.26, we also show the distribution as a function of the primary mass. As expected by looking at the observable horizon shown in Fig. 5.2, most of the distant binaries will have their primary mass around the peak of the experiment sensitivity, which is $m_1 \simeq 20 M_\odot$, not far from the characteristic PBH mass function scale $M_c$. Finally, we also computed the total number of observable distant events (i.e. integrated for $z > 30$) per year for both ET and CE detectors, finding

$$N_{\rm det}^{\rm ET}(z > 30) = 1315^{+305}_{-168}\,/{\rm yr}, \qquad\qquad N_{\rm det}^{\rm CE}(z > 30) = 12^{+22}_{-11}\,/{\rm yr}. \qquad (5.4.3)$$

We conclude that 3G detectors have the unique potential to unequivocally confirm the presence of a PBH subpopulation in the currently available dataset and to unveil a novel (primordial) family of BBHs.

Notice that any conclusion which is drawn on an event by event basis crucially depends on the measurement accuracy of the source redshift, which can be low for the most distant events even in the 3G era [579]. Another possibility is to focus on constraining the merger rate evolution at high redshift, thus increasing the information combining multiple detections [626].

In this section, we focused on confirming the origin of BH mergers using the difference between the number of events expected at high redshift between the ABH and PBH scenarios. It is interesting to mention that one could also try using the clustering properties of mergers to distinguish the two populations. As ABH and PBH are characterised by a different linear bias, it was proposed that cross-correlating the GW event catalogue or the SGWB with the large-scale structure probes [627–632] could help to distinguish between the two populations. Let us conclude by also noting that there may be other smoking-gun signatures of PBHs, one of which being the detection of a BH with a subsolar mass. This could be achieved at ground-based detectors searching for subsolar PBH binaries [423, 633–636] or at LISA by searching for Extreme Mass Ratio Inspirals (EMRI) involving a subsolar PBH merging with a supermassive BH [637–639]. A potential contamination from the astrophysical sector in this mass range could be the existence of a population of $\mathcal{O}(1)M_\odot$ BHs coming from transmuted neutron stars could form [640, 641] potentially contaminating a portion of the mass range below the Chandrasekhar's limit with non-primordial objects. However, this population may be distinguished from PBHs as it may posses a mass distribution correlated with the neutron star one [640, 641].

# Chapter 6

# Gravitational Waves as footprints of the PBH production

This chapter is devoted to the study of GW signatures related to the PBH production mechanism. We will primarily focus on the generation of a SGWB background induced at second order by the scalar perturbations which are responsible for the generation of overdensities regions in the early universe that can collapse to form PBHs. This SGWB, as we will see, can be used as a new tool to test the abundance of PBHs coming from the collapse of density perturbations, see Ref. [642] for a recent review.

The fundamental property relating the PBH formation models with the GW sector, is the relation between the PBH mass and GW frequency. As we will see in details, the GWs are emitted at the time of horizon crossing of the perturbation, that is when $k = 2\pi f = aH$. Since the PBH mass is a significant fraction of the horizon mass, see discussion in Sec. 2.1, one finds [643]

$$f \simeq 3\,\mathrm{mHz} \left( \frac{M_{\mathrm{PBH}}}{10^{-12} M_\odot} \right)^{-1/2}.$$ (6.0.1)

From this relation, we see that PBHs with masses of order $\approx M_\odot$ correspond to frequencies close to nHz, probed by Pulsar Timing Array experiments like PPTA [644], NANOGrav [645] and EPTA [646] and, in the future, SKA [647] (see also [648]). For PBHs with masses around $\approx 10^{-12} M_\odot$, instead, the frequency falls in the band observable by LISA [277]. [1]

It is important to stress that the emission of GWs is not related to the dynamics of collapse. Even though each gravitational collapse is a violent event, since PBHs are rare, their formation do not give the dominant contribution to the expected SGWB. A sizeable background is instead generated in each Hubble patch when large scalar perturbations cross the horizon [98, 643, 649–658]. As we showed, models producing PBHs have a characteristic enhancement of perturbations at small scales reaching $\mathcal{P}_\zeta(k_{\mathrm{PBH}}) \approx 10^{-2}$, which makes the SGWB potentially observable at current and future experiments like LIGO/Virgo and LISA. Given the potential relevance of these signals in upcoming GW experiments, it is important to characterise the SGWB produced by this mechanism [21, 22]. [2]

We stress that an additional contribution to the SGWB investigated in this chapter comes from the dynamics of overdensities produced in the PBH formation model which are not large enough to form PBHs. As they are inevitably contracting under the action of the gravitational force upon horizon crossing, they generate a time-dependent quadrupole leading to the emission of GWs with the same frequency as the one characterising the SGWB induced at second order. The amplitude of the SGWB produced with this mechanism is found to be, however, at least one order of magnitude smaller than the signal discussed in this chapter. We refer to Ref. [29] for more details about this mechanism.

---

[1] This frequency relation should not be confused with the one connecting GWs to the mass of PBH mergers, which follows completely different dynamics, discussed in details in Sec. 5.

[2] Other mechanisms generating a SGWB in the early universe can be found in Refs. [287, 659–666]. Also, we will not consider the GWs induced at second order by non-Gaussian scalar perturbations, as done in Ref. [228, 229, 667–669].

In the first section, we will focus on the properties of the induced SGWB spectrum, while moving to the characterisation of the GW three-point function and anisotropies in Secs. 6.2, 6.3 and 6.4, respectively. Finally, we will briefly discuss the potential gauge dependence of the SGWB before studying in details the expected signals at LISA and Pulsar Timing Array experiments coming from motivated PBH models.

## 6.1 SGWB induced at second order

In this section, we will describe the generation of a SGWB from scalar perturbations at second order in perturbation theory. We will show few results considering characteristic power spectra typically encountered in PBH formation models, by following closely Refs. [17, 21, 22].

### 6.1.1 SGWB frequency spectrum

We start by defining the second order tensor perturbation of the metric as

$$ds^2 = a^2(\eta) \left[ -d\eta^2 + \left( \delta_{ij} + \frac{1}{2} h_{ij} \right) dx^i dx^j \right],$$ (6.1.1)

where the spin-2 tensor $h_{ij}$ can be decomposed into its helicities in Fourier space as

$$h_{ij}(\eta, \vec{x}) = \int \frac{d^3k}{(2\pi)^3} \sum_{\lambda=R,L} h_\lambda(\eta, \vec{k}) \, e_{ij,\lambda}(\hat{k}) \, e^{i\vec{k}\cdot\vec{x}}.$$ (6.1.2)

By definition, the circular polarization operators $e_{ij,\lambda}$ are transverse and traceless, and satisfy the orthonormality condition

$$e_{ij,\lambda}(\vec{k}) e_{ij,\lambda'}^*(\vec{k}) = \delta_{\lambda\lambda'}.$$ (6.1.3)

The dynamics is dictated by Einstein's equation which, expanded up to second order in scalar perturbations in the Newtonian gauge, takes the form

$$h_{ij}'' + 2\mathcal{H} h_{ij}' - \nabla^2 h_{ij} = -4\mathcal{T}_{ij}{}^{\ell m} \mathcal{S}_{\ell m},$$ (6.1.4)

where the prime denotes the derivative with respect to conformal time $\eta$, $d\eta = dt/a$ and $\mathcal{H} = a'/a$ is the conformal Hubble parameter as a function of the scale factor $a(\eta)$.[3] The tensor $\mathcal{T}_{ij}{}^{\ell m}$ projects the right-hand side onto its transverse and traceless components. It is defined in Fourier space as

$$\mathcal{T}_{ij}^{lm} = e_{ij,L} \otimes e_L^{lm} + e_{ij,R} \otimes e_R^{lm},$$ (6.1.5)

where $e_{ij,\lambda}(\vec{k})$ are the polarisation tensors written in the chiral basis (L, R). In Eq. (6.1.4), the source term $\mathcal{S}_{\ell m}$ takes the form [649]

$$\mathcal{S}_{ij} = 4\Psi \partial_i \partial_j \Psi + 2\partial_i \Psi \partial_j \Psi - \partial_i \left( \frac{\Psi'}{\mathcal{H}} + \Psi \right) \partial_j \left( \frac{\Psi'}{\mathcal{H}} + \Psi \right)$$ (6.1.6)

in a radiation dominated (RD) universe. We do not consider the free-streaming effect of neutrinos on the GW amplitude [670], and therefore we have equated the Bardeen's potentials

$$\Psi = \Phi.$$ (6.1.7)

We discuss more details of the cosmological perturbation theory in App. E. The linear scalar perturbation $\Psi(\eta, \vec{k})$ defining the source can be related to the gauge-invariant comoving curvature perturbation using

$$\Psi(\eta, \vec{k}) = \frac{2}{3} T(k\eta) \zeta(\vec{k}),$$ (6.1.8)

---

[3]In Sec. 6.5, we will comment about differences in the source term encountered when different gauges are adopted and whether this issue is relevant as far as the predicted signal at present day experiments is concerned.

where the transfer function $T(k\eta)$ in the radiation-dominated era defined in Eq. (2.1.25).

We can already identify few properties of the GW source by investigating Eq. (6.1.4). As the source presents gradients of the scalar perturbations, it is suppressed at super-Hubble scales as

$$\mathcal{S} \sim \left(\frac{k}{\mathcal{H}}\right)^2 \qquad \text{for} \qquad k \ll \mathcal{H}. \tag{6.1.9}$$

On the other hand, at sub-Hubble scales, the curvature modes are damped by the radiation pressure, the source falls at a rate

$$\mathcal{S} \sim \left(\frac{k}{\mathcal{H}}\right)^{-2} \qquad \text{for} \qquad k \gg \mathcal{H} \tag{6.1.10}$$

due to the linear transfer function. It follows that GWs are mainly generated at the time of horizon crossing of the relevant scalar perturbations. In the PBH scenario, the relevant modes re-enter the horizon deep in the radiation dominated phase of the universe. For this reason, we will focus only on the GW emission in a radiation dominated universe. Secondly, as the source is intrinsically defined at second-order in the scalar perturbation $\Psi$, it will give rise to non-vanishing higher-order correlators of the tensor modes. This reveals the intrinsic non-Gaussian nature of the GWs generated through this mechanism. Finally, since the source contains two spatial derivatives, the resulting bispectrum is expected to peak in the equilateral configuration of the momenta in Fourier space.

Using the Green's function method and by defining the dimensionless variables $x = p/k$ and $y = |\vec{k} - \vec{p}|/k$, the solution of the equation of motion (6.1.4) is found to be

$$h_{\vec{k}}^\lambda(\eta) = \frac{4}{9} \int \frac{\mathrm{d}^3 p}{(2\pi)^3} \frac{1}{k^3\eta} e^\lambda(\vec{k},\vec{p}) \zeta(\vec{p}) \zeta(\vec{k}-\vec{p}) \Big[ \mathcal{I}_c(x,y) \cos(k\eta) + \mathcal{I}_s(x,y) \sin(k\eta) \Big], \tag{6.1.11}$$

where we have introduced the compact notation

$$e^\lambda(\vec{k},\vec{p}) = e_{ij}^\lambda(\vec{k}) p^i p^j \tag{6.1.12}$$

and the time integrated transfer functions [671, 672]

$$\begin{aligned}
\mathcal{I}_c(x,y) &= 4 \int_0^\infty \mathrm{d}\tau\, \tau(-\sin\tau) \left[ 2T(x\tau)T(y\tau) + \big(T(x\tau) + x\tau\, T'(x\tau)\big) \big(T(y\tau) + y\tau\, T'(y\tau)\big) \right], \\
\mathcal{I}_s(x,y) &= 4 \int_0^\infty \mathrm{d}\tau\, \tau(\cos\tau) \left[ 2T(x\tau)T(y\tau) + \big(T(x\tau) + x\tau\, T'(x\tau)\big) \big(T(y\tau) + y\tau\, T'(y\tau)\big) \right].
\end{aligned} \tag{6.1.13}$$

The exact analytical expressions of $\mathcal{I}_c(x,y)$ and $\mathcal{I}_s(x,y)$ can be found in App. D of Ref. [671] and reads

$$\mathcal{I}_c(x,y) = -36\pi \frac{(s^2+d^2-2)^2}{(s^2-d^2)^3} \Theta(s-1) \,, \tag{6.1.14}$$

$$\mathcal{I}_s(x,y) = -36 \frac{(s^2+d^2-2)}{(s^2-d^2)^2} \left[ \frac{(s^2+d^2-2)}{(s^2-d^2)} \log\frac{(1-d^2)}{|s^2-1|} + 2 \right], \tag{6.1.15}$$

with

$$d \equiv \frac{1}{\sqrt{3}}|x-y|, \qquad s \equiv \frac{1}{\sqrt{3}}(x+y), \qquad (d,s) \in [0, 1/\sqrt{3}] \times [1/\sqrt{3}, +\infty). \tag{6.1.16}$$

It is useful to define the GW power spectrum as

$$\left\langle h^{\lambda_1}(\eta, \vec{k}_1) h^{\lambda_2}(\eta, \vec{k}_2) \right\rangle' \equiv \delta^{\lambda_1\lambda_2} \frac{2\pi^2}{k_1^3} \mathcal{P}_h(\eta, k_1), \tag{6.1.17}$$

where, in the radiation-dominated era, $\mathcal{P}_h(\eta, k)$ takes the form [22]

$$\begin{aligned}
\mathcal{P}_h(\eta, k) = \frac{4}{81} \frac{1}{k^2\eta^2} \iint_{\mathcal{S}} \mathrm{d}x\, \mathrm{d}y \frac{x^2}{y^2} \left[ 1 - \frac{(1+x^2-y^2)^2}{4x^2} \right]^2 \mathcal{P}_\zeta(kx) \mathcal{P}_\zeta(ky) \\
\times \left[ \cos^2(k\eta) \mathcal{I}_c^2 + \sin^2(k\eta) \mathcal{I}_s^2 + \sin(2k\eta) \mathcal{I}_c \mathcal{I}_s \right].
\end{aligned} \tag{6.1.18}$$

In the previous equation, the integration domain $\mathcal{S}$ corresponds to the region in the $(x, y)$ plane satisfying the triangular inequality, see also Fig. 2 of Ref. [671].

We can compute the energy density of the SGWB emitted by using its general relativistic definition [286][4]

$$\rho_{\mathrm{GW}} = \frac{M_p^2}{16} \left\langle \dot{h}_{ab}(t, \vec{x}) \, \dot{h}_{ab}(t, \vec{x}) \right\rangle_T, \tag{6.1.19}$$

in terms of the time derivative of the tensor field, averaged over a time interval $T$ much longer than the GW period. In the following, we will use multiple times the same averaging procedure. When performing time averages, it is enough to take out the quickly oscillating pieces in the right hand side of Eqs. (6.1.11) or (6.1.18) by using

$$\langle \sin^2(k\eta) \rangle_T = \langle \cos^2(k\eta) \rangle_T = 1/2, \qquad \text{while} \qquad \langle \cos(k\eta) \sin(k\eta) \rangle_T = 0. \tag{6.1.20}$$

In terms of the GW power spectrum, the energy density takes the form

$$\Omega_{\mathrm{GW}}(\eta, k) = \frac{\rho_{\mathrm{GW}}(\eta, k)}{\rho_b(\eta)} = \frac{1}{24} \left( \frac{k}{\mathcal{H}(\eta)} \right)^2 \overline{\mathcal{P}_h(\eta, k)}, \tag{6.1.21}$$

where the overline denotes an average over conformal time $\eta$ and $\rho_b$ the background energy density of the universe.

As the GW density scales as $\propto 1/a^4$ since GWs are decoupled, we can relate $\rho_{\mathrm{GW}}$ to the radiation density $\rho_r(\eta_f)$ at the time all GWs have been emitted, up to a coefficient accounting for the change in the number of effective degrees of freedom. Indeed, the radiation density $\rho_r(\eta_f)$ is related its value today using conservation of entropy, giving

$$c_g \equiv \frac{a_f^4 \rho_r(\eta_f)}{\rho_r(\eta_0)} = \frac{g_*}{g_*^0} \left( \frac{g_{*S}^0}{g_{*S}} \right)^{4/3} \approx 0.4, \tag{6.1.22}$$

where $g_{*S}$ is the effective degrees of freedom for entropy density and we assume $g_{*S} \approx g_* \approx 106.75$ from the Standard Model at time $\eta_f$. The indicative value of 0.4 is found by assuming all the emission takes place before the universe cools below the temperature of decoupling of the Top quark. Therefore, the present density of GWs is found to be

$$\Omega_{\mathrm{GW}}(\eta_0, k) = \frac{a_f^4 \rho_{\mathrm{GW}}(\eta_f, k)}{\rho_r(\eta_0)} \Omega_{r,0} = c_g \frac{\Omega_{r,0}}{24} \frac{k^2}{\mathcal{H}(\eta_f)^2} \overline{\mathcal{P}_h(\eta_f, k)}, \tag{6.1.23}$$

where $\Omega_{r,0}$ is the present radiation energy density fraction. The time average of the oscillating terms in Eq. (6.1.18), together with the simplification of the time factor coming from the relation $\mathcal{H}^2(\eta_f) = 1/\eta_f^2$ (valid as long as $\eta_f < \eta_{\mathrm{eq}}$), gives the current abundance of GWs

$$\Omega_{\mathrm{GW}}(k) = \frac{c_g}{972} \Omega_{r,0} \iint_{\mathcal{S}} \mathrm{d}x \mathrm{d}y \frac{x^2}{y^2} \left[ 1 - \frac{(1 + x^2 - y^2)^2}{4x^2} \right]^2 \mathcal{P}_{\zeta}(kx) \, \mathcal{P}_{\zeta}(ky) \, \mathcal{I}^2(x, y), \tag{6.1.24}$$

where we defined $\mathcal{I}^2 \equiv \mathcal{I}_c^2 + \mathcal{I}_s^2$. Two important properties can be highlighted here:

- In the low $k$ tail (i.e. for modes much smaller than the one corresponding to the peak of the curvature power spectrum $k_*$), it can be shown that the GW spectrum is always characterised by a scaling

$$\Omega_{\mathrm{GW}}(k \ll k_\star) \propto k^3, \tag{6.1.25}$$

  which is entirely dictated by causality [671, 673, 674]. This behaviour is violated only in the case of an exactly monochromatic power spectrum, which is however an unphysical approximation of a narrow spectrum.

---

[4]We note that the additional factor $1/4$ in the definition of the energy density with respect to Ref. [286] comes from our normalisation of second order tensor modes with a pre-factor $1/2$ in Eq. (6.1.9).

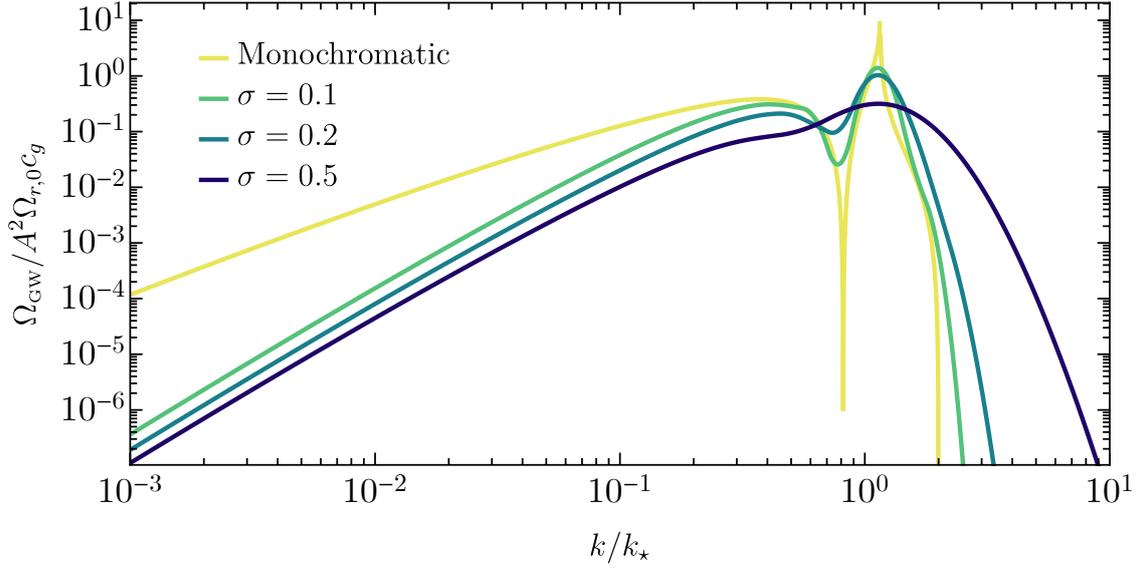

Figure 6.1: *Spectrum of the SGWB induced at second order by a power spectrum of the form* (6.1.27) *and* (6.1.29)*. In the case of the monochromatic power spectrum, a resonant effect at $k \sim 2k_\star/\sqrt{3}$ produces the observed spike, see for related discussion in Ref.* [651]*.*

- At large frequencies $k \gg k_\star$, the tail of the spectrum is scaling like

$$\Omega_{\mathrm{GW}}(k \gg k_\star) \propto \mathcal{P}_\zeta^2(k \gg k_\star), \tag{6.1.26}$$

  retaining information, therefore, about the shape of the curvature spectrum inducing the emission of GWs.

In the next subsection, we will show the resulting spectrum of SGWB for different choices of curvature power spectrum $\mathcal{P}_\zeta(k)$.

### 6.1.2   Monochromatic power spectrum

In this subsection, we make the idealized assumption that the scalar power spectrum is monochromatic and defined as

$$\mathcal{P}_\zeta(k) = A \, k_\star \delta \left( k - k_\star \right), \tag{6.1.27}$$

that can be considered as the limiting case of a very narrow lognormal power spectrum considered in the following section. This approximation allows us to obtain analytical results. Indeed, inserting Eq. (6.1.27) in the expression (6.5.54), one obtains (see also Refs. [643, 653])

$$\Omega_{\mathrm{GW}}(\eta_0, k) = \frac{c_g}{15552} \Omega_{r,0} A_s^2 \frac{k^2}{k_\star^2} \left( \frac{4k_\star^2}{k^2} - 1 \right)^2 \theta \left( 2k_\star - k \right) \left[ \mathcal{I}_c^2 \left( \frac{k_\star}{k}, \frac{k_\star}{k} \right) + \mathcal{I}_s^2 \left( \frac{k_\star}{k}, \frac{k_\star}{k} \right) \right]. \tag{6.1.28}$$

This result is shown as a yellow line in Fig. 6.1.

### 6.1.3   Lognormal power spectrum

In this section, we generalise the previous result to the case of a wider lognormal power spectrum of the form

$$\mathcal{P}_\zeta(k) = \frac{A}{\sqrt{2\pi\sigma^2}} \exp \left( -\frac{\ln^2(k/k_\star)}{2\sigma^2} \right). \tag{6.1.29}$$

It is trivial to show that in the limit of vanishing width $\sigma \to 0$, one exactly recovers the monochromatic scenario. While it is not possible to find a general analytical solution for $\Omega_{\mathrm{GW}}(k)$ (see Ref. [658] for some progress in this direction), we show few numerical results in Fig. 6.1. At high frequencies, at odds with the monochromatic case, there is no upper bound at $2k_\star$ because the scalar power spectrum presents a non-vanishing tail for $k > k_\star$. At lower frequencies, the spectral tilt is about $\simeq 3$, as expected from causality arguments, see details in Ref. [671].

## 6.2 Primordial bispectrum of GWs

We already highlighted that the induced GWs are non-Gaussian as they are produced at second order. Therefore, as we will show, their primordial three-point correlator is not vanishing. We will perform the computation as described in Ref. [21].

The computation of the three-point function is based on Eq. (6.1.11) and corresponds to evaluating

$$
\begin{aligned}
\mathcal{B}_{\lambda_i}\left(\vec{k}_i\right) &= \left\langle h_{\lambda_1}(\eta_1, \vec{k}_1) h_{\lambda_2}(\eta_2, \vec{k}_2) h_{\lambda_3}(\eta_3, \vec{k}_3)\right\rangle' \\
&= \left(\frac{8\pi}{9}\right)^3 \int \mathrm{d}^3 p_1 \frac{1}{k_1^3 k_2^3 k_3^3 \eta_1 \eta_2 \eta_3} \cdot e_{\lambda_1}^*(\vec{k}_1, \vec{p}_1) e_{\lambda_2}^*(\vec{k}_2, \vec{p}_2) e_{\lambda_3}^*(\vec{k}_3, \vec{p}_3) \frac{\mathcal{P}_\zeta(p_1)}{p_1^3} \frac{\mathcal{P}_\zeta(p_2)}{p_2^3} \frac{\mathcal{P}_\zeta(p_3)}{p_3^3} \\
&\quad \times \left[\left(\cos(k_1 \eta_1) \mathcal{I}_c\left(\frac{p_1}{k_1}, \frac{p_2}{k_1}\right) + \sin(k_1 \eta_1) \mathcal{I}_s\left(\frac{p_1}{k_1}, \frac{p_2}{k_1}\right)\right) \times \text{permutations}\right],
\end{aligned}
\tag{6.2.1}
$$

where we have set $\vec{p}_2 = \vec{p}_1 - \vec{k}_1$ and $\vec{p}_3 = \vec{p}_1 + \vec{k}_3$ and the prime indicates we have dropped the prefactor $(2\pi)^3$ and the Dirac delta enforcing the momentum conservation.

In the next pages, we will perform the computation first by assuming a monochromatic power spectrum of curvature perturbations, that will allow us to find analytical solutions. Then, we will generalise the result in the case of a lognormal spectrum.

### 6.2.1 Monochromatic power spectrum

We adopt Eq. (6.1.27) as the curvature power spectrum $\mathcal{P}_\zeta(k)$. Introducing it into Eq. (6.2.1), one obtains

$$
\begin{aligned}
\mathcal{B}_{\lambda_i}\left(\vec{k}_i\right) &= \left(\frac{8\pi}{9}\right)^3 \frac{A^3 k_\star^3}{k_1^3 k_2^3 k_3^3 \eta_1 \eta_2 \eta_3} \int \mathrm{d}^3 p_1\, e_{\lambda_1}^*\left(\vec{k}_1, \vec{p}_1\right) e_{\lambda_2}^*\left(\vec{k}_2, \vec{p}_1 - \vec{k}_1\right) e_{\lambda_3}^*\left(\vec{k}_3, \vec{p}_1 + \vec{k}_3\right) \\
&\quad \times \frac{\delta\left(p_1 - k_\star\right)}{k_\star^3} \frac{\delta\left(\left|\vec{p}_1 - \vec{k}_1\right| - k_\star\right)}{k_\star^3} \frac{\delta\left(\left|\vec{p}_1 + \vec{k}_3\right| - k_\star\right)}{k_\star^3} \\
&\quad \times \prod_{i=1}^3 \left[\cos\left(k_i \eta_i\right) \mathcal{I}_c\left(\frac{k_\star}{k_i}, \frac{k_\star}{k_i}\right) + \sin\left(k_i \eta_i\right) \mathcal{I}_s\left(\frac{k_\star}{k_i}, \frac{k_\star}{k_i}\right)\right].
\end{aligned}
\tag{6.2.2}
$$

As analysed in detail in Ref. [660], the bispectrum depends on the magnitude and orientation of the three vectors $\vec{k}_i$. We can fix, without loss of generality,

$$
\vec{k}_1 = k_1\, \hat{v}_1, \qquad \vec{k}_2 = k_2\, \hat{v}_2, \qquad \vec{k}_3 = -\vec{k}_1 - \vec{k}_2,
\tag{6.2.3}
$$

where

$$
\begin{aligned}
\hat{v}_1 &= \left(1\ ,\ 0\ ,\ 0\right), \\
\hat{v}_2 &= \left(\frac{k_3^2 - k_1^2 - k_2^2}{2k_1 k_2}\ ,\ \sqrt{1 - \left(\frac{k_3^2 - k_1^2 - k_2^2}{2k_1 k_2}\right)^2}\ ,\ 0\right).
\end{aligned}
\tag{6.2.4}
$$

One can show that the combination of various Dirac deltas has a non-vanishing support where the three momenta satisfy

$$
\mathcal{A}\left[k_1, k_2, k_3\right] > \frac{k_1 k_2 k_3}{4 k_\star},
\tag{6.2.5}
$$

where we introduced

$$\mathcal{A}\left[k_1,\, k_2,\, k_3\right] \equiv \frac{1}{4}\sqrt{\left(k_1+k_2+k_3\right)\left(-k_1+k_2+k_3\right)\left(k_1-k_2+k_3\right)\left(k_1+k_2-k_3\right)}. \tag{6.2.6}$$

With some algebraic passages, one can arrive at

$$\mathcal{B}_{\lambda_i}\left(\eta_i,\, \vec{k}_i\right) = \frac{A^3 \Theta\left(\mathcal{A}\left[r_1,\, r_2,\, r_3\right] - \frac{r_1 r_2 r_3}{4}\right)}{k_1^2 k_2^2 k_3^2\, k_\star^3\, \eta_1 \eta_2 \eta_3} \frac{1024 \pi^3}{729} \mathcal{D}_{\lambda_i}\left(\hat{k}_i,\, r_i\right)$$

$$\times \left(\frac{16\,\mathcal{A}^2\left[r_1,\, r_2,\, r_3\right]}{r_1^2 r_2^2 r_3^2} - 1\right)^{-1/2} \frac{r_1^4}{r_2^2 r_3^2} \prod_{i=1}^{3}\left[\frac{\mathcal{I}_i^*}{2}\, e^{i\eta_i k_i} + \frac{\mathcal{I}_i}{2}\, e^{-i\eta_i k_i}\right], \tag{6.2.7}$$

where we defined $r_i \equiv k_i/k_\star$, and introduced the combinations

$$\mathcal{I}_i \equiv \mathcal{I}\left(\frac{1}{r_i}\right) \equiv \mathcal{I}_c\left(\frac{1}{r_i},\, \frac{1}{r_i}\right) + i\,\mathcal{I}_s\left(\frac{1}{r_i},\, \frac{1}{r_i}\right). \tag{6.2.8}$$

Also, we defined the contractions

$$\mathcal{D}_{\lambda_i}\left(\hat{k}_i,\, r_i\right) \equiv e_{ab,\lambda_1}^*\left(\hat{k}_1\right) e_{cd,\lambda_2}^*\left(\hat{k}_2\right) e_{ef,\lambda_3}^*\left(\hat{k}_3\right) \left\{\left[\vec{q}_a\,\vec{q}_b\left(\vec{q}-\hat{k}_1\right)_c\left(\vec{q}-\hat{k}_1\right)_d \vec{q}_e\,\vec{q}_f\right]_I \right.$$

$$\left. + \left[\vec{q}_a\,\vec{q}_b\left(\vec{q}-\hat{k}_1\right)_c\left(\vec{q}-\hat{k}_1\right)_d \vec{q}_e\,\vec{q}_f\right]_{II}\right\}, \tag{6.2.9}$$

where we sum over the two points defined as

$$(\vec{p}_1)_{I,II} = k_1\left(\frac{1}{2},\, \frac{-k_1^2 + k_2^2 + k_3^2}{8\,\mathcal{A}\left[k_1,\, k_2,\, k_3\right]},\, \pm\frac{\sqrt{16\mathcal{A}^2\left[k_1,\, k_2,\, k_3\right]k_\star^2 - k_1^2 k_2^2 k_3^2}}{4\mathcal{A}\left[k_1,\, k_2,\, k_3\right]k_1}\right). \tag{6.2.10}$$

The explicit expressions for the contractions $\mathcal{D}_{\lambda_i}\left(\hat{k}_i,\, r_i\right)$ are rather cumbersome. For simplicity, we only report their explicit expression for the isosceles configuration while the interested reader can find more details in Ref. [21]. Setting $r_1 = r_2$ one finds

$$\mathcal{D}_{\text{RRR}} = \frac{1}{256}\left[\frac{32 r_1^3}{(2r_1 + r_3)^3} - \frac{24\left(3r_1^2 + 8\right)}{(2r_1 + r_3)^2} + \frac{32\left(r_1^2 + 4\right)r_3}{r_1^5} + \frac{32\left(r_1^2 - 1\right)^2}{r_1^3(2r_1 - r_3)}\right.$$

$$\left. + \frac{32\left(2\left(r_1^2 + 6\right)r_1^2 + 9\right)}{r_1^3(2r_1 + r_3)} - \frac{\left(r_1^4 + 24r_1^2 + 16\right)r_3^2}{r_1^6} - \frac{4\left(33r_1^4 + 24r_1^2 + 16\right)}{r_1^6} - 32\right],$$

$$\mathcal{D}_{\text{LRR}} = \frac{\left(r_1^2 - 4\right)^2\left(8r_1^4 - 4r_1^2\left(r_3^2 + 4\right) + r_3^4 + 4r_3^2\right)}{256 r_1^6\left(4r_1^2 - r_3^2\right)},$$

$$\mathcal{D}_{\text{RRL}} = \frac{1}{256}\left[-\frac{32 r_1^3}{(r_3 - 2r_1)^3} - \frac{24\left(3r_1^2 + 8\right)}{(r_3 - 2r_1)^2} - \frac{32\left(r_1^2 + 4\right)r_3}{r_1^5} - \frac{32\left(2\left(r_1^2 + 6\right)r_1^2 + 9\right)}{r_1^3(r_3 - 2r_1)}\right.$$

$$\left. + \frac{32\left(r_1^2 - 1\right)^2}{r_1^3(2r_1 + r_3)} - \frac{\left(r_1^4 + 24r_1^2 + 16\right)r_3^2}{r_1^6} - \frac{4\left(33r_1^4 + 24r_1^2 + 16\right)}{r_1^6} - 32\right]. \tag{6.2.11}$$

Recall that the contractions are invariant under the parity transformation (L $\leftrightarrow$ R). This follows from the fact that the full bispectrum is expected to be parity invariant.

In the equilateral (EQ) case $r_1 = r_2 = r_3$, the equal time bispectrum, averaged over the time oscillations of the amplitude, reads

$$\mathcal{B}_{\lambda_i}^{\text{EQ}}\left(\eta,\, |\vec{k}_i| = k\right) = \frac{A^3}{k_\star^3 \eta^3} \frac{1}{k^6} \frac{1024 \pi^3}{729} \frac{\Theta\left(\sqrt{3}\,k_\star - k\right)}{\sqrt{\frac{3k_\star^2}{k^2} - 1}} \left|\frac{1}{\sqrt{2}}\mathcal{I}\left(\frac{k_\star}{k}\right)\right|^3$$

$$\times \begin{cases} \frac{365}{6912} - \frac{61}{192}\frac{k_\star^2}{k^2} + \frac{9}{16}\frac{k_\star^4}{k^4} - \frac{1}{4}\frac{k_\star^6}{k^6} & \text{for} \qquad \text{RRR, LLL,} \\[2ex] \frac{\left[-4 + (k/k_\star)^2\right]^2\left[-12 + 5(k/k_\star)^2\right]}{768(k/k_\star)^6}, & \text{otherwise.} \end{cases} \tag{6.2.12}$$

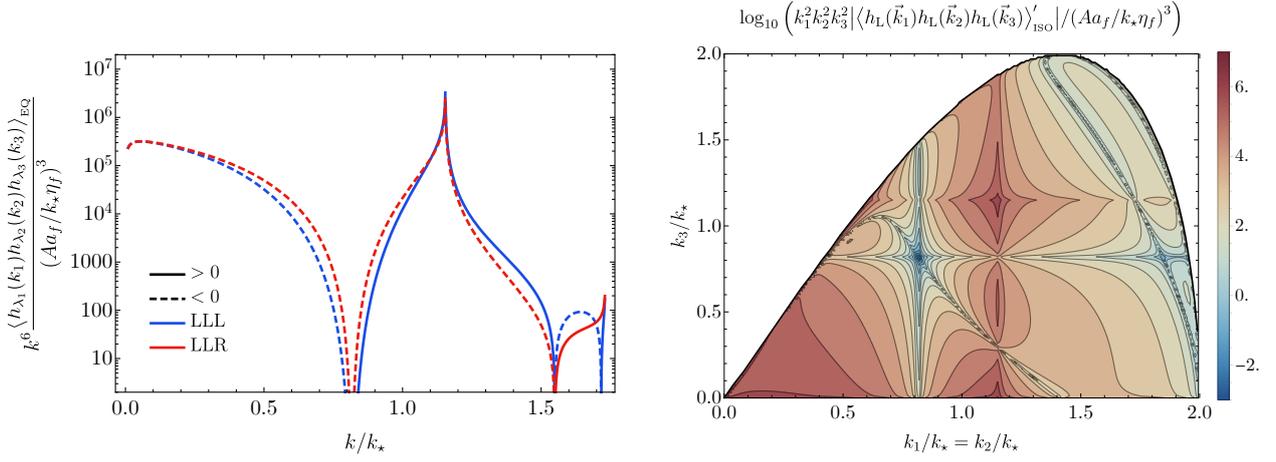

Figure 6.2: *Monochromatic power spectrum.* **Left:** *Plot of the rescaled primordial bispectrum in the equilateral configuration. Notice that the bispectrum vanishes for $k > \sqrt{3}k_\star$, as a consequence of the conditions imposed by the sharp curvature spectra.* **Right:** *Contour plot of the rescaled bispectrum in the isosceles configuration. The white portion of the plot indicates the region where condition* (6.2.5) *is violated and, therefore, the bispectrum vanishes.*

In Fig. 6.2 we show both the equilateral and isosceles primordial bispectrum generated by the monochromatic power spectrum. We see that the bispectrum is peaked in the equilateral configuration for $k_i = 2k_\star/\sqrt{3}$. This property will also be clear in the plot showing the bispectrum *shape*, as defined in Eq. (6.2.13) and shown in Fig. 6.4.

### 6.2.2  Lognormal power spectrum

We now turn to the results obtained in the case a lognormal power spectrum (6.1.29) is considered. Following Ref. [21], we adopt a power spectrum peaked at $k_\star \to 3k_\star/2$, resulting in a shift of the results to larger relative momenta $k_i/k_\star$. We also fix the width to be $\sigma = 0.5$.

When a lognormal $\mathcal{P}_\zeta$ is considered, it is not possible to find analytical solutions describing the bispectrum. In Fig. 6.3 we show both the result for an equilateral configuration, where we set $k_1 = k_2 = k_3$ and the isosceles configuration, for which $k_1 = k_2$. Firstly, compared to the monochromatic case, we see that the wider shape of $\mathcal{P}_\zeta$ results in a lower peak in the equilateral configuration, making the two peaks with opposite sign in the LLL configuration comparable. In the isosceles configuration, the more regular profile does not introduce a sharp cut-off present in the monochromatic case, with a non-vanishing result in all the parameter space of $(k_1, k_2, k_3)$ respecting the triangular inequality.

### 6.2.3  Shape of the tensor three-point function

For presentation purposes, it is interesting to define a rescaled three-point function, capturing the shape of the bispectrum, as

$$S_h^{\lambda_1\lambda_2\lambda_3}(\vec{k}_1, \vec{k}_2, \vec{k}_3) = k_1^2 k_2^2 k_3^2 \frac{\langle h_{\lambda_1}(\eta, \vec{k}_1) h_{\lambda_2}(\eta, \vec{k}_2) h_{\lambda_3}(\eta, \vec{k}_3) \rangle'}{\sqrt{\mathcal{P}_h(\eta, k_1)\mathcal{P}_h(\eta, k_2)\mathcal{P}_h(\eta, k_3)}}. \tag{6.2.13}$$

The shape, as defined in Eq. (6.2.13), is shown in Fig. 6.4 for both models previously considered.

These results show that the primordial bispectrum of GWs has its maximum at the equilateral configuration, $k_1 \simeq k_2 \simeq k_3$. This is expected due to the nature of the GW source, which is composed by gradients of curvature perturbations, giving the largest contribution at the time when the latter re-enters the horizon.

In principle, if such a SGWB was discovered, measuring the related shape would help to confirm the origin of the signal, while providing a consistency relation between the bispectrum and the power spectrum of GWs. One should not, however, have many expectations in the actual feasibility of this endeavour, as we are going to show how the propagation of GWs from the early epochs to the detectors in the present-day universe is expected to suppress the non-Gaussian signatures to an undetectable level. Therefore, it will turn out, the signal of a SGWB coming from the emission at second order in the early universe is predicted to be Gaussian, up to extremely high precision.

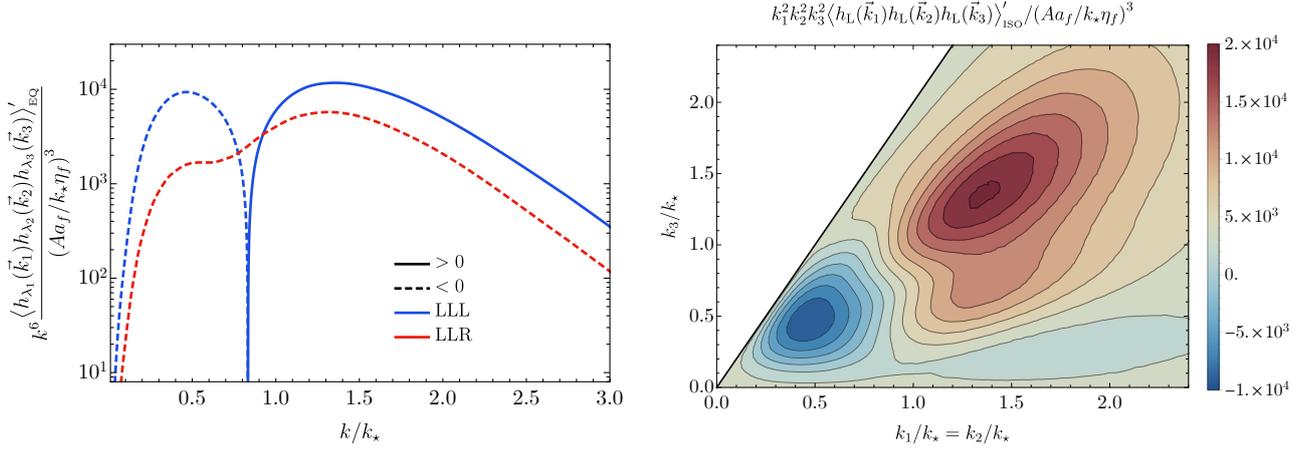

Figure 6.3: *Same as Fig. 6.2 for a lognormal power spectrum. The white region in the right plot indicates the parameter space where the momenta violate the triangular inequality. We recall that the power spectrum in Eq. (6.1.29) is, for presentation purposes, considered to peak at $k_\star \to 3k_\star/2$.*

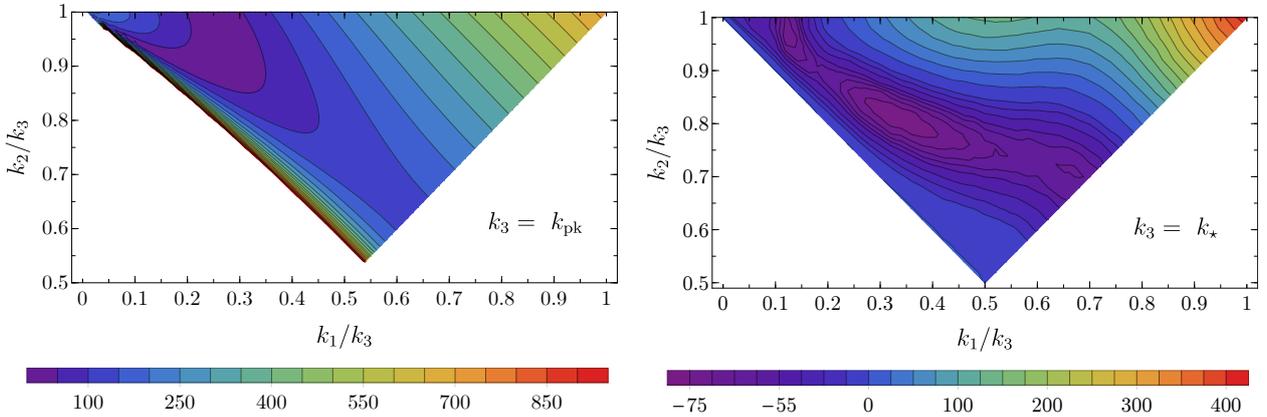

Figure 6.4: *Shape of the three-point function.* **Left:** *Monochromatic case, where $k_3$ was set at the scale corresponding to the peak of the three-point function. Notice that a portion of the lower triangular space of configurations in $(k_1, k_2)$ is vanishing identically due to the restrictive condition 6.2.5 imposed by the monochromatic spectrum.* **Right:** *Lognormal case with $k_3$ set at the reference scale $k_\star$.*

## 6.3 GW propagation in a perturbed universe

To discuss the actual detectability of the SGWB non-Gaussianity, one needs to account for propagation effects in the perturbed universe. By definition, a non-vanishing non-Gaussian signal requires the correlation among the different phases of the GWs at the time of detection [675]. In the scenario considered in this chapter, the GW phases are correlated at production since the source at second order comes from scalar perturbation generated during the inflationary stage. However, as we shall see, this coherence is destroyed as the detectable SGWB signal is made of contributions coming from all possible directions. Because of the Shapiro time delay caused by the presence of large-scale gravitational potentials along various lines of sight, different directions experience different delays, thus destroying the initial phase correlation.

In the next section, we will explicitly show that this effect does not change the power spectrum,

which can be thought of as propagating the information of the characteristic amplitude of the SGWB. It does, however, suppress the GW bispectrum. In simple terms, this effect can be viewed as a consequence of the central limit theorem: the time delay does not suppress the non-Gaussianity of the signal arriving from a single line of sight. However, as the SGWB is built from signals coming from all possible directions, the Shapiro-time delay makes the signal Gaussian.

We address this issue in three steps: in the first one, we compute the solution for the GW propagation in the geometrical optic limit, while, in the second and third ones, we will compute the effect on the power spectrum and bispectrum, respectively.

### 6.3.1   The propagation in the geometrical optic limit

Our starting point is the metric in the Newtonian longitudinal gauge (neglecting the shear, such that $\Phi = \Psi$, as done in the previous section), which can be written as

$$g_{00} = -(1 + 2\Phi) \quad , \quad g_{0i} = 0 \quad , \quad g_{ij} = a^2 \left[ (1 - 2\Phi)\, \delta_{ij} + h_{ij} \right]. \tag{6.3.1}$$

The spatial components of the Einstein tensor up to second-order in the metric perturbations written above are

$$
\begin{aligned}
G^i_j = {}& \delta_{ij} \left( -H^2 - \frac{2\ddot{a}}{a} \right) \\
& + \frac{1}{2}\ddot{h}_{ij} + \frac{3}{2}H\dot{h}_{ij} - \frac{1}{2a^2}h_{ij,kk} + \mathrm{O}\left( \Phi \right) \\
& + \left( \ddot{\Phi} + 3H\dot{\Phi} \right)h_{ij} - \frac{1}{a^2}\left[ \Phi h_{ij,kk} + (\Phi h_{ij})_{,kk} - (\Phi_{,k}h_{ik})_{,j} - (\Phi_{,k}h_{kj})_{,i} \right] \\
& + \delta_{ij}\left[ -\frac{1}{a^2}h_{mn}\Phi_{,mn} \right] + \mathcal{O}\left( \Phi^2 \right) + \mathcal{O}\left( h^2 \right).
\end{aligned}
\tag{6.3.2}
$$

We study the propagation of the GWs in the matter-dominated phase where we can safely neglect the tiny generation of the GWs caused by the $\mathcal{O}\left( \Phi^2 \right)$ during the propagation. We also adopt the geometrical optic limit which amounts to assuming that the SGWB's characteristic frequency is much larger than the typical momentum associated with the gravitational potential $\Phi$. In this limit, it is enough to consider the leading term in an expansion in gradients of $\Phi$. The corresponding equation reads

$$h''_{ij} + 2\mathcal{H}h'_{ij} - (1 + 4\Phi)\,h_{ij,kk} = 0, \tag{6.3.3}$$

where the projection on the transverse and traceless part will only be performed in a later stage of the computation. In momentum space, the GW (real in coordinate space) can be expanded in terms of the complex mode functions

$$h_{ij}(\vec{k}, \eta) = h^A_{ij}(\vec{k}, \eta) + h^{*A}_{ij}(-\vec{k}, \eta). \tag{6.3.4}$$

Then, by making the ansatz

$$h^A_{ij} = A_{ij}\mathrm{e}^{ik\eta}\mathrm{e}^{i\int^\eta d\eta' F_A(\eta')}, \tag{6.3.5}$$

we obtain the following equation for the leading and next-to-leading order solution

$$
\begin{cases}
A''_{ij} + 2ikA'_{ij} - k^2 A_{ij} + \frac{4}{\eta}\left[ A'_{ij} + ikA_{ij} \right] + k^2 A_{ij} = 0, \\
2iF_A A'_{ij} + \left[ iF'_A - 2kF_A - F^2_A + \frac{4}{\eta}iF_A + 4k^2\Phi \right]A_{ij} \simeq 0.
\end{cases}
\tag{6.3.6}
$$

In the previous equation, we have disregarded the spatial derivatives acting on $F_A$. The first equation is solved by

$$A_{ij} = \frac{C^A_{ij}}{k^2\eta^2}\left( 1 + \frac{i}{k\eta} \right), \tag{6.3.7}$$

where $C_{ij}^A$ are integration constants. As we are interested in the sub-horizon limit of these solutions, i.e. $k\eta \gg 1$, one has $A_{ij}' = -2A_{ij}/\eta$. Then, the second equation becomes

$$iF_A' - 2kF_A - F_A^2 + 4k^2\Phi \simeq 0, \tag{6.3.8}$$

which is solved by $F_A = 2k\Phi$, and therefore

$$h_{ij}^A = A_{ij}\mathrm{e}^{ik\eta + 2ik\int^\eta d\eta'\,\Phi(\eta')}. \tag{6.3.9}$$

The final step is performed by choosing the arbitrary constants such that one matches the solution (6.1.11) at early times. This procedure gives

$$h_\lambda\left(\vec{k}\right) = \frac{\eta_{\mathrm{eq}}}{\eta^2}\left[h_\lambda^{\mathrm{RD},c}\left(\vec{k}\right)\cos\Omega + h_\lambda^{\mathrm{RD},s}\left(\vec{k}\right)\sin\Omega\right]$$
$$= \frac{4}{9}\int\frac{d^3p}{(2\pi)^3}\frac{\eta_{\mathrm{eq}}}{k^3\eta^2}\,e_\lambda^*\left(\vec{k},\vec{p}\right)\zeta\left(\vec{p}\right)\zeta\left(\vec{k}-\vec{p}\right)\left[\mathcal{I}_c\left(x,y\right)\cos\Omega + \mathcal{I}_s\left(x,y\right)\sin\Omega\right], \tag{6.3.10}$$

where we defined the time and momentum dependent frequency as

$$\Omega = k\,\eta + 2k\int_{\eta_{\mathrm{eq}}}^\eta d\eta'\,\Phi\left(\eta',\,\vec{x}_0 + (\eta' - \eta_0)\,\hat{k}\right) = k\,\eta + \frac{6}{5}k\int_{\eta_{\mathrm{eq}}}^\eta d\eta'\,\zeta^L\left((\eta' - \eta_0)\,\hat{k}\right). \tag{6.3.11}$$

In Eq. (6.3.11), $\vec{x}_0$ is the location of the detector that can be set to zero without loss of generality, while $\hat{k}$ identifies the direction of motion of the GW. To stress that the gravitational potential $\Phi$ has a typical momentum much smaller than $k$, we have used the relation $\Phi = 3\zeta^L/5$. We take $\zeta^L$ to be Gaussian with its corresponding scale-invariant power spectrum $\mathcal{P}_\zeta^L$ having an amplitude matching CMB observations.[5] In the matter-dominated phase of the universe relevant for our considerations, $\zeta^L$ is time-independent on linear scales. This solution can also be written in a more convenient form as [21]

$$h_\lambda\left(\vec{k}\right) = \frac{2}{9}\int\frac{d^3p}{(2\pi)^3}\frac{\eta_{\mathrm{eq}}}{k^3\eta^2}\,e_{\lambda,ij}^*\left(\hat{k}\right)\vec{p}_i\vec{p}_j\zeta\left(\vec{p}\right)\zeta\left(\vec{k}-\vec{p}\right)$$
$$\times\left[\mathcal{I}^*\mathrm{e}^{ik\eta + i\frac{6}{5}k\int_{\eta_{\mathrm{eq}}}^\eta d\eta'\,\zeta^L\left((\eta'-\eta_0)\hat{k}\right)} + \mathcal{I}\,\mathrm{e}^{-ik\eta - i\frac{6}{5}k\int_{\eta_{\mathrm{eq}}}^\eta d\eta'\,\zeta^L\left(-(\eta'-\eta_0)\hat{k}\right)}\right], \tag{6.3.12}$$

where, to simplify the notation, we dropped the dependency on the momenta $\vec{k}$ and $\vec{p}$, that can be written explicitly as

$$\mathcal{I}^* = \mathcal{I}^*\left(\frac{p}{k},\frac{|\vec{k}-\vec{p}|}{k}\right), \qquad \mathcal{I} = \mathcal{I}\left(\frac{p}{k},\frac{|\vec{k}-\vec{p}|}{k}\right). \tag{6.3.13}$$

### 6.3.2 Impact on the tensor two-point function

In order to see the effect of the propagation onto the power spectrum, it is convenient to define the quantity

$$\hat{Z}\left(\eta,\vec{k}\right) \equiv \frac{6}{5}k\int_{\eta_{\mathrm{eq}}}^\eta d\eta'\,\zeta^L\left((\eta'-\eta_0)\,\hat{k}\right). \tag{6.3.14}$$

Then, one can compute the power spectrum after the GWs have propagated through the perturbed universe, again by considering the monochromatic case introduced in Sec. 6.1.2, as

$$\left\langle h_{\lambda_1}\left(\eta,\vec{k}\right)h_{\lambda_2}\left(\eta,\vec{k}'\right)\right\rangle' = \frac{\pi^2}{648}\frac{A^2}{k_*^2}\frac{\eta_{\mathrm{eq}}^2}{k^7\eta^4}\,(k-2k_*)^2\,(k+2k_*)^2\,\theta\,(2k_*-k)\,e_{ab,\lambda_1}^*\left(\vec{k}\right)e_{ab,\lambda_2}\left(\vec{k}\right)$$
$$\times\left\langle\left[\mathcal{I}^*\left(\frac{k_*}{k},\frac{k_*}{k}\right)\mathrm{e}^{ik\eta + i\hat{Z}(\eta,\vec{k})} + \mathcal{I}\left(\frac{k_*}{k},\frac{k_*}{k}\right)\mathrm{e}^{-ik\eta - i\hat{Z}(\eta,-\vec{k})}\right]\right.$$
$$\left.\times\left[\mathcal{I}^*\left(\frac{k_*}{k},\frac{k_*}{k}\right)\mathrm{e}^{ik\eta + i\hat{Z}(\eta,-\vec{k})} + \mathcal{I}\left(\frac{k_*}{k},\frac{k_*}{k}\right)\mathrm{e}^{-ik\eta - i\hat{Z}(\eta,\vec{k})}\right]\right\rangle, \tag{6.3.15}$$

---

[5]Notice that, in this section, we denote the long modes ($k_L \ll k_{\mathrm{GW}}$) with a capital letter $L$, contrarily to what was done in Sec. 3.3, to avoid confusion with the multipolar index $\ell$.

where we have exploited the fact that the short modes of $\zeta$, responsible for the GW production during the radiation-dominated era, and the long modes $\zeta^L$ are not correlated. This is the same procedure routinely used in the peak background split picture when primordial non-Gaussianities are absent. In the product within the angular brackets, terms proportional to $\mathcal{I}^2$ and $\mathcal{I}^{*2}$ are fast oscillating and can be dropped when taking the time average, while all phases are eliminated in those proportional to $|\mathcal{I}|^2$. This proves the propagation of the GWs in the perturbed universe does not affect the power spectrum, which brings information about the unaffected amplitude of the GW oscillations.

### 6.3.3  Impact on the tensor bispectrum

To compute the tensor three-point function, after the GWs have propagated through the universe, we proceed as in the previous section by starting from the solution in Eq. (6.3.12). For presentation purposes, and to simplify the computations, let us focus on the equilateral configuration

$$|\vec{k}_1| = |\vec{k}_2| = |\vec{k}_3| = k, \tag{6.3.16}$$

and consider again the case of a monochromatic spectrum of curvature perturbations. The resulting bispectrum takes the form

$$
\begin{aligned}
B_h^{\lambda_i}\left(\eta_i, \vec{k}_i\right) = \mathcal{N}\Bigg\langle &\left[\mathcal{I}^*\left(\frac{k_*}{k}, \frac{k_*}{k}\right)\mathrm{e}^{ik\eta_1 + i\hat{Z}(\eta_1, \vec{k}_1)} + \mathcal{I}\left(\frac{k_*}{k}, \frac{k_*}{k}\right)\mathrm{e}^{-ik\eta_1 - i\hat{Z}(\eta_1, -\vec{k}_1)}\right] \\
&\left[\mathcal{I}^*\left(\frac{k_*}{k}, \frac{k_*}{k}\right)\mathrm{e}^{ik\eta_2 + i\hat{Z}(\eta_2, \vec{k}_2)} + \mathcal{I}\left(\frac{k_*}{k}, \frac{k_*}{k}\right)\mathrm{e}^{-ik\eta_2 - i\hat{Z}(\eta_2, -\vec{k}_2)}\right] \\
&\left[\mathcal{I}^*\left(\frac{k_*}{k}, \frac{k_*}{k}\right)\mathrm{e}^{ik\eta_3 + i\hat{Z}(\eta_3, \vec{k}_3)} + \mathcal{I}\left(\frac{k_*}{k}, \frac{k_*}{k}\right)\mathrm{e}^{-ik\eta_3 - i\hat{Z}(\eta_3, -\vec{k}_3)}\right]\Bigg\rangle,
\end{aligned} \tag{6.3.17}
$$

where we have defined a normalisation pre-factor as

$$\mathcal{N} \equiv \frac{128\pi^3}{729}\frac{A_s^3\,\eta_{\mathrm{eq}}^3}{k_*^3\,\eta_1^2\eta_2^2\eta_3^2}\frac{\theta\left(\sqrt{3}k_* - k\right)}{k^6} \times \mathcal{D}_{\lambda_i}\left(\sqrt{\frac{3k_*^2}{k^2} - 1}\right)^{-1/2}. \tag{6.3.18}$$

We also set all times $\eta_i$ equal, since small relative variations of the order of the measurement timescale are irrelevant compared to the propagation time ($\simeq \eta_0$) and thus are not affecting our result. By defining

$$\mathcal{G}_{c_1,c_2,c_3}^{\vec{k}_1,\vec{k}_2,\vec{k}_3} \equiv \left\langle \mathrm{e}^{ic_1\hat{Z}(\eta,\vec{k}_1)}\mathrm{e}^{ic_2\hat{Z}(\eta,\vec{k}_2)}\mathrm{e}^{ic_3\hat{Z}(\eta,\vec{k}_3)}\right\rangle, \tag{6.3.19}$$

we can express Eq. (6.3.18) as

$$
\begin{aligned}
B_h^{\lambda_i}\left(\eta, \vec{k}_i\right) = \mathcal{N}\Bigg\{ &\mathcal{I}^{*3}\,\mathrm{e}^{3ik\eta}\,\mathcal{G}_{+++}^{\vec{k}_1,\vec{k}_2,\vec{k}_3} + \mathcal{I}^3\,\mathrm{e}^{-3ik\eta}\,\mathcal{G}_{---}^{-\vec{k}_1,-\vec{k}_2,-\vec{k}_3} \\
&+ \mathcal{I}^{*2}\,\mathcal{I}\,\mathrm{e}^{ik\eta}\left[\mathcal{G}_{++-}^{\vec{k}_1,\vec{k}_2,-\vec{k}_3} + \mathcal{G}_{++}^{\vec{k}_1,-\vec{k}_2,\vec{k}_3} + \mathcal{G}_{-++}^{-\vec{k}_1,\vec{k}_2,\vec{k}_3}\right] \\
&+ \mathcal{I}^*\,\mathcal{I}^2\,\mathrm{e}^{-ik\eta}\left[\mathcal{G}_{+--}^{\vec{k}_1,-\vec{k}_2,-\vec{k}_3} + \mathcal{G}_{-+-}^{-\vec{k}_1,\vec{k}_2,-\vec{k}_3} + \mathcal{G}_{--+}^{-\vec{k}_1,-\vec{k}_2,\vec{k}_3}\right]\Bigg\},
\end{aligned} \tag{6.3.20}
$$

where, having all the same argument, we indicate $\mathcal{I}\left(\frac{k_*}{k}, \frac{k_*}{k}\right) \to \mathcal{I}$. Next, to perform the expectation values, we use the identity

$$\left\langle \mathrm{e}^{\varphi_1}\mathrm{e}^{\varphi_2}\mathrm{e}^{\varphi_3}\right\rangle = \mathrm{e}^{\frac{\langle\varphi_1^2\rangle}{2} + \frac{\langle\varphi_2^2\rangle}{2} + \frac{\langle\varphi_3^2\rangle}{2} + \langle\varphi_1\varphi_2\rangle + \langle\varphi_1\varphi_3\rangle + \langle\varphi_2\varphi_3\rangle} \tag{6.3.21}$$

valid for a general Gaussian operator, and write

$$\mathcal{G}_{c_1,c_2,c_3}^{\vec{k}_1,\vec{k}_2,\vec{k}_3} = \exp^{-\frac{1}{2}c_1^2\mathcal{C}(\vec{k}_1,\vec{k}_1) - \frac{1}{2}c_2^2\mathcal{C}(\vec{k}_2,\vec{k}_2) - \frac{1}{2}c_3^2\mathcal{C}(\vec{k}_3,\vec{k}_3) - c_1c_2\mathcal{C}(\vec{k}_1,\vec{k}_2) - c_1c_3\mathcal{C}(\vec{k}_1,\vec{k}_3) - c_2c_3\mathcal{C}(\vec{k}_2,\vec{k}_3)}, \tag{6.3.22}$$

where we defined

$$
\begin{aligned}
\mathcal{C}\left(\vec{k}_1,\,\vec{k}_2\right) &\equiv \left\langle \hat{Z}\left(\eta,\,\vec{k}_1\right)\,\hat{Z}\left(\eta,\,\vec{k}_2\right)\right\rangle \\
&= \frac{36}{25}k_1 k_2 \int_{\eta_{\mathrm{eq}}}^{\eta} d\eta' \int_{\eta_{\mathrm{eq}}}^{\eta} d\eta'' \left\langle \zeta^L\left(\left(\eta'-\eta_0\right)\hat{k}_1\right)\zeta^L\left(\left(\eta''-\eta_0\right)\hat{k}_2\right)\right\rangle .
\end{aligned}
\tag{6.3.23}
$$

This correlator can be computed by using a decomposition in spherical harmonics. We refer to Ref. [21] for all the details about this computation. We follow the procedure of Ref. [676] and remove the monopole from the sum. For all the other multiples, we use the approximation (that is, strictly speaking, valid for $\ell \gg 1$) [677]

$$
\int_0^{\infty} dp\, p^2\, f\left(p\right) j_\ell\left(p\eta\right) j_\ell\left(p\eta'\right) \simeq \frac{\pi}{2\eta^2} f\left(\frac{\ell+1/2}{\eta}\right) \delta_D\left(\eta-\eta'\right),
\tag{6.3.24}
$$

giving (after some manipulations)

$$
\mathcal{C}\left(\vec{k}_1,\,\vec{k}_2\right) = \frac{36}{25}k_1 k_2 \times 2\pi^2 \int_{\eta_{\mathrm{eq}}}^{\eta} d\eta'\,\left(\eta_0-\eta'\right) \sum_{\ell=1}^{\infty}\sum_{m=-\ell}^{\ell} \frac{A_L}{\left(\ell+\frac{1}{2}\right)^3}\, Y_{\ell m}^*\left(\hat{k}_1\right) Y_{\ell m}\left(\hat{k}_2\right).
\tag{6.3.25}
$$

In the time integration, we put $\eta_0$ as the upper extremum and disregard the contribution at the time of equality. Furthermore, we can always orient $\hat{k}_1$ along the $z-$axis, so that the sum involves only the $m=0$ terms. We then note that $\cos \hat{k}_2$ becomes $\hat{k}_1 \cdot \hat{k}_2 \equiv \mu$. With this convention, then

$$
Y_{\ell m}^*\left(\hat{k}_1\right) Y_{\ell m}\left(\hat{k}_2\right) = \delta_{m0}\frac{2\ell+1}{4\pi}P_\ell\left(1\right)P_\ell\left(\mu\right) = \delta_{m0}\frac{2\ell+1}{4\pi}P_\ell\left(\mu\right),
\tag{6.3.26}
$$

where in the last step we used $P_\ell\left(1\right)=1$. Collecting all these passages, one arrives at

$$
\mathcal{C}\left(\vec{k}_1,\,\vec{k}_2\right) = A_L\,k_1\,k_2\,\eta_0^2 \times \frac{18\,\pi}{25}\sum_{\ell=1}^{\infty} \frac{P_\ell\left(\hat{k}_1\cdot\hat{k}_2\right)}{\left(\ell+\frac{1}{2}\right)^2} \equiv A_L\,k_1\,k_2\,\eta_0^2 \times \mathcal{S}\left(\hat{k}_1\cdot\hat{k}_2\right).
\tag{6.3.27}
$$

As we are interested in the equilateral configuration of the bispectrum, we only need to consider the following cases

$$
\begin{aligned}
i=j: \quad & \mathcal{C}\left(\vec{k}_i,\,\vec{k}_j\right) = A_L\,k^2\,\eta_0^2 \times \mathcal{S}\left(1\right) \simeq 2.11\,A_L\,\left(k\,\eta_0\right)^2, \\
i\neq j: \quad & \mathcal{C}\left(\pm\vec{k}_i,\,\pm\vec{k}_j\right) = A_L\,k^2\,\eta_0^2 \times \mathcal{S}\left(-1/2\right) \simeq -0.50\,A_L\,\left(k\,\eta_0\right)^2, \\
i\neq j: \quad & \mathcal{C}\left(\pm\vec{k}_i,\,\mp\vec{k}_j\right) = A_L\,k^2\,\eta_0^2 \times \mathcal{S}\left(1/2\right) \simeq 0.37\,A_L\,\left(k\,\eta_0\right)^2.
\end{aligned}
\tag{6.3.28}
$$

Inserting Eq. (6.3.28) into Eq. (6.3.20), one obtains

$$
\begin{aligned}
B_h^{\lambda_i}\left(\eta_0,\,\vec{k}_i\right) &= \mathcal{N}\Bigg\{ \mathcal{I}^{*3}\,\mathrm{e}^{3ik\eta_0}\,\mathrm{e}^{-1.67\,A_L\,\left(k\,\eta_0\right)^2} + \mathcal{I}^{*2}\,\mathcal{I}\,\mathrm{e}^{ik\eta_0}\left[\mathrm{e}^{-1.93\,A_L\,\left(k\,\eta_0\right)^2}\times 3\right] \\
&\quad + \mathcal{I}^*\,\mathcal{I}^2\,\mathrm{e}^{-ik\eta_0}\left[\mathrm{e}^{-1.93\,A_L\,\left(k\,\eta_0\right)^2}\times 3\right] + \mathcal{I}^3\,\mathrm{e}^{-3ik\eta_0}\,\mathrm{e}^{-1.67\,A_L\,\left(k\,\eta_0\right)^2}\Bigg\} \\
&\simeq \mathcal{N}\,\mathrm{e}^{-1.67\,A_L\,\left(k\,\eta_0\right)^2}\left[\mathcal{I}^{*3}\,\mathrm{e}^{3ik\eta_0} + \mathcal{I}^3\,\mathrm{e}^{-3ik\eta_0}\right].
\end{aligned}
\tag{6.3.29}
$$

In the last step, we only accounted for the least suppressed terms. One can also recast the result in terms of the bispectrum computed neglecting the effect of inhomogeneities as

$$
\frac{B_h^{\lambda_i}\left(\eta_0,\,\vec{k}_i\right)\big|_{\mathrm{inhom.}}}{B_h^{\lambda_i}\left(\eta_0,\,\vec{k}_i\right)\big|_{\mathrm{no\ inhom.}}} = \mathrm{e}^{-1.67\,A_L\,\left(k\,\eta_0\right)^2}.
\tag{6.3.30}
$$

We further define and compute the RMS of the (relative) time delay $d$, see Ref. [676], as

$$
\begin{aligned}
d_{\text{rms}}^2 &= \frac{4}{\eta_0^2} \left\langle \int d\eta' d\eta'' \Phi\left(\eta'\right) \Phi\left(\eta''\right) \right\rangle = \frac{4}{\eta_0^2 k^2} \times \frac{9}{25} k^2 \left\langle \int d\eta' d\eta'' \zeta_L\left(\eta'\right) \zeta_L\left(\eta''\right) \right\rangle \\
&= \frac{1}{\eta_0^2 k^2} \, \mathcal{C}\left(\vec{k}, \vec{k}\right) = 2.11 \, A_L,
\end{aligned}
\tag{6.3.31}
$$

where in the last step we have used the first of Eqs. (6.3.28). Finally, the above ratio can be rewritten as

$$
\frac{B_h^{\lambda_i}\left(\eta_0, \vec{k}_i\right)\big|_{\text{inhom.}}}{B_h^{\lambda_i}\left(\eta_0, \vec{k}_i\right)\big|_{\text{no inhom.}}} = e^{-1.67\, A_L\, (k\,\eta_0)^2\, \frac{d_{\text{rms}}^2}{2.11\, A_L}} = e^{-0.8\, k^2 \eta_0^2\, d_{\text{rms}}^2}.
\tag{6.3.32}
$$

Therefore, the bispectrum is exponentially suppressed by propagation effects. For example, the argument of the exponential factor is $k\eta_0 \, d_{\text{rms}} \sim 10^9$, if one considers the LISA characteristic frequency $k \sim 10^{-3}$ Hz.

It is interesting to notice that in the squeezed limit $k_1 \sim k_2 \gg k_3$, the bispectrum should reduce to an average of the short-mode power spectrum in a background modulated by the long mode $k_3$. Indeed, repeating the steps we have performed above for the squeezed limit, it is easy to see that the bispectrum reduces to

$$
B_h^{\lambda_i}\left(\eta, \vec{k}_i\right) \propto |\mathcal{I}_1|^2 \left(\mathcal{I}_3^* e^{i\eta k_3} + \mathcal{I}_3 e^{-i\eta k_3}\right) e^{-\frac{1}{2} k_3^2 \eta_0^2\, d_{\text{rms}}^2},
\tag{6.3.33}
$$

where we have, again, only kept terms that are least suppressed by propagation effects. The result is proportional to the power spectrum of the short modes times a suppression coming from the average over the long modes performed over many directions. Unfortunately, it turns out that the suppression also in this configuration is sizeable. We conclude that propagation effects are present for arbitrary shapes. The results presented in this section were also confirmed in Ref. [678].

## 6.4 Induced SGWB anisotropy

Another interesting property characterising a SGWB is its level of anisotropy. By definition, a signal is anisotropic if it presents features that are direction-dependent. Similarly to the case of CMB temperature anisotropies, just to mention a renowned example, the SGWB can possess an energy density varying across different sky directions.

It is important to keep in mind that, even assuming a completely homogeneous and isotropic SGWB at its production, the GWs propagate in a perturbed universe. As a consequence, on top of the aforementioned suppression of the tensor three-point function, there will be potential imprinting of angular anisotropies. This phenomenon was already studied in details in Refs. [679–685], while also carefully considering departures from Gaussianity [686]. In this section, we will focus on the anisotropy which is due to the mechanism of SGWB emission. We will follow the analysis presented in Ref. [17].

As we shall see in details, a necessary condition for having observable large-scale anisotropies is the presence of large-scale perturbations producing correlations on cosmological scales $k_L^{-1}$, much greater than the scale $k_*^{-1}$ associated with the typical regions forming PBHs. This is because the GW experiments searching for SGWB signatures will only be able to observe anisotropies at very large scales ($\ell \lesssim 15$ at LISA, see for example [687]).

The GW formation is a local event and, as a consequence of the Equivalence Principle, it cannot be affected by modes of wavelength much greater than the horizon at the emission time (roughly coinciding with the horizon at the epoch of PBH production). Therefore, the anisotropies expected in a vanilla production of GW from second-order perturbations in a PBH model are only sizeable at the scale $k_*^{-1}$, unobservable at GW experiments. Non-Gaussian features of the primordial density perturbations, however, can lead to a small-long scale cross talk, generating a potential large-scale modulation of the local power of the density perturbations and, consequently, of the amount of GWs produced within each region. In the following sections, we will quantify the anisotropies of the SGWB induced at second order by first addressing the Gaussian case. Then, we will show how non-Gaussianities can lead to the formation of large scale anisotropies.

### 6.4.1 SGWB anisotropy at formation: the Gaussian case

To quantify the level anisotropies, one needs to compute the two-point correlation function of the GW energy density field $\langle \rho_{\mathrm{GW}}(\vec{x}) \rho_{\mathrm{GW}}(\vec{y}) \rangle$, where, from Eq. (6.1.11), each factor takes the form

$$
\rho_{\mathrm{GW}}(\eta, \vec{x}) = \frac{M_p^2}{81 \eta^2 a^2} \int \frac{d^3 k_1 d^3 k_2 d^3 p_1 d^3 p_2}{(2\pi)^{12}} \frac{1}{k_1^2 k_2^2} \, \mathrm{e}^{i \vec{x} \cdot (\vec{k}_1 + \vec{k}_2)} T[\hat{k}_1, \hat{k}_2, \vec{p}_1, \vec{p}_2]
$$

$$
\times \zeta(\vec{p}_1) \zeta(\vec{k}_1 - \vec{p}_1) \zeta(\vec{p}_2) \zeta(\vec{k}_2 - \vec{p}_2) \Big\langle \prod_{i=1}^{2} \Big[ \mathcal{I}_s(\vec{k}_i, \vec{p}_i) \cos(k_i \eta) - \mathcal{I}_c(\vec{k}_i, \vec{p}_i) \sin(k_i \eta) \Big] \Big\rangle_T, \tag{6.4.1}
$$

and

$$
T\left[\hat{k}_1, \hat{k}_2, \vec{p}_1, \vec{p}_2\right] \equiv \sum_{\lambda_1, \lambda_2} e_{ij,\lambda_1}(\hat{k}_1) e^*_{ab,\lambda_1}(\hat{k}_1) e_{ij,\lambda_2}(\hat{k}_2) e^*_{cd,\lambda_2}(\hat{k}_2) \vec{p}_{1a} \vec{p}_{1b} \vec{p}_{2c} \vec{p}_{2d}. \tag{6.4.2}
$$

The explicit expression of the function $T$ in terms of the momenta $\vec{k}_i$ and $\vec{p}_i$ can be found in App. A of Ref. [17]. The energy density two-point correlator depends on space positions only through the difference $|\vec{x} - \vec{y}|$ as a consequence of statistical isotropy and homogeneity.

In practice, the computation requires the evaluation of a comoving curvature perturbation eight-point function $\langle \zeta^8 \rangle$ as

$$
\langle \rho_{\mathrm{GW}}(\eta_1, \vec{x}) \rho_{\mathrm{GW}}(\eta_2, \vec{y}) \rangle = \left( \frac{M_p^2}{81 a^2(\eta_1) a^2(\eta_2) \eta_1 \eta_2} \right)^2 \int \frac{d^3 k_1 d^3 k_2 d^3 p_1 d^3 p_2}{(2\pi)^{12}} \frac{1}{k_1^2 k_2^2} \, \mathrm{e}^{i \vec{x} \cdot (\vec{k}_1 + \vec{k}_2)}
$$

$$
\times \int \frac{d^3 k_3 d^3 k_4 d^3 p_3 d^3 p_4}{(2\pi)^{12}} \frac{1}{k_3^2 k_4^2} \, \mathrm{e}^{i \vec{y} \cdot (\vec{k}_3 + \vec{k}_4)} T\left[\hat{k}_1, \hat{k}_2, \vec{p}_1, \vec{p}_2\right] T\left[\hat{k}_3, \hat{k}_4, \vec{p}_3, \vec{p}_4\right]
$$

$$
\times \Big\langle \zeta_{\vec{p}_1} \zeta_{\vec{k}_1 - \vec{p}_1} \zeta_{\vec{p}_2} \zeta_{\vec{k}_2 - \vec{p}_2} \zeta_{\vec{p}_3} \zeta_{\vec{k}_3 - \vec{p}_3} \zeta_{\vec{p}_4} \zeta_{\vec{k}_4 - \vec{p}_4} \Big\rangle
$$

$$
\times \Big\langle \Big[ \mathcal{I}_s(\vec{k}_1, \vec{p}_1) \cos(k_1 \eta_1) - \mathcal{I}_c(\vec{k}_1, \vec{p}_1) \sin(k_1 \eta_1) \Big] \Big[ \mathcal{I}_s(\vec{k}_2, \vec{p}_2) \cos(k_2 \eta_1) - \mathcal{I}_c(\vec{k}_2, \vec{p}_2) \sin(k_2 \eta_1) \Big] \Big\rangle_T
$$

$$
\times \Big\langle \Big[ \mathcal{I}_s(\vec{k}_3, \vec{p}_3) \cos(k_3 \eta_2) - \mathcal{I}_c(\vec{k}_3, \vec{p}_3) \sin(k_3 \eta_2) \Big] \Big[ \mathcal{I}_s(\vec{k}_4, \vec{p}_4) \cos(k_4 \eta_2) - \mathcal{I}_c(\vec{k}_4, \vec{p}_4) \sin(k_4 \eta_2) \Big] \Big\rangle_T, \tag{6.4.3}
$$

where we have introduced the notation $\zeta_{\vec{q}} \equiv \zeta(\vec{q})$. Let us focus on the comoving curvature perturbation correlator involving eight Gaussian fields. Using the Wick theorem, we can express it as

$$
\Big\langle \zeta_{\vec{p}_1} \zeta_{\vec{k}_1 - \vec{p}_1} \zeta_{\vec{p}_2} \zeta_{\vec{k}_2 - \vec{p}_2} \zeta_{\vec{p}_3} \zeta_{\vec{k}_3 - \vec{p}_3} \zeta_{\vec{p}_4} \zeta_{\vec{k}_4 - \vec{p}_4} \Big\rangle = 4 \langle \zeta_{\vec{p}_1} \zeta_{\vec{p}_2} \rangle \Big\langle \zeta_{\vec{k}_1 - \vec{p}_1} \zeta_{\vec{k}_2 - \vec{p}_2} \Big\rangle \langle \zeta_{\vec{p}_3} \zeta_{\vec{p}_4} \rangle \Big\langle \zeta_{\vec{k}_3 - \vec{p}_3} \zeta_{\vec{k}_4 - \vec{p}_4} \Big\rangle
$$

$$
+ 8 \langle \zeta_{\vec{p}_1} \zeta_{\vec{p}_3} \rangle \Big\langle \zeta_{\vec{k}_1 - \vec{p}_1} \zeta_{\vec{k}_3 - \vec{p}_3} \Big\rangle \langle \zeta_{\vec{p}_2} \zeta_{\vec{p}_4} \rangle \Big\langle \zeta_{\vec{k}_2 - \vec{p}_2} \zeta_{\vec{k}_4 - \vec{p}_4} \Big\rangle
$$

$$
+ 32 \langle \zeta_{\vec{p}_1} \zeta_{\vec{p}_2} \rangle \langle \zeta_{\vec{p}_3} \zeta_{\vec{p}_4} \rangle \Big\langle \zeta_{\vec{k}_1 - \vec{p}_1} \zeta_{\vec{k}_3 - \vec{p}_3} \Big\rangle \Big\langle \zeta_{\vec{k}_2 - \vec{p}_2} \zeta_{\vec{k}_4 - \vec{p}_4} \Big\rangle
$$

$$
+ 16 \langle \zeta_{\vec{p}_1} \zeta_{\vec{p}_3} \rangle \Big\langle \zeta_{\vec{k}_1 - \vec{p}_1} \zeta_{\vec{k}_4 - \vec{p}_4} \Big\rangle \Big\langle \zeta_{\vec{k}_2 - \vec{p}_2} \zeta_{\vec{k}_3 - \vec{p}_3} \Big\rangle \langle \zeta_{\vec{p}_2} \zeta_{\vec{p}_4} \rangle, \tag{6.4.4}
$$

where the first line indicates the disconnected contribution (case A), while the remaining lines correspond to the connected pieces (case B, C, D, respectively). All cases are plotted diagrammatically in Figs. 6.5 and 6.6.

Case A gives the homogeneous contribution to the energy density of GWs and does not contribute to the anisotropies as it is space independent

$$
\langle \rho_{\mathrm{GW}}(\eta, \vec{x}) \rho_{\mathrm{GW}}(\eta, \vec{y}) \rangle_A = \langle \rho_{\mathrm{GW}}(\eta, \vec{x}) \rangle \langle \rho_{\mathrm{GW}}(\eta, \vec{y}) \rangle = \langle \rho_{\mathrm{GW}}(\eta) \rangle^2. \tag{6.4.5}
$$

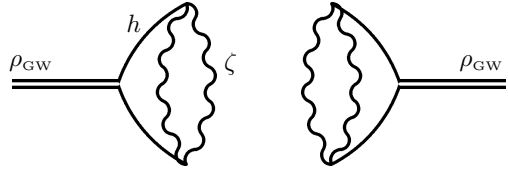

Figure 6.5: *Feynman diagram for the disconnected term (case A) in the energy density two-point function. The double lines identify the energy density field, the solid lines identify the GWs and the wiggly lines identify the curvature field.*

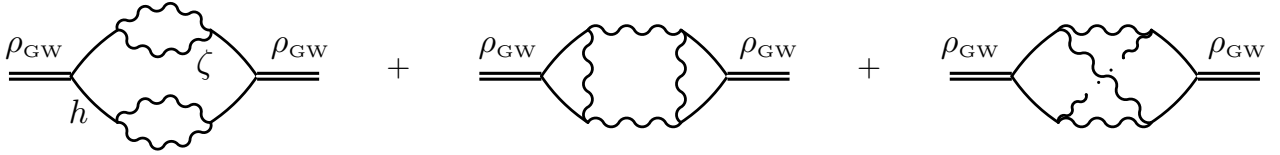

Figure 6.6: *Same as Fig. 6.5 for the connected diagrams in the energy density two-point function (case B, C and D respectively).*

This gives the monopole contribution as in Eq. (6.5.54). The remaining cases, in order B, C and D, correspond to different topologies of connected diagrams shown in Fig. 6.6.

As we will see, these contributions are completely negligible at distances $|\vec{x} - \vec{y}|$ of our interest. Since GW experiments are only able to capture angular distances corresponding to large fractions of the observable universe (i.e. small multipoles $\ell$), characteristic scales much shorter than the present horizon are not observable. In the PBH model, one always has $k_* |\vec{x} - \vec{y}| \gg 1$, and therefore, for Gaussian perturbations, no sizeable statistical correlations on cosmological scales are imprinted at formation. It is also easy to check that, as imposed by the Equivalence Principle, the anisotropies decay at large distances as $(k_* |\vec{x} - \vec{y}|)^{-2}$. In other words, as one can measure the GW energy density at some given angular scale, this effectively corresponds to coarse-graining the GW energy density with a resolution related to that scale. In practice, this is equivalent to averaging an extremely large number of patches of size $k_*^{-1}$, with the result of making the energy density extremely homogeneous due to the central limit theorem.

We will show that the connected contributions lead to anisotropies at the scale $k_*^{-1}$ by computing diagram B. We again assume a monochromatic power spectrum of curvature perturbations for simplicity. The qualitative result is expected to be the same for different choices of $P_\zeta$. In this setup, the contribution of case B to (6.4.3) becomes

$$
\langle \rho_{\text{GW}}(\eta, \vec{x}) \, \rho_{\text{GW}}(\eta, \vec{y}) \rangle_B = \frac{1}{4(2\pi)^3} \left( \frac{M_p^2 A_s^2}{81 a^2 \eta^2} \right)^2 \int dq \, q^2 j_0(q \, k_* \, |\vec{x} - \vec{y}|) \int d\Omega_s \int d\Omega_{p_1} \int d\Omega_{p_4}
$$

$$
\times \, \delta\left(|q \, \hat{e}_z + \hat{s} - \hat{p}_1 + \hat{p}_4| - 1\right) \frac{1}{|\hat{p}_1 - \hat{s}|^4} \frac{1}{|q \, \hat{e}_z + \hat{s} - \hat{p}_1|^4} T^2 \left[ \frac{\hat{p}_1 - \hat{s}}{|\hat{p}_1 - \hat{s}|}, \frac{q \, \hat{e}_z + \hat{s} - \hat{p}_1}{|q \, \hat{e}_z + \hat{s} - \hat{p}_1|}, \hat{p}_1, -\hat{p}_4 \right]
$$

$$
\times \left[ \mathcal{I}_s\left( \frac{1}{|\hat{p}_1 - \hat{s}|}, \frac{1}{|\hat{p}_1 - \hat{s}|} \right) \mathcal{I}_s\left( \frac{1}{|q \, \hat{e}_z + \hat{s} - \hat{p}_1|}, \frac{1}{|q \, \hat{e}_z + \hat{s} - \hat{p}_1|} \right) + \right.
$$

$$
\left. \mathcal{I}_c\left( \frac{1}{|\hat{p}_1 - \hat{s}|}, \frac{1}{|\hat{p}_1 - \hat{s}|} \right) \mathcal{I}_c\left( \frac{1}{|q \, \hat{e}_z + \hat{s} - \hat{p}_1|}, \frac{1}{|q \, \hat{e}_z + \hat{s} - \hat{p}_1|} \right) \right]^2 \left[ \frac{\sin(\Delta_{12} T)}{\Delta_{12} T} \right]^2, \tag{6.4.6}
$$

where we changed the integration variables into $\vec{q} = \left( \vec{k}_1 - \vec{k}_4 \right) / k_*$ and $\vec{s} = \vec{p}_1 - \vec{k}_1$ and

$$
\Delta_{12} T = \Delta_{34} T = (k_1 - k_4) T = \{|\hat{p}_1 - \hat{s}| - |q \, \hat{e}_z + \hat{s} - \hat{p}_1|\} \, k_* T. \tag{6.4.7}
$$

In this case, as before, $T$ represents the time interval over which one takes the average required by the definition of the energy density in Eq. (6.1.19). The spherical Bessel function $j_0(q \, k_* \, |\vec{x} - \vec{y}|)$ acts as a window function that forces its argument to be at most $\mathcal{O}(1)$, and therefore

$$q \approx \frac{1}{k_* \, |\vec{x} - \vec{y}|} \ll 1. \tag{6.4.8}$$

Following the notation adopted in Ref. [686], we can express the result in terms of the GW overdensity, conveniently defined as

$$\delta_{\mathrm{GW}}(\eta, \vec{x}, \vec{k}) = \frac{\Omega_{\mathrm{GW}}(\eta, \vec{x}, \vec{k}) - \bar{\Omega}_{\mathrm{GW}}(\eta, k)}{\bar{\Omega}_{\mathrm{GW}}(\eta, k)}. \tag{6.4.9}$$

Then, in the limit of small external momentum $q$, one can perform the integration numerically, and find

$$\langle \delta_{\mathrm{GW}}(\eta, \vec{x}) \, \delta_{\mathrm{GW}}(\eta, \vec{y}) \rangle_B \simeq 2 \cdot 10^2 \left( \frac{1}{k_* \, |\vec{x} - \vec{y}|} \right)^3. \tag{6.4.10}$$

This explicitly shows that the large scale anisotropies imprinted at formation in this model are highly suppressed. A similar contribution is found for the remaining two connected diagrams. We conclude that, as far as gaussian perturbations are concerned, this scenario predicts the emission of an extremely isotropic SGWB at large scales.

Let us stress again that there are additional effects involving the long-scale modes that can lead to sizeable anisotropies. The first effect is the propagation effect and was addressed in details in Refs. [680, 686]. A second possibility, that we are investigating in the following section, requires considering non-Gaussian scalar perturbations.

## 6.4.2 SGWB anisotropy at formation: the non-Gaussian case

Let us consider a local non-Gaussianity of the primordial scalar perturbations parametrised as

$$\zeta(\vec{k}) = \zeta_g(\vec{k}) + \frac{3}{5} f_{\mathrm{NL}} \int \frac{d^3 p}{(2\pi)^3} \zeta_g(\vec{p}) \, \zeta_g(\vec{k} - \vec{p}). \tag{6.4.11}$$

The specific ansatz used in Eq. (6.4.11) corresponds to a local expansion of the curvature field in real space. We recall that, as we are interested in very large scales, this shape of non-Gaussianity was constrained by CMB observations performed by the Planck collaboration [688]. The currently allowed range for the non-linear parameter is

$$-11.1 \le f_{\mathrm{NL}} \le 9.3 \quad , \quad \text{at } 95\% \text{ C.L.} \tag{6.4.12}$$

We consider the peak-background split picture, where one can expand the Gaussian comoving curvature perturbation field $\zeta_g$ as the sum of short $\zeta_s$ and long $\zeta_L$ components. In this case, one finds connected contributions to the energy density two-point function, where the long mode is bridging the two diagrams, see Fig. 6.7. In this case, the leading order term for $\langle \rho_{\mathrm{GW}}(\vec{x}) \, \rho_{\mathrm{GW}}(\vec{y}) \rangle$, not yet averaged over the long modes, results in a curvature perturbation four point function of the schematic form

$$\langle \zeta^4 \rangle = \left( 1 + \frac{24}{5} f_{\mathrm{NL}} \zeta_L \right) \left( \langle \zeta_s^2 \rangle \langle \zeta_s^2 \rangle + \langle \zeta_s^2 \rangle \langle \zeta_s^2 \rangle + \langle \zeta_s^2 \rangle \langle \zeta_s^2 \rangle \right), \tag{6.4.13}$$

where each term in the rightmost sum schematically corresponds to different ways to contract the long mode with the vertices in the diagram 6.7. In practice, one can write the resulting abundance, as a function of the long mode $\zeta_L$ as

$$\Omega_{\mathrm{GW}}(\eta, \vec{x}, k) = \bar{\Omega}_{\mathrm{GW}}(\eta, k) \left[ 1 + \frac{24}{5} f_{\mathrm{NL}} \int \frac{d^3 q}{(2\pi)^3} e^{i\vec{q}\cdot\vec{x}} \zeta_L(\vec{q}) \right], \tag{6.4.14}$$

where the term $\bar{\Omega}_{\mathrm{GW}}(\eta, k)$ identifies the contribution with the absence of the long mode, see Eq. (6.5.54).

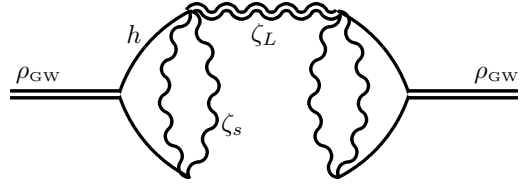

Figure 6.7: *Feynman diagram describing the energy density two-point function connected by a $f_{NL}$ bridge. The double wiggly line indicates the long mode $\zeta_L$.*

Following again the notation introduced in [686], we write the amount of anisotropy in the GW abundance by introducing the contrast

$$\delta_{\mathrm{GW}}(\eta,\vec{x},\vec{k}) = \frac{\Omega_{\mathrm{GW}}(\eta,\,\vec{x},\,\vec{k}) - \bar{\Omega}_{\mathrm{GW}}(\eta,\,k)}{\bar{\Omega}_{\mathrm{GW}}(\eta,\,k)} \equiv \Gamma_I(\eta,\vec{x},\vec{k})\left(4 - \frac{\partial \ln \bar{\Omega}_{\mathrm{GW}}(\eta,k)}{\partial \ln k}\right), \quad (6.4.15)$$

in terms of the quantity

$$\Gamma_I(\eta,\vec{x},\vec{k}) = \frac{3}{5}\tilde{f}_{\mathrm{NL}}(k)\int \frac{d^3 q}{(2\pi)^3}\,\mathrm{e}^{i\vec{q}\cdot\vec{x}}\,\zeta_L(\vec{q}), \quad (6.4.16)$$

where we defined a scale dependent coefficient

$$\tilde{f}_{\mathrm{NL}}(k) \equiv \frac{8\,f_{\mathrm{NL}}}{4 - \frac{\partial \ln \bar{\Omega}_{\mathrm{GW}}(\eta,k)}{\partial \ln k}}. \quad (6.4.17)$$

This term carries all the information about the amount of anisotropy imprinted at the formation and therefore acts as an initial condition (thus the suffix $I$). Fig. 6.8 shows the behaviour of the rescaled non-linear parameter as a function of the GW momentum for both a Dirac delta and lognormal power spectrum, as defined in Eqs. (6.1.27) and (6.1.29) respectively.

To work out the prediction for the anisotropies of the SGWB reaching the detectors today, we need to include the effect of propagation in the perturbed universe. This was studied in Ref. [686] by solving the free Boltzmann equation for a general SGWB (for a discussion on the graviton collisional corrections, see Ref. [689] and Refs. therein).

Without loss of generality, we set our location at the origin and define $\vec{k} = k\,\hat{n}$. In this reference frame, the position of the source is at $\vec{x} = \hat{n}(\eta_{\mathrm{in}} - \eta)$, where $\eta_{\mathrm{in}}$ indicates the emission time which we associate to the moment when the modes $k_*$ re-enter the horizon. It is also useful to perform an expansion in spherical harmonics, to get [6]

$$\Gamma_{\ell m, I}(k) = 4\pi\,(-i)^\ell\,\frac{3}{5}\tilde{f}_{\mathrm{NL}}(k)\int \frac{d^3 q}{(2\pi)^3}\zeta_L(\vec{q})\,Y_{\ell m}^*(\hat{q})\,j_\ell(q\,(\eta_0 - \eta_{\mathrm{in}})). \quad (6.4.18)$$

We account for the anisotropies induced by the GW propagation by introducing

$$\Gamma_S(\eta_0,\vec{q}) = \mathcal{T}^S(q,\,\eta_0,\,\eta_{\mathrm{in}})\,\zeta_L(\vec{q}), \quad (6.4.19)$$

where

$$\mathcal{T}^S(q,\,\eta_0,\,\eta_{\mathrm{in}}) = \int_{\eta_{\mathrm{in}}}^{\eta_0} d\eta'\,e^{-i\hat{k}\cdot\vec{q}q(\eta_0-\eta')}\left[T_\Phi(\eta',\,q)\,\delta(\eta'-\eta_{\mathrm{in}}) + \frac{\partial\left[T_\Psi(\eta',\,q) + T_\Phi(\eta',\,q)\right]}{\partial\eta'}\right], \quad (6.4.20)$$

and

$$\Phi(\eta,\vec{k}) \equiv T_\Phi(\eta,k)\zeta(\vec{k}), \qquad \Psi(\eta,\vec{k}) \equiv T_\Psi(\eta,k)\zeta(\vec{k}). \quad (6.4.21)$$

This factor account for the effect of large scale scalar perturbations on the SGWB. We will neglect the subdominant contribution coming from large-scale tensor modes. As we are looking for perturbing modes at very large scales, those have surely entered the horizon during matter domination and their transfer functions become $T_\Phi (\eta_{in}, q) = T_\Psi (\eta_{in}, q) = 3/5$. The effect of scalar perturbations can be thought of as being analogous to the well known Sachs-Wolfe and integrated Sachs-Wolfe effects notoriously encountered when describing the CMB anisotropies. As the latter is subdominant in the present case [686, 690], one can approximate the total contribution of the long mode affecting both the emission and propagation of the GWs, at leading order in the non-linear parameter, through the quantity

$$\Gamma_{\ell m, I+S} (k) \simeq 4\pi (-i)^\ell \int \frac{d^3 q}{(2\pi)^3} \zeta_L (\vec{q}) \, Y^*_{\ell m} (\hat{q}) \, \frac{3}{5} \left[ 1 + \tilde{f}_{NL} (k) \right] j_\ell (q (\eta_0 - \eta_{in})). \quad (6.4.22)$$

---

[6]We are using spherical harmonic functions obeying the normalisation condition $\int d\hat{n}\,Y_{\ell m}Y_{\ell' m'}^* = \delta_{\ell\ell'}\delta_{mm'}$.

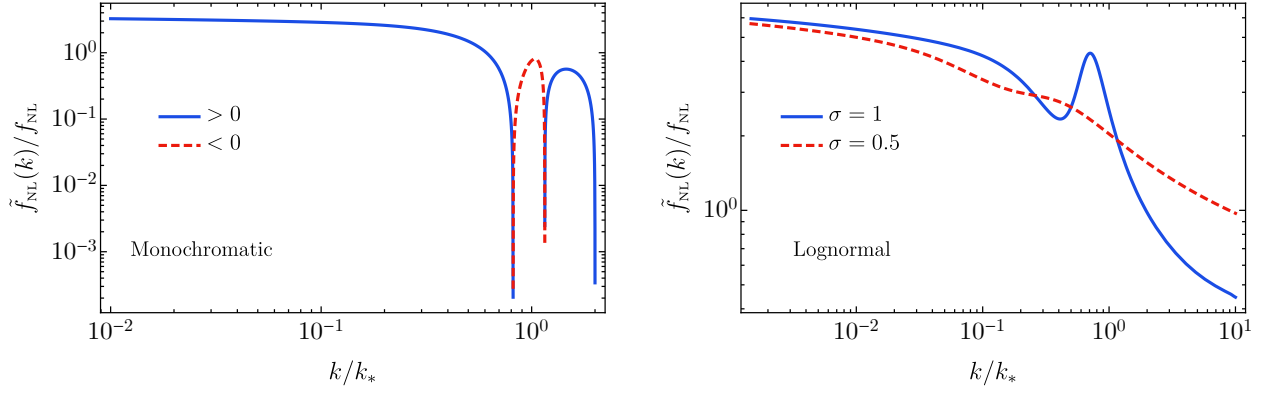

Figure 6.8: $\tilde{f}_{NL}/f_{NL}$ as a function of the ratio $k/k_*$ for both monochromatic (left) and lognormal (right) power spectrum.

In the following subsections, we will compute the two-point and three-point functions of the rescaled energy density as a function of the long modes power spectrum and the local non-linear parameter.

### 6.4.3  Energy density two-point function

Let us focus on the computation of the two point function of the term (6.4.22). This can be written as

$$\langle \Gamma_{\ell_1 m_1, I+S} (k) \, \Gamma^*_{\ell_2 m_2, I+S} (k) \rangle = (4\pi)^2 (-i)^{\ell_1 - \ell_2} \int \frac{d^3 q_1}{(2\pi)^3} \frac{d^3 q_2}{(2\pi)^3} Y^*_{\ell_1 m_1} (\hat{q}_1) \, Y_{\ell_2 m_2} (\hat{q}_2)$$

$$\times \left( \frac{3}{5} \right)^2 \left[ 1 + \tilde{f}_{NL} (k) \right]^2 j_{\ell_1} (q_1 (\eta_0 - \eta_{in})) \, j_{\ell_2} (q_2 (\eta_0 - \eta_{in})) \, \langle \zeta_L (\vec{q}_1) \, \zeta^*_L (\vec{q}_2) \rangle. \quad (6.4.23)$$

Then, using the property of orthonormality of the spherical harmonics and adopting a scale invariant power spectrum of the long modes $\mathcal{P}_{\zeta_L}(q) \simeq \mathcal{P}_{\zeta_L}$, the previous expression becomes

$$\langle \Gamma_{\ell_1 m_1, I+S} (k) \, \Gamma^*_{\ell_2 m_2, I+S} (k) \rangle = \delta_{\ell_1 \ell_2} \delta_{m_1 m_2} 4\pi \left( \frac{3}{5} \right)^2 \left[ 1 + \tilde{f}_{NL}(k) \right]^2 \frac{1}{2\ell_1 (\ell_1 + 1)} \mathcal{P}_{\zeta_L}. \quad (6.4.24)$$

Following the notation of [686], we can therefore express the two point function with the characteristic $C_\ell$ coefficients defined as

$$\langle \Gamma_{\ell_1 m_1, I+S} (k) \, \Gamma^*_{\ell_2 m_2, I+S} (k) \rangle = \delta_{\ell_1 \ell_2} \delta_{m_1 m_2} \, C_{\ell, I+S} (k). \quad (6.4.25)$$

Finally, one finds

$$\sqrt{\frac{\ell (\ell + 1)}{2\pi} C_{\ell, I+S} (k)} \simeq 2.8 \cdot 10^{-4} \left| \frac{1 + \tilde{f}_{NL} (k)}{10} \right| \left( \frac{\mathcal{P}_{\zeta_L}}{2.2 \cdot 10^{-9}} \right)^{1/2}. \quad (6.4.26)$$

We chose to normalise the quantities to the measured power spectrum amplitude at the CMB scales and the non-linear parameter close to its upper bound (6.4.12). We will come back to this result in Sec. 6.6, where we will focus the discussion on the potential signal observable at the LISA experiment.

### 6.4.4    Energy density three-point function

To compute the three-point function, accounting for both the anisotropy imprinted at emission and the one induced by propagation effects, we need to extend the quantities $\Gamma$ to the next-to-leading order in the non-linear parameter $f_{\text{NL}}$. Therefore, one gets [17]

$$
\begin{aligned}
\Gamma_{\ell m, I+S}\left(k\right) \;\simeq\; & 4\pi\left(-i\right)^{\ell} \int \frac{d^3 q}{\left(2\pi\right)^3}\, Y_{\ell m}^{*}\left(\hat{q}\right)\, j_{\ell}\left(q\left(\eta_0 - \eta_{\text{in}}\right)\right) \left\{ \frac{3}{5}\left[1 + \tilde{f}_{\text{NL}}\left(k\right)\right] \zeta_L\left(\vec{q}\right) \right. \\
& \left. + \; \frac{9}{25}\, f_{\text{NL}}\left[1 + 3\tilde{f}_{\text{NL}}\left(k\right)\right] \int \frac{d^3 p}{\left(2\pi\right)^3}\, \zeta_L\left(\vec{p}\right) \zeta_L\left(\vec{q} - \vec{p}\right) \right\},
\end{aligned}
\tag{6.4.27}
$$

where we stress once again that all the long modes $\zeta_L$ in this expression are Gaussian fields. The three point function is found by computing

$$
\begin{aligned}
\left\langle \prod_{i=1}^{3} \Gamma_{\ell_i m_i, I+S}\left(k\right) \right\rangle = & \;(4\pi)^3\left(-i\right)^{\ell_1 + \ell_2 + \ell_3}\, \frac{81}{625}\, f_{\text{NL}}\left[1 + \tilde{f}_{\text{NL}}\left(k\right)\right]^2 \left[1 + 3\tilde{f}_{\text{NL}}\left(k\right)\right] \\
& \times \int \frac{d^3 q_1}{\left(2\pi\right)^3} \int \frac{d^3 q_2}{\left(2\pi\right)^3} \int \frac{d^3 q_3}{\left(2\pi\right)^3} \int \frac{d^3 p}{\left(2\pi\right)^3} \left[\prod_{i=1}^{3} Y_{\ell_i m_i}^{*}\left(\hat{q}_i\right)\, j_{\ell_i}\left(q_i\left(\eta_0 - \eta_{\text{in}}\right)\right)\right] \\
& \times \left\langle \zeta_L\left(\vec{q}_1\right) \zeta_L\left(\vec{q}_2\right) \zeta_L\left(\vec{p}\right) \zeta_L\left(\vec{q}_3 - \vec{p}\right) \right\rangle + 2\,\text{perm}.
\end{aligned}
\tag{6.4.28}
$$

The tensorial structure of the three-point function is dictated by statistical isotropy and found to be proportional to the Gaunt integral [691], defined as

$$
\mathcal{G}_{\ell_1 \ell_2 \ell_3}^{m_1 m_2 m_3} = \int d\Omega_y\, Y_{\ell_1 m_1}^{*}\left(\Omega_y\right) Y_{\ell_2 m_2}^{*}\left(\Omega_y\right) Y_{\ell_3 m_3}^{*}\left(\Omega_y\right).
\tag{6.4.29}
$$

Therefore, after few manipulations, one finds [17]

$$
\left\langle \Gamma_{\ell_1 m_1, I+S}(k)\Gamma_{\ell_2 m_2, I+S}(k)\Gamma_{\ell_3 m_3, I+S}(k)\right\rangle = \mathcal{G}_{\ell_1 \ell_2 \ell_3}^{m_1 m_2 m_3}\, b_{\ell_1 \ell_2 \ell_3, I+S}\left(k\right),
\tag{6.4.30}
$$

where $b_{\ell_1 \ell_2 \ell_3, I+S}$ can be written in terms of the two-point functions found in Eq. (6.4.26) as

$$
b_{\ell_1 \ell_2 \ell_3, I+S}\left(k\right) \simeq \frac{2\, f_{\text{NL}}\left[1 + 3\, \tilde{f}_{\text{NL}}\left(k\right)\right]}{\left[1 + \tilde{f}_{\text{NL}}\left(k\right)\right]^2}\left[C_{\ell_1, I+S}\, C_{\ell_2, I+S} + C_{\ell_1, I+S}\, C_{\ell_3, I+S} + C_{\ell_2, I+S}\, C_{\ell_3, I+S}\right].
\tag{6.4.31}
$$

We will discuss this result further by focusing on the LISA detector in Sec. 6.6.

## 6.5    Induced SGWB gauge (in-)dependence

The GW signatures discussed in this chapter have the peculiar property of being produced by a second-order source. This forces us to consider the subtle issue of gauge dependence of the SGWB. In simple terms, it is a standard result of perturbation theory that the linear tensor perturbations of the metric are gauge-invariant. However, this is not true anymore when tensor modes at second order are considered, see details in App. E. Thus, one needs to check whether the observables computed are in fact gauge-invariant, as it should be for a well define observable. This issue was first pointed out in this context in Refs. [652, 692–695]

Any proper observation should be, by definition of gauge redundancy, invariant under gauge transformations. Therefore, we will start the discussion by considering how the GW measurements are performed in ground-based and space-based detectors. Then, we will describe a well-defined procedure to build gauge-invariant descriptions of second-order tensor perturbations and their equation of motion. Finally, we will show that the computation performed in the Newtonian gauge gives rise to the correct, gauge-invariant, *late-time* SGWB spectrum, confirming the result obtained in the previous sections. We will follow the discussion in Ref. [16].

### 6.5.1   The Measurement of GWs and the TT frame

In this section, we discuss how GWs are observed and what the experimental apparatus, in practice, can measure. We follow the steps described in the seminal book in Ref. [273] where many more details can be found.

The most advanced technology proposed so far to measure GWs makes use of laser interferometers. Let us sketch here the simplified procedure by which they can measure the passage of a GW signal. We consider the arms A and B of lengths $L_A \sim L_B \sim L$, aligned in the $\hat{x}$ and $\hat{y}$ directions, respectively. Without loss of generality, we fix the origin of our coordinate system with the position of the beam splitter at the initial time $t_0$. The experimental apparatus sends a beam of photons to mirrors placed at the end of the arms and collect the electromagnetic signal which comes back to the receiver. By measuring the modulation of the power observed at the receiver, one can infer the variation of the time $\Delta t_{A,B}$ elapsed in the two different travel paths. For presentation purposes, let us only focus on a single component of the electromagnetic vector field (of frequency $\omega_L$) which gets a phase shift in both arms given by

$$E_A(t) = -\frac{1}{2}E_0 e^{-i\omega_L(t - 2L_A) + i\Delta\phi_A(t)} \qquad \text{with} \qquad \Delta\phi_A = -\omega_L\Delta t_A, \tag{6.5.1}$$

$$E_B(t) = -\frac{1}{2}E_0 e^{-i\omega_L(t - 2L_B) + i\Delta\phi_B(t)} \qquad \text{with} \qquad \Delta\phi_B = -\omega_L\Delta t_B. \tag{6.5.2}$$

The observer records the total power of the electric field $P \sim |E_A + E_B|^2$ which is modulated by the passage of a GW signal as

$$P(t) = \frac{P_0}{2}\left\{1 - \cos\left[2\phi_0 + \Delta\phi_{GW}(t)\right]\right\}, \tag{6.5.3}$$

where we have conveniently defined $\phi_0 = k_L(L_A - L_B)$ and $\Delta\phi_{GW}(t) = \Delta\phi_A - \Delta\phi_B$. We note also that the description of the passage of a GW through the detector is frame dependent. One can characterise the GW passage by describing it as inducing a movement of both mirrors or as responsible for a modification of the photon geodesic (with the mirrors at fixed locations of the reference frame). We will discuss these two equivalent descriptions separately.

As the measurement of GWs takes place in time intervals that are much smaller than the typical rate of change of the cosmological background, in the following we will safely neglect the expansion of the universe and take flat spacetime limit, that is the limit in which we can set the scale factor $a = 1$.

There are fundamental differences between ground-based and space-based detectors that one needs to consider. For ground-based detectors, the typical GW frequency is of the order of $\omega_{GW}L \sim 10^{-2}$. Therefore, it is possible to expand in powers of $\omega_{GW}L$. At leading order, one can define an inertial reference frame that contains both sides of the experimental apparatus and where the metric is locally flat. This is called the proper detector frame. In this case, the effect of a GW can be described as a Newtonian force acting on both mirrors, whose movement is described by the geodesic deviation equation, thus changing the distance $L_A$ and $L_B$. For space-based observatories like LISA, on the contrary, one has $\omega_{GW}L \sim \pi/2$ and the expansion in powers of $\omega_{GW}L$ inevitably fails. In other words, it is not possible to define a single reference frame where the whole apparatus is described by an (approximately) flat metric in the presence of the GW. Therefore, the description must rely on a general relativistic framework, where the coordinate system is conveniently defined by the positions of the mirrors. This is called the TT frame (also known as the synchronous frame in a cosmological setting). In this setup, the GW affects the proper time measured by the observer at the origin, as we will see.

#### The measurement in the proper detector frame at first-order

If one assumes that $\omega_{GW}L \ll 1$, as we discussed above, the experimental apparatus can be described in the proper detector frame. In this case, the spacetime metric around the detector is flat, and the time differentials can be written as

$$dt^2 = d\tau^2\left(1 + \frac{dx^i}{d\tau}\frac{dx^i}{d\tau}\right) = d\tau^2\left(1 + \mathcal{O}(\delta g^2)\right), \tag{6.5.4}$$

since any movement of the mirrors must be at least at first order with respect to the metric perturbation $\delta g$. Therefore, time intervals can be identified with proper time intervals. One can also assume the velocity to be small with respect to $c = 1$ and $\mathrm{d}x^i/\mathrm{d}\tau \ll \mathrm{d}x^0/\mathrm{d}\tau$, to get the equation of motion for the infinitesimal displacement of the mirrors as

$$\frac{\mathrm{d}^2\xi^i}{\mathrm{d}\tau^2} = -R^i_{0j0}\xi^j\left(\frac{\mathrm{d}x^0}{\mathrm{d}\tau}\right)^2. \tag{6.5.5}$$

By focusing on the leading order in $\mathcal{O}(\delta g)$, and since in a flat background spacetime $R^i_{0j0} \sim \mathcal{O}(\delta g)$, one can rewrite Eq. (6.5.5) as

$$\ddot{\xi_i} = -R_{i0j0}\xi^j, \tag{6.5.6}$$

where the dot denotes the derivative with respect to the coordinate time $t = \tau$. This equation can be solved perturbatively to get

$$\xi_{\mathrm{A}}^i = \left[L_{\mathrm{A}} + \delta\xi_{\mathrm{A}}^i + \mathcal{O}(\delta g^2)\right]\hat{x}^i = \left[L_{\mathrm{A}} - L_{\mathrm{A}}\int^t \mathrm{d}t'\int^{t'}\mathrm{d}t''R_{1010} + \mathcal{O}(\delta g^2)\right]\hat{x}^i,$$

$$\xi_{\mathrm{B}}^i = \left[L_{\mathrm{B}} + \delta\xi_{\mathrm{B}}^i + \mathcal{O}(\delta g^2)\right]\hat{y}^i = \left[L_{\mathrm{B}} - L_{\mathrm{B}}\int^t \mathrm{d}t'\int^{t'}\mathrm{d}t''R_{2020} + \mathcal{O}(\delta g^2)\right]\hat{y}^i. \tag{6.5.7}$$

We follow the notation defined in App. E (where we also define our convention for indicating derivatives and index symmetrisation) and write a generic perturbation of the metric as

$$\mathrm{d}s^2 = -a^2(1 + 2\phi)\mathrm{d}\eta^2 + 2a^2B_i\mathrm{d}\eta\,\mathrm{d}x^i + a^2(\delta_{ij} + 2C_{ij})\mathrm{d}x^i\mathrm{d}x^j, \tag{6.5.8}$$

for which, at leading order and in the flat spacetime limit, the Riemann tensor $R^{(1)}_{i0j0}$ is

$$R^{(1)}_{i0j0} = \partial_i\partial_j\phi + \partial_{(i}B'_{j)} - C''_{ij}. \tag{6.5.9}$$

Using the scalar-vector-tensor (SVT) decomposition, the term $C_{ij}$ can be expanded as

$$C_{ij} = -\psi\delta_{ij} + E_{,ij} + F_{(i,j)} + \frac{1}{2}h_{ij}. \tag{6.5.10}$$

where $h_{ij}$ indicates the linear tensor perturbation, while $(\phi, \psi, E)$ and $(B_i, F_i)$ indicate the scalar and vector perturbations of the metric, respectively.

By looking at Eq. 6.5.7, it becomes evident that the passage of the GW affects the movement of the mirrors, and thus the roundtrip time of photons in each arm. One can also show, see Eq. (E.18), that $R^{(1)}_{i0j0}$ is explicitly gauge-invariant under a first-order coordinate transformation [273, 696]. Therefore, it can be computed in any reference frame. Finally, the proper time spent by a photon to complete a round trip ($\mathrm{d}t = \pm\mathrm{d}x$) is found to be

$$\Delta t_{\mathrm{A,B}} = 2\int_{t_0}^{L_{\mathrm{A,B}}+\delta\xi_{\mathrm{A,B}}}\mathrm{d}x = 2L_{\mathrm{A,B}} + 2\delta\xi_{\mathrm{A,B}}(t_0 + L_{\mathrm{A,B}}). \tag{6.5.11}$$

As, at first-order, this result is gauge-invariant, such a description is equivalent to the one in the TT frame at first-order in $\omega_{\mathrm{GW}}L \ll 1$, as we will show in the following.

### The measurement in the TT frame at first and second-order

In the TT frame, the coordinates are fixed by the locations of both mirrors, see Fig. 6.9, and one does not need to assume an expansion in powers of $\omega_{\mathrm{GW}}L$. One can show explicitly that in the TT frame the mirrors remain at rest also when perturbed by a passing GW.

Relying on the general relativistic description of the system, the test masses follows the geodesic equation of motion as

$$\frac{\mathrm{d}^2x^\mu}{\mathrm{d}\tau^2} = -\Gamma^\mu_{\nu\rho}\frac{\mathrm{d}x^\nu}{\mathrm{d}\tau}\frac{\mathrm{d}x^\rho}{\mathrm{d}\tau}, \tag{6.5.12}$$

**TT gauge**

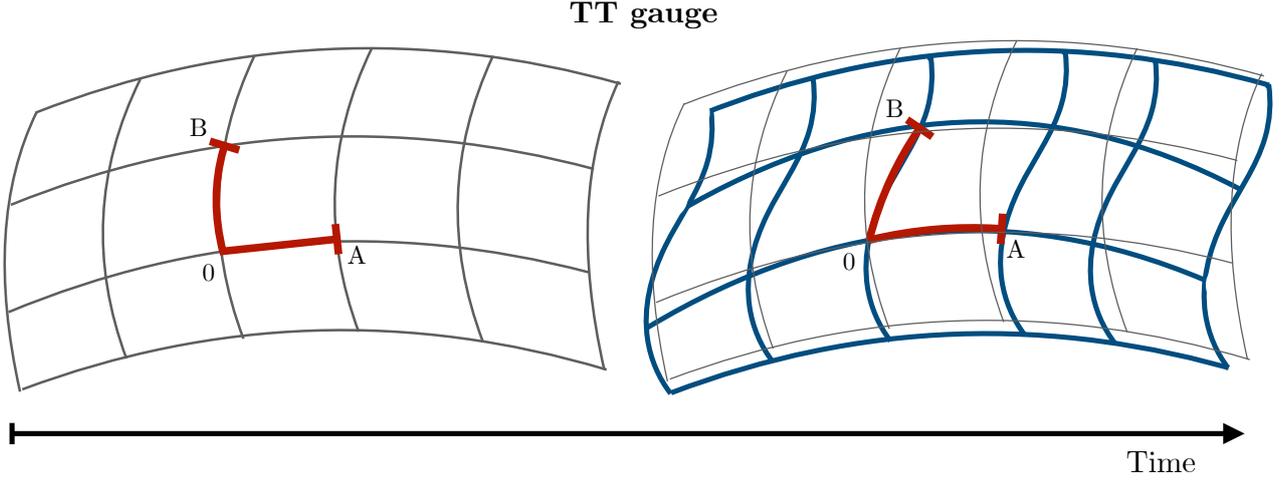

Time

Figure 6.9: *Pictorial representation of a GW passing through the detector in the TT frame. The coordinates are chosen such that the positions of the interferometer arms (in red) are fixed with the position of the mirrors and, consequently, they do not move even in the presence of a GW.*

where $\tau$ identifies the proper time. Assuming they are initially at rest, Eq. (6.5.12) becomes

$$\frac{\mathrm{d}^2 x^i}{\mathrm{d}\tau^2} = -\Gamma^i_{00} \left( \frac{\mathrm{d}x^0}{\mathrm{d}\tau} \right)^2 . \tag{6.5.13}$$

Now we impose the TT gauge, which is characterised by $\delta g_{00} = 0$ and $\delta g_{0i} = 0$, or in the notation defined in Eq. (6.5.8),

$$\widetilde{\phi} = 0 \qquad \text{and} \qquad \widetilde{B}_i = 0. \tag{6.5.14}$$

With this choice, the Christoffel symbols $\Gamma^i_{00}$ vanish, and therefore one finds

$$\frac{\mathrm{d}^2 x^i}{\mathrm{d}\tau^2} = 0, \tag{6.5.15}$$

showing that mirrors that are initially at rest, always remain fixed in their location in the TT frame.

The physical effect of a passing GW is captured by studying the proper times which are measured at the origin of the interferometer. The photons travelling to the mirror and back to the receiver follows the geodesic equation

$$\mathrm{d}s^2 = 0 = -\mathrm{d}t^2 + (\delta_{ij} + 2C_{1ij} + C_{2ij}) \, \mathrm{d}x^i \mathrm{d}x^j. \tag{6.5.16}$$

Focusing, for example, on the arm A of the detector, at second-order one finds

$$\mathrm{d}x = \pm \mathrm{d}t \left[ 1 - C_{1ij} + \frac{3}{2} C_{1ij}^2 - \frac{1}{2} C_{2ij} \right]_{i=j=1} + \dots \tag{6.5.17}$$

where the upper (lower) sign holds for the travel towards (back from) the mirrors at the end of the arms. Therefore, the time shifts are, up to second order perturbations,

$$\Delta t_{\mathrm{A,B}} = L_{\mathrm{A,B}} - \int_{t_0}^{t_0 + 2L_{\mathrm{A,B}}} \mathrm{d}t \left( -C_{1ij} + \frac{3}{2} C_{1ij}^2 - \frac{1}{2} C_{2ij} \right)_{i=j=1,2} . \tag{6.5.18}$$

One can explicitly show that, expanding at leading order in $\omega_{\mathrm{GW}} L \ll 1$ and $\delta g$ one recovers the same result for the time shift found in Eq. (6.5.11).

As the GWs reaching the detectors are tiny perturbations of the underlying spacetime and decoupled from the second-order source which is only sizeable in the early universe during the radiation dominated era, the first-order description of the measurement procedure will be sufficient for any practical purposes. Also, the TT frame provides a general framework to describe both ground-based and space-based detectors and therefore, this frame will be used in the following sections to compare the prediction of the SGWB generated in the PBH model with the sensitivity curves of GWs experiments.

### 6.5.2 Gauge-invariant description of second-order tensor perturbations

It is always possible to promote a gauge dependent quantity to a gauge-invariant one, following the explicit construction defined in Refs. [691, 697–700]. Let us stress, however, that this procedure is not unique. In other words, it only accounts for fixing the coordinate dependence of quantities and provides the possibility of working with explicitly gauge-invariant combinations, while it does not address the question of what is the physical observable, which is tightly related to the nature of the measurement performed. We will come back to this question in the following section.

Here, we show how this can be done for second order tensor modes. We start by considering a given gauge. In order to set such a gauge, one always performs a coordinate transformation of the form

$$x^\mu \to \widetilde{x}^\mu = x^\mu + \xi^\mu \qquad \text{with} \qquad \xi^\mu \equiv \left( \alpha, \xi^i \right),  \tag{6.5.19}$$

where the parameters $\alpha_1^{\mathrm{GC}}$ and $\xi_{1i}^{\mathrm{GC}}$ are functions of the perturbation fields. Then, the same functions can be used to perform an additional gauge transformation of the original fields yielding explicitly gauge-invariant quantities.

Let us show this procedure for the case of first-order scalar potentials $\phi_1$ and $\psi_1$. They transform as, see Eq. (E.18),

$$\widetilde{\phi}_1 \equiv \phi_1 + \mathcal{H}\alpha_1 + \alpha_1',  \tag{6.5.20}$$

$$\widetilde{\psi}_1 \equiv \psi_1 - \mathcal{H}\alpha_1.  \tag{6.5.21}$$

If we promote the gauge parameter $\alpha_1$ to be the field combination adopted to set a particular gauge, $\alpha_1^{\mathrm{GC}}(\delta g)$, one obtains explicitly gauge-invariant scalar perturbations. Adopting this procedure starting from the Poisson gauge leads to the definition of the gauge-invariant Bardeen's potentials, as we will see.

The same steps can be used to define gauge-invariant second-order tensor perturbations. Promoting the gauge transformation of the second order tensor as in Eq. (E.21g), one defines [697, 701]

$$h_{2ij}^{\mathrm{GI}} \equiv h_{2ij} + \mathcal{X}_{ij}^{\mathrm{GC}} + \frac{1}{2} \left( \nabla^{-2} \mathcal{X}_{,kl}^{\mathrm{GC}kl} - \mathcal{X}_{k}^{\mathrm{GC}k} \right) \delta_{ij} + \frac{1}{2} \nabla^{-2} \nabla^{-2} \mathcal{X}_{,klij}^{\mathrm{GC}kl}$$
$$+ \frac{1}{2} \nabla^{-2} \mathcal{X}_{k,ij}^{\mathrm{GC}k} - \nabla^{-2} \left( \mathcal{X}_{ik,\ j}^{\mathrm{GC}k} + \mathcal{X}_{jk,\ i}^{\mathrm{GC}k} \right),  \tag{6.5.22}$$

where

$$\mathcal{X}_{ij}^{\mathrm{GC}} \equiv 2\bigg[ \left( \mathcal{H}^2 + \frac{a''}{a} \right) \alpha_1^{\mathrm{GC}2} + \mathcal{H} \left( \alpha_1^{\mathrm{GC}} \alpha_1^{\mathrm{GC}\prime} + \alpha_{1,k}^{\mathrm{GC}} \xi_1^{\mathrm{GC}k} \right) \bigg] \delta_{ij} + 2 \left( B_{1i} \alpha_{1,j}^{\mathrm{GC}} + B_{1j} \alpha_{1,i}^{\mathrm{GC}} \right)$$
$$+ 4 \bigg[ \alpha_1^{\mathrm{GC}} \left( C_{1ij}' + 2\mathcal{H}C_{1ij} \right) + C_{1ij,k} \xi_1^{\mathrm{GC}k} + C_{1ik} \xi_{1\ ,j}^{\mathrm{GC}k} + C_{1kj} \xi_{1\ ,i}^{\mathrm{GC}k} \bigg]$$
$$+ 4\mathcal{H}\alpha_1^{\mathrm{GC}} \left( \xi_{1i,j}^{\mathrm{GC}} + \xi_{1j,i}^{\mathrm{GC}} \right) - 2\alpha_{1,i}^{\mathrm{GC}} \alpha_{1,j}^{\mathrm{GC}} + 2\xi_{1k,i}^{\mathrm{GC}} \xi_{1\ ,j}^{\mathrm{GC}k} + \alpha_1^{\mathrm{GC}} \left( \xi_{1i,j}^{\mathrm{GC}\prime} + \xi_{1j,i}^{\mathrm{GC}\prime} \right)$$
$$+ \left( \xi_{1i,jk}^{\mathrm{GC}} + \xi_{1j,ik}^{\mathrm{GC}} \right) \xi_1^{\mathrm{GC}k} + \xi_{1i,k}^{\mathrm{GC}} \xi_{1\ ,j}^{\mathrm{GC}k} + \xi_{1j,k}^{\mathrm{GC}} \xi_{1\ ,i}^{\mathrm{GC}k} + \xi_{1i}^{\mathrm{GC}\prime} \alpha_{1,j}^{\mathrm{GC}} + \xi_{1j}^{\mathrm{GC}\prime} \alpha_{1,i}^{\mathrm{GC}}  \tag{6.5.23}$$

in terms of the field combinations $\alpha_1^{\mathrm{GC}}(\delta g)$ and $\xi_{1i}^{\mathrm{GC}}(\delta g)$. The presence of non-local terms in the definition of the gauge-invariant second-order tensor modes is not a concern as these terms disappear in the equation of motion projected on the transverse and traceless components. This becomes more transparent when considering the helicity fields

$$h_\lambda(t, \boldsymbol{k}) = e_\lambda^{ij}(\boldsymbol{k}) \int \mathrm{d}^3 x e^{-i\boldsymbol{k}\cdot\boldsymbol{x}} h_{ij}(t, \boldsymbol{x}),  \tag{6.5.24}$$

for which

$$h_{2\lambda}^{\mathrm{GI}} = h_{2\lambda} + e_\lambda^{ij}(\boldsymbol{k}) \mathcal{X}_{ij}^{\mathrm{GC}}.  \tag{6.5.25}$$

We stress again that this construction of gauge-invariant tensor modes is not unique and, in the following, we will provide a couple of explicit examples.

**Explicit construction from the Poisson (Newtonian) gauge**

We clarify the preceding discussion by showing the explicit example of the construction of gauge-invariant combinations starting from the Poisson gauge. This gauge is commonly used in the literature in the context of GWs induced at second-order.

We recall here for clarity that the general perturbation of the metric can be written as, see App. E,

$$
\begin{aligned}
\delta g_{00} &= -2a^2\phi, \\
\delta g_{0i} &= a^2\left(B_{,i} - S_i\right), \\
\delta g_{ij} &= 2a^2\left(C_{ij} = -\psi\delta_{ij} + E_{,ij} + F_{(i,j)} + \frac{1}{2}h_{ij}\right).
\end{aligned}
\tag{6.5.26}
$$

Setting the Poisson gauge requires imposing

$$
\widetilde{E}^{\mathrm{P}} = 0, \qquad \widetilde{B}^{\mathrm{P}} = 0 \qquad \text{and} \qquad \widetilde{S}_i^{\mathrm{P}} = 0.
\tag{6.5.27}
$$

Therefore, using the gauge transformation property in appendix E.1.2, at first-order the gauge fixing is done by choosing

$$
\alpha_1^{\mathrm{P}} = B_1 - E_1', \qquad \beta_1^{\mathrm{P}} = -E_1, \qquad \gamma_{1i}^{\mathrm{P}} = \int^\eta S_{1i}\mathrm{d}\eta' + \hat{\mathcal{C}}_{1i}(\boldsymbol{x}),
\tag{6.5.28}
$$

up to an arbitrary constant 3-vector $\hat{\mathcal{C}}_{1i}$ which is set by choosing the spatial coordinates on an initial hypersurface. Now, using Eqs. (6.5.20) and (6.5.21), and promoting the gauge parameters to field combinations defined by (6.5.28), one finds

$$
\Phi_1 \equiv \phi_1 + \mathcal{H}\alpha_1^{\mathrm{P}} + \alpha_1^{\mathrm{P}\prime} = \phi_1 + \mathcal{H}(B_1 - E_1') + (B_1 - E_1')',
\tag{6.5.29}
$$
$$
\Psi_1 \equiv \psi_1 - \mathcal{H}\alpha_1^{\mathrm{P}} = \psi_1 - \mathcal{H}\left(B_1 - E_1'\right).
\tag{6.5.30}
$$

One can easily check that these combinations are explicitly gauge-invariant and equivalent to the Bardeen potentials [378].

With similar steps, one obtains the gauge-invariant second-order tensor perturbations as

$$
\begin{aligned}
h_{2ij}^{\mathrm{GI,P}} &\equiv h_{2ij} + \mathcal{X}_{ij}^{\mathrm{P}} + \frac{1}{2}\left(\nabla^{-2}\mathcal{X}^{\mathrm{P}kl}{}_{,kl} - \mathcal{X}^{\mathrm{P}k}{}_k\right)\delta_{ij} + \frac{1}{2}\nabla^{-2}\nabla^{-2}\mathcal{X}^{\mathrm{P}kl}{}_{,klij} \\
&\quad + \frac{1}{2}\nabla^{-2}\mathcal{X}^{\mathrm{P}k}{}_{k,ij} - \nabla^{-2}\left(\mathcal{X}_{ik,}^{\mathrm{P}\ k}{}_j + \mathcal{X}_{jk,}^{\mathrm{P}\ k}{}_i\right)
\end{aligned}
\tag{6.5.31}
$$

in terms of

$$
\begin{aligned}
\mathcal{X}_{ij}^{\mathrm{P}} &\equiv 2\left[\left(\mathcal{H}^2 + \frac{a''}{a}\right)\alpha_1^{\mathrm{P}2} + \mathcal{H}\left(\alpha_1^{\mathrm{P}}\alpha_1^{\mathrm{P}\prime} + \alpha_{1,k}^{\mathrm{P}}\xi_1^{\mathrm{P}k}\right)\right]\delta_{ij} \\
&\quad + 4\left[\alpha_1^{\mathrm{P}}\left(C_{1ij}' + 2\mathcal{H}C_{1ij}\right) + C_{1ij,k}\xi_1^{\mathrm{P}k} + C_{1ik}\xi_1^{\mathrm{P}k}{}_{,j} + C_{1kj}\xi_1^{\mathrm{P}k}{}_{,i}\right] + 2\left(B_{1i}\alpha_{1,j}^{\mathrm{P}} + B_{1j}\alpha_{1,i}^{\mathrm{P}}\right) \\
&\quad + 4\mathcal{H}\alpha_1^{\mathrm{P}}\left(\xi_{1i,j}^{\mathrm{P}} + \xi_{1j,i}^{\mathrm{P}}\right) - 2\alpha_{1,i}^{\mathrm{P}}\alpha_{1,j}^{\mathrm{P}} + 2\xi_{1k,i}^{\mathrm{P}}\xi_1^{\mathrm{P}k}{}_{,j} + \alpha_1^{\mathrm{P}}\left(\xi_{1i,j}^{\mathrm{P}\prime} + \xi_{1j,i}^{\mathrm{P}\prime}\right) + \left(\xi_{1i,jk}^{\mathrm{P}} + \xi_{1j,ik}^{\mathrm{P}}\right)\xi_1^{\mathrm{P}k} \\
&\quad + \xi_{1i,k}^{\mathrm{P}}\xi_1^{\mathrm{P}k}{}_{,j} + \xi_{1j,k}^{\mathrm{P}}\xi_1^{\mathrm{P}k}{}_{,i} + \xi_{1i}^{\mathrm{P}\prime}\alpha_{1,j}^{\mathrm{P}} + \xi_{1j}^{\mathrm{P}\prime}\alpha_{1,i}^{\mathrm{P}}.
\end{aligned}
\tag{6.5.32}
$$

The explicit expression of $\mathcal{X}_{ij}^{\mathrm{P}}$ can be found in App. E, see Eq. (E.22).

### 6.5.3   Gauge-invariant equation of motion for GWs

In the previous section, we have constructed gauge-invariant metric perturbations starting from the fields defined in a particular gauge. In the same manner, one can write down gauge-invariant equations of motion for the tensor field, as we will show.

The equation of motion for the transverse and traceless metric perturbation $h_{ij}$ at second-order can be written as (see, for example, Ref. [694])

$$
h_\lambda''(\eta, \boldsymbol{k}) + 2\mathcal{H}h_\lambda'(\eta, \boldsymbol{k}) + k^2 h_\lambda(\eta, \boldsymbol{k}) = 2a^2(\eta)e_\lambda^{ij}(\boldsymbol{k})s_{ij}(\boldsymbol{k}).
\tag{6.5.33}
$$

The second-order source gets contributions from three distinct pieces, which are

$$s_{ij} = s_{ij}^{(ss)} + s_{ij}^{(st)} + s_{ij}^{(tt)}, \tag{6.5.34}$$

and corresponds to the scalar-scalar, scalar-tensor and tensor-tensor terms. The piece at second order in tensor modes can be safely neglected, being a subdominant contribution. Then, it is easy to show that the scalar-scalar term is the one responsible for the GW emission, while the scalar-tensor affects the GW propagation, as discussed in Sec. 6.3.

We simplify the notation by introducing the shear potential $\sigma_1 = E_1' - B_1$. Then, each term in Eq. (6.5.34) can be expressed in full generality as [694]

$$
\begin{aligned}
s_{ij}^{(ss)} = & -\frac{1}{a^4}\frac{\mathrm{d}}{\mathrm{d}\eta}\Big[a^2\big(2\psi_1\sigma_{1,ij} + \psi_{1,i}\sigma_{1,j} + \psi_{1,j}\sigma_{1,i}\big)\Big] + \frac{1}{a^2}\big(3\mathcal{H}\phi_1 + 3\psi_1' - \sigma_{1,kk}\big)\sigma_{1,ij} \\
& + \frac{1}{a^2}\Big[2\phi_1\sigma_{1,ij}' + \mathcal{H}\phi_1\sigma_{1,ij} + \phi_1'\sigma_{1,ij} - 2(\phi_1 - \psi_1)\phi_{1,ij} - \phi_{1,i}\phi_{1,j} + 2\psi_{1,(i}\phi_{,j)}\Big] \\
& - \frac{1}{a^2}\big(4\psi_1\psi_{1,ij} + 3\psi_{1,i}\psi_{1,j}\big) + \frac{1}{a^2}\sigma_{1,i}^{,k}\sigma_{1,jk} + 8\pi G(\rho + P)v_{,i}v_{,j},
\end{aligned}
\tag{6.5.35}
$$

where $v$ is the scalar velocity potential, $G$ is the Newton's gravitational constant and

$$
\begin{aligned}
s_{ij}^{(st)} = & \frac{1}{2a}\frac{\mathrm{d}}{\mathrm{d}\eta}\Big[\frac{1}{a}h_{1ij}'\phi_1 - \frac{2}{a}\Big(\psi_1 h_{1ij}' + \psi_1'h_{1ij} - h_{1i}^k\sigma_{1,jk}\Big) + \frac{\sigma_1^{,k}}{a}\Big(h_{1ik,j} + h_{1jk,i} - h_{1ij,k}\Big)\Big] \\
& + \frac{3}{2}\frac{\mathcal{H}}{a^2}\Big[h_{1ij}'\phi_1 - 2\Big(\psi_1 h_{1ij}' + \psi_1'h_{1ij} - h_{1i}^k\sigma_{1,jk}\Big) + \sigma_1^{,k}\Big(h_{1ik,j} + h_{1jk,i} - h_{1ij,k}\Big)\Big] \\
& + \phi_1\frac{1}{2a}\frac{\mathrm{d}}{\mathrm{d}\eta}\Big(\frac{1}{a}h_{1ij}'\Big) - \frac{1}{2a^2}\sigma_1^{,k}h_{1ij,k}' + \frac{1}{2a^2}h_{1ij}'\big(3\mathcal{H}\phi_1 + 3\psi_1' - \sigma_{1,kk}\big) + \frac{1}{2a^2}\sigma_{1,i}^{,k}h_{1jk}' \\
& - \frac{1}{2a^2}\sigma_{1,j}^{,k}h_{1ik}' - \frac{1}{2a^2}\Big[2h_{1i}^k\phi_{1,jk} + \Big(h_{1ik,j} + h_{1jk,i} - h_{1ij,k}\Big)\phi_1^{,k}\Big] \\
& + \frac{1}{2a^2}\Big[2\Big(2\psi_1 h_{1ij,kk} - h_{1j}^k\psi_{1,ik} + h_{1ij}\psi_{1,kk}\Big) - \frac{1}{2}\psi_1^{,k}\Big(h_{1ik,j} + h_{1jk,i} - 3h_{1ij,k}\Big)\Big].
\end{aligned}
\tag{6.5.36}
$$

The equation of motion Eq. (6.5.33), expanded at second-order by keeping all the perturbations of the metric, is gauge-invariant by construction. This can be also checked explicitly by employing the second-order gauge transformation reviewed in appendix E.1.2.

**Gauge-invariant emission equation from the Poisson gauge**

One can also manipulate the general equation of motion (6.5.33) to identify the gauge-invariant combinations defined in the previous section. In particular, focusing on the gauge-invariant tensor modes defined starting from the Poisson gauge, one finds

$$
h_{2ij}^{\mathrm{GI,P}\prime\prime} + 2\mathcal{H}h_{2ij}^{\mathrm{GI,P}\prime} - h_{2ij,kk}^{\mathrm{GI,P}} = -4\mathcal{T}_{ij}^{lm}\Big[4\Psi_1\Psi_{1,lm} + 2\Psi_{1,l}\Psi_{1,m} - \partial_l\Big(\frac{\Psi_1'}{\mathcal{H}} + \Psi_1\Big)\partial_m\Big(\frac{\Psi_1'}{\mathcal{H}} + \Psi_1\Big)\Big],
\tag{6.5.37}
$$

where we introduced the transverse and traceless projector $\mathcal{T}_{ij}^{lm}$, see Eqs. (E.5) and (6.1.5). In practice, this equation is formally identical to the one we used to compute the GWs produced by second-order scalar perturbations in Sec. 6.1. However, this time it is written in terms of individually gauge-invariant quantities.

Notice that, however, other gauge-invariant definitions of tensor modes built starting from different gauges are possible. This would give rise to different forms for the gauge-invariant equations of motion and, therefore, potentially different predictions for the induced gravitational waves. Thus, the issue of which is the observable combination can only be addressed by focusing on the measurement process, as we will see in the next section.

**Gauge-invariant propagation equation from the Poisson gauge**

Focusing on the GW propagation equation, a procedure analogous to the one we followed in the previous section leads to

$$h_{2ij}^{\mathrm{GI,P}\prime\prime} + 2\mathcal{H}h_{2ij}^{\mathrm{GI,P}\prime} - h_{2ij,kk}^{\mathrm{GI,P}} = 4\mathcal{T}_{ij}^{lm}\left(2\Psi_1 h_{1lm,kk}\right), \tag{6.5.38}$$

where we neglected terms with a smaller number of derivatives in the tensor modes. This equation describes the Shapiro time-delay effect in the geometrical optics approximation, see Sec. 6.3.

### 6.5.4   The gauge-invariant observable SGWB

In this section, we address the issue of what is the actual physical combination which is observable at GW detectors. As discussed previously, the TT frame is the only one allowing for a tractable and consistent description of the GW measurement. In particular, the sensitivity curve for the upcoming space-based detector LISA is necessarily provided in the TT frame [273]. In this section, we are going to perform the computation of the SGWB induced at second order in the TT frame, and show that the late-time observable SGWB coincide with the one computed in the Poisson (or Newtonian) frame. This can be explained on physical grounds by noticing that, as the emission of gravitational waves takes place deep in the radiation dominated universe, the scalar perturbations responsible for the emission are quickly damped by their transfer functions. Therefore, the GW effectively decouples from the source and becomes a linear perturbation of the metric and, as such, gauge-invariant.

**Linear solutions in the TT gauge**

We start by computing the linear evolution of scalar perturbations in the TT gauge by performing a gauge transformation from the known results in the Poisson gauge. Using the gauge transformation defined in Eq. (E.18), one finds

$$\phi_1^{\mathrm{TT}} = 0, \qquad \psi_1^{\mathrm{TT}} = \psi_1^{\mathrm{P}} - \mathcal{H}\alpha_1^{\mathrm{TT}}, \qquad B_1^{\mathrm{TT}} = 0, \qquad E_1^{\mathrm{TT}} = \beta_1^{\mathrm{TT}},$$
$$S_{1i}^{\mathrm{TT}} = 0, \qquad F_{1i}^{\mathrm{TT}} = F_{1i}^{\mathrm{P}},$$
$$h_{1ij}^{\mathrm{TT}} = h_{1ij}^{\mathrm{P}}, \tag{6.5.39}$$

where the gauge parameters used to impose the TT gauge are

$$\alpha_1^{\mathrm{TT}} = -\frac{1}{a}\left[\int a\phi_1 \mathrm{d}\eta - \mathcal{C}_1(\boldsymbol{x})\right], \tag{6.5.40}$$

$$\beta_1^{\mathrm{TT}} = \int \left(\alpha_1^{\mathrm{TT}} - B_1\right)\mathrm{d}\eta + \hat{\mathcal{C}}_1(\boldsymbol{x}), \tag{6.5.41}$$

$$\gamma_{1i}^{\mathrm{TT}} = \int S_{1i}\mathrm{d}\eta + \hat{\mathcal{C}}_{1i}(\boldsymbol{x}). \tag{6.5.42}$$

Then, the scalar potential and shear potential $\sigma_1$ can be written as

$$\psi_1^{\mathrm{TT}}(\boldsymbol{k},\eta) = \psi_1^{\mathrm{P}}(\boldsymbol{k},\eta) - \mathcal{H}\left[\frac{1}{a(\eta)}\mathcal{C}_1(\boldsymbol{k}) - \frac{1}{a(\eta)}\int^\eta a(\eta')\psi_1^{\mathrm{P}}(\boldsymbol{k},\eta')\mathrm{d}\eta'\right], \tag{6.5.43}$$

$$\sigma_1^{\mathrm{TT}}(\boldsymbol{k},\eta) = E_1^{\mathrm{TT}\prime}(\boldsymbol{k},\eta) = -\frac{1}{\mathcal{H}}\left[\psi_1^{\mathrm{TT}}(\boldsymbol{k},\eta) - \psi_1^{\mathrm{P}}(\boldsymbol{k},\eta)\right]. \tag{6.5.44}$$

As long as we focus on the radiation-dominated epoch, we can use $a \sim \eta$ and $\mathcal{H} = 1/\eta$, to find

$$\psi_1^{\mathrm{TT}}(\boldsymbol{k},\eta) = \frac{2}{3}\zeta(\boldsymbol{k})3\left[\frac{j_1(z)}{z} - \frac{j_0(z)}{z^2}\right] - \frac{\mathcal{C}(\boldsymbol{k})}{\eta^2}, \tag{6.5.45}$$

$$\sigma_1^{\mathrm{TT}}(\boldsymbol{k},\eta) = \frac{2}{3}\zeta(\boldsymbol{k})3\eta\frac{j_0(z)}{z^2} + \frac{\mathcal{C}(\boldsymbol{k})}{\eta}, \tag{6.5.46}$$

where $\zeta(\boldsymbol{k})$ is the primordial comoving curvature perturbation and $z = k\eta/\sqrt{3}$. The arbitrary constant $\mathcal{C}(k)$ can be set by requiring a finite value of the perturbations in the super-horizon limit $k\eta \to 0$ in accordance with [702]. This procedure selects $\mathcal{C}(\boldsymbol{k}) = -6\zeta(\boldsymbol{k})/k^2$ and, finally, one can write

$$\psi_1^{\mathrm{TT}}(\boldsymbol{k}, \eta) = \frac{2}{3}\zeta(\boldsymbol{k})3\left[\frac{j_1(z)}{z} - \frac{j_0(z)}{z^2} + \frac{1}{z^2}\right] \equiv \frac{2}{3}\zeta(\boldsymbol{k})T_\psi(\eta, k), \tag{6.5.47}$$

$$\sigma_1^{\mathrm{TT}}(\boldsymbol{k}, \eta) = \frac{2}{3}\zeta(\boldsymbol{k})3\eta\left[\frac{j_0(z)}{z^2} - \frac{1}{z^2}\right] \equiv \frac{2}{3}\zeta(\boldsymbol{k})\frac{\sqrt{3}}{k}T_\sigma(\eta, k). \tag{6.5.48}$$

Those relations correspond to the solution of the scalar linear equation of motion in the absence of anisotropic stress.

**GWs emission in the TT gauge**

The source of GWs at second order can be read in Eq. (6.5.35), by specialising it to the TT gauge as

$$s_{ij,\mathrm{TT}}^{(ss)} = -\frac{1}{a^2(\eta)}\left[\psi_1\psi_{1,ij} + 2\psi_{1,(i}'\sigma_{1,j)} - \psi_1'\sigma_{1,ij} + \sigma_{1,ij}\sigma_{1,kk} - \sigma_{1,ik}\sigma_{1,jk} + \frac{2}{\mathcal{H}' - \mathcal{H}^2}\psi_{1,i}'\psi_{1,j}'\right], \tag{6.5.49}$$

which, in a radiation-dominated universe, give rise the the equation of motion for the gauge-invariant combinations defined starting from the TT gauge as

$$h_\lambda^{\mathrm{TT}\prime\prime}(\eta, \boldsymbol{k}) + 2\mathcal{H}h_\lambda^{\mathrm{TT}\prime}(\eta, \boldsymbol{k}) + k^2 h_\lambda^{\mathrm{TT}}(\eta, \boldsymbol{k}) =$$
$$-4e_\lambda^{ij}(\boldsymbol{k})\left[\psi_1^{\mathrm{TT}}\psi_{1,ij}^{\mathrm{TT}} + 2\psi_{1,(i}^{\mathrm{TT}\prime}\sigma_{1,j)}^{\mathrm{TT}} - \psi_1^{\mathrm{TT}\prime}\sigma_{1,ij}^{\mathrm{TT}} + \sigma_{1,ij}^{\mathrm{TT}}\sigma_{1,mm}^{\mathrm{TT}} - \sigma_{1,im}^{\mathrm{TT}}\sigma_{1,jm}^{\mathrm{TT}} - \frac{1}{\mathcal{H}^2}\psi_{1,i}^{\mathrm{TT}\prime}\psi_{1,j}^{\mathrm{TT}\prime}\right]. \tag{6.5.50}$$

We can solve this equation to find the SGWB by following the same steps adopted in Sec. 6.1. In practice, using the Green's function method, one arrives at the following form for the tensor modes

$$h_\lambda^{\mathrm{TT}}(\eta, \boldsymbol{k}) = \frac{4}{9}\int\frac{\mathrm{d}^3p}{(2\pi)^3}\frac{1}{k^3p}e_\lambda(\boldsymbol{k}, \boldsymbol{p})\zeta(\boldsymbol{k})\zeta(\boldsymbol{k} - \boldsymbol{p})\left[\mathcal{I}_c^{\mathrm{TT}}(x, y)\cos(k\eta) + \mathcal{I}_s^{\mathrm{TT}}(x, y)\sin(k\eta)\right], \tag{6.5.51}$$

where, in this case, the time integrated transfer functions are defined as

$$\mathcal{I}_c^{\mathrm{TT}}(x, y) = \int_0^\infty \mathrm{d}u\, u(-\sin u)f^{\mathrm{TT}}(x, y, u),$$
$$\mathcal{I}_s^{\mathrm{TT}}(x, y) = \int_0^\infty \mathrm{d}u\, u(\cos u)f^{\mathrm{TT}}(x, y, u), \tag{6.5.52}$$

as a function of

$$f^{\mathrm{TT}}(x, y, u) = \frac{54}{u^5 x^3 y^2}\left\{3\left[ux\left(u^2\left(x^2 - y^2 - 1\right) - 2\right) + \sqrt{3}\left(u^2\left(-x^2 + y^2 + 1\right) + 4\right)\sin\left(\frac{ux}{\sqrt{3}}\right)\right.\right.$$
$$\left.-2ux\cos\left(\frac{ux}{\sqrt{3}}\right)\right] + \frac{1}{y}\sin\left(\frac{uy}{\sqrt{3}}\right)\left[\sqrt{3}x\left(u^2\left(-3x^2 + y^2 + 3\right) + 24\right)\right.$$
$$\left.+u\left(y^2\left(2u^2x^2 - 15\right) - 3\left(x^2 + 3\right)\right)\sin\left(\frac{ux}{\sqrt{3}}\right) + 4\sqrt{3}x\left(u^2y^2 - 6\right)\cos\left(\frac{ux}{\sqrt{3}}\right)\right]$$
$$\left.+2\cos\left(\frac{uy}{\sqrt{3}}\right)\left[2\sqrt{3}\left(u^2x^2 - 3\right)\sin\left(\frac{ux}{\sqrt{3}}\right) - 9ux + 15ux\cos\left(\frac{ux}{\sqrt{3}}\right)\right]\right\}. \tag{6.5.53}$$

A plot of the functions 6.5.52 can be found in Ref. [16]. The resulting energy density power spectrum, analogously to Eq. (6.5.54), is found by computing the integral

$$\Omega_{\mathrm{GW}}^{\mathrm{TT}}(\eta_0, k) = c_g\frac{\Omega_{r,0}}{972}\iint_{\mathcal{S}}\mathrm{d}x\mathrm{d}y\,\frac{x^2}{y^2}\left[1 - \frac{\left(1 + x^2 - y^2\right)^2}{4x^2}\right]^2\mathcal{P}_\zeta(k\,x)\,\mathcal{P}_\zeta(k\,y)$$
$$\times\left[\mathcal{I}_c^{\mathrm{TT}2}(x, y) + \mathcal{I}_s^{\mathrm{TT}2}(x, y)\right]. \tag{6.5.54}$$

In the late time universe, after all the relevant GWs emission has taken place, one can see that the SGWB found starting from the tensor combination defined in the TT gauge, agrees with the result in the Poisson gauge reported in the previous sections. This is a consequence of the fact that, as mentioned in the introduction, once the GWs are generated and propagate freely inside our horizon, they can be treated as linear objects and all gauges must provide the same result.

This point is made manifest by the fact that, in the late time limit, the combination $\mathcal{I}_c^2 + \mathcal{I}_s^2$ is the same in all gauges where the scalar perturbations are suppressed in the subhorizon limit and the tensor modes decouple from them. This result also agrees with Refs. [703–705]. One can see this explicitly by comparing the results in the two gauges [703]

$$[\mathcal{I}^{\mathrm{p}}(k\eta) - \mathcal{I}^{\mathrm{TT}}(k\eta)] \propto \frac{1}{k\eta}, \tag{6.5.55}$$

which vanishes in the subhorizon limit. This is phenomenon is analogous to the way energy density perturbations lose their gauge dependency in the subhorizon limit [702].

**Propagation equation in the TT gauge**

The scalar-tensor source in the TT gauge is found to be

$$s_{ij,\mathrm{TT}}^{(st)} = -\frac{1}{a^2(\eta)} \Big[ -\psi_1 h_{1ij,kk} + 2h_{1ij}\mathcal{H}\psi_1' + \frac{1}{2} h_{1ij}'\psi_1' + 2h_{1\ i}^{\ k}\psi_{1,j)k} + h_{1ij}\psi_1'' - h_{1ij}\psi_{1,kk}$$
$$+ \frac{1}{2} h_{1ij}'\sigma_{1,kk} + \sigma_{1,k} \left( h_{1ij,k}' - h_{1k(i,j)}' \right) + 2\psi_{1,k} \left( h_{1k(i,j)} - h_{1ij,k} \right) - h_{1k(i}'\sigma_{1,j)k} \Big]. \tag{6.5.56}$$

We want to consider the geometrical optics approximation, as done in Sec. 6.3. This amounts to take the leading order in the derivatives of the tensor field as

$$h_\lambda^{\mathrm{TT}''}(\eta,\boldsymbol{k}) + 2\mathcal{H}h_\lambda^{\mathrm{TT}'}(\eta,\boldsymbol{k}) + k^2 h_\lambda^{\mathrm{TT}}(\eta,\boldsymbol{k}) = 4e_\lambda^{ij}(\boldsymbol{k}) \Big[ \psi_1^{\mathrm{TT}} h_{1ij,kk}^{\mathrm{TT}} - \sigma_{1,k}^{\mathrm{TT}} \left( h_{1ij,k}^{\mathrm{TT}\prime} - h_{1k(i,j)}^{\mathrm{TT}\prime} \right) \Big]. \tag{6.5.57}$$

As argued above, this is the equation that describes the propagation of GWs which are later on observable by ground-based and space-based observatories. The Shapiro time delay gives rise to a phase accumulated during the propagation in a perturbed universe which does not affect the power spectrum [21]. Therefore, even in a perturbed universe, the GW abundance is independent of the gauge.

## 6.6   Observability at LISA

Having discussed in details the properties of the GW signal induced at second order in models producing a sizeable population of PBHs, let us now turn our attention to GW experiments that would be able to detect it. Let us start by focusing on the future space-based detector LISA [37, 277].

### 6.6.1   PBH models observable at LISA

As we introduced at the beginning of this chapter, the fundamental property in this scenario is the relation between the PBH mass and the characteristic GW frequency (6.0.1). The LISA experiment will be able to measure GW with frequencies in the range

$$f \simeq (10^{-5} \div 10^{-1})\mathrm{Hz}, \tag{6.6.1}$$

with a peak sensitivity at $f_{\mathrm{LISA}} \simeq 3.4\mathrm{mHz}$. Those frequencies are able to probe the primordial curvature perturbation power spectrum on scales in the range $(10^{10} - 10^{15})\,\mathrm{Mpc}^{-1}$. Also, those frequencies are associated with the formation of PBHs with masses falling in the range

$$M_{\mathrm{PBH}} \sim \mathcal{O}\left(10^{-15} - 10^{-8}\right) M_\odot. \tag{6.6.2}$$

In Fig. 6.10, we show the SGWB spectrum associated with various illustrative PBH populations produced by a narrow curvature power spectrum peaking in the relevant range of masses. We parametrise $\mathcal{P}_\zeta$ with a lognormal function ($\sigma = 0.5$) and an amplitude fixed by requiring an overall PBH abundance ranging between $f_{\rm PBH} = (10^{-6}, 1)$. Also, for comparison, we show what the proposed design (4 yrs, 2.5 Gm of length, 6 links) is expected to provide in terms of $\Omega_{\rm GW}$ sensitivity. The line we show is taken as the mean of "C1" and "C2" in Ref. [706]. In the bottom panel of Fig. 6.10, we show the corresponding PBH abundance of the same PBH populations compared to the existing constraints available to date.

We highlight a few noteworthy features of this result:

- When PBHs constitute the totality of the dark matter $f_{\rm PBH} = 1$, the corresponding GW signal is well within the observability of LISA. Even though a careful determination of possible background/foreground signals by other sources was not carried out, the amplitude of such a spectrum is so large that it may easily dominate the spectrum of GWs reaching the experiment;

- As the PBH abundance scales exponentially with the amplitude of the curvature spectrum, while the GW abundance only quadratically, a change of orders of magnitude in the abundance only corresponds to a factor of order few in the SGWB energy density. This can be seen explicitly by comparing the coloured bands in both panels of Fig. 6.10;

- As one can appreciate, the LISA experiment will be able to probe the entirety of the window which is still allowing PBH to be 100% of the dark matter;

- Again due to the exponential sensitivity of the PBH abundance to the amplitude of the curvature power spectrum, a null detection of a SGWB at LISA would imply the PBH abundance in the corresponding mass range to be completely negligible. This conclusion may be reconsidered if one assumed PBHs come from highly non-Gaussian curvature perturbations [667] or non standard cosmologies like an early matter-domination era [111], even though a sizeable SGWB could be produced also in this case, see for example Refs. [707, 708].

To conclude, we have shown that if PBHs with masses in range $M_{\rm PBH} \sim \mathcal{O}\left(10^{-15} - 10^{-8}\right) M_\odot$ form the totality (or a fraction of) dark matter, LISA will be able to measure the GWs sourced during the PBH formation. The signal is predicted to have a scaling going like $\propto k^3$ in the tail at low frequencies, while is model dependent in the high-frequency tail. Also, the SGWB is expected to be gaussian with extremely high precision due to propagation effects. We will discuss the potential anisotropies of such a signal in the next section.

### 6.6.2  SGWB anisotropies at LISA

To have a more physical intuition of the amount of anisotropy in the GWs abundance, we express the results found in Sec. 6.4 in terms of the GW density contrast $\delta_{\rm GW}$. Following Ref. [17], we define the two and three point functions as

$$\langle \delta_{{\rm GW},\ell m} \delta^*_{{\rm GW},\ell' m'} \rangle = \delta_{\ell\ell'} \delta_{mm'} \hat{C}_\ell(k),$$
$$\langle \delta_{{\rm GW},\ell_1 m_1} \delta_{{\rm GW},\ell_2 m_2} \delta_{{\rm GW},\ell_3 m_3} \rangle = \mathcal{G}^{m_1 m_2 m_3}_{\ell_1 \ell_2 \ell_3} \hat{b}_{\ell_1 \ell_2 \ell_3}(k), \tag{6.6.3}$$

where we have again factorised the tensorial structures dictated by statistical isotropy, and the multipoles are

$$\sqrt{\frac{\ell(\ell+1)}{2\pi} \hat{C}_\ell(k)} \simeq \frac{3}{5} \left| 1 + \tilde{f}_{\rm NL}(k) \right| \left| 4 - \frac{\partial \ln \bar{\Omega}_{\rm GW}(\eta, k)}{\partial \ln k} \right| \mathcal{P}^{1/2}_{\zeta_L}, \tag{6.6.4}$$

$$\hat{b}_{\ell_1 \ell_2 \ell_3}(k) \simeq \frac{\tilde{f}_{\rm NL}\left[1 + 3\tilde{f}_{\rm NL}(k)\right]}{4\left[1 + \tilde{f}_{\rm NL}(k)\right]^2} \left( \hat{C}_{\ell_1} \hat{C}_{\ell_2} + \hat{C}_{\ell_1} \hat{C}_{\ell_3} + \hat{C}_{\ell_2} \hat{C}_{\ell_3} \right). \tag{6.6.5}$$

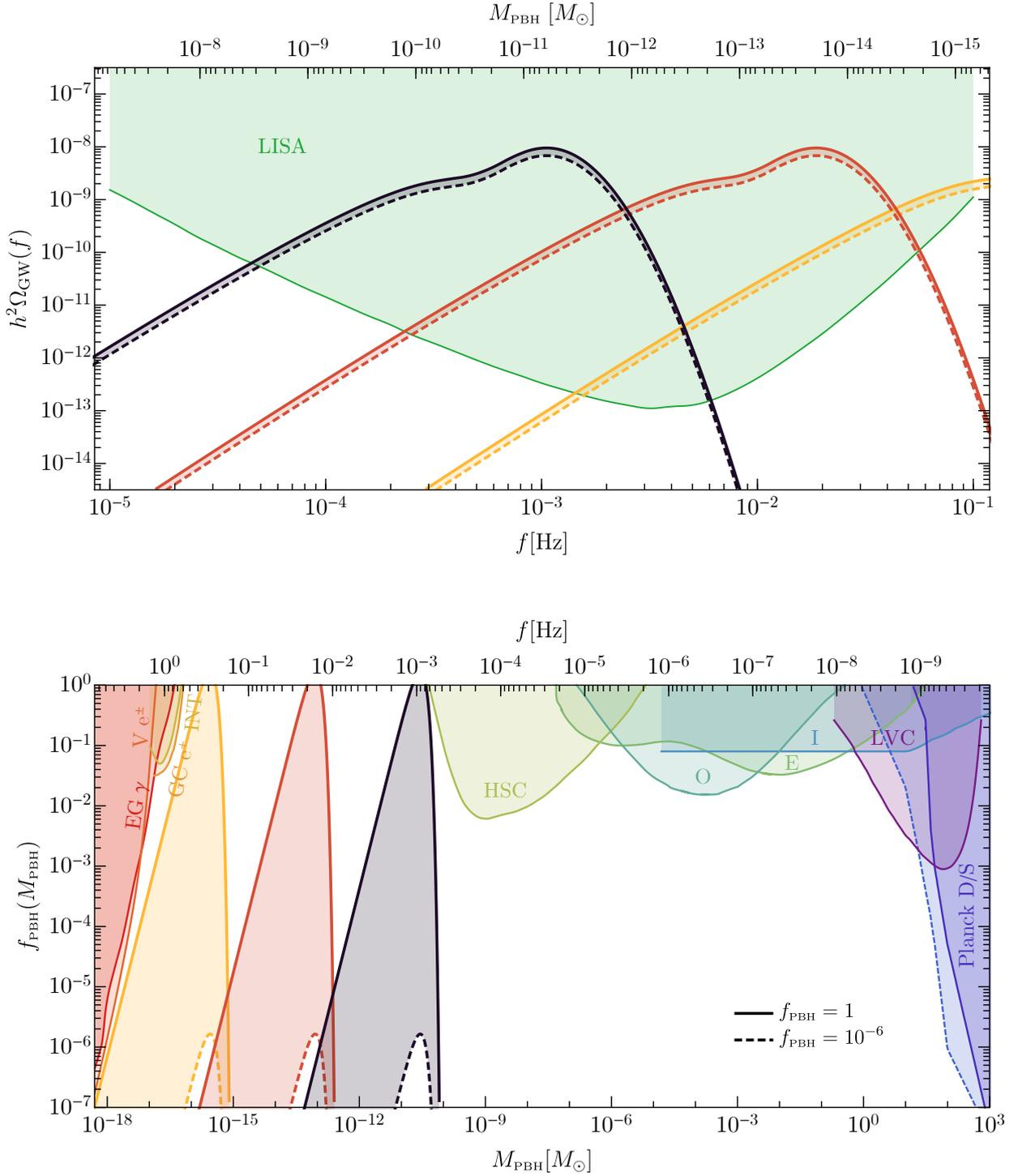

Figure 6.10: **Top:** *Comparison of the estimated sensitivity for LISA [277] with the GW abundance generated at second-order by the formation mechanism of PBHs for a lognormal power spectrum (6.1.29) with $\sigma = 0.5$. The spectra are centred at various scales, corresponding to the mass function shown below. The properties of the SGWB are described in Sec. 6.1.1.* **Bottom:** *PBH mass function produced by the same curvature power spectrum in Eq. (6.1.29). As we consider a narrow spectrum, for illustrative purposes, we approximated the mass function to the critical one, see Sec. 3.1. Each constraint is discussed in Sec. 1.2.2. In both plots, the solid (dashed) line corresponds to $f_{\rm PBH} = 1$ ($f_{\rm PBH} = 10^{-6}$).*

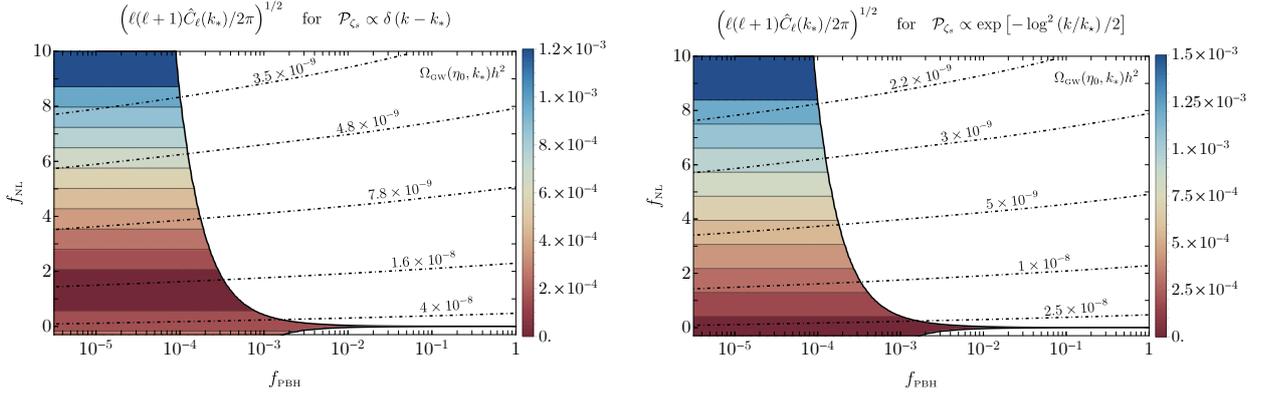

Figure 6.11: *Contour plot of* $\left(\ell(\ell+1)\hat{C}_\ell(k_*)/2\pi\right)^{1/2}$ *in the region permitted by the constraints of Planck on the combination* $f_{\rm PBH}$ *and* $f_{\rm NL}$ *for the choice of a Dirac delta (left) and lognormal (right) power spectrum of the short modes. The peak frequency* $k_*$ *has been chosen to correspond to a PBH population peaked at* $M_{\rm PBH} = 10^{-12} M_\odot$. *The dot-dashed lines identify the corresponding GWs abundance. Notice that the results shown here only hold for a local, and strictly scale-invariant, primordial non-Gaussianity of scalar perturbations.*

We recall that these results account for both the anisotropies imprinted at emission and due to propagation effects.

One can identify two characteristic regimes: the first when anisotropies are dominated by the propagation term while the second when they are controlled by the initial conditions. In the former case, which corresponds to the limit $\tilde{f}_{\rm NL} \to 0$, one finds

$$\text{Dominated by propagation}: \begin{cases} \sqrt{\frac{\ell(\ell+1)}{2\pi}\,\hat{C}_\ell\,(k)} \simeq \frac{3}{5}\,\left|4 - \frac{\partial \ln \bar{\Omega}_{\rm GW}(\eta, k)}{\partial \ln k}\right|\,\mathcal{P}^{1/2}_{\zeta_L}, \\[2ex] \hat{b}_{\ell_1\ell_2\ell_3}\,(k) \simeq \frac{1}{4}\tilde{f}_{\rm NL}\,\left[\hat{C}_{\ell_1}\,\hat{C}_{\ell_2} + \hat{C}_{\ell_1}\,\hat{C}_{\ell_3} + \hat{C}_{\ell_2}\,\hat{C}_{\ell_3}\right], \end{cases} \tag{6.6.6}$$

which agrees with Ref. [686]. The absence of anisotropies at formation when $\tilde{f}_{\rm NL} \to 0$ is interpreted as the consequence of the Equivalence Principle, for which different Hubble patches cannot be correlated on scales $R \gg 1/H$ in the absence of non-Gaussianities, see related discussion in Sec. 6.4. In the latter case, when the initial condition term dominates, one can instead consider the formal limit of large $\tilde{f}_{\rm NL}$. The correlators for $\delta_{\rm GW}$ then become

$$\text{Dominated by initial condition}: \begin{cases} \sqrt{\frac{\ell(\ell+1)}{2\pi}\,\hat{C}_\ell\,(k)} \simeq \frac{24}{5}\,|f_{\rm NL}|\,\mathcal{P}^{1/2}_{\zeta_L}, \\[2ex] \hat{b}_{\ell_1\ell_2\ell_3}\,(k) \simeq \frac{3}{4}\,\left(\hat{C}_{\ell_1}\,\hat{C}_{\ell_2} + \hat{C}_{\ell_1}\,\hat{C}_{\ell_3} + \hat{C}_{\ell_2}\,\hat{C}_{\ell_3}\right), \end{cases} \tag{6.6.7}$$

where we note that $f_{\rm NL}$ has disappeared from the last expression, since $\Gamma_I$ is maximally non-Gaussian (as opposite to $\Gamma_S$, that is Gaussian up to $\mathcal{O}\,(f_{\rm NL})$ non-Gaussianity).

We stress again here that observable anisotropies are imprinted at formation only when a cross-talk between the PBH scales and the large scales ($\ell \lesssim 15$) is present, as we parametrised in terms of the non-linear parameter $f_{\rm NL}$. This follows from the fact that the LISA experiment will only be able to measure anisotropies at very large scales [687]. If all, or a significant part, of the dark matter is composed of PBHs, then this non-Gaussianity is responsible for the production of isocurvature modes in the dark matter density fluid, which are strongly constrained by the CMB observations, as we discussed in Sec. 3.3.2. In practice, large $f_{\rm NL} \gtrsim 1$ can only be compatible with an $f_{\rm PBH} \lesssim 10^{-3}$.

In Fig. 6.11, we show the prediction for the multipoles $\hat{C}_\ell$ of the density contrast two-point function, for the choice of a Dirac delta and lognormal power spectrum of curvature perturbations on small scales. The peak frequency of this signal was chosen as the one corresponding to PBH masses given by $M_{\rm PBH} = 10^{-12} M_\odot$ for which PBHs can represent all the DM, also coinciding with the frequency of maximum sensitivity at LISA ($f_{\rm LISA} \simeq 3.4\,{\rm mHz}$). The region in white corresponds to the parameter space excluded by CMB constraint on isocurvature perturbations. Also, the dot-dashed lines identify the corresponding GWs abundance computed at present time and at the peak frequency. Finally, the results for similar PBH masses in the entire window observable by LISA do not change significantly. In summary, we conclude that the typical anisotropies expected for a SGWB produced by this mechanism are of the order of $\zeta_L \sim 10^{-4}$ and correspondingly, the reduced bispectrum is of the order of $\zeta_L^2 \sim 10^{-8}$. Also, if the SGWB possesses a large amount of anisotropies, only possible with a relatively large $f_{\rm NL}$, PBHs cannot be the dominant component of the dark matter in this mass range.

## 6.7 Pulsar timing array signals

In this section, we discuss the potential of Pulsar Timing Arrays (PTA) to discover and/or constrain the PBH scenario by searching for the corresponding SGWB. [7]

Pulsars are rapidly rotating neutron stars that continuously emit beams of electromagnetic waves in the radio spectrum. Those can be used to probe low-frequency GWs, see for example Ref. [714]. In particular, millisecond pulsars, rotating with a period of $\mathcal{O}({\rm ms})$, are characterised by a particularly stable orbital period. One can measure accurately the time of arrival of each pulse to construct a pulsar timing model and use the observations of arrival times to search for GWs.

The passage of a GW in between the pulsar and the observer on the earth can cause a variation of the time of arrival of a pulse [715, 716], similarly to what happens at ground-based interferometers. This induces residuals in the pulse arrival time $t$ (with respect to the reference $t = 0$) when compared to the pulsar timing model as

$$R(t) = -\int_0^t \frac{\delta\nu}{\nu} dt, \tag{6.7.1}$$

where $\nu$ is the pulse frequency, characteristic of each observed pulsar. In terms of the tensor perturbation $h_{ij}$ at the detector location $\vec{x}_{\rm d}$ and pulsar location $\vec{x}_{\rm p}$, the pulse frequency variation is given by [715]

$$\frac{\delta\nu}{\nu} = -H^{ij}\left[h_{ij}(t, \vec{x}_{\rm d}) - h_{ij}(t - D, \vec{x}_{\rm p})\right], \tag{6.7.2}$$

where $H^{ij}$ is a geometrical factor depending on the propagation direction of the GWs relative to the direction of the pulsar at a distance $D$.

It was shown by Hellings and Downs [717] that timing residuals induced by GW signals for pulsars separated by an angular distance $\theta$ follows the correlation pattern

$$C(\theta) = \frac{1 - \cos\theta}{2} \ln\left[\frac{1 - \cos\theta}{2}\right] - \frac{1 - \cos\theta}{12} + \frac{1}{3}. \tag{6.7.3}$$

This is known as the Hellings and Downs (HD) curve and it is shown in Fig. 6.12. By searching for such a signal at PTA experiments like Parkes Pulsar Timing Array (PPTA) [280], the North American Nanohertz Observatory for Gravitational Waves (NANOGrav) [281], and the European Pulsar Timing Array (EPTA) [282], one can constrain the abundance of the SGWB at frequencies close to nHz and, indirectly, the abundance of PBHs with masses around $M \sim 0.1 M_\odot$.

Up to now, no decisive detection of a SGWB has been reported (with the noticeable exception of NANOGrav 12.5 yr, which will be addressed in the following section). Therefore, using the limit on the SGWB one can derive a limit on $f_{\rm PBH}$. At the moment, they are still comparable with other constraints coming from lensing measurements, see discussion in Sec. 1.2.2.

---

[7] We will not discuss constraints that can be set with PTA experiments by measuring the effect of the Shapiro time-delay caused by PBHs and compact objects intervening between pulsars and the Earth. This was recently investigated in details in Refs. [709–713], where it was found that future Square Kilometer Array (SKA) experiment [279] will be able to constrain $f_{\rm PBH}$ and the abundance of compact subhalos to sub per cent values in the range of masses $M \simeq (10^{-11} \div 10^2) M_\odot$.

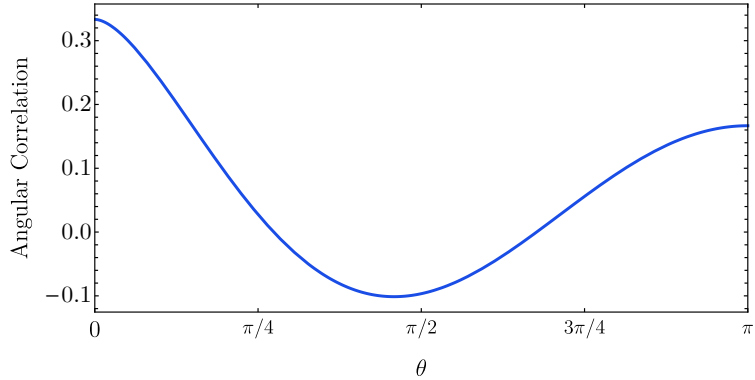

Figure 6.12: *Characteristic angular correlation between time residuals of pulsars separated by an angle θ induced by a GW signal.*

### 6.7.1   A potential PBH as DM explanation of the NANOGrav 12.5 yr data

The NANOGrav Collaboration has recently published the analysis of 12.5 yrs of pulsar timing data [231] reporting strong evidence for a stochastic common-spectrum process, see also Ref. [718]. This signal may be compatible with a SGWB with a characteristic strain amplitude $\sim 10^{-15}$ at a frequency $f \sim 3 \cdot 10^{-8}$ Hz with an almost flat spectrum, $\Omega_{\mathrm{GW}}(f) \sim f^{(-1.5 \div 0.5)}$ at $1\sigma$-level.

In particular, their analysis was performed on data coming from the observation of forty-five pulsars and showed the presence of a stochastic common process strongly preferred against an independent red-noise signal. It is important to note however that the NANOGrav Collaboration does not claim a firm detection of GWs since they have not yet found sufficient evidence for the characteristic HD quadrupole correlation, see discussion above. It is exciting that, more recently, the PPTA collaboration has confirmed the hint for a common-spectrum process in their data [719].

Following closely Ref. [8], we are going to show that the NANOGrav signal, if interpreted as a GW background, can be naturally explained by a flat spectrum of GWs generated at second-order during the formation of PBHs. [8] This scenario can explain the supposed SGWB signal and is characterised by the following interesting properties:

- PBHs formed in this scenario are the dark matter in our universe;

- The dominant contribution to the PBH mass function falls in the range $(10^{-15} \div 10^{-11}) M_{\odot}$ where no constraints on the PBH abundance are present;

- the SGWB spectrum reaches higher frequencies testable by future experiments, such as LISA [277].

Let us present this scenario in more details. We consider a class of models with a broad and flat power spectrum of the curvature perturbation of the form [15]

$$\mathcal{P}_{\zeta}(k) \approx A \, \Theta(k_s - k)\Theta(k - k_l), \qquad k_s \gg k_l \qquad (6.7.4)$$

where $\Theta$ is the Heaviside step function and $A$ is the amplitude of the power spectrum. This shape can naturally arise for modes exiting the Hubble radius during a non-attractor phase, obtained through an ultra slow-roll regime of the inflaton potential, as a result of a duality transformation which maps the non-attractor phase into a slow-roll phase [23, 739–741].

The power spectrum in Eq. (6.7.4) is responsible for both the generation of a population of PBHs with a broad mass function, see details in Sec. 3.1, and the emission of a SGWB with a flat

---

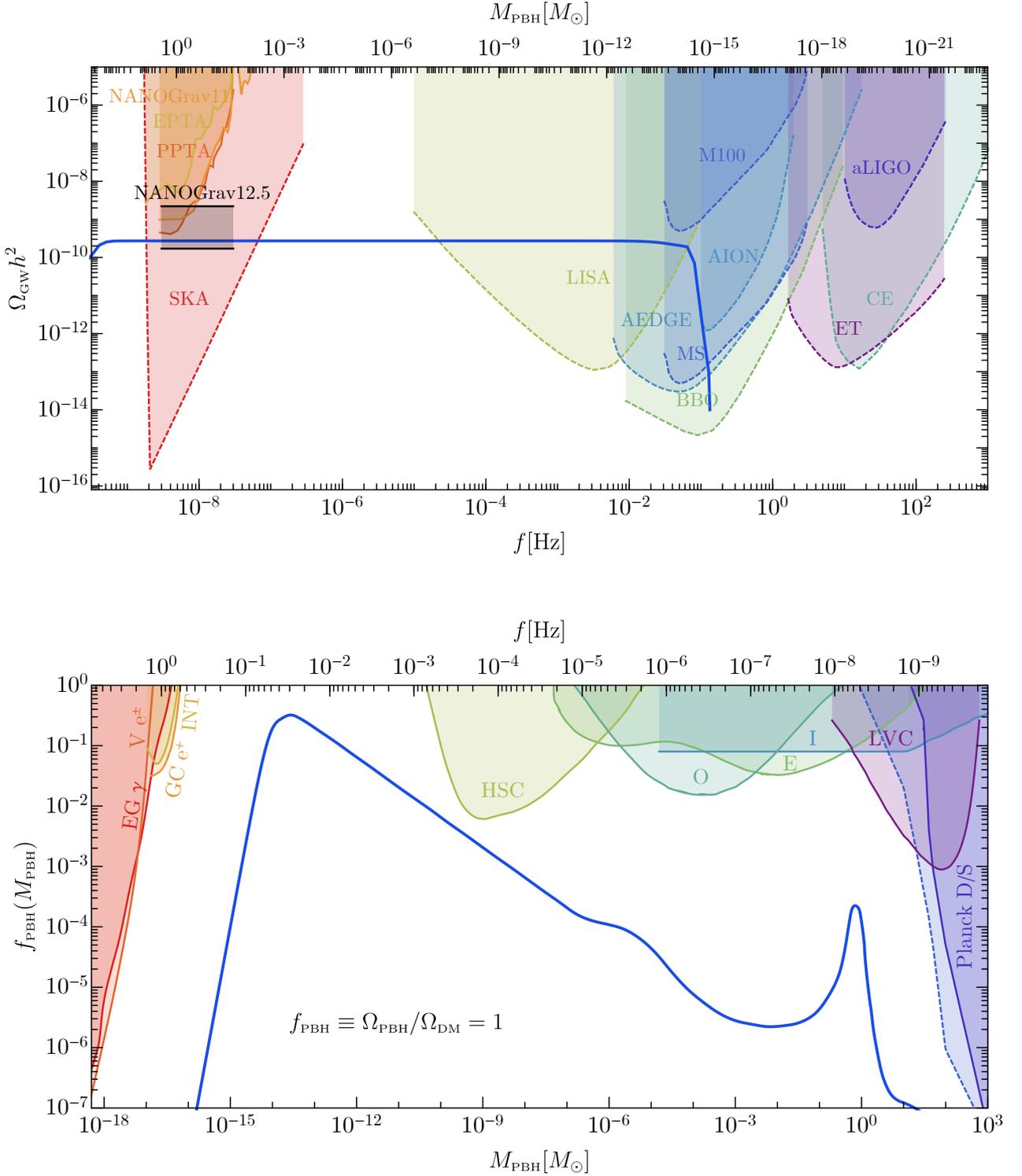

Figure 6.13: **Top:** *The abundance of GWs predicted in the scenario proposed in this section. The black band shows the 95% C.I. from the NANOGrav 12.5 yrs experiment. For more details about the projected sensitivities see the main text.* **Bottom:** *Mass function resulting from a flat power spectrum such that it peaks around $M_{PBH} \simeq 10^{-14} M_\odot$, with $A \simeq 5.8 \cdot 10^{-3}$ and $k_s = 10^9 k_l \simeq 1.6$ Hz, and PBHs comprise the totality of DM, i.e. $f_{PBH} = 1$. The bump in the tail of the population, around $M_{PBH} \sim M_\odot$, is due to the decrease of the threshold during the QCD epoch [133, 359].*

spectrum. We have collected both results in Fig. 6.13. In the bottom panel, we plot the mass function corresponding to the primordial curvature perturbation given in Eq. (6.7.4). As expected for a broad and flat spectrum of curvature perturbations, the peak of the mass function corresponds to the Hubble mass $M_H$ when the shortest scale $\sim 1/k_s$ re-enters the horizon. The requirement that the peak should fall in the range where PBH could be 100% of the dark matter select a range for smallest scale $k_s^{-1}$ in the model. At smaller masses, the mass function goes as $\sim (M_{\mathrm{PBH}})^{3.8}$ due to the dynamics of the critical collapse, while at larger masses falls as $\sim (M_{\mathrm{PBH}})^{-1/2}$ and has a sub-dominant peak around $\sim M_\odot$ due to the change of equation of state during the QCD phase transition [133, 359]. Given the absence of constraints [256, 257] in the mass range where most of the contribution to the PBH mass function comes from, the overall PBH abundance can be set to unity. This requirement dictates the choice of the power spectrum amplitude $A$. We find $A \simeq 5.8 \cdot 10^{-3}$ such that

$$f_{\mathrm{PBH}} = \int f_{\mathrm{PBH}}(M_{\mathrm{PBH}}) \mathrm{d} \ln M_{\mathrm{PBH}} = 1. \tag{6.7.5}$$

As a consequence, as one can appreciate by looking at the top panel of Fig. 6.13, the corresponding expected SGWB has an amplitude that is consistent with the 95% C.I. from the NANOGrav 12.5 yrs observation. Also, on the top panel of Fig. 6.13, we show the constraints coming from experiment EPTA [646], PPTA [644], NANOGrav 11 yrs [645, 742] and future sensitivity curves for planned experiments like SKA [279], LISA [277] (power-law integrated sensitivity curve expected to fall in between the designs "C1" and "C2" in Ref. [706]), DECIGO/BBO [743], CE [744], Einstein Telescope [648, 745], Advanced Ligo + Virgo collaboration [746], Magis-space and Magis-100 [747], AEDGE [748] and AION [749]. Notice that a portion of the 95% C.I. of NANOGrav 12.5 yrs is in tension with previous constraints from the NANOGrav 11 yrs and PPTA. However, according to the NANOGrav Collaboration [231], the improved priors for the intrinsic pulsar red noise used in the novel analysis relaxes the NANOGrav 11 yrs bound. Nevertheless, the predicted signal within our scenario falls below all bounds and in the 95% C.I. of the 12.5 yrs signal.

The SGWB spectrum is flat and extends up to larger frequencies, entering the LISA detectable region before decaying rapidly above the frequency corresponding to the shortest scale $1/k_s$. Another prediction of this scenario is that the second-order GWs seen by NANOGrav should also be detected by the forthcoming experiment LISA, and eventually by AEDGE, BBO and MS.

Finally, we stress that this scenario is also compatible with the candidate microlensing event observed by the Hyper-Supreme Camera (HSC) collaboration and discussed in details in Ref. [130] and can potentially be confirmed by future HSC observations as forecasted in Ref. [750].

To summarise, we showed there exists a scenario in which the NANOGrav observation can be explained as a SGWB generated at second order from the scalar perturbations responsible for the PBH production. By only requiring a flat spectrum of perturbations, and fixing the amplitude by requiring PBHs are the dark matter in our universe, the corresponding amplitude of $\Omega_{\mathrm{GW}}$ is compatible with NANOGrav results. This scenario can be tested with future experiments like LISA and HSC.

# Part IV

# Conclusions

# Chapter 7

# Conclusions and outlook

We are experiencing possibly the most exciting time to study PBHs. Following the widespread interest in the topic, just in the last few years tremendous steps forward were done in both the theory behind the PBH formation and the comparison with a variety of probes, with particular attention to the rising field of GW astronomy.

In this thesis, we addressed various aspects related to the physics of PBHs. Starting from the formation of PBHs in Chapter 2, we analysed the dependence of the PBH threshold on the statistical properties of the overdensity perturbations, while also extending the computation of the PBH abundance beyond the Gaussian paradigm by including both intrinsic and non-linearly induced non-Gaussianities. We then described, in Chapter 3, the fundamental properties of a PBH population at formation epoch by discussing the mass distribution, the spin of PBHs at birth and the spatial clustering of PBHs. Consequently, the evolution of the PBH population was addressed in Chapter 4, both focusing on isolated PBHs and PBHs in binaries. We identified and characterised accretion as the key player in modifying the PBH masses and spins, while clustering evolution was described through an analytical treatment that matches existing N-body simulations limited to high redshift. We finally discussed the formation and evolution of PBH binaries leading to the current state-of-the-art parametrisation of the PBH merger rate. The main source of uncertainty still affecting its computation comes from PBH clustering. Our results were inevitably based on extrapolations of the analytical descriptions of clustering at lower redshift compared to what was achieved so far with N-body simulations. As clustering can impact both the constraints on the PBH abundance and the merger rate, it deserves further intense investigations. Most of the work in the next years should be focused on refining the prediction of the PBH model to allow for a more solid comparison with the wealth of data that will soon become available (from GW experiments in particular).

In the third part of this thesis, we dedicated our focus to the comparison of the PBH scenario with GW observations. In Chapter 5, we focused on the latest catalog of GW events released by the LVC collaboration. First, we showed that PBHs can have a merger rate sufficient to produce a number of detectable signals compatible with observations without violating any bound on the PBH abundance. Then, we compared and mixed the PBH scenario with astrophysical channels to seize the potential PBH contribution to the current GW data. Sec. 5.2.4 can be considered as the culmination of the path followed in this thesis, started from sharpening the PBH predictions and arrived at the comparison with the most recent GW data. Overall, our work highlighted the following key take-home messages:

- Given current knowledge of both the astrophysical and primordial scenarios, PBHs are compatible with being a subpopulation of current GW data. In particular, they offer a competitive explanation of events in the mass gap such as GW190521. Our analysis not only proved that PBHs can play a major role in explaining some events in the catalog, but also that current knowledge of astrophysical models may still present shortcomings when trying to explain all the features of the GWTC-2 catalog.

- To robustly confirm the presence of PBHs in current and future data, a more extensive treatment of both astrophysical and primordial model uncertainties must be taken into account. This includes both extending the astrophysical models as well as going beyond the lognormal

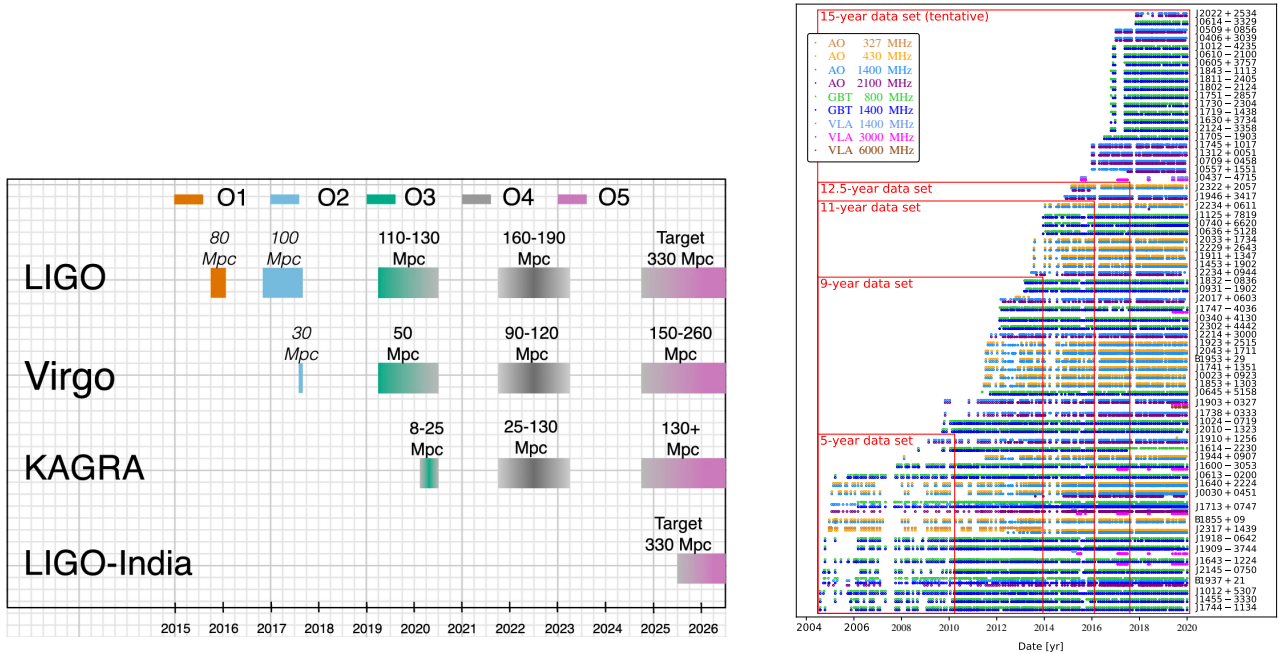

Figure 7.1: **Left:** *Timeline of the various observation runs of the LVCK collaboration. As one can see, within the next couple of years the fourth observation run will be completed, with an expected number of detections to fall in the $\mathcal{O}(10^2)$ ballpark. Image taken from Ref. [283].* **Right:** *Complete dataset collected by the NANOGrav collaboration to date. With the analysis of the 15 yr dataset ongoing, extending current results both in terms of observational time and number of pulsars traced, the nature of the signal recently reported by the collaboration is expected to be confirmed soon [754]. Figure taken from [755].*

parameterisation of the PBH mass function.

- We highlighted a few strategies to search for smoking-gun signatures of the PBH model. Those rely on searching for high redshift events at 3G detectors and either single-event or population studies searching for the key features of the PBH scenario, such as the $q - \chi_{\text{eff}}$ correlation introduced by accretion effects.

It is also important to stress that many more data are expected to come very soon, see Fig. 7.1. Recently, a new analysis of the first half of the O3 run provided few additional signals that were not previously identified [751]. In particular, two additional events were discovered with BH masses falling in the astrophysical mass gap (i.e. GW190403_051519 and GW190426_190642), possibly strengthening the case for a PBH contribution to current data. During the final stages of O3, a new detector was added to the network. KAGRA is a ground-based detector similar to LIGO and Virgo based in Japan [752] which allows for an improved capability of the network to detect and reconstruct GW signals. During the second phase of the O3 run, at least a couple of dozen of new detections were performed [283] and two neutron star-BH/BH-BH binaries has been already released [753]. During the O4 run, estimates predict several binary BH detections (i.e. $N_{\text{det}} = 79^{+89}_{-44}$) at the LIGO/Virgo/KAGRA facilities. This will more than double the number of events observed so far, paving the way to more robust population studies.

In Chapter 6, we focused on the characterisation of the SGWB induced at second order by the scalar perturbations responsible for the PBH formation. After a detailed discussion on the expected spectral shape, Gaussianity of the signal after GW propagation in the perturbed universe and gauge independence of the result, we compared the PBH formation signatures to current and future GW experiments. We found that LISA will be able to search for the signature of PBH production in the mass window where PBHs are still allowed to be the entirety of the dark matter. Also, assuming the NANOGrav data to be the first hint of a GW signal (to be confirmed with more data in the near future, see [754] and Fig. 7.1), we devised a scenario in which PBHs can comprise the totality of dark matter and, at the same time, being responsible for the generation of a SGWB compatible with the NANOGrav observations. In this direction, much work is still needed in order to understand if the signal coming from the PBH scenario could be distinguished from other astrophysical or cosmological sources, such as the expected SGWB from the merger of supermassive BH binaries.

With this in mind, as more GW data are certainly coming in the near future, a large amount of work is required to refine the predictions of the PBH model along the lines we just mentioned, to allow for solid and definitive comparison with increasingly statistically significant GW data. Depending on future analyses' outcomes, this ambitious line of research is expected to provide either a path towards first evidence of the existence of PBHs (telling us for the first time what a significant fraction of the dark matter is made of) or the most stringent constraints on the PBH abundance in the extensive mass range accessible by GW experiments. We contributed to this endeavour in this thesis, but this effort just scratches the surface of the amount of new knowledge humanity will acquire in the flourishing era of gravitational wave astronomy.

# Part V

# Appendices

# Appendix A

# Peak theory of gaussian fields

In this appendix, we report a short summary of the theory of peaks of gaussian random fields, introduced in the seminal paper in Ref. [297], with the aim of defining the basic properties of peaks routinely used in Chapters 2 and 3 and providing a justification of the commonly used assumption that extreme peaks tend to be spherical.

## A.1   Peak number density distribution

We consider a gaussian field $\delta$, which may represent either the curvature or the density perturbation. The joint gaussian probability distribution of the field $\delta$ and its first derivatives, defined as

$$\eta_i = \frac{\partial \delta}{\partial x_i}, \quad \zeta_{ij} = \frac{\partial^2 \delta}{\partial x_i \partial x_j},$$
(A.1)

can be written as

$$f(V_i) \mathrm{d}^{10} V_i = \frac{1}{(2\pi)^5 |\mathbf{M}|^{1/2}} e^{-\frac{1}{2}(V_i - \langle V_i \rangle) \mathbf{M}_{ij}^{-1} (V_j - \langle V_j \rangle)} \mathrm{d}^{10} V_i,$$
(A.2)

where we stacked all the variables in a single vector $V$ and where the covariance matrix $\mathbf{M}$ is defined as

$$\mathbf{M}_{ij} = \langle (V_i - \langle V_i \rangle)(V_j - \langle V_j \rangle) \rangle.$$
(A.3)

We stress that $\zeta_{ij}$ has only 6 independent components due to its symmetry. We compute the non-vanishing correlators as follows:

- The field variance:

$$\langle \delta \delta \rangle \equiv \langle \delta^2 \rangle = \frac{1}{(2\pi)^3} \int |\delta_{\vec{k}}|^2 4\pi k^2 \mathrm{d}k \equiv \sigma_0^2.$$
(A.4)

- The first derivatives two-point function:

$$\langle \eta_i \eta_j \rangle = \frac{1}{(2\pi)^3} \int |\delta_{\vec{k}}|^2 k_i k_j 4\pi k^2 \mathrm{d}k = \frac{1}{3} \delta_{ij} \frac{1}{(2\pi)^3} \int |\delta_{\vec{k}}|^2 k^2 4\pi k^2 \mathrm{d}k \equiv \frac{1}{3} \delta_{ij} \sigma_1^2.$$
(A.5)

- The second derivatives two-point function:

$$\langle \zeta_{ij} \zeta_{mn} \rangle = \frac{1}{(2\pi)^3} \int |\delta_{\vec{k}}|^2 k_i k_j k_m k_n 4\pi k^2 \mathrm{d}k$$
(A.6)

$$= \frac{1}{15} \left( \delta_{ij} \delta_{mn} + \delta_{im} \delta_{jn} + \delta_{in} \delta_{jm} \right) \frac{1}{(2\pi)^3} \int |\delta_{\vec{k}}|^2 k^4 4\pi k^2 \mathrm{d}k$$
(A.7)

$$\equiv \frac{1}{15} \left( \delta_{ij} \delta_{mn} + \delta_{im} \delta_{jn} + \delta_{in} \delta_{jm} \right) \sigma_2^2.$$
(A.8)

- The mixed term:

$$\langle \delta \zeta_{ij} \rangle = -\frac{1}{(2\pi)^3} \int |\delta_{\vec{k}}|^2 k_i k_j 4\pi k^2 \mathrm{d}k = -\frac{1}{3} \delta_{ij} \frac{1}{(2\pi)^3} \int |\delta_{\vec{k}}|^2 k^2 4\pi k^2 \mathrm{d}k \equiv -\frac{1}{3} \delta_{ij} \sigma_1^2.$$
(A.9)

To summarize, the non-vanishing entries of the correlation matrix are

$$\langle \delta\delta \rangle = \sigma_0^2, \qquad\qquad \langle \eta_i \eta_j \rangle = \frac{\sigma_1^2}{3}\delta_{ij}$$

$$\langle \delta\zeta_{ij} \rangle = -\frac{\sigma_1^2}{3}\delta_{ij}, \qquad\qquad \langle \zeta_{ij}\zeta_{kl} \rangle = \frac{\sigma_2^2}{15}\delta_{ij} \qquad\qquad (A.10)$$

One can show that the non-diagonal components of the matrix $\zeta_{ij}$ can be expressed in terms of the Euler angles defining the orientation of the eigenvectors of $\zeta_{ij}$. Marginalising over those, and defining both $\nu = \delta/\sigma_0$, and

$$\sigma_2 x = -(\zeta_{11} + \zeta_{22} + \zeta_{33}), \qquad \sigma_2 y = -\frac{\zeta_{11} - \zeta_{33}}{2}, \qquad \sigma_2 z = -\frac{\zeta_{11} - 2\zeta_{22} + \zeta_{33}}{2}, \qquad (A.11)$$

the joint Gaussian probability distribution becomes

$$P(\nu, \vec{\eta}, x, y, z)\mathrm{d}\nu\mathrm{d}^3\eta\mathrm{d}x\mathrm{d}y\mathrm{d}z = N|2y(y^2 - z^2)|e^{-Q/2}\mathrm{d}\nu\mathrm{d}x\mathrm{d}y\mathrm{d}z\frac{\mathrm{d}^3\eta}{\sigma_0^3}, \qquad (A.12)$$

as a function of

$$Q = \nu^2 + \frac{(x - x_*)^2}{(1 - \gamma^2)} + 15y^2 + 5z^2 + \frac{3\vec{\eta}\cdot\vec{\eta}}{\sigma_1^2} \qquad (A.13)$$

and

$$x_* = \gamma\nu, \quad \gamma = \frac{\sigma_1^2}{\sigma_0\sigma_2}, \quad N = \frac{(15)^{5/2}}{32\pi^3}\frac{6\sigma_0^3}{\sigma_1^3(1 - \gamma^2)^{1/2}}. \qquad (A.14)$$

We now focus on peaks of the field $\delta$. This means we require the first derivative of the field to vanish, while the eigenvalues of the Hessian matrix to be negative. We, therefore, expand the field around $r_{\mathrm{pk}}$ as

$$\delta(r) = \delta(r_{\mathrm{pk}}) + \frac{1}{2}\sum_{ij}\zeta_{ij}(r_{\mathrm{pk}})\left(r^i - r_{\mathrm{pk}}^i\right)\left(r^j - r_{\mathrm{pk}}^j\right) + \dots. \qquad (A.15)$$

The number density of peaks with field height between $\nu_0$ and $\nu_0 + \mathrm{d}\nu$ is found by computing

$$\mathcal{N}_{\mathrm{pk}}\mathrm{d}\nu = \left\langle \sum_{\mathrm{pk}}\delta^3(r - r_{\mathrm{pk}}) \right\rangle \simeq \delta^3(\vec{\eta})\Theta(-\zeta_{11})\Theta(-\zeta_{22})\Theta(-\zeta_{33})\delta(\nu - \nu_0)\mathrm{d}\nu, \qquad (A.16)$$

and its distribution can be written explicitly as [297]

$$\mathcal{N}_{\mathrm{pk}}(\nu, x, y, z)\mathrm{d}\nu\mathrm{d}x\mathrm{d}y\mathrm{d}z = \frac{5^{5/2}9}{(2\pi)^3}\frac{1}{R_*^3}\frac{\chi F(x, y, z)}{(1 - \gamma^2)^{1/2}}e^{-\frac{\nu^2}{2} - \frac{(x - x_*)^2}{2(1 - \gamma^2)} - \frac{5}{2}(3y^2 + z^2)}\mathrm{d}\nu\mathrm{d}x\mathrm{d}y\mathrm{d}z, \qquad (A.17)$$

where

$$F(x, y, z) = (x - 2z)\left[(x + z)^2 - (3y)^2\right]y\left(y^2 - z^2\right). \qquad (A.18)$$

In the previous expression the function $\chi$ is 1 in the allowed domain of the variables $x, y, z, \nu$ enforcing the eigenvalues of $\zeta_{ij}$ to be $\lambda_1 \geq \lambda_2 \geq \lambda_3 \geq 0$ and 0 elsewhere.

Integrating over $y$ and $z$, one finds

$$\mathcal{N}_{\mathrm{pk}}(\nu, x)\mathrm{d}\nu\mathrm{d}x = \frac{e^{-\nu^2/2}}{(2\pi)^2 R_*^3}f(x)\frac{\exp[-(x - x_*)^2/2(1 - \gamma^2)]}{[2\pi(1 - \gamma^2)]^{1/2}}\mathrm{d}\nu\mathrm{d}x, \qquad (A.19)$$

where we introduced

$$R_* = \sqrt{3}\frac{\sigma_1}{\sigma_2}, \qquad (A.20)$$

and

$$f(x) = \frac{(x^3 - 3x)}{2}\left[\mathrm{erf}\left(x\sqrt{\frac{5}{2}}\right) + \mathrm{erf}\left(\frac{x}{2}\sqrt{\frac{5}{2}}\right)\right] + \sqrt{\frac{2}{5\pi}}\left[\left(\frac{31x^2}{4} + \frac{8}{5}\right)e^{-\frac{5x^2}{8}} + \left(\frac{x^2}{2} - \frac{8}{5}\right)e^{-\frac{5x^2}{2}}\right]. \qquad (A.21)$$

The expression in Eq. (A.19) was used in the main text to compute the number density of peaks with given $x$ and $\nu$. We show the integrated distribution (A.19) in Fig. A.1 as a function of $\nu$ and for different values of $\gamma$.

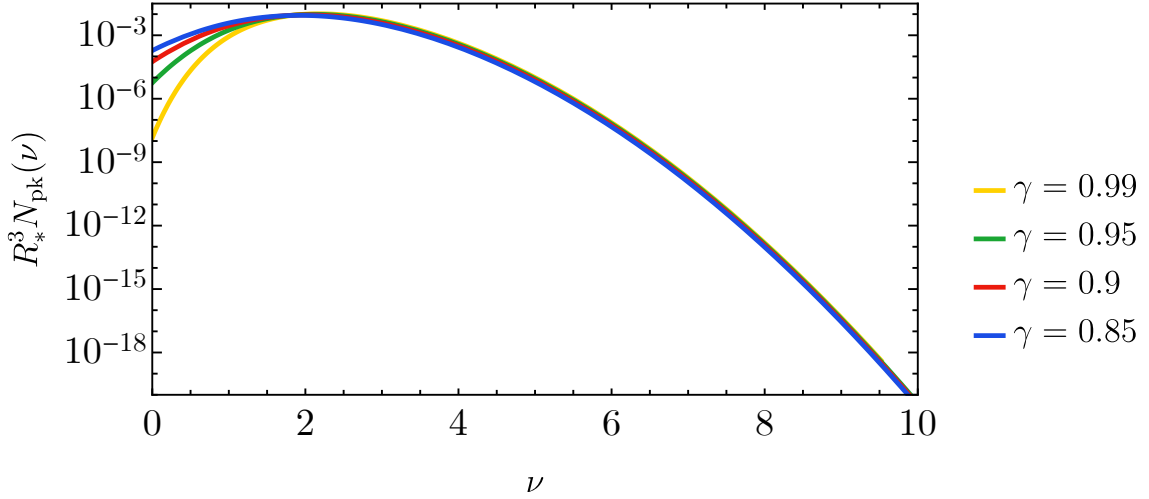

Figure A.1: *The rescaled comoving peak density for several choices of $\gamma$. The curves are very similar around $\nu \sim 8$, where this quantity becomes relevant for the computation of the PBH abundance.*

## A.2 Ellipticity and prolateness of extreme peaks

In order to study the geometrical properties of rare peaks, we can introduce the ellipticity $e$ and prolateness $p$ as

$$e = y/x, \qquad p = z/x. \tag{A.22}$$

The quantity $e\ (> 0)$ measures the ellipticity of peaks, while $p$ determined the oblateness (i.e. when $0 < p < e$) or prolateness (i.e. when $0 > p > -e$) of the ellipsoid. In terms of $e$ and $p$ one finds

$$\chi(e, p) = \begin{cases} 1 & \text{if} \quad 0 \leq e \leq 1/4 \quad \text{and} \quad -e \leq p \leq e \\ 1 & \text{if} \quad 1/4 \leq e \leq 1/2 \quad \text{and} \quad -(1 - 3e) \leq p \leq e \\ 0 & \text{elsewhere} \end{cases} \tag{A.23}$$

which is equivalent to the condition of having $\lambda_1 \geq \lambda_2 \geq \lambda_3 \geq 0$. This function is non-vanishing in a triangular region of the $(e, p)$ plane, see Fig. A.2.

The differential number density of peaks in terms of $\nu$, $x$, $e$ and $p$ is found after few manipulations of Eq. (A.17) as

$$\mathcal{N}_{\mathrm{pk}}(\nu, x, e, p)\mathrm{d}\nu\mathrm{d}x\mathrm{d}e\mathrm{d}p = \frac{225\sqrt{5}}{8\pi^3} \frac{1}{R_*^3} \frac{1}{\sqrt{1 - \gamma^2}} x^8 e(2p - 1)\left(e^2 - p^2\right)\left(9e^2 - (p + 1)^2\right)$$

$$\times \exp\left[-\frac{1}{2}\nu^2 - \frac{5}{2}(3e^2 + p^2)x^2 - \frac{(x - \gamma\nu)^2}{2(1 - \gamma^2)}\right]\mathrm{d}\nu\mathrm{d}x\mathrm{d}e\mathrm{d}p. \tag{A.24}$$

Therefore, one can write the probability distribution of peaks as a function of $(e, p)$ with given the height $\nu$ and curvature $x$. The result is independent of $\nu$ and can be found by computing

$$P(e, p|x)\mathrm{d}e\mathrm{d}p \equiv \frac{\mathcal{N}_{\mathrm{pk}}(\nu, x, e, p)}{\mathcal{N}_{\mathrm{pk}}(\nu, x)}. \tag{A.25}$$

In the high peak limit $\nu \gg 1$, one finds that the characteristic value of $x$ tends towards $x_* = \gamma\nu$, see Eq. (A.24). Therefore, we can expand in the large $x$ limit. In this setup, the distribution of $e$ and $p$ becomes nearly gaussian [297] as

$$P(e, p|x)\mathrm{d}e\mathrm{d}p \approx \exp\left[-\frac{(e - e_m)^2}{2\sigma_e^2} - \frac{(p - p_m)^2}{2\sigma_p^2}\right] \tag{A.26}$$

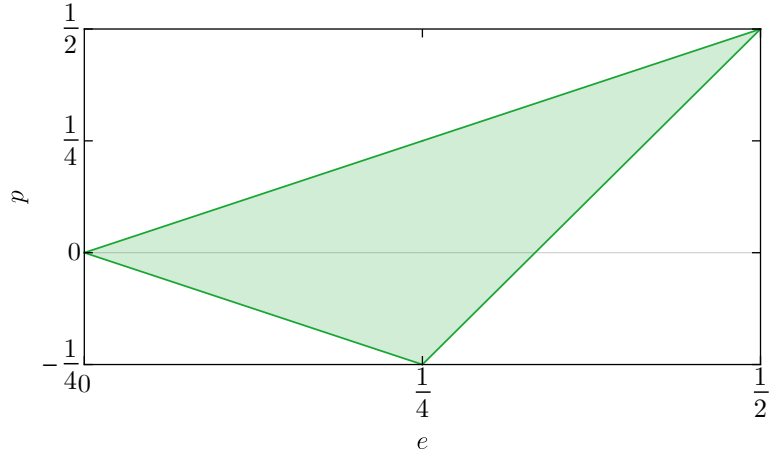

Figure A.2: *Support of $\chi(e,p)$ enforcing the condition $\lambda_1 \geq \lambda_2 \geq \lambda_3 \geq 0$.*

with

$$
e_m = \frac{1}{\sqrt{5}x\left[1 + 6/(5x^2)\right]^{1/2}}, \qquad \sigma_e = \frac{e_m}{\sqrt{6}},
$$
$$
p_m = \frac{6}{5x^4\left[1 + 6/(5x^2)\right]^2}, \qquad \sigma_p = \frac{p_m}{\sqrt{3}}. \tag{A.27}
$$

This shows that, in the limit of large $\nu$, for which $x \sim \gamma\nu$, the characteristic values of the deviations from sphericity rapidly decay as $e_m \sim 1/\gamma\nu$ and $p_m \sim 1/(\gamma\nu)^4$.

# Appendix B

# A general formula for PBH clustering induced by non-Gaussianities

In this appendix, we provide a general formula describing the PBH correlation function in the presence of non-Gaussianities. We will follow the same path integral approach used in Sec. 2.2.2, originally introduced in Ref. [324] in the context of large scale structure and applied to the PBH models in Ref. [24].

Using the peak theory model of bias, one can compute the PBH $n$-point correlation functions. For Gaussian fluctuations, this was performed in Ref. [756], where it was shown that the connected correlation function of peaks can be written as

$$\xi_\nu^{(2)}(\vec{x}_1, \vec{x}_2) = -1 + \exp\left(\nu^2 w^{(2)}(\vec{x}_1, \vec{x}_2)\right). \tag{B.1}$$

While the correlator $\xi_\nu^{(2)}(\vec{x}_1, \vec{x}_2)$ increments for higher values of $\nu$, as the distance between the two points is increased to super Hubble scales, the rescaled density two-point function $w^{(2)}$ is suppressed, leading to a small PBH correlation function, consistently with the previous section.

The technique explained in Ref. [324] allows to extend the gaussian formula with the inclusion of higher order correlation functions, capturing the non-Gaussian nature of the perturbations. To simplify the computations, we adopt the smoothing function $W(\vec{x}, R) = \delta^{(3)}(\vec{x})$. Then, analogously to Eq. (2.2.16), we can write

$$\Pi_\nu^{(N)}(\vec{x}_1, \cdots, \vec{x}_N) = \frac{e^{-N\nu^2}}{(2\pi\nu^2)^{N/2}} \prod_{i=1}^N Z[J_i], \tag{B.2}$$

which, by using Eq. (2.2.27), can be written as

$$\Pi_\nu^{(N)}(\vec{x}_1, \cdots, \vec{x}_N) = \langle \rho_\nu(\vec{x}) \rangle^N \exp\left\{ \sum_{n=2}^\infty \frac{\nu^n}{n!} \left[ \sum_{i_1, \cdots, i_n=1}^N w_{[i_n]}^{(n)} - N w^{(n)} \right] \right\}. \tag{B.3}$$

Therefore, we can define the $N$-point peak disconnected correlation function as

$$\xi_{\text{disc},\nu}^{(N)}(\vec{x}_1, \cdots, \vec{x}_N) = \left\langle \prod_{i=1}^N \frac{\rho_\nu(\vec{x}_i)}{\langle \rho_\nu \rangle} \right\rangle - 1, \tag{B.4}$$

which becomes

$$\xi_{\text{disc},\nu}^{(N)}(\vec{x}_1, \cdots, \vec{x}_N) = -1 + \exp\left\{ \sum_{n=2}^\infty \frac{\nu^n}{n!} \left[ \sum_{i_1, \cdots, i_n=1}^N w_{[i_n]}^{(n)} - N w_R^{(n)} \right] \right\}. \tag{B.5}$$

For instance, we can write explicitly the non-Gaussian two-peaks connected correlation function as

$$\xi_\nu^{(2)}(\vec{x}_1, \vec{x}_2) = = -1 + \exp\left\{ \sum_{n=2}^\infty \sum_{j=1}^{n-1} \frac{\nu^n \sigma^{-n}}{j!(n-1)!} \xi^{(n)}\left( \underbrace{\vec{x}_1, \cdots, \vec{x}_1}_{j-\text{times}}, \underbrace{\vec{x}_2, \cdots, \vec{x}_2}_{(n-j)-\text{times}} \right) \right\}. \tag{B.6}$$

Let us stress that this equation does not assume the variance to be small [324, 757, 758]. In the model with a local non-Gaussianity of the form (3.3.15), it can be shown that the expression (B.6) give rise to the scale dependent bias factor as presented in the previous section (see Ref. [759] for this result in the context of large scale structure).

# Appendix C

# Bayesian inference framework

In this Appendix, we present the statistical tools adopted to perform the confrontation of the PBH model to the LVC data.

## C.1 Event parameters and population parameters

In this section, we summarise, for clarity, the parameters used in the analysis discussed in Chapter 5. In Table C.1, we report the event parameters that characterise an individual merger event. Also, in the same table, we report the hyperparameters $\boldsymbol{\lambda}$ for all the population models considered. Those are, in particular, the PBH model (see Chapter 4), the phenomenological astrophysical model (see Sec. 5.2.3) and the astrophysical isolated and dynamical channels (see App. D).

In our analyses, reported in Chapter 5, we make the following assumptions:

- the prior $\pi(\boldsymbol{\lambda})$ on the hyperparameters $\boldsymbol{\lambda}$ of the model is considered flat. The prior ranges are reported in Tab. C.2. In the case of the astrophysical isolated and dynamical channels, only discrete points in the hyperparameter space are available from Ref. [577, 760].

- The binary BH parameters considered are taken to be $\boldsymbol{\theta} = \{m_1, m_2, z, \chi_{\mathrm{eff}}\}$, where $m_i$ is the source-frame mass of the $i$-th binary component, $z$ is the merger redshift, and

$$\chi_{\mathrm{eff}} \equiv \frac{\chi_1 \cos\alpha_1 + q\chi_2 \cos\alpha_2}{1+q} \tag{C.1}$$

is the effective spin parameter, which is a function of the mass ratio $q \equiv m_2/m_1 \leq 1$, of both BH spin magnitudes $\chi_j$ ($j = 1, 2$, with $0 \leq \chi_j \leq 1$), and of their orientation with respect to the orbital angular momentum parametrised by the angles $\alpha_j$. We neglect the precession spin component $\chi_{\mathrm{p}}$ in the inference, since this parameter is still poorly reconstructed for most of the GW events [507]. We stress that the PBH model predicts uncorrelated individual spins which are evenly distributed on the sphere. This property is similar to what is expected in the dynamical astrophysical channels.

- The assumption on the natal spin $\chi_{\mathrm{b}}$ (see discussion in App. D) only affects the astrophysical channels as the PBH spin is determined by a different physical process related to the collapse of radiation density perturbations in the early universe.

- The common envelope efficiency $\alpha_{\mathrm{CE}}$ is only assumed to affect the CE channel, see discussion in App. D.

- The astrophysical merger rates $N_i$ are assumed to be unconstrained and independent from both the remaining astrophysical hyperparameters $\alpha_{\mathrm{CE}}$ and $\chi_{\mathrm{b}}$.

- For values of $z_{\mathrm{cut\text{-}off}} \gtrsim 30$, accretion is negligible in the mass range of interest for LVC observations and PBHs always retain small spins. We therefore cut the PBH prior space at $z_{\mathrm{cut\text{-}off}} = 30$ as we checked that models with a larger cut-off $z_{\mathrm{cut\text{-}off}} > 30$ are degenerate and would produce a plateau in the posterior distribution.

Table C.1: *Relevant event parameters and hyperparameters in the various analyses.*

| | *Event parameters* $\boldsymbol{\theta}$ |
|---|---|
| $m_1$ | Source-frame primary mass |
| $m_2$ | Source-frame secondary mass |
| $\chi_{\text{eff}}$ | Effective spin |
| $z$ | Merger redshift |
| | *Hyperparameters* $\boldsymbol{\lambda}$ *of the PBH model* |
| $M_c$ | Peak reference mass of the lognormal distribution |
| $\sigma$ | Variance of the lognormal mass distribution |
| $f_{\text{PBH}}$ | Fraction of dark matter composed by PBHs at formation |
| $z_{\text{cut-off}}$ | Accretion cut-off redshift |
| | *Hyperparameters* $\boldsymbol{\lambda}$ *of the phenomenological ABH model* |
| $R_0$ | Integrated merger rate at $z = 0$ |
| $\kappa$ | Merger rate evolution in redshift, defined as $R(z) \propto (1+z)^{\kappa}$ |
| $\alpha$ | Exponent of the total mass factor in the differential rate |
| $\beta$ | Exponent of the symmetric mass ratio factor |
| $\zeta$ | ABH mass function power law scaling $\psi(m) \propto m^{-\zeta}$ |
| $m_{\text{min}}$ | Minimum ABH mass (step-like cut-off) |
| $m_{\text{max}}$ | Maximum ABH mass (step-like cut-off) |
| $\mu_{\chi_{\text{eff}}}$ | Mean value of $\chi_{\text{eff}}$ in the "Gaussian" spin model |
| $\sigma_{\chi_{\text{eff}}}$ | Width of the distribution of $\chi_{\text{eff}}$ in the "Gaussian" spin model |
| | *Hyperparameters* $\boldsymbol{\lambda}$ *of the isolated and dynamical ABH channels* |
| $\alpha_{\text{CE}}$ | Efficiency of the common envelope phase |
| $\chi_{\text{b}}$ | Astrophysical black hole natal spin |
| $N_{\text{CE}}$ | Intrinsic number of mergers from the CE channel |
| $N_{\text{SMT}}$ | Intrinsic number of mergers from the SMT channel |
| $N_{\text{GC}}$ | Intrinsic number of mergers from the GC channel |
| $N_{\text{NSC}}$ | Intrinsic number of mergers from the NSC channel |

## C.2   Hierarchical Bayesian inference

For population studies, the technique we use, following the recent literature, goes under the name of hierarchical bayesian framework. In full generality, a hierarchical Bayesian analysis determines the posterior distribution of a series of model hyperparameters $\boldsymbol{\lambda}$ based on the observed dataset $\boldsymbol{d}$. This is a powerful tool allowing for a comparison of physical models to GW observations fully accounting for selection effects and can be used to perform model selection, derive constraints on parameters and compare different formation channels. We refer the interested reader to more comprehensive descriptions that can be found in, for example, Refs. [535, 536, 761, 762].

GW detections are individual time series of the strain observed in the experimental apparatus which are identified as signals coming from compact objects and cleaned from the instrumental noise.

Table C.2: *Prior ranges for the hyperparameters of each model.*

| PBH model | | | | | |
|---|---|---|---|---|---|
| Parameter | $M_c[M_\odot]$ | $\sigma$ | | $\log f_{\mathrm{PBH}}$ | $z_{\mathrm{cut\text{-}off}}$ |
| Prior | $[5, 40]$ | $[0.1, 1.1]$ | | $[-4, -2]$ | $[10, 30]$ |
| Phenomenological ABH models | | | | | |
| Parameter | $R_0[\mathrm{Gpc}^{-3}\mathrm{yr}^{-1}]$ | $\beta$ | $\zeta$ | $m_{\mathrm{min}}[M_\odot]$ $m_{\mathrm{max}}[M_\odot]$ | $\mu_{\chi_{\mathrm{eff}}}$ $\sigma_{\chi_{\mathrm{eff}}}$ |
| Prior | $[1, 100]$ | [-4,12] | [-6,0] | $[2, 10]$ $[30, 100]$ | [-1,1] [0,1] |
| Isolated and dynamical ABH model | | | | | |
| Parameter | | | | $\alpha_{\mathrm{CE}}$ | $\chi_{\mathrm{b}}$ |
| Prior | | | | [0.2,0.5,1,2,5] | [0,0.1,0.2,0.5] |

Then, the data are analysed through a parameter estimation process [761] allowing to extract the physical parameters describing the binary BH merger, such as the BH masses, spins and redshift of the binary. The result of the analysis is made public by the LVC collaboration as a collection of posteriors characterising the expectation values and uncertainties on the observables related to the individual merger events, i.e. $p(\boldsymbol{\theta}|\boldsymbol{d}_i)$, where $\boldsymbol{\theta}$ is a vector of source parameters (e.g. the binary BH masses and spins), and $\boldsymbol{d}_i$ labels the time series of the $i$-th event in the catalog. We show in Fig. 5.1 a comprehensive view of the posteriors of each relevant quantity for all the binary BH events detected so far in the first three observing runs (O1/O2/O3a) of the LIGO and Virgo experiments.

By Bayes' theorem dictates that the posterior probability of the source parameters $\boldsymbol{\theta}$, given some data $\boldsymbol{d}$, is

$$p(\boldsymbol{\theta}|\boldsymbol{d}) \propto p(\boldsymbol{d}|\boldsymbol{\theta})p(\boldsymbol{\theta}), \tag{C.2}$$

where $p(\boldsymbol{d}|\boldsymbol{\theta})$ is the likelihood of observing the data given the model describing both the signal and detector, while $p(\boldsymbol{\theta})$ is the assumed prior on the source parameters. One can interpret the prior as capturing our previous knowledge of the underlying physics. Typically, in the LVC analysis, those are assumed to be uninformative. However, physically motivated prior choices may still impact on the interpretation of single events, see Refs. [10, 530–532, 607, 608].

A hierarchical Bayesian analysis parametrizes the priors as a function of the hyperparameters $\boldsymbol{\lambda}$ (which can be thought of as the channel rate, the PBH abundance, the characteristic scale of the mass function, etc., just to name few examples). Those hyperparameters uniquely determine the population model one wish to compare to the data. Then, the posterior distribution of $\boldsymbol{\lambda}$ can be inferred from the data using

$$p(\boldsymbol{\lambda}|\boldsymbol{d}) \propto p(\boldsymbol{\lambda}) \int p(\boldsymbol{d}|\boldsymbol{\theta})p_{\mathrm{pop}}(\boldsymbol{\theta}|\boldsymbol{\lambda})\mathrm{d}\boldsymbol{\theta}, \tag{C.3}$$

where $p(\boldsymbol{d}|\boldsymbol{\theta})$ is the single-event likelihood, $p(\boldsymbol{\lambda})$ is the prior on the hyperparameters $\boldsymbol{\lambda}$, and $p_{\mathrm{pop}}(\boldsymbol{\theta}|\boldsymbol{\lambda})$ is the *population likelihood*, equivalent to a prior parametrized by some hyperparameters. We present the various choices of both event and population parameters in the following section.

With the aim of shortening the computational time, the integration over $\boldsymbol{\theta}$ in Eq. (C.3) can be written as a weighted average over the event posterior's sample. The likelihood function is therefore computed according to

$$p(\boldsymbol{\lambda}|\boldsymbol{d}) = \pi(\boldsymbol{\lambda})e^{-N_{\mathrm{det}}(\boldsymbol{\lambda})}\left[N(\boldsymbol{\lambda})\right]^{N_{\mathrm{obs}}} \prod_{i=1}^{N_{\mathrm{obs}}} \frac{1}{\mathcal{S}_i} \sum_{j=1}^{\mathcal{S}_i} \frac{p_{\mathrm{pop}}(^j\boldsymbol{\theta}_i|\boldsymbol{\lambda})}{\pi(^j\boldsymbol{\theta}_i)}, \tag{C.4}$$

where $N_{\mathrm{obs}}$ is the number of events in the catalog, i.e. the number of detections, $N(\boldsymbol{\lambda})$ is the *total* number of events in the model characterised by the set of hyperparameters $\boldsymbol{\lambda}$, $N_{\mathrm{det}}(\boldsymbol{\lambda})$ is the expected number of *observable* events in the model characterised by the hyperparameters $\boldsymbol{\lambda}$ computed by accounting for the experimental selection bias, and $\mathcal{S}_i$ is the length of the sample taken from the posterior dataset of each event in the catalog. Finally, $\pi(\boldsymbol{\theta})$ is the prior on the binary parameters used by the LIGO/Virgo collaboration when performing the parameter estimation. As one can appreciate by looking at Eq. (C.4), this prior is removed to extract the values of the single-event likelihood, ensuring only the informative part of the event posterior is used. This analysis is therefore independent of any assumption on the priors adopted to infer the properties of each event. We remark that the last sum in Eq. (C.4) gives the average of the binary parameter distribution over posterior samples, while the factors proportional to

$$p(\boldsymbol{\lambda}|\boldsymbol{d}) \propto e^{-N_{\text{det}}(\boldsymbol{\lambda})} \left[N(\boldsymbol{\lambda})\right]^{N_{\text{obs}}} \tag{C.5}$$

characterize an inhomogeneous Poisson process [535, 574, 761, 763] and are responsible for introducing the rate information in the inference, as well as the experimental selection effects.

The analysis is then performed by sampling the likelihood function (C.4) in the hyperparameter space by using the Markov chain Monte Carlo software algorithm `emcee` [764].

## C.2.1   Population distributions and selection bias

The binary parameter distributions in a given model (either primordial or astrophysical), are computed starting from the differential merger rate $\mathrm{d}R$ as

$$p_{\text{pop}}(\boldsymbol{\theta}|\boldsymbol{\lambda}) \equiv \frac{1}{N(\boldsymbol{\lambda})} \left[T_{\text{obs}} \frac{1}{1+z} \frac{\mathrm{d}V}{\mathrm{d}z} \frac{\mathrm{d}R}{\mathrm{d}m_1 \mathrm{d}m_2}(\boldsymbol{\theta}|\boldsymbol{\lambda})\right], \tag{C.6}$$

in terms of the observation time $T_{\text{obs}}$, the prefactor $1/(1+z)$ accounting for the clock redshift at the source epoch and $\mathrm{d}V/\mathrm{d}z$, indicating the differential comoving volume factor, see for example [541],

$$\frac{\mathrm{d}V(z)}{\mathrm{d}z} = \frac{4\pi}{H_0} \frac{D_c^2(z)}{E(z)} = \frac{4\pi}{H_0^2} \frac{1}{E(z)} \left(\int_0^z \frac{\mathrm{d}z'}{E(z')}\right)^2. \tag{C.7}$$

In the relation above we introduced the comoving distance

$$D_c(z) = \frac{1}{H_0} \int_0^z \frac{\mathrm{d}z'}{E(z')}, \tag{C.8}$$

where

$$E(z') = \sqrt{\Omega_r(1+z')^4 + \Omega_m(1+z')^3 + \Omega_{\text{K}}(1+z')^2 + \Omega_\Lambda}, \tag{C.9}$$

and $\Omega_{\text{K}} = 0.0007$, $\Omega_r = 5.38 \times 10^{-5}$, $\Omega_\Lambda = 0.685$, $\Omega_m = 0.315$, $h = 0.674$, $H_0 = 1.0227 \times 10^{-10} h \, \text{yr}^{-1}$.

The number of expected detections for a given model with hyperparameter $\boldsymbol{\lambda}$ is defined as the integral of the differential merger rate, that is

$$N_{\text{det}}(\boldsymbol{\lambda}) \equiv T_{\text{obs}} \int \mathrm{d}m_1 \mathrm{d}m_2 \mathrm{d}z \, p_{\text{det}}(m_1, m_2, z) \frac{1}{1+z} \frac{\mathrm{d}V}{\mathrm{d}z} \frac{\mathrm{d}R}{\mathrm{d}m_1 \mathrm{d}m_2}(m_1, m_2, z|\boldsymbol{\lambda}). \tag{C.10}$$

The function $0 \leq p_{\text{det}}(\boldsymbol{\theta}) \leq 1$ introduces the selection effects caused by the sensitivity of the detectors, and can be interpreted as the probability that an event with parameters $\boldsymbol{\theta}$ would be detectable, see details in Sec. 5.1.2. The *total* number of events $N(\boldsymbol{\lambda})$ is computed in the same manner using Eq. (C.10) but without accounting for the experimental selection effects, i.e. by setting $p_{\text{det}} = 1$.

For the O1-O2 (O3a) observing run, there are $T_{\text{obs}} \approx 166.6$ (183.3) days of coincident data. The data released from the LVC collaboration can be found at the following repositories: "Combined" samples for the GWTC-1 events in Ref. [765]; "PublicationSamples" in Ref. [766] for the GWTC-2 events.

### C.2.2   Bayes factors

A quantitative way of comparing how much different models are supported by a given dataset is through the computation of the Bayes factors.

Given a model $\mathcal{M}$, the statistical evidence $Z$ is defined as the marginal population likelihood computed as the integral of the population posterior $p(\boldsymbol{\lambda}|\boldsymbol{d})$, i.e.

$$Z_{\mathcal{M}} \equiv \int \mathrm{d}\boldsymbol{\lambda}\, p(\boldsymbol{\lambda}|\boldsymbol{d}). \tag{C.11}$$

In simple terms, the evidence $Z_{\mathcal{M}}$ is a quantitative measure of the support for a given model given the data $\boldsymbol{d}$. Then, the so-called Bayes factor is defined as the ratio of the evidences of two different models, i.e.

$$\mathcal{B}_{\mathcal{M}_2}^{\mathcal{M}_1} \equiv \frac{Z_{\mathcal{M}_1}}{Z_{\mathcal{M}_2}}. \tag{C.12}$$

According to Jeffreys' scale criterion [580], a Bayes factor larger than $(10, 10^{1.5}, 10^2)$ would imply a strong, very strong, or decisive evidence in favour of model $\mathcal{M}_1$ with respect to model $\mathcal{M}_2$ given the available dataset.

## C.3   Parameter estimation for coalescing black hole binaries

Bayesian parameter estimation is also conventionally adopted to derive the posterior distributions $p(\boldsymbol{\theta}|\boldsymbol{d}, \mathcal{M})$ of parameters $\boldsymbol{\theta}$ assuming a model $\mathcal{M}$ from the observed data $\boldsymbol{d}$. This is done including information of the expectations for the parameter values via the prior $\pi = p(\boldsymbol{\theta}|\mathcal{M})$. The information contained in the data contributes through the likelihood function $p(\boldsymbol{d}|\boldsymbol{\theta}, \mathcal{M})$ and the posterior probability density function $p(\boldsymbol{\theta}|\boldsymbol{d}, \mathcal{M})$ is given by [767]

$$p(\boldsymbol{\theta}|\boldsymbol{d}, \mathcal{M}) = \frac{p(\boldsymbol{\theta}|\mathcal{M})p(\boldsymbol{d}|\boldsymbol{\theta}, \mathcal{M})}{Z}, \tag{C.13}$$

where $\boldsymbol{d} = \mathcal{M} + \mathcal{N}$, $\mathcal{N}$ is the noise, and the normalizing factor $Z = p(\boldsymbol{d}|\mathcal{M})$ is called the evidence. Assuming the noise model is Gaussian and stationary, the likelihood function can written as

$$L = p(\boldsymbol{d}|\boldsymbol{\theta}, \mathcal{M}) \propto e^{-\frac{1}{2}\langle \mathcal{N}|\mathcal{N}\rangle} = e^{-\frac{1}{2}\langle \boldsymbol{d}-\mathcal{M}|\boldsymbol{d}-\mathcal{M}\rangle}, \tag{C.14}$$

where $\langle .|. \rangle$ denotes the weighted inner product. In the context of estimating the source properties from a GW signal, $\mathcal{M}$ is the frequency-domain binary BH gravitational waveform $\widetilde{h(f)}$ and $\boldsymbol{d}$ is the output of the GW interferometers expressed in the frequency domain and

$$\langle \mathcal{M}|\boldsymbol{d}\rangle = \int_0^\infty df\, \frac{\widetilde{h}^*(f) \times \widetilde{\boldsymbol{d}}(f)}{S_n(f)}, \tag{C.15}$$

where $S_n(f)$ is the power spectral density (PSD) of the instrument. Finally, in order to quantify the goodness of fit to the data of a model, including information on the volume of the prior space (thereby penalizing over-fitting), one computes the evidence. This is obtained by completely marginalizing the posterior and can be written as

$$Z_{\mathcal{M}} = p(\boldsymbol{d}|\mathcal{M}) = \int d\boldsymbol{\theta}\, L(\boldsymbol{d}|\boldsymbol{\theta})\pi(\boldsymbol{\theta}). \tag{C.16}$$

We use the open science data for the binary BH coalescences as outlined in Ref. [768], using data from the two LIGO interferometers. We also use the BILBY Bayesian inference library [768, 769] to perform full Bayesian parameter estimations and infer all the 15 parameters of the GW coalescence waveform model, namely the source-frame masses $(m_1, m_2)$, the dimensionless BH spins magnitudes $(\chi_1, \chi_2)$, 4 angle variables $(\theta_{1,2}, \delta\phi, \delta_{JL})$ that describe the BH spin directions, the inclination angle $(\iota)$, the polarization angle $(\psi)$, the phase at coalescence $(\phi_c)$, the time of coalescence $(t_c)$, the right

Table C.3: *List of the non-informative priors we use for the waveform parameters. For the BH masses* $(m_1, m_2)$ *and spins* $(\chi_1, \chi_2)$ *we adopt either LVC or PBH-motivated priors.*

| Parameter | Prior |
|---|---|
| $\theta_{i,j}$ | Sin prior in $[0, \pi]$ |
| $\delta\phi$ | Uniform prior in $[0, 2\pi]$ |
| $\delta_{JL}$ | Uniform prior in $[0, 2\pi]$ |
| $t_c$ | Uniform prior around the event trigger time |
| $\phi_c$ | Uniform in $[0, 2\pi]$ |
| $\alpha$ | Uniform prior in $[0, 2\pi]$ |
| $\delta$ | Cosine prior in $[-\pi/2, \pi/2]$ |
| $d_L$ | Power law prior in $[50, 2000]$ Mpc |
| $\iota$ | Sin prior in $[0, \pi]$ |
| $\psi$ | Uniform in $[0, 2\pi]$ |

ascension ($\alpha$), the declination ($\delta$) and the luminosity distance ($d_L$). See details on their priors in Tab. C.3

For the analysis of the events presented in the main text, excluding GW190412, we use only the dominant $l = m = 2$ harmonic. On the other hand, for GW190412 (the most asymmetric binary BH system considered), the subdominant $l = m = 3$ harmonic has been measured [443]. Therefore, for this event we also included higher harmonics in the waveform model and discuss results with and without them separately. Indeed, while the inclusion of higher harmonics does not change the inferred parameter values for GW190412 significantly [443], it might change the evidence of a given model.

When considering only the dominant $l = m = 2$ harmonic, we used the "IMRPhenomPv2" frequency domain waveform implementation from LALSuite [770]. On the other hand, when higher harmonics are considered, we adopt "IMRPhenomHM" waveform approximant [771], including $(l, m) = [(2, 2), (2, 1), (3, 3), (3, 2), (4, 4), (4, 3)]$ in the parameter estimation, see more details in Ref. [10]. We stress that this waveform only describes systems with aligned spins. We leave the extension to a precessing waveform for future work. The PSD of the detectors is calculated from the data using the GWpy package [772, 773].

We stress that the parameter estimation is carried out for all the 15 binary BH parameters, but, for clarity, in the main text we only present marginalised posterior distributions of the relevant parameters.

# Appendix D

# Leading astrophysical models of black hole binaries

In this Appendix, we provide a brief summary of the astrophysical models adopted in the mixed ABH-PBH inference in Chapter 5. We follow the description given in Ref. [577] and references therein. The populations adopted in the analysis of Ref. [577] were made available at [760].

The astrophysical models can be broadly divided into two categories: the isolated formation channels, giving rise to BH binaries starting from stellar binaries evolving in isolation and the dynamical channels, where BHs are assembled in dense stellar environments. Both classes have different sub-models. In this work, we focused on the four main channels proposed in the literature (we refer to [774] for a recent review on the topic): Common Envelope (CE), Stable Mass Transfer (SMT), Globular Clusters (GC) and Nuclear Star Clusters (NSC). [1] Each one is characterised by specific predictions for the mass, spin and redshift distributions of the binary populations, while its characteristic merger rate is still affected by large uncertainties, but was shown to be able to reach at least a significant fraction of the one inferred by the LIGO/Virgo collaboration of 15–40 Gpc$^{-3}$ yr$^{-1}$ (90% C.I.) [507], see Ref. [780] for a recent review.

## D.1 Isolated formation channels

The main processes giving rise to the merger of a BH binary formed in isolation are the common envelope evolution and the stable mass transfer channel as discussed in details in Ref. [582].

### D.1.1 Common envelope and stable mass transfer

In the CE (SMT) scenario, a star binary initiates a phase of unstable (stable [776, 781]) mass transfer after one of its star components collapse to form a BH. In the first case, a common envelope forms and hardens (i.e. reduce the semi-major axis) the binary due to drag forces [782–784]. This hardening process allows the BHs to approach each other and reach a point where the gravitational-wave emission is strong enough to be able to make binary merge within less than the Hubble time [785–791] We note that Ref. [582] reported the SMT channel would become irrelevant in case of a highly super-Eddington accretion efficiency.

Both CE and SMT models were simulated in Ref. [577] with the POSYDON framework [583] combining the rapid population synthesis code COSMIC [584] with the binary evolution computation MESA [585, 792–795] as done in Ref. [582]. Then, the code COSMIC was used to evolve each binary in the time interval between the zero-age main sequence and the end of the second mass-transfer episode. Finally, MESA was adopted to determine the last phase of binary evolution which determines the second-born spin properties. These simulations take into account differential stellar rotation, tidal interactions, stellar winds, and the evolution of the Wolf–Rayet stellar structure. Finally, the synthetic binary BH population was distributed across cosmic time using the prescriptions presented in Sec. D.3.

---

[1]We will not address other scenarios such as the chemically homogeneous evolution [775–777], pop III population [616, 778, 779], binaries formed in AGN disks [599–601].

Uncertainties in the CE channel are parametrised using the common envelope efficiency parameter $\alpha_{CE}$. Large values of $\alpha_{CE} > 1$ correspond to an efficient CE evolution. This means that either additional energy (on top of the orbital one) helps to remove the envelope (see, for example, [796, 797]) or some of the envelope bounds to the stellar core (see, for example, Ref. [798]). The value of $\alpha_{CE}$ modifies the spin distribution, see Fig. D.1, since a lower $\alpha_{CE}$ (corresponding to tighter post-CE binaries) is associated with a more efficient tidal spin-up of the second-born BH. We note that Ref. [577] reported an estimated rate density for the CE channel at $17 \div 113 \, \mathrm{Gpc}^{-3} \, \mathrm{yr}^{-1}$, depending on the common envelope efficiency hyperparameter $\alpha_{CE}$. The range is delimited by the smallest value corresponding to $\alpha_{CE} = 5.0$ and the largest related to $\alpha_{CE} = 0.2$.

Another uncertainty, which is affecting all the channels discussed in this appendix, is related to the BH natal spin magnitude $\chi_b$. This depends on the presence of angular momentum transport in massive stars. If it is efficient, the core transfers angular momentum to the envelope, leading to a small spin at BH formation (see, for example, [799]). In Ref. [577], to standardise the assumptions on the natal spin across various channels, four values of spin at birth were considered $\chi_b \in [0, 0.1, 0.2, 0.5]$. In practice, components possessing a slow rotation $\chi < \chi_b$ at BBH formation were given a fixed spin magnitude $\chi_b$.

The populations produce in the CE and SMT channels for the choice of hyperparameters $\alpha_{CE} \in [0.2, 0.5, 1, 2, 5]$ along with the BH natal spin $\chi_b \in [0, 0.1, 0.2, 0.5]$ are available at [760]. In Fig. D.1, we show the observable distributions of binary parameters for the CE and SMT model during the O3 stage of the LIGO/Virgo experiment. The peculiar properties of the CE channel are the predominantly positive effective spin parameter when $\chi_b \neq 0$ and a chirp mass distribution which is not exceeding $\mathcal{M} \lesssim 40$ in all the hyperparameter space. Also, for a low CE efficiency, the binary tends to be more symmetric with a distribution of $q$ peaking close to unity. The SMT channel on the other hand, predicts more massive binaries (still below the upper mass gap) and a distribution of $q$ and $\chi_{eff}$ similar to the CE case.

## D.2 Dynamical formation channels

The dynamical formation channel predicts that BHs are formed in isolation after the death of massive stars in the dense stellar environments and only later on are assembled in binaries due to GW interactions. The binaries are then hardened through a series of gravitational interactions with other BHs or stars in the dense environment [800–805]. Those interactions are facilitated by the fact that heavier systems like BHs and BH binaries segregate towards the core of the clusters [802, 806]. It was shown that binaries formed and evolving in this channel can merge within a Hubble time [807–817].

The predictions of the dynamical channel depend on the environment in which BHs reside. In particular, the merger rate properties depend on the stellar density and cluster's escape velocity. In the following, we will present the models describing the dynamical binary formation in globular clusters and nuclear star clusters.

### D.2.1 Globular clusters

The GC models we adopt, following Ref. [577] are simulated using the Monte Carlo code CMC [813, 818–821]. To evolve stars and binaries in the cluster, the BSE package was used [822, 823], where the physical prescriptions describing the stellar winds, masses of compact objects, supernova natal kicks and the physics of pulsational-pair instability were adapted to be consistent with the code COSMIC, see Refs. [813, 824, 825] and Refs. therein. The three-body interactions with stars, allowing for the productions of many binary BHs were treated as prescribed in Ref. [826] while strong three- and four-body encounters were integrated using Fewbody [805], also accounting for relativistic corrections [825, 827, 828]. Finally, the BH remnants produced by binary mergers which are retained in the cluster are given new masses, spins, and GW recoil velocities taken from numerical relativity-based fitting formulae as in App. A of Ref. [825].

Consistently with the assumptions adopted for the other astrophysical channels, the BH spin at formation is dictated by the hyperparameter $\chi_b$. Also, the assumption of CE efficiency negligibly affects the evolution of the dynamical channels as the BHs are swapped multiple times throughout the binary evolution and are not tight enough for the tidal spin-up during the BH–Wolf–Rayet phase for $\alpha_{CE}$ to be relevant.

Finally, the time evolution of the merger rate for GC does not follow Appendix D.3, while instead rely on detailed modelling of GC formation across cosmic time [503, 829]. In Fig. D.1, we show the distribution of parameters in the GC model as observable during the O3 stage of the LIGO/Virgo experiment, i.e. including the observation bias.

The natal spin has a crucial role in determining the properties of binaries in the dynamical channel. This is because the large $\mathcal{M}$ and small $q$ portion of the distribution is populated by second-generation mergers. Hierarchical mergers happen when BH remnants are retained in the cluster. This crucially depends on both the initial BH spin [489, 555, 556] and escape velocity of the cluster in question [489, 556, 558, 576, 586, 814]. Indeed, large spins lead to an asymmetry in the merger dynamics, that is responsible for large relativistic recoil kicks due to the anisotropic emission of GWs [487, 830–839] and remnant BHs are more easily expelled from their host environments. Therefore, large values of $\chi_b$ are preventing a copious production of subsequent hierarchical mergers [555, 840, 841], leading to a smaller tail in the high portion of the chirp mass distribution and smaller support at small $q$. Finally, we note that the spins of BH in binaries assembled dynamically are uncorrelated and evenly oriented on the sphere. For this reason, the $\chi_{eff}$ distribution is symmetric with respect to the origin. Also, second-generation mergers can possess spins which are significantly larger than $\chi_b$ [842–844]. These properties can be observed in Fig. D.1.

### D.2.2　Nuclear star clusters

The dynamical process leading to the formation of binaries in the NSC channel is similar to the GC case. In Ref. [577], the evolution of NSCs was determined following the semi-analytical approach of Ref. [586] and neglecting the mass loss from stellar evolution and the escape of BH and stars (see [845] for caveats in this assumption). The binaries are formed dynamically in the cluster core and they are subsequently evolved until either they merge, they are ejected from the cluster or a time of 13 Gyr has passed. The properties of the remnant BHs are computed using Ref. [846]. If the merger remnant is retained inside the cluster, its spin and mass are computed using the prescriptions in Ref. [846].

As it was assumed for the GC channel, the impact of $\alpha_{CE}$ on the mergers was neglected. On the other hand, the star formation rate and metallicity evolution are assumed to be following the same prescriptions as the CE channel and described in Sec. D.3 (see, however, Ref. [847] for a discussion on the solidity of this assumption).

In Fig. D.1, we show the distribution of parameters in the NSC model as observable by the O3 stage of the LIGO/Virgo experiment. The impact of $\chi_b$ on the observable distributions of the NSC channel is qualitatively similar to the GC case.

## D.3　Formation rate and metallicity evolution

The evolution with redshift of each formation channel (apart from GC) is assumed to be following the Star Formation Rate (SFR) density in [848]

$$\psi(z) = 10^{-2} \, M_\odot \mathrm{yr}^{-1} \, \mathrm{Mpc}^{-3} \frac{(1+z)^{2.6}}{1 + [(1+z)/3.2]^{6.2}}, \tag{D.1}$$

determining the birth redshift of the BBH progenitor. Then, in the case of CE and SMT, the merger redshift is computed by including the inspiral time $t_{insp}$ computes using the orbital properties after the birth of the second BH using Ref. [267], see also Sec. 4.5.3. Therefore, one has

$$z_{merge} = z(t_{birth} + t_{BBH} + t_{insp}). \tag{D.2}$$

Finally, the metallicity distribution follows a truncated log-normal around the empirical median metallicity from [848] as

$$\log_{10} \langle Z/Z_\odot \rangle = 0.153 - 0.074 z^{1.34}, \tag{D.3}$$

where the solar metallicity is $Z_\odot = 0.017$ [849].

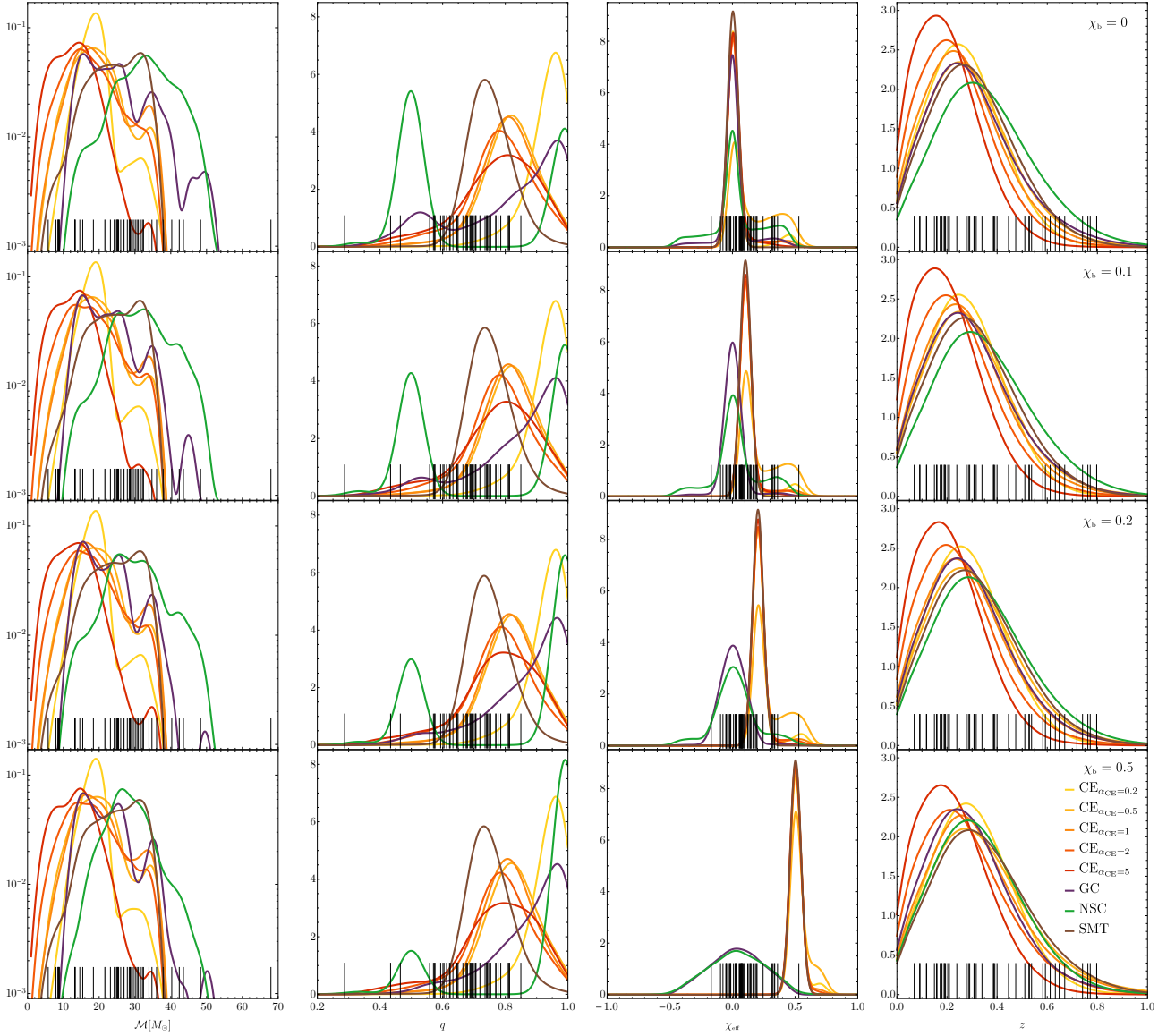

Figure D.1: *Plot of the "observable" distributions of the binary parameters in the CE, SMT, GC and NSC models for all the astro-hyperparameters $\alpha_{\text{CE}}$ (with different warm colours) and $\chi_{\text{b}}$ (each row from top to bottom) from [577]. The gridlines at the bottom of each plot indicate the mean values of $(\mathcal{M}, q, \chi_{\text{eff}}, z)$ for each detected event in the GWTC-2 catalog.*

# Appendix E

# Cosmological perturbation theory

In this appendix, we review the notation used to describe the cosmological perturbation theory in the previous chapters and, in particular, in Sec. 6.5. We adopted the notation used in Ref. [699]. We also explicitly provide some lengthy expression at second order such as the Riemann tensor and the tensor $\mathcal{X}_{ij}^{\mathrm{P}}$, as defined in Eq. (6.5.32).

## E.1 Metric perturbations and gauge transformations

First of all, we adopt the mostly plus sign notation for the spacetime metric signature. The ordinary and covariant derivatives are indicated as

$$\partial_\nu \mathcal{O}_\mu \equiv \mathcal{O}_{\mu,\nu}, \qquad D_\nu \mathcal{O}_\mu \equiv \partial_\nu \mathcal{O}_\mu - \Gamma_{\mu\nu}^\rho \mathcal{O}_\rho \equiv \mathcal{O}_{\mu;\nu}, \tag{E.1}$$

respectively. Finally, we use the compact notation for (anti-)symmetrisation, with the normalisation coefficient $1/2$, as

$$A_{(\mu\nu)} \equiv \frac{1}{2}\left(A_{\mu\nu} + A_{\nu\mu}\right), \qquad A_{[\mu\nu]} \equiv \frac{1}{2}\left(A_{\mu\nu} - A_{\nu\mu}\right). \tag{E.2}$$

Where not stated otherwise, we indicate with the prime symbol $'$ the ordinary derivative with respect to the conformal time $\eta$. The GW polarisation tensors $e_\lambda^{ij}$ are defined such that

$$e_L^{ij} e_{ij}^R = 0, \quad e_L^{ij} e_{ij}^L = e_R^{ij} e_{ij}^R = 1 \quad \text{and} \quad h_\lambda(t, \boldsymbol{k}) = e_\lambda^{ij}(\boldsymbol{k}) \int \mathrm{d}^3 x e^{-i\boldsymbol{k}\cdot\boldsymbol{x}} h_{ij}(t, \boldsymbol{x}), \tag{E.3}$$

with

$$e_\lambda^{ii}(\boldsymbol{k}) = 0 \qquad \text{and} \qquad k_i e_\lambda^{ij}(\boldsymbol{k}) = 0. \tag{E.4}$$

Finally, the transverse traceless projector $\mathcal{T}_{ij}^{lm}$ is written in terms of $e_\lambda^{ij}$ as

$$\mathcal{T}_{ij}^{lm} = e_{ij}^L \otimes e_L^{lm} + e_{ij}^R \otimes e_R^{lm}. \tag{E.5}$$

### E.1.1 Perturbations of the metric up to second-order

The background metric $\bar{g}_{\mu\nu}$ for an FRWL cosmology can be written as

$$\mathrm{d}s^2 \equiv \bar{g}_{\mu\nu}\mathrm{d}x^\mu \mathrm{d}x^\nu = -a^2(\eta)\left[\mathrm{d}\eta^2 + \delta_{ij}\,\mathrm{d}x^i\mathrm{d}x^j\right], \tag{E.6}$$

where $\eta$ is the conformal time. The background evolution is dictated by the Friedmann equations, i.e.

$$\mathcal{H}^2 = \frac{8\pi G}{3}a^2\rho, \qquad \mathcal{H}' = -\frac{4\pi G}{3}a^2(\rho + 3P), \tag{E.7}$$

where $\rho$ and $P$ correspond to the cosmic fluid's energy density and pressure. Assuming an equation of state of the form $P = w\rho$, one can write

$$\mathcal{H}' - \mathcal{H}^2 = -4\pi G a^2(\rho + P) \qquad \text{and} \qquad \mathcal{H}' = -\mathcal{H}^2\left(1 + 3w\right)/2. \tag{E.8}$$

The perturbed metric $g_{\mu\nu} \equiv \bar{g}_{\mu\nu} + \delta g_{\mu\nu}$ can be decomposed as

$$\delta g_{00} = -2a^2\phi, \qquad \delta g_{0i} = a^2 B_i, \qquad \delta g_{ij} = 2a^2 C_{ij}, \tag{E.9}$$

where, when expanded in the scalar-vector-tensor (SVT) decomposition, one can define

$$B_i = B_{,i} - S_i, \qquad C_{ij} = -\psi\delta_{ij} + E_{,ij} + F_{(i,j)} + \frac{1}{2}h_{ij}. \tag{E.10}$$

The vector and tensor degrees of freedom are transverse (divergence-free) and traceless, which means that they satisfy the conditions

$$S_{i,i} = 0, \qquad F_{i,i} = 0, \qquad \text{and} \qquad h^i_i = h_{ij,j} = 0. \tag{E.11}$$

One can single out each quantity at first and second-order as

$$\phi = \phi_1 + \frac{1}{2}\phi_2 + \dots \qquad\qquad \psi = \psi_1 + \frac{1}{2}\psi_2 + \dots \tag{E.12a}$$

$$B = B_1 + \frac{1}{2}B_2 + \dots \qquad\qquad E = E_1 + \frac{1}{2}E_2 + \dots \tag{E.12b}$$

$$S_i = S_{1i} + \frac{1}{2}S_{2i} + \dots \qquad\qquad F_i = F_{1i} + \frac{1}{2}F_{2i} + \dots \tag{E.12c}$$

$$h_{ij} = h_{1ij} + \frac{1}{2}h_{2ij} + \dots. \tag{E.12d}$$

The connection coefficients, defined as

$$\Gamma^\rho_{\mu\nu} = \frac{1}{2}g^{\rho\sigma}\left(g_{\mu\sigma,\nu} + g_{\nu\sigma,\mu} - g_{\mu\nu,\sigma}\right) \tag{E.13}$$

can be expressed, up to second-order, as [1]

$$\Gamma^0_{00} = \mathcal{H} + \epsilon\phi'_1 + \epsilon^2(B_{1b}B_1^b\mathcal{H} + B_1^bB'_{1b} - 2\phi_1\phi'_1 + \frac{1}{2}\phi'_2 + B_1^b\partial_b\phi_1) + \mathcal{O}(\epsilon^3),$$

$$\Gamma^0_{0i} = \epsilon(B_{1i}\mathcal{H} + \partial_i\phi_1) + \frac{1}{2}\epsilon^2(B_{2i}\mathcal{H} - 4B_{1i}\mathcal{H}\phi_1 + 2B_1^bC'_{1ib} - B_1^b\partial_bB_{1i} + B_1^b\partial_iB_{1b}$$
$$- 4\phi_1\partial_i\phi_1 + \partial_i\phi_2) + \mathcal{O}(\epsilon^3),$$

$$\Gamma^i_{00} = \epsilon(B_1^i\mathcal{H} + B_1^{i\prime} + \partial^i\phi_1) + \frac{1}{2}\epsilon^2(B_2^i\mathcal{H} - 4B_1^bC_{1b}{}^i\mathcal{H} - 4C_1^{ib}B'_{1b} + B_2^{i\prime} - 2B_1^i\phi'_1$$
$$- 4C_1^{ib}\partial_b\,\phi_1 + \partial^i\phi_2) + \mathcal{O}(\epsilon^3),$$

$$\Gamma^0_{ij} = g_{ij}\mathcal{H} + \epsilon(2C_{1ij}\mathcal{H} - 2g_{ij}\mathcal{H}\phi_1 + C'_{1ij} - \frac{1}{2}\partial_iB_{1j} - \frac{1}{2}\partial_jB_{1i})$$
$$+ \epsilon^2(C_{2ij}\mathcal{H} - B_{1b}B_1^bg_{ij}\mathcal{H} - 4C_{1ij}\mathcal{H}\phi_1 + 4g_{ij}\mathcal{H}\phi_1^2 - g_{ij}\mathcal{H}\phi_2 - 2\phi_1C'_{1ij} + \frac{1}{2}C'_{2ij}$$
$$- B_1^b\partial_bC_{1ij} + \phi_1\partial_iB_{1j} - \frac{1}{4}\partial_iB_{2j} + B_1^b\partial_iC_{1jb} + \phi_1\partial_jB_{1i} - \frac{1}{4}\partial_jB_{2i} + B_1^b\partial_jC_{1ib}) + \mathcal{O}(\epsilon^3),$$

$$\Gamma^i_{0j} = \delta^i{}_j\mathcal{H} + \epsilon(C_{1j}{}^{i\prime} - \frac{1}{2}\partial^iB_{1j} + \frac{1}{2}\partial_jB_1^i) + \frac{1}{4}\epsilon^2(-4B_1^iB_{1j}\mathcal{H} - 8C_1^{ib}C'_{1jb} + 2C_{2j}{}^{i\prime}$$
$$+ 4C_1^{ib}\partial_bB_{1j} - \partial^iB_{2j} - 4C_1^{ib}\partial_jB_{1b} + \partial_jB_2^i - 4B_1^i\partial_j\phi_1) + \mathcal{O}(\epsilon^3),$$

$$\Gamma^i_{jk} = \epsilon(-B_1^ig_{jk}\mathcal{H} - \partial^iC_{1jk} + \partial_jC_{1k}{}^i + \partial_kC_{1j}{}^i) + \frac{1}{2}\epsilon^2(-4B_1^iC_{1jk}\mathcal{H} - B_2^ig_{jk}\mathcal{H}$$
$$+ 4B_1^bC_{1b}{}^ig_{jk}\mathcal{H} + 4B_1^ig_{jk}\mathcal{H}\phi_1 - 2B_1^iC'_{1jk} + 4C_1^{ib}\partial_bC_{1jk} - \partial^iC_{2jk} + B_1^i\partial_jB_{1k}$$
$$- 4C_1^{ib}\partial_jC_{1kb} + \partial_jC_{2k}{}^i + B_1^i\partial_kB_{1j} - 4C_1^{ib}\partial_kC_{1jb} + \partial_kC_{2j}{}^i) + \mathcal{O}(\epsilon^3). \tag{E.14}$$

The bookkeeping parameter $\epsilon$ specifies each order in perturbation theory. Furthermore, the mixed components $\{i0j0\}$ of the Riemann tensor, defined as

$$R^\alpha{}_{\beta\mu\nu} = \Gamma^\alpha_{\beta\nu,\mu} - \Gamma^\alpha_{\beta\mu,\nu} + \Gamma^\alpha_{\lambda\mu}\Gamma^\lambda_{\beta\nu} - \Gamma^\alpha_{\lambda\nu}\Gamma^\lambda_{\beta\mu}, \tag{E.15}$$

[1]We acknowledge the usage of the Mathematica package xAct [850] for the symbolic tensor computations.

entering in the detector response (6.5.7) are given, up to second-order, by

$$
\begin{aligned}
R_{i0j0} = {}& -a^2 g_{ij}\partial_\tau\mathcal{H} + \frac{1}{4}\epsilon a^2\left[-4\mathcal{H}C'_{1ij} - 8C_{1ij}\mathcal{H}' + 4g_{ij}\mathcal{H}\phi'_1 + 2\partial_i B'_{1j} + 2\partial_j B'_{1i} - 4C''_{1ij}\right.\\
&\left. + 2\mathcal{H}\partial_i B_{1j} + 2\mathcal{H}\partial_j B_{1i} + 4\partial_j\partial_i\phi_1\right] + \frac{1}{4}\epsilon^2 a^2\left[4B_{1b}B^b_1 g_{ij}\mathcal{H}^2 + 4B^b_1 g_{ij}\mathcal{H}B'_{1b} + 4C_{1i}{}^{b'}C'_{1jb}\right.\\
&- 2\mathcal{H}C'_{2ij} - 4C_{2ij}\mathcal{H}' + 8C_{1ij}\mathcal{H}\phi'_1 - 8g_{ij}\mathcal{H}\phi_1\phi'_1 + 4C'_{1ij}\phi'_1 + 2g_{ij}\mathcal{H}\phi'_2 + \partial_i B'_{2j} + \partial_j B'_{2i}\\
&- 2C''_{2ij} + 4B^{b'}_1\mathcal{H}\partial_b C_{1ij} + 4B^{b'}_1\partial_b C_{1ij} + 4B^b_1 g_{ij}\mathcal{H}\partial_b\phi_1 - 2C'_{1jb}\partial^b B_{1i} + \partial_b B_{1j}\partial^b B_{1i}\\
&- 2C'_{1ib}\partial^b B_{1j} + 4\partial_b C_{1ij}\partial^b\phi_1 + 2C'_{1jb}\partial_i B^b_1 - \partial_b B_{1j}\partial_i B^b_1 - 2\phi'_1\partial_i B_{1j} + \mathcal{H}\partial_i B_{2j}\\
&- 4B^{b'}_1\mathcal{H}\partial_i C_{1jb} - 4B^{b'}_1\partial_i C_{1jb} - 4\partial^b\phi_1\partial_i C_{1jb} - 4B_{1j}\mathcal{H}\partial_i\phi_1 - \partial^b B_{1j}\partial_j B_{1b} + \partial_i B^b_1\partial_j B_{1b}\\
&+ 2C'_{1ib}\partial_j B^b_1 - 2\phi'_1\partial_j B_{1i} + \mathcal{H}\partial_j B_{2i} - 4B^{b'}_1\mathcal{H}\partial_j C_{1ib} - 4B^{b'}_1\partial_j C_{1ib} - 4\partial^b\phi_1\partial_j C_{1ib}\\
&\left. - 4\partial_i\phi_1\partial_j\phi_1 - 4B_{1i}\mathcal{H}(B_{1j}\mathcal{H} + \partial_j\phi_1) + 2\partial_j\partial_i\phi_2\right].
\end{aligned}
\tag{E.16}
$$

### E.1.2  Gauge transformations up to second-order

One can perform a second-order gauge transformation of the form $x^\mu \to \widetilde{x}^\mu = x^\mu + \xi^\mu$ with

$$
\xi^\mu = \left(\alpha_1 + \alpha_2, \xi^i_1 + \xi^i_2\right)\qquad\text{and}\qquad \xi^i_a = \beta^i_a + \gamma^i_a,
\tag{E.17}
$$

where $a = \{1, 2\}$ and where we defined a divergence-free vectorial parameter such that $\gamma^i_{,i} = 0$. The first-order gauge transformations are given by (see Ref. [699])

$$
\begin{aligned}
\widetilde{\phi}_1 &= \phi_1 + \mathcal{H}\alpha_1 + \alpha'_1\,, & \widetilde{\psi}_1 &= \psi_1 - \mathcal{H}\alpha_1\,, \tag{E.18a}\\
\widetilde{B}_1 &= B_1 - \alpha_1 + \beta'_1\,, & \widetilde{E}_1 &= E_1 + \beta_1\,, \tag{E.18b}\\
\widetilde{S_1}^i &= S_1^i - \gamma_1{}^{i'}\,, & \widetilde{F_1}^i &= F_1{}^i + \gamma_1{}^i\,, \tag{E.18c}\\
\widetilde{h}_{1ij} &= h_{1ij}\,. & & \tag{E.18d}
\end{aligned}
$$

These relations make explicit the gauge independence of the first-order tensor perturbations of the metric. At second-order, the gauge transformations can be written by defining the vector $\mathcal{X}_{\mathrm{B}i}$ and tensor $\mathcal{X}_{ij}$ (dependent only on first-order quantities squared) as

$$
\begin{aligned}
\mathcal{X}_{\mathrm{B}i} \equiv {}& 2\Big[\left(2\mathcal{H}B_{1i} + B'_{1i}\right)\alpha_1 + B_{1i,k}\xi^k_1 - 2\phi_1\alpha_{1,i} + B_{1k}\xi^k_{1,\,i} + B_{1i}\alpha'_1 + 2C_{1ik}\xi^{k'}_1\Big]\\
&+ 4\mathcal{H}\alpha_1\left(\xi'_{1i} - \alpha_{1,i}\right) + \alpha'_1\left(\xi'_{1i} - 3\alpha_{1,i}\right) + \alpha_1\left(\xi''_{1i} - \alpha'_{1,i}\right)\\
&+ \xi^{k'}_1\left(\xi_{1i,k} + 2\xi_{1k,i}\right) + \xi^k_1\left(\xi'_{1i,k} - \alpha_{1,ik}\right) - \alpha_{1,k}\xi^k_{1,\,i}\,,
\end{aligned}
\tag{E.19}
$$

and

$$
\begin{aligned}
\mathcal{X}_{ij} \equiv {}& 2\Big[\left(\mathcal{H}^2 + \frac{a''}{a}\right)\alpha^2_1 + \mathcal{H}\left(\alpha_1\alpha'_1 + \alpha_{1,k}\xi^k_1\right)\Big]\delta_{ij}\\
&+ 4\Big[\alpha_1\left(C'_{1ij} + 2\mathcal{H}C_{1ij}\right) + C_{1ij,k}\xi^k_1 + C_{1ik}\xi^k_1{}_{,j} + C_{1kj}\xi^k_1{}_{,i}\Big] + 2\left(B_{1i}\alpha_{1,j} + B_{1j}\alpha_{1,i}\right)\\
&+ 4\mathcal{H}\alpha_1\left(\xi_{1i,j} + \xi_{1j,i}\right) - 2\alpha_{1,i}\alpha_{1,j} + 2\xi_{1k,i}\xi^k_1{}_{,j} + \alpha_1\left(\xi'_{1i,j} + \xi'_{1j,i}\right) + \left(\xi_{1i,jk} + \xi_{1j,ik}\right)\xi^k_1\\
&+ \xi_{1i,k}\xi^k_1{}_{,j} + \xi_{1j,k}\xi^k_1{}_{,i} + \xi'_{1i}\alpha_{1,j} + \xi'_{1j}\alpha_{1,i}\,,
\end{aligned}
\tag{E.20}
$$

such that one finds

$$\widetilde{\phi}_2 = \phi_2 + \mathcal{H}\alpha_2 + \alpha_2' + \alpha_1 \left[ \alpha_1'' + 5\mathcal{H}\alpha_1' + \left( \mathcal{H}' + 2\mathcal{H}^2 \right) \alpha_1 + 4\mathcal{H}\phi_1 + 2\phi_1' \right]$$
$$+ 2\alpha_1' \left( \alpha_1' + 2\phi_1 \right) + \xi_{1k} \left( \alpha_1' + \mathcal{H}\alpha_1 + 2\phi_1 \right)^{,k} + \xi_{1k}' \left[ \alpha_1^{,k} - 2B_{1k} - \xi_1^{k'} \right], \tag{E.21a}$$

$$\widetilde{\psi}_2 = \psi_2 - \mathcal{H}\alpha_2 - \frac{1}{4}\mathcal{X}_k^k + \frac{1}{4}\nabla^{-2}\mathcal{X}^{ij}_{,ij}, \tag{E.21b}$$

$$\widetilde{B}_2 = B_2 - \alpha_2 + \beta_2' + \nabla^{-2}\mathcal{X}_{\mathrm{B}}{}^k_{,k}, \tag{E.21c}$$

$$\widetilde{E}_2 = E_2 + \beta_2 + \frac{3}{4}\nabla^{-2}\nabla^{-2}\mathcal{X}^{ij}_{,ij} - \frac{1}{4}\nabla^{-2}\mathcal{X}_k^k, \tag{E.21d}$$

$$\widetilde{S}_{2i} = S_{2i} - \gamma_{2i}' - \mathcal{X}_{\mathrm{B}i} + \nabla^{-2}\mathcal{X}_{\mathrm{B}}{}^k_{,ki}, \tag{E.21e}$$

$$\widetilde{F}_{2i} = F_{2i} + \gamma_{2i} + \nabla^{-2}\mathcal{X}_{ik,}{}^k - \nabla^{-2}\nabla^{-2}\mathcal{X}^{kl}_{,kli}, \tag{E.21f}$$

$$\widetilde{h}_{2ij} = h_{2ij} + \mathcal{X}_{ij} + \frac{1}{2}\left( \nabla^{-2}\mathcal{X}^{kl}_{,kl} - \mathcal{X}_k^k \right)\delta_{ij} + \frac{1}{2}\nabla^{-2}\nabla^{-2}\mathcal{X}^{kl}_{,klij} + \frac{1}{2}\nabla^{-2}\mathcal{X}^k_{,kij}$$
$$- \nabla^{-2}\left( \mathcal{X}_{ik,}{}^k{}_j + \mathcal{X}_{jk,}{}^k{}_i \right). \tag{E.21g}$$

### E.1.3   Explicit expression for $\mathcal{X}_{ij}^{\mathrm{P}}$

Here, for completeness, we report the explicit field combinations defining $\mathcal{X}_{ij}^{\mathrm{P}}$ as

$$\mathcal{X}_{ij}^{\mathrm{P}} = 4h_{1ij}\mathcal{H}(B_1 - E_1') + 2B_1 h_{1ij}' - 2E_1' h_{1ij}' + 2B_1 F_{1j,i}' - 2E_1' F_{1j,i}' + 2B_1 F_{1i,j}' - 2E_1' F_{1i,j}'$$
$$+ 4B_1 E_{1,ij}' - 2E_1' E_{1,ij}' - 2S1_j B_{1,i} + \mathcal{S}_{1j}' B_{1,i} - E_{1,j}' B_{1,i} + 4B_1 \mathcal{H} F_{1j,i} - 4\mathcal{H} E_1' F_{1j,i}$$
$$+ 4B_1 \mathcal{H} \mathcal{S}_{1j,i} - 4\psi_1 \mathcal{S}_{1j,i} - 4\mathcal{H} E_1' \mathcal{S}_{1j,i} + 2h_{1jk}\mathcal{S}_{1,i}^k + 2S1_j E_{1,i}' - \mathcal{S}_{1j}' E_{1,i}' + B_1 \mathcal{S}_{1j,i}'$$
$$- E_1' \mathcal{S}_{1j,i}' - 2B_1 E_{1,ij}' - 2S_{1i} B_{1,j} + \mathcal{S}_{1i}' B_{1,j} - E_{1,i}' B_{1,j} + 2B_{1,i} B_{1,j} + 4B_1 \mathcal{H} F_{1i,j}$$
$$- 4\mathcal{H} E_1' F_{1i,j} + 2\mathcal{S}_{1k,i} F_{1,j}^k + 4B_1 \mathcal{H} \mathcal{S}_{1i,j} - 4\psi_1 \mathcal{S}_{1i,j} - 4\mathcal{H} E_1' \mathcal{S}_{1i,j} + 2F_{1,i}^k \mathcal{S}_{1k,j}$$
$$+ 2\mathcal{S}_{1,i}^k \mathcal{S}_{1k,j} + 2h_{1ik}\mathcal{S}_{1,j}^k + 2S_{1i} E_{1,j}' - \mathcal{S}_{1i}' E_{1,j}' + B_1 \mathcal{S}_{1i,j}' - E_1' \mathcal{S}_{1i,j}' + 8\psi_1 E_{1,ij}$$
$$+ 2\mathcal{S}_1^k h_{1ij,k} - 2F_{1,j}^k E_{1,ik} + \mathcal{S}_{1,j}^k E_{1,ik} + 2\mathcal{S}_1^k F_{1ji,ik} + \mathcal{S}_1^k \mathcal{S}_{1ji,ik} - 2F_{1,i}^k E_{1,jk} + \mathcal{S}_{1,i}^k E_{1,jk}$$
$$+ 2\mathcal{S}_1^k F_{1i,jk} + \mathcal{S}_1^k \mathcal{S}_{1i,jk} + 2\mathcal{S}_1^k E_{1,ijk} - 2h_{1ij,k} E_{1,k} - 2F_{1j,ik} E_{1,k} - \mathcal{S}_{1ji,k} E_{1,k}$$
$$- 2F_{1i,jk} E_{1,k} - \mathcal{S}_{1ij,k} E_{1,k} - 2E_{1,ijk} E_{1,k} + 2g_{ij} \left[ 2B_1^2 \mathcal{H}^2 - 4B_1 \mathcal{H} \psi_1 + \mathcal{H} B_1' (B_1 - E_1') \right.$$
$$- 4B_1 \mathcal{H}^2 E_1' + 4\mathcal{H} \psi_1 E_1' + B_1^2 \mathcal{H}' - 2B_1 E_1' \mathcal{H}' + (E_1')^2 (2\mathcal{H}^2 + \mathcal{H}') - 2B_1 \psi_1' + 2E_1' \psi_1'$$
$$- B_1 \mathcal{H} E_1'' + \mathcal{H} E_1' E_1'' + \mathcal{S}_1^k \mathcal{H} B_{1,k} - 2\mathcal{S}_1^k \psi_{1,k} - \mathcal{S}_1^k \mathcal{H} E_{1,k}' - \mathcal{H} E_{1,k} B_{1,k} + 2\psi_{1,k} E_{1,k}$$
$$\left. + \mathcal{H} E_{1,k}' E_{1,k} \right] + 2\mathcal{S}_{1k,j} F_{1i,k} - 2E_{1,jk} F_{1i,k} + 2\mathcal{S}_{1k,i} F_{1j,k} - 2E_{1,ik} F_{1j,k} + \mathcal{S}_{1k,j} \mathcal{S}_{1i,k}$$
$$- E_{1,jk} \mathcal{S}_{1i,k} + \mathcal{S}_{1k,i} \mathcal{S}_{1j,k} - E_{1,ik} \mathcal{S}_{1j,k} - 2h_{1jk} E_{1,ik} - 4E_{1,jk} E_{1,ik} - 2h_{1ik} E_{1,jk}, \tag{E.22}$$

where we also introduced $\mathcal{S}_{1i} \equiv \int^{\eta} S_{1i}\mathrm{d}\eta' + \hat{\mathcal{C}}_{1i}(\boldsymbol{x})$.

## E.2   First-order dynamics in the Poisson gauge

In this section, we present the equations of motion for the perturbations in the Poisson gauge at first-order. The Poisson gauge is defined by setting

$$\widetilde{E}^{\mathrm{P}} = 0, \qquad \widetilde{B}^{\mathrm{P}} = 0 \qquad \text{and} \qquad \widetilde{S}_i^{\mathrm{P}} = 0. \tag{E.23}$$

Therefore, using the gauge transformation properties in Sec. E.1.2, at first-order the gauge is fixed by choosing

$$\alpha_1^{\mathrm{P}} = B_1 - E_1', \qquad \beta_1^{\mathrm{P}} = -E_1, \qquad \gamma_{1i}^{\mathrm{P}} = \int^{\eta} S_{1i}\mathrm{d}\eta' + \hat{\mathcal{C}}_{1i}(\boldsymbol{x}), \tag{E.24}$$

up to an arbitrary constant 3-vector $\hat{\mathcal{C}}_{1i}$ which depends on the choice of spatial coordinates on an initial hypersurface.

The equations of motion for the scalar, vector and tensor perturbations decouple at first order. Therefore we can study them separately.

**Scalar perturbations**

The Einstein equations at first-order provide two coupled equations for the scalar perturbations of the metric, given by [699]

$$\psi_1'' + 2\mathcal{H}\psi_1' + \mathcal{H}\phi_1' + \left(2\mathcal{H}' + \mathcal{H}^2\right)\phi_1 = 4\pi G a^2 \left(\delta P + \frac{2}{3}\nabla^2 \Pi_1\right), \tag{E.25}$$

$$\sigma_1' + 2\mathcal{H}\sigma_1 + \psi_1 - \phi_1 = 8\pi G a^2 \Pi_1, \tag{E.26}$$

where $\Pi_1$ is the scalar part of the (trace-free) first-order anisotropic stress. The second equation (E.26) can be written in terms of the gauge invariant Bardeen's potentials (see Eqs. (6.5.29) and (6.5.30) in the main text) as

$$\Psi_1 - \Phi_1 = 8\pi G a^2 \Pi_1. \tag{E.27}$$

Thus, in the absence of anisotropic stress, one finds $\Psi_1 = \Phi_1$.

When dealing with adiabatic perturbations one has $\delta P = c_s^2 \delta \rho$ and the remaining equation can be written as [699]

$$\Psi_1'' + 3(1 + c_s^2)\mathcal{H}\Psi_1' + [2\mathcal{H}' + (1 + 3c_s^2)\mathcal{H}^2 - c_s^2\nabla^2]\Psi_1 = 0, \tag{E.28}$$

which is solved, in a radiation-dominated universe, by

$$\Psi_1(\eta, k) \equiv \frac{2}{3}\zeta(k)T(k\eta) \tag{E.29}$$

where the transfer function can be written in terms of the first spherical Bessel function as

$$T(k\eta) \equiv 3\frac{j_1(k\eta/\sqrt{3})}{k\eta/\sqrt{3}} = 3\frac{\sin\left(k\eta/\sqrt{3}\right) - \left(k\eta/\sqrt{3}\right)\cos\left(k\eta/\sqrt{3}\right)}{\left(k\eta/\sqrt{3}\right)^3} \tag{E.30}$$

and $\zeta$ is the comoving curvature perturbation. Also, the equation of motion for the velocity potential $v$ in the Poisson gauge is written as

$$\Psi_1' + \mathcal{H}\Psi_1 = -4\pi G a^2 (\rho + P)v. \tag{E.31}$$

**Vector perturbations**

The gauge invariant vector metric perturbation is directly related to the divergence-free part of the momentum via the constraint equation

$$\nabla^2 \left(F_{1i}' + S_{1i}\right) = -16\pi G a^2 \delta q_i, \tag{E.32}$$

while the momentum conservation equation yields

$$\delta q_i' + 4\mathcal{H}\delta q_i = -\nabla^2 \Pi_{1i}. \tag{E.33}$$

Therefore, in the absence of anisotropic stress sourcing $\delta q_i$, the vector perturbations are rapidly redshifted with the Hubble expansion as

$$\delta q_i(\eta) = \delta q_i(\eta_{\text{in}})/a^4(\eta). \tag{E.34}$$

In particular, in both Poisson and TT gauge where one sets $S_{1i} = 0$, one reads directly from (E.32) that the vector perturbations are redshifted away in an expanding universe.

**Tensor perturbations**

From the linearised Einstein's equations, one finds

$$h_{1ij}'' + 2\mathcal{H}h_{1ij}' - \nabla^2 h_{1ij} = 8\pi G a^2 \Pi_{1ij}, \tag{E.35}$$

which becomes the standard homogeneous evolution equation for the linear tensor modes in the absence of the anisotropic stress.

# Acknowledgements

There is no other way to begin this list of acknowledgements than by expressing my deepest gratitude to *Toni*. You have been more than an excellent guide throughout this journey, as you showed to me, leading by example, what passion and dedication for physics really are. I learned from you more than I could have ever hoped to grasp in just four years. Among other things, you thought me the most important (unspoken) lesson of all: when driven by curiosity, one would leave no stone unturned. You constantly believed in me and pushed me to give my own best by always standing on my side, as friends do. I will always be indebted to you for all this.

I would like to thank *Alex*. I learned a lot by collaborating with you since the very beginning of my stay in Geneva and, most importantly, you always showed a relentless good spirit. Another kind word is for all the people I had the privilege to work with during the last four years: *Ameek, Angelo, Azadeh, Caner, Daniele, Dionysios, Emanuele, Enrico, Gian, Gianmassimo, Giulia, Ilia, Kaze, Ken, Konstantinos, Lam, Luca, Marco, Maresuke, Matteo, Nicola, Paolo, Riccardo, Sabino, Salvatore, Simone, Swetha, Vincent* and *Vishal*. There is not enough space here for me to thank you one by one. Rest assured, however, that I look forward to new exciting collaborations together.

A word of gratitude goes to *Bernard, Chris* and *Michele*. It is an honour for me to have you as members of the jury. I would like to thank all the professors, post-docs and students in the cosmology and astroparticle physics group at the University of Geneva. You have contributed to creating such a lively and challenging research environment, making my stay in Geneva exciting and extremely fruitful. A particular thank goes to *Adrian, Dmitry, Manuel* and *Tobias*, for sharing with me the experience of teaching and to *Charles* for constantly bearing such noisy Italian office mates and for helping me in revising the "résumé" of this thesis.

Turning slightly towards the personal sphere, I would like to express my gratitude to *Davide (R.)*. I clearly remember the moment you knocked at the door of my office to welcome me the first day I arrived. You have been a guide and a friend, even after you left Geneva. Another person who has been standing at my side is *Davide (L.)*. We first met during that GGI school in Florence and since then we have always had endless intriguing conversations and so much fun. Finally, a word for my partner in crime in all these years. *Valerio*, we have shared the apartment and the office, the excitements and the defeats. I thank you for all your restless cheerfulness and for being a very good friend.

I was lucky to meet, beyond the walls of my department, some wonderful friends, who have made me feel at home in Geneva, even during the hard time we faced since that infamous year 2020. *Anna, Chris, Clement, Francesco (I.), Francesco (L.), Julia, Kevin, Laura, Lucrezia, Marc* and *Solène*, you have been a fantastic group of friends. Also, it was incredible luck to unexpectedly find nearby an old friend from high school. *Diego*, I was fortunate to share this experience abroad together with you and new friends met along the way: *Lorenzo (C.), Lorenzo (R.)* and *Annalucia*.

Some last words are for those who have always been close to my heart. My *old friends*, among which I can identify my strongest supporters and who have always stood on my side. There is no need to state their name here, they know exactly who they are. *Corrado* and *Stefania*, for your ever encouraging words and the unconditioned esteem you showed for me. My absolute gratitude goes to my parents, *Bruno* and *Marta*. You have made all this possible by believing in my capabilities since I was only a child and by fostering my passions. A thank also goes to my brother *Matteo*, undoubtedly among my fiercest competitors but absolutely the person I look up to, along with his wife *Cristal*.

There is, lastly, one special person who more than anybody else has shared with me all the emotions of this journey, encouraging me even in the darkest hours. *Giulia*, my love for you remains one of the deepest mysteries I have ever encountered and I intend to contemplate it for the rest of my life.

# Bibliography

[1] V. De Luca, G. Franciolini, P. Pani, and A. Riotto, *The Minimum Testable Abundance of Primordial Black Holes at Future Gravitational-Wave Detectors*, (June 2021), arXiv: 2106.13769 [astro-ph.CO] (p. v).

[2] M. Biagetti, V. De Luca, G. Franciolini, A. Kehagias, and A. Riotto, *The Formation Probability of Primordial Black Holes*, (May 2021), arXiv: 2105.07810 [astro-ph.CO] (p. v).

[3] G. Franciolini, V. Baibhav, V. De Luca, K. K. Y. Ng, K. W. K. Wong, E. Berti, et al., *Quantifying the evidence for primordial black holes in LIGO/Virgo gravitational-wave data*, (May 2021), arXiv: 2105.03349 [gr-qc] (pp. v, 8, 140).

[4] V. De Luca, G. Franciolini, and A. Riotto, *Constraining the Initial Primordial Black Hole Clustering with CMB-distortion*, (Mar. 2021), arXiv: 2103.16369 [astro-ph.CO] (pp. v, 69, 70).

[5] V. De Luca, G. Franciolini, P. Pani, and A. Riotto, *Bayesian Evidence for Both Astrophysical and Primordial Black Holes: Mapping the GWTC-2 Catalog to Third-Generation Detectors*, JCAP 05 (2021), p. 003, arXiv: 2102.03809 [astro-ph.CO] (pp. v, 117, 140, 154, 155).

[6] I. Musco, V. De Luca, G. Franciolini, and A. Riotto, *The Threshold for Primordial Black Hole Formation: a Simple Analytic Prescription*, (Nov. 2020), arXiv: 2011.03014 [astro-ph.CO] (pp. v, 19, 20, 21, 22, 26, 27).

[7] K. W. K. Wong, G. Franciolini, V. De Luca, V. Baibhav, E. Berti, P. Pani, and A. Riotto, *Constraining the primordial black hole scenario with Bayesian inference and machine learning: the GWTC-2 gravitational wave catalog*, Phys. Rev. D 103.2 (2021), p. 023026, arXiv: 2011.01865 [gr-qc] (pp. v, 12, 125, 126).

[8] V. De Luca, G. Franciolini, and A. Riotto, *NANOGrav Data Hints at Primordial Black Holes as Dark Matter*, Phys. Rev. Lett. 126.4 (2021), p. 041303, arXiv: 2009.08268 [astro-ph.CO] (pp. v, 190).

[9] V. De Luca, V. Desjacques, G. Franciolini, and A. Riotto, *The clustering evolution of primordial black holes*, JCAP 11 (2020), p. 028, arXiv: 2009.04731 [astro-ph.CO] (pp. v, 97, 105, 114, 115).

[10] S. Bhagwat, V. De Luca, G. Franciolini, P. Pani, and A. Riotto, *The importance of priors on LIGO-Virgo parameter estimation: the case of primordial black holes*, JCAP 01 (2021), p. 037, arXiv: 2008.12320 [astro-ph.CO] (pp. v, 124, 134, 149, 206, 209).

[11] V. De Luca, V. Desjacques, G. Franciolini, P. Pani, and A. Riotto, *GW190521 Mass Gap Event and the Primordial Black Hole Scenario*, Phys. Rev. Lett. 126.5 (2021), p. 051101, arXiv: 2009.01728 [astro-ph.CO] (pp. v, 131, 133).

[12] V. De Luca, G. Franciolini, P. Pani, and A. Riotto, *Primordial Black Holes Confront LIGO/Virgo data: Current situation*, JCAP 06 (2020), p. 044, arXiv: 2005.05641 [astro-ph.CO] (pp. v, 72, 77, 80, 93, 105, 118, 126, 129, 132, 140, 141, 144, 149, 154).

[13] V. De Luca, G. Franciolini, P. Pani, and A. Riotto, *Constraints on Primordial Black Holes: the Importance of Accretion*, Phys. Rev. D 102.4 (2020), p. 043505, arXiv: 2003.12589 [astro-ph.CO] (pp. v, 11, 13, 85, 86, 133).

[14] V. De Luca, G. Franciolini, P. Pani, and A. Riotto, *The evolution of primordial black holes and their final observable spins*, JCAP 04 (2020), p. 052, arXiv: 2003.02778 [astro-ph.CO] (pp. v, 12, 72, 77, 85, 89, 90, 93, 96, 129, 144, 149).

[15] V. De Luca, G. Franciolini, and A. Riotto, *On the Primordial Black Hole Mass Function for Broad Spectra*, Phys. Lett. B 807 (2020), p. 135550, arXiv: 2001.04371 [astro-ph.CO] (pp. v, 24, 49, 51, 98, 190).

[16] V. De Luca, G. Franciolini, A. Kehagias, and A. Riotto, *On the Gauge Invariance of Cosmological Gravitational Waves*, JCAP 03 (2020), p. 014, arXiv: 1911.09689 [gr-qc] (pp. v, 176, 184).

[17] N. Bartolo, D. Bertacca, V. De Luca, G. Franciolini, S. Matarrese, M. Peloso, et al., *Gravitational wave anisotropies from primordial black holes*, JCAP 02 (2020), p. 028, arXiv: 1909.12619 [astro-ph.CO] (pp. v, 158, 170, 171, 176, 186).

[18] A. Moradinezhad Dizgah, G. Franciolini, and A. Riotto, *Primordial Black Holes from Broad Spectra: Abundance and Clustering*, JCAP 11 (2019), p. 001, arXiv: 1906.08978 [astro-ph.CO] (pp. v, 24, 38, 65, 67).

[19] V. De Luca, G. Franciolini, A. Kehagias, M. Peloso, A. Riotto, and C. Ünal, *The Ineludible non-Gaussianity of the Primordial Black Hole Abundance*, JCAP 07 (2019), p. 048, arXiv: 1904.00970 [astro-ph.CO] (pp. v, 24, 27, 32, 35, 38).

[20] V. De Luca, V. Desjacques, G. Franciolini, A. Malhotra, and A. Riotto, *The initial spin probability distribution of primordial black holes*, JCAP 05 (2019), p. 018, arXiv: 1903.01179 [astro-ph.CO] (pp. vi, 52, 57, 59, 60, 64, 141).

[21] N. Bartolo, V. De Luca, G. Franciolini, M. Peloso, D. Racco, and A. Riotto, *Testing primordial black holes as dark matter with LISA*, Phys. Rev. D 99.10 (2019), p. 103521, arXiv: 1810.12224 [astro-ph.CO] (pp. vi, 157, 158, 162, 163, 164, 167, 169, 185).

[22] N. Bartolo, V. De Luca, G. Franciolini, A. Lewis, M. Peloso, and A. Riotto, *Primordial Black Hole Dark Matter: LISA Serendipity*, Phys. Rev. Lett. 122.21 (2019), p. 211301, arXiv: 1810.12218 [astro-ph.CO] (pp. vi, 157, 158, 159).

[23] M. Biagetti, G. Franciolini, A. Kehagias, and A. Riotto, *Primordial Black Holes from Inflation and Quantum Diffusion*, JCAP 07 (2018), p. 032, arXiv: 1804.07124 [astro-ph.CO] (pp. vi, 28, 44, 46, 190).

[24] G. Franciolini, A. Kehagias, S. Matarrese, and A. Riotto, *Primordial Black Holes from Inflation and non-Gaussianity*, JCAP 03 (2018), p. 016, arXiv: 1801.09415 [astro-ph.CO] (pp. vi, 27, 28, 29, 30, 31, 202).

[25] S. S. Bavera, G. Franciolini, G. Cusin, A. Riotto, M. Zevin, and T. Fragos, *Stochastic gravitational-wave background as a tool to investigate multi-channel astrophysical and primordial black-hole mergers*, (Sept. 2021), arXiv: 2109.05836 [astro-ph.CO] (p. vi).

[26] V. De Luca, G. Franciolini, A. Kehagias, and A. Riotto, *Standard Model Baryon Number Violation Seeded by Black Holes*, Phys. Lett. B 819 (2021), p. 136454, arXiv: 2102.07408 [astro-ph.CO] (p. vi).

[27] K. Kritos, V. De Luca, G. Franciolini, A. Kehagias, and A. Riotto, *The Astro-Primordial Black Hole Merger Rates: a Reappraisal*, (Dec. 2020), arXiv: 2012.03585 [gr-qc] (p. vi).

[28] V. De Luca, G. Franciolini, A. Kehagias, A. Riotto, and M. Shiraishi, *Constraining graviton non-Gaussianity through the CMB bispectra*, Phys. Rev. D 100.6 (2019), p. 063535, arXiv: 1908.00366 [astro-ph.CO] (p. vi).

[29] V. De Luca, V. Desjacques, G. Franciolini, and A. Riotto, *Gravitational Waves from Peaks*, JCAP 09 (2019), p. 059, arXiv: 1905.13459 [astro-ph.CO] (pp. vi, 157).

[30] D. Anninos, V. De Luca, G. Franciolini, A. Kehagias, and A. Riotto, *Cosmological Shapes of Higher-Spin Gravity*, JCAP 04 (2019), p. 045, arXiv: 1902.01251 [hep-th] (p. vi).

[31] G. Franciolini, G. F. Giudice, D. Racco, and A. Riotto, *Implications of the detection of primordial gravitational waves for the Standard Model*, JCAP 05 (2019), p. 022, arXiv: 1811.08118 [hep-ph] (p. vi).

[32] G. Franciolini, L. Hui, R. Penco, L. Santoni, and E. Trincherini, *Stable wormholes in scalar-tensor theories*, JHEP 01 (2019), p. 221, arXiv: 1811.05481 [hep-th] (p. vi).

[33] G. Franciolini, L. Hui, R. Penco, L. Santoni, and E. Trincherini, *Effective Field Theory of Black Hole Quasinormal Modes in Scalar-Tensor Theories*, JHEP 02 (2019), p. 127, arXiv: 1810.07706 [hep-th] (p. vi).

[34] A. Moradinezhad Dizgah, G. Franciolini, A. Kehagias, and A. Riotto, *Constraints on long-lived, higher-spin particles from galaxy bispectrum*, Phys. Rev. D 98.6 (2018), p. 063520, arXiv: 1805.10247 [astro-ph.CO] (pp. vi, 43).

[35] G. Franciolini, A. Kehagias, A. Riotto, and M. Shiraishi, *Detecting higher spin fields through statistical anisotropy in the CMB bispectrum*, Phys. Rev. D 98.4 (2018), p. 043533, arXiv: 1803.03814 [astro-ph.CO] (p. vi).

[36] G. Franciolini, A. Kehagias, and A. Riotto, *Imprints of Spinning Particles on Primordial Cosmological Perturbations*, JCAP 02 (2018), p. 023, arXiv: 1712.06626 [hep-th] (p. vi).

[37] E. Barausse et al., *Prospects for Fundamental Physics with LISA*, Gen. Rel. Grav. 52.8 (2020), p. 81, arXiv: 2001.09793 [gr-qc] (pp. vi, 185).

[38] A. Kashlinsky et al., *Electromagnetic probes of primordial black holes as dark matter*, (Mar. 2019), arXiv: 1903.04424 [astro-ph.CO] (p. vi).

[39] G. Galilei, *Dialogue Concerning the Two Chief World Systems, Ptolemaic and Copernican*. Translated by S. Drake. Foreword by Albert Einstein. University of California Press, 1967. (Original 1632.) ISBN: 9780375757662, URL: https://books.google.ch/books?id=c-nIrKjBqOwC (p. vi).

[40] R. Sanders, *The Dark Matter Problem: A Historical Perspective*. Cambridge University Press, 2010. ISBN: 9781139485739, URL: https://books.google.ch/books?id=RpuAoqSOWQIC (p. 2).

[41] G. Bertone and D. Hooper, *History of dark matter*, Rev. Mod. Phys. 90.4 (2018), p. 045002, arXiv: 1605.04909 [astro-ph.CO] (pp. 2, 3).

[42] Kelvin, B., *Baltimore lectures on molecular dynamics and the wave theory of light.* 1904 (p. 2).

[43] J. H. Oort, *The force exerted by the stellar system in the direction perpendicular to the galactic plane and some related problems, Bull. Astron. Inst. Netherlands* 6 (Aug. 1932), p. 249 (p. 2).

[44] H. W. Babcock, *The rotation of the Andromeda Nebula, Lick Observatory Bulletin* 498 (Jan. 1939), pp. 41–51 (p. 2).

[45] V. C. Rubin and J. Ford W. Kent, *Rotation of the Andromeda Nebula from a Spectroscopic Survey of Emission Regions, "Astrophys. J."* 159 (Feb. 1970), p. 379 (p. 3).

[46] M. S. Roberts and R. N. Whitehurst, *The rotation curve and geometry of M31 at large galactocentric distances. "Astrophys. J."* 201 (Oct. 1975), pp. 327–346 (p. 3).

[47] C. Carignan, L. Chemin, W. K. Huchtmeier, and F. J. Lockman, *Extended hi rotation curve and mass distribution of m31, Astrophys. J. Lett.* 641 (2006), pp. L109–L112, arXiv: astro-ph/0603143 (p. 3).

[48] K. C. Freeman, *On the disks of spiral and SO Galaxies, Astrophys. J.* 160 (1970), p. 811 (p. 3).

[49] D. H. Rogstad and G. S. Shostak, *Gross Properties of Five Scd Galaxies as Determined from 21-centimeter Observations, "Astrophys. J."* 176 (Sept. 1972), p. 315 (p. 3).

[50] J. Einasto, A. Kaasik, and E. Saar, *Dynamic evidence on massive coronas of galaxies, "Nature"* 250.5464 (July 1974), pp. 309–310 (p. 3).

[51] J. P. Ostriker, P. J. E. Peebles, and A. Yahil, *The Size and Mass of Galaxies, and the Mass of the Universe, "Astrophys. J. Lett."* 193 (Oct. 1974), p. L1 (p. 3).

[52] F. Zwicky, *Die Rotverschiebung von extragalaktischen Nebeln, Helv. Phys. Acta* 6 (1933), pp. 110–127 (p. 3).

[53] M. Markevitch, A. H. Gonzalez, L. David, A. Vikhlinin, S. Murray, W. Forman, et al., *A Textbook example of a bow shock in the merging galaxy cluster 1E0657-56, Astrophys. J. Lett.* 567 (2002), p. L27, arXiv: astro-ph/0110468 (p. 3).

[54] D. Clowe, A. Gonzalez, and M. Markevitch, *Weak lensing mass reconstruction of the interacting cluster 1E0657-558: Direct evidence for the existence of dark matter, Astrophys. J.* 604 (2004), pp. 596–603, arXiv: astro-ph/0312273 (p. 3).

[55] D. Clowe, M. Bradac, A. H. Gonzalez, M. Markevitch, S. W. Randall, C. Jones, and D. Zaritsky, *A direct empirical proof of the existence of dark matter, Astrophys. J. Lett.* 648 (2006), pp. L109–L113, arXiv: astro-ph/0608407 (pp. 3, 4).

[56] M. Bradac, S. W. Allen, T. Treu, H. Ebeling, R. Massey, R. G. Morris, et al., *Revealing the properties of dark matter in the merging cluster MACSJ0025.4-1222, Astrophys. J.* 687 (2008), p. 959, arXiv: 0806.2320 [astro-ph] (p. 4).

[57] D. Harvey, R. Massey, T. Kitching, A. Taylor, and E. Tittley, *The non-gravitational interactions of dark matter in colliding galaxy clusters, Science* 347 (2015), pp. 1462–1465, arXiv: 1503.07675 [astro-ph.CO] (p. 4).

[58] M. Colless et al., *The 2dF Galaxy Redshift Survey: Spectra and redshifts, Mon. Not. Roy. Astron. Soc.* 328 (2001), p. 1039, arXiv: astro-ph/0106498 (pp. 4, 5).

[59] M. Tegmark et al., *The 3-D power spectrum of galaxies from the SDSS, Astrophys. J.* 606 (2004), pp. 702–740, arXiv: astro-ph/0310725 (pp. 4, 5).

[60] V. Springel et al., *Simulating the joint evolution of quasars, galaxies and their large-scale distribution, Nature* 435 (2005), pp. 629–636, arXiv: astro-ph/0504097 (pp. 4, 5).

[61] https://wwwmpa.mpa-garching.mpg.de/galform/millennium/. 2005 (p. 5).

[62] S. Dodelson, *Modern cosmology.* 2003 (pp. 4, 5).

[63] R. K. Sachs and A. M. Wolfe, *Perturbations of a cosmological model and angular variations of the microwave background, Astrophys. J.* 147 (1967), pp. 73–90 (p. 5).

[64] W. T. Hu, *"Wandering in the Background: A CMB Explorer"*. MA thesis. Aug. 1995, arXiv: astro-ph/9508126 (p. 5).

[65] N. Aghanim et al., *Planck 2018 results. VI. Cosmological parameters, Astron. Astrophys.* 641 (2020), A6, arXiv: 1807.06209 [astro-ph.CO] (p. 6).

[66] Y. B. N. Zel'dovich I. D., *The Hypothesis of Cores Retarded during Expansion and the Hot Cosmological Model, Soviet Astron. AJ (Engl. Transl.)*, 10 (1967), p. 602 (p. 6).

[67] S. Hawking, *Gravitationally collapsed objects of very low mass, Mon. Not. Roy. Astron. Soc.* 152 (1971), p. 75 (p. 6).

[68] S. W. Hawking, *Particle Creation by Black Holes, Commun. Math. Phys.* 43 (1975). Ed. by G. W. Gibbons and S. W. Hawking. [Erratum: Commun.Math.Phys. 46, 206 (1976)], pp. 199–220 (p. 6).

[69] S. W. Hawking, *Black hole explosions*, Nature 248 (1974), pp. 30–31 (p. 6).

[70] B. J. Carr and S. W. Hawking, *Black holes in the early Universe*, Mon. Not. Roy. Astron. Soc. 168 (1974), pp. 399–415 (p. 6).

[71] B. J. Carr, *The Primordial black hole mass spectrum*, Astrophys. J. 201 (1975), pp. 1–19 (pp. 6, 8, 19, 20, 51).

[72] G. F. Chapline, *Cosmological effects of primordial black holes*, Nature 253.5489 (1975), pp. 251–252 (p. 6).

[73] P. Meszaros, *Primeval black holes and galaxy formation. "Astronomy and Astrophysics"* 38.1 (Jan. 1975), pp. 5–13 (p. 6).

[74] B. J. Carr and M. J. Rees, *Can pregalactic objects generate galaxies?, "Mon. Not. Roy. Astron. Soc."* 206 (Feb. 1984), pp. 801–818 (pp. 6, 8).

[75] D. N. Page, *Particle emission rates from a black hole: Massless particles from an uncharged, nonrotating hole*, Phys. Rev. D 13 (2 1976), pp. 198–206, URL: https://link.aps.org/doi/10.1103/PhysRevD.13.198 (p. 6).

[76] J. H. MacGibbon, B. J. Carr, and D. N. Page, *Do Evaporating Black Holes Form Photospheres?*, Phys. Rev. D 78 (2008), p. 064043, arXiv: 0709.2480 [astro-ph] (p. 6).

[77] J. Silk and M. S. Turner, *Double Inflation*, Phys. Rev. D 35 (1987), p. 419 (p. 7).

[78] A. Dolgov and J. Silk, *Baryon isocurvature fluctuations at small scales and baryonic dark matter*, Phys. Rev. D 47 (1993), pp. 4244–4255 (pp. 7, 51).

[79] B. J. Carr and J. E. Lidsey, *Primordial black holes and generalized constraints on chaotic inflation*, Phys. Rev. D 48 (1993), pp. 543–553 (p. 7).

[80] B. J. Carr, J. H. Gilbert, and J. E. Lidsey, *Black hole relics and inflation: Limits on blue perturbation spectra*, Phys. Rev. D 50 (1994), pp. 4853–4867, arXiv: astro-ph/9405027 (p. 7).

[81] P. Ivanov, P. Naselsky, and I. Novikov, *Inflation and primordial black holes as dark matter*, Phys. Rev. D 50 (1994), pp. 7173–7178 (pp. 7, 8).

[82] L. Randall, M. Soljacic, and A. H. Guth, *Supernatural inflation: Inflation from supersymmetry with no (very) small parameters*, Nucl. Phys. B 472 (1996), pp. 377–408, arXiv: hep-ph/9512439 (p. 7).

[83] J. Yokoyama, *Formation of MACHO primordial black holes in inflationary cosmology*, Astron. Astrophys. 318 (1997), p. 673, arXiv: astro-ph/9509027 (p. 7).

[84] J. Garcia-Bellido, A. D. Linde, and D. Wands, *Density perturbations and black hole formation in hybrid inflation*, Phys. Rev. D 54 (1996), pp. 6040–6058, arXiv: astro-ph/9605094 (p. 7).

[85] H. I. Kim and C. H. Lee, *Constraints on the spectral index from primordial black holes*, Phys. Rev. D 54 (1996), pp. 6001–6007 (p. 7).

[86] M. Kawasaki, N. Sugiyama, and T. Yanagida, *Primordial black hole formation in a double inflation model in supergravity*, Phys. Rev. D 57 (1998), pp. 6050–6056, arXiv: hep-ph/9710259 (p. 7).

[87] A. M. Green and A. R. Liddle, *Constraints on the density perturbation spectrum from primordial black holes*, Phys. Rev. D 56 (1997), pp. 6166–6174, arXiv: astro-ph/9704251 (p. 7).

[88] P. Ivanov, *Nonlinear metric perturbations and production of primordial black holes*, Phys. Rev. D 57 (1998), pp. 7145–7154, arXiv: astro-ph/9708224 (pp. 7, 28).

[89] E. Kotok and P. Naselsky, *Blue spectra and induced formation of primordial black holes*, Phys. Rev. D 58 (1998), p. 103517, arXiv: astro-ph/9806139 (p. 7).

[90] S. M. Leach, I. J. Grivell, and A. R. Liddle, *Black hole constraints on the running mass inflation model*, Phys. Rev. D 62 (2000), p. 043516, arXiv: astro-ph/0004296 (p. 7).

[91] A. M. Green and K. A. Malik, *Primordial black hole production due to preheating*, Phys. Rev. D 64 (2001), p. 021301, arXiv: hep-ph/0008113 (p. 7).

[92] B. A. Bassett and S. Tsujikawa, *Inflationary preheating and primordial black holes*, Phys. Rev. D 63 (2001), p. 123503, arXiv: hep-ph/0008328 (p. 7).

[93] D. H. Lyth and D. Wands, *Generating the curvature perturbation without an inflaton*, Phys. Lett. B 524 (2002), pp. 5–14, arXiv: hep-ph/0110002 (p. 7).

[94] K. Kohri, D. H. Lyth, and A. Melchiorri, *Black hole formation and slow-roll inflation*, JCAP 04 (2008), p. 038, arXiv: 0711.5006 [hep-ph] (p. 7).

[95] L. Alabidi and K. Kohri, *Generating Primordial Black Holes Via Hilltop-Type Inflation Models*, Phys. Rev. D 80 (2009), p. 063511, arXiv: 0906.1398 [astro-ph.CO] (p. 7).

[96] M. Kawasaki, N. Kitajima, and T. T. Yanagida, *Primordial black hole formation from an axionlike curvaton model*, Phys. Rev. D 87.6 (2013), p. 063519, arXiv: 1207.2550 [hep-ph] (p. 7).

[97]  S. Clesse and J. García-Bellido, *Massive Primordial Black Holes from Hybrid Inflation as Dark Matter and the seeds of Galaxies*, Phys. Rev. D 92.2 (2015), p. 023524, arXiv: 1501.07565 [astro-ph.CO] (p. 7).

[98]  J. Garcia-Bellido, M. Peloso, and C. Unal, *Gravitational waves at interferometer scales and primordial black holes in axion inflation*, JCAP 12 (2016), p. 031, arXiv: 1610.03763 [astro-ph.CO] (pp. 7, 157).

[99]  K. Inomata, M. Kawasaki, K. Mukaida, Y. Tada, and T. T. Yanagida, *Inflationary primordial black holes for the LIGO gravitational wave events and pulsar timing array experiments*, Phys. Rev. D 95.12 (2017), p. 123510, arXiv: 1611.06130 [astro-ph.CO] (p. 7).

[100] J. Garcia-Bellido, M. Peloso, and C. Unal, *Gravitational Wave signatures of inflationary models from Primordial Black Hole Dark Matter*, JCAP 09 (2017), p. 013, arXiv: 1707.02441 [astro-ph.CO] (pp. 7, 12).

[101] J. Garcia-Bellido and E. Ruiz Morales, *Primordial black holes from single field models of inflation*, Phys. Dark Univ. 18 (2017), pp. 47–54, arXiv: 1702.03901 [astro-ph.CO] (p. 7).

[102] K. Kannike, L. Marzola, M. Raidal, and H. Veermäe, *Single Field Double Inflation and Primordial Black Holes*, JCAP 09 (2017), p. 020, arXiv: 1705.06225 [astro-ph.CO] (p. 7).

[103] G. Ballesteros and M. Taoso, *Primordial black hole dark matter from single field inflation*, Phys. Rev. D 97.2 (2018), p. 023501, arXiv: 1709.05565 [hep-ph] (p. 7).

[104] M. P. Hertzberg and M. Yamada, *Primordial Black Holes from Polynomial Potentials in Single Field Inflation*, Phys. Rev. D 97.8 (2018), p. 083509, arXiv: 1712.09750 [astro-ph.CO] (p. 7).

[105] H. Motohashi and W. Hu, *Primordial Black Holes and Slow-Roll Violation*, Phys. Rev. D 96.6 (2017), p. 063503, arXiv: 1706.06784 [astro-ph.CO] (pp. 7, 44).

[106] J. Martin, T. Papanikolaou, and V. Vennin, *Primordial black holes from the preheating instability in single-field inflation*, JCAP 01 (2020), p. 024, arXiv: 1907.04236 [astro-ph.CO] (p. 7).

[107] G. Ballesteros, J. Rey, M. Taoso, and A. Urbano, *Primordial black holes as dark matter and gravitational waves from single-field polynomial inflation*, JCAP 07 (2020), p. 025, arXiv: 2001.08220 [astro-ph.CO] (p. 7).

[108] M. Y. Khlopov and A. G. Polnarev, *Primordial black holes as a cosmological test of grand unification*, Phys. Lett. B 97 (1980), pp. 383–387 (p. 7).

[109] A. G. Polnarev and M. Y. Khlopov, *Cosmology, primordial black holes, and supermassive particles*, Sov. Phys. Usp. 28 (1985), pp. 213–232 (p. 7).

[110] A. M. Green, A. R. Liddle, and A. Riotto, *Primordial black hole constraints in cosmologies with early matter domination*, Phys. Rev. D 56 (1997), pp. 7559–7565, arXiv: astro-ph/9705166 (p. 7).

[111] T. Harada, C.-M. Yoo, K. Kohri, K.-i. Nakao, and S. Jhingan, *Primordial black hole formation in the matter-dominated phase of the Universe*, Astrophys. J. 833.1 (2016), p. 61, arXiv: 1609.01588 [astro-ph.CO] (pp. 7, 186).

[112] B. Carr, T. Tenkanen, and V. Vaskonen, *Primordial black holes from inflaton and spectator field perturbations in a matter-dominated era*, Phys. Rev. D 96.6 (2017), p. 063507, arXiv: 1706.03746 [astro-ph.CO] (p. 7).

[113] J. D. Barrow and B. J. Carr, *Formation and evaporation of primordial black holes in scalar - tensor gravity theories*, Phys. Rev. D 54 (1996), pp. 3920–3931 (p. 7).

[114] M. Khlopov, B. A. Malomed, and I. B. Zeldovich, *Gravitational instability of scalar fields and formation of primordial black holes*, Mon. Not. Roy. Astron. Soc. 215 (1985), pp. 575–589 (p. 7).

[115] E. Cotner and A. Kusenko, *Primordial black holes from supersymmetry in the early universe*, Phys. Rev. Lett. 119.3 (2017), p. 031103, arXiv: 1612.02529 [astro-ph.CO] (p. 7).

[116] E. Cotner, A. Kusenko, M. Sasaki, and V. Takhistov, *Analytic Description of Primordial Black Hole Formation from Scalar Field Fragmentation*, JCAP 10 (2019), p. 077, arXiv: 1907.10613 [astro-ph.CO] (p. 7).

[117] S. W. Hawking, *Black Holes From Cosmic Strings*, Phys. Lett. B 231 (1989), pp. 237–239 (p. 7).

[118] A. Polnarev and R. Zembowicz, *Formation of Primordial Black Holes by Cosmic Strings*, Phys. Rev. D 43 (1991), pp. 1106–1109 (p. 7).

[119] J. Garriga and M. Sakellariadou, *Effects of friction on cosmic strings*, Phys. Rev. D 48 (1993), pp. 2502–2515, arXiv: hep-th/9303024 (p. 7).

[120] R. R. Caldwell and P. Casper, *Formation of black holes from collapsed cosmic string loops*, Phys. Rev. D 53 (1996), pp. 3002–3010, arXiv: gr-qc/9509012 (p. 7).

[121] J. H. MacGibbon, R. H. Brandenberger, and U. F. Wichoski, *Limits on black hole formation from cosmic string loops*, Phys. Rev. D 57 (1998), pp. 2158–2165, arXiv: astro-ph/9707146 (p. 7).

[122] T. Helfer, J. C. Aurrekoetxea, and E. A. Lim, *Cosmic String Loop Collapse in Full General Relativity*, Phys. Rev. D 99.10 (2019), p. 104028, arXiv: 1808.06678 [gr-qc] (p. 7).

[123] A. C. Jenkins and M. Sakellariadou, *Primordial black holes from cusp collapse on cosmic strings*, (June 2020), arXiv: 2006.16249 [astro-ph.CO] (p. 7).

[124] S. G. Rubin, M. Y. Khlopov, and A. S. Sakharov, *Primordial black holes from nonequilibrium second order phase transition*, Grav. Cosmol. 6 (2000). Ed. by M. Y. Khlopov, M. E. Prokhorov, and A. A. Starobinsky, pp. 51–58, arXiv: hep-ph/0005271 (p. 7).

[125] S. G. Rubin, A. S. Sakharov, and M. Y. Khlopov, *The Formation of primary galactic nuclei during phase transitions in the early universe*, J. Exp. Theor. Phys. 91 (2001), pp. 921–929, arXiv: hep-ph/0106187 (p. 7).

[126] J. Garriga, A. Vilenkin, and J. Zhang, *Black holes and the multiverse*, JCAP 02 (2016), p. 064, arXiv: 1512.01819 [hep-th] (p. 7).

[127] H. Deng, J. Garriga, and A. Vilenkin, *Primordial black hole and wormhole formation by domain walls*, JCAP 04 (2017), p. 050, arXiv: 1612.03753 [gr-qc] (p. 7).

[128] H. Deng and A. Vilenkin, *Primordial black hole formation by vacuum bubbles*, JCAP 12 (2017), p. 044, arXiv: 1710.02865 [gr-qc] (p. 7).

[129] J. Liu, Z.-K. Guo, and R.-G. Cai, *Primordial Black Holes from Cosmic Domain Walls*, Phys. Rev. D 101.2 (2020), p. 023513, arXiv: 1908.02662 [astro-ph.CO] (p. 7).

[130] A. Kusenko, M. Sasaki, S. Sugiyama, M. Takada, V. Takhistov, and E. Vitagliano, *Exploring Primordial Black Holes from the Multiverse with Optical Telescopes*, Phys. Rev. Lett. 125 (2020), p. 181304, arXiv: 2001.09160 [astro-ph.CO] (pp. 7, 192).

[131] M. Crawford and D. N. Schramm, *Spontaneous Generation of Density Perturbations in the Early Universe*, Nature 298 (1982), pp. 538–540 (p. 7).

[132] H. Kodama, M. Sasaki, and K. Sato, *Abundance of Primordial Holes Produced by Cosmological First Order Phase Transition*, Prog. Theor. Phys. 68 (1982), p. 1979 (p. 7).

[133] K. Jedamzik, *Primordial black hole formation during the QCD epoch*, Phys. Rev. D 55 (1997), pp. 5871–5875, arXiv: astro-ph/9605152 (pp. 7, 191, 192).

[134] S. W. Hawking, I. G. Moss, and J. M. Stewart, *Bubble Collisions in the Very Early Universe*, Phys. Rev. D 26 (1982), p. 2681 (p. 7).

[135] I. G. Moss, *Singularity formation from colliding bubbles*, Phys. Rev. D 50 (1994), pp. 676–681 (p. 7).

[136] N. Kitajima and F. Takahashi, *Primordial Black Holes from QCD Axion Bubbles*, JCAP 11 (2020), p. 060, arXiv: 2006.13137 [hep-ph] (p. 7).

[137] J. R. Espinosa, D. Racco, and A. Riotto, *Cosmological Signature of the Standard Model Higgs Vacuum Instability: Primordial Black Holes as Dark Matter*, Phys. Rev. Lett. 120.12 (2018), p. 121301, arXiv: 1710.11196 [hep-ph] (p. 7).

[138] J. R. Espinosa, D. Racco, and A. Riotto, *Primordial Black Holes from Higgs Vacuum Instability: Avoiding Fine-tuning through an Ultraviolet Safe Mechanism*, Eur. Phys. J. C 78.10 (2018), p. 806, arXiv: 1804.07731 [hep-ph] (p. 7).

[139] M. W. Choptuik, *Universality and scaling in gravitational collapse of a massless scalar field*, Phys. Rev. Lett. 70 (1993), pp. 9–12 (pp. 7, 49).

[140] J. C. Niemeyer and K. Jedamzik, *Near-critical gravitational collapse and the initial mass function of primordial black holes*, Phys. Rev. Lett. 80 (1998), pp. 5481–5484, arXiv: astro-ph/9709072 (pp. 7, 49, 50).

[141] J. Yokoyama, *Formation of primordial black holes in the inflationary universe*, Phys. Rept. 307 (1998), pp. 133–139 (p. 7).

[142] J. C. Niemeyer and K. Jedamzik, *Dynamics of primordial black hole formation*, Phys. Rev. D 59 (1999), p. 124013, arXiv: astro-ph/9901292 (p. 7).

[143] M. Shibata and M. Sasaki, *Black hole formation in the Friedmann universe: Formulation and computation in numerical relativity*, Phys. Rev. D 60 (1999), p. 084002, arXiv: gr-qc/9905064 (pp. 7, 19, 20, 21, 56).

[144] C. Gundlach, *Critical phenomena in gravitational collapse*, Living Rev. Rel. 2 (1999), p. 4, arXiv: gr-qc/0001046 (p. 7).

[145] C. Gundlach, *Critical phenomena in gravitational collapse*, Phys. Rept. 376 (2003), pp. 339–405, arXiv: gr-qc/0210101 (p. 7).

[146] I. Musco, J. C. Miller, and L. Rezzolla, *Computations of primordial black hole formation*, Class. Quant. Grav. 22 (2005), pp. 1405–1424, arXiv: gr-qc/0412063 (pp. 7, 19, 49).

[147] A. G. Polnarev and I. Musco, *Curvature profiles as initial conditions for primordial black hole formation*, Class. Quant. Grav. 24 (2007), pp. 1405–1432, arXiv: gr-qc/0605122 (pp. 7, 19, 20).

[148] I. Musco, J. C. Miller, and A. G. Polnarev, *Primordial black hole formation in the radiative era: Investigation of the critical nature of the collapse*, Class. Quant. Grav. 26 (2009), p. 235001, arXiv: 0811.1452 [gr-qc] (pp. 7, 49).

[149] I. Musco and J. C. Miller, *Primordial black hole formation in the early universe: critical behaviour and self-similarity*, Class. Quant. Grav. 30 (2013), p. 145009, arXiv: 1201.2379 [gr-qc] (pp. 7, 49).

[150] T. Harada, C.-M. Yoo, and K. Kohri, *Threshold of primordial black hole formation*, Phys. Rev. D 88.8 (2013). [Erratum: Phys.Rev.D 89, 029903 (2014)], p. 084051, arXiv: 1309.4201 [astro-ph.CO] (pp. 7, 19).

[151] F. Kühnel, C. Rampf, and M. Sandstad, *Effects of Critical Collapse on Primordial Black-Hole Mass Spectra*, Eur. Phys. J. C 76.2 (2016), p. 93, arXiv: 1512.00488 [astro-ph.CO] (p. 7).

[152] E. Aubourg et al., *Evidence for gravitational microlensing by dark objects in the galactic halo*, Nature 365 (1993), pp. 623–625 (p. 7).

[153] C. Alcock et al., *The MACHO project LMC microlensing results from the first two years and the nature of the galactic dark halo*, Astrophys. J. 486 (1997), pp. 697–726, arXiv: astro-ph/9606165 (p. 7).

[154] P. Tisserand et al., *Limits on the Macho Content of the Galactic Halo from the EROS-2 Survey of the Magellanic Clouds*, Astron. Astrophys. 469 (2007), pp. 387–404, arXiv: astro-ph/0607207 (p. 7).

[155] L. Wyrzykowski et al., *The OGLE View of Microlensing towards the Magellanic Clouds. II. OGLE-II SMC data*, Mon. Not. Roy. Astron. Soc. 407 (2010), pp. 189–200, arXiv: 1004.5247 [astro-ph.GA] (p. 7).

[156] L. Wyrzykowski et al., *The OGLE View of Microlensing towards the Magellanic Clouds. III. Ruling out sub-solar MACHOs with the OGLE-III LMC data*, Mon. Not. Roy. Astron. Soc. 413 (2011), p. 493, arXiv: 1012.1154 [astro-ph.GA] (p. 7).

[157] L. Wyrzykowski et al., *The OGLE View of Microlensing towards the Magellanic Clouds. IV. OGLE-III SMC Data and Final Conclusions on MACHOs*, Mon. Not. Roy. Astron. Soc. 416 (2011), p. 2949, arXiv: 1106.2925 [astro-ph.GA] (p. 7).

[158] S. Calchi Novati, S. Mirzoyan, P. Jetzer, and G. Scarpetta, *Microlensing towards the SMC: a new analysis of OGLE and EROS results*, Mon. Not. Roy. Astron. Soc. 435 (2013), p. 1582, arXiv: 1308.4281 [astro-ph.GA] (p. 7).

[159] B. P. Abbott et al., *Observation of Gravitational Waves from a Binary Black Hole Merger*, Phys. Rev. Lett. 116.6 (2016), p. 061102, arXiv: 1602.03837 [gr-qc] (pp. 7, 14, 149, 150).

[160] S. Bird, I. Cholis, J. B. Muñoz, Y. Ali-Haïmoud, M. Kamionkowski, E. D. Kovetz, et al., *Did LIGO detect dark matter?*, Phys. Rev. Lett. 116.20 (2016), p. 201301, arXiv: 1603.00464 [astro-ph.CO] (pp. 7, 14, 107, 115).

[161] S. Clesse and J. García-Bellido, *The clustering of massive Primordial Black Holes as Dark Matter: measuring their mass distribution with Advanced LIGO*, Phys. Dark Univ. 15 (2017), pp. 142–147, arXiv: 1603.05234 [astro-ph.CO] (p. 7).

[162] M. Sasaki, T. Suyama, T. Tanaka, and S. Yokoyama, *Primordial Black Hole Scenario for the Gravitational-Wave Event GW150914*, Phys. Rev. Lett. 117.6 (2016). [Erratum: Phys.Rev.Lett. 121, 059901 (2018)], p. 061101, arXiv: 1603.08338 [astro-ph.CO] (pp. 7, 14, 106, 108).

[163] https://www.benty-fields.com/trending (p. 7).

[164] P. S. Cole and C. T. Byrnes, *Extreme scenarios: the tightest possible constraints on the power spectrum due to primordial black holes*, JCAP 02 (2018), p. 019, arXiv: 1706.10288 [astro-ph.CO] (p. 7).

[165] G. Sato-Polito, E. D. Kovetz, and M. Kamionkowski, *Constraints on the primordial curvature power spectrum from primordial black holes*, Phys. Rev. D 100.6 (2019), p. 063521, arXiv: 1904.10971 [astro-ph.CO] (pp. 7, 10).

[166] A. Kalaja, N. Bellomo, N. Bartolo, D. Bertacca, S. Matarrese, I. Musco, et al., *From Primordial Black Holes Abundance to Primordial Curvature Power Spectrum (and back)*, JCAP 10 (2019), p. 031, arXiv: 1908.03596 [astro-ph.CO] (pp. 7, 27).

[167] A. D. Gow, C. T. Byrnes, P. S. Cole, and S. Young, *The power spectrum on small scales: Robust constraints and comparing PBH methodologies*, JCAP 02 (2021), p. 002, arXiv: 2008.03289 [astro-ph.CO] (pp. 7, 10, 12, 50, 51, 148).

[168] C. T. Byrnes, E. J. Copeland, and A. M. Green, *Primordial black holes as a tool for constraining non-Gaussianity*, Phys. Rev. D 86 (2012), p. 043512, arXiv: 1206.4188 [astro-ph.CO] (pp. 8, 28).

[169] Y. Ali-Haïmoud, E. D. Kovetz, and M. Kamionkowski, *Merger rate of primordial black-hole binaries*, Phys. Rev. D 96.12 (2017), p. 123523, arXiv: 1709.06576 [astro-ph.CO] (pp. 8, 12, 72, 76, 84, 106, 112, 115, 154).

[170] M. Raidal, C. Spethmann, V. Vaskonen, and H. Veermäe, *Formation and Evolution of Primordial Black Hole Binaries in the Early Universe*, JCAP 02 (2019), p. 018, arXiv: 1812.01930 [astro-ph.CO] (pp. 8, 12, 102, 103, 107, 111, 112, 113, 114, 118, 126, 140, 141, 154).

[171] J. S. Bullock and M. Boylan-Kolchin, *Small-Scale Challenges to the ΛCDM Paradigm*, Ann. Rev. Astron. Astrophys. 55 (2017), pp. 343–387, arXiv: 1707.04256 [astro-ph.CO] (p. 8).

[172] P. Boldrini, Y. Miki, A. Y. Wagner, R. Mohayaee, J. Silk, and A. Arbey, *Cusp-to-core transition in low-mass dwarf galaxies induced by dynamical heating of cold dark matter by primordial black holes*, Mon. Not. Roy. Astron. Soc. 492.4 (2020), pp. 5218–5225, arXiv: 1909.07395 [astro-ph.CO] (p. 8).

[173] J. Silk, *Feedback by Massive Black Holes in Gas-rich Dwarf Galaxies*, "Astrophys. J. Lett." 839.1, L13 (Apr. 2017), p. L13, arXiv: 1703.08553 [astro-ph.GA] (p. 8).

[174] B. C. Lacki and J. F. Beacom, *Primordial Black Holes as Dark Matter: Almost All or Almost Nothing*, Astrophys. J. Lett. 720 (2010), pp. L67–L71, arXiv: 1003.3466 [astro-ph.CO] (p. 8).

[175] Y. N. Eroshenko, *Dark matter density spikes around primordial black holes*, Astron. Lett. 42.6 (2016), pp. 347–356, arXiv: 1607.00612 [astro-ph.HE] (p. 8).

[176] J. Adamek, C. T. Byrnes, M. Gosenca, and S. Hotchkiss, *WIMPs and stellar-mass primordial black holes are incompatible*, Phys. Rev. D 100.2 (2019), p. 023506, arXiv: 1901.08528 [astro-ph.CO] (pp. 8, 75).

[177] G. Bertone, A. M. Coogan, D. Gaggero, B. J. Kavanagh, and C. Weniger, *Primordial Black Holes as Silver Bullets for New Physics at the Weak Scale*, Phys. Rev. D 100.12 (2019), p. 123013, arXiv: 1905.01238 [hep-ph] (p. 8).

[178] B. Carr, F. Kuhnel, and L. Visinelli, *Black Holes and WIMPs: All or Nothing or Something Else*, (Nov. 2020), arXiv: 2011.01930 [astro-ph.CO] (p. 8).

[179] R. Bean and J. a. Magueijo, *Could supermassive black holes be quintessential primordial black holes?*, Phys. Rev. D 66 (6 2002), p. 063505, URL: https://link.aps.org/doi/10.1103/PhysRevD.66.063505 (p. 8).

[180] N. Duechting, *Supermassive black holes from primordial black hole seeds*, Phys. Rev. D 70 (2004), p. 064015, arXiv: astro-ph/0406260 (p. 8).

[181] M. Kawasaki, A. Kusenko, and T. T. Yanagida, *Primordial seeds of supermassive black holes*, Phys. Lett. B 711 (2012), pp. 1–5, arXiv: 1202.3848 [astro-ph.CO] (p. 8).

[182] J. L. Bernal, A. Raccanelli, L. Verde, and J. Silk, *Signatures of primordial black holes as seeds of supermassive black holes*, JCAP 05 (2018). [Erratum: JCAP 01, E01 (2020)], p. 017, arXiv: 1712.01311 [astro-ph.CO] (p. 8).

[183] P. D. Serpico, V. Poulin, D. Inman, and K. Kohri, *Cosmic microwave background bounds on primordial black holes including dark matter halo accretion*, Phys. Rev. Res. 2.2 (2020), p. 023204, arXiv: 2002.10771 [astro-ph.CO] (pp. 8, 12, 72, 84, 87, 88, 133).

[184] M. Volonteri, *Formation of Supermassive Black Holes*, Astron. Astrophys. Rev. 18 (2010), pp. 279–315, arXiv: 1003.4404 [astro-ph.CO] (p. 8).

[185] B. J. Carr, "Primordial black holes: Do they exist and are they useful?", *59th Yamada Conference on Inflating Horizon of Particle Astrophysics and Cosmology*. Nov. 2005, arXiv: astro-ph/0511743 (p. 9).

[186] A. Riotto, *Inflation and the theory of cosmological perturbations*, ICTP Lect. Notes Ser. 14 (2003). Ed. by G. Dvali, A. Perez-Lorenzana, G. Senjanovic, G. Thompson, and F. Vissani, pp. 317–413, arXiv: hep-ph/0210162 (p. 9).

[187] D. J. Fixsen, E. S. Cheng, J. M. Gales, J. C. Mather, R. A. Shafer, and E. L. Wright, *The Cosmic Microwave Background spectrum from the full COBE FIRAS data set*, Astrophys. J. 473 (1996), p. 576, arXiv: astro-ph/9605054 (pp. 10, 70).

[188] J. C. Mather et al., *Measurement of the Cosmic Microwave Background spectrum by the COBE FIRAS instrument*, Astrophys. J. 420 (1994), pp. 439–444 (p. 10).

[189] C. T. Byrnes, P. S. Cole, and S. P. Patil, *Steepest growth of the power spectrum and primordial black holes*, JCAP 06 (2019), p. 028, arXiv: 1811.11158 [astro-ph.CO] (pp. 10, 51).

[190] K. Inomata and T. Nakama, *Gravitational waves induced by scalar perturbations as probes of the small-scale primordial spectrum*, Phys. Rev. D 99.4 (2019), p. 043511, arXiv: 1812.00674 [astro-ph.CO] (p. 10).

[191] Z.-C. Chen, C. Yuan, and Q.-G. Huang, *Pulsar Timing Array Constraints on Primordial Black Holes with NANOGrav 11-Year Dataset*, Phys. Rev. Lett. 124.25 (2020), p. 251101, arXiv: 1910.12239 [astro-ph.CO] (pp. 10, 12, 130).

[192] A. S. Josan, A. M. Green, and K. A. Malik, *Generalised constraints on the curvature perturbation from primordial black holes*, Phys. Rev. D 79 (2009), p. 103520, arXiv: 0903.3184 [astro-ph.CO] (p. 10).

[193] M. Y. Khlopov, *Primordial Black Holes*, Res. Astron. Astrophys. 10 (2010), pp. 495–528, arXiv: 0801.0116 [astro-ph] (p. 10).

[194] B. J. Carr, K. Kohri, Y. Sendouda, and J. Yokoyama, *New cosmological constraints on primordial black holes*, Phys. Rev. D 81 (2010), p. 104019, arXiv: 0912.5297 [astro-ph.CO] (pp. 10, 11, 51).

[195] B. Carr, F. Kuhnel, and M. Sandstad, *Primordial Black Holes as Dark Matter*, Phys. Rev. D 94.8 (2016), p. 083504, arXiv: 1607.06077 [astro-ph.CO] (p. 10).

[196] M. Sasaki, T. Suyama, T. Tanaka, and S. Yokoyama, *Primordial black holes—perspectives in gravitational wave astronomy*, Class. Quant. Grav. 35.6 (2018), p. 063001, arXiv: 1801.05235 [astro-ph.CO] (pp. 10, 20, 49, 69, 105, 107, 108).

[197] B. Carr and F. Kuhnel, *Primordial Black Holes as Dark Matter: Recent Developments*, Ann. Rev. Nucl. Part. Sci. 70 (2020), pp. 355–394, arXiv: 2006.02838 [astro-ph.CO] (pp. 10, 88).

[198] A. M. Green and B. J. Kavanagh, *Primordial Black Holes as a dark matter candidate*, J. Phys. G 48.4 (2021), p. 4, arXiv: 2007.10722 [astro-ph.CO] (pp. 10, 12, 130).

[199] B. Carr, K. Kohri, Y. Sendouda, and J. Yokoyama, *Constraints on Primordial Black Holes*, (Feb. 2020), arXiv: 2002.12778 [astro-ph.CO] (pp. 11, 70, 75, 87, 132).

[200] A. Arbey, J. Auffinger, and J. Silk, *Constraining primordial black hole masses with the isotropic gamma ray background*, Phys. Rev. D 101.2 (2020), p. 023010, arXiv: 1906.04750 [astro-ph.CO] (p. 11).

[201] M. Boudaud and M. Cirelli, *Voyager 1 $e^{\pm}$ Further Constrain Primordial Black Holes as Dark Matter*, Phys. Rev. Lett. 122.4 (2019), p. 041104, arXiv: 1807.03075 [astro-ph.HE] (p. 11).

[202] W. DeRocco and P. W. Graham, *Constraining Primordial Black Hole Abundance with the Galactic 511 keV Line*, Phys. Rev. Lett. 123.25 (2019), p. 251102, arXiv: 1906.07740 [astro-ph.CO] (p. 11).

[203] R. Laha, J. B. Muñoz, and T. R. Slatyer, *INTEGRAL constraints on primordial black holes and particle dark matter*, Phys. Rev. D 101.12 (2020), p. 123514, arXiv: 2004.00627 [astro-ph.CO] (p. 11).

[204] G. Ballesteros, J. Coronado-Blázquez, and D. Gaggero, *X-ray and gamma-ray limits on the primordial black hole abundance from Hawking radiation*, Phys. Lett. B 808 (2020), p. 135624, arXiv: 1906.10113 [astro-ph.CO] (p. 11).

[205] R. Laha, *Primordial Black Holes as a Dark Matter Candidate Are Severely Constrained by the Galactic Center 511 keV $\gamma$ -Ray Line*, Phys. Rev. Lett. 123.25 (2019), p. 251101, arXiv: 1906.09994 [astro-ph.HE] (p. 11).

[206] H. Poulter, Y. Ali-Haïmoud, J. Hamann, M. White, and A. G. Williams, *CMB constraints on ultra-light primordial black holes with extended mass distributions*, (July 2019), arXiv: 1907.06485 [astro-ph.CO] (p. 11).

[207] B. Dasgupta, R. Laha, and A. Ray, *Neutrino and positron constraints on spinning primordial black hole dark matter*, Phys. Rev. Lett. 125.10 (2020), p. 101101, arXiv: 1912.01014 [hep-ph] (p. 11).

[208] R. Laha, P. Lu, and V. Takhistov, *Gas Heating from Spinning and Non-Spinning Evaporating Primordial Black Holes*, (Sept. 2020), arXiv: 2009.11837 [astro-ph.CO] (p. 11).

[209] A. Ray, R. Laha, J. B. Muñoz, and R. Caputo, *Closing the gap: Near future MeV telescopes can discover asteroid-mass primordial black hole dark matter*, (Feb. 2021), arXiv: 2102.06714 [astro-ph.CO] (p. 11).

[210] K. Inomata, M. Kawasaki, K. Mukaida, T. Terada, and T. T. Yanagida, *Gravitational Wave Production right after a Primordial Black Hole Evaporation*, Phys. Rev. D 101.12 (2020), p. 123533, arXiv: 2003.10455 [astro-ph.CO] (p. 11).

[211] M. H. Chan and C. M. Lee, *Constraining Primordial Black Hole Fraction at the Galactic Centre using radio observational data*, Mon. Not. Roy. Astron. Soc. 497.1 (2020), pp. 1212–1216, arXiv: 2007.05677 [astro-ph.HE] (p. 11).

[212] H. Kim, *A constraint on light primordial black holes from the interstellar medium temperature*, (July 2020), arXiv: 2007.07739 [hep-ph] (p. 11).

[213] C. M. Lee and M. Ho Chan, *The Evaporating Primordial Black Hole Fraction in Cool-core Galaxy Clusters*, Astrophys. J. 912.1 (2021), p. 24, arXiv: 2103.12354 [astro-ph.HE] (p. 11).

[214] J. Iguaz, P. D. Serpico, and T. Siegert, *Isotropic X-ray bound on Primordial Black Hole Dark Matter*, (Apr. 2021), arXiv: 2104.03145 [astro-ph.CO] (p. 11).

[215] V. De Romeri, P. Martínez-Miravé, and M. Tórtola, *Signatures of primordial black hole dark matter at DUNE and THEIA*, (June 2021), arXiv: 2106.05013 [hep-ph] (p. 11).

[216] S. Mittal, A. Ray, G. Kulkarni, and B. Dasgupta, *Constraining primordial black holes as dark matter using the global 21-cm signal with X-ray heating and excess radio background*, (July 2021), arXiv: 2107.02190 [astro-ph.CO] (p. 11).

[217] H. Niikura et al., *Microlensing constraints on primordial black holes with Subaru/HSC Andromeda observations*, Nature Astron. 3.6 (2019), pp. 524–534, arXiv: 1701.02151 [astro-ph.CO] (pp. 11, 13).

[218] N. Smyth, S. Profumo, S. English, T. Jeltema, K. McKinnon, and P. Guhathakurta, *Updated Constraints on Asteroid-Mass Primordial Black Holes as Dark Matter*, Phys. Rev. D 101.6 (2020), p. 063005, arXiv: 1910.01285 [astro-ph.CO] (pp. 11, 13).

[219] C. Alcock et al., *The MACHO project: microlensing detection efficiency*, Astrophys. J. Suppl. 136 (2001), pp. 439–462, arXiv: astro-ph/0003392 (pp. 11, 87).

[220] R. A. Allsman et al., *MACHO project limits on black hole dark matter in the 1-30 solar mass range*, Astrophys. J. Lett. 550 (2001), p. L169, arXiv: astro-ph/0011506 (pp. 11, 87).

[221] M. Oguri, J. M. Diego, N. Kaiser, P. L. Kelly, and T. Broadhurst, *Understanding caustic crossings in giant arcs: characteristic scales, event rates, and constraints on compact dark matter*, Phys. Rev. D 97.2 (2018), p. 023518, arXiv: 1710.00148 [astro-ph.CO] (pp. 11, 87).

[222] H. Niikura, M. Takada, S. Yokoyama, T. Sumi, and S. Masaki, *Constraints on Earth-mass primordial black holes from OGLE 5-year microlensing events*, Phys. Rev. D 99.8 (2019), p. 083503, arXiv: 1901.07120 [astro-ph.CO] (p. 12).

[223] M. Zumalacarregui and U. Seljak, *Limits on stellar-mass compact objects as dark matter from gravitational lensing of type Ia supernovae*, Phys. Rev. Lett. 121.14 (2018), p. 141101, arXiv: 1712.02240 [astro-ph.CO] (pp. 12, 87).

[224] V. Vaskonen and H. Veermäe, *Lower bound on the primordial black hole merger rate*, Phys. Rev. D 101.4 (2020), p. 043015, arXiv: 1908.09752 [astro-ph.CO] (pp. 12, 114, 115).

[225] R. Saito and J. Yokoyama, *Gravitational wave background as a probe of the primordial black hole abundance*, Phys. Rev. Lett. 102 (2009). [Erratum: Phys.Rev.Lett. 107, 069901 (2011)], p. 161101, arXiv: 0812.4339 [astro-ph] (p. 12).

[226] V. Atal, J. Cid, A. Escrivà, and J. Garriga, *PBH in single field inflation: the effect of shape dispersion and non-Gaussianities*, JCAP 05 (2020), p. 022, arXiv: 1908.11357 [astro-ph.CO] (pp. 12, 130).

[227] T. Nakama, J. Silk, and M. Kamionkowski, *Stochastic gravitational waves associated with the formation of primordial black holes*, Phys. Rev. D 95.4 (2017), p. 043511, arXiv: 1612.06264 [astro-ph.CO] (p. 12).

[228] R.-g. Cai, S. Pi, and M. Sasaki, *Gravitational Waves Induced by non-Gaussian Scalar Perturbations*, Phys. Rev. Lett. 122.20 (2019), p. 201101, arXiv: 1810.11000 [astro-ph.CO] (pp. 12, 157).

[229] C. Unal, *Imprints of Primordial Non-Gaussianity on Gravitational Wave Spectrum*, Phys. Rev. D 99.4 (2019), p. 041301, arXiv: 1811.09151 [astro-ph.CO] (pp. 12, 157).

[230] R.-G. Cai, S. Pi, S.-J. Wang, and X.-Y. Yang, *Pulsar Timing Array Constraints on the Induced Gravitational Waves*, JCAP 10 (2019), p. 059, arXiv: 1907.06372 [astro-ph.CO] (p. 12).

[231] Z. Arzoumanian et al., *The NANOGrav 12.5 yr Data Set: Search for an Isotropic Stochastic Gravitational-wave Background*, Astrophys. J. Lett. 905.2 (2020), p. L34, arXiv: 2009.04496 [astro-ph.HE] (pp. 12, 190, 192).

[232] Y. Ali-Haïmoud and M. Kamionkowski, *Cosmic microwave background limits on accreting primordial black holes*, Phys. Rev. D 95.4 (2017), p. 043534, arXiv: 1612.05644 [astro-ph.CO] (pp. 12, 72, 84, 87).

[233] V. Poulin, P. D. Serpico, F. Calore, S. Clesse, and K. Kohri, *CMB bounds on disk-accreting massive primordial black holes*, Phys. Rev. D 96.8 (2017), p. 083524, arXiv: 1707.04206 [astro-ph.CO] (pp. 12, 72).

[234] M. Ricotti, J. P. Ostriker, and K. J. Mack, *Effect of Primordial Black Holes on the Cosmic Microwave Background and Cosmological Parameter Estimates*, Astrophys. J. 680 (2008), p. 829, arXiv: 0709.0524 [astro-ph] (pp. 12, 72, 73, 74, 76, 77, 80, 84, 89, 90, 130).

[235] D. Inman and Y. Ali-Haïmoud, *Early structure formation in primordial black hole cosmologies*, Phys. Rev. D 100.8 (2019), p. 083528, arXiv: 1907.08129 [astro-ph.CO] (pp. 12, 75, 84, 98, 99, 100, 101, 103, 105, 118).

[236] G. Hütsi, M. Raidal, and H. Veermäe, *Small-scale structure of primordial black hole dark matter and its implications for accretion*, Phys. Rev. D 100.8 (2019), p. 083016, arXiv: 1907.06533 [astro-ph.CO] (pp. 12, 84, 102).

[237] D. Gaggero, G. Bertone, F. Calore, R. M. T. Connors, M. Lovell, S. Markoff, and E. Storm, *Searching for Primordial Black Holes in the radio and X-ray sky*, Phys. Rev. Lett. 118.24 (2017), p. 241101, arXiv: 1612.00457 [astro-ph.HE] (pp. 13, 87).

[238] J. Manshanden, D. Gaggero, G. Bertone, R. M. T. Connors, and M. Ricotti, *Multiwavelength astronomical searches for primordial black holes*, JCAP 06 (2019), p. 026, arXiv: 1812.07967 [astro-ph.HE] (pp. 13, 87).

[239] A. Hektor, G. Hütsi, and M. Raidal, *Constraints on primordial black hole dark matter from Galactic center X-ray observations*, Astron. Astrophys. 618 (2018), A139, arXiv: 1805.06513 [astro-ph.CO] (p. 13).

[240] Y. Inoue and A. Kusenko, *New X-ray bound on density of primordial black holes*, JCAP 10 (2017), p. 034, arXiv: 1705.00791 [astro-ph.CO] (pp. 13, 87).

[241] P. Lu, V. Takhistov, G. B. Gelmini, K. Hayashi, Y. Inoue, and A. Kusenko, *Constraining Primordial Black Holes with Dwarf Galaxy Heating*, Astrophys. J. Lett. 908.2 (2021), p. L23, arXiv: 2007.02213 [astro-ph.CO] (p. 13).

[242] V. Takhistov, P. Lu, G. B. Gelmini, K. Hayashi, Y. Inoue, and A. Kusenko, *Interstellar Gas Heating by Primordial Black Holes*, (May 2021), arXiv: 2105.06099 [astro-ph.GA] (p. 13).

[243] B. J. Carr and M. Sakellariadou, *Dynamical constraints on dark compact objects*, Astrophys. J. 516 (1999), pp. 195–220 (p. 13).

[244] B. Carr and J. Silk, *Primordial Black Holes as Generators of Cosmic Structures*, Mon. Not. Roy. Astron. Soc. 478.3 (2018), p. 3756–3775, arXiv: 1801.00672 [astro-ph.CO] (p. 13).

[245] D. P. Quinn, M. I. Wilkinson, M. J. Irwin, J. Marshall, A. Koch, and V. Belokurov, *On the Reported Death of the MACHO Era*, Mon. Not. Roy. Astron. Soc. 396 (2009), p. 11, arXiv: 0903.1644 [astro-ph.GA] (pp. 13, 87).

[246] T. D. Brandt, *Constraints on MACHO Dark Matter from Compact Stellar Systems in Ultra-Faint Dwarf Galaxies*, Astrophys. J. Lett. 824.2 (2016), p. L31, arXiv: 1605.03665 [astro-ph.GA] (pp. 13, 87).

[247] S. M. Koushiappas and A. Loeb, *Dynamics of Dwarf Galaxies Disfavor Stellar-Mass Black Holes as Dark Matter*, Phys. Rev. Lett. 119.4 (2017), p. 041102, arXiv: 1704.01668 [astro-ph.GA] (pp. 13, 87).

[248] J. Stegmann, P. R. Capelo, E. Bortolas, and L. Mayer, *Improved constraints from ultra-faint dwarf galaxies on primordial black holes as dark matter*, Mon. Not. Roy. Astron. Soc. 492.4 (2020), pp. 5247–5260, arXiv: 1910.04793 [astro-ph.GA] (p. 13).

[249] K. Inomata, M. Kawasaki, and Y. Tada, *Revisiting constraints on small scale perturbations from big-bang nucleosynthesis*, Phys. Rev. D 94.4 (2016), p. 043527, arXiv: 1605.04646 [astro-ph.CO] (p. 13).

[250] T. Nakama, B. Carr, and J. Silk, *Limits on primordial black holes from μ distortions in cosmic microwave background*, Phys. Rev. D 97.4 (2018), p. 043525, arXiv: 1710.06945 [astro-ph.CO] (p. 13).

[251] N. Afshordi, P. McDonald, and D. N. Spergel, *Primordial black holes as dark matter: The Power spectrum and evaporation of early structures*, Astrophys. J. Lett. 594 (2003), pp. L71–L74, arXiv: astro-ph/0302035 (p. 13).

[252] R. Murgia, G. Scelfo, M. Viel, and A. Raccanelli, *Lyman-α Forest Constraints on Primordial Black Holes as Dark Matter*, Phys. Rev. Lett. 123.7 (2019), p. 071102, arXiv: 1903.10509 [astro-ph.CO] (pp. 13, 87).

[253] A. Barnacka, J. F. Glicenstein, and R. Moderski, *New constraints on primordial black holes abundance from femtolensing of gamma-ray bursts*, Phys. Rev. D 86 (2012), p. 043001, arXiv: 1204.2056 [astro-ph.CO] (p. 13).

[254] P. W. Graham, S. Rajendran, and J. Varela, *Dark Matter Triggers of Supernovae*, Phys. Rev. D 92.6 (2015), p. 063007, arXiv: 1505.04444 [hep-ph] (p. 13).

[255] F. Capela, M. Pshirkov, and P. Tinyakov, *Constraints on primordial black holes as dark matter candidates from capture by neutron stars*, Phys. Rev. D 87.12 (2013), p. 123524, arXiv: 1301.4984 [astro-ph.CO] (p. 13).

[256] A. Katz, J. Kopp, S. Sibiryakov, and W. Xue, *Femtolensing by Dark Matter Revisited*, JCAP 12 (2018), p. 005, arXiv: 1807.11495 [astro-ph.CO] (pp. 13, 192).

[257] P. Montero-Camacho, X. Fang, G. Vasquez, M. Silva, and C. M. Hirata, *Revisiting constraints on asteroid-mass primordial black holes as dark matter candidates*, JCAP 08 (2019), p. 031, arXiv: 1906.05950 [astro-ph.CO] (pp. 13, 192).

[258] B. Carr, M. Raidal, T. Tenkanen, V. Vaskonen, and H. Veermäe, *Primordial black hole constraints for extended mass functions*, Phys. Rev. D 96.2 (2017), p. 023514, arXiv: 1705.05567 [astro-ph.CO] (pp. 13, 51, 86, 87).

[259]  N. Bellomo, J. L. Bernal, A. Raccanelli, and L. Verde, *Primordial Black Holes as Dark Matter: Converting Constraints from Monochromatic to Extended Mass Distributions*, *JCAP* 01 (2018), p. 004, arXiv: 1709.07467 [astro-ph.CO] (pp. 13, 86).

[260]  A. Einstein, *Näherungsweise Integration der Feldgleichungen der Gravitation*. German, *Berl. Ber.* 1916 (1916), pp. 688–696 (p. 14).

[261]  M. Coleman Miller and N. Yunes, *The new frontier of gravitational waves*, *Nature* 568.7753 (2019), pp. 469–476 (p. 14).

[262]  A. S. Eddington, *The Propagation of Gravitational Waves*, *Proceedings of the Royal Society of London Series A* 102.716 (Dec. 1922), pp. 268–282 (p. 14).

[263]  J. Weber and J. A. Wheeler, *Reality of the Cylindrical Gravitational Waves of Einstein and Rosen*, *Reviews of Modern Physics* 29.3 (July 1957), pp. 509–515 (p. 14).

[264]  A. Hewish, S. J. Bell, J. D. H. Pilkington, P. F. Scott, and R. A. Collins, *Observation of a Rapidly Pulsating Radio Source, "Nature"* 217.5130 (Feb. 1968), pp. 709–713 (p. 14).

[265]  T. Gold, *Rotating Neutron Stars as the Origin of the Pulsating Radio Sources, "Nature"* 218.5143 (May 1968), pp. 731–732 (p. 14).

[266]  P. C. Peters and J. Mathews, *Gravitational radiation from point masses in a Keplerian orbit*, *Phys. Rev.* 131 (1963), pp. 435–439 (pp. 14, 108).

[267]  P. C. Peters, *Gravitational Radiation and the Motion of Two Point Masses*, *Phys. Rev.* 136 (1964), B1224–B1232 (pp. 14, 108, 212).

[268]  R. A. Hulse and J. H. Taylor, *Discovery of a pulsar in a binary system. "Astrophys. J. Lett."* 195 (Jan. 1975), pp. L51–L53 (p. 14).

[269]  J. M. Weisberg and J. H. Taylor, *Gravitational radiation from an orbiting pulsar. General Relativity and Gravitation* 13.1 (Jan. 1981), pp. 1–6 (p. 14).

[270]  S. Vitale, *The first five years of gravitational-wave astrophysics*, (Nov. 2020), arXiv: 2011.03563 [gr-qc] (pp. 14, 122).

[271]  M. Bailes et al., *Gravitational-wave physics and astronomy in the 2020s and 2030s*, *Nature Rev. Phys.* 3.5 (2021), pp. 344–366 (p. 14).

[272]  A. Einstein, *Über Gravitationswellen, Sitzungsberichte der Königlich Preußischen Akademie der Wissenschaften (Berlin* (Jan. 1918), pp. 154–167 (p. 15).

[273]  M. Maggiore, *Gravitational Waves. Vol. 1: Theory and Experiments*. Oxford Master Series in Physics. Oxford University Press, 2007. ISBN: 978-0-19-857074-5 (pp. 15, 107, 125, 177, 178, 183).

[274]  R. Abbott et al., *GWTC-2: Compact Binary Coalescences Observed by LIGO and Virgo During the First Half of the Third Observing Run*, *Phys. Rev. X* 11 (2021), p. 021053, arXiv: 2010.14527 [gr-qc] (pp. 16, 52, 122, 123, 125, 153).

[275]  S. Hild et al., *Sensitivity Studies for Third-Generation Gravitational Wave Observatories*, *Class. Quant. Grav.* 28 (2011), p. 094013, arXiv: 1012.0908 [gr-qc] (pp. 16, 88, 92, 149, 153).

[276]  C. Explorer, (2020), URL: https://cosmicexplorer.org/researchers.html (pp. 16, 153).

[277]  P. Amaro-Seoane et al., *Laser Interferometer Space Antenna*, (Feb. 2017), arXiv: 1702.00786 [astro-ph.IM] (pp. 15, 16, 109, 157, 185, 187, 190, 192).

[278]  Z. Arzoumanian et al., *The NANOGrav 11-year Data Set: High-precision timing of 45 Millisecond Pulsars*, *Astrophys. J. Suppl.* 235.2 (2018), p. 37, arXiv: 1801.01837 [astro-ph.HE] (p. 16).

[279]  W. Zhao, Y. Zhang, X.-P. You, and Z.-H. Zhu, *Constraints of relic gravitational waves by pulsar timing arrays: Forecasts for the FAST and SKA projects*, *Phys. Rev. D* 87.12 (2013), p. 124012, arXiv: 1303.6718 [astro-ph.CO] (pp. 16, 189, 192).

[280]  R. N. Manchester et al., *The Parkes Pulsar Timing Array Project*, *Publ. Astron. Soc. Austral.* 30 (2013), p. 17, arXiv: 1210.6130 [astro-ph.IM] (pp. 15, 189).

[281]  M. A. McLaughlin, *The North American Nanohertz Observatory for Gravitational Waves*, *Class. Quant. Grav.* 30 (2013), p. 224008, arXiv: 1310.0758 [astro-ph.IM] (pp. 15, 189).

[282]  M. Kramer and D. J. Champion, *The European Pulsar Timing Array and the Large European Array for Pulsars*, *Class. Quant. Grav.* 30 (2013), p. 224009 (pp. 15, 189).

[283]  B. P. Abbott et al., *Prospects for Observing and Localizing Gravitational-Wave Transients with Advanced LIGO, Advanced Virgo and KAGRA*, *Living Rev. Rel.* 21.1 (2018), p. 3, arXiv: 1304.0670 [gr-qc] (pp. 15, 125, 195).

[284]  M. Punturo et al., *The Einstein Telescope: A third-generation gravitational wave observatory*, *Class. Quant. Grav.* 27 (2010). Ed. by F. Ricci, p. 194002 (p. 15).

[285]  D. Reitze et al., *Cosmic Explorer: The U.S. Contribution to Gravitational-Wave Astronomy beyond LIGO*, *Bull. Am. Astron. Soc.* 51.7 (2019), p. 035, arXiv: 1907.04833 [astro-ph.IM] (p. 15).

[286] M. Maggiore, *Gravitational wave experiments and early universe cosmology*, Phys. Rept. 331 (2000), pp. 283–367, arXiv: gr-qc/9909001 (pp. 16, 160).

[287] C. Caprini and D. G. Figueroa, *Cosmological Backgrounds of Gravitational Waves*, Class. Quant. Grav. 35.16 (2018), p. 163001, arXiv: 1801.04268 [astro-ph.CO] (pp. 16, 157).

[288] N. Christensen, *Stochastic Gravitational Wave Backgrounds*, Rept. Prog. Phys. 82.1 (2019), p. 016903, arXiv: 1811.08797 [gr-qc] (p. 16).

[289] K. Jedamzik and J. C. Niemeyer, *Primordial black hole formation during first order phase transitions*, Phys. Rev. D 59 (1999), p. 124014, arXiv: astro-ph/9901293 (p. 19).

[290] I. Hawke and J. M. Stewart, *The dynamics of primordial black hole formation*, Class. Quant. Grav. 19 (2002), pp. 3687–3707 (p. 19).

[291] I. Musco, *Threshold for primordial black holes: Dependence on the shape of the cosmological perturbations*, Phys. Rev. D 100.12 (2019), p. 123524, arXiv: 1809.02127 [gr-qc] (pp. 19, 20, 21, 22, 59).

[292] A. Escrivà, C. Germani, and R. K. Sheth, *Universal threshold for primordial black hole formation*, Phys. Rev. D 101.4 (2020), p. 044022, arXiv: 1907.13311 [gr-qc] (pp. 19, 22, 23).

[293] D. H. Lyth, K. A. Malik, and M. Sasaki, *A General proof of the conservation of the curvature perturbation*, JCAP 05 (2005), p. 004, arXiv: astro-ph/0411220 (p. 19).

[294] T. Harada, C.-M. Yoo, T. Nakama, and Y. Koga, *Cosmological long-wavelength solutions and primordial black hole formation*, Phys. Rev. D 91.8 (2015), p. 084057, arXiv: 1503.03934 [gr-qc] (pp. 19, 20, 56).

[295] C. Germani and I. Musco, *Abundance of Primordial Black Holes Depends on the Shape of the Inflationary Power Spectrum*, Phys. Rev. Lett. 122.14 (2019), p. 141302, arXiv: 1805.04087 [astro-ph.CO] (pp. 19, 24, 27).

[296] C.-M. Yoo, T. Harada, J. Garriga, and K. Kohri, *Primordial black hole abundance from random Gaussian curvature perturbations and a local density threshold*, PTEP 2018.12 (2018), 123E01, arXiv: 1805.03946 [astro-ph.CO] (pp. 19, 27, 32).

[297] J. M. Bardeen, J. R. Bond, N. Kaiser, and A. S. Szalay, *The Statistics of Peaks of Gaussian Random Fields*, Astrophys. J. 304 (1986), pp. 15–61 (pp. 19, 28, 35, 53, 59, 65, 70, 100, 198, 199, 200).

[298] C.-M. Yoo, T. Harada, and H. Okawa, *Threshold of Primordial Black Hole Formation in Nonspherical Collapse*, Phys. Rev. D 102.4 (2020), p. 043526, arXiv: 2004.01042 [gr-qc] (p. 19).

[299] K. Tomita, *Evolution of Irregularities in a Chaotic Early Universe*, Prog. Theor. Phys. 54 (1975), p. 730 (p. 20).

[300] D. S. Salopek and J. R. Bond, *Nonlinear evolution of long wavelength metric fluctuations in inflationary models*, Phys. Rev. D 42 (1990), pp. 3936–3962 (pp. 20, 67).

[301] A. Kehagias, I. Musco, and A. Riotto, *Non-Gaussian Formation of Primordial Black Holes: Effects on the Threshold*, JCAP 12 (2019), p. 029, arXiv: 1906.07135 [astro-ph.CO] (pp. 24, 27, 38).

[302] C.-M. Yoo, T. Harada, S. Hirano, and K. Kohri, *Abundance of Primordial Black Holes in Peak Theory for an Arbitrary Power Spectrum*, PTEP 2021.1 (2021), 013E02, arXiv: 2008.02425 [astro-ph.CO] (p. 27).

[303] S. Young, C. T. Byrnes, and M. Sasaki, *Calculating the mass fraction of primordial black holes*, JCAP 07 (2014), p. 045, arXiv: 1405.7023 [gr-qc] (pp. 27, 28, 66).

[304] M. Kawasaki and H. Nakatsuka, *Effect of nonlinearity between density and curvature perturbations on the primordial black hole formation*, Phys. Rev. D 99.12 (2019), p. 123501, arXiv: 1903.02994 [astro-ph.CO] (p. 27).

[305] S. Young, I. Musco, and C. T. Byrnes, *Primordial black hole formation and abundance: contribution from the non-linear relation between the density and curvature perturbation*, JCAP 11 (2019), p. 012, arXiv: 1904.00984 [astro-ph.CO] (p. 27).

[306] C. Germani and R. K. Sheth, *Nonlinear statistics of primordial black holes from Gaussian curvature perturbations*, Phys. Rev. D 101.6 (2020), p. 063520, arXiv: 1912.07072 [astro-ph.CO] (p. 27).

[307] S. Young and M. Musso, *Application of peaks theory to the abundance of primordial black holes*, JCAP 11 (2020), p. 022, arXiv: 2001.06469 [astro-ph.CO] (p. 27).

[308] F. Riccardi, M. Taoso, and A. Urbano, *Solving peak theory in the presence of local non-gaussianities*, (Feb. 2021), arXiv: 2102.04084 [astro-ph.CO] (p. 27).

[309] M. Taoso and A. Urbano, *Non-gaussianities for primordial black hole formation*, (Feb. 2021), arXiv: 2102.03610 [astro-ph.CO] (pp. 27, 37, 38).

[310] S. Young and C. T. Byrnes, *Primordial black holes in non-Gaussian regimes*, JCAP 08 (2013), p. 052, arXiv: 1307.4995 [astro-ph.CO] (pp. 27, 28).

[311] S. Young, D. Regan, and C. T. Byrnes, *Influence of large local and non-local bispectra on primordial black hole abundance*, JCAP 02 (2016), p. 029, arXiv: 1512.07224 [astro-ph.CO] (pp. 27, 28).

[312] C.-M. Yoo, J.-O. Gong, and S. Yokoyama, *Abundance of primordial black holes with local non-Gaussianity in peak theory*, JCAP 09 (2019), p. 033, arXiv: 1906.06790 [astro-ph.CO] (p. 27).

[313] D. Blais, T. Bringmann, C. Kiefer, and D. Polarski, *Accurate results for primordial black holes from spectra with a distinguished scale*, Phys. Rev. D 67 (2003), p. 024024, arXiv: astro-ph/0206262 (p. 27).

[314] P. Pina Avelino, *Primordial black hole constraints on non-gaussian inflation models*, Phys. Rev. D 72 (2005), p. 124004, arXiv: astro-ph/0510052 (p. 28).

[315] E. V. Bugaev and P. A. Klimai, *Primordial black hole constraints for curvaton models with predicted large non-Gaussianity*, Int. J. Mod. Phys. D 22 (2013), p. 1350034, arXiv: 1303.3146 [astro-ph.CO] (p. 28).

[316] J. S. Bullock and J. R. Primack, *NonGaussian fluctuations and primordial black holes from inflation*, Phys. Rev. D 55 (1997), p. 7423–7439, arXiv: astro-ph/9611106 (p. 28).

[317] J. Yokoyama, *Chaotic new inflation and formation of primordial black holes*, Phys. Rev. D 58 (1998), p. 083510, arXiv: astro-ph/9802357 (p. 28).

[318] R. Saito, J. Yokoyama, and R. Nagata, *Single-field inflation, anomalous enhancement of superhorizon fluctuations, and non-Gaussianity in primordial black hole formation*, JCAP 06 (2008), p. 024, arXiv: 0804.3470 [astro-ph] (p. 28).

[319] M. Kawasaki and Y. Tada, *Can massive primordial black holes be produced in mild waterfall hybrid inflation?*, JCAP 08 (2016), p. 041, arXiv: 1512.03515 [astro-ph.CO] (p. 28).

[320] C. Pattison, V. Vennin, H. Assadullahi, and D. Wands, *Quantum diffusion during inflation and primordial black holes*, JCAP 10 (2017), p. 046, arXiv: 1707.00537 [hep-th] (pp. 28, 45).

[321] V. Atal and C. Germani, *The role of non-gaussianities in Primordial Black Hole formation*, Phys. Dark Univ. 24 (2019), p. 100275, arXiv: 1811.07857 [astro-ph] (p. 28).

[322] Y. Tada and S. Yokoyama, *Primordial black holes as biased tracers*, Phys. Rev. D 91.12 (2015), p. 123534, arXiv: 1502.01124 [astro-ph.CO] (pp. 28, 65, 68, 69).

[323] S. Young and C. T. Byrnes, *Signatures of non-gaussianity in the isocurvature modes of primordial black hole dark matter*, JCAP 04 (2015), p. 034, arXiv: 1503.01505 [astro-ph.CO] (pp. 28, 68, 69).

[324] S. Matarrese, F. Lucchin, and S. A. Bonometto, *A path integral approach to large scale matter distribution originated by non-gaussian fluctuations*, Astrophys. J. Lett. 310 (1986), pp. L21–L26 (pp. 29, 202, 203).

[325] J. R. Bond, S. Cole, G. Efstathiou, and N. Kaiser, *Excursion set mass functions for hierarchical Gaussian fluctuations*, Astrophys. J. 379 (1991), p. 440 (pp. 38, 40, 41).

[326] A. R. Zentner, *The Excursion Set Theory of Halo Mass Functions, Halo Clustering, and Halo Growth*, Int. J. Mod. Phys. D 16 (2007), pp. 763–816, arXiv: astro-ph/0611454 (p. 39).

[327] M. Maggiore and A. Riotto, *The Halo Mass Function from Excursion Set Theory. I. Gaussian fluctuations with non-Markovian dependence on the smoothing scale*, Astrophys. J. 711 (2010), pp. 907–927, arXiv: 0903.1249 [astro-ph.CO] (p. 40).

[328] M. Maggiore and A. Riotto, *The Halo mass function from excursion set theory. II. The diffusing barrier*, Astrophys. J. 717 (2010), pp. 515–525, arXiv: 0903.1250 [astro-ph.CO] (p. 40).

[329] M. Maggiore and A. Riotto, *The Halo mass function from excursion set theory. III. Non-Gaussian fluctuations*, Astrophys. J. 717 (2010), pp. 526–541, arXiv: 0903.1251 [astro-ph.CO] (p. 40).

[330] C. G. Lacey and S. Cole, *Merger rates in hierarchical models of galaxy formation*, Mon. Not. Roy. Astron. Soc. 262 (1993), pp. 627–649 (pp. 41, 43, 104).

[331] A. M. Green and A. R. Liddle, *Critical collapse and the primordial black hole initial mass function*, Phys. Rev. D 60 (1999), p. 063509, arXiv: astro-ph/9901268 (p. 41).

[332] J. M. Ezquiaga, J. García-Bellido, and V. Vennin, *The exponential tail of inflationary fluctuations: consequences for primordial black holes*, JCAP 03 (2020), p. 029, arXiv: 1912.05399 [astro-ph.CO] (p. 44).

[333] C. Pattison, V. Vennin, H. Assadullahi, and D. Wands, *Stochastic inflation beyond slow roll*, JCAP 07 (2019), p. 031, arXiv: 1905.06300 [astro-ph.CO] (p. 44).

[334] H. Firouzjahi, A. Nassiri-Rad, and M. Noorbala, *Stochastic Ultra Slow Roll Inflation*, JCAP 01 (2019), p. 040, arXiv: 1811.02175 [hep-th] (p. 44).

[335] D. Cruces, C. Germani, and T. Prokopec, *Failure of the stochastic approach to inflation beyond slow-roll*, JCAP 03 (2019), p. 048, arXiv: 1807.09057 [gr-qc] (p. 44).

[336] J. M. Ezquiaga and J. García-Bellido, *Quantum diffusion beyond slow-roll: implications for primordial black-hole production*, JCAP 08 (2018), p. 018, arXiv: 1805.06731 [astro-ph.CO] (p. 44).

[337] G. Ballesteros, J. Rey, M. Taoso, and A. Urbano, *Stochastic inflationary dynamics beyond slow-roll and consequences for primordial black hole formation*, JCAP 08 (2020), p. 043, arXiv: 2006.14597 [astro-ph.CO] (p. 44).

[338] K. Ando and V. Vennin, *Power spectrum in stochastic inflation*, JCAP 04 (2021), p. 057, arXiv: 2012.02031 [astro-ph.CO] (p. 44).

[339] C. Pattison, V. Vennin, D. Wands, and H. Assadullahi, *Ultra-slow-roll inflation with quantum diffusion*, (Jan. 2021), arXiv: 2101.05741 [astro-ph.CO] (p. 44).

[340] D. H. Lyth and A. Riotto, *Particle physics models of inflation and the cosmological density perturbation*, Phys. Rept. 314 (1999), pp. 1–146, arXiv: hep-ph/9807278 (p. 44).

[341] W. H. Kinney, *Horizon crossing and inflation with large eta*, Phys. Rev. D 72 (2005), p. 023515, arXiv: gr-qc/0503017 (p. 44).

[342] C. Dvorkin and W. Hu, *Generalized Slow Roll for Large Power Spectrum Features*, Phys. Rev. D 81 (2010), p. 023518, arXiv: 0910.2237 [astro-ph.CO] (p. 44).

[343] M. H. Namjoo, H. Firouzjahi, and M. Sasaki, *Violation of non-Gaussianity consistency relation in a single field inflationary model*, EPL 101.3 (2013), p. 39001, arXiv: 1210.3692 [astro-ph.CO] (p. 44).

[344] J. Martin, H. Motohashi, and T. Suyama, *Ultra Slow-Roll Inflation and the non-Gaussianity Consistency Relation*, Phys. Rev. D 87.2 (2013), p. 023514, arXiv: 1211.0083 [astro-ph.CO] (p. 44).

[345] X. Chen, H. Firouzjahi, M. H. Namjoo, and M. Sasaki, *A Single Field Inflation Model with Large Local Non-Gaussianity*, EPL 102.5 (2013), p. 59001, arXiv: 1301.5699 [hep-th] (p. 44).

[346] S. Mooij and G. A. Palma, *Consistently violating the non-Gaussian consistency relation*, JCAP 11 (2015), p. 025, arXiv: 1502.03458 [astro-ph.CO] (p. 44).

[347] C. Germani and T. Prokopec, *On primordial black holes from an inflection point*, Phys. Dark Univ. 18 (2017), pp. 6–10, arXiv: 1706.04226 [astro-ph.CO] (p. 44).

[348] K. Dimopoulos, *Ultra slow-roll inflation demystified*, Phys. Lett. B 775 (2017), pp. 262–265, arXiv: 1707.05644 [hep-ph] (p. 44).

[349] B. Finelli, G. Goon, E. Pajer, and L. Santoni, *Soft Theorems For Shift-Symmetric Cosmologies*, Phys. Rev. D 97.6 (2018), p. 063531, arXiv: 1711.03737 [hep-th] (p. 44).

[350] Y.-F. Cai, X. Chen, M. H. Namjoo, M. Sasaki, D.-G. Wang, and Z. Wang, *Revisiting non-Gaussianity from non-attractor inflation models*, JCAP 05 (2018), p. 012, arXiv: 1712.09998 [astro-ph.CO] (p. 44).

[351] A. D. Linde, *Particle physics and inflationary cosmology*. Vol. 5. 1990, arXiv: hep-th/0503203 (p. 45).

[352] H. Risken and T. Frank, *The Fokker-Planck Equation*. Vol. 16. Springer-Verlag Berlin Heidelberg, 1996 (p. 45).

[353] M. Cicoli, V. A. Diaz, and F. G. Pedro, *Primordial Black Holes from String Inflation*, JCAP 06 (2018), p. 034, arXiv: 1803.02837 [hep-th] (pp. 46, 48).

[354] C. R. Evans and J. S. Coleman, *Observation of critical phenomena and selfsimilarity in the gravitational collapse of radiation fluid*, Phys. Rev. Lett. 72 (1994), pp. 1782–1785, arXiv: gr-qc/9402041 (p. 49).

[355] T. Koike, T. Hara, and S. Adachi, *Critical behavior in gravitational collapse of radiation fluid: A Renormalization group (linear perturbation) analysis*, Phys. Rev. Lett. 74 (1995), pp. 5170–5173, arXiv: gr-qc/9503007 (p. 49).

[356] A. Escrivà, *Simulation of primordial black hole formation using pseudo-spectral methods*, Phys. Dark Univ. 27 (2020), p. 100466, arXiv: 1907.13065 [gr-qc] (p. 49).

[357] J. Yokoyama, *Cosmological constraints on primordial black holes produced in the near critical gravitational collapse*, Phys. Rev. D 58 (1998), p. 107502, arXiv: gr-qc/9804041 (p. 50).

[358] S. Young and C. T. Byrnes, *Initial clustering and the primordial black hole merger rate*, JCAP 03 (2020), p. 004, arXiv: 1910.06077 [astro-ph.CO] (pp. 50, 70).

[359] C. T. Byrnes, M. Hindmarsh, S. Young, and M. R. S. Hawkins, *Primordial black holes with an accurate QCD equation of state*, JCAP 08 (2018), p. 041, arXiv: 1801.06138 [astro-ph.CO] (pp. 51, 191, 192).

[360] A. M. Green, *Microlensing and dynamical constraints on primordial black hole dark matter with an extended mass function*, Phys. Rev. D 94.6 (2016), p. 063530, arXiv: 1609.01143 [astro-ph.CO] (p. 51).

[361] B. Horowitz, *Revisiting Primordial Black Holes Constraints from Ionization History*, (Dec. 2016), arXiv: 1612 . 07264 [astro-ph.CO] (pp. 51, 72).

[362] F. Kühnel and K. Freese, *Constraints on Primordial Black Holes with Extended Mass Functions*, *Phys. Rev. D* 95.8 (2017), p. 083508, arXiv: 1701.07223 [astro-ph.CO] (p. 51).

[363] A. D. Gow, C. T. Byrnes, and A. Hall, *Primordial black holes from narrow peaks and the skew-lognormal distribution*, (Sept. 2020), arXiv: 2009 . 03204 [astro-ph.CO] (pp. 51, 148).

[364] A. Doroshkevich, *Spatial structure of perturbations and origin of galactic rotation in fluctuation theory*, *Astrophysics* 6 (1970), pp. 320–330 (p. 52).

[365] A. Heavens and J. Peacock, *Tidal torques and local density maxima*, *Mon. Not. Roy. Astron. Soc.* 232.2 (May 1988), pp. 339–360 (pp. 52, 54, 59, 61, 62).

[366] S. D. M. White, *Angular momentum growth in protogalaxies*, *Astrophys. J.* 286 (1984), pp. 38–41 (p. 52).

[367] J. Barnes and G. Efstathiou, *Angular momentum from tidal torques*, *Astrophys. J.* 319 (1987), p. 575 (pp. 52, 58).

[368] B. M. Schaefer and P. Merkel, *Galactic angular momenta and angular momentum couplings in the large-scale structure*, *Mon. Not. Roy. Astron. Soc.* 421 (2012), pp. 2751–2762, arXiv: 1101.4584 [astro-ph.CO] (p. 52).

[369] P. J. E. Peebles, *Origin of the Angular Momentum of Galaxies*, *Astrophys. J.* 155 (1969), p. 393 (p. 52).

[370] Y. B. Zeldovich, *Gravitational instability: An Approximate theory for large density perturbations*, *Astron. Astrophys.* 5 (1970), pp. 84–89 (p. 52).

[371] L. Hui and E. Bertschinger, *Local approximations to the gravitational collapse of cold matter*, *Astrophys. J.* 471 (1996), p. 1, arXiv: astro-ph/9508114 (p. 52).

[372] J. R. Bond and S. T. Myers, *The Hierarchical peak patch picture of cosmic catalogs. 1. Algorithms*, *Astrophys. J. Suppl.* 103 (1996), p. 1 (p. 52).

[373] R. K. Sheth, H. J. Mo, and G. Tormen, *Ellipsoidal collapse and an improved model for the number and spatial distribution of dark matter haloes*, *Mon. Not. Roy. Astron. Soc.* 323 (2001), p. 1, arXiv: astro-ph/9907024 (p. 52).

[374] V. Desjacques, *Environmental dependence in the ellipsoidal collapse model*, *Mon. Not. Roy. Astron. Soc.* 388 (2008), p. 638, arXiv: 0707 . 4670 [astro-ph] (p. 52).

[375] E. Gourgoulhon, *3+1 formalism and bases of numerical relativity*, (Mar. 2007), arXiv: gr-qc/0703035 (pp. 52, 53).

[376] M. Mirbabayi, A. Gruzinov, and J. Noreña, *Spin of Primordial Black Holes*, *JCAP* 03 (2020), p. 017, arXiv: 1901 . 05963 [astro-ph.CO] (pp. 52, 141).

[377] A. Komar, *Covariant conservation laws in general relativity*, *Phys. Rev.* 113 (1959), pp. 934–936 (p. 52).

[378] J. M. Bardeen, *Gauge Invariant Cosmological Perturbations*, *Phys. Rev. D* 22 (1980), pp. 1882–1905 (pp. 57, 181).

[379] N. Bartolo, S. Matarrese, and A. Riotto, *CMB Anisotropies at Second-Order. 2. Analytical Approach*, *JCAP* 01 (2007), p. 019, arXiv: astro-ph/0610110 (p. 57).

[380] T. Chiba and S. Yokoyama, *Spin Distribution of Primordial Black Holes*, *PTEP* 2017.8 (2017), 083E01, arXiv: 1704 .06573 [gr-qc] (pp. 63, 64).

[381] T. W. Baumgarte and C. Gundlach, *Critical collapse of rotating radiation fluids*, *Phys. Rev. Lett.* 116.22 (2016), p. 221103, arXiv: 1603 . 04373 [gr-qc] (p. 63).

[382] M. He and T. Suyama, *Formation threshold of rotating primordial black holes*, *Phys. Rev. D* 100.6 (2019), p. 063520, arXiv: 1906 . 10987 [astro-ph.CO] (p. 64).

[383] M. M. Flores and A. Kusenko, *Primordial Black Holes from Long-Range Scalar Forces and Scalar Radiative Cooling*, *Phys. Rev. Lett.* 126.4 (2021), p. 041101, arXiv: 2008 . 12456 [astro-ph.CO] (p. 65).

[384] M. M. Flores and A. Kusenko, *Spins of primordial black holes formed in different cosmological scenarios*, (June 2021), arXiv: 2106.03237 [astro-ph.CO] (p. 65).

[385] T. Harada, C.-M. Yoo, K. Kohri, and K.-I. Nakao, *Spins of primordial black holes formed in the matter-dominated phase of the Universe*, *Phys. Rev. D* 96.8 (2017). [Erratum: Phys.Rev.D 99, 069904 (2019)], p. 083517, arXiv: 1707.03595 [gr-qc] (p. 65).

[386] J. H. MacGibbon, *Can Planck-mass relics of evaporating black holes close the universe?*, *Nature* 329 (1987), pp. 308–309 (p. 65).

[387] J. A. de Freitas Pacheco and J. Silk, *Primordial Rotating Black Holes*, *Phys. Rev. D* 101.8 (2020), p. 083022, arXiv: 2003 . 12072 [astro-ph.CO] (p. 65).

[388] E. Cotner and A. Kusenko, *Primordial black holes from scalar field evolution in the early universe*, *Phys. Rev. D* 96.10 (2017), p. 103002, arXiv: 1706.09003 [astro-ph.CO] (p. 65).

[389] N. Kaiser, *On the Spatial correlations of Abell clusters*, Astrophys. J. Lett. 284 (1984), pp. L9–L12 (pp. 65, 66).

[390] T. Baldauf, U. Seljak, R. E. Smith, N. Hamaus, and V. Desjacques, *Halo stochasticity from exclusion and nonlinear clustering*, Phys. Rev. D 88.8 (2013), p. 083507, arXiv: 1305.2917 [astro-ph.CO] (p. 65).

[391] V. Desjacques, D. Jeong, and F. Schmidt, *Large-Scale Galaxy Bias*, Phys. Rept. 733 (2018), pp. 1–193, arXiv: 1611.09787 [astro-ph.CO] (p. 65).

[392] Y. Ali-Haïmoud, *Correlation Function of High-Threshold Regions and Application to the Initial Small-Scale Clustering of Primordial Black Holes*, Phys. Rev. Lett. 121.8 (2018), p. 081304, arXiv: 1805.05912 [astro-ph.CO] (p. 65).

[393] V. Desjacques and A. Riotto, *Spatial clustering of primordial black holes*, Phys. Rev. D 98.12 (2018), p. 123533, arXiv: 1806.10414 [astro-ph.CO] (pp. 65, 67).

[394] G. Ballesteros, P. D. Serpico, and M. Taoso, *On the merger rate of primordial black holes: effects of nearest neighbours distribution and clustering*, JCAP 10 (2018), p. 043, arXiv: 1807.02084 [astro-ph.CO] (p. 65).

[395] A. Gangui, F. Lucchin, S. Matarrese, and S. Mollerach, *The Three point correlation function of the cosmic microwave background in inflationary models*, Astrophys. J. 430 (1994), pp. 447–457, arXiv: astro-ph/9312033 (p. 67).

[396] L. Verde, R. Jimenez, M. Kamionkowski, and S. Matarrese, *Tests for primordial nonGaussianity*, Mon. Not. Roy. Astron. Soc. 325 (2001), p. 412, arXiv: astro-ph/0011180 (p. 67).

[397] E. Komatsu and D. N. Spergel, *Acoustic signatures in the primary microwave background bispectrum*, Phys. Rev. D 63 (2001), p. 063002, arXiv: astro-ph/0005036 (p. 67).

[398] N. Dalal, O. Dore, D. Huterer, and A. Shirokov, *The imprints of primordial non-gaussianities on large-scale structure: scale dependent bias and abundance of virialized objects*, Phys. Rev. D 77 (2008), p. 123514, arXiv: 0710.4560 [astro-ph] (p. 68).

[399] Y. Akrami et al., *Planck 2018 results. X. Constraints on inflation*, Astron. Astrophys. 641 (2020), A10, arXiv: 1807.06211 [astro-ph.CO] (p. 69).

[400] J. Chluba and R. A. Sunyaev, *The evolution of CMB spectral distortions in the early Universe*, Mon. Not. Roy. Astron. Soc. 419 (2012), pp. 1294–1314, arXiv: 1109.6552 [astro-ph.CO] (p. 70).

[401] J. Chluba, J. Hamann, and S. P. Patil, *Features and New Physical Scales in Primordial Observables: Theory and Observation*, Int. J. Mod. Phys. D 24.10 (2015), p. 1530023, arXiv: 1505.01834 [astro-ph.CO] (p. 70).

[402] A. Kogut et al., *The Primordial Inflation Explorer (PIXIE): A Nulling Polarimeter for Cosmic Microwave Background Observations*, JCAP 07 (2011), p. 025, arXiv: 1105.2044 [astro-ph.CO] (p. 70).

[403] J. Chluba et al., *New Horizons in Cosmology with Spectral Distortions of the Cosmic Microwave Background*, (Sept. 2019), arXiv: 1909.01593 [astro-ph.CO] (p. 70).

[404] J. Chluba et al., *Spectral Distortions of the CMB as a Probe of Inflation, Recombination, Structure Formation and Particle Physics: Astro2020 Science White Paper*, Bull. Am. Astron. Soc. 51.3 (2019), p. 184, arXiv: 1903.04218 [astro-ph.CO] (p. 70).

[405] W. Hu, D. Scott, and J. Silk, *Power spectrum constraints from spectral distortions in the cosmic microwave background*, Astrophys. J. Lett. 430 (1994), pp. L5–L8, arXiv: astro-ph/9402045 (p. 70).

[406] J. B. Dent, D. A. Easson, and H. Tashiro, *Cosmological constraints from CMB distortion*, Phys. Rev. D 86 (2012), p. 023514, arXiv: 1202.6066 [astro-ph.CO] (p. 70).

[407] F. Hofmann, E. Barausse, and L. Rezzolla, *The final spin from binary black holes in quasi-circular orbits*, Astrophys. J. Lett. 825.2 (2016), p. L19, arXiv: 1605.01938 [gr-qc] (pp. 72, 94).

[408] M. Ricotti, *Bondi accretion in the early universe*, Astrophys. J. 662 (2007), pp. 53–61, arXiv: 0706.0864 [astro-ph] (pp. 73, 74, 76, 84).

[409] Y. Suto, R.-y. Cen, and J. P. Ostriker, *Statistics of the cosmic Mach number from numerical simulations of a CDM universe*, (Dec. 1991) (p. 73).

[410] H. Bondi and F. Hoyle, *On the mechanism of accretion by stars*, Mon. Not. Roy. Astron. Soc. 104 (1944), p. 273 (p. 74).

[411] H. Bondi, *Spherically symmetrical models in general relativity*, Mon. Not. Roy. Astron. Soc. 107 (1947), pp. 410–425 (p. 74).

[412] H. Bondi, *On spherically symmetrical accretion*, Mon. Not. Roy. Astron. Soc. 112 (1952), p. 195 (p. 74).

[413] S. L. Shapiro and S. A. Teukolsky, *Black holes, white dwarfs, and neutron stars: The physics of compact objects*. 1983. ISBN: 978-0-471-87316-7 (p. 74).

[414] E. Barausse, V. Cardoso, and P. Pani, *Can environmental effects spoil precision gravitational-wave astrophysics?*, Phys. Rev. D 89.10 (2014), p. 104059, arXiv: 1404.7149 [gr-qc] (pp. 75, 108).

[415] E. Bertschinger, *Self - similar secondary infall and accretion in an Einstein-de Sitter universe*, Astrophys. J. Suppl. 58 (1985), p. 39 (p. 75).

[416] K. J. Mack, J. P. Ostriker, and M. Ricotti, *Growth of structure seeded by primordial black holes*, Astrophys. J. 665 (2007), pp. 1277–1287, arXiv: astro-ph/0608642 (p. 75).

[417] G. Hasinger, *Illuminating the dark ages: Cosmic backgrounds from accretion onto primordial black hole dark matter*, JCAP 07 (2020), p. 022, arXiv: 2003.05150 [astro-ph.CO] (pp. 76, 84).

[418] E. Poisson and C. Will, *Gravity: Newtonian, Post-Newtonian, Relativistic.* Cambridge University Press, 2014 (p. 78).

[419] S. P. Oh and Z. Haiman, *Fossil HII regions: Self-limiting star formation at high redshift*, Mon. Not. Roy. Astron. Soc. 346 (2003), p. 456, arXiv: astro-ph/0307135 (p. 84).

[420] V. Bosch-Ramon and N. Bellomo, *Mechanical feedback effects on primordial black hole accretion*, Astron. Astrophys. 638 (2020), A132, arXiv: 2004.11224 [astro-ph.CO] (p. 84).

[421] J. M. Bardeen, W. H. Press, and S. A. Teukolsky, *Rotating black holes: Locally nonrotating frames, energy extraction, and scalar synchrotron radiation*, Astrophys. J. 178 (1972), p. 347 (pp. 85, 90).

[422] C. F. Gammie, S. L. Shapiro, and J. C. McKinney, *Black hole spin evolution*, Astrophys. J. 602 (2004), pp. 312–319, arXiv: astro-ph/0310886 (pp. 85, 91).

[423] B. P. Abbott et al., *Search for Subsolar Mass Ultracompact Binaries in Advanced LIGO's Second Observing Run*, Phys. Rev. Lett. 123.16 (2019), p. 161102, arXiv: 1904.08976 [astro-ph.CO] (pp. 87, 88, 156).

[424] P. N. Wilkinson, D. R. Henstock, I. W. A. Browne, A. G. Polatidis, P. Augusto, A. C. S. Readhead, et al., *Limits on the cosmological abundance of supermassive compact objects from a search for multiple imaging in compact radio sources*, Phys. Rev. Lett. 86 (2001), pp. 584–587, arXiv: astro-ph/0101328 (p. 87).

[425] K. Inayoshi, E. Visbal, and Z. Haiman, *The Assembly of the First Massive Black Holes*, Ann. Rev. Astron. Astrophys. 58 (2020), pp. 27–97, arXiv: 1911.05791 [astro-ph.GA] (p. 88).

[426] E. Berti and M. Volonteri, *Cosmological black hole spin evolution by mergers and accretion*, Astrophys. J. 684 (2008), pp. 822–828, arXiv: 0802.0025 [astro-ph] (pp. 89, 97).

[427] P. Pani and A. Loeb, *Constraining Primordial Black-Hole Bombs through Spectral Distortions of the Cosmic Microwave Background*, Phys. Rev. D 88 (2013), p. 041301, arXiv: 1307.5176 [astro-ph.CO] (p. 89).

[428] R. Brito, V. Cardoso, and P. Pani, *Superradiance: New Frontiers in Black Hole Physics*, Lect. Notes Phys. 906 (2015), pp.1–237, arXiv: 1501.06570 [gr-qc] (p. 89).

[429] J. P. Conlon and C. A. R. Herdeiro, *Can black hole superradiance be induced by galactic plasmas?*, Phys. Lett. B 780 (2018), pp. 169–173, arXiv: 1701.02034 [astro-ph.HE] (p. 89).

[430] A. Dima and E. Barausse, *Numerical investigation of plasma-driven superradiant instabilities*, Class. Quant. Grav. 37.17 (2020), p. 175006, arXiv: 2001.11484 [gr-qc] (p. 89).

[431] D. Blas and S. J. Witte, *Quenching Mechanisms of Photon Superradiance*, Phys. Rev. D 102.12 (2020), p. 123018, arXiv: 2009.10075 [hep-ph] (p. 89).

[432] R. Narayan and I. Yi, *Advection dominated accretion: Underfed black holes and neutron stars*, Astrophys. J. 452 (1995), p. 710, arXiv: astro-ph/9411059 (p. 90).

[433] N. I. Shakura and R. A. Sunyaev, *Black holes in binary systems. Observational appearance*, Astron. Astrophys. 24 (1973), pp. 337–355 (p. 90).

[434] I. D. Novikov and K. S. Thorne, *Astrophysics of black holes.* Jan. 1973, pp. 343–450 (p. 90).

[435] K. S. Thorne, *Disk accretion onto a black hole. 2. Evolution of the hole.* Astrophys. J. 191 (1974), pp. 507–520 (pp. 90, 91).

[436] R. Brito, V. Cardoso, and P. Pani, *Black holes as particle detectors: evolution of superradiant instabilities*, Class. Quant. Grav. 32.13 (2015), p. 134001, arXiv: 1411.0686 [gr-qc] (p. 90).

[437] M. Volonteri, P. Madau, E. Quataert, and M. J. Rees, *The Distribution and cosmic evolution of massive black hole spins*, Astrophys. J. 620 (2005), pp. 69–77, arXiv: astro-ph/0410342 (p. 90).

[438] D. A. Hemberger, G. Lovelace, T. J. Loredo, L. E. Kidder, M. A. Scheel, B. Szilágyi, et al., *Final spin and radiated energy in numerical simulations of binary black holes with equal masses and equal, aligned or anti-aligned spins*, Phys. Rev. D 88 (2013), p. 064014, arXiv: 1305.5991 [gr-qc] (pp. 93, 118).

[439] B. P. Abbott et al., *GWTC-1: A Gravitational-Wave Transient Catalog of Compact Binary Mergers Observed by LIGO and Virgo during the First and Second Observing Runs*, Phys. Rev. X 9.3 (2019), p. 031040, arXiv: 1811. 12907 [astro-ph.HE] (pp. 95, 123, 149, 151).

[440] B. Zackay, T. Venumadhav, L. Dai, J. Roulet, and M. Zaldarriaga, *Highly spinning and aligned binary black hole merger in the Advanced LIGO first observing run*, Phys. Rev. D 100.2 (2019), p. 023007, arXiv: 1902.10331 [astro-ph.HE] (p. 95).

[441] T. Venumadhav, B. Zackay, J. Roulet, L. Dai, and M. Zaldarriaga, *New binary black hole mergers in the second observing run of Advanced LIGO and Advanced Virgo*, Phys. Rev. D 101.8 (2020), p. 083030, arXiv: 1904.07214 [astro-ph.HE] (p. 95).

[442] Y. Huang, C.-J. Haster, S. Vitale, A. Zimmerman, J. Roulet, T. Venumadhav, et al., *Source properties of the lowest signal-to-noise-ratio binary black hole detections*, Phys. Rev. D 102.10 (2020), p. 103024, arXiv: 2003.04513 [gr-qc] (p. 95).

[443] R. Abbott et al., *GW190412: Observation of a Binary-Black-Hole Coalescence with Asymmetric Masses*, Phys. Rev. D 102.4 (2020), p. 043015, arXiv: 2004.08342 [astro-ph.HE] (pp. 95, 122, 149, 150, 151, 209).

[444] B. P. Abbott et al., *Binary Black Hole Mergers in the first Advanced LIGO Observing Run*, Phys. Rev. X 6.4 (2016). [Erratum: Phys.Rev.X 8, 039903 (2018)], p. 041015, arXiv: 1606.04856 [gr-qc] (p. 94).

[445] E. Barausse and L. Rezzolla, *Predicting the direction of the final spin from the coalescence of two black holes*, Astrophys. J. Lett. 704 (2009), pp. L40–L44, arXiv: 0904.2577 [gr-qc] (p. 94).

[446] M. Kesden, U. Sperhake, and E. Berti, *Final spins from the merger of precessing binary black holes*, Phys. Rev. D 81 (2010), p. 084054, arXiv: 1002.2643 [astro-ph.GA] (p. 94).

[447] E. Barausse, *The evolution of massive black holes and their spins in their galactic hosts*, Mon. Not. Roy. Astron. Soc. 423 (2012), pp. 2533–2557, arXiv: 1201.5888 [astro-ph.CO] (p. 94).

[448] W. Tichy and P. Marronetti, *The Final mass and spin of black hole mergers*, Phys. Rev. D 78 (2008), p. 081501, arXiv: 0807.2985 [gr-qc] (p. 96).

[449] D. Gerosa and E. Berti, *Are merging black holes born from stellar collapse or previous mergers?*, Phys. Rev. D 95.12 (2017), p. 124046, arXiv: 1703.06223 [gr-qc] (pp. 97, 140, 153).

[450] M. Fishbach, D. E. Holz, and B. Farr, *Are LIGO's Black Holes Made From Smaller Black Holes?*, Astrophys. J. Lett. 840.2 (2017), p. L24, arXiv: 1703.06869 [astro-ph.HE] (pp. 97, 131, 134, 139).

[451] T. Padmanabhan, *Aspects of gravitational clustering*, (Nov. 1999), arXiv: astro-ph/9911374 (p. 99).

[452] P. J. E. Peebles, *Principles of physical cosmology*. 1994 (pp. 100, 101).

[453] F. Bernardeau, S. Colombi, E. Gaztanaga, and R. Scoccimarro, *Large scale structure of the universe and cosmological perturbation theory*, Phys. Rept. 367 (2002), pp. 1–248, arXiv: astro-ph/0112551 (p. 100).

[454] P. J. E. Peebles, *The Nature of the Distribution of Galaxies*, Astronomy and Astrophysics 32 (May 1974), p. 197 (p. 102).

[455] J. McClelland and J. Silk, *The correlation function for density perturbations in an expanding universe. II. Nonlinear theory*, Astrophysical Journal 217 (Oct. 1977), pp. 331–352 (p. 102).

[456] W. H. Press and P. Schechter, *Formation of galaxies and clusters of galaxies by selfsimilar gravitational condensation*, Astrophys. J. 187 (1974), pp. 425–438 (p. 102).

[457] A. Cooray and R. K. Sheth, *Halo Models of Large Scale Structure*, Phys. Rept. 372 (2002), pp. 1–129, arXiv: astro-ph/0206508 (p. 102).

[458] R. K. Sheth and B. Jain, *The Nonlinear correlation function and the shapes of virialized halos*, Mon. Not. Roy. Astron. Soc. 285 (1997), p. 231, arXiv: astro-ph/9602103 (pp. 102, 103).

[459] M. Trashorras, J. García-Bellido, and S. Nesseris, *The clustering dynamics of primordial black boles in N-body simulations*, Universe 7.1 (2021), p. 18, arXiv: 2006.15018 [astro-ph.CO] (p. 103).

[460] K. Jedamzik, *Primordial Black Hole Dark Matter and the LIGO/Virgo observations*, JCAP 09 (2020), p. 022, arXiv: 2006.11172 [astro-ph.CO] (pp. 103, 113).

[461] J. Binney and S. Tremaine, *Galactic dynamics*. Princeton University Press, 1987 (p. 104).

[462] T. Nakamura, M. Sasaki, T. Tanaka, and K. S. Thorne, *Gravitational waves from coalescing black hole MACHO binaries*, Astrophys. J. Lett. 487 (1997), pp. L139–L142, arXiv: astro-ph/9708060 (p. 106).

[463] K. Ioka, T. Chiba, T. Tanaka, and T. Nakamura, *Black hole binary formation in the expanding universe: Three body problem approximation*, Phys. Rev. D 58 (1998), p. 063003, arXiv: astro-ph/9807018 (p. 106).

[464] G. D. Quinlan and S. L. Shapiro, *Dynamical Evolution of Dense Clusters of Compact Stars*, Astrophysical Journal 343 (Aug. 1989), p. 725 (p. 107).

[465] H. Mouri and Y. Taniguchi, *Runaway merging of black holes: analytical constraint on the timescale*, Astrophys. J. Lett. 566 (2002), pp. L17–L20, arXiv: astro-ph/0201102 (p. 107).

[466] I. Cholis, E. D. Kovetz, Y. Ali-Haïmoud, S. Bird, M. Kamionkowski, J. B. Muñoz, and A. Raccanelli, *Orbital eccentricities in primordial black hole binaries*, Phys. Rev. D 94.8 (2016), p. 084013, arXiv: 1606.07437 [astro-ph.HE] (p. 107).

[467] R. M. O'Leary, B. Kocsis, and A. Loeb, *Gravitational waves from scattering of stellar-mass black holes in galactic nuclei*, Mon. Not. Roy. Astron. Soc. 395.4 (2009), pp. 2127–2146, arXiv: 0807.2638 [astro-ph] (p. 107).

[468] J. Samsing, D. J. D'Orazio, K. Kremer, C. L. Rodriguez, and A. Askar, *Single-single gravitational-wave captures in globular clusters: Eccentric deci-Hertz sources observable by DECIGO and Tian-Qin*, Phys. Rev. D 101.12 (2020), p. 123010, arXiv: 1907.11231 [astro-ph.HE] (p. 107).

[469] A. Caputo, L. Sberna, A. Toubiana, S. Babak, E. Barausse, S. Marsat, and P. Pani, *Gravitational-wave detection and parameter estimation for accreting black-hole binaries and their electromagnetic counterpart*, Astrophys. J. 892.2 (2020), p. 90, arXiv: 2001.03620 [astro-ph.HE] (pp. 108, 110).

[470] C. F. B. Macedo, P. Pani, V. Cardoso, and L. C. B. Crispino, *Into the lair: gravitational-wave signatures of dark matter*, Astrophys. J. 774 (2013), p. 48, arXiv: 1302.2646 [gr-qc] (p. 108).

[471] M. Maggiore et al., *Science Case for the Einstein Telescope*, JCAP 03 (2020), p. 050, arXiv: 1912.02622 [astro-ph.CO] (pp. 109, 117, 130, 149).

[472] L. Landau and E. Lifshitz, *Mechanics: Volume 1*. v. 1. Elsevier Science, 1982. ISBN: 9780080503479, URL: https://books.google.it/books?id=bE-9tUH2J2wC (pp. 109, 110).

[473] Y. N. Eroshenko, *Gravitational waves from primordial black holes collisions in binary systems*, J. Phys. Conf. Ser. 1051.1 (2018), p. 012010, arXiv: 1604.04932 [astro-ph.CO] (p. 112).

[474] L. Liu, Z.-K. Guo, and R.-G. Cai, *Effects of the surrounding primordial black holes on the merger rate of primordial black hole binaries*, Phys. Rev. D 99.6 (2019), p. 063523, arXiv: 1812.05376 [astro-ph.CO] (p. 112).

[475] J. Garriga and N. Triantafyllou, *Enhanced cosmological perturbations and the merger rate of PBH binaries*, JCAP 09 (2019), p. 043, arXiv: 1907.01455 [astro-ph.CO] (p. 112).

[476] G. Hütsi, M. Raidal, V. Vaskonen, and H. Veermäe, *Two populations of LIGO-Virgo black holes*, JCAP 03 (2021), p. 068, arXiv: 2012.02786 [astro-ph.CO] (pp. 113, 117, 118, 124, 126, 135, 136, 140).

[477] D. Lynden-Bell and R. Wood, *The gravothermal catastrophe in isothermal spheres and the onset of red-giant structure for stellar systems*, Mon. Not. Roy. Astron. Soc. 138.4 (1968), p. 495 (p. 114).

[478] G. D. Quinlan, *The Time scale for core collapse in spherical star clusters*, New Astron. 1 (1996), p. 255, arXiv: astro-ph/9606182 (p. 114).

[479] R. I. Epstein, *Proto-galactic perturbations*, Mon. Not. Roy. Astron. Soc. 205 (1983), pp. 207–229 (p. 114).

[480] R. K. Sheth, *The Generalized Poisson distribution and a model of clustering from Poisson initial conditions*, Mon. Not. Roy. Astron. Soc. 299 (1998), pp. 207–217, arXiv: astro-ph/9803045 (p. 114).

[481] K. Jedamzik, (2021). In preparation (p. 115).

[482] V. Korol, I. Mandel, M. C. Miller, R. P. Church, and M. B. Davies, *Merger rates in primordial black hole clusters without initial binaries*, Mon. Not. Roy. Astron. Soc. 496.1 (2020), pp. 994–1000, arXiv: 1911.03483 [astro-ph.HE] (p. 115).

[483] S. Clesse and J. García-Bellido, *Seven Hints for Primordial Black Hole Dark Matter*, Phys. Dark Univ. 22 (2018), pp. 137–146, arXiv: 1711.10458 [astro-ph.CO] (p. 116).

[484] L. Liu, Z.-K. Guo, and R.-G. Cai, *Effects of the merger history on the merger rate density of primordial black hole binaries*, Eur. Phys. J. C 79.8 (2019), p. 717, arXiv: 1901.07672 [astro-ph.CO] (p. 116).

[485] Y. Wu, *Merger history of primordial black-hole binaries*, Phys. Rev. D 101.8 (2020), p. 083008, arXiv: 2001.03833 [astro-ph.CO] (pp. 116, 126).

[486] W. B. Bonnor and M. A. Rotenberg, *Transport of Momentum by Gravitational Waves: The Linear Approximation*, Proceedings of the Royal Society of London Series A 265.1320 (1961), pp. 109–116 (p. 116).

[487] A. Peres, *Classical Radiation Recoil*, Phys. Rev. D 128 (1962), pp. 2471–2475 (pp. 116, 212).

[488] D. Gerosa, F. Hébert, and L. C. Stein, *Black-hole kicks from numerical-relativity surrogate models*, Phys. Rev. D 97.10 (2018), p. 104049, arXiv: 1802.04276 [gr-qc] (p. 116).

[489] D. Gerosa and E. Berti, *Escape speed of stellar clusters from multiple-generation black-hole mergers in the upper mass gap*, Phys. Rev. D 100.4 (2019), p. 041301, arXiv: 1906.05295 [astro-ph.HE] (pp. 116, 131, 134, 139, 212).

[490] V. Baibhav, D. Gerosa, E. Berti, K. W. K. Wong, T. Helfer, and M. Mould, *The mass gap, the spin gap, and the origin of merging binary black holes*, Phys. Rev. D 102.4 (2020), p. 043002, arXiv: 2004.00650 [astro-ph.HE] (pp. 116, 131, 134, 139).

[491] D. Gerosa and M. Fishbach, *Hierarchical mergers of stellar-mass black holes and their gravitational-wave signatures*, (May 2021), arXiv: 2105.03439 [astro-ph.HE] (p. 116).

[492] B. S. Sathyaprakash et al., *Cosmology and the Early Universe*, (Mar. 2019), arXiv: 1903.09260 [astro-ph.HE] (p. 117).

[493] X.-J. Zhu, E. Howell, T. Regimbau, D. Blair, and Z.-H. Zhu, *Stochastic Gravitational Wave Background from Coalescing Binary Black Holes*, Astrophys. J. 739 (2011), p. 86, arXiv: 1104.3565 [gr-qc] (p. 118).

[494] S. Wang, Y.-F. Wang, Q.-G. Huang, and T. G. F. Li, *Constraints on the Primordial Black Hole Abundance from the First Advanced LIGO Observation Run Using the Stochastic Gravitational-Wave Background*, Phys. Rev. Lett. 120.19 (2018), p. 191102, arXiv: 1610.08725 [astro-ph.CO] (p. 118).

[495] M. Raidal, V. Vaskonen, and H. Veermäe, *Gravitational Waves from Primordial Black Hole Mergers*, JCAP 09 (2017), p. 037, arXiv: 1707.01480 [astro-ph.CO] (p. 118).

[496] Z.-C. Chen, F. Huang, and Q.-G. Huang, *Stochastic Gravitational-wave Background from Binary Black Holes and Binary Neutron Stars and Implications for LISA*, Astrophys. J. 871.1 (2019), p. 97, arXiv: 1809.10360 [gr-qc] (pp. 118, 119).

[497] R. Abbott et al., *Upper Limits on the Isotropic Gravitational-Wave Background from Advanced LIGO's and Advanced Virgo's Third Observing Run*, (Jan. 2021), arXiv: 2101.12130 [gr-qc] (p. 119).

[498] P. Ajith et al., *Inspiral-merger-ringdown waveforms for black-hole binaries with non-precessing spins*, Phys. Rev. Lett. 106 (2011), p. 241101, arXiv: 0909.2867 [gr-qc] (p. 118).

[499] P. Ajith et al., *A Template bank for gravitational waveforms from coalescing binary black holes. I. Non-spinning binaries*, Phys. Rev. D 77 (2008). [Erratum: Phys.Rev.D 79, 129901 (2009)], p. 104017, arXiv: 0710.2335 [gr-qc] (p. 118).

[500] M. Dominik, K. Belczynski, C. Fryer, D. E. Holz, E. Berti, T. Bulik, et al., *Double Compact Objects II: Cosmological Merger Rates*, Astrophys. J. 779 (2013), p. 72, arXiv: 1308.1546 [astro-ph.HE] (p. 119).

[501] K. Belczynski, D. E. Holz, T. Bulik, and R. O'Shaughnessy, *The first gravitational-wave source from the isolated evolution of two 40-100 Msun stars*, Nature 534 (2016), p. 512, arXiv: 1602.04531 [astro-ph.HE] (p. 119).

[502] M. Mapelli, N. Giacobbo, F. Santoliquido, and M. C. Artale, *The properties of merging black holes and neutron stars across cosmic time*, Mon. Not. Roy. Astron. Soc. 487.1 (2019), pp. 2–13, arXiv: 1902.01419 [astro-ph.HE] (p. 119).

[503] C. L. Rodriguez and A. Loeb, *Redshift Evolution of the Black Hole Merger Rate from Globular Clusters*, Astrophys. J. Lett. 866.1 (2018), p. L5, arXiv: 1809.01152 [astro-ph.HE] (pp. 119, 212).

[504] J. Aasi et al., *Advanced LIGO*, Class. Quant. Grav. 32 (2015), p. 074001, arXiv: 1411.4547 [gr-qc] (p. 121).

[505] F. Acernese et al., *Advanced Virgo: a second-generation interferometric gravitational wave detector*, Class. Quant. Grav. 32.2 (2015), p. 024001, arXiv: 1408.3978 [gr-qc] (p. 121).

[506] J. R. Bond and B. J. Carr, *Gravitational waves from a population of binary black holes, "Mon. Not. Roy. Astron. Soc."* 207 (Apr. 1984), pp. 585–609 (p. 121).

[507] R. Abbott et al., *Population Properties of Compact Objects from the Second LIGO-Virgo Gravitational-Wave Transient Catalog*, (Oct. 2020), arXiv: 2010.14533 [astro-ph.HE] (pp. 121, 122, 124, 128, 129, 130, 134, 135, 139, 141, 204, 210).

[508] B. P. Abbott et al., *GW150914: The Advanced LIGO Detectors in the Era of First Discoveries*, Phys. Rev. Lett. 116.13 (2016), p. 131103, arXiv: 1602.03838 [gr-qc] (p. 122).

[509] A. H. Nitz, T. Dent, G. S. Davies, S. Kumar, C. D. Capano, I. Harry, et al., *2-OGC: Open Gravitational-wave Catalog of binary mergers from analysis of public Advanced LIGO and Virgo data*, Astrophys. J. 891 (Mar. 2020), p. 123, arXiv: 1910.05331 [astro-ph.HE] (p. 122).

[510] B. Zackay, L. Dai, T. Venumadhav, J. Roulet, and M. Zaldarriaga, *Detecting Gravitational Waves With Disparate Detector Responses: Two New Binary Black Hole Mergers*, (Oct. 2019), arXiv: 1910.09528 [astro-ph.HE] (p. 122).

[511] J. Roulet, T. Venumadhav, B. Zackay, L. Dai, and M. Zaldarriaga, *Binary Black Hole Mergers from LIGO/Virgo O1 and O2: Population Inference Combining Confident and Marginal Events*, Phys. Rev. D 102.12 (2020), p. 123022, arXiv: 2008.07014 [astro-ph.HE] (p. 122).

[512] R. Abbott et al., *GW190521: A Binary Black Hole Merger with a Total Mass of 150$M_\odot$*, Phys. Rev. Lett. 125.10 (2020), p. 101102, arXiv: 2009.01075 [gr-qc] (pp. 122, 123, 131, 133, 135, 142).

[513] G. Rakavy and G. Shaviv, *Instabilities in Highly Evolved Stellar Models*, "Astrophys. J." 148 (June 1967), p. 803 (pp. 122, 131, 139).

[514] Z. Barkat, G. Rakavy, and N. Sack, *Dynamics of Supernova Explosion Resulting from Pair Formation*, Phys. Rev. Lett. 18 (1967), pp. 379–381 (pp. 122, 131, 139).

[515] G. S. Fraley, *Supernovae Explosions Induced by Pair-Production Instability*, "Astrophys. & Space Sciences" 2.1 (Aug. 1968), pp. 96–114 (pp. 122, 131, 139).

[516] J. R. Bond, W. D. Arnett, and B. J. Carr, *The evolution and fate of Very Massive Objects*, "Astrophys. J." 280 (May 1984), pp. 825–847 (p. 122).

[517] A. Heger and S. E. Woosley, *The nucleosynthetic signature of population III*, Astrophys. J. 567 (2002), pp. 532–543, arXiv: astro-ph/0107037 (pp. 122, 131, 139).

[518] S. E. Woosley, S. Blinnikov, and A. Heger, *Pulsational pair instability as an explanation for the most luminous supernovae*, Nature 450 (2007), p. 390, arXiv: 0710.3314 [astro-ph] (pp. 122, 131, 139).

[519] K. Belczynski et al., *The Effect of Pair-Instability Mass Loss on Black Hole Mergers*, Astron. Astrophys. 594 (2016), A97, arXiv: 1607.03116 [astro-ph.HE] (pp. 122, 131, 139).

[520] S. E. Woosley, *Pulsational Pair-Instability Supernovae*, Astrophys. J. 836.2 (2017), p. 244, arXiv: 1608.08939 [astro-ph.HE] (pp. 122, 131, 139).

[521] S. Stevenson, M. Sampson, J. Powell, A. Vigna-Gómez, C. J. Neijssel, D. Szécsi, and I. Mandel, *The impact of pair-instability mass loss on the binary black hole mass distribution*, (Apr. 2019), arXiv: 1904.02821 [astro-ph.HE] (pp. 122, 131, 139).

[522] R. Farmer, M. Renzo, S. E. de Mink, P. Marchant, and S. Justham, *Mind the gap: The location of the lower edge of the pair instability supernova black hole mass gap*, (Oct. 2019), arXiv: 1910.12874 [astro-ph.SR] (pp. 122, 131, 139).

[523] M. Renzo, R. J. Farmer, S. Justham, S. E. de Mink, Y. Götberg, and P. Marchant, *Sensitivity of the lower-edge of the pair instability black hole mass gap to the treatment of time dependent convection*, Mon. Not. Roy. Astron. Soc. 493.3 (2020), pp. 4333–4341, arXiv: 2002.08200 [astro-ph.SR] (pp. 122, 131, 139).

[524] M. Mapelli, M. Spera, E. Montanari, M. Limongi, A. Chieffi, N. Giacobbo, et al., *Impact of the Rotation and Compactness of Progenitors on the Mass of Black Holes*, Astrophys. J. 888 (2020), p. 76, arXiv: 1909.01371 [astro-ph.HE] (pp. 122, 139).

[525] D. Croon, S. D. McDermott, and J. Sakstein, *New physics and the black hole mass gap*, Phys. Rev. D 102.11 (2020), p. 115024, arXiv: 2007.07889 [gr-qc] (pp. 122, 139).

[526] P. Marchant and T. Moriya, *The impact of stellar rotation on the black hole mass-gap from pair-instability supernovae*, Astron. Astrophys. 640 (2020), p. L18, arXiv: 2007.06220 [astro-ph.HE] (pp. 122, 139).

[527] J. Ziegler and K. Freese, *Filling the Black Hole Mass Gap: Avoiding Pair Instability in Massive Stars through Addition of Non-Nuclear Energy*, (Oct. 2020), arXiv: 2010.00254 [astro-ph.HE] (pp. 122, 139).

[528] K. Belczynski, *The most ordinary formation of the most unusual double black hole merger*, Astrophys. J. Lett. 905.2 (2020), p. L15, arXiv: 2009.13526 [astro-ph.HE] (pp. 122, 131, 139).

[529] R. Abbott et al., *GW190814: Gravitational Waves from the Coalescence of a 23 Solar Mass Black Hole with a 2.6 Solar Mass Compact Object*, Astrophys. J. Lett. 896.2 (2020), p. L44, arXiv: 2006.12611 [astro-ph.HE] (pp. 122, 124, 142).

[530] C. Pankow, L. Sampson, L. Perri, E. Chase, S. Coughlin, M. Zevin, and V. Kalogera, *Astrophysical Prior Information and Gravitational-wave Parameter Estimation*, Astrophys. J. 834.2 (2017), p. 154, arXiv: 1610.05633 [astro-ph.HE] (pp. 122, 124, 149, 206).

[531] S. Vitale, D. Gerosa, C.-J. Haster, K. Chatziioannou, and A. Zimmerman, *Impact of Bayesian Priors on the Characterization of Binary Black Hole Coalescences*, Phys. Rev. Lett. 119.25 (2017), p. 251103, arXiv: 1707.04637 [gr-qc] (pp. 122, 124, 149, 206).

[532] M. Zevin, C. P. L. Berry, S. Coughlin, K. Chatziioannou, and S. Vitale, *You Can't Always Get What You Want: The Impact of Prior Assumptions on Interpreting GW190412*, Astrophys. J. Lett. 899.1 (2020), p. L17, arXiv: 2006.11293 [astro-ph.HE] (pp. 122, 124, 149, 151, 206).

[533] B. P. Abbott et al., *Multi-messenger Observations of a Binary Neutron Star Merger*, Astrophys. J. Lett. 848.2 (2017), p. L12, arXiv: 1710.05833 [astro-ph.HE] (p. 124).

[534] B. P. Abbott et al., *GW190425: Observation of a Compact Binary Coalescence with Total Mass ∼ 3.4$M_\odot$*, Astrophys. J. Lett. 892.1 (2020), p. L3, arXiv: 2001.01761 [astro-ph.HE] (p. 124).

[535] I. Mandel, W. M. Farr, and J. R. Gair, *Extracting distribution parameters from multiple uncertain observations with selection biases*, Mon. Not. Roy. Astron. Soc. 486.1 (2019), pp. 1086–1093, arXiv: 1809.02063 [physics.data-an] (pp. 124, 205, 207).

[536] S. Vitale, D. Gerosa, W. M. Farr, and S. R. Taylor, *Inferring the properties of a population of compact binaries in presence of selection effects*, (July 2020), arXiv: 2007.05579 [astro-ph.IM] (pp. 124, 205).

[537] L. S. Finn and D. F. Chernoff, *Observing binary inspiral in gravitational radiation: One interferometer*, Phys. Rev. D 47 (1993), pp. 2198–2219, arXiv: gr-qc/9301003 (p. 124).

[538] L. S. Finn, *Binary inspiral, gravitational radiation, and cosmology*, Phys. Rev. D 53 (1996), pp. 2878–2894, arXiv: gr-qc/9601048 (p. 124).

[539] B. P. Abbott et al., *The Rate of Binary Black Hole Mergers Inferred from Advanced LIGO Observations Surrounding GW150914*, Astrophys. J. Lett. 833.1 (2016), p. L1, arXiv: 1602.03842 [astro-ph.HE] (p. 124).

[540] B. P. Abbott et al., *Binary Black Hole Population Properties Inferred from the First and Second Observing Runs of Advanced LIGO and Advanced Virgo*, Astrophys. J. Lett. 882.2 (2019), p. L24, arXiv: 1811.12940 [astro-ph.HE] (pp. 124, 134).

[541] M. Dominik, E. Berti, R. O'Shaughnessy, I. Mandel, K. Belczynski, C. Fryer, et al., *Double Compact Objects III: Gravitational Wave Detection Rates*, Astrophys. J. 806.2 (2015), p. 263, arXiv: 1405.7016 [astro-ph.HE] (pp. 125, 207).

[542] A. Nitz, I. Harry, D. Brown, C. M. Biwer, J. Willis, T. D. Canton, et al., *gwastro/pycbc: 1.18.0 release of PyCBC*. Version v1.18.0. Feb. 2021, URL: https://doi.org/10.5281/zenodo.4556907 (p. 125).

[543] S. Husa, S. Khan, M. Hannam, M. Pürrer, F. Ohme, X. Jiménez Forteza, and A. Bohé, *Frequency-domain gravitational waves from nonprecessing black-hole binaries. I. New numerical waveforms and anatomy of the signal*, Phys. Rev. D 93.4 (2016), p. 044006, arXiv: 1508.07250 [gr-qc] (pp. 125, 126).

[544] S. Khan, S. Husa, M. Hannam, F. Ohme, M. Pürrer, X. Jiménez Forteza, and A. Bohé, *Frequency-domain gravitational waves from nonprecessing black-hole binaries. II. A phenomenological model for the advanced detector era*, Phys. Rev. D 93.4 (2016), p. 044007, arXiv: 1508.07253 [gr-qc] (pp. 125, 126).

[545] K. W. K. Wong, K. K. Y. Ng, and E. Berti, *Gravitational-wave signal-to-noise interpolation via neural networks*, (July 2020), arXiv: 2007.10350 [astro-ph.HE] (p. 125).

[546] D. Gerosa, G. Pratten, and A. Vecchio, *Gravitational-wave selection effects using neural-network classifiers*, Phys. Rev. D 102.10 (2020), p. 103020, arXiv: 2007.06585 [astro-ph.HE] (p. 125).

[547] A. D. Dolgov, A. G. Kuranov, N. A. Mitichkin, S. Porey, K. A. Postnov, O. S. Sazhina, and I. V. Simkin, *On mass distribution of coalescing black holes*, JCAP 12 (2020), p. 017, arXiv: 2005.00892 [astro-ph.CO] (p. 126).

[548] A. Hall, A. D. Gow, and C. T. Byrnes, *Bayesian analysis of LIGO-Virgo mergers: Primordial vs. astrophysical black hole populations*, Phys. Rev. D 102 (2020), p. 123524, arXiv: 2008.13704 [astro-ph.CO] (pp. 126, 135, 136, 140).

[549] P. Madau and M. Dickinson, *Cosmic Star Formation History*, Ann. Rev. Astron. Astrophys. 52 (2014), pp. 415–486, arXiv: 1403.0007 [astro-ph.CO] (p. 129).

[550] B. S. Sathyaprakash et al., *Extreme Gravity and Fundamental Physics*, (Mar. 2019), arXiv: 1903.09221 [astro-ph.HE] (p. 130).

[551] J. Sakstein, D. Croon, S. D. McDermott, M. C. Straight, and E. J. Baxter, *Beyond the Standard Model Explanations of GW190521*, Phys. Rev. Lett. 125.26 (2020), p. 261105, arXiv: 2009.01213 [gr-qc] (p. 131).

[552] G. Costa, A. Bressan, M. Mapelli, P. Marigo, G. Iorio, and M. Spera, *Formation of GW190521 from stellar evolution: the impact of the hydrogen-rich envelope, dredge-up and $^{12}C(\alpha, \gamma)^{16}O$ rate on the pair-instability black hole mass gap*, Mon. Not. Roy. Astron. Soc.

[553] S. E. Woosley and A. Heger, *The Pair-Instability Mass Gap for Black Holes*, (Mar. 2021), arXiv: 2103 . 07933 [astro-ph.SR] (p. 131).

[554] E. J. Baxter, D. Croon, S. D. Mcdermott, and J. Sakstein, *Find the Gap: Black Hole Population Analysis with an Astrophysically Motivated Mass Function*, (Apr. 2021), arXiv: 2104.02685 [astro-ph.CO] (p. 131).

[555] C. L. Rodriguez, M. Zevin, P. Amaro-Seoane, S. Chatterjee, K. Kremer, F. A. Rasio, and C. S. Ye, *Black holes: The next generation—repeated mergers in dense star clusters and their gravitational-wave properties*, Phys. Rev. D 100.4 (2019), p. 043027, arXiv: 1906. 10260 [astro-ph.HE] (pp. 131, 134, 139, 141, 212).

[556] C. Kimball, C. Talbot, C. P. L. Berry, M. Carney, M. Zevin, E. Thrane, and V. Kalogera, *Black Hole Genealogy: Identifying Hierarchical Mergers with Gravitational Waves*, Astrophys. J. 900.2 (2020), p. 177, arXiv: 2005 . 00023 [astro-ph.HE] (pp. 131, 134, 139, 212).

[557] J. Samsing and K. Hotokezaka, *Populating the Black Hole Mass Gaps In Stellar Clusters: General Relations and Upper Limits*, (June 2020), arXiv: 2006 . 09744 [astro-ph.HE] (pp. 131, 139).

[558] M. Mapelli, F. Santoliquido, Y. Bouffanais, M. A. Sedda, N. Giacobbo, M. C. Artale, and A. Ballone, *Hierarchical mergers in young, globular and nuclear star clusters: black hole masses and merger rates*, (July 2020), arXiv: 2007 . 15022 [astro-ph.HE] (pp. 131, 134, 139, 212).

[559] U. N. Di Carlo, M. Mapelli, Y. Bouffanais, N. Giacobbo, F. Santoliquido, A. Bressan, et al., *Binary black holes in the pair-instability mass gap*, Mon. Not. Roy. Astron. Soc. 497.1 (2020), pp. 1043–1049, arXiv: 1911 . 01434 [astro-ph.HE] (pp. 131, 134).

[560] U. N. Di Carlo et al., *Binary black holes in young star clusters: the impact of metallicity*, Mon. Not. Roy. Astron. Soc. 498.1 (2020), pp. 495–506, arXiv: 2004 . 09525 [astro-ph.HE] (pp. 131, 134).

[561] Z. Roupas and D. Kazanas, *Generation of massive stellar black holes by rapid gas accretion in primordial dense clusters*, Astron. Astrophys. 632 (2019), p. L8, arXiv: 1911.03915 (p. 131).

[562] R. Abbott et al., *Properties and Astrophysical Implications of the 150 $M_\odot$ Binary Black Hole Merger GW190521*, Astrophys. J. Lett. 900.1 (2020), p. L13, arXiv: 2009 . 01190 [astro-ph.HE] (pp. 132, 133).

[563] A. Cruz-Osorio, F. D. Lora-Clavijo, and C. Herdeiro, *GW190521 formation scenarios via relativistic accretion*, (Jan. 2021), arXiv: 2101. 01705 [astro-ph.HE] (p. 132).

[564] C. Cutler and E. E. Flanagan, *Gravitational waves from merging compact binaries: How accurately can one extract the binary's parameters from the inspiral wave form?*, Phys. Rev. D 49 (1994), pp. 2658–2697, arXiv: gr - qc / 9402014 (pp. 135, 152).

[565] E. Poisson and C. M. Will, *Gravitational waves from inspiraling compact binaries: Parameter estimation using second postNewtonian wave forms*, Phys. Rev. D 52 (1995), pp. 848–855, arXiv: gr-qc/9502040 (pp. 135, 152).

[566] E. Baird, S. Fairhurst, M. Hannam, and P. Murphy, *Degeneracy between mass and spin in black-hole-binary waveforms*, Phys. Rev. D 87.2 (2013), p. 024035, arXiv: 1211 . 0546 [gr-qc] (pp. 135, 152).

[567] M. Pürrer, M. Hannam, P. Ajith, and S. Husa, *Testing the validity of the single-spin approximation in inspiral-merger-ringdown waveforms*, Phys. Rev. D 88 (2013), p. 064007, arXiv: 1306.2320 [gr-qc] (pp. 135, 152).

[568] K. Chatziioannou, N. Cornish, A. Klein, and N. Yunes, *Detection and Parameter Estimation of Gravitational Waves from Compact Binary Inspirals with Analytical Double-Precessing Templates*, Phys. Rev. D 89.10 (2014), p. 104023, arXiv: 1404.3180 [gr-qc] (pp. 135, 152).

[569] K. K. Y. Ng, S. Vitale, A. Zimmerman, K. Chatziioannou, D. Gerosa, and C.-J. Haster, *Gravitational-wave astrophysics with effective-spin measurements: asymmetries and selection biases*, Phys. Rev. D 98.8 (2018), p. 083007, arXiv: 1805.03046 [gr-qc] (pp. 135, 152).

[570] S. Vitale, R. Lynch, R. Sturani, and P. Graff, *Use of gravitational waves to probe the formation channels of compact binaries*, Class. Quant. Grav. 34.3 (2017), 03LT01, arXiv: 1503.04307 [gr-qc] (p. 137).

[571] S. Stevenson, F. Ohme, and S. Fairhurst, *Distinguishing compact binary population synthesis models using gravitational-wave observations of coalescing binary black holes*, Astrophys. J. 810.1 (2015), p. 58, arXiv: 1504.07802 [astro-ph.HE] (p. 137).

[572] M. Zevin, C. Pankow, C. L. Rodriguez, L. Sampson, E. Chase, V. Kalogera, and F. A. Rasio, *Constraining Formation Models of Binary Black Holes with Gravitational-Wave Observations*, Astrophys. J. 846.1 (2017), p. 82, arXiv: 1704.07379 [astro-ph.HE] (p. 137).

[573] C. Talbot and E. Thrane, *Determining the population properties of spinning black holes*, Phys. Rev. D 96.2 (2017), p. 023012, arXiv: 1704. 08370 [astro-ph.HE] (p. 137).

[574] S. R. Taylor and D. Gerosa, *Mining Gravitational-wave Catalogs To Understand Binary Stellar Evolution: A New Hierarchical Bayesian Framework*, Phys. Rev. D 98.8 (2018), p. 083017, arXiv: 1806 . 08365 [astro-ph.HE] (pp. 137, 207).

[575] Y. Bouffanais, M. Mapelli, D. Gerosa, U. N. Di Carlo, N. Giacobbo, E. Berti, and V. Baibhav, *Constraining the fraction of binary black holes formed in isolation and young star clusters with gravitational-wave data*, Astrophys. J. 886.1 (2019), arXiv: 1905 . 11054 [astro-ph.HE] (p. 137).

[576] C. Kimball et al., *Evidence for hierarchical black hole mergers in the second LIGO–Virgo gravitational-wave catalog*, (Nov. 2020), arXiv: 2011.05332 [astro-ph.HE] (pp. 137, 212).

[577] M. Zevin, S. S. Bavera, C. P. L. Berry, V. Kalogera, T. Fragos, P. Marchant, et al., *One Channel to Rule Them All? Constraining the Origins of Binary Black Holes Using Multiple Formation Pathways*, Astrophys. J. 910.2 (2021), p. 152, arXiv: 2011 . 10057 [astro-ph.HE] (pp. 137, 140, 141, 142, 146, 148, 204, 210, 211, 212, 213).

[578] K. W. K. Wong, K. Breivik, K. Kremer, and T. Callister, *Joint constraints on the field-cluster mixing fraction, common envelope efficiency, and globular cluster radii from a population of binary hole mergers via deep learning*, (Nov. 2020), arXiv: 2011 . 03564 [astro-ph.HE] (p. 137).

[579] K. K. Y. Ng, S. Vitale, W. M. Farr, and C. L. Rodriguez, *Probing multiple populations of compact binaries with third-generation gravitational-wave detectors*, (Dec. 2020), arXiv: 2012 . 09876 [astro-ph.CO] (pp. 137, 154, 155).

[580] H. Jeffreys, *The Theory of Probability (3rd ed.), Oxford, England.* 1998 (pp. 138, 208).

[581] D. Gerosa, N. Giacobbo, and A. Vecchio, *High mass but low spin: an exclusion region to rule out hierarchical black-hole mergers as a mechanism to populate the pair-instability mass gap*, (Apr. 2021), arXiv: 2104 . 11247 [astro-ph.HE] (p. 140).

[582] S. S. Bavera et al., *The impact of mass-transfer physics on the observable properties of field binary black hole populations*, Astron. Astrophys. 647 (2021), A153, arXiv: 2010 . 16333 [astro-ph.HE] (pp. 141, 210).

[583] T. Fragos et al, (2021). In preparation (pp. 141, 210).

[584] K. Breivik et al., *COSMIC Variance in Binary Population Synthesis*, Astrophys. J. 898.1 (2020), p. 71, arXiv: 1911 . 00903 [astro-ph.HE] (pp. 141, 210).

[585] B. Paxton, R. Smolec, J. Schwab, A. Gautschy, L. Bildsten, M. Cantiello, et al., *Modules for Experiments in Stellar Astrophysics (MESA): Pulsating Variable Stars, Rotation, Convective Boundaries, and Energy Conservation*, The Astrophysical Journal Supplement Series 243.1, 10 (July 2019), p. 10, arXiv: 1903.01426 [astro-ph.SR] (pp. 141, 210).

[586] F. Antonini, M. Gieles, and A. Gualandris, *Black hole growth through hierarchical black hole mergers in dense star clusters: implications for gravitational wave detections*, Mon. Not. Roy. Astron. Soc. 486.4 (2019), pp. 5008–5021, arXiv: 1811 . 03640 [astro-ph.HE] (pp. 141, 212).

[587] V. Gayathri, J. Healy, J. Lange, B. O'Brien, M. Szczepanczyk, I. Bartos, et al., *GW190521 as a Highly Eccentric Black Hole Merger*, (Sept. 2020), arXiv: 2009 . 05461 [astro-ph.HE] (p. 142).

[588] I. M. Romero-Shaw, P. D. Lasky, E. Thrane, and J. C. Bustillo, *GW190521: orbital eccentricity and signatures of dynamical formation in a binary black hole merger signal*, Astrophys. J. Lett. 903.1 (2020), p. L5, arXiv: 2009.04771 [astro-ph.HE] (p. 142).

[589] J. C. Bustillo, N. Sanchis-Gual, A. Torres-Forné, J. A. Font, A. Vajpeyi, R. Smith, et al., *GW190521 as a Merger of Proca Stars: A Potential New Vector Boson of $8.7 \times 10^{-13}$ eV*, Phys. Rev. Lett. 126.8 (2021), p. 081101, arXiv: 2009.05376 [gr-qc] (p. 142).

[590] A. H. Nitz and C. D. Capano, *GW190521 may be an intermediate mass ratio inspiral*, Astrophys. J. Lett. 907.1 (2021), p. L9, arXiv: 2010. 12558 [astro-ph.HE] (p. 142).

[591] M. Shibata, K. Kiuchi, S. Fujibayashi, and Y. Sekiguchi, *Alternative possibility of GW190521: Gravitational waves from high-mass black hole-disk systems*, Phys. Rev. D 103.6 (2021), p. 063037, arXiv: 2101 . 05440 [astro-ph.HE] (p. 142).

[592] H. Estellés et al., *A detailed analysis of GW190521 with phenomenological waveform models*, (May 2021), arXiv: 2105 . 06360 [gr-qc] (p. 142).

[593] R. Gamba, M. Breschi, G. Carullo, P. Rettegno, S. Albanesi, S. Bernuzzi, and A. Nagar, *GW190521: A dynamical capture of two black holes*, (June 2021), arXiv: 2106 . 05575 [gr-qc] (p. 142).

[594] B. O'Brien, M. Szczepanczyk, V. Gayathri, I. Bartos, G. Vedovato, G. Prodi, et al., *Detection of LIGO-Virgo binary black holes in the pair-instability mass gap*, (June 2021), arXiv: 2106.00605 [gr-qc] (p. 142).

[595] A. K. Mehta, A. Buonanno, J. Gair, M. C. Miller, E. Farag, R. J. deBoer, et al., *Observing intermediate-mass black holes and the upper–stellar-mass gap with LIGO and Virgo*, (May 2021), arXiv: 2105.06366 [gr-qc] (p. 142).

[596] C. D. Capano, M. Cabero, J. Westerweck, J. Abedi, S. Kastha, A. H. Nitz, et al., *Observation of a multimode quasi-normal spectrum from a perturbed black hole*, (May 2021), arXiv: 2105.05238 [gr-qc] (p. 142).

[597] S. Olsen, J. Roulet, H. S. Chia, L. Dai, T. Venumadhav, B. Zackay, and M. Zaldarriaga, *Mapping the Likelihood of GW190521 with Diverse Mass and Spin Priors*, (June 2021), arXiv: 2106.13821 [astro-ph.HE] (p. 142).

[598] A. Olejak, K. Belczynski, and N. Ivanova, *The impact of common envelope development criteria on the formation of LIGO/Virgo sources*, (Feb. 2021), arXiv: 2102.05649 [astro-ph.HE] (p. 142).

[599] B. McKernan, K. E. S. Ford, R. O'Shaughnessy, and D. Wysocki, *Monte Carlo simulations of black hole mergers in AGN discs: Low $\chi_{eff}$ mergers and predictions for LIGO*, Mon. Not. Roy. Astron. Soc. 494.1 (2020), pp. 1203–1216, arXiv: 1907.04356 [astro-ph.HE] (pp. 144, 210).

[600] H. Tagawa, Z. Haiman, and B. Kocsis, *Formation and Evolution of Compact Object Binaries in AGN Disks*, Astrophys. J. 898.1 (2020), p. 25, arXiv: 1912.08218 [astro-ph.GA] (pp. 144, 210).

[601] Y. Yang, I. Bartos, Z. Haiman, B. Kocsis, Z. Marka, N. Stone, and S. Marka, *AGN Disks Harden the Mass Distribution of Stellar-mass Binary Black Hole Mergers*, Astrophys. J. 876.2 (2019), p. 122, arXiv: 1903.01405 [astro-ph.HE] (pp. 144, 210).

[602] N. Fernandez and S. Profumo, *Unraveling the origin of black holes from effective spin measurements with LIGO-Virgo*, JCAP 08 (2019), p. 022, arXiv: 1905.13019 [astro-ph.HE] (p. 149).

[603] M. Safarzadeh, *The branching ratio of LIGO binary black holes*, Astrophys. J. Lett. 892.1 (2020), p. L8, arXiv: 2003.02764 [astro-ph.HE] (p. 149).

[604] J. García-Bellido, J. F. Nuño Siles, and E. Ruiz Morales, *Bayesian analysis of the spin distribution of LIGO/Virgo black holes*, Phys. Dark Univ. 31 (2021), p. 100791, arXiv: 2010.13811 [astro-ph.CO] (p. 149).

[605] S. Dwyer, D. Sigg, S. W. Ballmer, L. Barsotti, N. Mavalvala, and M. Evans, *Gravitational wave detector with cosmological reach*, Phys. Rev. D 91.8 (2015), p. 082001, arXiv: 1410.0612 [astro-ph.IM] (p. 149).

[606] B. P. Abbott et al., *GW170608: Observation of a 19-solar-mass Binary Black Hole Coalescence*, Astrophys. J. Lett. 851 (2017), p. L35, arXiv: 1711.05578 [astro-ph.HE] (pp. 149, 151).

[607] I. Mandel and T. Fragos, *An alternative interpretation of GW190412 as a binary black hole merger with a rapidly spinning secondary*, Astrophys. J. Lett. 895.2 (2020), p. L28, arXiv: 2004.09288 [astro-ph.HE] (pp. 151, 206).

[608] D. Gerosa, S. Vitale, and E. Berti, *Astrophysical implications of GW190412 as a remnant of a previous black-hole merger*, Phys. Rev. Lett. 125.10 (2020), p. 101103, arXiv: 2005.04243 [astro-ph.HE] (pp. 151, 153, 206).

[609] S. Vitale and M. Evans, *Parameter estimation for binary black holes with networks of third generation gravitational-wave detectors*, Phys. Rev. D 95.6 (2017), p. 064052, arXiv: 1610.06917 [gr-qc] (p. 153).

[610] M. Fishbach, D. E. Holz, and W. M. Farr, *Does the Black Hole Merger Rate Evolve with Redshift?*, Astrophys. J. Lett. 863.2 (2018), p. L41, arXiv: 1805.10270 [astro-ph.HE] (p. 154).

[611] T. Callister, M. Fishbach, D. Holz, and W. Farr, *Shouts and Murmurs: Combining Individual Gravitational-Wave Sources with the Stochastic Background to Measure the History of Binary Black Hole Mergers*, Astrophys. J. Lett. 896.2 (2020), p. L32, arXiv: 2003.12152 [astro-ph.HE] (p. 154).

[612] M. Fishbach, Z. Doctor, T. Callister, B. Edelman, J. Ye, R. Essick, et al., *When are LIGO/Virgo's Big Black-Hole Mergers?*, (Jan. 2021), arXiv: 2101.07699 [astro-ph.HE] (p. 154).

[613] S. Vitale, W. M. Farr, K. Ng, and C. L. Rodriguez, *Measuring the star formation rate with gravitational waves from binary black holes*, Astrophys. J. Lett. 886.1 (2019), p. L1, arXiv: 1808.00901 [astro-ph.HE] (p. 154).

[614] X. Ding, K. Liao, M. Biesiada, and Z.-H. Zhu, *Black hole mass function and its evolution – the first prediction for the Einstein Telescope*, (Feb. 2020), arXiv: 2002.02981 [astro-ph.GA] (p. 154).

[615] Z.-C. Chen and Q.-G. Huang, *Distinguishing Primordial Black Holes from Astrophysical Black Holes by Einstein Telescope and Cosmic Explorer*, JCAP 08 (2020), p. 039, arXiv: 1904.02396 [astro-ph.CO] (p. 154).

[616] T. Kinugawa, K. Inayoshi, K. Hotokezaka, D. Nakauchi, and T. Nakamura, *Possible Indirect Confirmation of the Existence of Pop III Massive Stars by Gravitational Wave*, Mon. Not. Roy. Astron. Soc. 442.4 (2014), pp. 2963–2992, arXiv: 1402.6672 [astro-ph.HE] (pp. 154, 210).

[617] T. Kinugawa, A. Miyamoto, N. Kanda, and T. Nakamura, *The detection rate of inspiral and quasi-normal modes of Population III binary black holes which can confirm or refute the general relativity in the strong gravity region*, Mon. Not. Roy. Astron. Soc. 456.1 (2016), pp. 1093–1114, arXiv: 1505.06962 [astro-ph.SR] (p. 154).

[618] T. Hartwig, M. Volonteri, V. Bromm, R. S. Klessen, E. Barausse, M. Magg, and A. Stacy, *Gravitational Waves from the Remnants of the First Stars*, Mon. Not. Roy. Astron. Soc. 460.1 (2016), pp. L74–L78, arXiv: 1603.05655 [astro-ph.GA] (p. 154).

[619] R. Schneider, A. Ferrara, B. Ciardi, V. Ferrari, and S. Matarrese, *Gravitational waves signals from the collapse of the first stars*, Mon. Not. Roy. Astron. Soc. 317 (2000), p. 385, arXiv: astro-ph/9909419 (p. 154).

[620] R. Schneider, A. Ferrara, P. Natarajan, and K. Omukai, *First stars, very massive black holes and metals*, Astrophys. J. 571 (2002), pp. 30–39, arXiv: astro-ph/0111341 (p. 154).

[621] R. Schneider, A. Ferrara, R. Salvaterra, K. Omukai, and V. Bromm, *Low-mass relics of early star formation*, Nature 422 (2003), pp. 869–871, arXiv: astro-ph/0304254 (p. 154).

[622] B. Liu and V. Bromm, *Gravitational waves from Population III binary black holes formed by dynamical capture*, Mon. Not. Roy. Astron. Soc. 495.2 (2020), pp. 2475–2495, arXiv: 2003.00065 [astro-ph.CO] (p. 154).

[623] R. Valiante, M. Colpi, R. Schneider, A. Mangiagli, M. Bonetti, G. Cerini, et al., *Unveiling early black hole growth with multifrequency gravitational wave observations*, Mon. Not. Roy. Astron. Soc. 500.3 (2020), pp. 4095–4109, arXiv: 2010.15096 [astro-ph.GA] (pp. 154, 155).

[624] S. M. Koushiappas and A. Loeb, *Maximum redshift of gravitational wave merger events*, Phys. Rev. Lett. 119.22 (2017), p. 221104, arXiv: 1708.07380 [astro-ph.CO] (p. 155).

[625] T. Nakamura et al., *Pre-DECIGO can get the smoking gun to decide the astrophysical or cosmological origin of GW150914-like binary black holes*, PTEP 2016.9 (2016), 093E01, arXiv: 1607.00897 [astro-ph.HE] (p. 155).

[626] K. K. Ng, V. Baibhav, E. Berti, V. De Luca, G. Franciolini, P. Pani, et al., (). in preparation (p. 155).

[627] T. Namikawa, A. Nishizawa, and A. Taruya, *Detecting Black-Hole Binary Clustering via the Second-Generation Gravitational-Wave Detectors*, Phys. Rev. D 94.2 (2016), p. 024013, arXiv: 1603.08072 [astro-ph.CO] (p. 155).

[628] A. Raccanelli, E. D. Kovetz, S. Bird, I. Cholis, and J. B. Munoz, *Determining the progenitors of merging black-hole binaries*, Phys. Rev. D 94.2 (2016), p. 023516, arXiv: 1605.01405 [astro-ph.CO] (p. 155).

[629] G. Scelfo, N. Bellomo, A. Raccanelli, S. Matarrese, and L. Verde, *GW×LSS: chasing the progenitors of merging binary black holes*, JCAP 09 (2018), p. 039, arXiv: 1809.03528 [astro-ph.CO] (p. 155).

[630] G. Cañas Herrera, O. Contigiani, and V. Vardanyan, *Cross-correlation of the astrophysical gravitational-wave background with galaxy clustering*, Phys. Rev. D 102.4 (2020), p. 043513, arXiv: 1910.08353 [astro-ph.CO] (p. 155).

[631] F. Calore, A. Cuoco, T. Regimbau, S. Sachdev, and P. D. Serpico, *Cross-correlating galaxy catalogs and gravitational waves: a tomographic approach*, Phys. Rev. Res. 2 (2020), p. 023314, arXiv: 2002.02466 [astro-ph.CO] (p. 155).

[632] G. Scelfo, M. Spinelli, A. Raccanelli, L. Boco, A. Lapi, and M. Viel, *Gravitational waves × HI intensity mapping: cosmological and astrophysical applications*, (June 2021), arXiv: 2106.09786 [astro-ph.CO] (p. 155).

[633] B. P. Abbott et al., *Search for Subsolar-Mass Ultracompact Binaries in Advanced LIGO's First Observing Run*, Phys. Rev. Lett. 121.23 (2018), p. 231103, arXiv: 1808.04771 [astro-ph.CO] (p. 156).

[634] Y.-F. Wang and A. H. Nitz, *Prospects for detecting gravitational waves from eccentric subsolar mass compact binaries*, Astrophys. J. 912.1 (2021), p. 53, arXiv: 2101.12269 [astro-ph.HE] (p. 156).

[635] A. H. Nitz and Y.-F. Wang, *Search for gravitational waves from the coalescence of sub-solar mass and eccentric compact binaries*, (Feb. 2021), arXiv: 2102.00868 [astro-ph.HE] (p. 156).

[636] A. H. Nitz and Y.-F. Wang, *Search for gravitational waves from the coalescence of sub-solar mass binaries in the first half of Advanced LIGO and Virgo's third observing run*, (June 2021), arXiv: 2106.08979 [astro-ph.HE] (p. 156).

[637] H.-K. Guo, J. Shu, and Y. Zhao, *Using LISA-like Gravitational Wave Detectors to Search for Primordial Black Holes*, Phys. Rev. D 99.2 (2019), p. 023001, arXiv: 1709 . 03500 [astro-ph.CO] (p. 156).

[638] F. Kuhnel, A. Matas, G. D. Starkman, and K. Freese, *Waves from the Centre: Probing PBH and other Macroscopic Dark Matter with LISA*, Eur. Phys. J. C 80.7 (2020), p. 627, arXiv: 1811.06387 [gr-qc] (p. 156).

[639] Y.-F. Wang, Q.-G. Huang, T. G. F. Li, and S. Liao, *Searching for primordial black holes with stochastic gravitational-wave background in the space-based detector frequency band*, Phys. Rev. D 101.6 (2020), p. 063019, arXiv: 1910.07397 [astro-ph.CO] (p. 156).

[640] B. Dasgupta, R. Laha, and A. Ray, *Low Mass Black Holes from Dark Core Collapse*, Phys. Rev. Lett. 126.14 (2021), p. 141105, arXiv: 2009.01825 [astro-ph.HE] (p. 156).

[641] V. Takhistov, G. M. Fuller, and A. Kusenko, *Test for the Origin of Solar Mass Black Holes*, Phys. Rev. Lett. 126.7 (2021), p. 071101, arXiv: 2008.12780 [astro-ph.HE] (p. 156).

[642] C. Yuan and Q.-G. Huang, *A topic review on probing primordial black hole dark matter with scalar induced gravitational waves*, (Mar. 2021), arXiv: 2103 . 04739 [astro-ph.GA] (p. 157).

[643] R. Saito and J. Yokoyama, *Gravitational-Wave Constraints on the Abundance of Primordial Black Holes*, Prog. Theor. Phys. 123 (2010). [Erratum: Prog.Theor.Phys. 126, 351–352 (2011)], pp. 867–886, arXiv: 0912 . 5317 [astro-ph.CO] (pp. 157, 161).

[644] R. M. Shannon et al., *Gravitational waves from binary supermassive black holes missing in pulsar observations*, Science 349.6255 (2015), pp. 1522–1525, arXiv: 1509 . 07320 [astro-ph.CO] (pp. 157, 192).

[645] Z. Arzoumanian et al., *The NANOGrav 11-year Data Set: Pulsar-timing Constraints On The Stochastic Gravitational-wave Background*, Astrophys. J. 859.1 (2018), p. 47, arXiv: 1801.02617 [astro-ph.HE] (pp. 157, 192).

[646] L. Lentati et al., *European Pulsar Timing Array Limits On An Isotropic Stochastic Gravitational-Wave Background*, Mon. Not. Roy. Astron. Soc. 453.3 (2015), pp. 2576–2598, arXiv: 1504.03692 [astro-ph.CO] (pp. 157, 192).

[647] P. E. Dewdney, P. J. Hall, R. T. Schilizzi, and T. J. L. W. Lazio, *The Square Kilometre Array*, IEEE Proceedings 97.8 (2009), pp. 1482–1496 (p. 157).

[648] C. J. Moore, R. H. Cole, and C. P. L. Berry, *Gravitational-wave sensitivity curves*, Class. Quant. Grav. 32.1 (2015), p. 015014, arXiv: 1408.0740 [gr-qc] (pp. 157, 192).

[649] V. Acquaviva, N. Bartolo, S. Matarrese, and A. Riotto, *Second order cosmological perturbations from inflation*, Nucl. Phys. B 667 (2003), pp. 119–148, arXiv: astro - ph / 0209156 [astro-ph] (pp. 157, 158).

[650] S. Mollerach, D. Harari, and S. Matarrese, *CMB polarization from secondary vector and tensor modes*, Phys. Rev. D 69 (2004), p. 063002, arXiv: astro-ph/0310711 (p. 157).

[651] K. N. Ananda, C. Clarkson, and D. Wands, *The Cosmological gravitational wave background from primordial density perturbations*, Phys. Rev. D 75 (2007), p. 123518, arXiv: gr-qc/0612013 (pp. 157, 161).

[652] D. Baumann, P. J. Steinhardt, K. Takahashi, and K. Ichiki, *Gravitational Wave Spectrum Induced by Primordial Scalar Perturbations*, Phys. Rev. D 76 (2007), p. 084019, arXiv: hep-th/0703290 (pp. 157, 176).

[653] E. Bugaev and P. Klimai, *Induced gravitational wave background and primordial black holes*, Phys. Rev. D 81 (2010), p. 023517, arXiv: 0908.0664 [astro-ph.CO] (pp. 157, 161).

[654] H. Assadullahi and D. Wands, *Constraints on primordial density perturbations from induced gravitational waves*, Phys. Rev. D 81 (2010), p. 023527, arXiv: 0907.4073 [astro-ph.CO] (p. 157).

[655] T. Nakama and T. Suyama, *Primordial black holes as a novel probe of primordial gravitational waves*, Phys. Rev. D 92.12 (2015), p. 121304, arXiv: 1506 . 05228 [gr-qc] (p. 157).

[656] T. Nakama and T. Suyama, *Primordial black holes as a novel probe of primordial gravitational waves. II: Detailed analysis*, Phys. Rev. D 94.4 (2016), p. 043507, arXiv: 1605.04482 [gr-qc] (p. 157).

[657] G. Domènech, *Induced gravitational waves in a general cosmological background*, Int. J. Mod. Phys. D 29.03 (2020), p. 2050028, arXiv: 1912. 05583 [gr-qc] (p. 157).

[658] S. Pi and M. Sasaki, *Gravitational Waves Induced by Scalar Perturbations with a Lognormal Peak*, JCAP 09 (2020), p. 037, arXiv: 2005.12306 [gr-qc] (pp. 157, 161).

[659] M. C. Guzzetti, N. Bartolo, M. Liguori, and S. Matarrese, *Gravitational waves from inflation*, Riv. Nuovo Cim. 39.9 (2016), pp. 399–495, arXiv: 1605.01615 [astro-ph.CO] (p. 157).

[660] N. Bartolo et al., *Science with the space-based interferometer LISA. IV: Probing inflation with gravitational waves*, JCAP 12 (2016), p. 026, arXiv: 1610.06481 [astro-ph.CO] (pp. 157, 162).

[661] M. Geller, A. Hook, R. Sundrum, and Y. Tsai, *Primordial Anisotropies in the Gravitational Wave Background from Cosmological Phase Transitions*, Phys. Rev. Lett. 121.20 (2018), p. 201303, arXiv: 1803.10780 [hep-ph] (p. 157).

[662] A. Ricciardone and G. Tasinato, *Anisotropic tensor power spectrum at interferometer scales induced by tensor squeezed non-Gaussianity*, JCAP 02 (2018), p. 011, arXiv: 1711.02635 [astro-ph.CO] (p. 157).

[663] E. Dimastrogiovanni, M. Fasiello, and G. Tasinato, *Searching for Fossil Fields in the Gravity Sector*, Phys. Rev. Lett. 124.6 (2020), p. 061302, arXiv: 1906.07204 [astro-ph.CO] (p. 157).

[664] J. L. Cook and L. Sorbo, *Particle production during inflation and gravitational waves detectable by ground-based interferometers*, Phys. Rev. D 85 (2012). [Erratum: Phys.Rev.D 86, 069901 (2012)], p. 023534, arXiv: 1109.0022 [astro-ph.CO] (p. 157).

[665] C. Caprini, D. G. Figueroa, R. Flauger, G. Nardini, M. Peloso, M. Pieroni, et al., *Reconstructing the spectral shape of a stochastic gravitational wave background with LISA*, JCAP 11 (2019), p. 017, arXiv: 1906.09244 [astro-ph.CO] (p. 157).

[666] N. Bartolo, V. Domcke, D. G. Figueroa, J. García-Bellido, M. Peloso, M. Pieroni, et al., *Probing non-Gaussian Stochastic Gravitational Wave Backgrounds with LISA*, JCAP 11 (2018), p. 034, arXiv: 1806.02819 [astro-ph.CO] (p. 157).

[667] C. Yuan and Q.-G. Huang, *Gravitational waves induced by the local-type non-Gaussian curvature perturbations*, (July 2020), arXiv: 2007.10686 [astro-ph.CO] (pp. 157, 186).

[668] P. Adshead, K. D. Lozanov, and Z. J. Weiner, *Non-Gaussianity and the induced gravitational wave background*, (May 2021), arXiv: 2105.01659 [astro-ph.CO] (p. 157).

[669] V. Atal and G. Domènech, *Probing non-Gaussianities with the high frequency tail of induced gravitational waves*, (Mar. 2021), arXiv: 2103.01056 [astro-ph.CO] (p. 157).

[670] S. Weinberg, *Damping of tensor modes in cosmology*, Phys. Rev. D 69 (2004), p. 023503, arXiv: astro-ph/0306304 (p. 158).

[671] J. R. Espinosa, D. Racco, and A. Riotto, *A Cosmological Signature of the SM Higgs Instability: Gravitational Waves*, JCAP 09 (2018), p. 012, arXiv: 1804.07732 [hep-ph] (pp. 159, 160, 162).

[672] K. Kohri and T. Terada, *Semianalytic calculation of gravitational wave spectrum nonlinearly induced from primordial curvature perturbations*, Phys. Rev. D 97.12 (2018), p. 123532, arXiv: 1804.08577 [gr-qc] (p. 159).

[673] R.-G. Cai, S. Pi, and M. Sasaki, *Universal infrared scaling of gravitational wave background spectra*, Phys. Rev. D 102.8 (2020), p. 083528, arXiv: 1909.13728 [astro-ph.CO] (p. 160).

[674] A. Hook, G. Marques-Tavares, and D. Racco, *Causal gravitational waves as a probe of free streaming particles and the expansion of the Universe*, JHEP 02 (2021), p. 117, arXiv: 2010.03568 [hep-ph] (p. 160).

[675] T. Matsubara, *Statistics of Fourier Modes in Non-Gaussian Fields*, Astrophys. J. Suppl. 170 (2007), p. 1, arXiv: astro-ph/0610536 (p. 165).

[676] W. Hu and A. Cooray, *Gravitational time delay effects on cosmic microwave background anisotropies*, Phys. Rev. D 63 (2001), p. 023504, arXiv: astro-ph/0008001 (pp. 169, 170).

[677] Gorbunov, D.S. and Rubakov, V.A., *Introduction to the Theory of the Early Universe: Cosmological Perturbations and Inflationary Theory*. v. 2. World Scientific, 2011. ISBN: 9789814343787, URL: https://books.google.it/books?id=scgqbaKeeCEC (p. 169).

[678] A. Margalit, C. R. Contaldi, and M. Pieroni, *Phase decoherence of gravitational wave backgrounds*, Phys. Rev. D 102.8 (2020), p. 083506, arXiv: 2004.01727 [astro-ph.CO] (p. 170).

[679] V. Alba and J. Maldacena, *Primordial gravity wave background anisotropies*, JHEP 03 (2016), p. 115, arXiv: 1512.01531 [hep-th] (p. 170).

[680] C. R. Contaldi, *Anisotropies of Gravitational Wave Backgrounds: A Line Of Sight Approach*, Phys. Lett. B 771 (2017), pp. 9–12, arXiv: 1609.08168 [astro-ph.CO] (pp. 170, 173).

[681] D. Bertacca, A. Raccanelli, N. Bartolo, and S. Matarrese, *Cosmological perturbation effects on gravitational-wave luminosity distance estimates*, Phys. Dark Univ. 20 (2018), pp. 32–40, arXiv: 1702.01750 [gr-qc] (p. 170).

[682] G. Cusin, C. Pitrou, and J.-P. Uzan, *Anisotropy of the astrophysical gravitational wave background: Analytic expression of the angular power spectrum and correlation with cosmological observations*, Phys. Rev. D 96.10 (2017), p. 103019, arXiv: 1704 . 06184 [astro-ph.CO] (p. 170).

[683] A. C. Jenkins and M. Sakellariadou, *Anisotropies in the stochastic gravitational-wave background: Formalism and the cosmic string case*, Phys. Rev. D 98.6 (2018), p. 063509, arXiv: 1802.06046 [astro-ph.CO] (p. 170).

[684] G. Cusin, R. Durrer, and P. G. Ferreira, *Polarization of a stochastic gravitational wave background through diffusion by massive structures*, Phys. Rev. D 99.2 (2019), p. 023534, arXiv: 1807.10620 [astro-ph.CO] (p. 170).

[685] A. Renzini and C. Contaldi, *Improved limits on a stochastic gravitational-wave background and its anisotropies from Advanced LIGO O1 and O2 runs*, Phys. Rev. D 100.6 (2019), p. 063527, arXiv: 1907.10329 [gr-qc] (p. 170).

[686] N. Bartolo, D. Bertacca, S. Matarrese, M. Peloso, A. Ricciardone, A. Riotto, and G. Tasinato, *Anisotropies and non-Gaussianity of the Cosmological Gravitational Wave Background*, Phys. Rev. D 100.12 (2019), p. 121501, arXiv: 1908 . 00527 [astro-ph.CO] (pp. 170, 173, 174, 175, 188).

[687] C. R. Contaldi, M. Pieroni, A. I. Renzini, G. Cusin, N. Karnesis, M. Peloso, et al., *Maximum likelihood map-making with the Laser Interferometer Space Antenna*, Phys. Rev. D 102.4 (2020), p. 043502, arXiv: 2006 . 03313 [astro-ph.CO] (pp. 170, 188).

[688] Y. Akrami et al., *Planck 2018 results. IX. Constraints on primordial non-Gaussianity*, Astron. Astrophys. 641 (2020), A9, arXiv: 1905. 05697 [astro-ph.CO] (p. 173).

[689] N. Bartolo, A. Hoseinpour, G. Orlando, S. Matarrese, and M. Zarei, *Photon-graviton scattering: A new way to detect anisotropic gravitational waves?*, Phys. Rev. D 98.2 (2018), p. 023518, arXiv: 1804 . 06298 [gr-qc] (p. 174).

[690] N. Bartolo, D. Bertacca, S. Matarrese, M. Peloso, A. Ricciardone, A. Riotto, and G. Tasinato, *Characterizing the cosmological gravitational wave background: Anisotropies and non-Gaussianity*, Phys. Rev. D 102.2 (2020), p. 023527, arXiv: 1912.09433 [astro-ph.CO] (p. 175).

[691] N. Bartolo, E. Komatsu, S. Matarrese, and A. Riotto, *Non-Gaussianity from inflation: Theory and observations*, Phys. Rept. 402 (2004), pp. 103–266, arXiv: astro - ph / 0406398 (pp. 176, 180).

[692] J.-C. Hwang, D. Jeong, and H. Noh, *Gauge dependence of gravitational waves generated from scalar perturbations*, Astrophys. J. 842.1 (2017), p. 46, arXiv: 1704 . 03500 [astro-ph.CO] (p. 176).

[693] G. Domènech and M. Sasaki, *Hamiltonian approach to second order gauge invariant cosmological perturbations*, Phys. Rev. D 97.2 (2018), p. 023521, arXiv: 1709.09804 [gr-qc] (p. 176).

[694] J.-O. Gong, *Analytic integral solutions for induced gravitational waves*, (Sept. 2019), arXiv: 1909.12708 [gr-qc] (pp. 176, 181, 182).

[695] K. Tomikawa and T. Kobayashi, *Gauge dependence of gravitational waves generated at second order from scalar perturbations*, Phys. Rev. D 101.8 (2020), p. 083529, arXiv: 1910.01880 [gr-qc] (p. 176).

[696] E. E. Flanagan and S. A. Hughes, *The Basics of gravitational wave theory*, New J. Phys. 7 (2005), p. 204, arXiv: gr-qc/0501041 (p. 178).

[697] K. A. Malik and D. Wands, *Gauge invariant variables on cosmological hypersurfaces*, (Apr. 1998), arXiv: gr-qc/9804046 (p. 180).

[698] K. A. Malik and D. Wands, *Evolution of second-order cosmological perturbations*, Class. Quant. Grav. 21 (2004), pp. L65–L72, arXiv: astro-ph/0307055 (p. 180).

[699] K. A. Malik and D. Wands, *Cosmological perturbations*, Phys. Rept. 475 (2009), pp. 1–51, arXiv: 0809.4944 [astro-ph] (pp. 180, 214, 216, 218).

[700] F. Arroja, H. Assadullahi, K. Koyama, and D. Wands, *Cosmological matching conditions for gravitational waves at second order*, Phys. Rev. D 80 (2009), p. 123526, arXiv: 0907.3618 [astro-ph.CO] (p. 180).

[701] S. Matarrese, S. Mollerach, and M. Bruni, *Second order perturbations of the Einstein-de Sitter universe*, Phys. Rev. D 58 (1998), p. 043504, arXiv: astro-ph/9707278 (p. 180).

[702] C.-P. Ma and E. Bertschinger, *Cosmological perturbation theory in the synchronous and conformal Newtonian gauges*, Astrophys. J. 455 (1995), pp. 7–25, arXiv: astro - ph / 9506072 (pp. 184, 185).

[703] K. Inomata and T. Terada, *Gauge Independence of Induced Gravitational Waves*, Phys. Rev. D 101.2 (2020), p. 023523, arXiv: 1912. 00785 [gr-qc] (p. 185).

[704] C. Yuan, Z.-C. Chen, and Q.-G. Huang, *Scalar induced gravitational waves in different gauges*, Phys. Rev. D 101.6 (2020), p. 063018, arXiv: 1912.00885 [astro-ph.CO] (p. 185).

[705] G. Domènech and M. Sasaki, *Approximate gauge independence of the induced gravitational wave spectrum*, Phys. Rev. D 103.6 (2021), p. 063531, arXiv: 2012.14016 [gr-qc] (p. 185).

[706] C. Caprini et al., *Science with the space-based interferometer eLISA. II: Gravitational waves from cosmological phase transitions*, JCAP 04 (2016), p. 001, arXiv: 1512.06239 [astro-ph.CO] (pp. 186, 192).

[707] T. Nakama, *Stochastic gravitational waves associated with primordial black holes formed during an early matter era*, Phys. Rev. D 101.6 (2020), p. 063519 (p. 186).

[708] I. Dalianis and C. Kouvaris, *Gravitational Waves from Density Perturbations in an Early Matter Domination Era*, (Dec. 2020), arXiv: 2012.09255 [astro-ph.CO] (p. 186).

[709] K. Kashiyama and N. Seto, *Enhanced exploration for primordial black holes using pulsar timing arrays*, Monthly Notices of the Royal Astronomical Society 426.2 (2012), 1369–1373. ISSN: 0035-8711, URL: http://dx.doi.org/10.1111/j.1365-2966.2012.21935.x (p. 189).

[710] K. Schutz and A. Liu, *Pulsar timing can constrain primordial black holes in the LIGO mass window*, Phys. Rev. D 95.2 (2017), p. 023002, arXiv: 1610.04234 [astro-ph.CO] (p. 189).

[711] J. A. Dror, H. Ramani, T. Trickle, and K. M. Zurek, *Pulsar Timing Probes of Primordial Black Holes and Subhalos*, Phys. Rev. D 100.2 (2019), p. 023003, arXiv: 1901.04490 [astro-ph.CO] (p. 189).

[712] H. Ramani, T. Trickle, and K. M. Zurek, *Observability of Dark Matter Substructure with Pulsar Timing Correlations*, JCAP 12 (2020), p. 033, arXiv: 2005.03030 [astro-ph.CO] (p. 189).

[713] V. S. H. Lee, S. R. Taylor, T. Trickle, and K. M. Zurek, *Bayesian Forecasts for Dark Matter Substructure Searches with Mock Pulsar Timing Data*, (Apr. 2021), arXiv: 2104.05717 [astro-ph.CO] (p. 189).

[714] G. Hobbs and S. Dai, *Gravitational wave research using pulsar timing arrays*, Natl. Sci. Rev. 4.5 (2017), pp. 707–717, arXiv: 1707.01615 [astro-ph.IM] (p. 189).

[715] S. L. Detweiler, *Pulsar timing measurements and the search for gravitational waves*, Astrophys. J. 234 (1979), pp. 1100–1104 (p. 189).

[716] B. Bertotti, B. J. Carr, and M. J. Rees, *Limits from the timing of pulsars on the cosmic gravitational wave background. "Mon. Not. Roy. Astron. Soc."* 203 (June 1983), pp. 945–954 (p. 189).

[717] R. W. Hellings and G. S. Downs, *Upper Limits On The Isotropic Gravitational Radiation Background From Pulsar Timing Analysis*, Astrophys. J. Lett. 265 (1983), pp. L39–L42 (p. 189).

[718] N. S. Pol et al., *Astrophysics Milestones For Pulsar Timing Array Gravitational Wave Detection*, (Oct. 2020), arXiv: 2010.11950 [astro-ph.HE] (p. 190).

[719] B. Goncharov et al., *On the evidence for a common-spectrum process in the search for the nanohertz gravitational wave background with the Parkes Pulsar Timing Array*, (July 2021), arXiv: 2107.12112 [astro-ph.HE] (p. 190).

[720] A. Sesana, F. Haardt, P. Madau, and M. Volonteri, *Low-frequency gravitational radiation from coalescing massive black hole binaries in hierarchical cosmologies*, Astrophys. J. 611 (2004), pp. 623–632, arXiv: astro-ph/0401543 (p. 190).

[721] H. Middleton, A. Sesana, S. Chen, A. Vecchio, W. Del Pozzo, and P. A. Rosado, *Massive black hole binary systems and the NANOGrav 12.5 year results*, Mon. Not. Roy. Astron. Soc. 502.1 (2021), pp. L99–L103, arXiv: 2011.01246 [astro-ph.HE] (p. 190).

[722] K. Kohri and T. Terada, *Solar-Mass Primordial Black Holes Explain NANOGrav Hint of Gravitational Waves*, Phys. Lett. B 813 (2021), p. 136040, arXiv: 2009.11853 [astro-ph.CO] (p. 190).

[723] G. Domènech and S. Pi, *NANOGrav Hints on Planet-Mass Primordial Black Holes*, (Oct. 2020), arXiv: 2010.03976 [astro-ph.CO] (p. 190).

[724] K. Inomata, M. Kawasaki, K. Mukaida, and T. T. Yanagida, *NANOGrav results and LIGO-Virgo primordial black holes in axion-like curvaton model*, Phys. Rev. Lett. 126.13 (2021), p. 131301, arXiv: 2011.01270 [astro-ph.CO] (p. 190).

[725] V. Atal, A. Sanglas, and N. Triantafyllou, *NANOGrav signal as mergers of Stupendously Large Primordial Black Holes*, (Dec. 2020), arXiv: 2012.14721 [astro-ph.CO] (p. 190).

[726] J. Ellis and M. Lewicki, *Cosmic String Interpretation of NANOGrav Pulsar Timing Data*, Phys. Rev. Lett. 126.4 (2021), p. 041304, arXiv: 2009.06555 [astro-ph.CO] (p. 190).

[727] S. Blasi, V. Brdar, and K. Schmitz, *Has NANOGrav found first evidence for cosmic strings?*, Phys. Rev. Lett. 126.4 (2021), p. 041305, arXiv: 2009.06607 [astro-ph.CO] (p. 190).

[728] W. Buchmuller, V. Domcke, and K. Schmitz, *From NANOGrav to LIGO with metastable cosmic strings*, Phys. Lett. B **811** (2020), p. 135914, arXiv: 2009.10649 [astro-ph.CO] (p. 190).

[729] A. Addazi, Y.-F. Cai, Q. Gan, A. Marciano, and K. Zeng, *NANOGrav results and Dark First Order Phase Transitions*, (Sept. 2020), arXiv: 2009.10327 [hep-ph] (p. 190).

[730] Y. Nakai, M. Suzuki, F. Takahashi, and M. Yamada, *Gravitational Waves and Dark Radiation from Dark Phase Transition: Connecting NANOGrav Pulsar Timing Data and Hubble Tension*, Phys. Lett. B **816** (2021), p. 136238, arXiv: 2009.09754 [astro-ph.CO] (p. 190).

[731] A. Neronov, A. Roper Pol, C. Caprini, and D. Semikoz, *NANOGrav signal from magnetohydrodynamic turbulence at the QCD phase transition in the early Universe*, Phys. Rev. D **103**.4 (2021), p. 041302, arXiv: 2009.14174 [astro-ph.CO] (p. 190).

[732] A. Paul, U. Mukhopadhyay, and D. Majumdar, *Gravitational Wave Signatures from Domain Wall and Strong First-Order Phase Transitions in a Two Complex Scalar extension of the Standard Model*, (Oct. 2020), arXiv: 2010.03439 [hep-ph] (p. 190).

[733] S.-L. Li, L. Shao, P. Wu, and H. Yu, *NANOGrav Signal from First-Order Confinement/Deconfinement Phase Transition in Different QCD Matters*, (Jan. 2021), arXiv: 2101.08012 [astro-ph.CO] (p. 190).

[734] Z. Arzoumanian et al., *Searching For Gravitational Waves From Cosmological Phase Transitions With The NANOGrav 12.5-year dataset*, (Apr. 2021), arXiv: 2104.13930 [astro-ph.CO] (p. 190).

[735] C. J. Moore and A. Vecchio, *Ultra-low frequency gravitational waves: distinguishing cosmological backgrounds from astrophysical foregrounds*, (Apr. 2021), arXiv: 2104.15130 [astro-ph.CO] (p. 190).

[736] D. Borah, A. Dasgupta, and S. K. Kang, *Gravitational waves from a dark $U(1)_D$ phase transition in the light of NANOGrav 12.5 yr data*, (May 2021), arXiv: 2105.01007 [hep-ph] (p. 190).

[737] S. Bhattacharya, S. Mohanty, and P. Parashari, *Implications of the NANOGrav result on primordial gravitational waves in nonstandard cosmologies*, Phys. Rev. D **103**.6 (2021), p. 063532, arXiv: 2010.05071 [astro-ph.CO] (p. 190).

[738] J. Yokoyama, *Implication of Pulsar Timing Array Experiments on Cosmological Gravitational Wave Detection*, (May 2021), arXiv: 2105.07629 [gr-qc] (p. 190).

[739] D. Wands, *Duality invariance of cosmological perturbation spectra*, Phys. Rev. D **60** (1999), p. 023507, arXiv: gr-qc/9809062 (p. 190).

[740] S. M. Leach and A. R. Liddle, *Inflationary perturbations near horizon crossing*, Phys. Rev. D **63** (2001), p. 043508, arXiv: astro-ph/0010082 (p. 190).

[741] S. M. Leach, M. Sasaki, D. Wands, and A. R. Liddle, *Enhancement of superhorizon scale inflationary curvature perturbations*, Phys. Rev. D **64** (2001), p. 023512, arXiv: astro-ph/0101406 (p. 190).

[742] K. Aggarwal et al., *The NANOGrav 11-Year Data Set: Limits on Gravitational Waves from Individual Supermassive Black Hole Binaries*, Astrophys. J. **880** (2019), p. 2, arXiv: 1812.11585 [astro-ph.GA] (p. 192).

[743] K. Yagi and N. Seto, *Detector configuration of DECIGO/BBO and identification of cosmological neutron-star binaries*, Phys. Rev. D **83** (2011). [Erratum: Phys.Rev.D 95, 109901 (2017)], p. 044011, arXiv: 1101.3940 [astro-ph.CO] (p. 192).

[744] B. P. Abbott et al., *Exploring the Sensitivity of Next Generation Gravitational Wave Detectors*, Class. Quant. Grav. **34**.4 (2017), p. 044001, arXiv: 1607.08697 [astro-ph.IM] (p. 192).

[745] B. S. Sathyaprakash and B. F. Schutz, *Physics, Astrophysics and Cosmology with Gravitational Waves*, Living Rev. Rel. **12** (2009), p. 2, arXiv: 0903.0338 [gr-qc] (p. 192).

[746] B. P. Abbott et al., *Upper Limits on the Stochastic Gravitational-Wave Background from Advanced LIGO's First Observing Run*, Phys. Rev. Lett. **118**.12 (2017). [Erratum: Phys.Rev.Lett. 119, 029901 (2017)], p. 121101, arXiv: 1612.02029 [gr-qc] (p. 192).

[747] J. Coleman, *Matter-wave Atomic Gradiometer InterferometricSensor (MAGIS-100) at Fermilab*, PoS ICHEP2018 (2019), p. 021, arXiv: 1812.00482 [physics.ins-det] (p. 192).

[748] Y. A. El-Neaj et al., *AEDGE: Atomic Experiment for Dark Matter and Gravity Exploration in Space*, EPJ Quant. Technol. **7** (2020), p. 6, arXiv: 1908.00802 [gr-qc] (p. 192).

[749] L. Badurina et al., *AION: An Atom Interferometer Observatory and Network*, JCAP **05** (2020), p. 011, arXiv: 1911.11755 [astro-ph.CO] (p. 192).

[750] S. Sugiyama, V. Takhistov, E. Vitagliano, A. Kusenko, M. Sasaki, and M. Takada, *Testing Stochastic Gravitational Wave Signals from Primordial Black Holes with Optical Telescopes*, Phys. Lett. B **814** (2021), p. 136097, arXiv: 2010.02189 [astro-ph.CO] (p. 192).

[751] R. Abbott et al., *GWTC-2.1: Deep Extended Catalog of Compact Binary Coalescences Observed by LIGO and Virgo During the First Half of the Third Observing Run*, (Aug. 2021), arXiv: 2108.01045 [gr-qc] (p. 195).

[752] T. Akutsu et al., *KAGRA: 2.5 Generation Interferometric Gravitational Wave Detector*, *Nature Astron.* 3.1 (2019), pp. 35–40, arXiv: 1811.08079 [gr-qc] (p. 195).

[753] R. Abbott et al., *Observation of gravitational waves from two neutron star-black hole coalescences*, *Astrophys. J. Lett.* 915 (2021), p. L5, arXiv: 2106.15163 [astro-ph.HE] (p. 195).

[754] N. S. Pol et al., *Astrophysics Milestones for Pulsar Timing Array Gravitational-wave Detection*, *Astrophys. J. Lett.* 911.2 (2021), p. L34, arXiv: 2010.11950 [astro-ph.HE] (p. 195).

[755] http://nanograv.org. 2021 (p. 195).

[756] H. D. Politzer and M. B. Wise, *Relations Between Spatial Correlations of Rich Clusters of Galaxies*, *Astrophys. J. Lett.* 285 (1984), pp. L1–L3 (p. 202).

[757] B. Grinstein and M. B. Wise, *Nongaussian Fluctuations and the Correlations of Galaxies or Rich Clusters of Galaxies*, *Astrophys. J.* 310 (1986), pp. 19–22 (p. 203).

[758] F. Lucchin, S. Matarrese, and N. Vittorio, *Scale Invariant Clustering and Primordial Biasing*, *Astrophys. J. Lett.* 330 (1988), pp. L21–L23 (p. 203).

[759] S. Matarrese and L. Verde, *The effect of primordial non-Gaussianity on halo bias*, *Astrophys. J. Lett.* 677 (2008), pp. L77–L80, arXiv: 0801.4826 [astro-ph] (p. 203).

[760] M. Zevin, (2021), URL: https://doi.org/10.5281/zenodo.4448170 (pp. 204, 210, 211).

[761] E. Thrane and C. Talbot, *An introduction to Bayesian inference in gravitational-wave astronomy: parameter estimation, model selection, and hierarchical models*, *Publ. Astron. Soc. Austral.* 36 (2019). [Erratum: Publ.Astron.Soc.Austral. 37, e036 (2020)], e010, arXiv: 1809.02293 [astro-ph.IM] (pp. 205, 206, 207).

[762] S. M. Gaebel, J. Veitch, T. Dent, and W. M. Farr, *Digging the population of compact binary mergers out of the noise*, *Mon. Not. Roy. Astron. Soc.* 484.3 (2019), pp. 4008–4023, arXiv: 1809.03815 [astro-ph.IM] (p. 205).

[763] T. J. Loredo, *Accounting for source uncertainties in analyses of astronomical survey data*, *AIP Conf. Proc.* 735.1 (2004). Ed. by R. Fischer, R. Preuss, and U. von Toussaint, pp. 195–206, arXiv: astro-ph/0409387 (p. 207).

[764] D. Foreman-Mackey, D. W. Hogg, D. Lang, and J. Goodman, *emcee: The MCMC Hammer*, *Publ. Astron. Soc. Pac.* 125 (2013), pp. 306–312, arXiv: 1202.3665 [astro-ph.IM] (p. 207).

[765] https://dcc.ligo.org/LIGO-P1800370/public, 2018 (p. 207).

[766] https://dcc.ligo.org/LIGO-P2000223/public, 2020 (p. 207).

[767] L. Xiang, *A Review of: "Introduction to Bayesian Statistics"*, *IIE Transactions* 39.8 (2007), pp. 829–829, eprint: https://doi.org/10.1080/07408170600941656, URL: https://doi.org/10.1080/07408170600941656 (p. 208).

[768] I. M. Romero-Shaw et al., *Bayesian inference for compact binary coalescences with bilby: validation and application to the first LIGO–Virgo gravitational-wave transient catalogue*, *Mon. Not. Roy. Astron. Soc.* 499.3 (2020), pp. 3295–3319, arXiv: 2006.00714 [astro-ph.IM] (p. 208).

[769] G. Ashton et al., *BILBY: A user-friendly Bayesian inference library for gravitational-wave astronomy*, *Astrophys. J. Suppl.* 241.2 (2019), p. 27, arXiv: 1811.02042 [astro-ph.IM] (p. 208).

[770] LIGO Scientific Collaboration, *LIGO Algorithm Library - LALSuite*. free software (GPL). 2018 (p. 209).

[771] C. Kalaghatgi, M. Hannam, and V. Raymond, *Parameter estimation with a spinning multimode waveform model*, *Phys. Rev. D* 101.10 (2020), p. 103004, arXiv: 1909.10010 [gr-qc] (p. 209).

[772] M. Walker, A. F. Agnew, J. Bidler, A. Lundgren, A. Macedo, D. Macleod, et al., *Identifying correlations between LIGO's astronomical range and auxiliary sensors using lasso regression*, *Class. Quant. Grav.* 35.22 (2018), p. 225002, arXiv: 1807.02592 [astro-ph.IM] (p. 209).

[773] S. Coughlin, S. Bahaadini, N. Rohani, M. Zevin, O. Patane, M. Harandi, et al., *Classifying the unknown: Discovering novel gravitational-wave detector glitches using similarity learning*, *Phys. Rev. D* 99 (8 2019), p. 082002, URL: https://link.aps.org/doi/10.1103/PhysRevD.99.082002 (p. 209).

[774] I. Mandel and A. Farmer, *Merging stellar-mass binary black holes*, (June 2018), arXiv: 1806.05820 [astro-ph.HE] (p. 210).

[775] S. E. de Mink and I. Mandel, *The chemically homogeneous evolutionary channel for binary black hole mergers: rates and properties of gravitational-wave events detectable by advanced LIGO*, Monthly Notices of the Royal Astronomical Society 460.4 (Aug. 2016), pp. 3545–3553, arXiv: 1603.02291 [astro-ph.HE] (p. 210).

[776] C. J. Neijssel, A. Vigna-Gómez, S. Stevenson, J. W. Barrett, S. M. Gaebel, F. S. Broekgaarden, et al., *The effect of the metallicity-specific star formation history on double compact object mergers*, Monthly Notices of the Royal Astronomical Society 490.3 (Dec. 2019), pp. 3740–3759, arXiv: 1906.08136 [astro-ph.SR] (p. 210).

[777] P. Marchant, N. Langer, P. Podsiadlowski, T. M. Tauris, and T. J. Moriya, *A new route towards merging massive black holes*, Astronomy & Astrophysics 588, A50 (Apr. 2016), A50, arXiv: 1601.03718 [astro-ph.SR] (p. 210).

[778] P. Madau and M. J. Rees, *Massive Black Holes as Population III Remnants*, The Astrophysical Journal Letters 551.1 (Apr. 2001), pp. L27–L30, arXiv: astro-ph/0101223 [astro-ph] (p. 210).

[779] K. Inayoshi, R. Hirai, T. Kinugawa, and K. Hotokezaka, *Formation pathway of Population III coalescing binary black holes through stable mass transfer*, Monthly Notices of the Royal Astronomical Society 468.4 (July 2017), pp. 5020–5032, arXiv: 1701.04823 [astro-ph.HE] (p. 210).

[780] I. Mandel and F. S. Broekgaarden, *Rates of Compact Object Coalescences*, (July 2021), arXiv: 2107.14239 [astro-ph.HE] (p. 210).

[781] E. P. J. van den Heuvel, S. F. Portegies Zwart, and S. E. de Mink, *Forming short-period Wolf-Rayet X-ray binaries and double black holes through stable mass transfer*, Monthly Notices of the Royal Astronomical Society 471.4 (Nov. 2017), pp. 4256–4264, arXiv: 1701.02355 [astro-ph.SR] (p. 210).

[782] B. Paczynski, "Common Envelope Binaries", *Structure and Evolution of Close Binary Systems*. Ed. by P. Eggleton, S. Mitton, and J. Whelan. Vol. 73. Jan. 1976, p. 75 (p. 210).

[783] E. P. J. van den Heuvel, "Late Stages of Close Binary Systems", *Structure and Evolution of Close Binary Systems*. Ed. by P. Eggleton, S. Mitton, and J. Whelan. Vol. 73. Jan. 1976, p. 35 (p. 210).

[784] A. V. Tutukov and L. R. Yungelson, *The merger rate of neutron star and black hole binaries*. Monthly Notices of the Royal Astronomical Society 260 (Feb. 1993), pp. 675–678 (p. 210).

[785] H. A. Bethe and G. E. Brown, *Evolution of binary compact objects which merge*, Astrophys. J. 506 (1998), pp. 780–789, arXiv: astro-ph/9802084 (p. 210).

[786] K. Belczynski, V. Kalogera, and T. Bulik, *A Comprehensive study of binary compact objects as gravitational wave sources: Evolutionary channels, rates, and physical properties*, Astrophys. J. 572 (2001), pp. 407–431, arXiv: astro-ph/0111452 (p. 210).

[787] M. Dominik, K. Belczynski, C. Fryer, D. Holz, E. Berti, T. Bulik, et al., *Double Compact Objects I: The Significance of the Common Envelope on Merger Rates*, Astrophys. J. 759 (2012), p. 52, arXiv: 1202.4901 [astro-ph.HE] (p. 210).

[788] K. Belczynski, S. Repetto, D. E. Holz, R. O'Shaughnessy, T. Bulik, E. Berti, et al., *Compact Binary Merger Rates: Comparison with LIGO/Virgo Upper Limits*, Astrophys. J. 819.2 (2016), p. 108, arXiv: 1510.04615 [astro-ph.HE] (p. 210).

[789] J. J. Eldridge and E. R. Stanway, *BPASS predictions for Binary Black-Hole Mergers*, Mon. Not. Roy. Astron. Soc. 462.3 (2016), pp. 3302–3313, arXiv: 1602.03790 [astro-ph.HE] (p. 210).

[790] S. Stevenson, A. Vigna-Gómez, I. Mandel, J. W. Barrett, C. J. Neijssel, D. Perkins, and S. E. de Mink, *Formation of the first three gravitational-wave observations through isolated binary evolution*, Nature Commun. 8 (2017), p. 14906, arXiv: 1704.01352 [astro-ph.HE] (p. 210).

[791] N. Giacobbo and M. Mapelli, *The progenitors of compact-object binaries: impact of metallicity, common envelope and natal kicks*, Mon. Not. Roy. Astron. Soc. 480.2 (2018), pp. 2011–2030, arXiv: 1806.00001 [astro-ph.HE] (p. 210).

[792] B. Paxton, L. Bildsten, A. Dotter, F. Herwig, P. Lesaffre, and F. Timmes, *Modules for Experiments in Stellar Astrophysics (MESA)*, Astrophys. J. Suppl. 192 (2011), p. 3, arXiv: 1009.1622 [astro-ph.SR] (p. 210).

[793] B. Paxton et al., *Modules for Experiments in Stellar Astrophysics (MESA): Planets, Oscillations, Rotation, and Massive Stars*, Astrophys. J. Suppl. 208 (2013), p. 4, arXiv: 1301.0319 [astro-ph.SR] (p. 210).

[794] B. Paxton et al., *Modules for Experiments in Stellar Astrophysics (MESA): Binaries, Pulsations, and Explosions*, Astrophys. J. Suppl. 220.1 (2015), p. 15, arXiv: 1506.03146 [astro-ph.SR] (p. 210).

[795] B. Paxton et al., *Modules for Experiments in Stellar Astrophysics (MESA): Convective Boundaries, Element Diffusion, and Massive Star Explosions*, Astrophys. J. Suppl. 234.2 (2018), p. 34, arXiv: 1710.08424 [astro-ph.SR] (p. 210).

[796] N. Ivanova et al., *Common Envelope Evolution: Where we stand and how we can move forward*, Astron. Astrophys. Rev. 21 (2013), p. 59, arXiv: 1209.4302 [astro-ph.HE] (p. 211).

[797] J. L. A. Nandez and N. Ivanova, *Common envelope events with low-mass giants: understanding the energy budget*, Monthly Notices of the Royal Astronomical Society 460.4 (Aug. 2016), pp. 3992–4002, arXiv: 1606.04922 [astro-ph.SR] (p. 211).

[798] T. Fragos, J. J. Andrews, E. Ramirez-Ruiz, G. Meynet, V. Kalogera, R. E. Taam, and A. Zezas, *The Complete Evolution of a Neutron-star Binary through a Common Envelope Phase Using 1D Hydrodynamic Simulations*, The Astrophysical Journal Letters 883.2, L45 (Oct. 2019), p. L45, arXiv: 1907.12573 [astro-ph.HE] (p. 211).

[799] J. Fuller and L. Ma, *Most Black Holes are Born Very Slowly Rotating*, Astrophys. J. Lett. 881.1 (2019), p. L1, arXiv: 1907.03714 [astro-ph.SR] (p. 211).

[800] S. McMillan, P. Hut, and J. Makino, *Star Cluster Evolution with Primordial Binaries. II. Detailed Analysis*, Astrophysical Journal 372 (May 1991), p. 111 (p. 211).

[801] P. Hut, S. McMillan, J. Goodman, M. Mateo, E. S. Phinney, C. Pryor, et al., *Binaries in Globular Clusters*, Publications of the Astronomical Society of the Pacific 104 (Nov. 1992), p. 981 (p. 211).

[802] S. Sigurdsson and L. Hernquist, *Primordial black holes in globular clusters*, Nature 364 (1993), pp. 423–425 (p. 211).

[803] M. C. Miller and D. P. Hamilton, *Four-Body Effects in Globular Cluster Black Hole Coalescence*, Astrophysical Journal 576.2 (Sept. 2002), pp. 894–898, arXiv: astro-ph/0202298 [astro-ph] (p. 211).

[804] K. Gültekin, M. C. Miller, and D. P. Hamilton, *Three-Body Dynamics with Gravitational Wave Emission*, Astrophysical Journal 640.1 (Mar. 2006), pp. 156–166, arXiv: astro-ph/0509885 [astro-ph] (p. 211).

[805] J. M. Fregeau and F. A. Rasio, *Monte Carlo Simulations of Globular Cluster Evolution. 4. Direct Integration of Strong Interactions*, Astrophys. J. 658 (2007), p. 1047, arXiv: astro-ph/0608261 (p. 211).

[806] A. P. Lightman and S. L. Shapiro, *The dynamical evolution of globular clusters*, Rev. Mod. Phys. 50 (1978), pp. 437–481 (p. 211).

[807] S. F. Portegies Zwart and S. McMillan, *Black hole mergers in the universe*, Astrophys. J. Lett. 528 (2000), p. L17, arXiv: astro-ph/9910061 (p. 211).

[808] R. M. O'Leary, F. A. Rasio, J. M. Fregeau, N. Ivanova, and R. W. O'Shaughnessy, *Binary mergers and growth of black holes in dense star clusters*, Astrophys. J. 637 (2006), pp. 937–951, arXiv: astro-ph/0508224 (p. 211).

[809] J. M. B. Downing, M. J. Benacquista, M. Giersz, and R. Spurzem, *Compact Binaries in Star Clusters I - Black Hole Binaries Inside Globular Clusters*, Mon. Not. Roy. Astron. Soc. 407 (2010), p. 1946, arXiv: 0910.0546 [astro-ph.SR] (p. 211).

[810] J. Samsing, M. MacLeod, and E. Ramirez-Ruiz, *The Formation of Eccentric Compact Binary Inspirals and the Role of Gravitational Wave Emission in Binary-Single Stellar Encounters*, Astrophys. J. 784 (2014), p. 71, arXiv: 1308.2964 [astro-ph.HE] (p. 211).

[811] B. M. Ziosi, M. Mapelli, M. Branchesi, and G. Tormen, *Dynamics of stellar black holes in young star clusters with different metallicities – II. Black hole–black hole binaries*, Mon. Not. Roy. Astron. Soc. 441.4 (2014), pp. 3703–3717, arXiv: 1404.7147 [astro-ph.GA] (p. 211).

[812] C. L. Rodriguez, M. Morscher, B. Pattabiraman, S. Chatterjee, C.-J. Haster, and F. A. Rasio, *Binary Black Hole Mergers from Globular Clusters: Implications for Advanced LIGO*, Phys. Rev. Lett. 115.5 (2015). [Erratum: Phys.Rev.Lett. 116, 029901 (2016)], p. 051101, arXiv: 1505.00792 [astro-ph.HE] (p. 211).

[813] C. L. Rodriguez, S. Chatterjee, and F. A. Rasio, *Binary Black Hole Mergers from Globular Clusters: Masses, Merger Rates, and the Impact of Stellar Evolution*, Phys. Rev. D 93.8 (2016), p. 084029, arXiv: 1602.02444 [astro-ph.HE] (p. 211).

[814] F. Antonini and F. A. Rasio, *Merging black hole binaries in galactic nuclei: implications for advanced-LIGO detections*, Astrophys. J. 831.2 (2016), p. 187, arXiv: 1606.04889 [astro-ph.HE] (pp. 211, 212).

[815] A. Askar, M. Szkudlarek, D. Gondek-Rosińska, M. Giersz, and T. Bulik, *MOCCA-SURVEY Database – I. Coalescing binary black holes originating from globular clusters*, Mon. Not. Roy. Astron. Soc. 464.1 (2017), pp. L36–L40, arXiv: 1608.02520 [astro-ph.HE] (p. 211).

[816] S. Banerjee, *Stellar-mass black holes in young massive and open stellar clusters and their role in gravitational-wave generation – II*, Mon. Not. Roy. Astron. Soc. 473.1 (2018), pp. 909–926, arXiv: 1707 . 00922 [astro-ph.HE] (p. 211).

[817] J. Samsing and E. Ramirez-Ruiz, *On the Assembly Rate of Highly Eccentric Binary Black Hole Mergers*, Astrophys. J. Lett. 840.2 (2017), p. L14, arXiv: 1703 . 09703 [astro-ph.HE] (p. 211).

[818] M. H. Hénon, *The Monte Carlo Method (Papers appear in the Proceedings of IAU Colloquium No. 10 Gravitational N-Body Problem (ed. by Myron Lecar), R. Reidel Publ. Co. , Dordrecht-Holland.)* Astrophysics and Space Science 14.1 (Nov. 1971), pp. 151–167 (p. 211).

[819] M. Hénon, *Monte Carlo Models of Star Clusters (Part of the Proceedings of the IAU Colloquium No. 10, held in Cambridge, England, August 12-15, 1970.)* Astrophysics and Space Science 13.2 (Oct. 1971), pp. 284–299 (p. 211).

[820] K. Joshi, F. Rasio, and S. F. Portegies Zwart, *Monte Carlo simulations of globular cluster evolution - 1. Method and test calculations*, Astrophys. J. 540 (2000), p. 969, arXiv: astro-ph/9909115 (p. 211).

[821] B. Pattabiraman, S. Umbreit, W.-K. Liao, A. Choudhary, V. Kalogera, G. Memik, and F. A. Rasio, *A Parallel Monte Carlo Code for Simulating Collisional N-body Systems*, Astrophys. J. Suppl. 204 (2013), p. 15, arXiv: 1206.5878 [astro-ph.IM] (p. 211).

[822] J. R. Hurley, O. R. Pols, and C. A. Tout, *Comprehensive analytic formulae for stellar evolution as a function of mass and metallicity*, Mon. Not. Roy. Astron. Soc. 315 (2000), p. 543, arXiv: astro-ph/0001295 (p. 211).

[823] J. R. Hurley, C. A. Tout, and O. R. Pols, *Evolution of binary stars and the effect of tides on binary populations*, Mon. Not. Roy. Astron. Soc. 329 (2002), p. 897, arXiv: astro-ph/0201220 (p. 211).

[824] S. Chatterjee, J. M. Fregeau, S. Umbreit, and F. A. Rasio, *Monte Carlo Simulations of Globular Cluster Evolution. V. Binary Stellar Evolution*, Astrophys. J. 719 (2010), pp. 915–930, arXiv: 0912.4682 [astro-ph.GA] (p. 211).

[825] C. L. Rodriguez, P. Amaro-Seoane, S. Chatterjee, and F. A. Rasio, *Post-Newtonian Dynamics in Dense Star Clusters: Highly-Eccentric, Highly-Spinning, and Repeated Binary Black Hole Mergers*, Phys. Rev. Lett. 120.15 (2018), p. 151101, arXiv: 1712.04937 [astro-ph.HE] (p. 211).

[826] M. Morscher, S. Umbreit, W. M. Farr, and F. A. Rasio, *Retention of Stellar-Mass Black Holes in Globular Clusters*, Astrophys. J. Lett. 763 (2013), p. L15, arXiv: 1211 . 3372 [astro-ph.GA] (p. 211).

[827] J. M. Antognini, B. J. Shappee, T. A. Thompson, and P. Amaro-Seoane, *Rapid Eccentricity Oscillations and the Mergers of Compact Objects in Hierarchical Triples*, Mon. Not. Roy. Astron. Soc. 439.1 (2014), pp. 1079–1091, arXiv: 1308.5682 [astro-ph.HE] (p. 211).

[828] P. Amaro-Seoane and X. Chen, *Relativistic mergers of black hole binaries have large, similar masses, low spins and are circular*, Mon. Not. Roy. Astron. Soc. 458.3 (2016), pp. 3075–3082, arXiv: 1512 . 04897 [astro-ph.CO] (p. 211).

[829] K. El-Badry, E. Quataert, D. R. Weisz, N. Choksi, and M. Boylan-Kolchin, *The formation and hierarchical assembly of globular cluster populations*, Monthly Notices of the Royal Astronomical Society 482.4 (Feb. 2019), pp. 4528–4552, arXiv: 1805 . 03652 [astro-ph.GA] (p. 212).

[830] J. D. Bekenstein, *Gravitational-Radiation Recoil and Runaway Black Holes*, Astrophys. J. 183 (1973), pp. 657–664 (p. 212).

[831] A. G. Wiseman, *Coalescing binary systems of compact objects to (post)5/2 Newtonian order. 2. Higher order wave forms and radiation recoil*, Phys. Rev. D 46 (1992), pp. 1517–1539 (p. 212).

[832] M. Favata, S. A. Hughes, and D. E. Holz, *How black holes get their kicks: Gravitational radiation recoil revisited*, Astrophys. J. Lett. 607 (2004), pp. L5–L8, arXiv: astro-ph/0402056 (p. 212).

[833] J. G. Baker, J. Centrella, D.-I. Choi, M. Koppitz, J. R. van Meter, and M. C. Miller, *Getting a kick out of numerical relativity*, Astrophys. J. Lett. 653 (2006), pp. L93–L96, arXiv: astro-ph/0603204 (p. 212).

[834] M. Koppitz, D. Pollney, C. Reisswig, L. Rezzolla, J. Thornburg, P. Diener, and E. Schnetter, *Recoil Velocities from Equal-Mass Binary-Black-Hole Mergers*, Phys. Rev. Lett. 99 (2007), p. 041102, arXiv: gr-qc/0701163 (p. 212).

[835] D. Pollney et al., *Recoil velocities from equal-mass binary black-hole mergers: A Systematic investigation of spin-orbit aligned configurations*, Phys. Rev. D 76 (2007), p. 124002, arXiv: 0707.2559 [gr-qc] (p. 212).

[836]  K. Holley-Bockelmann, K. Gultekin, D. Shoe-
       maker, and N. Yunes, *Gravitational Wave Re-
       coil and the Retention of Intermediate Mass
       Black Holes*, Astrophys. J. 686 (2008), p. 829,
       arXiv: 0707.1334 [astro-ph] (p. 212).

[837]  C. O. Lousto, M. Campanelli, Y. Zlochower,
       and H. Nakano, *Remnant Masses, Spins and
       Recoils from the Merger of Generic Black-Hole
       Binaries*, Class. Quant. Grav. 27 (2010). Ed.
       by S. Husa and B. Krishnan, p. 114006, arXiv:
       0904.3541 [gr-qc] (p. 212).

[838]  L. Blanchet, *Gravitational Radiation from
       Post-Newtonian Sources and Inspiralling Com-
       pact Binaries*, Living Rev. Rel. 17 (2014), p. 2,
       arXiv: 1310.1528 [gr-qc] (p. 212).

[839]  U. Sperhake, *The numerical relativity break-
       through for binary black holes*, Class. Quant.
       Grav. 32.12 (2015), p. 124011, arXiv: 1411.
       3997 [gr-qc] (p. 212).

[840]  S. Banerjee, *Stellar-mass black holes in young
       massive and open stellar clusters – IV. Up-
       dated stellar-evolutionary and black hole spin
       models and comparisons with the LIGO-Virgo
       O1/O2 merger-event data*, Mon. Not. Roy. As-
       tron. Soc. 500.3 (2020), pp. 3002–3026, arXiv:
       2004.07382 [astro-ph.HE] (p. 212).

[841]  G. Fragione and J. Silk, *Repeated mergers
       and ejection of black holes within nuclear star
       clusters*, Mon. Not. Roy. Astron. Soc. 498.4
       (2020), pp. 4591–4604, arXiv: 2006.01867
       [astro-ph.GA] (p. 212).

[842]  F. Pretorius, *Evolution of binary black hole
       spacetimes*, Phys. Rev. Lett. 95 (2005),
       p. 121101, arXiv: gr-qc/0507014 (p. 212).

[843]  J. A. Gonzalez, U. Sperhake, B. Bruegmann,
       M. Hannam, and S. Husa, *Total recoil: The
       Maximum kick from nonspinning black-hole
       binary inspiral*, Phys. Rev. Lett. 98 (2007),
       p. 091101, arXiv: gr-qc/0610154 (p. 212).

[844]  A. Buonanno, L. E. Kidder, and L. Lehner, *Es-
       timating the final spin of a binary black hole co-
       alescence*, Phys. Rev. D 77 (2008), p. 026004,
       arXiv: 0709.3839 [astro-ph] (p. 212).

[845]  F. Antonini and M. Gieles, *Population syn-
       thesis of black hole binary mergers from star
       clusters*, Mon. Not. Roy. Astron. Soc. 492.2
       (2020), pp. 2936–2954, arXiv: 1906.11855
       [astro-ph.HE] (p. 212).

[846]  L. Rezzolla, E. Barausse, E. N. Dorband, D.
       Pollney, C. Reisswig, J. Seiler, and S. Husa,
       *On the final spin from the coalescence of two
       black holes*, Phys. Rev. D 78 (2008), p. 044002,
       arXiv: 0712.3541 [gr-qc] (p. 212).

[847]  N. Neumayer, A. Seth, and T. Boeker, *Nuclear
       Star Clusters*, (Jan. 2020), arXiv: 2001.03626
       [astro-ph.GA] (p. 212).

[848]  P. Madau and T. Fragos, *Radiation Back-
       grounds at Cosmic Dawn: X-Rays from Com-
       pact Binaries*, Astrophys. J. 840.1 (2017), p. 39,
       arXiv: 1606.07887 [astro-ph.GA] (p. 212).

[849]  N. Grevesse and A. J. Sauval, *Standard So-
       lar Composition*, Space Sci. Rev. 85 (1998),
       pp. 161–174 (p. 212).

[850]  J. M. Martin-Garcia, *xAct: Efficient tensor
       computer algebra for the Wolfram Language*,
       (2021) (p. 215).